\documentclass{book}

\usepackage{color}
\usepackage{tikz}
\usepackage{ifthen}
\usepackage{animate}
\usetikzlibrary{calc}

\usepackage{amssymb,amsmath,amsthm}
\usepackage{makeidx,mathrsfs}         
\usepackage{graphicx}        
\usepackage{upgreek}        

\usepackage{hyperref}

\usepackage{psfrag}
\usepackage{a4wide}

\usepackage{imakeidx}

\makeindex

\theoremstyle{plain}
\newtheorem{thm}{Theorem}[chapter]
\newtheorem{prop}[thm]{Proposition}
\newtheorem{lemma}[thm]{Lemma}
\newtheorem{cor}[thm]{Corollary}
\theoremstyle{definition}
\newtheorem{definition}[thm]{Definition}

\newtheorem{remark}[thm]{Remark}
\newtheorem{example}[thm]{Example}
\newtheorem{remarks}[thm]{Remarks}

\newcommand{\tn}[1]{\ensuremath{\mathbb{T}^{#1}}}

\newcommand{\rn}[1]{\ensuremath{\mathbb{R}^{#1}}}
\newcommand{\nn}[1]{\ensuremath{\mathbb{N}^{#1}}}
\newcommand{\zn}[1]{\ensuremath{\mathbb{Z}^{#1}}}
\newcommand{\sn}[1]{\ensuremath{\mathbb{S}^{#1}}}
\newcommand{\cn}[1]{\ensuremath{\mathbb{C}^{#1}}}

\newcommand{\Mn}[2]{\ensuremath{\mathbf{M}_{#1}(#2)}}

\newcommand{\Spe}{\mathrm{Sp}}
\newcommand{\roF}{\mathrm{F}}

\newcommand{\roKG}{\mathrm{KG}}
\newcommand{\fK}{\mathrm{fK}}
\newcommand{\rohyb}{\mathrm{hyb}}
\newcommand{\rohs}{\mathrm{hs}}
\newcommand{\rosob}{\mathrm{Sob}}
\newcommand{\mKG}{m_{\mathrm{KG}}}
\newcommand{\rohy}{\mathrm{hy}}
\newcommand{\rohc}{\mathrm{hc}}
\newcommand{\roHom}{\mathrm{Hom}}
\newcommand{\roG}{\mathrm{G}}
\newcommand{\rond}{\mathrm{nd}}
\newcommand{\romat}{\mathrm{mat}}
\newcommand{\rode}{\mathrm{de}}
\newcommand{\mrod}{\mathrm{od}}

\newcommand{\mrI}{\mathrm{I}}
\newcommand{\mrII}{\mathrm{II}}
\newcommand{\mrIII}{\mathrm{III}}
\newcommand{\mrIV}{\mathrm{IV}}
\newcommand{\mrV}{\mathrm{V}}
\newcommand{\mrVI}{\mathrm{VI}}
\newcommand{\mrVII}{\mathrm{VII}}
\newcommand{\mrVIII}{\mathrm{VIII}}
\newcommand{\bRic}{\overline{\mathrm{Ric}}}
\newcommand{\grad}{\mathrm{grad}}

\newcommand{\zo}{\mathbb{Z}}
\newcommand{\ro}{\mathbb{R}}

\newcommand{\roK}{\mathrm{K}}
\newcommand{\roQ}{\mathrm{Q}}

\newcommand{\rovar}{\mathrm{var}}

\newcommand{\roc}{\mathrm{c}}

\newcommand{\rocor}{\mathrm{cor}}
\newcommand{\rocoeff}{\mathrm{coeff}}
\newcommand{\rodiv}{\mathrm{div}}
\newcommand{\rocau}{\mathrm{cau}}
\newcommand{\roRe}{\mathrm{Re}}
\newcommand{\rorel}{\mathrm{rel}}
\newcommand{\rolas}{\mathrm{las}}

\newcommand{\co}{\mathbb{C}}
\newcommand{\so}{\mathbb{S}^{1}}
\newcommand{\sfL}{\mathsf{L}}

\newcommand{\sfR}{\mathsf{R}}

\newcommand{\sfx}{\mathsf{x}}
\newcommand{\sfX}{\mathsf{X}}
\newcommand{\ONSF}{\mathsf{X}}
\newcommand{\ONCS}{\mathsf{\bar{\mathsf{\Gamma}}}}
\newcommand{\sfY}{\mathsf{Y}}

\newcommand{\bsfx}{\bar{\mathsf{x}}}
\newcommand{\hsfx}{\hat{\mathsf{x}}}
\newcommand{\hell}{\hat{\ell}}

\newcommand{\msL}{\mathscr{L}}
\newcommand{\msA}{\mathscr{A}}
\newcommand{\msX}{\mathscr{X}}
\newcommand{\msT}{\mathscr{T}}
\newcommand{\msS}{\mathscr{S}}
\newcommand{\msU}{\mathscr{U}}
\newcommand{\msE}{\mathscr{E}}

\newcommand{\msfP}{\mathsf{P}}
\newcommand{\msfD}{\mathsf{D}}
\newcommand{\mfP}{\mathfrak{P}}
\newcommand{\mfS}{\mathfrak{S}}
\newcommand{\mfC}{\mathfrak{C}}
\newcommand{\mfE}{\mathfrak{E}}
\newcommand{\mfT}{\mathfrak{T}}
\newcommand{\mfPb}{\bar{\mathfrak{P}}}
\newcommand{\mfL}{\mathfrak{L}}
\newcommand{\mfQ}{\mathfrak{Q}}

\newcommand{\bna}{\bar{\nabla}}

\newcommand{\rest}{\mathrm{rest}}

\newcommand{\rem}{\mathrm{rem}}
\newcommand{\refer}{\mathrm{ref}}

\newcommand{\rosh}{\mathrm{sh}}

\newcommand{\rotot}{\mathrm{tot}}

\newcommand{\coeff}{\mathrm{coeff}}
\newcommand{\coe}{\mathrm{co}}

\newcommand{\roloc}{\mathrm{loc}}

\newcommand{\rol}{\mathrm{l}}

\newcommand{\ros}{\mathrm{s}}

\newcommand{\rodS}{\mathrm{dS}}

\newcommand{\roacc}{\mathrm{acc}}
\newcommand{\rocom}{\mathrm{com}}

\newcommand{\roode}{\mathrm{ode}}

\newcommand{\robas}{\mathrm{bas}}
\newcommand{\rowe}{\mathrm{we}}
\newcommand{\robal}{\mathrm{bal}}
\newcommand{\ropar}{\mathrm{par}}

\newcommand{\rocon}{\mathrm{con}}

\newcommand{\rood}{\mathrm{od}}

\newcommand{\rodiff}{\mathrm{diff}}

\newcommand{\tr}{\mathrm{tr}}

\newcommand{\pre}{\mathrm{pre}}

\newcommand{\Id}{\mathrm{Id}}

\newcommand{\fin}{\mathrm{fin}}

\newcommand{\bd}{\bar{d}}

\newcommand{\ba}{\bar{a}}

\newcommand{\bka}{\bar{\kappa}}

\newcommand{\mutgc}{\mu_{\tilde{g};c}}

\newcommand{\tx}{\tilde{x}}

\newcommand{\tX}{\tilde{X}}

\newcommand{\tg}{\tilde{g}}

\newcommand{\bk}{\bar{k}}
\newcommand{\bK}{\bar{K}}
\newcommand{\hK}{\hat{K}}
\newcommand{\hE}{\hat{E}}
\newcommand{\hT}{\hat{T}}
\newcommand{\hpi}{\hat{\pi}}
\newcommand{\hX}{\hat{X}}

\newcommand{\chth}{\check{\theta}}

\newcommand{\chN}{\check{N}}

\newcommand{\hmu}{\hat{\mu}}
\newcommand{\hml}{\hat{\ml}}

\newcommand{\tc}{\tilde{c}}
\newcommand{\tchi}{\tilde{\chi}}
\newcommand{\dotchi}{\dot{\chi}}

\newcommand{\hY}{\hat{Y}}

\newcommand{\bge}{\bar{g}}
\newcommand{\bS}{\bar{S}}

\newcommand{\bD}{\bar{D}}
\newcommand{\bX}{\bar{X}}

\newcommand{\bR}{\bar{R}}
\newcommand{\bM}{\bar{M}}
\newcommand{\tM}{\tilde{M}}

\newcommand{\bC}{\bar{C}}
\newcommand{\bJ}{\bar{J}}
\newcommand{\bmJ}{\bar{\mathcal{J}}}

\newcommand{\bmu}{\bar{\mu}}
\newcommand{\bla}{\bar{\lambda}}
\newcommand{\tla}{\tilde{\lambda}}

\newcommand{\bnabla}{\overline{\nabla}}

\newcommand{\bareta}{\bar{\eta}}

\newcommand{\bfI}{\mathbf{I}}

\newcommand{\bfJ}{\mathbf{J}}
\newcommand{\bfK}{\mathbf{K}}
\newcommand{\bfL}{\mathbf{L}}
\newcommand{\bfM}{\mathbf{M}}
\newcommand{\bfN}{\mathbf{N}}
\newcommand{\bfA}{\mathbf{A}}
\newcommand{\bbA}{\mathbb{A}}
\newcommand{\bbE}{\mathbb{E}}
\newcommand{\bbW}{\mathbb{W}}
\newcommand{\bfB}{\mathbf{B}}
\newcommand{\bfbC}{\mathbf{C}}
\newcommand{\bfX}{\mathbf{X}}
\newcommand{\bfE}{\mathbf{E}}
\newcommand{\bfW}{\mathbf{W}}
\newcommand{\bfz}{\mathbf{0}}

\newcommand{\bga}{\bar{\gamma}}
\newcommand{\bG}{\bar{\Gamma}}
\renewcommand{\a}{\alpha}
\newcommand{\e}{\epsilon}
\newcommand{\vare}{\varepsilon}

\newcommand{\de}{\delta}

\renewcommand{\b}{\beta}
\newcommand{\teta}{\tilde{\eta}}
\newcommand{\g}{\gamma}
\newcommand{\G}{\Gamma}
\renewcommand{\d}{\partial}

\newcommand{\me}{\mathcal{E}}
\newcommand{\mK}{\mathcal{K}}

\newcommand{\eSpe}{\varepsilon_{\Spe}}
\newcommand{\emK}{\varepsilon_{\mK}}
\newcommand{\mKsup}{C_{\mK}}
\newcommand{\bDlnhNsup}{C_{\rorel}}
\newcommand{\mM}{\mathcal{M}}
\newcommand{\mL}{\mathcal{L}}
\newcommand{\bmK}{\bar{\mK}}

\newcommand{\ma}{\mathcal{A}}
\newcommand{\mA}{\mathcal{A}}

\newcommand{\mO}{\mathcal{O}}

\newcommand{\ml}{\mathcal{L}}

\newcommand{\mt}{\mathcal{T}}
\newcommand{\ms}{\mathcal{S}}
\newcommand{\mU}{\mathcal{U}}

\newcommand{\mS}{\mathcal{S}}
\newcommand{\mSb}{\bar{\mathcal{S}}}
\newcommand{\mN}{\mathcal{N}}
\newcommand{\mF}{\mathcal{F}}
\newcommand{\mW}{\mathcal{W}}
\newcommand{\omW}{\overline{\mW}}

\newcommand{\mfF}{\mathfrak{F}}
\newcommand{\mfX}{\mathfrak{X}}
\newcommand{\mfY}{\mathfrak{Y}}
\newcommand{\mfV}{\mathfrak{V}}

\newcommand{\Weight}{\mathfrak{V}}
\newcommand{\mfA}{\mathfrak{A}}
\newcommand{\Index}{\mathfrak{I}}
\newcommand{\bfl}{\mathbf{l}}
\newcommand{\bfm}{\mathbf{m}}
\newcommand{\bfk}{\mathbf{k}}

\newcommand{\utc}{\mathfrak{W}}
\newcommand{\weight}{\mathfrak{v}}

\newcommand{\cweight}{\mathfrak{u}}
\newcommand{\bcweight}{\bar{\mathfrak{u}}}
\newcommand{\zweight}{\mathfrak{w}}

\newcommand{\mfR}{\mathfrak{R}}
\newcommand{\mfG}{\mathfrak{G}}
\newcommand{\mfD}{\mathfrak{D}}

\newcommand{\mH}{\mathcal{H}}
\newcommand{\mG}{\mathcal{G}}
\newcommand{\mP}{\mathcal{P}}

\newcommand{\mR}{\mathcal{R}}

\newcommand{\bmR}{\bar{\mathcal{R}}}
\newcommand{\bmS}{\bar{\mathcal{S}}}
\newcommand{\bmN}{\bar{\mathcal{N}}}
\newcommand{\mQ}{\mathcal{Q}}
\newcommand{\mc}{\mathcal{C}}
\newcommand{\mcX}{\mathcal{X}}
\newcommand{\mcY}{\mathcal{Y}}
\newcommand{\mcV}{\mathcal{V}}
\newcommand{\tmcV}{\tilde{\mcV}}
\newcommand{\hmcV}{\hat{\mcV}}
\newcommand{\hmcX}{\hat{\mcX}}
\newcommand{\hmcY}{\hat{\mcY}}
\newcommand{\md}{\mathcal{D}}

\newcommand{\mfr}{\mathfrak{r}}
\newcommand{\mfs}{\mathfrak{s}}
\newcommand{\mft}{\mathfrak{t}}
\newcommand{\mfq}{\mathfrak{q}}
\newcommand{\mfp}{\mathfrak{p}}
\newcommand{\mcP}{\mathcal{P}}
\newcommand{\mcR}{\mathcal{R}}
\newcommand{\mfg}{\mathfrak{g}}

\newcommand{\mfe}{\mathfrak{E}}
\newcommand{\mfse}{\mathfrak{e}}

\newcommand{\mfH}{\mathfrak{H}}

\newcommand{\hN}{\hat{N}}
\newcommand{\tN}{\tilde{N}}

\newcommand{\hF}{\hat{F}}
\newcommand{\hGe}{\hat{G}}
\newcommand{\hU}{\hat{U}}
\newcommand{\hD}{\hat{D}}
\newcommand{\bGe}{\bar{G}}
\newcommand{\bF}{\bar{F}}
\newcommand{\fb}{\bar{f}}
\newcommand{\hf}{\hat{f}}
\newcommand{\hk}{\hat{k}}

\newcommand{\hg}{\hat{g}}
\newcommand{\chg}{\check{g}}

\newcommand{\tga}{\tilde{\gamma}}
\newcommand{\chvarphi}{\check{\varphi}}
\newcommand{\tvarphi}{\tilde{\varphi}}
\newcommand{\chvarrho}{\check{\varrho}}
\newcommand{\tvarrho}{\tilde{\varrho}}
\newcommand{\chD}{\check{D}}
\newcommand{\chk}{\check{k}}
\newcommand{\chK}{\check{K}}
\newcommand{\chmK}{\check{\mathcal{K}}}
\newcommand{\chG}{\check{G}}

\newcommand{\hH}{\hat{H}}
\newcommand{\hG}{\hat{\G}}

\newcommand{\hal}{\hat{\a}}

\newcommand{\hA}{\hat{A}}

\newcommand{\cha}{\check{a}}

\newcommand{\hL}{\hat{L}}

\newcommand{\htau}{\hat{\tau}}

\newcommand{\hPhi}{\hat{\Phi}}

\newcommand{\hchi}{\hat{\chi}}

\newcommand{\hna}{\hat{\nabla}}

\newcommand{\bx}{\bar{x}}
\newcommand{\by}{\bar{y}}

\newcommand{\bfw}{\mathbf{w}}

\newcommand{\bp}{\bar{p}}

\newcommand{\bchi}{\bar{\chi}}
\newcommand{\bc}{\bar{c}}
\newcommand{\bb}{\bar{b}}

\newcommand{\bfC}{\mathbf{C}}
\newcommand{\bfD}{\mathbf{D}}

\newcommand{\tPsi}{\tilde{\Psi}}
\newcommand{\bPsi}{\bar{\Psi}}
\newcommand{\ldr}[1]{\langle #1\rangle}

\begin{document}

\author{Hans Ringstr\"{o}m}
\title{Wave equations on silent big bang backgrounds}

\maketitle

\frontmatter

\include{dedicbook}

\setcounter{tocdepth}{1}
\tableofcontents

\mainmatter





\part{Introduction}

\chapter{Introduction}

The subject of these notes is the asymptotic behaviour of solutions to linear systems of wave equations in the vicinity of big bang singularities.
In particular, we are interested in the case of crushing singularities (cf. Definition~\ref{def:crushingsingularity} below) with
silent and anisotropic asymptotics. Beyond studying wave equations, we here develop a geometric framework for understanding such singularities, and
in a companion article \cite{RinGeometry}, we combine this framework with Einstein's equations in order to deduce additional information. Due to the
length of these notes, we, in the present chapter, wish to give an overview of the context of this study, as well as of the motivation, goals, assumptions
and results. In the following chapter, we introduce additional terminology and justify the importance of the anisotropic setting. We also provide
quite a detailed overview of previous results. This material serves as a background for the formal assumptions, stated in Chapter~\ref{chapter:assumptions}.
A detailed formulation of the results is then to be found in Chapter~\ref{chapter:results}. For an outline of these notes, the reader is referred to
Section~\ref{section:FullOutline}. 

\section{Big bang singularities}\label{section:bigbang}
Soon after the formulation of the general theory of relativity, the spatially homogeneous and isotropic Friedman-Lema\^{i}tre-Robertson-Walker (FLRW)
spacetimes, cf. (\ref{eq:RWmetric}) below, became the dominant models when describing the universe. In spite of the fact that the corresponding solutions
typically contain a big bang singularity, and in spite of the observations by, e.g., Hubble indicating that our universe expands, the existence of a
cosmological singularity only became accepted much later. Hawking's singularity theorem, providing robust conditions that guarantee the presence of
incomplete causal geodesics, combined with the discovery of the cosmic microwave background radiation by Penzias and Wilson, made it difficult to avoid
the conclusion that our universe began with a big bang.

The currently preferred $\Lambda$CDM models of the universe can be demonstrated to be future globally non-linearly stable; cf., e.g., \cite{stab} and
references cited therein. However, spatially homogeneous and isotropic solutions are typically unstable in the direction of the singularity; cf.
Section~\ref{section:anisotropy} below. There are some exceptions, correponding to matter models (such as stiff fluids and scalar fields) that give rise
to so-called quiescent asymptotics; see Chapter~\ref{chapter:basnotpreresults} below for more details. However, even in these cases, the isotropic
solutions are stable but not asymptotically stable, and there is no reason to expect the asymptotics to be isotropic; cf. Section~\ref{section:anisotropy}
below.

Since there is observational support for the spatial homogeneity and isotropy of the universe (even though the degree of this support can be questioned),
there is a tension between the observations and the instability. One way to resolve it is to say that the universe may be approximately spatially
homogeneous and isotropic back to some time (say, e.g., the surface of last scattering or the end of inflation, assuming that there
is an inflationary phase in the universe), but that it before that could be substantially different.
Another way is to say that the ``initial data'' for our universe are very special. However, regardless of perspective, it is of interest to have a more
general understanding of big bang singularities, in order to see if there are classes of solutions which are far from spatially homogeneous and isotropic
before some time which are still consistent with observations; or, alternatively, to see how special the initial data have to be in order to be consistent
with observations.

\section{Motivation}
This paper is the first in a series of two in which we develop a geometric framework for understanding highly anisotropic and oscillatory
big bang singularities. The observations of the previous section constitute the main motivation for doing so. However, an additional motivation is that
understanding highly anisotropic and oscillatory singularities is the natural next step in a hierarchy of difficulty in the study of the asymptotics of
cosmological solutions to Einstein's
equations. The hierarchy is determined by several features of the asymptotics: isotropic/anisotropic; silent/not silent; quiescent/oscillatory. We discuss
these notions in greater detail in the following chapter, but for the purposes of the present discussion, assume that there is a crushing singularity; cf.
Definition~\ref{def:crushingsingularity} below. Let $\mK$ denote the \textit{expansion normalised Weingarten map} associated with the foliation, i.e., the
Weingarten map of the leaves of the foliation divided by the mean curvature;
cf. Definition~\ref{def:normalisedWeingartenmap} below for a formal definition. Then (local) isotropy corresponds to $\mK$ being a multiple of the
identity. Moreover, for the purposes of the present discussion, the asymptotics are said to be \textit{quiescent} if the eigenvalues of $\mK$ converge
along causal curves going into the singularity and \textit{oscillatory} if they do not. Heuristically, the condition of \textit{silence} should be
interpreted as saying that different observers (i.e., causal curves) going into the singularity typically lose the ability to communicate (i.e., close
enough to the singularity, there is no past pointing causal curve from one observer to the other); cf. Section~\ref{section:silence} below for a more
formal discussion. Isotropic situations are easier to analyse than anisotropic ones; silent situations are easier to handle than non-silent ones; and
quiescent situations are less difficult than oscillatory ones.

Until recently, the results on future and past global non-linear stability were, at least to the best of our knowledge, all concerned with the
near isotropic setting. In the expanding direction, there
is by now a vast literature of stability results in the case of accelerated expansion. However, in that setting, the solutions isotropise asymptotically.
There are also results concerning the future stability of the Milne model and similar solutions. Again, these solutions exhibit isotropic asymptotics. In
the direction of the singularity, there are proofs of stable big bang formation; cf. Subsection~\ref{ssection:quiescent} below for further details.
Until recently, these results concerned solutions that are close to isotropic or moderately anisotropic. However, in late 2020, the authors of
\cite{FRS} achieved a breakthrough by demonstrating stable big bang formation for the Einstein-vacuum and Einstein-scalar field equations in the full
regime expected on the basis of heuristic arguments. In particular, the results cover the highly anisotropic setting. On a general level, it is
therefore of interest to investigate the asymptotics in highly anisotropic and oscillatory settings, since it represents a new
level of difficulty and might yield insights concerning the dynamics in unexplored regimes. On the other hand, to simplify the setting, while still
allowing oscillations and substantial anisotropies, it is natural to assume silence. 

An additional important observation is that for large classes of cosmological singularities, the expansion normalised Weingarten map is bounded. This
bound holds for examples with quiescent asymptotics; examples with
oscillatory asymptotics; for examples that are spatially homogeneous; and for examples that are spatially inhomogeneous. In fact, we only know of one
exception: In the case of so-called non-degenerate true spikes in $\tn{3}$-Gowdy symmetric vacuum solutions, the expansion normalised Weingarten map
is unbounded along causal geodesics that end up on the tip of a non-degenerate true spike. However, for generic $\tn{3}$-Gowdy symmetric vacuum solutions,
there are only finitely many non-degenerate true spikes. It is therefore to be expected that a generic causal geodesic going into the singularity does
not end up on the tip of such a spike; cf. Section~\ref{section:t3Gowdy} and, more specifically, Subsection~\ref{ssection:nondegeneratetruespikes}
below for more details on this topic. To conclude, it is of interest to analyse what can be deduced from the assumption that the expansion normalised
Weingarten map is bounded in the direction of the singularity, since such an assumption can be expected to be a natural bootstrap assumption in the
context of a non-linear stability argument. In some respects, this is the main motivation for writing these notes. 

\section{Goals}
In these notes, we formulate the assumptions of the geometric framework. However, our main goal is to analyse the asymptotic behaviour of solutions
to linear systems of wave equations on the corresponding backgrounds. An important step in achieving both this goal and the goal of obtaining a clear
picture of the asymptotic geometry is to find an appropriate frame. We do so here, and we deduce the central properties of the frame. 
Beyond stating results concerning the asymptotics of solutions to linear systems of wave equations, we, in the present paper, formulate some of the
conclusions concerning the geometry. However, we devote a separate paper to the conclusions that follow from combining the geometric framework
with Einstein's equations. In particular, we there demonstrate that the so-called Kasner map appears naturally.

In the present paper, we do not formulate non-linear results. One of the reasons is that we expect the geometric framework developed here to be only one,
albeit important, ingredient in a bootstrap argument. However, as is illustrated by the results and methods of the present paper, controlling the geometry
comes at the price of losing derivatives. It is therefore to be expected that the geometric framework will have to be combined with methods to
obtain crude estimates without a derivative loss in order to obtain non-linear results. Moreover, we expect the particular form of the methods to obtain
crude estimates to depend on the context. 

\section{Assumptions}

We formulate the assumptions of these notes in Chapter~\ref{chapter:assumptions} below. However, as a part of the introduction, we wish to give an outline
of the results. This necessitates providing a rough description of the assumptions, which is the purpose of the present section. 

\textbf{The expansion normalised Weingarten map.} The main assumptions are formulated in terms of the \textit{expansion normalised Weingarten map},
denoted $\mK$ and defined as follows. If $(M,g)$ is a spacetime with a crushing singularity (cf. Definition~\ref{def:crushingsingularity} below) with
corresponding foliation $M=\bM\times I$ (where $I$ is an open interval), then the expansion normalised Weingarten map of $\bM_{t}:=\bM\times\{t\}$ is
defined to be the Weingarten map (or shape operator) of $\bM_{t}$ divided by the mean curvature $\theta$ of $\bM_{t}$; cf.
Definition~\ref{def:normalisedWeingartenmap} below. The notion of (local) isotropy can be interpreted in terms of $\mK$; at a given spacetime point,
isotropy corresponds to $\mK$ being a multiple of the identity. 

\textbf{The logarithmic volume density.} 
For the assumptions to be general enough, it is important that some quantities are allowed to diverge in the direction of the singularity. Moreover, we
need to quantify the rate of divergence. One way of doing so is by introducing the \textit{volume density} $\varphi$ by demanding that the relation
$\mu_{\bge}=\varphi\mu_{\bge_{\refer}}$ hold. Here $\bge$ is the metric induced on $\bM_{t}$ (considered as a Riemannian metric on $\bM$), $\bge_{\refer}$ is a
fixed reference metric on $\bM$ and $\mu_{h}$ is the volume form associated with a given Riemannian metric $h$ on $\bM$. Here we assume $\varphi$ to
converge to zero in the direction of the singularity. The \textit{logarithmic volume density} $\varrho:=\ln\varphi$ can therefore be used as a measure
of proximity to the singularity. 

\textbf{Non-degeneracy.} Since we are interested in the highly anisotropic setting, we assume the eigenvalues of $\mK$ to be distinct, and the absolute
value of the differences of the different eigenvalues to have a positive lower bound. Since $\mK$ is symmetric with respect to $\bge$, there are thus
$n$ distinct real eigenvalues $\ell_{1}<\dots<\ell_{n}$ (and, by assumption, $|\ell_{i}-\ell_{j}|$ has a positive lower bound for $i\neq j$). By taking
a finite covering space of $\bM$, if necessary, there is an associated frame $\{X_{A}\}$, $A=1,\dots,n$, such that $\mK X_{A}=\ell_{A}X_{A}$ (no summation)
and such that $\bge_{\refer}(X_{A},X_{A})=1$. Note also that the frame $\{X_{A}\}$ is orthogonal with respect to $\bge$. 

\textbf{Silence.} One important assumption in our framework is that the causal structure of the singularity is silent; cf. \cite{EUW,PastAttr} for the
origin of the terminology. Heuristically, the condition of silence should be
interpreted as saying that different observers (i.e., causal curves) going into the singularity typically lose the ability to communicate (i.e.,
close enough to the singularity, there is no past pointing causal curve from one observer to the other). One way to express the condition of silence
formally is via the Weingarten map, say $\chK$, of the conformally rescaled metric $\hg:=\theta^{2}g$. The condition of silence we impose here is that
$\chK$ is negative definite in the sense that there is a constant $\e_{\Spe}>0$ such that $\chK\leq -\e_{\Spe}\mathrm{Id}$; cf.
Definition~\ref{def:silenceintro} below.

\textbf{Frame.} If $U$ is the future pointing unit normal to the leaves of the foliation and $\hU:=\theta^{-1}U$, then combining $\hU$ with the $X_{A}$
yields an orthogonal frame of $g$ (and $\hg$). Moreover, $\hU$ is a future pointing unit vector field with respect to $\hg$ and
$\hg(X_{A},X_{A})=e^{2\mu_{A}}$ for some functions $\mu_{A}$. 

\textbf{Sobolev norms.} If $\bM$ is closed and $\mt(\cdot,t)$ is a tensorfield on $\bM_{t}$ for each $t\in I$, let 
\[
\|\mt(\cdot,t)\|_{H^{\bfl}_{\weight}(\bM)}:=\left(\int_{\bM}\textstyle{\sum}_{m=l_{0}}^{l_{1}}\ldr{\varrho(\cdot,t)}^{-2\weight_{a}-2m\weight_{b}}
|\bD^{m}\mt(\cdot,t)|_{\bge_{\refer}}^{2}\mu_{\bge_{\refer}}\right)^{1/2},
\]
where $\bfl=(l_{0},l_{1})$; $\weight=(\weight_{a},\weight_{b})$; $\weight_{a}$ and $\weight_{b}$ are non-negative real numbers; $l_{0},l_{1}$ are non-negative
integers; and $l_{0}\leq l_{1}$. Here $\bD$ is the Levi-Civita connection of $(\bM,\bge_{\refer})$ and $\ldr{\xi}:=(1+|\xi|^{2})^{1/2}$. We introduce
similar notation when imposing control in $C^{k}$; cf. (\ref{eq:mtClbS}) below. Note that the norms and the covariant derivative are defined using a
\textit{fixed} Riemannian metric on $\bM$, not the induced metric $\bge$.

\textbf{Boundedness of the expansion normalised Weingarten map.} It is a remarkable fact that for large classes of big bang singularities, $\mK$ and its
covariant derivatives are uniformly bounded with respect to a fixed metric on $\bM$. Here, we assume $\mK$ to be bounded with respect to weighted $C^{k}$
and Sobolev spaces. For example, we assume $\|\mK(\cdot,t)\|_{H^{\bfl}_{\weight}(\bM)}$ to be uniformly bounded for some $\bfl=(0,l)$, $l\in\nn{}$,
$\weight=(0,\cweight)$ and $0\leq \cweight\in\ro$. Note that this bound is consistent with the pointwise norms of the covariant derivatives of $\mK$
diverging. It is of interest to allow faster blow up of the derivatives. However, in order to obtain results in such a setting, we expect it to be
necessary to make more detailed assumptions concerning the eigenvalues $\ell_{A}$, and, potentially, to make the weights dependent on the tangential
directions of the derivatives. Nevertheless, we expect the methods developed in these notes to be of interest under such circumstances as well. 

Next, consider the \textit{expansion normalised normal derivative} of $\mK$, denoted $\hml_{U}\mK$. This quantity is essentially an expansion normalised
Lie derivative of $\mK$ with respect to $U$; cf. Section~\ref{section:timederivativeofmK} below for a formal definition.
In this case, we impose bounds on the covariant derivatives similar to those imposed on $\mK$. In particular, we assume
$\|\hml_{U}\mK(\cdot,t)\|_{H^{\bfl}_{\weight}(\bM)}$
to be uniformly bounded, where $\bfl=(0,l)$, $\weight:=(\cweight,\cweight)$, $l\in\nn{}$ and $0\leq \cweight\in\ro$. It is important to note that
such a bound is consistent with the pointwise norm of the expansion normalised normal derivative of $\mK$ diverging in the direction of the singularity. 

Finally, we impose bounds on the components of $\hml_{U}\mK$ with respect to the eigenspaces of $\mK$. To be more precise, if $\{Y^{A}\}$ is the frame dual
to $\{X_{A}\}$, then we impose decay conditions on $(\hml_{U}\mK)(Y^{A},X_{B})$ for $B>1$ and $A\neq B$; cf. Definition~\ref{def:offdiagonalexpdec} below for
further details. Note that since the $\ell_{A}$ are ordered, and since the $X_{A}$ are ordered accordingly, it matters if $A>B$ or $B>A$. A posteriori, it
is possible to improve the bounds for $A<B$. However, in the case of $3+1$-dimensions, the case that $B=2$ and $A=3$ remains, and this constitutes the main
assumption. Nevertheless, in the companion article \cite[Corollary~52, p.~35]{RinGeometry}, we demonstrate that, when combining the assumptions with
Einstein's equations, the
estimate in this remaining case can also be improved a posteriori. That the above conditions are satisfied for large classes of spacetimes is justified
below; cf., in particular, Appendix~\ref{chapter:examples}. 

\textbf{The mean curvature.} Since information concerning the mean curvature cannot be extracted from the expansion normalised Weingarten map, we need to
impose conditions on the mean curvature separately. The assumptions take two forms. First, we impose a uniform bound on $\|\ln\theta\|_{H^{\bfl}_{\weight}(\bM)}$,
where $\bfl=(1,l)$, $l\in\nn{}$, $\weight=(0,\cweight)$ and $0\leq\cweight\in\ro$. Note in particular, that such a bound does not impose any restrictions
on the rate of blow up
of $\ln\theta$. Moreover, it is consistent with the covariant derivatives of $\ln\theta$ blowing up. We also impose restrictions on the expansion
normalised normal derivative of $\ln\theta$. It is convenient to express the corresponding conditions in terms of the \textit{deceleration parameter} $q$,
defined by the equality $\hU(n\ln\theta)=-1-q$. Concerning the deceleration parameter, we, e.g., impose uniform bounds on $\|q\|_{H^{\bfl}_{\weight}(\bM)}$,
where $\bfl=(0,l)$, $l\in\nn{}$, $\weight=(0,\cweight)$ and $0\leq\cweight\in\ro$. 

\textbf{Lapse and shift.} We also impose bounds on the \textit{shift vector field} $\chi$ and the relative spatial variation of the \textit{lapse function}
$N$, defined by $\d_{t}=NU+\chi$. The conditions imposed on the lapse function are similar to those imposed on the mean curvature. The shift vector field is
the only quantity on which we impose a smallness condition. However, we also need to impose boundedness conditions on higher covariant derivatives (with
appropriate weights). We refer the reader interested in the details to Chapter~\ref{chapter:assumptions} below. 

\textbf{Equations.} In these notes, we are interested in analysing the asymptotics of solutions to linear systems of wave equations taking the following
form:
\begin{equation}\label{eq:theequation}
\Box_{g}u+\mcX (u)+\a u=f,
\end{equation}
\index{$\a$Aa@Notation!Coefficients of the equation!$\mcX$}%
\index{$\a$Aa@Notation!Coefficients of the equation!$\a$}%
where $u$ is an $\rn{m_{\ros}}$ valued function on $M$, $\mcX$ is an $m_{\ros}\times m_{\ros}$-matrix of vector fields on $M$,
$\a\in C^{\infty}[M,\Mn{m_{\ros}}{\ro}]$ and $\Mn{m_{\ros}}{\ro}$ denotes the set of real valued $m_{\ros}\times m_{\ros}$-matrices. Moreover,
$f\in C^{\infty}(M,\rn{m_{\ros}})$. Due to the assumed silence, the global topology of the manifold is not of importance. In particular, $u$
could equally well be assumed to take its values in a vector bundle. 

\textbf{Coefficients of the equations.} In order to derive conclusions concerning solutions to linear systems of wave equations, we, needless to say,
also need to impose conditions on the coefficients of these systems. The conditions take the form of bounds on weighted norms of expansion normalised
versions of the coefficients, such as $\hal:=\theta^{-2}\a$. For example, we assume $\|\hal\|_{H^{\bfl}_{\weight}(\bM)}$ to be uniformly bounded, where
$\bfl=(0,l)$, $l\in\nn{}$, $\weight=(0,\cweight)$ and $0\leq\cweight\in\ro$. The expansion normalised version of $\mcX$ takes the form
\begin{equation}\label{eq:hmcXintrointro}
  \hmcX:=\theta^{-2}\mcX=\hmcX^{0}\hU+\hmcX^{\perp}=\hmcX^{0}\hU+\hmcX^{A}X_{A},
\end{equation}
where the components of $\hmcX^{\perp}$ are tangential to $\bM_{t}$. Here we require $\hmcX^{0}$ to satisfy weighted bounds similar to those imposed on
$\mK$. Concerning $\hmcX^{\perp}$, we demand that the components are bounded relative to the metric induced on the hypersurfaces $\bM_{t}$ by $\hg$.
However, we also impose bounds on weighted Sobolev norms etc. We refer the reader interested in the details concerning the different coefficients to
Chapter~\ref{chapter:assumptions} below. 

\textbf{Generality of the assumptions.} Below, we discuss the generality of the assumptions by comparing them with the properties of known solutions
to Einstein's equations; cf., in particular, Appendix~\ref{chapter:examples}.  

\section{Results}\label{section:resultsintrointo}
The main results of these notes concern the asymptotic behaviour of solutions to linear systems of wave equations under the assumptions described in
the previous section. In order to understand the asymptotics, it is convenient to write down the equation with respect to the frame introduced in the
previous section. It then takes the form
\begin{equation}\label{eq:equationintermsofcanonicalframeintro}
  -\hU^{2}u+\textstyle{\sum}_{A}e^{-2\mu_{A}}X_{A}^{2}u+Z^{0}\hU u+Z^{A}X_{A}u+\hal u=\hf.
\end{equation}
Here the coefficients $Z^{0}$ and $Z^{A}$ can be calculated in terms of $\hmcX$ and the geometry; cf. Subsection~\ref{ssection:reformofequationintro}
below. When analysing the asymptotics, the most important coefficients are $\hal$ and
\begin{equation}\label{eq:Zzdefintrointro}
  Z^{0} := \frac{1}{n}[q-(n-1)]\Id+\hmcX^{0}.
\end{equation}
Due to this formula, it is clear that the difference $q-(n-1)$ is of importance. In many quiescent settings, this quantity converges to zero
exponentially; cf. Appendix~\ref{chapter:examples} below.

\textbf{Energies.} To begin with, we derive energy estimates for energies such as 
\[
\hE[u](t):=\frac{1}{2}\int_{\bM_{t}}\left(|\hU(u)|^{2}+\textstyle{\sum}_{A}e^{-2\mu_{A}}|X_{A}(u)|^{2}+|u|^{2}\right)\theta\mu_{\bge}
\]
and higher order versions thereof; using the volume form $\theta\mu_{\bge}$ turns out to simplify the derivation of the estimates. When formulating
the results, it is convenient to change the time coordinate to $\tau(t):=\varrho(\bx_{0},t)$ for some reference point $\bx_{0}\in \bM$. The exact
estimate will depend on the choice of $\bx_{0}$. However, the main observation is that the energy could, potentially, grow exponentially (in terms of the
$\tau$-time) in the direction of the singularity, but that the rate of exponential growth does not depend on the number of derivatives. Conclusions of
this nature do not depend on the choice of $\bx_{0}$. The resulting estimates may not seem to be very useful. However, they are an essential first step
in making it possible to derive more detailed estimates in localised regions. 

\textbf{Localising the estimates.} In order to obtain more detailed information, it is necessary to localise the analysis. If $\g$ is an inextendible
future pointing causal curve, it is natural to focus on the behaviour of solutions in regions such as $J^{+}(\g)$, the causal future of the range of $\g$;
note that we are here interested in the asymptotic behaviour of solutions towards the past. Due to the silence, the spatial component of $\g$, say
$\bga$, converges in the direction of the singularity. Assume, from now on, that the limit point is $\bx_{0}$. Again, due to the silence, the 
variation of $\varrho$ in spatial slices of $J^{+}(\g)$ decays exponentially in the direction of the singularity. This means that in $J^{+}(\g)$,
$\varrho$ and $\tau$ are essentially the same. On the other hand, it can be demonstrated that $\hU(\varrho)$ is essentially equal to $1$. From this
perspective, it is therefore natural
to think of $\hU$ as $\d_{\tau}$. In the spirit of the BKL conjecture (cf. Subsection~\ref{subsection:BKLConjecture} below), it should also be possible
to ignore the spatial derivatives. Applying these ideas to (\ref{eq:equationintermsofcanonicalframeintro}) leads (assuming $f=0$) to the following model
equation for the asymptotic behaviour in $J^{+}(\g)$:
\begin{equation}\label{eq:modelintrointro}
  -u_{\tau\tau}+Z^{0}_{\roloc}u_{\tau}+\hal_{\roloc}u=0.
\end{equation}
Here $Z^{0}_{\roloc}(t):=Z^{0}(\bx_{0},t)$ and $\hal_{\roloc}(t):=\hal(\bx_{0},t)$, though we could  just as well localise the coefficients along $\g$.

At this point, the crucial question is: how do solutions to the model equation (\ref{eq:modelintrointro}) compare with solutions to the actual
equation? In order to answer that question, we need to know something about how solutions to the model equation behave. However, the assumptions
are such that we only know $Z^{0}_{\roloc}$ and $\hal_{\roloc}$ to be bounded. In particular, we do not know that they converge. On
the other hand, since the coefficients of the model equation are bounded, solutions cannot grow faster than exponentially. This indicates one way
of quantifying the asymptotic behaviour of solutions to the model equation: assuming a specific estimate for the flow associated with the model
equation. The hope would then be that solutions to the actual equation can be demonstrated to satisfy the same estimate. In order to be more specific,
note that (\ref{eq:modelintrointro}) can be written as a first order system of ODE's: $\Psi_{\tau}=A\Psi$; cf. (\ref{eq:PsiAdefintro}) below. Let $\Phi$
be the flow associated with this first order system; cf. (\ref{eq:Phidefintro}) below. Let $C_{A}$, $d_{A}$ and $\varpi_{A}$ be constants such that if
$s_{1}\leq s_{2}\leq 0$, then 
\begin{equation}\label{eq:Phinormbasasslocintrointro}
 \|\Phi(s_{1};s_{2})\|\leq C_{A}\ldr{s_{2}-s_{1}}^{d_{A}}e^{\varpi_{A}(s_{1}-s_{2})}. 
\end{equation}
Then one of the main results of these notes is that if (\ref{eq:Phinormbasasslocintrointro}) holds and $u$ is a solution to
(\ref{eq:equationintermsofcanonicalframeintro}) with $f=0$, then 
\begin{equation}\label{eq:hUuanduoptest}
  |\hU u|+|u|\leq C\ldr{\varrho}^{d_{A}}e^{\varpi_{A}\varrho}
\end{equation}
in $J^{+}(\g)$. In other words, the solution satisfies the best estimate one could hope for. 
Note that $\varpi_{A}$ and $d_{A}$ are determined by $A$; i.e., by $\hal_{\roloc}$ and $Z^{0}_{\roloc}$. In particular, these constants depend
on $\bx_{0}$, i.e. on $\g$. We also obtain higher order versions of the estimate (\ref{eq:hUuanduoptest}). 

\textbf{Asymptotics.} In order to derive asymptotics, we need to make more detailed assumptions concerning the coefficients. Say, for the sake of
argument, that $Z^{0}_{\roloc}$ and $\hal_{\roloc}$ converge exponentially (in $\tau$-time) to limits $Z^{0}_{\infty}$ and $\hal_{\infty}$ respectively.
Then we replace $Z^{0}_{\roloc}$ and $\hal_{\roloc}$ with $Z^{0}_{\infty}$ and $\hal_{\infty}$ respectively in the model equation (\ref{eq:modelintrointro}). 
This results in a linear system of second order constant coefficient ODE's which can be rewritten in first order form as $\Psi_{\tau}=A_{0}\Psi$, where
$A_{0}$ is given by (\ref{eq:AzeroAremdefintro}) below. In this setting, $d_{A}$ and $\varpi_{A}$ can be calculated in terms of $A_{0}$. Moreover,
given a solution $u$ to (\ref{eq:equationintermsofcanonicalframeintro}), there is a vector $V_{\infty}$ and a $\b>0$ such that
\begin{equation}\label{eq:uhUuasymptoticsintrointro}
\left|\left(\begin{array}{c} u \\ \hU u\end{array}\right)-e^{A_{0}\varrho}V_{\infty}\right|
  \leq  Ce^{(\varpi_{A}+\b)\varrho}
\end{equation}
in $J^{+}(\g)$. In other words, the solution to the actual equation behaves as a solution to the model equation. The estimate
(\ref{eq:uhUuasymptoticsintrointro}) also holds with $\hU u$ replaced by $u_{\tau}$. Additionally, detailed asymptotics for the higher order derivatives
can be derived; cf. Subsection~\ref{ssection:higherorderderivativesintro} below. It is also possible to specify the leading order asymptotics; cf.
Section~\ref{section:specifyingasymptoticsintro}. Due to this fact, it is possible to prove that estimates such as (\ref{eq:hUuanduoptest}) are optimal.
Note, however, that these estimates are associated with substantial losses in derivatives.

\textbf{Lack of uniformity.} In addition to the above, there are results of the following nature. Given a finite number of distinct points, say
$\bx_{i}\in\bM$, $i=1,\dots,l$; a finite set of real numbers (characterising the growth/decay rate), say $a_{i}\in\ro$, $i=1,\dots,l$; and future pointing
inextendible causal curves $\g_{i}$, $i=1,\dots,l$ such that the spatial component of $\g_{i}$ converges to $\bx_{i}$ in the direction of the singularity;
there is an equation and a corresponding solution such that the (exponential) growth rate of the energy density of the solution in
$J^{+}(\g_{i})$ is given by $a_{i}$ for $i=1,\dots,l$, and for causal curves $\g$ such that the spatial component of $\g$ converges to a point
$\bx\notin\{\bx_{1},\dots,\bx_{l}\}$, the solution decays at a fixed prespecified rate. Note, in particular, that the optimal rate in general depends
discontinuously on the endpoint of the spatial component of the causal curve. The above observations make it clear that it is not reasonable to hope
a general energy estimate to yield detailed information, since the behaviour of the solution in $J^{+}(\g)$ can be expected to depend strongly (and
discontinuously) on the choice of causal curve. 

It is of interest to compare the results mentioned above with the BKL proposal, which we discuss in Subsection~\ref{subsection:BKLConjecture}
below. One of the key ideas of this proposal is that, with respect to suitable foliations, solutions to Einstein's equations should be well approximated
by solutions to the equations obtained by dropping the spatial derivatives. The results mentioned above yield conclusions of
this nature. However, it is important to note that in the BKL proposal, it is \textit{assumed} that the spatial derivatives can be ignored, whereas we
here formulate conditions that make it possible to \textit{prove} that the spatial derivatives can be ignored. On the other hand, these notes are only
concerned with linear systems of wave equations on given backgrounds, as opposed to the Einstein equations. 

\section{Outline}
In addition to the present chapter, the introductory part consists of three chapters. In Chapter~\ref{chapter:basnotpreresults},
we introduce some of the basic notions. Moreover, we justify the importance of considering the highly anisotropic and oscillatory setting
and give an overview of mathematical results concerning big bang singularities.
In Chapter~\ref{chapter:assumptions}, we then describe the assumptions, as well as some of the basic conclusions. Finally, in
Chapter~\ref{chapter:results}, we describe the results and give an outline of the contents of these notes. 

\section*{Acknowledgments}

This research was funded by the Swedish Research Council, dnr. 2017-03863. It was also supported by the Swedish Research Council
under grant no. 2016-06596 while the author was in residence at Institut Mittag-Leffler in Djursholm, Sweden during the fall of 2019.

\chapter{Basic notions and previous results}\label{chapter:basnotpreresults}

The purpose of the present chapter is to justify why it is natural to consider highly anisotropic and oscillatory solutions in the direction of the
singularity; to introduce some basic terminology; to briefly describe existing conjectures concerning big bang singularities; and to give examples of
previous results. In other words, beyond the terminology, the present chapter is largely motivational. The examples of previous results also serve the
purpose of providing a frame of reference for the assumptions we make in these notes. However, it should be mentioned that, logically, the present chapter
could largely be skipped by the reader only interested in the formal statements and proofs. 

\section{Anisotropy}\label{section:anisotropy}

As noted in Section~\ref{section:bigbang}, spatially homogeneous and isotropic solutions are typically unstable in the direction of the big bang
singularity. In the present section, we justify this statement. However, before doing so, we need to introduce notation allowing us to quantify the
anisotropies of solutions. This naturally leads to the introduction of the expansion normalised Weingarten map, the central object in these notes.

\subsection{The expansion normalised Weingarten map}

In these notes, we restrict our attention to crushing singularities. 

\begin{definition}\label{def:crushingsingularity}
  A spacetime $(M,g)$ is said to have a \textit{crushing singularity}
  \index{Crushing singularity}%
  \index{Singularity!Crushing}%
  if the following conditions are satisfied. First, $(M,g)$ is foliated by spacelike
  Cauchy hypersurfaces in the sense that $M=\bM\times I$, where $\bM$ is an $n$-dimensional manifold, $I=(t_{-},t_{+})$ is an interval, the metric $\bge$
  induced on the leaves $\bM_{t}:=\bM\times\{t\}$ of the foliation is Riemannian, and $\bM_{t}$ is a Cauchy hypersurface in $(M,g)$ for all $t\in I$. Second,
  the mean curvature, say $\theta$, of the leaves of the foliation tends to infinity as $t\rightarrow t_{-}+$.
\end{definition}
\begin{remark}
  A spacetime is a time oriented Lorentz manifold. And given a foliation as in the statement of the definition, $\d_{t}$ is always assumed to be future
  pointing. 
\end{remark}
Given a crushing singularity, let $\bK$ be the Weingarten map (shape operator) of the leaves of the foliation. In other words, $\bK$ is the second
fundamental form, considered
as a linear map from the tangent space of the leaves of the foliation to itself (or, alternately, $\bK$ is obtained from the second fundamental form by
raising one index). Then the expansion normalised Weingarten map, in many ways the central object in these notes, is defined as follows.

\begin{definition}\label{def:normalisedWeingartenmap}
  Let $(M,g)$ be a spacetime with a crushing singularity. Let $\theta$ be the mean curvature and $\bK$ be the Weingarten map of the leaves of the foliation.
  Assume $\theta$ to always be strictly positive. Then the \textit{expansion normalised Weingarten map}
  \index{Expansion normalised Weingarten map}%
  \index{Weingarten map!Expansion normalised}%
  is defined by $\mK:=\bK/\theta$.
  \index{$\a$Aa@Notation!Tensor fields!$\mK$}%
\end{definition}
\begin{remark}
  In these notes, we are interested in the asymptotics in the direction of a crushing singularity. For that reason, the assumption that $\theta$ be strictly
  positive is not a substantial restriction, since limiting one's attention to a region of the spacetime close enough to the singularity ensures that this
  condition is satisfied. 
\end{remark}
\begin{remark}\label{remark:ellAdefinitionintro}
  Since $\mK$ is symmetric with respect to $\bge$, the eigenvalues of $\mK$, say $\ell_{A}$,
  \index{$\a$Aa@Notation!Eigenvalues!$\ell_{A}$}%
  are real, and, due to the normalisation, their sum equals one.
  In what follows, we order them so that $\ell_{1}\leq\ell_{2}\leq\cdots\leq\ell_{n}$. In the case of $3+1$-dimensions, it is convenient to summarise the
  information contained in the $\ell_{A}$ by $\ell_{\pm}$, defined as follows:
  \begin{align}
    \ell_{+} := & \frac{3}{2}\left(\ell_{2}+\ell_{3}-\frac{2}{3}\right)=\frac{3}{2}\left(\frac{1}{3}-\ell_{1}\right),\label{eq:ellplus}\\
    \ell_{-} := & \frac{\sqrt{3}}{2}(\ell_{2}-\ell_{3}).\label{eq:ellminus}
  \end{align}
  \index{$\a$Aa@Notation!Eigenvalues!$\ell_{\pm}$}%
\end{remark}
\begin{remark}
  If the eigenvalues $\ell_{A}$ are all equal, then $\mK=\mathrm{Id}/n$. A solution is said to be \textit{asymptotically isotropic} if the eigenvalues
  $\ell_{A}$ asymptotically become equal (since the sum of the eigenvalues equals $1$, this means that the eigenvalues all have to converge to $1/n$). In
  the case of $3+1$-dimensions this requirement is equivalent to $(\ell_{+},\ell_{-})$ converging to $(0,0)$. 
\end{remark}
With the above terminology, the distinction between quiescent and oscillatory asymptotics can be defined as follows.
\begin{definition}\label{def:oscandquiescent}
  Assume the conditions of Definition~\ref{def:normalisedWeingartenmap} to be satisfied and let $\{\ell_{A}\}$ be defined by
  Remark~\ref{remark:ellAdefinitionintro}. Then the singularity is said to be \textit{quiescent}
  \index{Singularity!Quiescent}%
  \index{Quiescent singularity}%
  if, for every future pointing and past inextendible causal
  curve $\g:(s_{-},s_{+})\rightarrow M$, and for every $A\in \{1,\dots,n\}$, $\ell_{A}\circ\g(s)$ converges as $s\rightarrow s_{-}+$. If the singularity is not
  quiescent, it is said to be \textit{oscillatory}.
  \index{Singularity!Oscillatory}%
  \index{Oscillatory singularity}%
\end{definition}
Before proceeding, it is convenient to introduce some classes of solutions that can be used to illustrate general definitions etc. in the discussions to
follow.

\begin{figure}
  \begin{center}
    \includegraphics{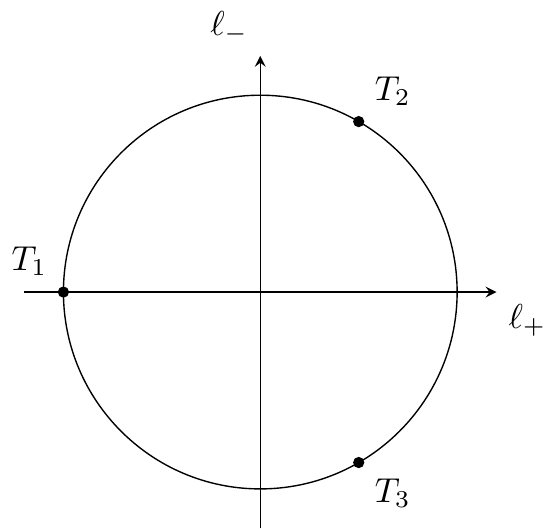}
  \end{center}
  \caption{The Kasner circle with the special points $T_{i}$, $i=1,2,3$, indicated.}\label{fig:Kasnercircle}
\end{figure}

\begin{example}[The Kasner solutions]\label{example:Kasnersolutions}
The \textit{Kasner solutions} to Einstein's vacuum equations are the metrics
\begin{equation}\label{eq:groKdef}
g_{\roK}:=-dt\otimes dt+\textstyle{\sum}_{i=1}^{n}t^{2p_{i}}dx^{i}\otimes dx^{i}
\end{equation}
\index{Solutions!Kasner}%
\index{Kasner!Solutions}%
on the manifold $M_{\roK}:=\rn{n}\times (0,\infty)$, where $p_{i}$ are constants satisfying the so-called \textit{Kasner relations}:
\begin{equation}\label{eq:Kasnerrelations}
  \textstyle{\sum}_{i=1}^{n}p_{i}=1,\ \ \
  \textstyle{\sum}_{i=1}^{n}p_{i}^{2}=1.
\end{equation}
\index{Kasner!Relations}%
In this case the constant-$t$ hypersurfaces constitute a natural foliation, and the mean curvature of $\rn{n}\times \{t\}$ satisfies $\theta=t^{-1}$.
In particular, $(M_{\roK},g_{\roK})$ has a crushing singularity corresponding to $t\rightarrow 0+$. Next, note that $\mK^{i}_{\phantom{i}j}=p_{i}\de^{i}_{j}$
(no summation on $i$), where we calculate the components of $\mK$ using the frame $\{\d_{i}\}$ and its dual. In particular, the $p_{i}$ are the
eigenvalues of $\mK$ so that $\ell_{i}=p_{i}$. In case $n=3$, we can define $\ell_{\pm}$ as in (\ref{eq:ellplus}) and (\ref{eq:ellminus}). With this
terminology, the Kasner relations (\ref{eq:Kasnerrelations}) can be summarised by one equality: $\ell_{+}^{2}+\ell_{-}^{2}=1$. The corresponding set
is referred to as the \textit{Kasner circle},
\index{Kasner!Circle}%
and plays a central role in what follows; cf. Figure~\ref{fig:Kasnercircle}. If one of the $p_{i}=1$ and
all the others equal $0$, then the corresponding spacetime is flat (as opposed to Ricci flat). These conditions define the \textit{flat Kasner solutions},
\index{Flat Kasner solutions}%
\index{Kasner!Flat solutions}%
and they correspond to subsets of Minkowski space (or quotients of subsets, in case the spatial topology is different from $\rn{n}$). On the Kasner circle,
the flat Kasner solutions correspond to three points, $T_{1}=(-1,0)$, $T_{2}=(1/2,\sqrt{3}/2)$ and $T_{3}=(1/2,-\sqrt{3}/2)$,
\index{$\a$Aa@Notation!Points!$T_{i}$, Special points on the Kasner circle}%
referred to as the
\textit{special points}; cf. Figure~\ref{fig:Kasnercircle}.
\index{Special points}%
\index{Kasner!Circle!Special points}%
\end{example}
\begin{remark}
  Note that, except for Minkowski space, all maximal globally hyperbolic developments (MGHD's) of left invariant vacuum initial data on $\rn{n}$ (with
  respect to the standard Lie group structure) can be written in the form (\ref{eq:groKdef}). Moreover, all of these solutions can be considered to be
  solutions on $\tn{n}\times (0,\infty)$. Note, however, that when taking the quotient, the edges of the corresponding fundamental domains need not be
  aligned with the $\d_{i}$ appearing in (\ref{eq:groKdef}). Moreover, the sizes of the fundamental domains are variable. Note also that Minkowski space,
  considered as a solution to Einstein's vacuum equations on $\tn{n}\times\ro$, is unstable. 
\end{remark}

\subsection{Instability of spatially homogeneous and isotropic solutions}

As already mentioned in Section~\ref{section:bigbang}, cosmologists normally use FLRW spacetimes to model the universe. They take the form
$(M_{\roF},g_{\roF})$, where
\begin{equation}\label{eq:RWmetric}
  g_{\roF}=-dt\otimes dt+a^{2}(t)\bge,
\end{equation}
\index{Spacetimes!FLRW}%
\index{FLRW!Spacetimes}%
\index{FLRW!Metrics}%
\index{Metrics!FLRW}%
$M_{\roF}:=\Sigma\times I$, $I$ is an open interval, $a\in C^{\infty}[I,(0,\infty)]$ and $(\Sigma,\bge)$ is a complete Riemannian manifold of constant curvature
$0$, $1$ or $-1$; i.e., $(\Sigma,\bge)$ is a quotient of Euclidean, spherical or hyperbolic space. Since we are interested in crushing singularities, we
here assume $\dot{a}/a$ to tend to infinity as
$t\rightarrow t_{-}+$ (assuming the range of the foliation to be given by $I=(t_{-},t_{+})$). This does not necessarily mean that $a\rightarrow 0$ as
$t\rightarrow t_{-}+$. However, for the spacetimes of interest here, this condition is satisfied, and we, in what follows, tacitly assume it. In order to
connect the Lorentz manifolds of the form $(M_{\roF},g_{\roF})$ with cosmology, we have to make a choice of matter model and impose Einstein's equations. 
In the standard models of the universe, the matter content is normally modeled by perfect fluids, defined as follows. 

\textbf{Perfect fluids.} On a spacetime $(M,g)$, the stress energy tensor associated with a \textit{perfect fluid}
\index{Perfect fluid}%
takes the form
\begin{equation}\label{eq:setperfectfluid}
  T=(\rho+p)U^{\flat}\otimes U^{\flat}+pg.
\end{equation}
\index{Perfect fluid!Stress energy tensor}%
Here $U$ is the \textit{flow vector field of the fluid}.
\index{Perfect fluid!Flow vector field}%
In particular, it is a future pointing unit timelike vector field. Moreover, $U^{\flat}$ is the 
metrically equivalent one-form field. Finally, $\rho$ and $p$ are the \textit{energy density}
\index{Perfect fluid!Energy density}%
and \textit{pressure}
\index{Perfect fluid!Energy density}%
of the fluid. In particular, they
are smooth functions on $M$. In order to be able to deduce how the fluid evolves, we here, in addition, impose a linear equation of state $p=(\g-1)\rho$,
where $\g$ is a constant. Here $\g=1$ corresponds to \textit{dust}
\index{Dust}%
\index{Perfect fluid!Dust}%
(this is used to model ordinary and dark matter), $\g=4/3$ corresponds to a
\textit{radiation fluid}
\index{Radiation fluid}%
\index{Perfect fluid!Radiation}%
(describing radiation and highly relativistic particles) and $\g=2$ corresponds to a \textit{stiff fluid}.
\index{Stiff fluid}%
\index{Perfect fluid!Stiff}%
Note that a positive
cosmological constant can be interpreted as as a perfect fluid with $p=-\rho$: i.e., $\g=0$. When taking this perspective, the cosmological constant
can be thought of as a particular form of dark energy. The equations that have to be satisfied by the matter are summarised by the requirement that the
stress energy tensor be divergence free. Note that, in the case of $\g=0$, this requirement implies that $\rho$ is constant (assuming $M$ to be connected),
and this constant is then the cosmological constant. 

\textbf{Perfect fluids in the spatially homogeneous and isotropic setting.} In the spatially homogeneous and isotropic setting, $U$ has to be orthogonal to
the spatial hypersurfaces of homogeneity $\Sigma_{t}:=\Sigma\times\{t\}$ and $p$ and $\rho$ have to be independent of the spatial variable. This means, in
particular, that $U=\d_{t}$ and that $p$ and $\rho$ only depend on $t$.  In the case of the metric (\ref{eq:RWmetric}), it can then be deduced that 
$\dot{\rho}=-3(\rho+p)\dot{a}/a$; cf. \cite[Corollary~13, p.~346]{oneill}. Due to the equation of state, this equality is equivalent to the statement that
$a^{3\g}\rho$ is constant. In
particular, when $a\rightarrow 0+$, the energy density of dust tends to infinity as $a^{-3}$; the energy density of a radiation fluid tends to
infinity as $a^{-4}$; the energy density of a stiff fluid tends to infinity as $a^{-6}$; and the energy density of dark energy remains constant. 

\textbf{The $\Lambda$CDM models.} The currently preferred models of the universe are spatially flat, include cold dark matter, ordinary matter, radiation
and a positive cosmological constant $\Lambda$. The different matter components can be modeled in different ways. However, one specific choice is that
$\bge$ is Euclidean, that $g_{\roF}$ is a solution to
\[
G+\Lambda g=T,
\]
where $G$ is the Einstein tensor, $\Lambda$ is the cosmological constant and $T$ is the sum of three contributions: dust corresponding to ordinary matter,
dust corresponding to dark matter and a radiation fluid corresponding to radiation and highly relativistic particles. When analysing the asymptotics in
the direction of the singularity, physicists normally ignore the contribution from the dark energy and from the ordinary and dark matter. The reason for
this is quite simple: the energy density of the radiation fluid grows as $a^{-4}$, whereas the energy density of the remaining components of the matter
is bounded by $Ca^{-3}$. Thus the radiation fluid will dominate asymptotically. For that reason, we, for the rest of this subsection, restrict our attention
to solutions to Einstein's equations with a vanishing cosmological constant and matter consisting of a radiation fluid.

\begin{figure}
  \begin{center}
    \includegraphics{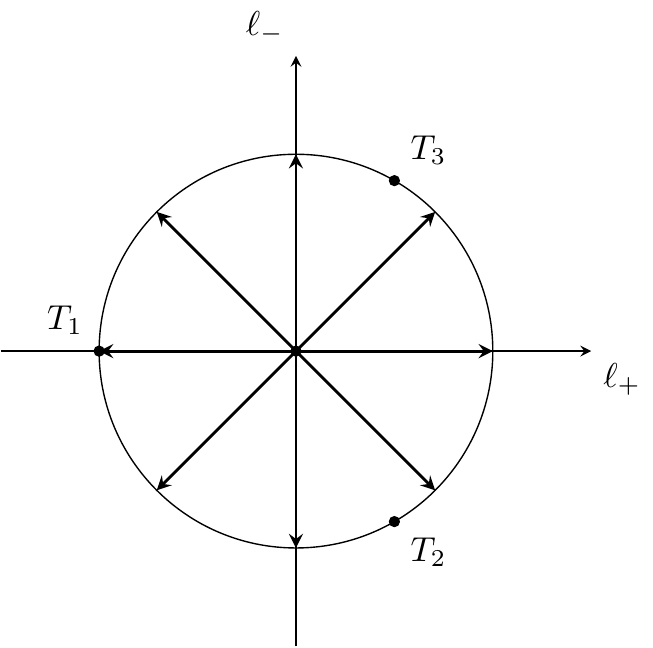}    
  \end{center}
  \caption{A projection of the dynamics of Bianchi type I radiation fluid solutions to the $\ell_{+}\ell_{-}$-plane. In fact, all Bianchi type I perfect
    fluid solutions exhibit these dynamics if $2/3<\g<2$.}\label{fig:Bianchi-I-Fluid-projection}
\end{figure}
\begin{figure}
  \begin{center}
    \includegraphics{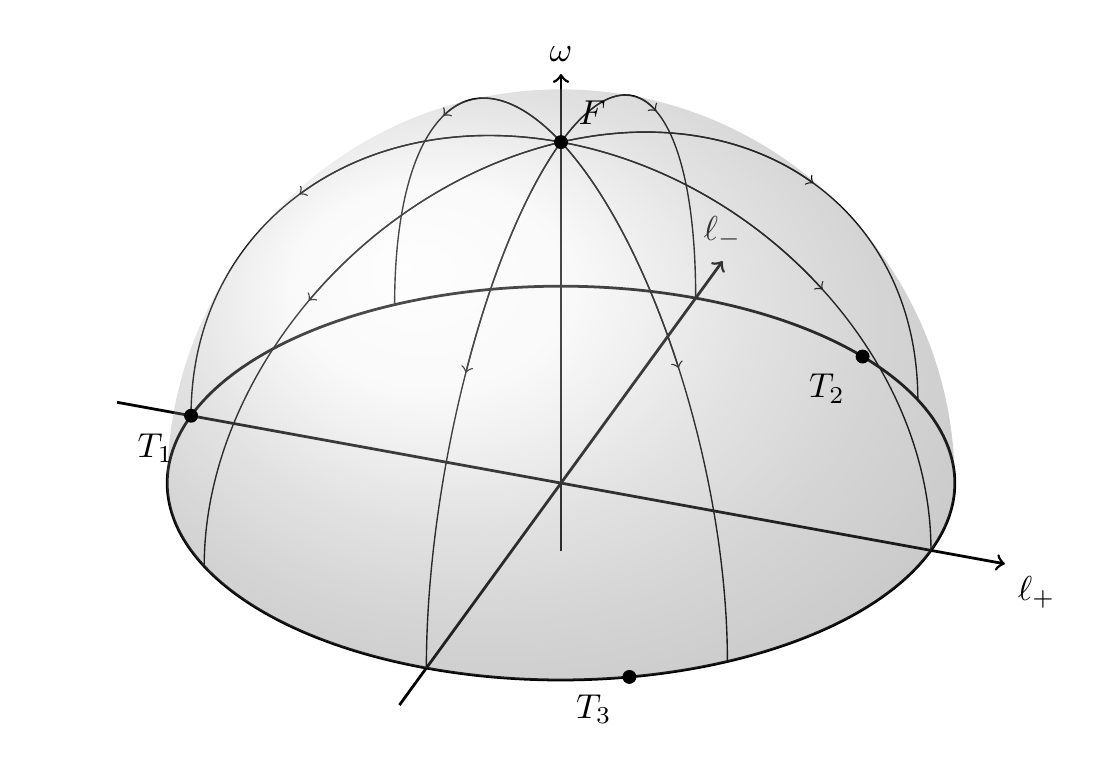}    
  \end{center}
  \caption{The dynamics of Bianchi type I radiation fluid solutions in the full state space. Here $F$ denotes the fixed point corresponding to the
  isotropic solutions. Moreover, $\omega$ corresponds to the square root of a rescaled version of the energy density. All Bianchi type I fluid
  solutions exhibit these dynamics if $2/3<\g<2$.}\label{fig:Bianchi-I-Fluid-full}
\end{figure}

\textbf{Instability to anisotropic perturbations.} In order to determine the stability of the above solutions in the direction of the singularity with respect
to anisotropic perturbations, it is natural to begin by addressing the stability in the simplest setting possible, namely that of Bianchi type I solutions.
The Bianchi type I solutions are the maximal globally hyperbolic developments (MGHD's)
\index{MGHD}%
of left invariant initial data on $\rn{3}$ or a quotient thereof.
\index{Bianchi!Type I}%
In the Bianchi type I state space, the
isotropic solutions coincide with a single fixed point (assuming one uses, e.g., the expansion normalised variables introduced by Wainwright and Hsu, cf.
\cite{wah}). We here denote it $F$. The full Bianchi type I state space corresponds to a hemisphere and the equator corresponds to the
Kasner circle. In particular, the north pole and the equator consist of fixed points. Moreover, the dynamics can be summarised as saying that, in the
direction of the singularity, $(\ell_{+},\ell_{-})$ moves radially towards the Kasner circle; and in the expanding direction $(\ell_{+},\ell_{-})$  moves
radially towards the origin; cf. Figure~\ref{fig:Bianchi-I-Fluid-projection} for an illustration of the projected dynamics. The dynamics in the full state
space are illustrated in Figure~\ref{fig:Bianchi-I-Fluid-full}. 
For a justification of the above statements, cf., e.g., \cite[Section~8, p.~428]{BianchiIXattr}.

Given the above observations, it is of interest to ask if the Kasner solutions are stable. This is not to be expected, for the following reason.
First, the Bianchi type I solutions are on the boundary of the state space of Bianchi type IX solutions (with respect to the Wainwright Hsu variables),
where Bianchi type IX solutions are the MGHD's of left invariant initial data on $\mathrm{SU}(2)$.
\index{Bianchi!Type IX}%
Perturbing into the Bianchi type IX state space, the
Kasner solutions are unstable, and the dynamics are expected to be well approximated by the Kasner map (cf. Figure~\ref{fig:TheKasnerMap} below); cf.
\cite[Proposition~6.1, p.~421]{BianchiIXattr} and its proof for a justification. The topologies of the spatial hypersurfaces of
homogeneity are of course different in the Bianchi type I and IX settings. For this reason,
global perturbations from Bianchi type I to Bianchi type IX are not meaningful. However, local perturbation are, and they indicate the instability of
the Kasner solutions.

\begin{figure}
  \begin{center}
    \includegraphics{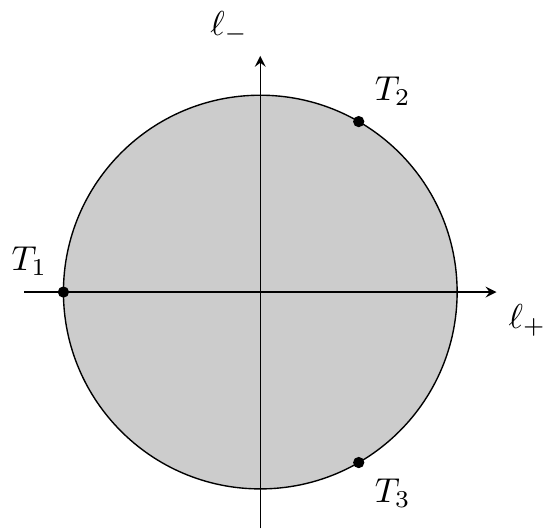}    
  \end{center}
  \caption{A projection of the Bianchi type I stiff fluid state space ($\g=2$) to the $\ell_{+}\ell_{-}$-plane. The state space consists of fixed
    points.}\label{fig:Kasner-Bianchi-I-Quiescent-Regime}
\end{figure}

\textbf{Stiff fluids.} The dynamics in the Bianchi type I setting are illustrated by Figure~\ref{fig:Bianchi-I-Fluid-full} for all perfect fluids
satisfying $2/3<\g<2$. However, for stiff fluids the dynamics are different. In that case, the hemisphere illustrated in 
Figure~\ref{fig:Bianchi-I-Fluid-full} consists of fixed points; i.e., there are no dynamics. Projecting the state space to the $\ell_{+}\ell_{-}$-plane
yields Figure~\ref{fig:Kasner-Bianchi-I-Quiescent-Regime}.
Again, the question arises if these fixed points are stable. It turns out that when perturbing initial data corresponding to the fixed points belonging
to the full disc in Figure~\ref{fig:Kasner-Bianchi-I-Quiescent-Regime} into the Bianchi type VIII and IX state spaces, then only the fixed points belonging
to the shaded area in Figure~\ref{fig:QuiescentConvergentRegime} are stable. More specifically, all Bianchi type VIII and IX stiff fluid solutions with
a non-vanishing energy density converge to a point in the shaded area of Figure~\ref{fig:QuiescentConvergentRegime} below; cf.
\cite[Theorem~19.1, p.~478]{BianchiIXattr}. Here the Bianchi type VIII solutions are the MGHD's corresponding to left invariant initial data
on the universal covering group of $\mathrm{Sl}(2,\ro)$.
\index{Bianchi!Type VIII}%

Considering a solution which is similar to a $\Lambda$CDM model but with a small stiff fluid component, it is reasonable to expect the stiff fluid component
to dominate asymptotically, so that spatially homogeneous and isotropic solutions are stable. On the other hand, for this to be true, the stiff fluid
component has to be large enough in comparison with the anisotropic perturbations. Since there is no stiff fluid component at all in the standard models,
it is not obvious that such a condition is satisfied. In that setting, it may therefore be more reasonable to expect anisotropic perturbations, combined
with, say, a radiation fluid, to, initially, generate significant anisotropies. At a later stage, the stiff fluid then begins to dominate, leading to
quiescent behaviour. However, since the solution is already anisotropic by that time, and since isotropic solutions are not asymptotically stable in the
stiff fluid setting, there is no reason to prefer a specific subset of the stable regime depicted in Figure~\ref{fig:QuiescentConvergentRegime} below.

\textbf{Inflation.} Inflation is an important ingredient of the standard models of the universe. However, since it is supposed to begin and end at times
which are determined in a somewhat ad hoc fashion, and since the relevant times are both distinct from the asymptotic regime, we do not discuss this topic
further here.

\begin{example}[Bianchi type I stiff fluids]\label{example:BianchiTypeIStiffFluids}
  As is clear from the above discussion, the Bianchi type I stiff fluid solutions are of particular interest.
  \index{Bianchi!Type I!Stiff fluid}%
  The corresponding metrics can be written
  \begin{equation}\label{eq:groQdef}
    g_{\roQ}:=-dt\otimes dt+\textstyle{\sum}_{i=1}^{n}t^{2p_{i}}dx^{i}\otimes dx^{i}
  \end{equation}
  on the manifold $M_{\roQ}:=\rn{n}\times (0,\infty)$, where $p_{i}$ and $p_{\phi}$ are constants satisfying
  \begin{equation}\label{eq:KasnerQuiescentRelations}
    \textstyle{\sum}_{i=1}^{n}p_{i}=1,\ \ \
    \textstyle{\sum}_{i=1}^{n}p_{i}^{2}+p_{\phi}^{2}=1.
  \end{equation}
  Defining $\rho_{\roQ}:=p_{\phi}^{2}/(2t^{2})$, $(M_{\roQ},g_{\roQ},\rho_{\roQ})$ is a solution to the Einstein stiff fluid equations. Moreover, fixing
  $\phi_{0}\in\ro$ and defining $\phi_{\roQ}=p_{\phi}\ln t+\phi_{0}$, $(M_{\roQ},g_{\roQ},\phi_{\roQ})$ is a solution to the Einstein scalar field equations.
  The mean curvature and the expansion normalised Weingarten map can be calculated as in Example~\ref{example:Kasnersolutions}. In particular, $t=0$
  represents a crushing singularity in $(M_{\roQ},g_{\roQ})$. 
\end{example}

\section{Silence}\label{section:silence}

An extremely important notion in these notes is that of silence; cf. \cite{EUW}, in particular \cite[Subsection~5.3]{EUW}, and \cite{PastAttr},
in particular \cite[Subsection~III.B.3]{PastAttr} and \cite[Section~IV]{PastAttr}, for the origin of the terminology. There are various ways of defining
it. On a heuristic level, the idea is that observers
going into the singularity typically lose the ability to communicate. On the weakest level, there should be points $p,q\in M$ such that
$J^{-}(p)\cap J^{-}(q)=\varnothing$. Another indication of silence is the presence of particle horizons. Here, a \textit{particle horizon}
\index{Particle horizon}%
is a set which is
non-empty and which can be written as the boundary of $J^{+}[J^{-}(p)]$ for some $p\in M$. However, in practice it is often convenient to formulate the
property of silence in terms of a foliation, even though the resulting notion is foliation dependent. Given a foliation $M=\bM\times I$ of the spacetime,
the idea is then that the spatial component of past intextentible causal curves should converge with respect to some reference metric on $\bM$. However, in
these notes we make an even stronger assumption.

\begin{definition}\label{def:silenceintro}
  Let $(M,g)$ be a spacetime with a crushing singularity. Let $\theta$ be the mean curvature of the leaves of the corresponding foliation and assume
  $\theta$ to always be strictly positive. Let $\hg:=\theta^{2}g$
  \index{$\a$Aa@Notation!Metrics!$\hg$}%
  \index{$\a$Aa@Notation!Tensor fields!$\hg$}%
  and let $\chK$ be the Weingarten map of the leaves of the foliation with respect to $\hg$.
  \index{Weingarten map!Conformally rescaled metric}%
  \index{$\a$Aa@Notation!Tensor fields!$\chK$}%
  If there is a constant $\e_{\Spe}>0$ such that
  \index{$\a$Aa@Notation!Constants!$\e_{\Spe}$}%
  \begin{equation}\label{eq:chKeSpeintro}
    \chK\leq -\e_{\Spe}\Id
  \end{equation}
  on $M$, then $\chK$ is said to satisfy a \textit{silent upper bound} on $M$.
\end{definition}
\begin{remark}\label{remark:interpretationnegativedefinite}
  The inequality (\ref{eq:chKeSpeintro}) should be interpreted as saying that
  \[
  \bge(\chK v,v)\leq -\e_{\Spe}\bge(v,v)
  \]
  for all tangent vectors $v$ to the leaves of the foliation. Here $\bge$ is the metric induced on the leaves of the foliation by $g$.
\end{remark}

\begin{example}\label{example:chKKasnersolutions}
  In the case of the Kasner solutions introduced in Example~\ref{example:Kasnersolutions}, $\chK$ takes the form
  \[
  \chK^{i}_{\phantom{i}j}=(p_{i}-1)\de^{i}_{j}
  \]
  (no summation on $i$), where we calculate the components of $\chK$ using the frame $\{\d_{i}\}$ and its dual. Note, in particular, that
  for all Kasner solutions except the flat ones, $\chK$ satisfies a silent upper bound on $M_{\roK}$. The above calculation is also valid  
  for Bianchi type I stiff fluids; cf. Example~\ref{example:BianchiTypeIStiffFluids}. In case the fluid is non-vanishing, it follows that
  $p_{\phi}\neq 0$ and that $p_{i}<1$ for all $i$, with the consequence that $\chK$ satisfies a silent upper bound on $M_{\roQ}$.
\end{example}

\section{Conjectures and results concerning big bang singularities}\label{section:conjecturesandresults}

In these notes, we develop a framework for analysing anisotropic big bang singularities. For this framework to be of interest, it, of course, has to be
consistent with large classes of solutions whose asymptotics are understood. In the present section, we therefore first formulate a general conjecture
concerning big bang singularities and then give an overview of known results. The organising principle in our description of the results is that of a
symmetry hierarchy. However, the interested reader is also referred to, e.g., \cite{LUW} for a discussion of more specific solutions illustrating silence
breaking, different types of curvature dominance etc. 

\subsection{The BKL conjecture}\label{subsection:BKLConjecture}
In the physics literature, the dominant conjecture concerning the generic behaviour in the direction of the singularity is due to Belinski\v{\i}, Khalatnikov
and Lifschitz (BKL); cf. \cite{bkl1} and \cite{bkl2}, as well as, e.g., \cite{dhn,dah,HUL,huar} for recent refinements. The idea of the corresponding
\textit{BKL conjecture}
\index{BKL!Conjecture}%
\index{Conjecture!BKL}%
is that the singularity should be spacelike, in the sense that there is silence asymptotically, and oscillatory. Moreover, the matter
content should not play a role asymptotically, so that it is sufficient to focus on vacuum solutions. More specifically, for an appropriately chosen
foliation of the spacetime, the simplified equations obtained by dropping the spatial derivatives in the original equations should yield a good model of the
asymptotic behaviour. Dropping the spatial derivatives, one is left with a system of ODE's for each spatial point. According to the BKL picture, the relevant
ODE's are the equations for the spatially homogeneous vacuum solutions with the maximal number of degrees of freedom; i.e., vacuum Bianchi type VIII, IX or
VI${}_{-1/9}$ solutions.  Finally, the asymptotic behaviour of solutions to the model ODE's is oscillatory and described by the Kasner map
\index{Kasner!Map}%
(essentially a
chaotic billiard); cf. Figure~\ref{fig:TheKasnerMap} for an illustration. The BKL picture is conjectured to be valid for Einstein's equations coupled to
large families of matter sources in $3+1$-dimensions. However, in the presence of a scalar field or a stiff fluid, e.g., the matter
plays a role asymptotically, the model ODE's are different, and instead of being well approximated by the Kasner map, the asymptotics are quiescent. In
higher dimensions, the picture is also different. The statements are in many ways quite vague, and the BKL perspective should not be thought of as a
mathematical conjecture. However, it is a very useful perspective to have in mind when studying solutions. Spikes is a phenomenon which is not discussed
by BKL, but which has turned out to be important in numerical and analytical studies. The spikes were first discovered numerically, see
\cite{BM}, and were later constructed analytically; cf. \cite{raw} and \cite{lim}. See also \cite{HUL} for a generalisation of the BKL picture involving
spikes and spike oscillations.

\subsection{Spatially homogeneous solutions}\label{ssection:sphomsol}
\begin{figure}
  \begin{center}
    \includegraphics{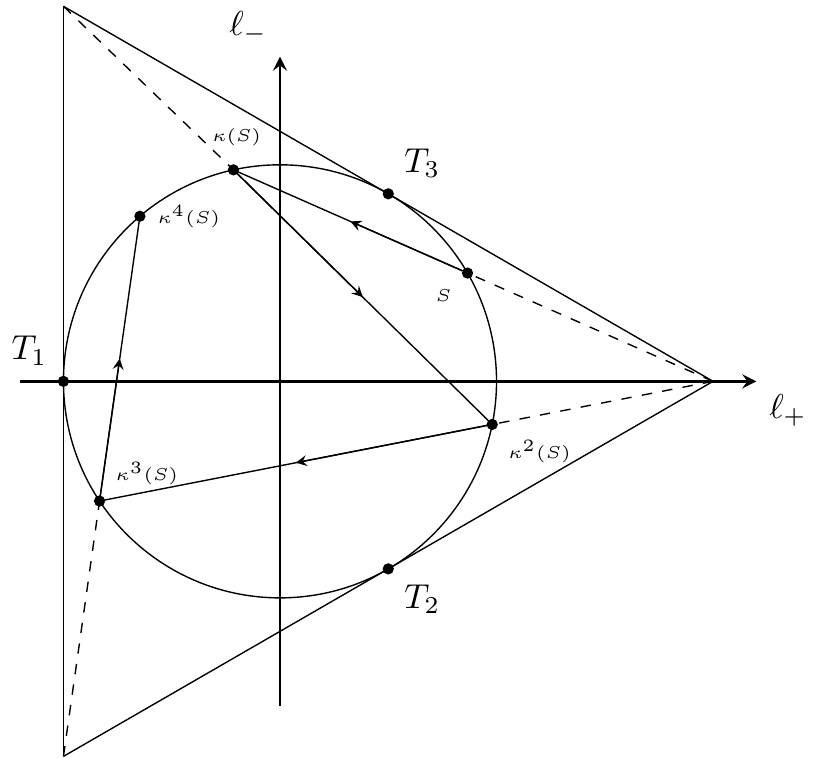}    
  \end{center}
  \caption{The Kasner map, here denoted $\kappa$, is a map from the Kasner circle to itself. Given a point $S$ on the circle, $\kappa(S)$ is obtained by
    taking the nearest corner of the triangle, drawing a straight line from this corner to $S$, and then continuing this straight line to the next
    intersection with the circle. This next intersection defines $\kappa(S)$. Above we illustrate four iterations of the map. That the dynamics associated
    with the Kasner map are chaotic follows from the fact that the Kasner map is topologically conjugate to the map $\theta\mapsto -2\theta$ on
  $\ro/\zo$; cf. \cite[Section~8, p.~22]{beguin}.}\label{fig:TheKasnerMap}
\end{figure}
Due to the central role spatially homogeneous solutions
\index{Solutions!Spatially homogeneous}%
\index{Symmetry!Spatial homogeneity}%
play in the BKL conjecture, it is of importance to analyse their asymptotics. These solutions are
classified as being of Bianchi class A,
\index{Bianchi!Class A}%
Bianchi class B
\index{Bianchi!Class B}%
or Kantowski-Sachs type.
\index{Kantowski-Sachs}%
The Bianchi class A (B) solutions are the MGHD's of left invariant initial
data on $3$-dimensional unimodular (non-unimodular) Lie groups; and the Kantowski-Sachs solutions are the MGHD's of initial data invariant under the isometry
group of the standard metric on $\mathbb{S}^{2}\times\mathbb{R}$. The Bianchi A and B classes are further divided into types according to a classification of
the corresponding Lie algebras. Since the Kantowski-Sachs solutions typically exhibit simpler dynamics, it is natural to focus on Bianchi class A and B. In
\cite{eam}, the authors develop a general perspective on the Bianchi class A and B setting. Building on these ideas, scale invariant versions of the
equations (for all Bianchi types except VI${}_{-1/9}$) are developed in \cite{wah,haw}. The importance of developing a scale invariant perspective is
due to the fact that the mean curvature (and many other geometric quantities) diverge in the direction of the singularity. However, using the mean
curvature to extract a scale and to change the time coordinate leads to a dynamical system with a state space which is either compact or such that the
solution is asymptotically contained in a compact subset of the state space. Moreover, extracting a scale yields a clearer picture of the dynamics.

\textbf{Mechanisms causing oscillatory and quiescent asymptotics.} Turning to results, it is convenient to classify them according to whether the
asymptotics are quiescent or oscillatory; cf. Definition~\ref{def:oscandquiescent}. In the companion article \cite{RinGeometry}, we provide a 
systematic way to predict whether the asymptotics will be quiescent or oscillatory (in the vacuum and scalar field settings). However, for the purposes
of the present discussion, let us just note that there are two main aspects that influence the outcome. To begin with,
symmetry assumptions and particular matter models can suppress the oscillations. Moreover, certain matter models can also reactivate oscillations under
symmetry assumptions that would otherwise have suppressed them. Turning to specific examples, Bianchi type I vacuum solutions (i.e., the Kasner solutions,
cf. Figure~\ref{fig:Kasnercircle}) are clearly quiescent, contrary to the BKL expectation concerning generic vacuum solutions. However, in this case, the
oscillations are suppressed by the symmetry assumption that the initial data be invariant under left translations in the Lie group $\rn{n}$. Generic
Bianchi type VIII and IX vacuum spacetimes exhibit oscillatory behaviour; cf. \cite{cbu}. However, adding a non-vanishing stiff fluid eliminates the
oscillations; cf. \cite{BianchiIXattr}. In fact, in the case of Bianchi type VIII and IX stiff fluid spacetimes, $(\ell_{+},\ell_{-})$ converges to a
point in the interior of the shaded triangle in Figure~\ref{fig:QuiescentConvergentRegime}; cf. \cite[Theorem~19.1, p.~478]{BianchiIXattr}. Finally,
Bianchi type VI${}_{0}$ vacuum and generic orthogonal perfect fluid solutions with $\g\in (2/3,2)$ are quiescent; cf., e.g.,
\cite[Proposition~22.16, p. 239]{minbok} and \cite[Theorem~1.6, p.~3076]{hog}. However, magnetic Bianchi type VI${}_{0}$ solutions are oscillatory; cf.
\cite[Theorem, p.~426]{wea}.

\textbf{Results concerning spatially homogeneous solutions with quiescent asymptotics.} There is a vast literature of results in the spatially homogeneous
and quiescent setting and, as a consequence, it is not realistic to describe them all. Some examples can be found in
\cite{wah,haw,wae,BianchiIXattr,HU,RadermacherNonStiff,RadermacherStiff,hog}. These results include conclusions for all Bianchi types except VIII, IX and
VI${}_{-1/9}$
in the orthogonal perfect fluid settings. However, the exact restrictions on the equation of state differ between the references. Concerning the stiff
fluid setting, there are results for all Bianchi types except VI${}_{-1/9}$; cf. \cite{BianchiIXattr,RadermacherStiff}. Beyond being quiescent, spatially
homogeneous solutions with quiescent asymptotics typically have the property that all the expansion normalised variables parametrising the relevant state
space converge. Moreover, $\chK$ typically satisfies a silent upper bound asymptotically. 

\textbf{Results concerning spatially homogeneous solutions with oscillatory asymptotics.} As already mentioned, generic Bianchi type VIII and IX vacuum
spacetimes exhibit oscillatory asymptotics, and the same is true of magnetic Bianchi type VI${}_{0}$ solutions. That generic Bianchi type IX solutions (in
the orthogonal and non-stiff perfect fluid setting) converge to an attractor on which the dynamics are described by the Kasner map (cf.
Figure~\ref{fig:TheKasnerMap}) is demonstrated in \cite{BianchiIXattr}. Lebesgue generic Bianchi type VIII and IX vacuum solutions have silent asymptotics in
the sense that the spatial component of causal curves (with respect to the uniquely determined foliation by constant mean curvature hypersurfaces) converges
in the direction of the singularity. This is demonstrated in \cite{brehm}. Finally, one can specify orbits of the Kasner map and then prove that there
are stable manifolds of solutions to the full Bianchi type VIII and IX dynamics that shadow these orbits. In the case of periodic orbits, this is
demonstrated in \cite{lea}. In the case of aperiodic orbits that stay away from the special points (cf. Figure~\ref{fig:Kasnercircle}) this is demonstrated
in \cite{beguin}. In \cite{du}, Dutilleul proves that for Lebesgue almost every point $p$ of the Kasner circle, the heteroclinic chain $H$ starting at $p$
(i.e., the orbit of the Kasner map starting at $p$) is such that the union of all the type IX orbits shadowing $H$ contains a $3$-dimensional Lipschitz
immersed submanifold. Moreover, for every subset $E$ of the Kasner circle with positive $1$-dimensional Lebesgue measure, the union of all the type IX orbits
shadowing some heteroclinic chain starting at a point of $E$ has positive 4-dimensional Lebesgue measure. Concerning Bianchi type VI${}_{-1/9}$ solutions,
there is a qualitative description of the expected dynamics, cf. \cite{hhw}, but, to the best of our knowledge, no mathematical results. 

\subsection{$\tn{3}$-Gowdy symmetry}\label{subsection:t3gowdy}
Proceeding beyond spatial homogeneity, it is natural to consider Gowdy and $\mathbb{T}^{2}$-symmetry.
\index{Symmetry!Gowdy}%
\index{Symmetry!$\tn{2}$}%
\index{Gowdy!Symmetry}%
In these cases, there is a
$2$-dimensional isometry group, so that the equations are effectively a system of $1+1$-dimensional wave equations. In the vacuum Gowdy setting, the symmetry
is such that the oscillations are suppressed. However, this is not expected to be the case for general $\mathbb{T}^{2}$-symmetric solutions. In the
$\mathbb{T}^{3}$-Gowdy symmetric vacuum setting, there is an analysis of the asymptotics for generic initial data, as well as a proof of generic curvature
blow up (and, thereby, strong cosmic censorship); cf., e.g., \cite{asvelGowdy,SCCGowdy} and references cited therein. Even though the methods used in
\cite{asvelGowdy,SCCGowdy} cannot be expected to carry over to the general setting, the conclusions of the analysis do have important implications. In
order to formulate the conclusions, note that the metric can be assumed to take the form
\begin{equation}\label{eq:Gowdymetricareal}
  g=t^{-1/2}e^{\lambda/2}(-dt^{2}+d\vartheta^{2})+te^{P}(dx+Qdy)^{2}+te^{-P}dy^{2}
\end{equation}
\index{Metrics!Gowdy}%
\index{Gowdy!Metric}%
on $\tn{3}\times (0,\infty)$. Here the functions $P$, $Q$ and $\lambda$ only depend on $t$ and $\vartheta$, so that the metric is invariant under the
action of $\tn{2}$ corresponding to translations in $x$ and $y$. In what follows, it is also convenient to use the time coordinate $\tau=-\ln t$. With
this choice, the big bang singularity corresponds to $\tau\rightarrow\infty$.

Let $\g$ be a past inextendible causal curve. Then, due to the causal structure of the metric $g$ given by (\ref{eq:Gowdymetricareal}), the
$\vartheta$-component of $\g$ converges in the direction of the singularity. Denote the limit by $\vartheta_{0}$. Letting
$\kappa=P_{\tau}^{2}+e^{2P}Q_{\tau}^{2}$, it can then be demonstrated that $\kappa$ converges (in the direction of the singularity) uniformly in $J^{+}(\g)$ to
a limit. We denote this limit by $v_{\infty}^{2}(\vartheta_{0})$ and refer to the function $v_{\infty}\geq 0$ as the \textit{asymptotic velocity}.
\index{Asymptotic velocity}%
\index{Gowdy!Asymptotic velocity}%
A proof of these statements is provided in \cite{asvelGowdy}; cf. Subsection~\ref{subsection:aslimiteigenvalues} below for a more
detailed discussion and more detailed references. The eigenvalues, $\ell_{A}$, $A=1,2,3$, of $\mK$ can be calculated; cf.
Remark~\ref{remark:eigenvaluesintthreegowdy} below. The corresponding eigenvector fields $X_{A}$, $A=1,2,3$, can be chosen such that $X_{1}=\d_{\vartheta}$,
and $X_{A}=X_{A}^{x}\d_{x}+X_{A}^{y}\d_{y}$ for $A=2,3$, where $X_{A}^{x}$ and $X_{A}^{y}$ only depend on $t$ and $\vartheta$. Note, in particular, that
$[X_{2},X_{3}]=0$. Next, it can be demonstrated that the eigenvalues $\ell_{1}$, $\ell_{2}$ and $\ell_{3}$ converge uniformly to
\begin{equation}\label{eq:aseigenvTthreeGowdy}
  \frac{v_{\infty}^{2}(\vartheta_{0})-1}{v_{\infty}^{2}(\vartheta_{0})+3},\ \ \
  2\frac{1-v_{\infty}(\vartheta_{0})}{v_{\infty}^{2}(\vartheta_{0})+3},\ \ \
  2\frac{1+v_{\infty}(\vartheta_{0})}{v_{\infty}^{2}(\vartheta_{0})+3}
\end{equation}
respectively in $J^{+}(\g)$; cf. (\ref{eq:ellonelim})--(\ref{eq:ellthreelim}) below. Denoting the limits by $\ell_{i,\infty}(\vartheta_{0})$, it can be
verified that they satisfy the Kasner relations; cf. (\ref{eq:elliinfKasnerrelations}) below. It can also be verified that the deceleration parameter $q$
converges to $2$ uniformly in $J^{+}(\g)$; cf. Lemma~\ref{lemma:quniformbdGowdy} below. This means that the eigenvalues of $\chK$ converge uniformly to
\[
-\frac{4}{v_{\infty}^{2}(\vartheta_{0})+3},\ \ \
-\frac{[v_{\infty}(\vartheta_{0})-1]^{2}}{v_{\infty}^{2}(\vartheta_{0})+3},\ \ \
-\frac{[v_{\infty}(\vartheta_{0})+1]^{2}}{v_{\infty}^{2}(\vartheta_{0})+3}
\]
in $J^{+}(\g)$; cf. (\ref{eq:lambdaonelim})--(\ref{eq:lambdathreelim}) below. In particular, $\chK$ is asymptotically negative definite
unless $v_{\infty}(\vartheta_{0})=1$. That $v_{\infty}(\vartheta_{0})=1$ is, potentially, an obstruction to silence is illustrated by the fact that $P=\tau$,
$Q=0$ and $\lambda=-\tau$ is a solution to the $\tn{3}$-Gowdy symmetric vacuum equations. Moreover, this solution is a flat Kasner solution (which has a Cauchy
horizon).

\textbf{Generic solutions.}
\index{Gowdy!Generic solutions}%
\index{Generic!Gowdy solutions}%
The above observations hold for all $\tn{3}$-Gowdy symmetric vacuum solutions. However, some values of $v_{\infty}$ are not
stable under perturbations. In fact, generic solutions are such that $0<v_{\infty}<1$ for all but a finite number of points. Moreover, the exceptional
points are so-called non-degenerate true spikes, for which, in particular, $1<v_{\infty}<2$. These statements are justified in \cite{SCCGowdy};
cf. Section~\ref{section:t3Gowdy} and Subsection~\ref{ssection:nondegeneratetruespikes} below for a more detailed discussion and more detailed references.
In particular, it is clear that there is something special about the regime $0<v_{\infty}<1$. This can be understood from (\ref{eq:aseigenvTthreeGowdy}).
Due to (\ref{eq:aseigenvTthreeGowdy}), it is clear that $\ell_{1}$ is asymptotically negative and $\ell_{2}$, $\ell_{3}$ are asymptotically positive
if $0<v_{\infty}<1$. In particular, the one negative eigenvalue corresponds to an eigenvector field which is orthogonal to two commuting eigenvector fields.
Note that the fact that this combination is possible is due to the particular structure of $\tn{3}$-Gowdy symmetry. In the companion article
\cite{RinGeometry}, we, moreover, argue that this particular combination is related to the suppression of oscillations and the appearance of a convergent
regime (for $0<v_{\infty}<1$) in $\tn{3}$-Gowdy symmetric vacuum spacetimes. 

\textit{The low velocity regime.}
\index{Gowdy!Low velocity regime}%
\index{Low velocity regime, Gowdy}%
Consider a solution and a $\vartheta_{0}\in\so$ such that $0<v_{\infty}(\vartheta_{0})<1$. Then there is an open neighbourhood
$I$ containing $\vartheta_{0}$ such that the conditions of these notes are satisfied in $I$. In fact, $\mK$ converges exponentially in any $C^{k}$ norm
on $I$; $\hml_{U}\mK$ converges exponentially to zero with respect to any $C^{k}$-norm etc. The justification for these statements is given in
Subsections~\ref{ssection:convergentsettingGowdy} and \ref{subsection:falsespikes} below.

\textit{Non-degenerate true spikes.} Next, consider a non-degenerate true spike; cf. Subsection~\ref{ssection:nondegeneratetruespikes} below for a precise
definition of this notion. Given that $\vartheta_{0}$ corresponds to the tip of the spike, assume $\g$ to be a past inextendible causal curve such that
the $\vartheta$-component of $\g$ converges to $\vartheta_{0}$ in the direction of the singularity. Then, with respect to suitable local coordinates on
$\tn{3}$, all the components of $\mK$ but one converge in $J^{+}(\g)$ in the direction of the singularity. However, the remaining component tends to infinity.
Moreover, the eigenvector fields $X_{2}$ and $X_{3}$ asymptotically point in the same direction, even though the eigenvalues corresponding to $X_{2}$ and
$X_{3}$ converge to distinct values. In other words, the span of the limits of the eigenvector
fields $X_{2}$ and $X_{3}$ is a one dimensional subspace. This is clearly not the case when $0<v_{\infty}(\vartheta_{0})<1$, since $\mK$ converges and the
limits of the eigenvalues are distinct in that case. In other words, for a generic solution, the non-degenerate true spikes are characterised by the
property that the span of the limits of the eigenvector fields $X_{2}$ and $X_{3}$ is a one dimensional subspace. The above statments are justified in
Subsection~\ref{ssection:nondegeneratetruespikes}. 

\begin{figure}
  \includegraphics{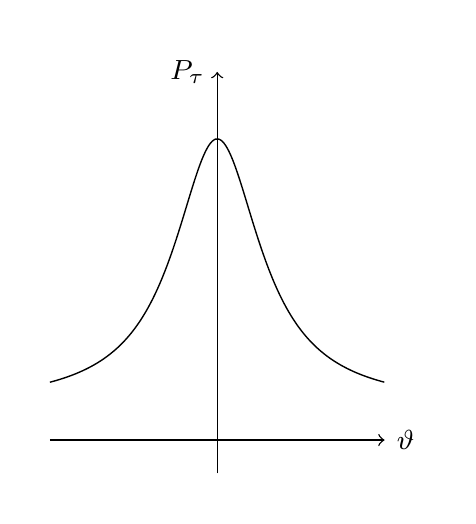}
  \includegraphics{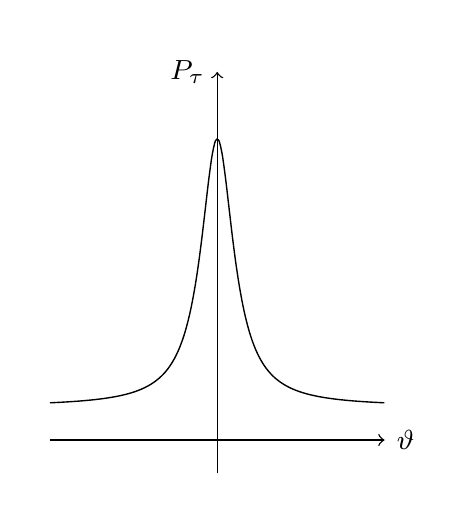}
  \includegraphics{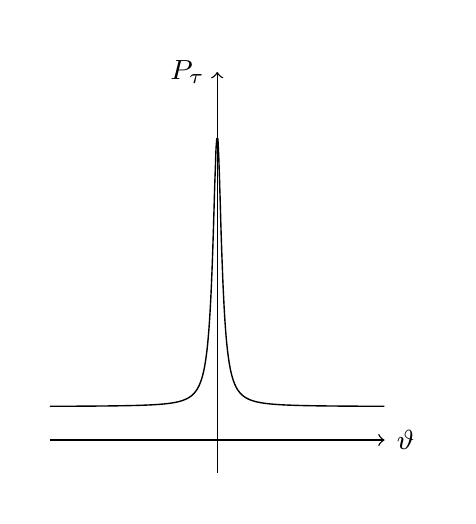}
  \caption{In a neighbourhood of a true spike, the asymptotic velocity is the limit of $P_{\tau}$. The plots are of $P_{\tau}$ at three different
    times. The limit, i.e. the asymptotic velocity, is discontinuous.}\label{fig:TrueSpikes}
\end{figure}

\textbf{Localisations.} An important lesson to be learnt from the study of $\tn{3}$-Gowdy symmetric spacetimes is that focussing on regions
of the form $J^{+}(\g)$ substantially simplifies the analysis. In order to justify this statement, it is useful to consider the spikes in greater
detail. Figure~\ref{fig:TrueSpikes} illustrates a non-degenerate true spike. Note, in particular, that the tip of the spike is a point of discontinuity
for $v_{\infty}$. 
If one abandons the requirement of non-degeneracy, there can be infinitely many spikes, and the corresponding asymptotic behaviour is very complicated. On
the other hand, following a causal curve, say $\gamma$, into the singularity, then intersecting the leaves of the natural foliation with $J^{+}(\gamma)$,
the spatial variation of, e.g., the eigenvalues of $\mK$, in the corresponding sets decays to zero in the direction of the singularity. And this is true
even if the spatial component of $\gamma$ converges to a point on the singularity which is an accumulation point of spikes. The important observation here
is that
\begin{itemize}
\item in order to prove, e.g., generic curvature blow up, it is sufficient to consider the behaviour of solutions along causal curves, 
\item in order to predict the behaviour of the solution along a causal curve going into the singularity, it is, from a PDE perspective, sufficient to
  control the behaviour in $J^{+}(\gamma)$,
\item the behaviour in $J^{+}(\gamma)$ is in general much less complicated; e.g., the eigenvalues of $\mK$ converge and their spatial variation dies out, 
\item considering larger regions that intersect the singularity in a subset containing an open set, the behaviour can be extremely complicated; there
  can be infinitely many spikes and infinitely many discontinuity points of the asymptotic velocity.
\end{itemize}
In short: it is sufficient to focus on sets of the form $J^{+}(\gamma)$, and considering the solution in larger regions in general takes the degree of
difficulty to a completely different level.

\subsection{Quiescent singularities}\label{ssection:quiescent}
In spite of the central role of the BKL proposal in cosmology, there is no construction of a spatially inhomogeneous solution with the properties stated in
the BKL conjecture. There is not even a construction of a spatially inhomogeneous solution with an oscillatory singularity. However, according to the BKL
proposal, the presence of a scalar field or a stiff fluid is expected to suppress the oscillations and produce a quiescent singularity. In addition, as
noted in \cite{Henneauxetal}, even for Einstein's vacuum equations, there are quiescent regimes in the case of $n+1$-dimensions for $n\geq 10$. Moreover,
as already discussed above, the presence of symmetries can suppress oscillations. 

\textbf{Specifying data on the singularity.}
\index{Specifying data on the singularity}%
The vacuum $\mathbb{T}^{3}$-Gowdy setting is the most general cosmological setting in which
the generic behaviour of solutions in the direction of the singularity has been analysed. There are Gowdy settings with different spatial
topologies ($\mathbb{S}^{3}$ and $\mathbb{S}^{1}\times\mathbb{S}^{2}$) as well as the so-called polarised $\mathbb{T}^{2}$-symmetric solutions, all of which
are expected to be quiescent and for which the asymptotics could potentially be analysed. However, due to the difficulty, the results going
beyond these classes largely consist of specifying data on the singularity. The idea here is to specify the asymptotic behaviour of solutions, and
then to prove that there are solutions with the prescribed asymptotics. This point of view is applied to the $\mathbb{T}^{3}$-Gowdy symmetric setting
in \cite{kar}, an article which generated substantial activity in the area; cf., e.g., \cite{iak,ren,aarendall,sta,iam,daetal}. Even though results of
this nature allow for the correct number of free functions, it is unclear how large a subset of regular initial data the constructed solutions correspond
to. In particular, it is unclear if they correspond to an open set. As mentioned before, in order to obtain quiescent behaviour in a situation without
symmetries, it is necessary to introduce matter (such as a scalar field or a stiff fluid), or to consider higher dimensions; e.g., the Einstein vacuum
equations in $n+1$ dimensions, where $n\geq 10$. In \cite{aarendall,daetal}, results are derived in these contexts in the class of real analytic solutions,
using Fuchsian techniques.
\index{Fuchsian techniques}%
Two more recent results on specifying data on the singularity are \cite{aeta,fal}.
The results of \cite{fal} (cf. also \cite{klinger}) are of particular importance, in that they apply to the Einstein vacuum equations in $3+1$-dimensions
in the absence of symmetries. In particular, the authors construct a class of solutions such that for each ``point on the singularity'', the asymptotics are
approximately those of a Kasner solution; cf. Example~\ref{example:Kasnersolutions}. This may seem to contradict the BKL proposal. However, in spite of the
fact that the solutions are not symmetric, they are still expected to be highly non-generic; cf. the companion article \cite{RinGeometry} for a discussion.
On the other hand, the results of \cite{fal} are in the $C^{\infty}$-setting.

\begin{figure}
  \begin{center}
    \includegraphics{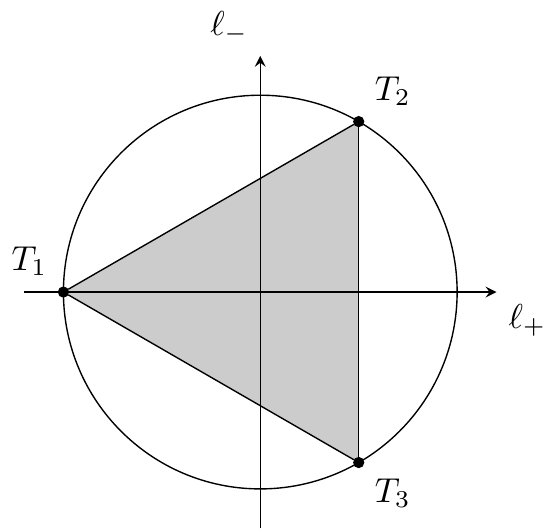}
  \end{center}
  \caption{The Kasner disc. The gray area indicates the subset in which stable big bang formation is expected in the stiff fluid/scalar field setting.
    Note that all Bianchi type VIII and IX stiff fluid spacetimes (with non-vanishing energy density) asymptotically converge to a point in the gray
    region; cf. \cite{BianchiIXattr}.}
  \label{fig:QuiescentConvergentRegime}
\end{figure}

In spite of the weaknesses described above, the results allowing the specification of data on the singularity are very important, in that they (in particular
\cite{aarendall,daetal}) indicate that there are regimes for which one could hope for stable big bang formation. In particular, in the $3+1$-dimensional
stiff fluid and scalar field setting, the initial data are, essentially, freely specifiable under the constraint that the pointwise asymptotic limits of
$(\ell_{+},\ell_{-})$ belong to the shaded region in Figure~\ref{fig:QuiescentConvergentRegime}. 

\textbf{Stable big bang formation.}
\index{Stable big bang formation}%
In \cite{rasql,rasq,rsh,specks3}, the authors accomplish an important breakthrough in the study of big bang
singularities. In particular, they demonstrate stable big bang formation in the case of stiff fluids, in the case of scalar fields, and in the case of
higher dimensions. One drawback is that the results only yield solutions that are close to isotropic or whose anisotropy has a definite bound which
excludes the full range of possibilities one would expect on the basis of \cite{aarendall,daetal}. In order to explain the discrepancy, consider first
the $3+1$-dimensional setting. Due to \cite{aarendall}, the expectation in the scalar field/stiff fluid setting is that stable big bang formation should be
obtained for $(\ell_{+},\ell_{-})$ belonging to the interior of the equilateral triangle with vertices given by the special points $T_{i}$, $i=1,2,3$,
introduced in Example~\ref{example:Kasnersolutions}; cf.
Figure~\ref{fig:QuiescentConvergentRegime}.
The results obtained in \cite{rasq} yield stable big bang formation in a neighbourhood of $\ell_{+}=\ell_{-}=0$. In that sense, there is a large region
missing for which one expects to be able to prove stable big bang formation. In \cite{rsh}, the authors prove stable big bang formation for Einstein's
vacuum equations in $n+1$-dimensions for $n\geq 38$. However, as noted above, $n+1$-dimensions with $n\geq 10$ should be enough. For these reasons, the
paper \cite{FRS} constitutes an important breakthrough. In \cite{FRS}, the authors prove stable big bang formation in the Einstein-scalar field setting
(for $n\geq 3$) and for the Einstein vacuum equations (for $n\geq 10$). Moreover, their results cover the full regimes expected on the basis of
\cite{aarendall,daetal}. In order to prove their results, the authors use Fermi-Walker transported frames. This gives a more geometric perspective
on the behaviour of solutions than the coordinate based analysis of \cite{rasql,rasq,rsh,specks3}. Related results are obtained in
\cite{AF}. In this paper, the authors demonstrate a stability result for the Schwarzschild singularity. The result concerns the interior of the
black hole for solutions to Einstein's vacuum equations with polarized axial symmetry.

One potential problem with the methodology used in \cite{rasql,rasq,rsh,specks3,FRS}, is that the gauge is non-local. As pointed out concerning the vacuum
$\mathbb{T}^{3}$-Gowdy setting, it can in general be expected to be of central importance to localise the analysis to sets of the form $J^{+}(\gamma)$ for a
causal curve $\gamma$ going into the singularity. In case the gauge is non-local, this might be problematic.

\chapter{Assumptions}\label{chapter:assumptions}

\section{Equations and basic terminology}

\textbf{Equation.} Many of the fundamental questions in general relativity can be phrased in terms of the asymptotic behaviour of solutions to Einstein's
equations. There are various ways of defining an asymptotic regime, but here we use a foliation. This is a somewhat non-geometric approach. However, given
information along a foliation, it is typically possible to draw geometric conclusions. In the present paper, we are interested in a toy problem associated
with the Einstein equations, namely that of analysing the asymptotic behaviour of solutions to systems of linear wave equations of the form
(\ref{eq:theequation}). 

\textbf{Induced metric and second fundamental form.} In these notes, we focus on spacetimes $(M,g)$ with a crushing singularity; cf.
Definition~\ref{def:crushingsingularity}. The justification for this is that for large classes of solutions with big bang singularities, such as the ones
discussed in Section~\ref{section:conjecturesandresults}, the singularity is crushing; cf. Appendix~\ref{chapter:examples} below. 
We use the notation $\bge$ and $\bk$
\index{$\a$Aa@Notation!Tensor fields!$\bge$}%
\index{$\a$Aa@Notation!Metrics!$\bge$}%
\index{$\a$Aa@Notation!Tensor fields!$\bk$}%
\index{Induced!Metric}%
\index{Induced!Second fundamental form}%
for the metric and second fundamental form induced on the leaves of the associated foliation. We think of
$\bge$ and $\bk$ as families of symmetric covariant $2$-tensor fields on $\bM$ (here and below we use the notation introduced in
Definition~\ref{def:crushingsingularity}). The \textit{mean curvature} is of particular interest, and we denote it
$\theta:=\tr_{\bge}\bk$.
\index{Induced!Mean curvature}%
\index{Mean curvature}%
\index{$\a$Aa@Notation!Functions!$\theta$}%
Next, the \textit{volume density} $\varphi$
\index{Volume density}%
\index{$\a$Aa@Notation!Functions!$\varphi$}%
is defined by the requirement that 
\begin{equation}\label{eq:varphidefitobmubgedp}
  \mu_{\bge}=\varphi\mu_{\bge_{\refer}}.
\end{equation}
Here $\mu_{\bge}$ and $\mu_{\bge_{\refer}}$
\index{$\a$Aa@Notation!Volume forms!$\mu_{\bge}$}%
\index{$\a$Aa@Notation!Volume forms!$\mu_{\bge_{\refer}}$}%
are the volume forms with respect to $\bge$ and $\bge_{\refer}$ respectively. Moreover, $\bge_{\refer}$
can be chosen to be any reference (Riemannian) metric on $\bM$. However, for the sake of convenience, we here assume $\bge_{\refer}$
\index{Reference metric $\bge_{\refer}$}%
\index{$\a$Aa@Notation!Metrics!$\bge_{\refer}$}%
\index{$\a$Aa@Notation!Tensor fields!$\bge_{\refer}$}%
to equal the metric induced on $\bM_{t_{0}}$ for some fixed reference time $t_{0}\in I$; this means that $\varphi(\bx,t_{0})=1$ for all $\bx\in\bM$. It is
also convenient to introduce the \textit{logarithmic volume density}:
\index{Logarithmic volume density}%
\index{$\a$Aa@Notation!Functions!$\varrho$}%
\begin{equation}\label{eq:varrhodefdp}
\varrho:=\ln\varphi. 
\end{equation}
In the case of a big bang singularity, it is natural to assume $\varphi$ to converge to zero as $t\rightarrow t_{-}$ (this is satisfied for
the spacetimes discussed in Section~\ref{section:conjecturesandresults}; cf. Appendix~\ref{chapter:examples} below). Then $\varrho\rightarrow -\infty$
as $t\rightarrow t_{-}$. Finally, we assume that $\theta>0$ on the entire foliation. Since we are interested in the asymptotic regime where
$\theta\rightarrow\infty$ uniformly, this is not a restriction; if it is not fulfilled, we can restrict $I$ in such a way that it is. 

\textbf{Terminology.} Sometimes, it is of interest to consider more general situations than the one discussed above. We then
use the following terminology. 
\begin{definition}\label{def:basicnotions}
  Let $(M,g)$ be a time oriented Lorentz manifold. A \textit{partial pointed foliation}
  \index{Partial pointed foliation}%
  \index{Foliation!Partial pointed}%
  of $(M,g)$
  \index{Partial pointed foliation!Spacetime $(M,g)$}%
  \index{$\a$Aa@Notation!Manifolds!$M$}%
  is a triple $\bM$,
  \index{Partial pointed foliation!Spatial slice $\bM$}%
  \index{$\a$Aa@Notation!Manifolds!$\bM$}%
  $I$
  \index{$\a$Aa@Notation!Intervals!$I$}%
  \index{Partial pointed foliation!Interval $I$}%
  and $t_{0}\in I$,
  \index{$\a$Aa@Notation!Points!$t_{0}$}%
  where $\bM$
  is a closed $n$-dimensional manifold; $I$ is an interval with left end point $t_{-}$ and right end point $t_{+}$; and there is an open interval $J$
  containing $I$ and a diffeomorphism from $\bM\times J$ to an open subset of $M$. Moreover, the hypersurfaces $\bM_{t}:=\bM\times \{t\}$ are required
  to be spacelike Cauchy hypersurfaces (in $(\bM\times J,g)$) and $\d_{t}$ is required to be future pointing timelike with respect to $g$ (where $\d_{t}$
  represents differentiation with respect to the variable on $I$). Given a partial pointed foliation,
  the \textit{associated induced metric, second fundamental form, mean curvature and future pointing unit normal}
  \index{Metrics!Partial pointed foliation}%
  \index{Partial pointed foliation!Induced metric}%
  \index{Second fundamental form!Partial pointed foliation}%
  \index{Partial pointed foliation!Induced seccond fundamental form}%
  are denoted $\bge$, $\bk$,
  \index{$\a$Aa@Notation!Tensor fields!$\bge$}%
  \index{$\a$Aa@Notation!Metrics!$\bge$}%
  \index{$\a$Aa@Notation!Tensor fields!$\bk$}%
  \index{Induced!Metric}%
  \index{Induced!Second fundamental form}%
  $\theta$
  \index{Mean curvature}%
  \index{Induced!Mean curvature}%
  \index{Partial pointed foliation!Induced mean curvature}%
  \index{$\a$Aa@Notation!Functions!$\theta$}%
  and $U$
  \index{Unit normal!Future pointing}%
  \index{Partial pointed foliation!Induced future pointing unit normal}%
  \index{$\a$Aa@Notation!Vector fields!$U$}%
  respectively;
  the \textit{associated Weingarten map}
  \index{Partial pointed foliation!Induced Weingarten map}%
  $\bK$
  \index{Weingarten map!Partial pointed foliation}%
  \index{$\a$Aa@Notation!Tensor fields!$\bK$}%
  is the family of $(1,1)$ tensor fields on $\bM$ obtained by raising one of the indices of $\bk$ with $\bge$;
  the \textit{associated reference metric}
  \index{Partial pointed foliation!Induced reference metric}%
  is the metric induced on $\bM_{t_{0}}$ by $g$ (it is denoted by $\bge_{\refer}$
  \index{$\a$Aa@Notation!Metrics!$\bge_{\refer}$}%
  \index{$\a$Aa@Notation!Tensor fields!$\bge_{\refer}$}%
  with associated Levi-Civita connection $\bD$);
  \index{Partial pointed foliation!Levi-Civita connection of induced reference metric}%
  \index{$\a$Aa@Notation!Levi-Civita connection!$\bD$}%
  and the \textit{volume density $\varphi$ and logarithmic volume density $\varrho$ associated with the partial pointed foliation} are defined by
  (\ref{eq:varphidefitobmubgedp}) and (\ref{eq:varrhodefdp}) respectively.

  An \textit{expanding partial pointed foliation}
  \index{Expanding partial pointed foliation}%
  \index{Folitation!Expanding partial pointed}%
  is a partial pointed foliation such that the mean curvature $\theta$ of the leaves of the foliation
  $\bM_{t}$, $t\in I$, is always strictly positive. Given an expanding partial pointed foliation,
  the \textit{associated expansion normalised Weingarten map}
  \index{Expanding partial pointed foliation!Induced expansion normalised Weingarten map}%
  \index{Weingarten map!Expansion normalised}%
  \index{$\a$Aa@Notation!Tensor fields!$\mK$}%
  $\mK$ is the family of $(1,1)$ tensor fields on $\bM$ given by $\mK:=\bK/\theta$;
  the \textit{associated conformal metric} is $\hg:=\theta^{2}g$;
  \index{Expanding partial pointed foliation!Induced conformal metric}%
  \index{$\a$Aa@Notation!Tensor fields!$\hg$}%
  \index{$\a$Aa@Notation!Metrics!$\hg$}%
  the \textit{associated induced conformal metric, second fundamental form, mean curvature and future pointing unit normal}
  \index{Metrics!Induced by conformal metric}%
  \index{Expanding partial pointed foliation!Metric induced by conformal metric}%
  \index{Second fundamental form!Conformal metric}%
  \index{Induced!Second fundamental form of conformal metric}%
  \index{Expanding partial pointed foliation!Induced second fundamental form of conformal metric}%
  \index{Mean curvature!Conformal metric}%
  \index{Induced!Mean curvature of conformal metric}%
  \index{Expanding partial pointed foliation!Induced mean curvature of conformal metric}%
  \index{Unit normal!Future pointing, conformal metric}%
  \index{Expanding partial pointed foliation!Future pointing unit normal induced by conformal metric}%
  are denoted $\chg$, $\chk$, $\chth$ and $\hU$
  \index{$\a$Aa@Notation!Metrics!$\chg$}%
  \index{$\a$Aa@Notation!Tensor fields!$\chg$}%
  \index{$\a$Aa@Notation!Tensor fields!$\chk$}%
  \index{$\a$Aa@Notation!Functions!$\chth$}%
  \index{$\a$Aa@Notation!Vector fields!$\hU$}%
  respectively, and they are the objects induced on the hypersurfaces $\bM_{t}$ by the conformal metric $\hg$;
  and the \textit{associated conformal Weingarten map}
  \index{Expanding partial pointed foliation!Conformal Weingarten map}%
  $\chK$
  \index{Weingarten map!Conformally rescaled metric}%
  \index{$\a$Aa@Notation!Tensor fields!$\chK$}%
  is the family of $(1,1)$ tensor fields on $\bM$ obtained by raising one of the indices of $\chk$ with $\chg$.
\end{definition}
\begin{remark}
  We consider the family $\bge$ of Riemannian metrics to be defined on $\bM$ (in other words, we identify $\bM_{t}$ and $\bM$). Similar
  comments apply to $\bk$, $\chg$ etc. We also consider $\bge_{\refer}$ to be defined on $\bM$. 
\end{remark}
\begin{remark}
Given a partial pointed foliation of a spacetime, we, in what follows, speak of $M$, $g$, $n$, $\bge$, $U$, $\bk$, $\theta$, $\bK$, $\bM$, $I$, 
$t_{\pm}$, $t_{0}$, $\bge_{\refer}$, $\bD$, $\varphi$ and $\varrho$ without further comment. Given an expanding partial pointed foliation, 
we, in addition, speak of $\hg$, $\chg$, $\hU$, $\chk$, $\chth$, $\mK$ and $\chK$ without further comment.
\end{remark}
\begin{remark}
The assumption that $\bM$ be closed is mainly for convenience. With slightly modified assumptions, the arguments presented below should also work
for non-compact $\bM$. The reason we do not assume $\bM\times I$ to be diffeomorphic to $M$ is that we wish to be able to use the results presented 
below in the context of a bootstrap argument. Then $I$ is an interval the size of which increases in the course of the argument. 
\end{remark}
It is of interest to relate $\bK$, $\mK$ and $\chK$. Note, to this end, that 
\begin{equation}\label{eq:chKmKthetarelation}
\chK=\theta^{-1}\bK+\hU(\ln\theta)\Id=\mK+\hU(\ln\theta)\Id.
\end{equation}
In particular, $\chK$, $\bK$ and $\mK$ have the same eigenspaces. On the other hand, the eigenvalues are quite different. 

\subsection{Deceleration parameter}

We are interested in situations where the mean curvature of the leaves of the foliation tends to infinity. We can therefore not impose bounds on
$\theta$. However, in many applications, $\hU(\ln\theta)$ is bounded. For that reason, it is of interest to introduce the notion of a
deceleration parameter, defined as follows.
\begin{definition}\label{def:decelerationpar}
  Let $(M,g)$ be a time oriented Lorentz manifold. Assume that it has an expanding partial pointed foliation. Then the \textit{deceleration parameter}
  \index{Deceleration parameter}%
  $q$
  \index{$\a$Aa@Notation!Functions!$q$}%
  is defined by
  \begin{equation}\label{eq:hUnlnthetamomqbas}
    \hU(n\ln\theta)=-1-q.
  \end{equation}
\end{definition}
\begin{remark}
  For an FLRW spacetime with scale factor $a(t)$, cf. (\ref{eq:RWmetric}), it can be computed that $q=-a\ddot{a}/\dot{a}^{2}$. In this sense, $q$ measures
  the deceleration. In more general situations, the Raychaudhuri equation can be used to compute $q$. Moreover, the Hamiltonian constraint can be used to
  draw conclusions concerning the boundedness of $q$; cf. \cite{RinGeometry} for further details. 
\end{remark}
For future reference, it is of interest to note that taking the trace of (\ref{eq:chKmKthetarelation}) yields
\begin{equation}\label{eq:chthexpressionhUlntheta}
  \chth=1+\hU(n\ln\theta)=-q,
\end{equation}
where we appealed to (\ref{eq:hUnlnthetamomqbas}) in the last step. 

\subsection{Lapse and shift}\label{ssection:lapseandshift}

Two important objects associated with a foliation are the lapse function and the shift vector field. They are defined as follows. 

\begin{definition}\label{def:lapseandshift}
  Let $(M,g)$ be a time oriented Lorentz manifold. Assume that it has an expanding partial pointed foliation. Then the \textit{lapse function}
  \index{Lapse function}%
  $N$
  \index{$\a$Aa@Notation!Functions!$N$}%
  and the
  \textit{shift vector field}
  \index{Shift vector field}%
  $\chi$
  \index{$\a$Aa@Notation!Vector fields!$\chi$}%
  associated with the foliation are defined by the condition that
  \begin{equation}\label{eq:dtNUchispl}
    \d_{t}=NU+\chi
  \end{equation}
  and the condition that $\chi$ is tangential to the constant $t$ hypersurfaces. In the case of $\hg$, the lapse function and shift vector field
  are defined by $\d_{t}=\hN\hU+\hat{\chi}$. In particular, $\hN=\theta N$
  \index{$\a$Aa@Notation!Functions!$\hN$}%
  and $\hat{\chi}=\chi$. 
\end{definition}
\begin{remark}
  Since $\d_{t}$ is future pointing timelike, $N$ is a strictly positive function. Moreover,
  \begin{equation}\label{eq:futurenormalNchiexpr}
    U=N^{-1}(\d_{t}-\chi).
  \end{equation}
\end{remark}
\begin{remark}
Since the shift vector field is the same for $g$ and $\hg$, we, from now on, only speak of $\chi$.
\end{remark}
In the process of constructing a spacetime via a foliation, it is necessary to make a choice of lapse and shift. They are defined, explicitly or 
implicitly, via gauge conditions. What gauge conditions are appropriate to impose depends on the situation. However, we are mainly interested in
situations in which the shift vector field is small. Note, in particular, that in all the examples discussed in Section~\ref{section:conjecturesandresults},
$\chi=0$. Moreover, except for the results concerning $\tn{3}$-Gowdy symmetric solutions and stable big bang formation, $N=1$. However, in the case of the
results on stable big bang formation, $N$ converges to $1$. 

\section{Basic assumptions}

To begin with, we make assumptions concerning the eigenvalues of $\mK$ and $\chK$. 

\subsection{Silence and non-degeneracy}

Two fundamental assumptions concerning the geometry are silence and non-degeneracy. They can be formulated purely in terms of $\mK$ and $\chK$,
and are the basis for drawing conclusions concerning the causal structure.
\begin{definition}\label{def:silenceandnondegeneracy}
  Let $(M,g)$ be a time oriented Lorentz manifold. Assume that it has an expanding partial pointed foliation. If there is a constant $\e_{\Spe}>0$
  \index{$\a$Aa@Notation!Constants!$\e_{\Spe}$}%
  such that
  \begin{equation}\label{eq:chKeSpe}
    \chK\leq -\e_{\Spe}\Id
  \end{equation}
  (i.e., if $\chK$ is negative definite) on $\bM\times I$, then $\chK$ is said to have a \textit{silent upper bound} on $I$.
  \index{Silent upper bound}%
  \index{Upper bound!Silent}%
  \index{Weingarten map!Conformally rescaled metric!Silent upper bound}%
  In what follows, $\e_{\Spe}$
  is assumed to satisfy $\e_{\Spe}\leq 2$. If the eigenvalues of $\mK$ are distinct and there is an $\e_{\rond}>0$ such that the distance between different
  eigenvalues is bounded from below by $\e_{\rond}$ on $I$, then $\mK$ is said to be \textit{non-degenerate} on $I$.
  \index{Non-degenerate!Expansion normalised Weingarten map}%
  \index{Weingarten map!Expansion normalised!Non-degenerate}%
\end{definition}
\begin{remark}
  Remark~\ref{remark:interpretationnegativedefinite} is equally relevant here. Note also that the inequality (\ref{eq:chKeSpe}) is equivalent to the
  statement that the eigenvalues of $\chK$ are bounded from above by $-\e_{\Spe}$. 
\end{remark}
\begin{remark}\label{remark:qlwbd}
  If (\ref{eq:chKeSpe}) holds, then $q\geq n\e_{\Spe}$, where $q$ is introduced in Definition~\ref{def:decelerationpar}; cf.
  (\ref{eq:chthexpressionhUlntheta}). 
\end{remark}

The quiescent examples discussed in Section~\ref{section:conjecturesandresults} are generally such that $\chK$ has
a silent upper bound; cf. Appendix~\ref{chapter:examples} below for a more detailed discussion. In the oscillatory setting, the situation is more
complicated. For large periods of time, an estimate such as (\ref{eq:chKeSpe}) holds. However, there will, at the very least, be short periods of
time during which this inequality is violated. Moreover, if the solution is such that its $\a$-limit set contains one of the special points on the
Kasner circle, then there will also be long periods of time during which the largest eigenvalue of $\chK$ is close to zero; cf.
Example~\ref{example:chKKasnersolutions}. Nevertheless, regions in which (\ref{eq:chKeSpe}) is satisfied are essential when analysing the asymptotics of
solutions.

Turning to the condition of non-degeneracy, one would expect it to be satisfied generically. However, there will be periods of time where it is
violated. In the oscillatory setting, the violations can mainly be expected to take place during short periods of time. However, in either case, if
there are violations during longer periods of time, the situation in some sense simplifies. The reason for this is that if two eigenvalues are roughly
equal, then there is no reason to distinguish the corresponding eigenspaces and it should (with, presumably, somewhat different methods) be possible
to treat the direct sum of the eigenspaces on the same footing as the eigenspaces of the distinct eigenvalues. 

\subsection{Frame}

In order to formulate the next assumptions, we need to introduce a frame on the constant $t$ hypersurfaces. 

\begin{definition}\label{def:XAellA}
  Let $(M,g)$ be a time oriented Lorentz manifold. Assume it to have an expanding partial pointed foliation and $\mK$ to be non-degenerate on
  $I$. By assumption, the eigenvalues, say $\ell_{1}<\cdots<\ell_{n}$,
  \index{$\a$Aa@Notation!Eigenvalues!$\ell_{A}$}%
  of $\mK$ are distinct. Locally, there is, for each $A\in \{1,\dots,n\}$
  an eigenvector $X_{A}$ of $\mK$ corresponding to $\ell_{A}$ such that
  \begin{equation}\label{eq:XAbgenormcond}
    |X_{A}|_{\bge_{\refer}}=1.
  \end{equation}
  If there is a global smooth frame with this property, say $\{X_{A}\}$,
  \index{$\a$Aa@Notation!Vector fields!$X_{A}$}%
  \index{$\a$Aa@Notation!Frames!$\{X_{A}\}$}%
  then $\mK$ is said to have a \textit{global frame}
  \index{Frame!Global}%
  \index{Global frame}%
  and $\{Y^{A}\}$
  \index{$\a$Aa@Notation!One form fields!$Y^{A}$}%
  \index{$\a$Aa@Notation!Coframes!$\{Y^{A}\}$}%
  denotes the frame dual to $\{X_{A}\}$.  
\end{definition}
\begin{remark}
  Since $\mK$ is smooth, the eigenvalues $\ell_{A}$ are smooth. 
\end{remark}
\begin{remark}\label{remark:existenceglobalframe}
  Note that, once we have fixed the $X_{A}$ at one point of $M$, they are uniquely determined in a neighbourhood by the conditions that $X_{A}$ be an
  eigenvector of $\mK$ corresponding to $\ell_{A}$; (\ref{eq:XAbgenormcond}); and the condition that the $X_{A}$ be smooth vector fields. On the other hand,
  there may be global topological obstructions to extending this local frame to a global one. Nevertheless, by taking a finite cover of $\bM$, if
  necessary, it can be ensured that there is a global frame; cf. Section~\ref{section:globalframefinitecover} below. The local geometry of this finite
  cover is of course identical to the original geometry. In other words, no geometric understanding is lost by going to the finite cover. Moreover, as
  will become clear, since we are
  interested in the silent setting, we can localise the analysis asymptotically, so that the issue of the existence of a global frame is, in practice,
  not a problem. For these reasons, we below restrict our attention to the case that $\mK$ has a global frame. In what follows, if $\mK$ is non-degenerate
  and has a global frame, we speak of $\{X_{A}\}$ and $\{Y^{A}\}$ without further comment. 
\end{remark}
\begin{remark}
  The assumptions imply that $\bM$ is parallelisable, which, in general, is a topological restriction. Note, however, that in the case of $n=3$, $\bM$
  is parallelisable as long as it is orientable; cf. \cite{bal} and references cited therein. Nevertheless, allowing degeneracy is, in general, of
  interest. However, degeneracy is in some respects
  associated with a higher degree of symmetry; e.g., all the eigenvalues coinciding corresponds to isotropy. Moreover, many of the complications in the
  analysis of the dynamics of cosmological solutions are associated with different rates of expansion in different spatial directions (which, in its turn,
  corresponds to non-degeneracy). If there is complete degeneracy (in the sense that all the eigenvalues are similar), different methods
  should be applicable (since there is no reason to distinguish the different spatial directions, due to the similar rates of expansion/contraction). If
  there is partial degeneracy in the sense that two or more eigenvalues are similar (or that there are pairs of similar eigenvalues etc.), it should be
  possible to divide the tangent space of $\bM$ into a finite sum of vector spaces (which are not necessarily one-dimensional), in which the eigenvalues
  are similar. The analysis in the present notes should suffice to analyse the distinct eigenspaces, and methods similar to those of, e.g., Rodnianski
  and Speck should suffice to analyse the behaviour in one of the vector spaces. Nevertheless, in order to obtain a clear picture of the geometry, we
  here insist on non-degeneracy. 
\end{remark}
\begin{remark}\label{remark:framenondegenerate}
  If all the assumptions of the definition are satisfied, there is a global orthonormal frame $\{E_{i}\}$
  \index{$\a$Aa@Notation!Vector fields!$E_{i}$}%
  \index{$\a$Aa@Notation!Frames!$\{E_{i}\}$}%
  on $\bM$ with respect to the metric
  $\bge_{\refer}$, with dual frame $\{\omega^{i}\}$.
  \index{$\a$Aa@Notation!One form fields!$\omega^{i}$}%
  \index{$\a$Aa@Notation!Coframes!$\{\omega^{i}\}$}%
\end{remark}

Given that the assumptions of the definition are satisfied, a standard argument implies that $\{X_{A}\}$ is an orthogonal frame with respect to $\bge$; cf. 
(\ref{eq:XAorthogonal}) below. This naturally leads to the following definition. 

\begin{definition}\label{def:muAbmuA}
  Let $(M,g)$ be a time oriented Lorentz manifold. Assume it to have an expanding partial pointed foliation and $\mK$ to be non-degenerate on
  $I$ and to have a global frame. Let the frame $\{X_{A}\}$ be given by Definition~\ref{def:XAellA}. Then $\mu_{A}$ and $\bmu_{A}$ are defined by
  \begin{align}
    \chg(X_{A},X_{A}) = & e^{2\mu_{A}},\label{eq:muAdef}\\
    \bge(X_{A},X_{A}) = & e^{2\bmu_{A}}.\label{eq:bgenormXA}
  \end{align}
  \index{$\a$Aa@Notation!Functions!$\mu_{A}$}%
  \index{$\a$Aa@Notation!Functions!$\bmu_{A}$}%
  In particular, $\mu_{A}=\bmu_{A}+\ln\theta$. 
\end{definition}

\subsection{Off-diagonal exponential decay/growth}

Most of our assumptions take the form of bounds. However, we need to impose additional conditions on the off-diagonal components of the expansion
normalised normal derivative of $\mK$. By the normal derivative of $\mK$, we here mean the Lie derivative of $\mK$ with respect to the future pointing
unit normal $U$, denoted $\ml_{U}\mK$, and the expansion normalised normal derivative of $\mK$ is defined by $\hml_{U}\mK:=\theta^{-1}\ml_{U}\mK$. 
However, it is not completely obvious how to define $\ml_{U}\mK$: $\mK$ is a family of $(1,1)$-tensor fields on $\bM$, and $\ml_{U}\mK$
should be an object of the same type. On the other hand, $U$ is clearly not tangential to $\bM$. The precise definition is straightforward
but somewhat lengthy. For that reason, we only provide it in Section~\ref{section:timederivativeofmK} below. If Einstein's equations are satisfied,
$\hml_{U}\mK$ can be calculated in terms of the stress energy tensor, $\mK$, the lapse function and the spatial geometry; cf. \cite{RinGeometry}. However,
we here do not assume Einstein's equations to be satisfied, and therefore we impose bounds directly on $\hml_{U}\mK$. 

\begin{definition}\label{def:offdiagonalexpdec}
  Let $(M,g)$ be a time oriented Lorentz manifold. Assume it to have an expanding partial pointed foliation and $\mK$ to be non-degenerate on
  $I$ and to have a global frame. Then $\hml_{U}\mK$ is said to satisfy an \textit{off-diagonal exponential bound}
  \index{Off-diagonal exponential bound}%
  if there are constants $C_{\mK,\mrod}>0$,
  \index{$\a$Aa@Notation!Constants!$C_{\mK,\mrod}$}%
  $G_{\mK,\mrod}>0$,
  \index{$\a$Aa@Notation!Constants!$G_{\mK,\mrod}$}%
  $M_{\mK,\mrod}>0$
  \index{$\a$Aa@Notation!Constants!$M_{\mK,\mrod}$}%
  and $0<\e_{\mK}\leq 2$
  \index{$\a$Aa@Notation!Constants!$\e_{\mK}$}%
  such that
  \begin{equation}\label{eq:hmlUhmlUsqmKoffdiagonalexpbd}
    |(\hml_{U}\mK)(Y^{A},X_{B})|\leq C_{\mK,\mrod}e^{\e_{\mK}\varrho}+G_{\mK,\mrod}e^{-\e_{\mK}\varrho}
  \end{equation}
  on $\bM\times I$ for all $A\neq B$, where
  \begin{equation}\label{eq:expgrowthwithupperbound}
    G_{\mK,\mrod}e^{-\e_{\mK}\varrho}\leq M_{\mK,\mrod}
  \end{equation}
  on $\bM\times I$. If there are constants $C_{\mK,\mrod}>0$, $G_{\mK,\mrod}>0$, $M_{\mK,\mrod}>0$ and $0<\e_{\mK}\leq 2$ such that
  (\ref{eq:hmlUhmlUsqmKoffdiagonalexpbd}) and (\ref{eq:expgrowthwithupperbound}) hold on $\bM\times I$ for all $A,B$ such that $A\neq B$ and $B>1$,
  then $\hml_{U}\mK$ is said to satisfy a \textit{weak off-diagonal exponential bound}.
  \index{Off-diagonal exponential bound!Weak}%
  \index{Weak off-diagonal exponential bound}%
\end{definition}
\begin{remark}
  We have ordered the eigenvalues of $\mK$ so that $\ell_{1}<\cdots<\ell_{n}$. For this reason, the order of $A$ and $B$ in
  (\ref{eq:hmlUhmlUsqmKoffdiagonalexpbd}) is potentially important. In fact, it turns out that the condition (\ref{eq:hmlUhmlUsqmKoffdiagonalexpbd})
  is much stronger if $A>B$ than if $A<B$. Moreover, the estimate
  (\ref{eq:hmlUhmlUsqmKoffdiagonalexpbd}) can, under quite general circumstances, be improved in the case that $A<B$; cf.
  Proposition~\ref{prop:strengthenassumpAltB} below. For this reason, it is of interest to note that we here only assume that
  the estimates (\ref{eq:hmlUhmlUsqmKoffdiagonalexpbd}) and (\ref{eq:expgrowthwithupperbound}) hold in the case that $B>1$; cf., e.g.,
  Lemma~\ref{lemma:lowerbdonmumin}, Corollary~\ref{cor:expdecmWABmWAAetc} and Proposition~\ref{prop:strengthenassumpAltB} below. Note also
  that in the case of $3+1$-dimensions, the only $A,B$ satisfying $B>1$ and $A>B$ are $A=3$ and $B=2$. The only condition that cannot be improved
  by appealing to Proposition~\ref{prop:strengthenassumpAltB} is thus when $A=3$ and $B=2$. However, if we impose Einstein's equations, and make
  suitable assumptions concerning the matter, the estimate for this remaining component can also, a posteriori, be improved; cf.
  \cite[Corollary~52]{RinGeometry}. 
\end{remark}
\begin{remark}
  The conditions are only imposed for $A\neq B$. As an illustration of the importance of this observation,
  note that Bianchi type VIII and IX vacuum spacetimes are such that there is a time independent frame with respect to which $\mK$ is diagonal.
  Thus, in that case, the left hand side of (\ref{eq:hmlUhmlUsqmKoffdiagonalexpbd}) vanishes identically for all $A\neq B$. In this respect,
  (\ref{eq:hmlUhmlUsqmKoffdiagonalexpbd}) is consistent with an oscillatory singularity. Note also that, for generic Bianchi type
  VIII and IX vacuum spacetimes, $(\hml_{U}\mK)(Y^{A},X_{A})$ (no summation on $A$) does not converge to zero in the direction of the singularity. 
\end{remark}
\begin{remark}
  The estimates (\ref{eq:hmlUhmlUsqmKoffdiagonalexpbd}) and (\ref{eq:expgrowthwithupperbound}) may seem like a curious combination of conditions.
  However, there are two reasons to impose them. First, if you consider oscillatory spatially homogeneous solutions, then there are typically
  exponentially decaying terms and exponentially growing terms. On the other hand, the exponentially growing terms are typically bounded.
  This combination is captured by (\ref{eq:hmlUhmlUsqmKoffdiagonalexpbd}) and (\ref{eq:expgrowthwithupperbound}). Second, integrating a non-negative
  function $f$ over an interval $[a,b]$ on which $f(t)\leq Ce^{\e t}\leq M$ yields an estimate
  \[
  \int_{a}^{b}f(t)dt\leq \e^{-1}M.
  \]
  In particular, there is a bound on the integral which is independent of the length of the interval, a property which is very useful when
  deriving estimates. 
\end{remark}

Returning to the results discussed in Section~\ref{section:conjecturesandresults}, note that, generally speaking, quiescent singularities are
such that $\hml_{U}\mK$ decays to zero exponentially (in $\varrho$); cf. Appendix~\ref{chapter:examples} below for a more precise statement and a
justification. In particular, the off-diagonal components converge to zero exponentially. In the case of Bianchi type VIII and IX orthogonal perfect
fluids, the off-diagonal components vanish identically. 

\subsection{Weighted norms and assumptions concerning the expansion normalised Weingarten map}

A remarkable feature of many, if not all, of the big bang singularities for which the asymptotics are understood is that $\mK$ is bounded with respect to a
fixed metric on $\bM$; cf. Appendix~\ref{chapter:examples} below for a more detailed discussion. Since this is the case, it is of interest to analyse what
conclusions can be drawn from the assumption that this bound holds. In some respects, this is the main motivation for writing these notes. In order to
obtain conclusions concerning, e.g., solutions to partial differential equations, it is, however, not sufficient to only assume bounds on $\mK$. We also need
to impose bounds on its derivatives. For many singularities, the derivatives of $\mK$ are bounded; cf. Appendix~\ref{chapter:examples} below. In fact,
in the case of quiescent singularities, $\mK$ typically converges exponentially. For the spatially homogeneous and oscillatory spacetimes, $\mK$ does not
converge, but it and its derivatives are bounded. However, in the case of non-degenerate true spikes in $\tn{3}$-Gowdy symmetric vacuum solutions,
$\mK$ is not bounded; cf. Subsection~\ref{ssection:nondegeneratetruespikes} below. On the other hand, a generic $\tn{3}$-Gowdy symmetric vacuum solution only
has a finite number of non-degenerate true spikes, and for every other point on the singularity, there is an open neighbourhood thereof such that
$\mK$ converges exponentially in any $C^{k}$-norm in that neighbourhood; cf. Section~\ref{section:t3Gowdy} below.

Here, we are going to impose bounds with respect to weighted Sobolev and
$C^{k}$-norms. The bounds are consistent with the derivatives of $\mK$ growing polynomially in $\varrho$, but not exponentially. However, that is not to say
that the methods developed in these notes are not useful in the latter context. On the other hand,
if we allow a faster rate of blow up of the spatial derivatives, we expect it to be necessary to impose more detailed assumptions concerning the eigenvalues
$\ell_{A}$, in fact to relate the rate of blow up of derivatives in specific directions with corresponding eigenvalues $\ell_{A}$. In short: in order to analyse
this situation, we expect it to be necessary to make very specific and interconnected assumptions concerning the eigenvalues and the rate of blow up. Here we
wish to avoid doing so. We therefore make stronger assumptions concerning the bounds on $\mK$. 

We also need to impose bounds on $\hml_{U}\mK$. We do not assume $\hml_{U}\mK$ to be bounded with respect to a fixed metric, but we assume it not to blow up
faster than polynomially in $\varrho$. We also impose weighted Sobolev and $C^{k}$-bounds. In the quiescent setting, such bounds are satisfied with a margin
since $\hml_{U}\mK$ typically converges to zero exponentially in this setting; cf. Appendix~\ref{chapter:examples} below. In the spatially homogeneous
orthogonal perfect fluid setting (including the oscillatory Bianchi
type VIII and IX solutions), $\hml_{U}\mK$ and its spatial derivatives are bounded but do not, in general, converge to zero. In the $\tn{3}$-Gowdy symmetric
setting, the spikes can be expected to cause complications.

As noted above, in the context of Einstein's equations, $\hml_{U}\mK$ can be calculated in terms of the stress energy tensor, $\mK$, the lapse function and the
spatial geometry. However, since we do not assume Einstein's equations to be satisfied here, we impose conditions on $\hml_{U}\mK$ directly. 

In order to define the weighted Sobolev and $C^{k}$-norms used to phrase the assumptions, let
\[
\Weight:=\{(\weight_{a},\weight_{b})\in\rn{2}:\weight_{a}\geq 0, \weight_{b}\geq 0\}.
\]
\index{$\a$Aa@Notation!Sets!$\Weight$}%
Let, moreover, 
\[
\Index:=\{(l_{0},l_{1})\in\zn{2}:0\leq l_{0}\leq l_{1}\}.
\]
\index{$\a$Aa@Notation!Sets!$\Index$}%
Then, if $(\weight_{a},\weight_{b})=\weight\in\Weight$, $(l_{0},l_{1})=\bfl\in\Index$ and $\mt$ is a family of tensor fields on $\bM$ for $t\in I$, 
\begin{align}
\|\mt(\cdot,t)\|_{C^{\bfl}_{\weight}(\bM)} := & 
\textstyle{\sup}_{\bx\in\bM}\left(\textstyle{\sum}_{j=l_{0}}^{l_{1}}\ldr{\varrho(\bx,t)}^{-2\weight_{a}-2j\weight_{b}}
|\bD^{j}\mt(\bx,t)|_{\bge_{\refer}}^{2}\right)^{1/2},\label{eq:mtClbS}\\
\|\mt(\cdot,t)\|_{H^{\bfl}_{\weight}(\bM)} := & 
\left(\int_{\bM}\textstyle{\sum}_{j=l_{0}}^{l_{1}}\ldr{\varrho(\cdot,t)}^{-2\weight_{a}-2j\weight_{b}}|\bD^{j}\mt(\cdot,t)|_{\bge_{\refer}}^{2}\mu_{\bge_{\refer}}\right)^{1/2}.
\label{eq:mtHlbS}
\end{align}
\index{$\a$Aa@Notation!Norms!$C^{\bfl}_{\weight}(\bM)$}%
\index{$\a$Aa@Notation!Norms!$H^{\bfl}_{\weight}(\bM)$}%
Here $\ldr{\xi}:=(1+|\xi|^{2})^{1/2}$.
\index{$\a$Aa@Notation!Functions!$\ldr{\xi}$}%
In case $\weight=0$, we write $C^{\bfl}(\bM)$ and $H^{\bfl}(\bM)$ for the spaces and correspondingly for the norms. In case $\bfl=(0,l)$, then we replace 
$\bfl$ with $l$ (in practice, this will be signalled by the fact that the superscript is not in boldface) in the names of the spaces and the notation for
the norms. Note that the norms are calculated with respect to the time independent Riemannian reference metric $\bge_{\refer}$, and not with respect to the 
induced metric $\bge$. 

\begin{remark}
In order to justify the above, somewhat cumbersome, notation, note that we wish $\mK$ to be bounded. For the norms of $\mK$, it is therefore natural to 
assume that there is no weight in front of the zeroth order term in the sum on the right hand sides of (\ref{eq:mtClbS}) and (\ref{eq:mtHlbS}). For other 
tensor fields, it might be natural to include a weight also in front of the zeroth order term. The reason for introducing the terminology $\Index$ is that 
in the case of, e.g., $\theta$, we wish to impose conditions on the derivatives of $\ln\theta$, but not on the $C^{0}$- or $L^{2}$-norm of $\ln\theta$. 
\end{remark}
\begin{remark}
Throughout these notes, we assume that there is a constant $\mKsup$ such that 
\begin{equation}\label{eq:mKsupbasest}
\|\mK(\cdot,t)\|_{C^{0}(\bM)}\leq \mKsup
\end{equation}
\index{$\a$Aa@Notation!Constants!$\mKsup$}%
for all $t\in I_{-}$, where
\begin{equation}\label{eq:Iminusdef}
  I_{-}:=\{t\in I:t\leq t_{0}\}.
\end{equation}
\index{$\a$Aa@Notation!Intervals!$I_{-}$}%
\end{remark}
\begin{remark}
We are mainly interested in imposing conditions on the Sobolev norms of $\mK$ and its normal derivative. However, the assumptions yielding the basic 
conclusions concerning the geometry are most naturally formulated using lower order supremum norms. It is of course also possible to deduce estimates
for the supremum norms using Sobolev embedding. 
\end{remark}

\subsection{Assumptions concerning the mean curvature}

We are interested in situations where the mean curvature of the leaves of the foliation tends to infinity. We can therefore not impose boundedness
conditions on $\theta$. However, in the case of many big bang singularities, the deceleration parameter $q$ introduced in
Definition~\ref{def:decelerationpar} is bounded. For example, the $3+1$-dimensional quiescent singularities discussed in
Section~\ref{section:conjecturesandresults} are typically such that $q$ converges to $2$ exponentially; cf. Appendix~\ref{chapter:examples} below.
In the case of the oscillatory and spatially homogeneous solutions discussed in Section~\ref{section:conjecturesandresults}, $q$ and its derivatives
are bounded, but $q$ does not converge. For these reasons, it is natural to impose bounds on $q$, and we do so in what follows. We also need to impose
bounds on the relative spatial variation of the mean curvature. In order to develop a feeling for what bounds are natural to impose, note that we are
here interested in singularities such that the mean curvature tends to infinity
in a synchronised way. In other words, if $t_{-}$ represents the singularity, then, for all $\bx\in\bM$, $\theta(\bx,t)\rightarrow\infty$ as 
$t\rightarrow t_{-}$. Combining this assumption with weighted bounds on $q$ and $\ln N$, and assuming that $\chi=0$, we deduce that weighted norms 
of $\bD\ln\theta$ are bounded; cf. Section~\ref{section:syncblowupmc} below for a more detailed justification. For this reason, we typically demand
that weighted norms of $\bD\ln\theta$ are bounded. Note also that most of the examples mentioned in Section~\ref{section:conjecturesandresults} are
such that $\theta$ is constant over the leaves of the foliation or such that the relative spatial variation decays in the direction of the singularity.
However, the $\tn{3}$-Gowdy setting is somewhat different; cf. Section~\ref{section:t3Gowdy} below. 

\begin{remark}
  In what follows, we always assume that there is a constant $\bDlnhNsup$ such that
  \begin{equation}\label{eq:bDlnNbDlnthetabd}
    |\bD \ln \hN|_{\bge_{\refer}} \leq  \bDlnhNsup
  \end{equation}
  \index{$\a$Aa@Notation!Constants!$\bDlnhNsup$}%
  on $\bM\times I_{-}$. 
\end{remark}

\subsection{Assumptions concerning the lapse function and the shift vector field}\label{ssection:asslash}

The conditions on the lapse function are imposed implicitly, since we impose weighted bounds on derivatives of $\ln\hN$ and $\ln\theta$. However,
they are analogous to those imposed on $\theta$. For reference, note that $N=1$ or $N\rightarrow 1$ for most of the examples in
Section~\ref{section:conjecturesandresults}. 

Turning to the shift vector field, we assume $\chi$ to be small. In order to develop a feeling for which norms are appropriate to use concerning
$\chi$, note that (\ref{eq:dtNUchispl}) implies that
\[
g(\d_{t},\d_{t})=-N^{2}+|\chi|_{\bge}^{2}. 
\]
Here, we are interested in foliations such that $\d_{t}$ is timelike; i.e., such that $N^{-1}|\chi|_{\bge}<1$. In what follows, we therefore assume that
\begin{equation}\label{eq:dttimelikecondition}
\frac{1}{N}|\chi|_{\bge}\leq \frac{1}{2}.
\end{equation}
This inequality ensures that $\d_{t}$ is timelike, with a margin. We also need to impose conditions on derivatives of $\chi$. However, we 
wish to measure the size of the derivatives with respect to a fixed metric, in analogy with the conditions imposed on $\mK$. To this end, we 
introduce the following hybrid measure: if $\xi$ is a vector field on $M$ which is tangential to the leaves of the foliation, let 
\begin{equation}\label{eq:bDkchirohydef}
|\bD^{k}\xi|_{\rohy}:=N^{-1}\left(\bge_{\refer}^{i_{1}j_{1}}\cdots \bge_{\refer}^{i_{k}j_{k}}\bge_{lm}\bD_{i_{1}}\cdots\bD_{i_{k}}\xi^{l}
\bD_{j_{1}}\cdots\bD_{j_{k}}\xi^{m}\right)^{1/2}. 
\end{equation}
With this notation, the inequality (\ref{eq:dttimelikecondition}) can be written $|\chi|_{\rohy}\leq 1/2$. Given 
$(\weight_{a},\weight_{b})=\weight\in\Weight$ and $(l_{0},l_{1})=\bfl\in\Index$, it is also convenient to introduce the notation
\begin{align}
\|\xi(\cdot,t)\|_{H^{\bfl,\weight}_{\rohy}(\bM)} := & \left(\int_{\bM}\textstyle{\sum}_{k=l_{0}}^{l_{1}}
\ldr{\varrho(\cdot,t)}^{-2\weight_{a}-2k\weight_{b}}|\bD^{k}\xi(\cdot,t)|_{\rohy}^{2}\mu_{\bge_{\refer}}\right)^{1/2},
\label{eq:Hlrohydef}\\
\|\xi(\cdot,t)\|_{C^{\bfl,\weight}_{\rohy}(\bM)} := & \sup_{\bx\in\bM}\textstyle{\sum}_{k=l_{0}}^{l_{1}}
\ldr{\varrho(\bx,t)}^{-\weight_{a}-k\weight_{b}}|\bD^{k}\xi(\bx,t)|_{\rohy}. \label{eq:Clrohydef}
\end{align}
\index{$\a$Aa@Notation!Norms!$C^{\bfl,\weight}_{\rohy}(\bM)$}%
\index{$\a$Aa@Notation!Norms!$H^{\bfl,\weight}_{\rohy}(\bM)$}%
In case $\bfl=(0,l)$, we replace $\bfl$ with $l$ (in practice, this will be signalled by the fact that the superscript is not in boldface) in the
names of the spaces and the notation for the norms. In case $\weight=0$, we also use the notation $H^{\bfl}_{\rohy}(\bM)$ and $C^{\bfl}_{\rohy}(\bM)$.
In what follows, we also need to impose bounds on 
\begin{equation}\label{eq:dotchidef}
  \dotchi:=\overline{\ml_{\hU}\chi}. 
\end{equation}
Here the overline represents orthogonal projection to the tangent spaces of $\bM_{t}$; i.e., $\dotchi-\ml_{\hU}\chi$ is parallel to $U$. Note that
$\dotchi$ can also be written
\[
\dotchi=\overline{\hml_{U}\chi},
\]
\index{$\a$Aa@Notation!Vector fields!$\dotchi$}%
where $\hml_{U}:=\theta^{-1}\ml_{U}$. 

In the case of the examples mentioned in Section~\ref{section:conjecturesandresults}, the shift vector field vanishes, so that the conditions concerning
$\chi$ are trivially satisfied. 

\subsection{Assumptions concerning the coefficients}

Turning to the assumptions concerning the coefficients of the equation, it is useful to take an expansion normalised perspective. Effectively, this means
that we multiply (\ref{eq:theequation}) by $\theta^{-2}$ (or, alternately, that we rephrase the wave operator in terms of the wave operator associated with
the conformally rescaled metric $\hg$; cf. Subsection~\ref{ssection:confeq} below). In particular, we therefore need to impose conditions on
\begin{equation}\label{eq:hmcXhaldef}
  \hmcX:=\theta^{-2}\mcX=\hmcX^{0}\hU+\hmcX^{\perp},\ \ \
  \hal:=\theta^{-2}\a,
\end{equation}
\index{$\a$Aa@Notation!Coefficients of the equation!$\hmcX$}%
\index{$\a$Aa@Notation!Coefficients of the equation!$\hmcX^{0}$}%
\index{$\a$Aa@Notation!Coefficients of the equation!$\hmcX^{\perp}$}%
\index{$\a$Aa@Notation!Coefficients of the equation!$\hal$}%
where the components of $\hmcX^{\perp}$ consist of vector fields that are perpendicular to $\hU$ with respect to $g$. Concerning $\hal$ and $\hmcX^{0}$, we
impose bounds with respect to norms such as (\ref{eq:mtClbS}) and (\ref{eq:mtHlbS}). However, when it comes to $\hmcX^{\perp}$, we need to proceed
differently. To begin with, if $\xi$ is a vector field on $M$ which is tangential to the leaves of the foliation, let 
\begin{equation}\label{eq:bDkchirohcdef}
|\bD^{k}\xi|_{\rohc}:=\left(\bge_{\refer}^{i_{1}j_{1}}\cdots \bge_{\refer}^{i_{k}j_{k}}\chg_{lm}\bD_{i_{1}}\cdots\bD_{i_{k}}\xi^{l}
\bD_{j_{1}}\cdots\bD_{j_{k}}\xi^{m}\right)^{1/2}. 
\end{equation}
Given $(\weight_{a},\weight_{b})=\weight\in\Weight$ and $(l_{0},l_{1})=\bfl\in\Index$, it is also convenient to introduce the notation
\begin{align}
\|\xi(\cdot,t)\|_{H^{\bfl,\weight}_{\rohc}(\bM)} := & \left(\int_{\bM}\textstyle{\sum}_{k=l_{0}}^{l_{1}}
\ldr{\varrho(\cdot,t)}^{-2\weight_{a}-2k\weight_{b}}|\bD^{k}\xi(\cdot,t)|_{\rohc}^{2}\mu_{\bge_{\refer}}\right)^{1/2},
\label{eq:Hlrohcdef}\\
\|\xi(\cdot,t)\|_{C^{\bfl,\weight}_{\rohc}(\bM)} := & \sup_{\bx\in\bM}\textstyle{\sum}_{k=l_{0}}^{l_{1}}
\ldr{\varrho(\bx,t)}^{-\weight_{a}-k\weight_{b}}|\bD^{k}\xi(\bx,t)|_{\rohc}. \label{eq:Clrohcdef}
\end{align}
\index{$\a$Aa@Notation!Norms!$C^{\bfl,\weight}_{\rohc}(\bM)$}%
\index{$\a$Aa@Notation!Norms!$H^{\bfl,\weight}_{\rohc}(\bM)$}%
In case $\bfl=(0,l)$, we replace $\bfl$ with $l$ (in practice, this will be signalled by the fact that the superscript is not in boldface) in the
names of the spaces and the notation for the norms. In case $\weight=0$, we also use the notation $H^{\bfl}_{\rohc}(\bM)$ and $C^{\bfl}_{\rohc}(\bM)$. Below,
we impose boundedness of $\hmcX^{\perp}$ with respect to norms such as the ones introduced in (\ref{eq:Hlrohcdef}) and (\ref{eq:Clrohcdef}).

It is of interest to analyse how strong the assumptions are by considering a specific example, such as the Klein-Gordon equation. In that case $\mcX=0$ and
$\a$ is constant. In the context of interest here, it can be demonstrated that $\theta$ tends to infinity exponentially (with respect to $\varrho$). Since
$\a$ is constant, this means that $\hal$ converges to zero exponentially. In particular, it is in that setting trivial to prove that $\hal$ is bounded with
respect to norms such as (\ref{eq:mtClbS}) and (\ref{eq:mtHlbS}). 

\section{Assumptions}

Since it is cumbersome to repeat all the assumptions in the statement of every lemma, we here formulate the basic assumptions.

\begin{definition}\label{def:basicassumptions}
  Let $(M,g)$ be a time oriented Lorentz manifold. Assume it to have an expanding partial pointed foliation, $\mK$ to be non-degenerate on $I$, $\mK$ to
  have a global frame and $\chK$ to have a silent upper bound on $I$; cf. Definition~\ref{def:silenceandnondegeneracy}. Assume, moreover, $\hml_{U}\mK$ to
  satisfy a weak off-diagonal exponential bound; cf. Definition~\ref{def:offdiagonalexpdec}. Next, let $\weight_{0}=(0,\cweight)\in\Weight$ and assume that
  there is a constant $K_{\cweight}$ such that
  \begin{equation}\label{eq:mKComfwbd}
    \|\mK(\cdot,t)\|_{C^{1}_{\weight_{0}}(\bM)}\leq K_{\cweight}
  \end{equation}
  \index{$\a$Aa@Notation!Constants!$K_{\cweight}$}%
  for all $t\in I_{-}$; in particular, there is a constant $\mKsup$ such that (\ref{eq:mKsupbasest}) holds. Assume, finally, that
  (\ref{eq:bDlnNbDlnthetabd}) holds; and that
  \begin{equation}\label{eq:basbaschibd}
    \|\chi(\cdot,t)\|_{C^{0}_{\rohy}(\bM)}\leq \frac{1}{2}
  \end{equation}
  for all $t\in I_{-}$. Then the \textit{basic assumptions}
  \index{Assumptions!Basic}%
  \index{Basic assumptions}%
  are said to be fulfilled. The associated constants are denoted by
  \begin{equation}\label{eq:crobasdef}
    c_{\robas}:=(n,\e_{\Spe},\e_{\mK},\e_{\rond},C_{\mK},C_{\mK,\mrod},M_{\mK,\mrod},\cweight,K_{\cweight},\bDlnhNsup).
  \end{equation}
  \index{$\a$Aa@Notation!Constants!$c_{\robas}$}%
\end{definition}

\subsection{Higher order Sobolev assumptions}\label{subsection:higherordersobolevassumptions}

In Definition~\ref{def:basicassumptions} we state the basic assumptions. However, in many contexts, it is of interest to make assumptions
concerning higher order derivatives. In the corresponding definitions, and in what is to follow, it is convenient to use the following notation
\begin{equation}\label{eq:thetazdef}
  \theta_{0,-}:=\inf_{\bx\in \bM}\theta(\bx,t_{0}),\ \ \
  \theta_{0,+}:=\sup_{\bx\in \bM}\theta(\bx,t_{0}). 
\end{equation}
\index{$\a$Aa@Notation!Constants!$\theta_{\pm}$}%

\begin{definition}\label{def:sobklassumptions}
  Given that the basic assumptions, cf. Definition~\ref{def:basicassumptions}, are satisfied, let $1\leq l\in\zo$, $\bfl_{0}:=(1,1)$,
  $\bfl:=(1,l)$ and $\bfl_{1}:=(1,l+1)$. Let $\cweight$ and $\weight_{0}$ be defined as in the statement of Definition~\ref{def:basicassumptions}.
  Let, moreover, $\weight:=(\cweight,\cweight)$. Then the $(\cweight,l)$-\textit{Sobolev assumptions}
  \index{Assumptions!$(\cweight,l)$-Sobolev}%
  \index{Assumptionst@$(\cweight,l)$-Sobolev assumptions}%
  are said to be satisfied if there are constants
  $S_{\rorel,l}$, $S_{\chi,l}$, $S_{\mK,l}$, $S_{\theta,l}$, $C_{\rorel,1}$, $C_{\mK,1}$, $C_{\chi,1}$ and $C_{\theta,1}$ such that
  \begin{align*}
    \|\ln\hN\|_{H^{\bfl_{1}}_{\weight_{0}}(\bM)}+\|\hU(\ln\hN)\|_{H^{l}_{\weight}(\bM)} \leq & S_{\rorel,l},\\
    \theta_{0,-}^{-1}\|\chi\|_{H^{l+2,\weight_{0}}_{\rohy}(\bM)}+\theta_{0,-}^{-1}\|\dotchi\|_{H^{l,\weight}_{\rohy}(\bM)} \leq & S_{\chi,l},\\
    \|\mK\|_{H^{l+1}_{\weight_{0}}(\bM)}+\|\hml_{U}\mK\|_{H^{l+1}_{\weight}(\bM)} \leq & S_{\mK,l},\\
    \|\ln\theta\|_{H^{\bfl_{1}}_{\weight_{0}}(\bM)}+\|q\|_{H^{l}_{\weight_{0}}(\bM)} \leq & S_{\theta,l}
  \end{align*}
  for all $t\in I_{-}$, where $I_{-}$ is defined by (\ref{eq:Iminusdef}), and
  \begin{align*}
    \|\ln\hN\|_{C^{\bfl_{0}}_{\weight_{0}}(\bM)}+\|\hU(\ln\hN)\|_{C^{0}_{\weight}(\bM)} \leq & C_{\rorel,1},\\
    \theta_{0,-}^{-1}\|\chi\|_{C^{2,\weight_{0}}_{\rohy}(\bM)}+\theta_{0,-}^{-1}\|\dotchi\|_{C^{1,\weight}_{\rohy}(\bM)} \leq & C_{\chi,1},\\
    \|\mK\|_{C^{1}_{\weight_{0}}(\bM)}+\|\hml_{U}\mK\|_{C^{0}_{\weight}(\bM)} \leq & C_{\mK,1},\\
    \|\ln\theta\|_{C^{\bfl_{0}}_{\weight_{0}}(\bM)}+\|q\|_{C^{0}_{\weight_{0}}(\bM)} \leq & C_{\theta,1}
  \end{align*}
  for all $t\in I_{-}$. Given that the $(\cweight,l)$-Sobolev assumptions hold, let
  \[
  s_{\cweight,l}:=(c_{\robas},l,S_{\rorel,l},S_{\chi,l},S_{\mK,l},S_{\theta,l},C_{\rorel,1},C_{\mK,1},C_{\chi,1},C_{\theta,1}).
  \]
  \index{$\a$Aa@Notation!Constants!$s_{\cweight,l}$}%
\end{definition}
\begin{remark}\label{remark:toogeneral}
  In specific situations, we typically do not need to make all these assumptions. However, in order to avoid stating distinct and detailed
  assumptions in every lemma, and in order to avoid listing dependence on a large number of constants, we here prefer to make all the needed
  assumptions in one place.
\end{remark}
\begin{remark}\label{remark:lossofderivativestogeometry}
  There are two undesirable assumptions in the above definition. First, we bound $\hml_{U}\mK$ in $H^{l+1}$ instead of in $H^{l}$.
  Second, we bound $\chi$ in $H^{l+2}$ instead of in $H^{l+1}$. Both of these anomalies have the same origin, namely the fact that 
  we need to bound $\mu_{A}$, defined by (\ref{eq:muAdef}), in $H^{l+1}$. Moreover, we only control $\mu_{A}$ via $\hml_{U}\mK$ and $\chi$.
  In short, the reason for these anomalies is that we wish to express the spatial derivatives in the equation with respect to a geometric
  frame. But the geometric frame is defined using the second fundamental form, which, in the end, leads to a loss of derivatives. In other
  words, we are losing derivatives in order to obtain a clear geometric picture. 
\end{remark}

The above assumptions concern the geometry. However, it is also necessary to make assumptions concerning the coefficients of the equation.
The conditions we impose here are of the following form. For a suitable choice of $0\leq l\in \zo$, we assume the existence of a constant
$s_{\coeff,l}$ such that 
\begin{equation}\label{eq:Sobcoefflassumptions}
  \|\hmcX^{0}(\cdot,t)\|_{H^{l}_{\weight_{0}}(\bM)}+\textstyle{\sum}_{i,j}\|\hmcX^{\perp}_{ij}(\cdot,t)\|_{H^{l,\weight_{0}}_{\rohc}(\bM)}
  +\|\hal(\cdot,t)\|_{H^{l}_{\weight_{0}}(\bM)}\leq s_{\coeff,l}
\end{equation}
\index{$\a$Aa@Notation!Constants!$s_{\coeff,l}$}%
for all $t\in I_{-}$, where $\weight_{0}$ and $\weight$ are given in Definition~\ref{def:sobklassumptions}. Since $\hmcX^{0}$ and $\hal$ are matrix
valued, the meaning of the left hand side of (\ref{eq:Sobcoefflassumptions}) needs to be clarified. Here, we take it to be understood that 
\begin{equation}\label{eq:halHlmatrixconv}
  \|\hal(\cdot,t)\|_{H^{l}_{\weight_{0}}(\bM)}:=\textstyle{\sum}_{i,j=1}^{m_{\ros}}\|\hal_{ij}(\cdot,t)\|_{H^{l}_{\weight_{0}}(\bM)}
\end{equation}
and similarly for the norm of $\hmcX^{0}$. 
  
\subsection{Higher order $C^{k}$-assumptions}\label{subsection:higherorderckassumptions}

Next, we introduce the $C^{k}$-terminology analogous to Definition~\ref{def:sobklassumptions}. 

\begin{definition}\label{def:supmfulassumptions}
  Given that the basic assumptions, cf. Definition~\ref{def:basicassumptions}, are satisfied, let $0\leq l\in\zo$
  and $\bfl_{1}:=(1,l+1)$. Let $\cweight$ and $\weight_{0}$ be defined as in the statement of Definition~\ref{def:basicassumptions}.
  Let, moreover, $\weight:=(\cweight,\cweight)$. Then the $(\cweight,l)$-\textit{supremum assumptions}
  \index{Assumptions!$(\cweight,l)$-Supremum}%
  \index{Assumptionst@$(\cweight,l)$-Supremum assumptions}%
  are said to be satisfied if there are constants
  $C_{\rorel,l}$, $C_{\chi,l}$, $C_{\mK,l}$, $C_{\theta,l}$ such that
  \begin{align*}
    \|\ln\hN\|_{C^{\bfl_{1}}_{\weight_{0}}(\bM)}+\|\hU(\ln\hN)\|_{C^{l}_{\weight}(\bM)} \leq & C_{\rorel,l},\\
    \theta_{0,-}^{-1}\|\chi\|_{C^{l+2,\weight_{0}}_{\rohy}(\bM)}+\theta_{0,-}^{-1}\|\dotchi\|_{C^{l,\weight}_{\rohy}(\bM)} \leq & C_{\chi,l},\\
    \|\mK\|_{C^{l+1}_{\weight_{0}}(\bM)}+\|\hml_{U}\mK\|_{C^{l+1}_{\weight}(\bM)} \leq & C_{\mK,l},\\
    \|\ln\theta\|_{C^{\bfl_{1}}_{\weight_{0}}(\bM)}+\|q\|_{C^{l}_{\weight_{0}}(\bM)} \leq & C_{\theta,l}
  \end{align*}
  for all $t\in I_{-}$. Given that the $(\cweight,l)$-supremum assumptions hold, let
  \[
  c_{\cweight,l}:=(c_{\robas},l,C_{\rorel,l},C_{\chi,l},C_{\mK,l},C_{\theta,l}).
  \]
  \index{$\a$Aa@Notation!Constants!$c_{\coeff,l}$}%
\end{definition}
\begin{remark}
  Remarks~\ref{remark:toogeneral} and \ref{remark:lossofderivativestogeometry} are equally relevant in the present setting. 
\end{remark}

Again, the above assumptions concern the geometry, but we also need to make assumptions concerning the coefficients of the equation.
For a suitable choice of $0\leq l\in \zo$, we assume the existence of a constant $c_{\coeff,l}$ such that 
\begin{equation}\label{eq:coefflassumptions}
  \|\hmcX^{0}(\cdot,t)\|_{C^{l}_{\weight_{0}}(\bM)}+\textstyle{\sum}_{i,j}\|\hmcX^{\perp}_{ij}(\cdot,t)\|_{C^{l,\weight_{0}}_{\rohc}(\bM)}
  +\|\hal(\cdot,t)\|_{C^{l}_{\weight_{0}}(\bM)}\leq c_{\rocoeff,l}
\end{equation}
\index{$\a$Aa@Notation!Constants!$c_{\coeff,l}$}%
for all $t\in I_{-}$, where $\weight_{0}$ and $\weight$ are given in Definition~\ref{def:supmfulassumptions}. Moreover, the notation is
analogous to that introduced in (\ref{eq:halHlmatrixconv}).

\section{Smallness of the shift vector field}

In these notes, we only make one smallness assumption, namely that the shift vector field is small.

\begin{lemma}\label{lemma:smallnessshiftconsequences}
  Assume the conditions of Definition~\ref{def:basicassumptions} to be fulfilled; i.e., the basic assumptions to hold. Assume, moreover, that
  there is a constant $c_{\chi,2}$ such that
  \[
  \theta_{0,-}^{-1}\|\chi\|_{C^{2,\weight_{0}}_{\rohy}(\bM)}\leq c_{\chi,2}
  \]
  \index{$\a$Aa@Notation!Constants!$c_{\chi,2}$}%
  holds for all $t\in I_{-}$, where $\weight_{0}$ is the same as in Definition~\ref{def:basicassumptions}. Then there is an $\e_{\chi}>0$, depending only
  on $c_{\robas}$, and a $\de_{\chi}$, depending only on $c_{\robas}$, $c_{\chi,2}$ and $(\bM,\bge_{\refer})$, such that if
  \begin{align}
    n^{1/2}\theta_{0,-}^{-1}|\chi|_{\rohy} \leq & \de_{\chi},\label{eq:dechicontrol}\\
    n^{1/2}\theta_{0,-}^{-1}|\bD\chi|_{\rohy} \leq & \e_{\chi}\label{eq:echicontrol}
  \end{align}
  \index{$\a$Aa@Notation!Constants!$\de_{\chi}$}%
  \index{$\a$Aa@Notation!Constants!$\e_{\chi}$}%
  hold on $M_{-}:=\bM\times I_{-}$,
  \index{$\a$Aa@Notation!Manifolds!$M_{-}$}%
  then
  \begin{equation}\label{eq:muminmainlowerboundintro}
    \mu_{\min} \geq  -\e_{\Spe}\varrho+\ln\theta_{0,-}-M_{\min}
  \end{equation}
  on $M_{-}$, where $M_{\min}$ only depends on $c_{\robas}$. Here $\mu_{\min}:=\min_{A}\mu_{A}$.
  \index{$\a$Aa@Notation!Functions!$\mu_{\min}$}%
  Moreover, there is a constant $C_{\varrho}$, depending only
  on $c_{\robas}$, $c_{\chi,2}$ and $(\bM,\bge_{\refer})$, such that $|\bD\varrho|_{\bge_{\refer}}\leq C_{\varrho}\ldr{\varrho}$. Next, there is a constant
  $K_{\rovar}$, depending only on $\bDlnhNsup$ and $(\bM,\bge_{\refer})$, such that if $\bx_{1},\bx_{2}\in\bM$ and $t_{1},t_{2}\in I_{-}$ are such that
  $t_{1}<t_{2}$, then 
  \begin{equation}\label{eq:DeltavarrhorelvariationEiintro}
    \frac{1}{3K_{\rovar}}\leq \frac{\varrho(\bx_{2},t_{2})-\varrho(\bx_{2},t_{1})}{\varrho(\bx_{1},t_{2})-\varrho(\bx_{1},t_{1})}\leq 3K_{\rovar}.
  \end{equation}
  \index{$\a$Aa@Notation!Constants!$K_{\rovar}$}%
  Finally  
  \begin{equation}\label{eq:hNinvdtvarrhoestEiintro}
    1/2\leq \hN^{-1}\d_{t}\varrho\leq 3/2
  \end{equation}
  holds on $M_{-}$. 
\end{lemma}
\begin{remark}
  The fact that (\ref{eq:muminmainlowerboundintro}) holds can be interpreted as saying that the conformally rescaled spacetime
  exhibits exponential expansion in the direction towards the singularity. The estimate (\ref{eq:DeltavarrhorelvariationEiintro}) yields a bound
  on the relative spatial variation of $\varrho$. Finally, (\ref{eq:hNinvdtvarrhoestEiintro}) allows us to, roughly speaking, introduce $\varrho$
  as a time coordinate. 
\end{remark}
\begin{remark}
  The values of the constants $\e_{\chi}$ and $\de_{\chi}$ can be deduced from the statements of Lemmas~\ref{lemma:lowerbdonmumin} and
  \ref{lemma:taurelvaryingbxEi} respectively.
\end{remark}
\begin{proof}
  The statement follows by combining Lemmas~\ref{lemma:lowerbdonmumin}, \ref{lemma:respvarvarrhoEi} and \ref{lemma:taurelvaryingbxEi} below.
\end{proof}

In most of the arguments and results presented in these notes, it will be important to know that the conclusions of 
Lemma~\ref{lemma:smallnessshiftconsequences} hold. For this reason, it is convenient to introduce the following terminology.

\begin{definition}\label{def:standardassumptions}
  Assume that the conditions of Definition~\ref{def:basicassumptions} are fulfilled. If, in addition, the conditions of
  Lemma~\ref{lemma:smallnessshiftconsequences} are satisfied, then the \textit{standard assumptions}
  \index{Assumptions!Standard}%
  \index{Standard assumptions}%
  are said to be satisfied. 
\end{definition}

\textbf{Time coordinate.} Given that the standard assumptions hold, it is convenient to introduce a new time coordinate by fixing a reference
point $\bx_{0}\in\bM$ and defining 
\begin{equation}\label{eq:taudefinitionintro}
  \tau(t):=\varrho(t,\bx_{0});
\end{equation}
\index{$\a$Aa@Notation!Functions!$\tau$}%
cf. (\ref{eq:taudefinitionEi}) below. Moreover, several conclusions concerning this time coordinate can be deduced; cf.
Lemma~\ref{lemma:epsilonlowdefEi} below.

\chapter{Results and outline}\label{chapter:results}

Given the terminology introduced in the previous chapter, we are in a position to formulate the conclusions. There are several types of results:
general energy estimates; localised energy estimates (in regions of the form $J^{+}(\g)$ for causal curves $\g$ going into the singularity);
a derivation of the leading order asymptotics and the corresponding asymptotic data; and a specification of the leading order asymptotics (leading to
a proof of optimality of the localised energy estimates). The corresponding theorems are formulated in 
Sections~\ref{section:energyestimates}--\ref{section:specifyingasymptoticsintro} below. It is of interest to compare the results of these notes with
the ones obtained in previous work, and we do so in Section~\ref{section:previousresults} below. We also provide an outlook in
Section~\ref{section:outlook}. Finally, we provide an outline of these notes in Section~\ref{section:FullOutline}. 

\section{Energy estimates}\label{section:energyestimates}

Before formulating the results, it is convenient to introduce some terminology. 

\subsection{Reformulation of the equation}\label{ssection:reformofequationintro}
The subject of these notes is the asymptotic behaviour of solutions to (\ref{eq:theequation}). We begin by stating energy estimates.
Before doing so, it is convenient to rewrite the equation in terms of the global frame introduced in Definition~\ref{def:XAellA}. It then takes the
form
\begin{equation}\label{eq:equationintermsofcanonicalframe}
  -\hU^{2}u+\textstyle{\sum}_{A}e^{-2\mu_{A}}X_{A}^{2}u+Z^{0}\hU u+Z^{A}X_{A}u+\hal u=\hf.
\end{equation}
Here $\hU$ and $X_{A}$ are introduced in Definitions~\ref{def:basicnotions} and \ref{def:XAellA} respectively; and $\hal$ is defined by
(\ref{eq:hmcXhaldef}). Moreover,
\begin{align}
  Z^{0} := & \frac{1}{n}[q-(n-1)]\Id+\hmcX^{0},\label{eq:Zzdefintro}\\
  Z^{A} := & \hmcY^{A}\Id+\hmcX^{A};\label{eq:ZAdefintro}
\end{align}
\index{$\a$Aa@Notation!Functions!$Z^{0}$}%
\index{$\a$Aa@Notation!Functions!$Z^{A}$}%
cf. (\ref{eq:theeqreformEi})--(\ref{eq:hmcYAdefEi}) below, as well as (\ref{eq:chthexpressionhUlntheta}). Note that here $\hmcY^{A}$ is given by
(\ref{eq:hmcYAdefEi}), (\ref{eq:hGAdef}) and (\ref{eq:aAdef}). Moreover, $\hmcX^{0}$ is defined by (\ref{eq:hmcXhaldef}) and $\hmcX^{A}=Y^{A}(\hmcX^{\perp})$,
where $Y^{A}$ is given by Definition~\ref{def:XAellA} and $\hmcX^{\perp}$ is given by (\ref{eq:hmcXhaldef}). In what follows, it is also convenient
to use the notation
\begin{equation}\label{eq:hmcXperpchgnormintro}
  \|\hmcX^{\perp}\|_{\chg}:=\left(\textstyle{\sum}_{A}e^{2\mu_{A}}\|\hmcX^{A}\|^{2}\right)^{1/2}.
\end{equation}

\subsection{Basic energy}
How the energy is defined depends on the coefficients of the equation. In order to separate the different cases, fix $\tau_{c}\leq 0$.
If there is a constant $d_{\a}$ such that
\begin{equation}\label{eq:haldecayintro}
  \|\hal(\cdot,t)\|_{C^{0}(\bM)}\leq d_{\a}\ldr{\tau(t)-\tau_{c}}^{-3}
\end{equation}
for all $t\leq t_{c}$, where $\tau_{c}=\tau(t_{c})$, we choose $\iota_{a}=0$ and $\iota_{b}=1$; here $\tau$ is the time coordinate introduced in
(\ref{eq:taudefinitionintro}). Otherwise, we choose $\iota_{a}=1$ and $\iota_{b}=0$. Let
\begin{equation}\label{eq:medefintro}
  \me[u]:=\frac{1}{2}\left(|\hU(u)|^{2}+\textstyle{\sum}_{A}e^{-2\mu_{A}}|X_{A}(u)|^{2}+\iota_{a}|u|^{2}+\iota_{b}\ldr{\tau-\tau_{c}}^{-3}|u|^{2}\right).
\end{equation}
\index{$\a$Aa@Notation!Energy densities!$\me[u]$}%
This expression represents the energy density. In order to use $\me$ to define an $L^{2}$-based energy, we need to fix a measure on $\bM$. Three naive
choices are $\mu_{\bge_{\refer}}$, $\mu_{\bge}$ and $\mu_{\chg}$. However, considering the identities that appear when deriving energy estimates, it turns
out that $\theta\mu_{\bge}=\theta\varphi\mu_{\bge_{\refer}}$ is a more promising candidate. Nevertheless, this measure also has a deficiency. In fact, it
is sometimes of interest to express the estimates in terms of a starting time, say $t_{c}$, different from $t_{0}$. In that context, it is natural
to express the control at $t_{c}$ in terms of a measure which does not depend on $t_{c}$, such as $\mu_{\bge_{\refer}}$. On the other hand, if $t_{c}$ is
close to the singularity, then the constants relating $\mu_{\bge_{\refer}}$ and $\theta\mu_{\bge}$ diverge. For this reason, it is convenient to introduce
$\tvarphi:=\theta\varphi$, $\tvarphi_{c}(\bx,t):=\tvarphi(\bx,t_{c})$ and
\begin{equation}\label{eq:hEubasicdefinition}
  \hE[u](\tau;\tau_{c}):=\int_{\bM_{\tau}}\me[u]\mutgc,
\end{equation}
\index{$\a$Aa@Notation!Energies!$\hE[u]$}%
where
\[
\mutgc=\tvarphi_{c}^{-1}\theta^{-(n-1)}\mu_{\chg}=\tvarphi_{c}^{-1}\theta\mu_{\bge}=\tvarphi_{c}^{-1}\tvarphi\mu_{\bge_{\refer}}. 
\]
\index{$\a$Aa@Notation!Volume forms!$\mutgc$}%
However, in many situations it is of interest to relate this energy to
\begin{equation}\label{eq:hGedefintro}
  \hGe[u](\tau):=\int_{\bM_{\tau}}\me[u]\mu_{\bge_{\refer}}.
\end{equation}
\index{$\a$Aa@Notation!Energies!$\hGe[u]$}%
One special situation of interest is the following.
\begin{lemma}\label{lemma:hEhGequivalenceintro}
  Assume that the standard assumptions are satisfied (cf. Definition~\ref{def:standardassumptions});
  that there is a constant $c_{\theta,1}$ such that
  \begin{equation}\label{eq:cthetaoneestimateintro}
    \|(\ln\theta)(\cdot,t)\|_{C^{\bfl_{0}}_{\weight_{0}}(\bM)}\leq c_{\theta,1}
  \end{equation}
  holds for all $t\leq t_{c}$, where $\bfl_{0}=(1,1)$; and that there is a constant $d_{q}$ such that
  \begin{equation}\label{eq:qconvergenceintro}
    \|\ldr{\varrho(\cdot,t)}^{3/2}[q(\cdot,t)-(n-1)]\|_{C^{0}(\bM)} \leq d_{q}
  \end{equation}
  for all $t\leq t_{c}$. Then there is a constant $c_{G}\geq 1$, depending only on $c_{\robas}$, $c_{\theta,1}$, $c_{\chi,2}$, $d_{q}$ and $(\bM,\bge_{\refer})$
  such that
  \[
  c_{G}^{-1}\hGe[u](\tau)\leq\hE[u](\tau;\tau_{c})\leq c_{G}\hGe[u](\tau)
  \]
  for all $t\leq t_{c}$. 
\end{lemma}
\begin{remark}
  As mentioned in the previous chapter, the $3+1$-dimensional quiescent singularities discussed in Section~\ref{section:conjecturesandresults} are
  typically such that $q$ converges to $2$ exponentially; cf. Appendix~\ref{chapter:examples} below. They are also such that
  (\ref{eq:cthetaoneestimateintro}) holds. 
\end{remark}
\begin{proof}
  The statement is an immediate consequence of Lemma~\ref{lemma:thetavarrhorelqconvtonmotwo} below. Note that $K_{\rovar}$ appearing in the statement of
  Lemma~\ref{lemma:thetavarrhorelqconvtonmotwo} is given by (\ref{eq:KrovarEi}). 
\end{proof}
The following result represents the basic energy estimate.

\begin{prop}\label{prop:basicenergyestimateintro}
  Assume the standard assumptions to be fulfilled; cf. Definition~\ref{def:standardassumptions}. Assume, moreover, (\ref{eq:coefflassumptions})
  to hold for $l=0$; $q$ to be bounded on $M_{-}$; and assume that there is a constant $c_{\theta,1}$ such that (\ref{eq:cthetaoneestimateintro})
  holds for all $t\in I_{-}$, where $\bfl_{0}=(1,1)$. Then, if $u$ is a solution to (\ref{eq:theequation}) with vanishing right hand side, 
  \begin{equation}\label{eq:basicenergyestimateintro}
    \begin{split}
      \hE(\tau_{a};\tau_{c}) \leq &  \hE(\tau_{b};\tau_{c})+\int_{\tau_{a}}^{\tau_{b}}[c_{0}+\kappa_{\rem}(\tau)]\hE(\tau;\tau_{c})d\tau
    \end{split}
  \end{equation}
  for all $\tau_{a}\leq\tau_{b}\leq \tau_{c}\leq 0$, where $c_{0}$ is a constant and  $\kappa_{\rem}\in L^{1}(-\infty,\tau_{c}]$. Moreover, the $L^{1}$-norm of
  $\kappa_{\rem}$ only depends on $c_{\robas}$, $c_{\chi,2}$, $c_{\theta,1}$, $(\bM,\bge_{\refer})$, $d_{\a}$ (in case $\iota_{b}=1$) and a lower bound on
  $\theta_{0,-}$.

  Assume, in addition to the above, that (\ref{eq:haldecayintro}) holds and that there are constants $d_{q}$ and $d_{\coeff}$ such that
  (\ref{eq:qconvergenceintro}) and
  \begin{align}      
    \sup_{\bx\in\bM}[\|\hmcX^{0}(\bx,t)\|+\|\hmcX^{\perp}(\bx,t)\|_{\chg}] \leq & d_{\coeff}\ldr{\tau(t)-\tau_{c}}^{-3/2}\label{eq:coeffconvergenceintro}
  \end{align}
  hold for all $t\leq t_{c}$. Then (\ref{eq:basicenergyestimateintro}) holds with $c_{0}=0$. Moreover, the $L^{1}$-norm of $\kappa_{\rem}$
  is bounded by a constant depending only on $c_{\robas}$, $c_{\chi,2}$, $c_{\theta,1}$, $(\bM,\bge_{\refer})$, $d_{\a}$, $d_{q}$, $d_{\coeff}$ and a lower bound on
  $\theta_{0,-}$. Finally,
  \begin{equation}\label{eq:basicenergyboundedness}
    \hGe[u](\tau)\leq C\hGe[u](\tau_{c})
  \end{equation}
  for all $\tau\leq\tau_{c}$, where $C$ only depends on $c_{\robas}$, $c_{\chi,2}$, $c_{\theta,1}$, $(\bM,\bge_{\refer})$, $d_{\a}$, $d_{q}$, $d_{\coeff}$ and a
  lower bound on $\theta_{0,-}$.
\end{prop}
\begin{remarks}
  Due to (\ref{eq:basicenergyestimateintro}), $\hE$ does not grow faster than exponentially.
  It is important to note that if estimates such as (\ref{eq:coefflassumptions}) do not hold for $l=0$, then the energy could grow superexponentially.
  For a justification of this statement, see \cite[Chapter~2]{finallinsys}. 
\end{remarks}
\begin{remark}
  The constant $c_{0}$ can be calculated in terms of $q$ and the coefficients of the equation; cf. (\ref{eq:czCbdef}) below.
\end{remark}
\begin{remark}\label{remark:KGboundedbasicenergyintro}
  In the case of the Klein-Gordon equation, (\ref{eq:haldecayintro}) and (\ref{eq:coeffconvergenceintro}) are automatically satisfied. The reason for this
  is that then $\hmcX=0$ and $\hal=-\theta^{-2}\mKG^{2}$, where $\mKG$ is a constant. Moreover, due to (\ref{eq:lnthetalowbd}), $\theta$ tends to infinity
  exponentially as $\tau\rightarrow-\infty$. Beyond the main assumptions in Proposition~\ref{prop:basicenergyestimateintro},
  it is thus sufficient to assume (\ref{eq:qconvergenceintro}) to be satisfied in order to conclude that (\ref{eq:basicenergyboundedness}) holds. 
\end{remark}
\begin{proof}
  The statement is an immediate consequence of Corollary~\ref{cor:basicenergyestimate} (a result which also gives conclusions in the case that
  $f\neq 0$), Lemma~\ref{lemma:hEhGequivalenceintro} and Gr\"{o}nwall's lemma. Note also that the notion of $C^{0}$-balance is introduced in
  Definition~\ref{def:Czerobalance} and that the equation is $C^{0}$-balanced on $I_{-}$ due to Remark~\ref{remark:Czbalancediffcond}. 
\end{proof}

\subsection{Higher order energies}
In order to define the higher order energies, it is convenient to recall that there is a global orthonormal frame $\{E_{i}\}$ on
$(\bM,\bge_{\refer})$; cf. Remark~\ref{remark:framenondegenerate}. We also use the following terminology. 
\begin{definition}\label{def:multiindexnotation}
  Let $(M,g)$ be a time oriented Lorentz manifold. Assume that it has an expanding partial pointed foliation. Assume, moreover, $\mK$ to be
  non-degenerate on $I$ and to have a global frame. Then a \textit{vector field multiindex}
  \index{Vector field multiindex}%
  \index{Multiindex!Vector field}%
  is a vector, say $\bfI=(I_{1},\dots,I_{l})$,
  \index{$\a$Aa@Notation!Vector field multiindices!$\bfI$}%
  where $I_{j}\in \{1,\dots,n\}$. The number $l$ is said to be the \textit{order}
  \index{Order!Vector field multiindex}%
  \index{Vector field multiindex!Order}%
  of the vector field multiindex, and it is denoted by $|\bfI|$. The
  vector field multiindex corresponding to the empty set is denoted by $\bfz$. Moreover, $|\bfz|=0$. Given that the letter used for the vector
  field multiindex is $\bfI$, $\bfJ$ etc.,
  \begin{align*}
    \bfE_{\bfI} := & (E_{I_{1}},\dots,E_{I_{l}}),\ \ \
    \bD_{\bfI}:=\bD_{E_{I_{1}}}\cdots \bD_{E_{I_{l}}},\ \ \
    E_{\bfI} := E_{I_{1}}\cdots E_{I_{l}}
  \end{align*}
  \index{$\a$Aa@Notation!Vector field multiindices!$\bfE_{\bfI}$}%
  \index{$\a$Aa@Notation!Vector field multiindices!$\bD_{\bfI}$}%
  \index{$\a$Aa@Notation!Vector field multiindices!$E_{\bfI}$}%
  etc. where $\bfI=(I_{1},\dots,I_{l})$, with the special convention that $\bD_{\bfz}$ and $E_{\bfz}$ are the identity operators,
  and $\bfE_{\bfz}$ is the empty argument.
\end{definition}
Given this notation, the higher order energies are defined as follows:
\begin{equation}\label{eq:hEkdefintro}
  \hE_{k}[u](\tau;\tau_{c}):=\textstyle{\sum}_{|\bfI|\leq k}\hE[E_{\bfI}u](\tau;\tau_{c}).
\end{equation}
\index{$\a$Aa@Notation!Energies!$\hE_{k}[u]$}%
In analogy with (\ref{eq:hGedefintro}), we also introduce
\begin{equation}\label{eq:hGekdefintro}
  \hGe_{k}[u](\tau):=\textstyle{\sum}_{|\bfI|\leq k}\hGe[E_{\bfI}u](\tau).
\end{equation}
\index{$\a$Aa@Notation!Energies!$\hGe_{k}[u]$}%
In case the conditions of Lemma~\ref{lemma:hEhGequivalenceintro} are satisfied, we then have
\begin{equation}\label{eq:hEkhGekequivalenceintroduction}
  c_{G}^{-1}\hGe_{k}[u](\tau)\leq\hE_{k}[u](\tau;\tau_{c})\leq c_{G}\hGe_{k}[u](\tau)
\end{equation}
for all $t\leq t_{c}$. The basic estimate of the higher order energies takes the following form. 

\begin{prop}\label{prop:EnergyEstimateSobolevAssumptionsintro}
  Let $0\leq \cweight\in\ro$, $\weight_{0}=(0,\cweight)$ and $\weight=(\cweight,\cweight)$. Assume that the standard assumptions are fulfilled (cf.
  Definition~\ref{def:standardassumptions}) and let $\kappa_{1}$ be the smallest integer strictly larger than $n/2+1$. Assume the
  $(\cweight,\kappa_{1})$-supremum assumptions to be satisfied; and that there is a constant $c_{\coeff,\kappa_{1}}$ such that (\ref{eq:coefflassumptions})
  holds with $l$ replaced by $\kappa_{1}$. Fix $l\geq\kappa_{1}$ as in Definition~\ref{def:sobklassumptions} and assume the
  $(\cweight,l)$-Sobolev assumptions to be satisfied. Assume, moreover, that there is a constant $s_{\coeff,l}$ such that (\ref{eq:Sobcoefflassumptions})
  holds. Assume, finally, (\ref{eq:theequation}) to be satisfied with vanishing right hand side. Then
  \begin{equation}\label{eq:TheEnergyEstimateintro}
    \begin{split}
      \hE_{l}(\tau_{a};\tau_{c}) \leq & C_{a}\ldr{\tau_{a}}^{2\a_{l,n}\cweight}\ldr{\tau_{a}-\tau_{c}}^{2\b_{l,n}}e^{c_{0}(\tau_{c}-\tau_{a})}\hE_{l}(\tau_{c};\tau_{c})
    \end{split}    
  \end{equation}
  for all $\tau_{a}\leq\tau_{c}\leq 0$. Here $c_{0}$ is the constant appearing in the statement of Proposition~\ref{prop:basicenergyestimateintro};
  $\a_{l,n}$ and $\b_{l,n}$ only depend on $n$ and $l$; and $C_{a}$ only depends on $s_{\cweight,l}$, $s_{\coeff,l}$, $c_{\cweight,\kappa_{1}}$, $c_{\coeff,\kappa_{1}}$,
  $m_{\ros}$, $d_{\a}$ (in case $\iota_{b}\neq 0$), $(\bM,\bge_{\refer})$ and a lower bound on $\theta_{0,-}$. If,
  in addition to the above assumptions, (\ref{eq:haldecayintro}), (\ref{eq:qconvergenceintro}) and (\ref{eq:coeffconvergenceintro}) hold for all
  $t\leq t_{c}$, then (\ref{eq:TheEnergyEstimateintro}) holds with $c_{0}=0$ and $\hE_{j}$ replaced by $\hGe_{j}$. However, in this case, the constant
  $C_{a}$, additionally, depends on $d_{q}$, $d_{\a}$ and $d_{\coeff}$. 
\end{prop}
\begin{remark}
  The combination of $C^{k}$ and Sobolev assumptions may seem somewhat strange. However, the logic is that the $C^{k}$ assumptions allow the deduction of
  energy estimates up to a certain order. Combining these energy estimates with Sobolev embedding yields $C^{m}$ control of the solution up to the order
  necessary for the combination of Sobolev assumptions, energy arguments and Moser-type estimates to yield control of the the higher order energies. 
\end{remark}
\begin{proof}
  The statement of the lemma is an immediate consequence of Proposition~\ref{prop:EnergyEstimateSobolevAssumptions},
  Remark~\ref{remark:EnergyEstimateSobolevAssumptionsImpCon} and (\ref{eq:hEkhGekequivalenceintroduction}).
\end{proof}

In some respects, the result is not very impressive, since it only states that the energy does not grow faster than exponentially, and since the rate
of exponential growth is quite rough. However, an estimate of this form is very valuable, and it can be used to derive much more detailed information.
The reason for this is that the rate of exponential growth is \textit{independent of the order of the energy}; in general, one might expect the
rate of exponential growth of the $l$'th energy to depend on $l$. Combining this independence with the assumed silence,
cf. Definition~\ref{def:silenceandnondegeneracy}, the asymptotic estimates can gradually be improved in order to obtain more detailed information. 

\subsection{The Klein-Gordon equation}

It is of interest to draw more detailed conclusions in the case of the Klein-Gordon equation
\begin{equation}\label{eq:KGeqdefintro}
  \Box_{g}u-\mKG^{2}u=0,
\end{equation}
\index{Klein-Gordon!Equation}%
\index{Equation!Klein-Gordon}%
where $\mKG$ is a constant. 

\begin{prop}  
  Let $0\leq \cweight\in\ro$, $\weight_{0}=(0,\cweight)$ and $\weight=(\cweight,\cweight)$. Assume the standard assumptions (cf.
  Definition~\ref{def:standardassumptions}) and the $(\cweight,\kappa_{1})$-supremum assumptions to be fulfilled, where $\kappa_{1}$ is the smallest
  integer strictly larger than $n/2+1$. Assume, additionally, that there are constants $\de_{q}$ and $\e_{q}>0$ such that 
  \begin{equation}\label{eq:qmnmoexpdecestintro}
    \|[q(\cdot,t)-(n-1)]\|_{C^{0}(\bM)} \leq \de_{q}e^{\e_{q}\tau(t)}
  \end{equation}
  for all $t\in I_{-}$. Let $\e_{\roKG}:=\min\{\e_{q},\eSpe\}$ and $u$ be a solution to (\ref{eq:KGeqdefintro}). Here $\eSpe=\e_{\Spe}/(3K_{\rovar})$, where
  $K_{\rovar}$ is the constant appearing in (\ref{eq:DeltavarrhorelvariationEiintro}). Then there is a $\psi_{\infty}\in C^{0}(\bM)$ such that
  \begin{align}
    \|(\hU u)(\cdot,\tau)-\psi_{\infty}\|_{C^{0}(\bM)} \leq & C_{\roKG}\ldr{\tau}^{\a_{n}\cweight+\b_{n}}e^{\e_{\roKG}\tau}\hGe_{\kappa_{1}}^{1/2}(0),\label{eq:hUvinflimintro}\\
    \|\psi_{\infty}\|_{C^{0}(\bM)}\leq & C_{\roKG}\hGe_{\kappa_{1}}^{1/2}(0),\label{eq:vinfCzbdintro}
  \end{align}
  for all $\tau\leq 0$, where $C_{\roKG}$ only depends on $c_{\cweight,\kappa_{1}}$, $\de_{q}$, $\e_{q}$, $\mKG$, 
  $(\bM,\bge_{\refer})$ and a lower bound on $\theta_{0,-}$. Moreover, $\a_{n}$ and $\b_{n}$ only depend on $n$.
\end{prop}
\begin{remark}
  Similar conclusions hold for more general classes of equations; cf. Proposition~\ref{prop:asvelocityKGlikeeq} below.
\end{remark}
\begin{remark}
  Making stronger assumptions, it might be possible to derive stronger conclusions. In particular, it might be possible to prove that there is,
  additionally, a function $u_{\infty}\in C^{0}(\bM)$ such that $u-\psi_{\infty}\varrho-u_{\infty}$ becomes small asymptotically; cf.
  Remarks~\ref{remark:higherorderasvel} and \ref{remark:vinftyuinftyexistence} for a discussion. However, we do not prove such estimates here.
  Nevertheless, in the context of the Einstein-scalar field equations, we do derive such estimates in \cite{RinGeometry} (as well as higher order
  versions thereof). 
\end{remark}
\begin{proof}
  Since the $(\cweight,\kappa_{1})$-supremum assumptions are fulfilled, the $(\cweight,\kappa_{1})$-Sobolev assumptions are fulfilled. Turning to the
  coefficients of the equation, note that $\mcX=0$ and that $\hal=-\theta^{-2}\mKG^{2}$. Due to the proof of Lemma~\ref{lemma:halestimateKG}, it follows
  that for $j\leq\kappa_{1}$,
  \[
  \|\hal(t,\cdot)\|_{C^{j}_{\weight_{0}}(\bM)}\leq C\theta_{0,-}^{-2}e^{2\eSpe\tau}\ldr{\tau}^{j\cweight}
  \]
  for all $\tau\leq 0$, where $C$ only depends on $\mKG$, $c_{\cweight,\kappa_{1}}$ and $(\bM,\bge_{\refer})$. Here $\eSpe=\e_{\Spe}/(3K_{\rovar})$ is defined in
  the statement of the proposition. In particular, (\ref{eq:Sobcoefflassumptions})
  and (\ref{eq:coefflassumptions}) are satisfied with $l=\kappa_{1}$. Moreover, since $\tau_{c}=0$, (\ref{eq:haldecayintro}) is satisfied with $d_{\a}$
  only depending on $\mKG$, $c_{\cweight,\kappa_{1}}$, $(\bM,\bge_{\refer})$ and a lower bound on $\theta_{0,-}$. Finally, note that (\ref{eq:qconvergenceintro})
  holds with $d_{q}$ depending only on $c_{\robas}$, $\e_{q}$ and $(\bM,\bge_{\refer})$; in order to obtain this conclusion, we appeal to
  (\ref{eq:DeltavarrhorelvariationEiintro}). Due to these observations, Proposition~\ref{prop:asvelocityKGlikeeq} applies and yields the statement of
  the proposition. 
\end{proof}

\section{Energy estimates in causally localised regions}\label{section:energyestimatesincausallylocalisedregions}

The estimates obtained in Propositions~\ref{prop:basicenergyestimateintro} and \ref{prop:EnergyEstimateSobolevAssumptionsintro} are crude in
that they only state that the energies do not grow faster than exponentially. However, there is one very important advantage of
these estimates, namely that the exponential rate does not depend on the number of derivatives. Due to this fact and the fact that the geometry is
silent, it is possible to improve the estimates in causally localised regions. In order to state the results, we first need to define the regions
in which the estimates hold. 

\begin{lemma}\label{lemma:pastinextendiblecausalcurvelocintro}
  Given that the standard assumptions are satisfied, cf. Definition~\ref{def:standardassumptions}, let $\tau$ be defined by
  (\ref{eq:taudefinitionintro}). Let $\g:(s_{-},s_{+})\rightarrow \bM\times I$ be a future pointing and past inextendible causal curve. Writing
  $\g(s)=[\bga(s),\g^{0}(s)]$, where $\bga(s)\in \bM$, there is an $\bx_{\g}\in\bM$ such that
  \[
  \lim_{s\rightarrow s_{-}+}d(\bga(s),\bx_{\g})=0,
  \]
  where $d$ is the topological metric induced on $\bM$ by $\bge_{\refer}$. Moreover, there is a constant $K_{A}$ such that if $\bx_{\g}=\bx_{0}$
  (where $\bx_{0}\in \bM$ is the reference point introduced in connection with (\ref{eq:taudefinitionintro})), then
  \begin{equation}\label{eq:Aplusdefintroduction}
    A^{+}(\g):=\{(\bx,t)\in \bM\times I: d(\bx,\bx_{\g})\leq K_{A}e^{\e_{\Spe}\tau(t)}\}
  \end{equation}
  \index{$\a$Aa@Notation!Sets!$A^{+}(\g)$}%
  has the property that $J^{+}(\g)\cap J^{-}(\bM_{t_{0}})\subset A^{+}(\g)$. Here $K_{A}$ only depends on $c_{\robas}$, $c_{\chi,2}$, $(\bM,\bge_{\refer})$
  and a lower bound on $\theta_{0,-}$.
\end{lemma}
\begin{remark}
  In what follows, it is also, given a $t_{c}\leq t_{0}$, convenient to use the notation \index{$\a$Aa@Notation!Sets!$A^{+}_{c}(\g)$}%
  \[
  A^{+}_{c}(\g):=\{(\bx,t)\in A^{+}(\g):t\leq t_{c}\}.
  \]  
\end{remark}
\begin{proof}
  The statement of the lemma follows from Lemma~\ref{lemma:pastinextendiblecausalcurveloc}, Remark~\ref{remark:bgaconvtobxbgaloc} and the observations
  made in connection with (\ref{eq:Aplusgammadef}).
\end{proof}
There is no restriction in assuming $\bx_{\g}=\bx_{0}$, and therefore we do so in what follows. Moreover, we focus on deriving estimates in
regions of the form $A^{+}_{c}(\g)$. Before stating the result concerning the evolution of the energy in $A^{+}_{c}(\g)$, it is of interest to develop
some intuition. Considering (\ref{eq:equationintermsofcanonicalframe}) and keeping in mind that the geometry is silent (which implies that $e^{-\mu_{A}}$
converges to zero exponentially in $\tau$-time), it is natural to discard the $X_{A}$-derivatives; i.e., to omit the spatial derivatives. Note that this
idea is in accordance with the BKL conjecture (which we briefly describe in Subsection~\ref{subsection:BKLConjecture}). In case $f=0$, the corresponding
(preliminary) model equation is
\begin{equation}\label{eq:modelequationfirststepintro}
  -\hU^{2}u+Z^{0}\hU u+\hal u=0.
\end{equation}
On the other hand, due to (\ref{eq:hUvarrhoident}) and (\ref{eq:rodivchiestimpr}), $\hU(\varrho)$ equals $1$ up to an exponentially small error. Moreover,
$\tau=\varrho(\bx_{0},t)$ so that, in $A^{+}(\g)$, $\tau$ and $\varrho$ should be comparable. Naively, it should thus be possible to replace $\hU$ with
$\d_{\tau}$. Finally, since the region $A^{+}(\g)$ shrinks exponentially, it should be possible to replace $Z^{0}$ and $\hal$ with localised versions of the
coefficients, defined as follows:
\begin{equation}\label{eq:Zzerolochallocintro}
Z^{0}_{\roloc}(t):=Z^{0}(\bx_{0},t),\ \ \
\hal_{\roloc}(t):=\hal(\bx_{0},t). 
\end{equation}
\index{$\a$Aa@Notation!Functions!$Z^{0}_{\roloc}$}%
\index{$\a$Aa@Notation!Functions!$\hal_{\roloc}$}%
In some respects, it would be more intuitive to evaluate the coefficients along the causal curve $\g$, and we could equally well do so. The above ideas lead
to the model equation
\[
-u_{\tau\tau}+Z^{0}_{\roloc}u_{\tau}+\hal_{\roloc}u=0.
\]
This is a system of ODE's which can be written in first order form as:
\begin{equation}\label{eq:PsiAdefintro}
  \Psi_{\tau}=A\Psi,\ \ \
  \Psi:=\left(\begin{array}{c} u \\ u_{\tau}\end{array}\right),\ \ \
  A:=\left(\begin{array}{cc} 0 & \Id \\ \hal_{\roloc} & Z^{0}_{\roloc}\end{array}\right).
\end{equation}
The naive expectation concerning the growth/decay of the solution is then that it should be determined by the flow associated with $\Psi_{\tau}=A\Psi$. To
be more specific, define the matrix valued function $\Phi$ by
\begin{equation}\label{eq:Phidefintro}
  \Phi_{\tau}=A\Phi,\ \ \
  \Phi(\tau;\tau)=\Id.
\end{equation}
Assume now that there are constants $C_{A}$, $d_{A}$ and $\varpi_{A}$ such that if $s_{1}\leq s_{2}\leq 0$, then 
\begin{equation}\label{eq:Phinormbasasslocintro}
 \|\Phi(s_{1};s_{2})\|\leq C_{A}\ldr{s_{2}-s_{1}}^{d_{A}}e^{\varpi_{A}(s_{1}-s_{2})}. 
\end{equation}
The assumptions we make in these notes are such that $\|A\|$ is bounded; cf. Definition~\ref{def:supmfulassumptions}, (\ref{eq:coefflassumptions})
and (\ref{eq:Zzdefintro}). For this reason, there are $C_{A}$, $d_{A}$ and $\varpi_{A}$ such that (\ref{eq:Phinormbasasslocintro}) holds. However, how
well the corresponding numbers reflect the actual behaviour of solutions is unclear. In practice, it is natural to take the supremum of all the
$\varpi_{A}$ such that there is a $C_{A}$ and a $d_{A}$ with the properties that (\ref{eq:Phinormbasasslocintro}) holds for all $s_{1}\leq s_{2}\leq 0$.
Any number strictly smaller than this supremum would then be a valid choice of $\varpi_{A}$. Note also that $C_{A}$, $d_{A}$ and $\varpi_{A}$ depend on
$\bx_{0}$, and as examples below will illustrate, the optimal choice of $\varpi_{A}$ can typically be expected to depend discontinuously on $\bx_{0}$.

\begin{thm}\label{thm:asgrowthofenergyintro}
  Let $0\leq \cweight\in\ro$, $\weight_{0}=(0,\cweight)$ and $\weight=(\cweight,\cweight)$. Assume that the standard assumptions, cf.
  Definition~\ref{def:standardassumptions}, are satisfied. Let $\kappa_{0}$ be the smallest integer which is strictly larger than $n/2$;
  $\kappa_{1}=\kappa_{0}+1$; $\kappa_{1}\leq k\in\zo$; and $l=k+\kappa_{0}$. Assume the $(\cweight,k)$-supremum and the $(\cweight,l)$-Sobolev
  assumptions to be satisfied; and that there are constants $c_{\coeff,k}$ and $s_{\coeff,l}$ such that (\ref{eq:Sobcoefflassumptions}) holds and
  such that (\ref{eq:coefflassumptions}) holds with $l$ replaced by $k$. Let $\g$ and $\bx_{\g}$ be as in
  Lemma~\ref{lemma:pastinextendiblecausalcurvelocintro}, and assume that $\bx_{0}=\bx_{\g}$. Assume, finally, that
  (\ref{eq:equationintermsofcanonicalframe}) is satisfied with vanishing right hand side; and that if $A$ is defined by (\ref{eq:PsiAdefintro})
  and $\Phi$ is defined by (\ref{eq:Phidefintro}), then there are constants $C_{A}$, $d_{A}$ and $\varpi_{A}$ such that (\ref{eq:Phinormbasasslocintro})
  holds. Let $c_{0}$ be the constant appearing in the statement of Proposition~\ref{prop:basicenergyestimateintro} and $\tc_{0}$ be defined by
  \begin{equation}\label{eq:tczdefintro}
    \tc_{0}:=c_{0}+1-1/n-\e_{\Spe}.
  \end{equation}
  Let $m_{0}$ be the smallest integer greater than or equal to 
  \begin{equation}\label{eq:mzdefintro}
    \max\left\{1,\frac{2\varpi_{A}+\tc_{0}}{2\e_{\Spe}}+\frac{1}{2}\right\}. 
  \end{equation}
  Assuming $k\geq m_{0}$ and letting $m_{1}:=m_{0}+\kappa_{0}$, the estimate 
  \begin{equation}\label{eq:melindassfslocfinalstmtintro}
    \begin{split}
      \me_{m}^{1/2} \leq & C_{m,a}\ldr{\tau-\tau_{c}}^{\kappa_{m,a}}\ldr{\tau}^{\lambda_{m,a}}e^{\varpi_{A}(\tau-\tau_{c})}\hGe_{m+m_{1}}^{1/2}(\tau_{c})
    \end{split}    
  \end{equation}
  holds on $A^{+}_{c}(\g)$ for $0\leq m\leq k-m_{0}$, where $C_{m,a}$ only depends on $s_{\cweight,l}$, $s_{\coeff,l}$, $c_{\cweight,k}$, $c_{\coeff,k}$,
  $m_{\ros}$, $d_{\a}$ (in case $\iota_{b}\neq 0$), $C_{A}$, $d_{A}$, $(\bM,\bge_{\refer})$ and a lower bound on $\theta_{0,-}$; $\kappa_{m,a}$ only
  depends on $d_{A}$, $n$, $m$ and $k$; $\lambda_{m,a}$ only depends on $\cweight$, $n$, $m$ and $k$; and we use the notation introduced in
  (\ref{eq:hGekdefintro}). Moreover, $\kappa_{0,a}=d_{A}$ and $\lambda_{0,a}=0$. 
\end{thm}
\begin{remark}
  Note, in particular, that $\me_{0}^{1/2}\leq C\ldr{\tau-\tau_{c}}^{d_{A}}e^{\varpi_{A}(\tau-\tau_{c})}$ on $A^{+}_{c}(\g)$, which, given
  (\ref{eq:Phinormbasasslocintro}), is the best estimate one could hope for. 
\end{remark}
\begin{proof}
  The statement is a direct consequence of Theorem~\ref{thm:asgrowthofenergy}.
\end{proof}

It is important to note that the above result is associated with a substantial loss of derivatives. Moreover, considering (\ref{eq:mzdefintro}), it is clear
that the loss tends to infinity as $\e_{\Spe}\rightarrow 0+$. In other words, in the limit that the causal structure is no longer silent, the  loss of
derivatives tends to infinity. This could be a deficiency of the method. However, it is of interest to note that a similar phenomenon appears in at
least two other contexts. In \cite{sta}, the author specifies smooth data on the singularity in the $\sn{3}$- and $\sn{2}\times\sn{1}$-Gowdy vacuum
settings. However, the closer the data are to those of a solution with a horizon, the higher the order of the correction terms that need to be added
to the unknowns in order to construct a solution; cf., in particular, \cite[(52)--(54), p.~4501]{sta} and the adjacent text. In \cite{olp}, the author
specifies
initial data on compact Cauchy horizons for wave equations. Again, the results are in the smooth setting. Moreover, the arguments use families of
approximate solutions that are defined using gradually higher numbers of derivatives of the data on the horizon. Due to these examples, it is tempting to
suggest that horizons are associated with a possibly infinite loss of derivatives. Moreover, since generic solutions are, according to the BKL proposal,
expected to behave locally like Bianchi type IX solutions; since Bianchi type IX solutions are supposed to be well approximated by the Kasner map;
and since generic orbits of the Kasner map have the special points (which correspond to solutions with compact Cauchy horizons) as limit points, it is
tempting to conjecture that the loss of derivatives is a generic phenomenon, so that, in the generic setting, it is necessary to restrict one's attention
to the smooth setting. 

On the other hand, the results \cite{sta,olp} are concerned with specifying data on the singularity. This could, potentially, be the cause of the loss of
derivatives in these settings. Moreover, the loss of derivatives in the above result could perhaps be avoided if more detailed assumptions are made
concerning the asymptotic geometry; note, e.g., that optimal energy estimates without a loss of derivatives are obtained in \cite{finallinsys} (on the
other hand, the optimal energy estimates without a loss of derivatives can, in general, be expected to be worse (in terms of growth/decay) than the
optimal energy estimates with a loss of derivatives). 

\subsection{Coefficients converging along a causal curve}
The case that the matrix valued function $A$, introduced in (\ref{eq:PsiAdefintro}), converges is of particular interest. In order to state the
corresponding results, we need to introduce the following terminology. 
\begin{definition}\label{def:SpRspdefintro}
  Given $A\in\Mn{k}{\co}$, let $\Spe A$
  \index{$\a$Aa@Notation!Sets!$\Spe A$}%
  denote the set of eigenvalues of $A$. Moreover, let
  \[
  \varpi_{\max}(A):=\sup\{\mathrm{Re}\lambda\ |\ \lambda\in \Spe A\},\ \ \
  \varpi_{\min}(A):=\inf\{\mathrm{Re}\lambda\ |\ \lambda\in \Spe A\}.
  \]
  \index{$\a$Aa@Notation!Functions!$\varpi_{\max}$}%
  \index{$\a$Aa@Notation!Functions!$\varpi_{\min}$}%
  In addition, if $\varpi\in \{\mathrm{Re}\lambda\ |\ \lambda\in \Spe A\}$, then $d_{\max}(A,\varpi)$
  \index{$\a$Aa@Notation!Functions!$d_{\max}$}%
  is defined to be the largest dimension of a Jordan block corresponding to an eigenvalue of $A$ with real part $\varpi$. 
\end{definition}
\begin{remark}
  Here $\Mn{k}{\mathbb{K}}$
  \index{$\a$Aa@Notation!Sets!$\Mn{k}{\mathbb{K}}$}%
  denotes the set of $k\times k$-matrices with coefficients in the field $\mathbb{K}$. 
\end{remark}

\begin{cor}\label{cor:Asymptoticallyconstantintro}
  Assume that the conditions of Theorem~\ref{thm:asgrowthofenergyintro} are satisfied. Let $A$ be the matrix defined by (\ref{eq:PsiAdefintro}) and
  consider it to be a function of $\tau$. Assume that there is an $A_{0}\in \Mn{2m_{\ros}}{\ro}$ such that $A(\tau)\rightarrow A_{0}$ as
  $\tau\rightarrow -\infty$. Let $\varpi_{A}=\varpi_{\min}(A_{0})$ and $d_{A}:=d_{\max}(A_{0},\varpi_{A})-1$. Let $\xi(\tau):=\ldr{\tau}^{d_{A}}\|A(\tau)-A_{0}\|$.
  If $\|\xi\|_{1}:=\|\xi\|_{L^{1}(-\infty,0]}<\infty$, then there is a constant $C_{A}$, depending only on $A_{0}$ and $\|\xi\|_{1}$, such that
  (\ref{eq:Phinormbasasslocintro}) holds. In particular, (\ref{eq:melindassfslocfinalstmtintro}) holds with $\varpi_{A}=\varpi_{\min}(A_{0})$.
\end{cor}
\begin{remark}
  One particular consequence of the corollary is that the energy growth is determined by the limit of the coefficients, assuming this limit exists and
  the convergence is sufficiently fast. Note also that the limit could equally well be calculated along $\g$, since the spatial variation of the
  coefficients in $A^{+}(\g)$ is exponentially small.
\end{remark}
\begin{remark}
  It is important to note that we only assume the coefficients to converge as $\tau\rightarrow-\infty$ for one fixed $\bx_{0}\in\bM$. In particular,
  the coefficients need not converge, even pointwise, in a punctured neighbourhood of $\bx_{0}$, and even if they do converge, the limiting function need
  not be continuous. 
\end{remark}
\begin{remark}
  It is of interest to ask if $\varpi_{A}$ and $d_{A}$ obtained in the corollary are optimal. Below, we demonstrate that if the rate of convergence of $A$
  to $A_{0}$ is exponential, then the rate is optimal.
\end{remark}
\begin{proof}
  The statement follows from Theorem~\ref{thm:asgrowthofenergy} and Corollary~\ref{cor:Asymptoticallyconstant}. 
\end{proof}

\section{Asymptotics in causally localised regions}

In Theorem~\ref{thm:asgrowthofenergyintro}, we assume neither $Z^{0}_{\roloc}$ nor $\hal_{\roloc}$ to converge. In
Corollary~\ref{cor:Asymptoticallyconstantintro} we assume them to converge at a specific polynomial rate. This allows us to estimate the growth/decay of the
energies in terms of the growth/decay associated with an asymptotic system of ODE's. In order to obtain more detailed asymptotic information, it is, however,
convenient to assume the coefficients to converge exponentially. In order to state the relevant results, we first need to introduce additional terminology;
cf. \cite[Definition~4.7, p.~48]{finallinsys}.

\begin{definition}\label{def:defofgeneigenspintro}
  Let $1\leq k\in\zo$, $B\in\Mn{k}{\co}$ and $P_{B}(X)$ be the characteristic polynomial of $B$. Then
  \[
  P_{B}(X)=\prod_{\lambda\in\Spe B}(X-\lambda)^{k_{\lambda}},
  \]
  where $1\leq k_{\lambda}\in\zo$. Moreover, given $\lambda\in \Spe B$, the \textit{generalised eigenspace of $B$ corresponding
    to $\lambda$},
  \index{Generalised eigenspace of a matrix}%
  denoted $E_{\lambda}$, is defined by
  \begin{equation}\label{eq:Elambdadefintro}
    E_{\lambda}:=\ker (B-\lambda\Id_{k})^{k_{\lambda}},
  \end{equation}
  where $\Id_{k}$ denotes the $k\times k$-dimensional identity matrix. If $J\subseteq\ro$ is an interval, then the
  $J$-\textit{generalised eigenspace of}
  \index{J@$J$-generalised eigenspace of a matrix}%
  $B$, denoted $E_{B,J}$, is the subspace of $\cn{k}$ defined to be the direct sum of the generalised eigenspaces
  of $B$ corresponding to eigenvalues with real parts belonging to $J$ (in case there are no eigenvalues with real part belonging to $J$, then $E_{B,J}$
  is defined to be $\{0\}$). Finally, given $0<\b\in\ro$, the \textit{first generalised eigenspace in the $\b$, $B$-decomposition of $\cn{k}$}, denoted
  $E_{B,\b}$, is defined to be $E_{B,J_{\b}}$, where $J_{\b}:=(\varpi-\b,\varpi]$ and $\varpi:=\varpi_{\max}(B)$; cf. Definition~\ref{def:SpRspdefintro}.
\end{definition}
\begin{remark}\label{remark:EBJrealintro}
  In case $B\in\Mn{k}{\ro}$, the vector spaces $E_{B,J}$ have bases consisting of vectors in $\rn{k}$. The reason for this is that if $\lambda$ is an
  eigenvalue of $B$ with $\mathrm{Re}\lambda\in J$, then $\lambda^{*}$ (the complex conjugate of $\lambda$) is an eigenvalue of $B$  with
  $\mathrm{Re}\lambda^{*}\in J$. Moreover, if $v\in E_{\lambda}$, then $v^{*}\in E_{\lambda^{*}}$. Combining the bases of $E_{\lambda}$ and $E_{\lambda^{*}}$, we
  can thus construct a basis of the direct sum of these two vector spaces which consists of vectors in $\rn{k}$. 
\end{remark}

\begin{thm}\label{thm:leadingorderasymptoticszerointro}
  Let $0\leq \cweight\in\ro$, $\weight_{0}=(0,\cweight)$ and $\weight=(\cweight,\cweight)$. Assume that the standard assumptions, cf.
  Definition~\ref{def:standardassumptions}, are satisfied. Let $\kappa_{0}$ be the smallest integer which is strictly larger than $n/2$;
  $\kappa_{1}=\kappa_{0}+1$; $\kappa_{1}\leq k\in\zo$; and $l=k+\kappa_{0}$. Assume the $(\cweight,k)$-supremum and the $(\cweight,l)$-Sobolev
  assumptions to be satisfied; and that there are constants $c_{\coeff,k}$ and $s_{\coeff,l}$ such that (\ref{eq:Sobcoefflassumptions}) holds and
  such that (\ref{eq:coefflassumptions}) holds with $l$ replaced by $k$. Assume, moreover, that (\ref{eq:equationintermsofcanonicalframe}) is satisfied
  with vanishing right hand side and that $\varrho(\bx_{0},t)\rightarrow -\infty$ as $t$ tends to the left endpoint of $I_{-}$; cf. (\ref{eq:Iminusdef}).
  Let $\g$ and $\bx_{\g}$ be as in Lemma~\ref{lemma:pastinextendiblecausalcurvelocintro}, and assume
  that $\bx_{0}=\bx_{\g}$. Assume, finally, that there are $Z^{0}_{\infty},\hal_{\infty}\in\Mn{m_{\ros}}{\ro}$ and constants $\e_{A}>0$, $c_{\rem}\geq 0$
  such that
  \begin{equation}\label{eq:Zzerohalexpconvergenceintro}
    [\|Z^{0}_{\roloc}(\tau)-Z^{0}_{\infty}\|^{2}+\|\hal_{\roloc}(\tau)-\hal_{\infty}\|^{2}]^{1/2}\leq c_{\rem}e^{\e_{A}\tau}
  \end{equation}
  for all $\tau\leq 0$, where $Z^{0}_{\roloc}$ and $\hal_{\roloc}$ are introduced in (\ref{eq:Zzerolochallocintro}). Let 
  \begin{equation}\label{eq:AzeroAremdefintro}
    A_{0}:=\left(\begin{array}{cc} 0 & \Id \\ \hal_{\infty} & Z^{0}_{\infty}\end{array}\right),
  \end{equation}
  $\varpi_{A}:=\varpi_{\min}(A_{0})$ and $d_{A}:=d_{\max}(A_{0},\varpi_{A})-1$. Let $m_{0}$ be defined as in the statement of
  Theorem~\ref{thm:asgrowthofenergyintro} and assume $k>m_{0}$. Let, moreover, $\b:=\min\{\e_{A},\e_{\Spe}\}$, $J_{a}:=[\varpi_{A},\varpi_{A}+\b)$,
  $E_{a}:=E_{A_{0},J_{a}}$ and 
  \begin{equation}\label{eq:Vdefintro}
    V:=\left(\begin{array}{c} u \\ \hU u\end{array}\right).
  \end{equation}
  Then, given $\tau_{c}\leq 0$, there is a unique $V_{\infty,a}\in E_{a}$ with $V_{\infty,a}\in\rn{2m_{s}}$ such that
  \begin{equation}\label{eq:VAplusestimateintro}
    \left|V-e^{A_{0}(\tau-\tau_{c})}V_{\infty,a}\right|
    \leq  C_{a}\ldr{\tau_{c}}^{\eta_{b}}\hGe_{l}^{1/2}(\tau_{c})\ldr{\tau-\tau_{c}}^{\eta_{a}}e^{(\varpi_{A}+\b)(\tau-\tau_{c})}
  \end{equation}
  on $A^{+}_{c}(\g)$, where $C_{a}$ only depends on $s_{\cweight,l}$, $s_{\coeff,l}$, $c_{\cweight,k}$, $c_{\coeff,k}$, $d_{\a}$ (in case $\iota_{b}\neq 0$), $A_{0}$,
  $c_{\rem}$, $\e_{A}$, $(\bM,\bge_{\refer})$ and a lower bound on $\theta_{0,-}$; and $\eta_{a}$, $\eta_{b}$ only depend on $\cweight$, $A_{0}$, $n$ and $k$.
  Moreover,
  \begin{equation}\label{eq:Vinfestimatefinalotintro}
    |V_{\infty,a}| \leq C_{a}\ldr{\tau_{c}}^{\eta_{b}}\hGe_{l}^{1/2}(\tau_{c}),
  \end{equation}
  where $C_{a}$ and $\eta_{b}$ have the same dependence as in the case of (\ref{eq:VAplusestimateintro}). 
\end{thm}
\begin{remark}
  Note that $e^{A_{0}(\tau-\tau_{c})}V_{\infty,a}$ is a solution to the model equation
  \begin{equation}\label{eq:asymptoticmodelequation}
    -u_{\tau\tau}+Z^{0}_{\infty}u_{\tau}+\hal_{\infty}u=0
  \end{equation}
  written in first order form. On a heuristic level, the estimate (\ref{eq:VAplusestimateintro}) thus says that the leading order behaviour
  of the solution in $A^{+}_{c}(\g)$ is given by a solution to the model equation (\ref{eq:asymptoticmodelequation}). 
\end{remark}
\begin{remark}
  Due to the proof, the function $V$ appearing in (\ref{eq:VAplusestimateintro}) can be replaced by $\Psi$ introduced in (\ref{eq:PsiAdefintro}).
\end{remark}
\begin{remark}\label{remark:Vinftyimprovementintro}
  The estimate (\ref{eq:VAplusestimateintro}) can be improved in that there is a $V_{\infty}\in \rn{2m_{s}}$ such that
  \begin{equation}\label{eq:VAplusestimateimprovedintro}
    \left|V-e^{A_{0}(\tau-\tau_{c})}V_{\infty}\right|
    \leq  C_{a}\ldr{\tau_{c}}^{\eta_{b}}e^{\b\tau_{c}}\hGe_{l}^{1/2}(\tau_{c})\ldr{\tau-\tau_{c}}^{\eta_{a}}e^{(\varpi_{A}+\b)(\tau-\tau_{c})}
  \end{equation}
  on $A^{+}_{c}(\g)$, where $C_{a}$, $\eta_{a}$ and $\eta_{b}$ have the same dependence as in the case of (\ref{eq:VAplusestimateintro}). However, the
  corresponding $V_{\infty}$ is not unique. Nevertheless, $V_{\infty}$ can be chosen so that it satisfies (\ref{eq:Vinfestimatefinalotintro}) with $V_{\infty,a}$
  replaced by $V_{\infty}$. On the other hand, letting $\tau_{c}$ be close enough to $-\infty$, the factor $C_{a}\ldr{\tau_{c}}^{\eta_{b}}e^{\b\tau_{c}}$ appearing
  on the right hand side of (\ref{eq:VAplusestimateimprovedintro}) can be chosen to be as small as we wish. 
\end{remark}
\begin{proof}
  The statement is an immediate consequence of Theorem~\ref{thm:leadingorderasymptoticszero}.
\end{proof}

\subsection{Asymptotics of the higher order derivatives}\label{ssection:higherorderderivativesintro}

Due to the fact that the causal structure is silent, (\ref{eq:modelequationfirststepintro}) is a natural model equation for the asymptotic behaviour.
This equation is the basis for the localised energy estimates obtained in Theorem~\ref{thm:asgrowthofenergyintro} and the
asymptotics derived in Theorem~\ref{thm:leadingorderasymptoticszerointro}. However, it is also of interest to derive the asymptotic behaviour for
the higher order derivatives; i.e., for $E_{\bfI}u$ and $\hU E_{\bfI}u$. In order to do so, we first need to commute (\ref{eq:modelequationfirststepintro})
with $E_{\bfI}$. However, commuting $E_{i}$ with $\hU$ leads to terms that cannot be neglected. Nevertheless, in the general spirit of neglecting spatial
derivatives, it is possible to derive a model equation of the form
\begin{equation}\label{eq:modelequationhigherorderderintro}
  -\d_{\tau}^{2}E_{\bfI}u+Z^{0}_{\infty}\d_{\tau} E_{\bfI}u+\hal_{\infty}E_{\bfI}u=L_{\pre,\bfI}u,
\end{equation}
where $L_{\pre,\bfI}u$ can, roughly speaking, be written in the form
\begin{equation}\label{eq:LbfIcompdefinitionpreintro}
  L_{\pre,\bfI}u=\textstyle{\sum}_{|\bfJ|<|\bfI|}\textstyle{\sum}_{m=0}^{2}L_{\pre,\bfI,\bfJ}^{m}\d_{\tau}^{m}E_{\bfJ}u.
\end{equation}
We refer the reader to Section~\ref{section:asymptoticshigherorderderivatives} below for a more detailed discussion and justification. A simplifying
feature of the system given by (\ref{eq:modelequationhigherorderderintro}) and (\ref{eq:LbfIcompdefinitionpreintro}) is that it is hierarchical
in the following sense. In case $|\bfI|=0$, the right hand side of (\ref{eq:modelequationhigherorderderintro}) vanishes, and it is sufficient to solve
the model equation (\ref{eq:asymptoticmodelequation}). This yields $u$, $u_{\tau}$ and, via (\ref{eq:asymptoticmodelequation}), $u_{\tau\tau}$. Thus
$L_{\pre,\bfI}u$ can be calculated for $|\bfI|=1$, so that the right hand side of (\ref{eq:modelequationhigherorderderintro}) can be considered to be
given for $|\bfI|=1$. Thus $E_{\bfI}u$, $E_{\bfI}u_{\tau}$ and $E_{\bfI}u_{\tau\tau}$ can be calculated by solving (\ref{eq:modelequationhigherorderderintro})
where the right hand side is given. This process can be continued to any order.

When deriving asymptotics, the above perspective is sufficient. However, below we are also interested in specifying asymptotics. In that context, the fact
that the different $E_{\bfI}u$ are not independent causes problems. In fact, $E_{\bfI}u$ can be expressed in terms of $E_{\omega}u$ for $\rn{n}$-multiindices
$\omega$ satisfying $|\omega|\leq |\bfI|$; if $\omega$ is an $\rn{n}$-multiindex, we here use the notation
\[
E_{\omega}u:=E_{1}^{\omega_{1}}\cdots E_{n}^{\omega_{n}}u.
\]
Again, we refer the reader to Section~\ref{section:asymptoticshigherorderderivatives} below for details. This leads, roughly speaking, to the model system
\begin{equation}\label{eq:modelequationhigherorderderivativesintro}
  -\d_{\tau}^{2}U_{\bfI}+Z^{0}_{\infty}\d_{\tau}U_{\bfI}+\hal_{\infty}U_{\bfI}=\hL_{\bfI},
\end{equation}
where 
\begin{equation}\label{eq:LbfIcompdefinitionintro}
  \hL_{\bfI}(\tau):=\textstyle{\sum}_{|\omega|<|\bfI|}\textstyle{\sum}_{m=0}^{2}L_{\bfI,\omega}^{m}(\bx_{0},\tau)\d_{\tau}^{m}U_{\omega}(\tau)
\end{equation}
and $\omega$ are $\rn{n}$-multiindices. Here $L_{\bfI,\omega}^{m}(\bx_{0},\cdot)$ can be calculated in terms of the geometry, the coefficients of the equation
and the structure constants of the frame $\{E_{i}\}$; cf. Section~\ref{section:asymptoticshigherorderderivatives} below. Moreover,
$U_{\bfI}$ should be thought of as $(E_{\bfI}u)(\bx_{0},\cdot)$ and $U_{\omega}$ should be thought of as $(E_{\omega}u)(\bx_{0},\cdot)$. Again, the system given by
(\ref{eq:modelequationhigherorderderivativesintro}) and (\ref{eq:LbfIcompdefinitionintro}) is hierarchical in the above sense. The solutions can be written
\[
\left(\begin{array}{c} U_{\bfI}(\tau) \\ (\d_{\tau}U_{\bfI})(\tau) \end{array}\right)
=e^{A_{0}(\tau-\tau_{c})}X_{\bfI}+\int_{\tau}^{\tau_{c}}e^{A_{0}(\tau-s)}\left(\begin{array}{c} 0 \\ \hL_{\bfI}(s) \end{array}\right)ds,
\]
where $X_{\bfI}\in\rn{2m_{\ros}}$. For this reason, the goal is to prove that for a suitable choice of $X_{\bfI}$, the difference 
\[
\left(\begin{array}{c} E_{\bfI}u \\ \hU E_{\bfI}u \end{array}\right)
-e^{A_{0}(\tau-\tau_{c})}X_{\bfI}-\int_{\tau}^{\tau_{c}}e^{A_{0}(\tau-s)}\left(\begin{array}{c} 0 \\ \hL_{\bfI}(s) \end{array}\right)ds
\]
is small in $A^{+}_{c}(\g)$.

\begin{thm}\label{thm:leadingorderasymptoticszerohointro}
  Let $0\leq \cweight\in\ro$, $\weight_{0}=(0,\cweight)$ and $\weight=(\cweight,\cweight)$. Assume that the standard assumptions, cf.
  Definition~\ref{def:standardassumptions}, are satisfied. Let $\kappa_{0}$ be the smallest integer which is strictly larger than $n/2$;
  $\kappa_{1}=\kappa_{0}+1$; $\kappa_{1}\leq k\in\zo$; and $l=k+\kappa_{0}$. Assume the $(\cweight,k)$-supremum and the $(\cweight,l)$-Sobolev
  assumptions to be satisfied; and that there are constants $c_{\coeff,k}$ and $s_{\coeff,l}$ such that (\ref{eq:Sobcoefflassumptions}) holds and
  such that (\ref{eq:coefflassumptions}) holds with $l$ replaced by $k$. Assume that (\ref{eq:equationintermsofcanonicalframe}) is
  satisfied with vanishing right hand side and that $\varrho(\bx_{0},t)\rightarrow -\infty$ as $t$ tends to the left endpoint of $I_{-}$;
  cf. (\ref{eq:Iminusdef}). Let $\g$ and $\bx_{\g}$ be as in Lemma~\ref{lemma:pastinextendiblecausalcurvelocintro}, and assume
  that $\bx_{0}=\bx_{\g}$. Assume, finally, that there are $Z^{0}_{\infty},\hal_{\infty}\in\Mn{m_{\ros}}{\ro}$ and constants $\e_{A}>0$, $c_{\rem}\geq 0$
  such that (\ref{eq:Zzerohalexpconvergenceintro}) holds for all $\tau\leq 0$. Let $A_{0}$ be defined by (\ref{eq:AzeroAremdefintro}). Let,
  moreover, $\varpi_{A}:=\varpi_{\min}(A_{0})$ and $d_{A}:=d_{\max}(A_{0},\varpi_{A})-1$. Let $m_{0}$ be defined as in the statement of
  Theorem~\ref{thm:asgrowthofenergyintro} and assume $k>m_{0}+1$. Let, moreover, $\b:=\min\{\e_{A},\e_{\Spe}\}$, $J_{a}:=[\varpi_{A},\varpi_{A}+\b)$,
  $E_{a}:=E_{A_{0},J_{a}}$, $V$ be defined by (\ref{eq:Vdefintro}) and 
  \[
  V_{\bfI}:=\left(\begin{array}{c} E_{\bfI}u \\ \hU E_{\bfI}u\end{array}\right).
  \]
  Fix $\tau_{c}\leq 0$, let $V_{\infty,a}$ be defined as in the statement of Theorem~\ref{thm:leadingorderasymptoticszerointro} and define
  $U_{0,m}\in C^{\infty}(\ro,\rn{m_{\ros}})$, $m=0,1,2$, by
  \begin{equation}\label{eq:Uzeromdefintro}
    \left(\begin{array}{c} U_{0,0}(\tau) \\ U_{0,1}(\tau)\end{array}\right):=e^{A_{0}(\tau-\tau_{c})}V_{\infty,a},\ \ \
    U_{0,2}(\tau):=Z^{0}_{\infty}U_{0,1}(\tau)+\hal_{\infty}U_{0,0}(\tau).
  \end{equation}
  Let $1\leq j\leq k-m_{0}-1$ and assume that $U_{\bfJ,m}$ has been defined for $|\bfJ|<j$ and $m=0,1,2$ (for $\bfJ=0$, these functions are defined by
  (\ref{eq:Uzeromdefintro}) and for $|\bfJ|>0$, they are defined inductively by (\ref{eq:UbfIzeroandonedefintro}) and (\ref{eq:UbfItwodefintro}) below).
  Let $\bfI$ be such that $|\bfI|=j$ and define $\sfL_{\bfI}$ by
  \[
  \sfL_{\bfI}(\tau):=\textstyle{\sum}_{|\omega|<|\bfI|}\textstyle{\sum}_{m=0}^{2}L_{\bfI,\omega}^{m}(\bx_{0},\tau)U_{\omega,m}(\tau).
  \]
  Then there is a unique $V_{\bfI,\infty,a}\in E_{a}$ with
  $V_{\bfI,\infty,a}\in\rn{2m_{s}}$ such that
  \begin{equation}\label{eq:VAplusestimatehointro}
    \begin{split}
      & \left|V_{\bfI}-e^{A_{0}(\tau-\tau_{c})}V_{\bfI,\infty,a}-\int_{\tau}^{\tau_{c}}e^{A_{0}(\tau-s)}\left(\begin{array}{c} 0 \\ \sfL_{\bfI}(s)\end{array}\right)ds\right|\\
      \leq & C_{a}\ldr{\tau_{c}}^{\eta_{b}}\hGe_{l}^{1/2}(\tau_{c})\ldr{\tau-\tau_{c}}^{\eta_{a}}e^{(\varpi_{A}+\b)(\tau-\tau_{c})}
    \end{split}    
  \end{equation}
  on $A^{+}_{c}(\g)$, where $C_{a}$ only depends on $s_{\cweight,l}$, $s_{\coeff,l}$, $c_{\cweight,k}$, $c_{\coeff,k}$, $d_{\a}$ (in case $\iota_{b}\neq 0$), $A_{0}$,
  $c_{\rem}$, $\e_{A}$, $(\bM,\bge_{\refer})$ and a lower bound on $\theta_{0,-}$; and $\eta_{a}$ and $\eta_{b}$ only depend on $\cweight$, $A_{0}$, $n$ and 
  $k$. Moreover,
  \begin{equation}\label{eq:Vinfestimatefinalothointro}
    |V_{\bfI,\infty,a}| \leq C_{a}\ldr{\tau_{c}}^{\eta_{b}}\hGe_{l}^{1/2}(\tau_{c}),
  \end{equation}
  where $C_{a}$ and $\eta_{b}$ have the same dependence as in the case of (\ref{eq:VAplusestimatehointro}). Given $V_{\bfI,\infty,a}$ as above,
  define $U_{\bfI,m}$, $m=0,1,2$, by
  \begin{align}
    \left(\begin{array}{c} U_{\bfI,0}(\tau) \\ U_{\bfI,1}(\tau)\end{array}\right)
    := & e^{A_{0}(\tau-\tau_{c})}V_{\bfI,\infty,a}+\int_{\tau}^{\tau_{c}}e^{A_{0}(\tau-s)}
    \left(\begin{array}{c} 0 \\ \sfL_{\bfI}(s)\end{array}\right)ds,\label{eq:UbfIzeroandonedefintro}\\
    U_{\bfI,2}(\tau) := & Z^{0}_{\infty}U_{\bfI,1}(\tau)+\hal_{\infty}U_{\bfI,0}(\tau)-\sfL_{\bfI}(\tau).\label{eq:UbfItwodefintro}
  \end{align}
  Proceeding inductively as above yields $U_{\bfI,m}$ and $V_{\bfI,\infty,a}$ for $|\bfI|\leq k-m_{0}-1$ and $m=0,1,2$ such that (\ref{eq:VAplusestimatehointro})
  holds. 
\end{thm}
\begin{remark}\label{remark:higherorderestimatesimprovedestimatesintro}
  It is possible to improve the estimates. First, define $V_{\infty}$ as in Remark~\ref{remark:Vinftyimprovementintro}. This yields
  (\ref{eq:VAplusestimateimprovedintro}). Defining $U_{0,m}$, $m=0,1,2$, by (\ref{eq:Uzeromdefintro}) with $V_{\infty,a}$ replaced by $V_{\infty}$, we can proceed
  inductively as in the statement of the theorem. In particular, a $V_{\bfI,\infty}\in\rn{2m_{s}}$ can be constructed such that (\ref{eq:VAplusestimatehointro})
  is improved to
  \begin{equation}\label{eq:VAplusestimatehoimprovedintro}
    \begin{split}
      & \left|V_{\bfI}-e^{A_{0}(\tau-\tau_{c})}V_{\bfI,\infty}-\int_{\tau}^{\tau_{c}}e^{A_{0}(\tau-s)}\left(\begin{array}{c} 0 \\ \sfL_{\bfI}(s)\end{array}\right)ds\right|\\
      \leq & C_{a}\ldr{\tau_{c}}^{\eta_{b}}e^{\b\tau_{c}}\hGe_{l}^{1/2}(\tau_{c})\ldr{\tau-\tau_{c}}^{\eta_{a}}e^{(\varpi_{A}+\b)(\tau-\tau_{c})}
    \end{split}    
  \end{equation}
  on $A^{+}_{c}(\g)$, where $C_{a}$, $\eta_{a}$ and $\eta_{b}$ have the same dependence as in (\ref{eq:VAplusestimatehointro}). Defining $U_{\bfI,m}$ as in
  (\ref{eq:UbfIzeroandonedefintro}) and (\ref{eq:UbfItwodefintro}) with $V_{\bfI,\infty,a}$ replaced by $V_{\bfI,\infty}$, and modifying $\sfL_{\bfI}$ accordingly, it
  can be demonstrated that (\ref{eq:VAplusestimatehoimprovedintro}) holds for $|\bfI|\leq k-m_{0}-1$. Note that the advantage here is that by taking
  $\tau_{c}$ close enough to $-\infty$, the factor $C_{a}\ldr{\tau_{c}}^{\eta_{b}}e^{\b\tau_{c}}$ can be chosen to be as small as we wish. The disadvantage of
  the estimate is that $V_{\bfI,\infty}$ is not unique. However, $V_{\bfI,\infty}$ satisfies (\ref{eq:Vinfestimatefinalothointro}) with $V_{\bfI,\infty,a}$ replaced
  by $V_{\bfI,\infty}$. 
\end{remark}
\begin{proof}
  The statements of the theorem and of the remark follow from Theorem~\ref{thm:leadingorderasymptoticszeroho} and
  Remark~\ref{remark:higherorderestimatesimprovedestimates}.
\end{proof}

\section{Specifying asymptotics}\label{section:specifyingasymptoticsintro}

Theorems~\ref{thm:leadingorderasymptoticszerointro} and \ref{thm:leadingorderasymptoticszerohointro} yield the leading order asymptotics. However,
the statement of Theorem~\ref{thm:leadingorderasymptoticszerointro}, e.g., does not guarantee that $V_{\infty,a}\neq 0$. If, for the sake of argument,
$V_{\infty,a}$ always vanishes, irrespective of the solution, then the energy estimate obtained in Theorem~\ref{thm:asgrowthofenergyintro} is not optimal
and Theorem~\ref{thm:leadingorderasymptoticszerointro} does not yield the leading order asymptotics of solutions. It is therefore of interest to ask if
it is possible to specify the asymptotic data. This turns out to be possible, but before stating the corresponding result, it is convenient to introduce
the following terminology.

\begin{definition}\label{def:upomegadefintro}
  Given a vector field multiindex $\bfI=(I_{1},\dots,I_{p})$, let $\upomega(\bfI)\in\nn{n}$
  \index{$\a$Aa@Notation!Functions!$\upomega$}%
  be the vector whose components, written
  $\upomega_{i}(\bfI)$, $i=1,\dots,n$, are given as follows: $\upomega_{i}(\bfI)$ equals the number of times $I_{q}=i$, $q=1,\dots,p$. 
\end{definition}

\begin{thm}\label{thm:specifyingtheasymptoticsintro}
  Assume that the conditions of Theorem~\ref{thm:leadingorderasymptoticszerohointro} are satisfied. Then, using the notation of
  Theorem~\ref{thm:leadingorderasymptoticszerohointro}, the following holds. Fix vectors $v_{\omega}\in E_{a}$ for $\rn{n}$-multiindices $\omega$
  satisfying $|\omega|\leq k-m_{0}-1$. Then, given $\tau_{c}$ close enough to $-\infty$, there is a solution to (\ref{eq:equationintermsofcanonicalframe})
  with vanishing right hand side such that if $V_{\bfI,\infty,a}$ are the vectors uniquely determined by the solution as in the statement of
  Theorem~\ref{thm:leadingorderasymptoticszerohointro}, then $V_{\bfI_{\omega},\infty,a}=v_{\omega}$, where $\bfI_{\omega}=(I_{1},\dots,I_{p})$ is the vector field
  multiindex such that $I_{j}\leq I_{j+1}$ for $j=1,\dots,p-1$ and such that $\upomega(\bfI_{\omega})=\omega$. 
\end{thm}
\begin{remark}
  The bound $\tau_{c}$ has to satisfy in order for the conclusions to hold is of the form $\tau_{c}\leq T_{c}$, where $T_{c}$ only depends on $s_{\cweight,l}$,
  $s_{\coeff,l}$, $c_{\cweight,k}$, $c_{\coeff,k}$, $d_{\a}$ (in case $\iota_{b}\neq 0$), $A_{0}$, $c_{\rem}$, $\e_{A}$, $(\bM,\bge_{\refer})$, a lower bound on
  $\theta_{0,-}$, a choice of local coordinates on $\bM$ around $\bx_{0}$ and a choice of a cut-off function near $\bx_{0}$.
\end{remark}
\begin{remark}
  The solutions constructed in the theorem are such that
  \begin{equation}\label{eq:VAplusestimatehoimprovedconstrintro}
    \begin{split}
      & \sum_{|\bfI|\leq k-m_{0}-1}\left|V_{\bfI}-e^{A_{0}(\tau-\tau_{c})}V_{\bfI,\infty,a}-\int_{\tau}^{\tau_{c}}
      e^{A_{0}(\tau-s)}\left(\begin{array}{c} 0 \\ \sfL_{\bfI}(s)\end{array}\right)ds\right|\\
      \leq & C_{a}\ldr{\tau_{c}}^{\eta_{b}}e^{\b\tau_{c}}\ldr{\tau-\tau_{c}}^{\eta_{a}}e^{(\varpi_{A}+\b)(\tau-\tau_{c})}
      \textstyle{\sum}_{|\omega|\leq k-m_{0}-1}|v_{\omega}|
    \end{split}    
  \end{equation}
  on $A^{+}_{c}(\g)$, where $C_{a}$ only depends on $s_{\cweight,l}$, $s_{\coeff,l}$, $c_{\cweight,k}$, $c_{\coeff,k}$, $d_{\a}$ (in case $\iota_{b}\neq 0$), $A_{0}$,
  $c_{\rem}$, $\e_{A}$, $(\bM,\bge_{\refer})$, a lower bound on $\theta_{0,-}$, a choice of local coordinates on $\bM$ around $\bx_{0}$ and a choice of a cut-off
  function near $\bx_{0}$; and $\eta_{a}$ and $\eta_{b}$ only depend on $\cweight$, $A_{0}$, $n$, and $k$. Note, in particular, that by choosing $\tau_{c}$
  close enough to $-\infty$, the factor $C_{a}\ldr{\tau_{c}}^{\eta_{b}}e^{\b\tau_{c}}$ appearing on the right hand side of
  (\ref{eq:VAplusestimatehoimprovedconstr}) can be chosen to be as small as we wish. 
\end{remark}
\begin{proof}
  The statement is an immediate consequence of Theorem~\ref{thm:specifyingtheasymptoticdata}.
\end{proof}

Due to this result, it is clear that Theorem~\ref{thm:asgrowthofenergyintro} yields optimal energy estimates and that
Theorems~\ref{thm:leadingorderasymptoticszerointro} and \ref{thm:leadingorderasymptoticszerohointro} yield the leading order asymptotics of solutions. 
Assuming the geometry and the equation to be such that for every $\bx\in\bM$, $Z^{0}(\bx,\cdot)$ and $\hal(\bx,\cdot)$ converge exponentially, we can
therefore, with each $\bx\in\bM$, associate $\varpi_{A}(\bx)$ and $d_{A}(\bx)$ such that the following holds. Let $\g$ be a causal curve with the
properties stated in Lemma~\ref{lemma:pastinextendiblecausalcurvelocintro}, and let $\bx_{\g}$ be the associated limit point on $\bM$. Then, if
$u$ is a solution to (\ref{eq:equationintermsofcanonicalframe}) with vanishing right hand side, there is a constant $C$ such that
\[
|(\hU u)\circ\g(s)|+|u\circ\g(s)|\leq C\ldr{\varrho\circ\g(s)}^{d_{A}(\bx_{\g})}e^{\varpi_{A}(\bx_{\g})\cdot \varrho\circ\g(s)}.
\]
Moreover, this estimate is optimal in the sense that there is a solution and a $C>0$ such that the reverse estimate holds asymptotically. The functions
$\varpi_{A}$ and $d_{A}$ need not be continuous. The following example illustrates some of the possibilities.

\begin{example}\label{example:discontofasymptotics}
  Consider a non-flat Kasner solution to Einstein's vacuum equations, say $(M,g_{K})$, where $M=\tn{n}\times (0,\infty)$ and
  \[
  g_{K}=-dt\otimes dt+\textstyle{\sum}_{i=1}^{n}t^{2p_{i}}dx^{i}\otimes dx^{i}.
  \]
  Here $p_{i}$ are constants such that $p_{i}<1$, $\sum p_{i}=1$ and $\sum p_{i}^{2}=1$. We also assume the $p_{i}$ to be distinct. Choosing $t_{0}=1$, the
  metric $\bge_{\refer}$ becomes the standard metric on $\tn{n}$. Moreover, $\varphi=t$, so that $\varrho=\ln t$ and $\tau=\ln t$. Additionally,
  $\theta=t^{-1}$, $N=1$, $\chi=0$, $U=\d_{t}$ and $\hU=t\d_{t}=\d_{\tau}$. Moreover,
  \[
  \mK=\textstyle{\sum}_{i=1}^{n}p_{i}\d_{x^{i}}\otimes dx^{i}.
  \]
  In particular, $p_{i}$ are the eigenvalues of $\mK$ and the $\d_{x^{i}}$ are the corresponding eigenvectors. Moreover, if the $p_{i}$ are distinct, then
  $\mK$ is non-degenerate. Note also that $\hml_{U}\mK=0$, cf. (\ref{eq:hmlUmtinfixedspatialcoord}), and that
  \[
  \hg_{K}=-d\tau\otimes d\tau+\textstyle{\sum}_{i=1}^{n}e^{2\b_{i}\tau} dx^{i}\otimes dx^{i},\ \ \
  \hK=\textstyle{\sum}_{i=1}^{n}\b_{i}\d_{x^{i}}\otimes dx^{i}
  \]
  where $\b_{i}=p_{i}-1<0$. In particular, $\hK$ is negative definite and $\e_{\Spe}=1-p_{\max}$, where $p_{\max}:=\max\{p_{1},\dots,p_{n}\}$. Moreover, the
  $\mu_{A}$'s correspond to the functions $\b_{i}\tau$. Next, note that
  \[
  -1-q=\hU(n\ln\theta)=\d_{\tau}(n\ln t^{-1})=n\d_{\tau}(-\tau)=-n,
  \]
  so that $q=n-1$. Consider the homogeneous version of the equation (\ref{eq:theequation}), where $g$ is given by $g_{K}$. It can be rewritten as
  (\ref{eq:equationintermsofcanonicalframe}) with $\hf=0$; i.e.
  \[
  -u_{\tau\tau}+\textstyle{\sum}_{i}e^{-2\b_{i}\tau}\d_{i}^{2}u+\hmcX^{0}u_{\tau}+\hmcX^{i}\d_{i}u+\hal u=0
  \]
  in the current setting, where we appealed to (\ref{eq:Zzdefintro}); the fact that $q=n-1$; (\ref{eq:ZAdefintro}), (\ref{eq:hmcYAdefEi}),
  (\ref{eq:hGAdef}) and (\ref{eq:aAdef}); the fact that $\mu_{A}$, $\mu_{\rotot}$, $\hN$ only depend on time; and the fact that the structure
  constants $\g^{A}_{BC}$ associated with the frame $\{\d_{x^{i}}\}$ vanish. Here, the coefficients of $u_{\tau}$, $\d_{i}u$ and $u$ are freely
  specifiable. As long as $\mcX$ is such that the second terms on the left hand sides of (\ref{eq:Sobcoefflassumptions}) and (\ref{eq:coefflassumptions})
  are bounded for all $l$, what $\hmcX^{i}$ is does not affect the asymptotics. From now on, we therefore only assume $\hmcX^{i}$ to satisfy
  these bounds. Let $\phi\in C^{\infty}_{0}(\rn{n})$ be such that $\phi=1$ in an open neighbourhood of $0$ and such that $\phi(\bx)=0$ for $|\bx|\geq 1$.
  Let $0<\cweight\in\ro$ and $\bx_{i}\in\tn{n}$, $i=1,\dots,m$, be distinct. Then we can think of
  \[
  \psi_{i}(\bx,t):=\phi\left[\ldr{\ln t}^{\cweight}(\bx-\bx_{i})\right]
  \]
  as being defined on $M$. Let $a_{j},b_{j}\in\ro$, $j=0,\dots,m$, and let
  \[
  \hmcX^{0}=a_{0}+\textstyle{\sum}_{i=1}^{m}(a_{i}-a_{0})\psi_{i},\ \ \
  \hal=b_{0}+\textstyle{\sum}_{i=1}^{m}(b_{i}-b_{0})\psi_{i}. 
  \]
  Then (\ref{eq:Sobcoefflassumptions}) and (\ref{eq:coefflassumptions}) are satisfied to any order. Note also that the standard assumptions are
  satisfied. Moreover, the $(\cweight,l)$-supremum and the
  $(\cweight,k)$-Sobolev assumptions are satisfied to any order. Finally, note that if $\bx\neq\bx_{i}$ for all $i$, then, for $t$ close enough to
  $0$, $Z^{0}(\bx,t)=a_{0}$ and $\hal(\bx,t)=b_{0}$. In particular, (\ref{eq:Zzerohalexpconvergenceintro}) is satisfied for $\bx_{0}=\bx$ and any choice
  of $\e_{A}$. Moreover, for $t$ close enough to $0$, $Z^{0}(\bx_{i},t)=a_{i}$ and $\hal(\bx_{i},t)=b_{i}$. Thus (\ref{eq:Zzerohalexpconvergenceintro}) is
  again satisfied for $\bx_{0}=\bx_{i}$, $i=1,\dots,m$, and any choice of $\e_{A}$. To conclude, the assumptions of
  Theorem~\ref{thm:leadingorderasymptoticszerohointro} are satisfied for all $\bx\in\tn{n}$. Let
  \[
  A_{0}:=\left(\begin{array}{cc} 0 & 1 \\ b_{0} & a_{0}\end{array}\right),\ \ \
  A_{i}:=\left(\begin{array}{cc} 0 & 1 \\ b_{i} & a_{i}\end{array}\right).
  \]
  Then $\varpi_{A}(\bx)=\varpi_{\max}(A_{0})$ and $d_{A}(\bx)=d_{\max}[A_{0},\varpi_{A}(\bx)]-1$ for $\bx\notin\{\bx_{1},\dots,\bx_{m}\}$, where we used
  the notation introduced in Definition~\ref{def:SpRspdefintro}. Similarly, $\varpi_{A}(\bx_{i})=\varpi_{\max}(A_{i})$ and
  $d_{A}(\bx_{i})=d_{\max}[A_{i},\varpi_{A}(\bx_{i})]-1$ for $i=1,\dots,m$. In particular, we can specify the $a_{i}$ and $b_{i}$ so that
  the solution decays at any given rate along causal curves $\g$ with $\bx_{\g}\notin\{\bx_{1},\dots,\bx_{m}\}$ and such that the solution
  grows at any given rate along causal curves $\g$ with $\bx_{\g}\in\{\bx_{1},\dots,\bx_{m}\}$. Here the latter statement requires an application
  of Theorem~\ref{thm:specifyingtheasymptoticsintro}. However, Theorem~\ref{thm:specifyingtheasymptoticsintro} does apply and can be used to
  not only demonstrate that the decay/growth rate is the expected one along causal curves $\g$ with $\bx_{\g}=\bx_{i}$, but also to demonstrate
  that the solution, to leading order, coincides with a solution to $\xi_{\tau}=A_{i}\xi$ in $A^{+}_{c}(\g)$. 
\end{example}
\begin{remark}
  Due to Example~\ref{example:discontofasymptotics}, it is clear that uniform growth rates such as those derived in
  Propositions~\ref{prop:basicenergyestimateintro} and
  \ref{prop:EnergyEstimateSobolevAssumptionsintro} cannot be expected to be very informative, since the asymptotic behaviour can be substantially
  different along different causal curves. In particular, given $\varpi_{1}>0$, $\varpi_{2}<0$ and $\bx_{2}\in\tn{n}$, we can construct equations
  with solutions such that along causal curves $\g$ with $\bx_{\g}\neq\bx_{2}$, the energy density of the solution decays at the rate $\varpi_{1}$
  and along causal curves $\g$ with $\bx_{\g}=\bx_{2}$, the energy density of the solution grows at the rate $\varpi_{2}$. Since a uniform estimate 
  is worse than the worst causally localised estimate, any uniform estimate will be misleading when it comes to describing the asymptotic behaviour
  along most causal curves. 
\end{remark}

\section{Previous results}\label{section:previousresults}

The subject of these notes is linear systems of wave equations on cosmological backgrounds. There are several previous results on this topic; cf., e.g.,
\cite{aren,pet,rasql,finallinsys,afaf,gns,bac,KGCos} and references cited therein. As far as the study of the singularity is concerned, the assumptions
made in these notes are less restrictive than the ones made in most of these references. However, let us briefly relate the results of these notes with
those of \cite{KGCos,finallinsys}.

In \cite{KGCos}, we consider solutions to the Klein-Gordon equation on Bianchi backgrounds. In particular, we analyse the asymptotic behaviour of solutions
in the direction of the big bang singularity. Since the background geometries are spatially homogeneous, and since we only consider the Klein-Gordon
equation, several of the results of \cite{KGCos} are corollaries of the results of these notes. However, \cite{KGCos} also yields results in the
degenerate setting, and, more importantly, in the case of generic Bianchi type VIII and IX vacuum solutions. Note that for generic Bianchi type VIII and
IX vacuum solutions, the expectation is that there is no $\e_{\Spe}>0$ such that the estimate (\ref{eq:chKeSpeintro}) holds.

In \cite{finallinsys}, we analyse the asymptotics of solutions to systems of wave equations both in the direction of the singularity and in the expanding
direction. However, the equations studied in \cite{finallinsys} are
assumed to be separable. This is a very strong assumption which we do not make here. On the other hand, in \cite{finallinsys} we obtain optimal energy
estimates without a loss of derivatives. Moreover, given suitable assumptions, we essentially control every mode of the solution for all times. We are
very far from doing so here; the results of these notes typically entail a substantial loss of derivatives, cf. the text below
Theorem~\ref{thm:asgrowthofenergyintro}. Concerning the map from initial data to asymptotic data, the results of these notes involve a derivative loss,
but in the results of \cite{finallinsys}, the regularity of the asymptotic data is sometimes higher than that of the initial data; cf.,
e.g., the discussion in \cite[Section~8, pp.~618--620]{KGCos}. In particular, if $u$ is a solution to the Klein-Gordon equation on a non-flat Kasner
background, then the limit of $u_{\tau}$ is half a derivative more regular than the initial data for $u_{\tau}$; here $\tau$ is the time coordinate introduced
in Example~\ref{example:discontofasymptotics}. Turning to Einstein's equations, one can naively think of the metric components as the unknown. This means
that if one could prove that the normal derivative of the unknown has better regularity asymptotically, one would obtain improved asymptotic knowledge
concerning the second fundamental form. In view of the central role played by the expansion normalised Weingarten map in these notes, such an improvement
could potentially be very important. 

\section{Outlook}\label{section:outlook}

As mentioned in the introduction, this article is the first in a series of two. In the present paper, we focus on analysing the asymptotics of solutions
to linear systems of wave equations. In the companion paper \cite{RinGeometry}, we consider the geometric consequences of the assumptions. In particular,
we combine the
assumptions made here with Einstein's equations in order to derive conclusions concerning, e.g., how $\ell_{\pm}$ evolve (in fact, we recover the Kasner map
from the assumptions). We also demonstrate that the combination yields improvements of some of the assumptions. Making stronger assumptions concerning
$\ell_{\pm}$ (such as demanding, e.g., that they belong to the triangle depicted in Figure~\ref{fig:QuiescentConvergentRegime}), we deduce, moreover,
exponential decay of $\hml_{U}\mK$ and convergence of $\mK$. 

Needless to say, the purpose of these notes is to develop methods that can ultimately be used in a non-linear setting. Here the assumptions concerning
the foliation and the geometry are quite general (we do not make any specific gauge choices) and the purpose is to illustrate the features that are
general and, hopefully, common to several different settings. Exactly what gauge choices and additional simplifications will be useful can be expected
to depend on the situation one wishes to study.

\section{Outline}\label{section:FullOutline}

These notes are divided into four parts: an introductory part, a geometry part, a PDE part, and appendices. The present section ends the introductory
part.

\subsection{Part II: Geometry}

\textbf{The frame.}
In Chapter~\ref{chapter:propertiesframe}, we begin by deriving the basic properties of the frame $\{X_{A}\}$, introduced in Definition~\ref{def:XAellA},
and its dual frame $\{Y^{A}\}$. To begin with, we need to estimate the norm of the elements of the dual frame. We are also interested in estimating
the covariant derivatives of the eigenvalues $\ell_{A}$ as well as of the elements of the frame and the dual frame. The goal is to estimate these
quantities in terms of the covariant derivatives of $\mK$; cf., e.g., Lemma~\ref{lemma:bDbfabDlmjchKest} below. We end
Chapter~\ref{chapter:propertiesframe} by estimating products that we will need to bound in later arguments.

\textbf{Geometric formulae.}
In Chapter~\ref{chapter:liederivativesframe}, we derive formulae relating some of the basic geometric quantities. To begin with, we
express $\hU(\ell_{A})$ in terms of $\hml_{U}\mK$ and the frame $\{X_{A}\}$. Introducing $\mW^{A}_{B}$ by
\begin{equation}\label{eq:WABdefintroduction}
  \hml_{U}X_{A}=\mW^{B}_{A}X_{B}+\omW^{0}_{A}U,
\end{equation}
we express $\mW^{A}_{B}$ in terms of $\hml_{U}\mK$, the frame $\{X_{A}\}$, the eigenvalues $\ell_{A}$, the lapse function, the shift vector field, and the
reference metric. We end Chapter~\ref{chapter:liederivativesframe} by discussing the commutator between $\hU$ and $E_{i}$:
\begin{equation}\label{eq:Aikdefintroduction}
  [\hU,E_{i}]=A_{i}^{0}\hU+A_{i}^{k}E_{k}.
\end{equation}
We need to estimate $A_{i}^{0}$, $A_{i}^{k}$ and their expansion normalised normal derivatives. We take a first step in this direction in
Section~\ref{section:contributionfromshift}. 

\textbf{Lower bounds on $\mu_{A}$.}
The main point of Chapter~\ref{chapter:normofelementsofframe} is to derive a lower bound for the $\mu_{A}$ introduced in Definition~\ref{def:muAbmuA}.
In particular, we prove that $\mu_{A}$ grows at least as $-\e_{\Spe}\varrho$ in the direction of the singularity; cf. (\ref{eq:muminmainlowerbound})
below. An important secondary goal is to control the relative spatial variation of $\varrho$; cf. Lemmas~\ref{lemma:respvarvarrhoEi} and
\ref{lemma:taurelvaryingbxEi}. However, we begin the chapter by deriving estimates of Lie derivatives involving the shift vector field in terms of
the covariant derivatives. We also estimate the divergence of $\chi$.

Throughout these notes, $\varrho$ and $\mu_{A}$ play a central role. We largely
control these quantities via evolution equations. In fact, we derive expressions for $\hU(\varrho)$ and $\hU(\bmu_{A})$ in
Lemma~\ref{lemma:geometricidentitiesbmuAvarrho}. Following this derivation, we state and prove the basic estimates for $\mu_{A}$ in
Section~\ref{section:mainmuAestimate}. The main assumptions needed to obtain the corresponding result are non-degeneracy, silence, that
$\mK$ is $C^{0}$-bounded and that $\hml_{U}\mK$ satisfies a weak off-diagonal exponential bound; cf. Definition~\ref{def:offdiagonalexpdec}. However, we
also need to impose a smallness assumption on $\chi$. This is the only smallness assumption we impose in these notes. The proof of the bounds on $\mu_{A}$
consists of a bootstrap argument. The point is that if the contribution from the shift vector field is small, then $\mu_{A}$ can be demonstrated
to grow in the direction of the singularity. However, if the $\mu_{A}$ grow, then it can be demonstrated that the contribution from the shift
vector field not only remains small, but in fact is integrable along integral curves of $\hU$. Assuming an off-diagonal exponential bound, lower
bounds on all the $\mu_{A}$ can be deduced directly. However, it is preferable to only require a weak off-diagonal exponential bound. Under such
assumptions $\mu_{A}$ for $A>1$ and $\mu_{1}$ have to be treated differently. First, we derive estimates for $\mu_{A}$, $A>1$, and then we combine
these estimates with information concerning the sum of the $\bmu_{A}$ and the sum of the $\ell_{A}$ in order to obtain estimates for $\mu_{1}$. 
The conclusions are stated in Lemma~\ref{lemma:lowerbdonmumin}. It is also of interest to note that under the assumptions of 
Lemma~\ref{lemma:lowerbdonmumin} and a weighted $C^{0}$-bound on $\hml_{U}\mK$, some of the assumption corresponding to a weak off-diagonal exponential
bound can be improved; cf. Proposition~\ref{prop:strengthenassumpAltB}.

In Section~\ref{section:estimatingrelativespatialvariation}, we turn to the problem of estimating the relative spatial variation of $\varrho$. We derive
the estimates by commuting the evolution equation for $\varrho$ with a spatial vector field. We also derive estimates for the time derivative of $\varrho$
in order to demonstrate that $\tau(t):=\varrho(\bx_{0},t)$ can be used as a time coordinate. In order to obtain the desired estimates, we have to impose
bounds such as (\ref{eq:bDlnNbDlnthetabd}) as well as additional smallness assumptions concerning the shift vector field. 

In the remainder of the chapter, we derive consequences of the assumption that $q-(n-1)$ converges to zero at a suitable rate (in many quiescent
settings, this quantity converges to zero exponentially, and it is of interest to work out the consequences of such an estimate). The conclusions
we obtain are of importance when deriving energy estimates.

\textbf{Function spaces and estimates.} In Chapter~\ref{chapter:functionspacesandestimates}, we introduce several function spaces. We also relate
the corresponding norms and derive Moser type estimates. The proofs are partly based on Gagliardo-Nirenberg type estimates derived in
Appendix~\ref{chapter:gagnir}. In particular, we derive estimates for the shift vector field. We also estimate weighted Sobolev norms of $\ell_{A}$,
$X_{A}$ and $Y^{A}$ in terms of $\mK$. 

\textbf{Estimating Lie derivatives.} In the derivation of energy estimates, we need bounds on $\mW_{A}^{B}$, $A_{i}^{k}$ and $\hU(A_{i}^{k})$,
introduced in (\ref{eq:WABdefintroduction}) and (\ref{eq:Aikdefintroduction}), with respect to weighted Sobolev and $C^{k}$-norms. The purpose of
Chapter~\ref{chapter:higherorderestimatesnormsLieframe} is to derive such estimates. We end the chapter by recording the result of combining such
estimates with the assumptions stated in Subsections~\ref{subsection:higherordersobolevassumptions} and \ref{subsection:higherorderckassumptions}.

\textbf{Estimating the components of the metric.} Due to our choice of frame, the metric takes a very simple form; cf. (\ref{eq:muAdef}) and 
(\ref{eq:bgenormXA}). However, in order for this information to be of interest, we need to estimate $\mu_{A}$ with respect to weighted Sobolev
and $C^{k}$-norms. This is the purpose of Chapter~\ref{chapter:estimatingthecomponentsofmetric}. We use energy estimates to derive the desired
conclusion. In the Sobolev setting, we integrate over the leaves of the foliation, but in the $C^{k}$-setting, we consider the evolution along
integral curves of $\hU$. Due to the definition of the $\mu_{A}$ in terms of eigenvectors of $\mK$, the arguments involve a loss of derivatives;
cf. Remark~\ref{remark:lossofderivativestogeometry}. 

\subsection{Part III: Wave equations}

\textbf{Basic energy estimates.} We begin Chapter~\ref{chapter:systemsofwaveequations} by rewriting the equation in terms of the wave operator
of the conformally rescaled metric $\hg$. We also derive a basic energy identity in Lemma~\ref{lemma:basicenergyestimate}. Combining this
identity with $C^{0}$-assumptions concerning the coefficients results in a basic energy estimate; cf. Section~\ref{section:basicenergyestimates}.
We end the chapter by expressing the conformal wave operator in terms of the frame; cf. Lemma~\ref{lemma:Boxhguocalcoord}. This also allows
us to calculate the relation between $\hmcX^{0}$, $\hmcX^{A}$ appearing in, e.g., (\ref{eq:hmcXintrointro}) and $Z^{0}$ and $Z^{A}$ appearing in
(\ref{eq:equationintermsofcanonicalframeintro}).

\textbf{Commutators.} The equation (\ref{eq:equationintermsofcanonicalframeintro}) can be written $Lu=\hf$. In order to take the step from the
basic energy estimate to higher order energy estimates, we need to calculate the commutator $[E_{\bfI},L]$. This is the subject of
Chapter~\ref{chapter:commutators}. The higher order energy estimates will be derived in two steps. First we derive conclusions on the basis of
weighted $C^{k}$-assumptions. Due to the resulting estimates, we obtain bounds on the unknown and its first derivatives in $C^{0}$. Combining these bounds
with higher order Sobolev assumptions and Moser type estimates yields energy estimates with a lower loss of derivatives; this is the second step. 
However, what is the most convenient expression for $[E_{\bfI},L]$ depends on which of these steps one is taking. The reason for this is that in the
$C^{k}$-setting, it is of interest to extract the expressions arising from the geometry and the coefficients directly in $C^{0}$. However, in the
Sobolev setting, one wants to apply a Moser estimate. The expressions and estimates for the commutators derived in Chapter~\ref{chapter:commutators}
are the basis for both steps.

\textbf{Energy estimates, step I.} In Chapter~\ref{chapter:higherorderenergyestimatespartI}, we derive energy estimates on the
basis of weighted $C^{k}$-assumptions. Since we know the basic energy estimate to hold, it is sufficient to estimate $[L,E_{\bfI}]u$ in $L^{2}$.
We therefore begin by combining the conclusions of Chapter~\ref{chapter:commutators} with the $(\cweight,l)$-supremum assumptions and the equation
in order to bound $[L,E_{\bfI}]u$. The resulting estimate, the basic energy estimate and an inductive argument then together yield a higher order
energy estimate; cf. (\ref{eq:basichigherorderenergyestimate}). Combining the result with a weighted version of Sobolev embedding, we obtain estimates
of the weighted higher order energy densities in Section~\ref{section:Ckestimatesweightedenergydensity}. 

\textbf{Energy estimates, step II.} In Chapter~\ref{chapter:hoeepII}, we derive energy estimates based on a combination of
$(\cweight,l)$-supremum and $(\cweight,l)$-Sobolev assumptions. However, in this setting, we have to address the fact that the output of Moser
estimates is expressions of the form
\[
\int_{\bM_{\tau}}|E_{\bfI}(e^{-\mu_{A}}X_{A}u)|^{2}\mutgc.
\]
On the other hand, the expressions that naturally appear in the energies are of the form
\[
\int_{\bM_{\tau}}|e^{-\mu_{A}}X_{A}E_{\bfI}u|^{2}\mutgc. 
\]
For this reason, the first problem we have to address is that of reordering the derivatives. This is the subject of
Section~\ref{section:reorderingderivatives}. We then estimate $[E_{\bfI},L]u$ by appealing to the results of Chapter~\ref{chapter:commutators}, 
Moser estimates, and the results concerning reordering of derivatives. Once this has been done, we essentially immediately obtain higher order energy
estimates in Section~\ref{section:energyestimatesstepII}. We end the chapter by deriving energy estimates in the case of, e.g., the Klein-Gordon equation.
Combining the energy estimates with some additional assumptions (in particular, we assume that $q-(n-1)$ converges to zero exponentially) leads to
partial asymptotics of solutions to the Klein-Gordon equation; cf. Proposition~\ref{prop:asvelocityKGlikeeq}. 

\textbf{Localising the analysis.} The energy estimates derived in Chapters~\ref{chapter:higherorderenergyestimatespartI} and \ref{chapter:hoeepII}
are quite crude in the sense that they yield exponential growth of solutions, without providing detailed information concerning the rate. On the other
hand, it is very important to note that the rate of growth is independent of the order of the energy. Due to this fact and the silence, it is possible to
obtain more detailed information by localising the analysis. This is the subject of
Chapters~\ref{chapter:localisingtheanalysis}-\ref{chapter:specifyingtheasymptotics}. We begin, in Section~\ref{section:causalstructure}, by analysing the
causal structure in the direction
of the singularity. In particular, we wish to limit our attention to sets of the form $J^{+}(\g)$, where $\g$ is a past inextendible
causal curve. In order to obtain specific estimates, we demonstrate that, to the past of $t_{0}$, $J^{+}(\g)$ is contained in a set of the form
$A^{+}(\g)$; cf. (\ref{eq:Aplusdefintroduction}). We also estimate the distance between $\varrho$ and $\tau$ in $A^{+}(\g)$ and derive an expression
for the weight $w$ used in the energy estimates; cf. Lemma~\ref{lemma:lnwintqminusnminusoneestimate}. Once this preliminary analysis has been carried
out, the main goal is to estimate the error terms that arise when replacing $\hU$ with $\d_{\tau}$, omitting ``spatial derivatives'' and localising the
coefficients; cf. the heuristic discussions in Sections~\ref{section:resultsintrointo} and \ref{section:energyestimatesincausallylocalisedregions}. In
Section~\ref{section:localisingeqfirstder}, we begin by estimating expressions such as $\d_{\tau}\psi-\hU\psi$. We then proceed to estimate
$\d_{\tau}^{2}\psi-\hU^{2}\psi$. In the end, we conclude that if $Lu=0$, then $u$ satisfies the model equation (\ref{eq:modelintrointro}), up to
an error term which is estimated in Corollary~\ref{cor:rhsreplcfderfullylocal}. In fact, if $\tau_{c}=0$, an estimate of the form
\begin{equation}\label{eq:estrhsofmodelequ}
  \begin{split}
    |-\d_{\tau}^{2}E_{\bfI}u+Z^{0}_{\roloc}\d_{\tau} E_{\bfI}u+\hal_{\roloc} E_{\bfI}u|
    \leq & C_{a}\ldr{\tau}^{\eta_{a}}e^{\e_{\Spe}\tau}\me_{m+1}^{1/2}
    +C_{b}\ldr{\tau}^{\eta_{b}}\me_{m-1}^{1/2}
  \end{split}
\end{equation}
holds; cf. (\ref{eq:rhsreplcfderfullylocal}). Here $m=|\bfI|$ and the second term on the right hand side of (\ref{eq:estrhsofmodelequ}) should be omited in
case $m=0$. 

\textbf{Localised energy estimates.} Given the estimate (\ref{eq:estrhsofmodelequ}), we are in a position to compare solutions to the actual equation
with solutions to the model equation. Since we cannot, in general, determine the asymptotic behaviour of solutions to the model equation, we, in general,
have to make assumptions concerning the evolution associated with the model equation. These assumptions take the form of estimates such as
(\ref{eq:Phinormbasasslocintro}). In the end, we obtain estimates such as (\ref{eq:melindassfslocfinalstmt}). The way to prove this estimate is to
proceed by induction. In some sense, there are in fact two induction arguments. To begin with, we have estimates for all the energy densities $\me_{j}$,
with a degree of exponential growth that does not depend on the order. However, there is, a priori no relation between this exponential growth and the
estimate (\ref{eq:Phinormbasasslocintro}). Given the estimate for all the $\me_{j}$, $j\leq l_{0}$ (for some $l_{0}$), we begin by considering 
(\ref{eq:estrhsofmodelequ}) with $m=0$. Then the second term on the right hand side vanishes and in the first term, there is a factor $e^{\e_{\Spe}\tau}$
in front of $\me_{1}$. If $\me_{0}$ and $\me_{1}$ are not already known to satisfy estimates corresponding to (\ref{eq:Phinormbasasslocintro}), then
(\ref{eq:estrhsofmodelequ}) can be used to improve the estimate for $\me_{0}$. Once an improved estimate for $\me_{0}$ has been derived,
(\ref{eq:estrhsofmodelequ}) can be used to improve the estimate for $\me_{1}$ etc. Proceeding in this way, we can improve the estimates for $\me_{j}$
for $j\leq l_{0}-1$. In other words, we can improve the estimates by a factor of $e^{\e_{\Spe}\tau}$ at the loss of one derivative (in practice, we typically
also get a deterioration in terms of polynomial factors). This process can be iterated as long as the estimates for $\me_{j}$ are worse than the estimates
for the model equation. In the end, it leads to the desired estimate, and a loss of $m_{0}$ derivatives; cf. (\ref{eq:mzdefintro}) and the adjacent text. 
In particular, as $\e_{\Spe}$ tends to zero, the number of derivatives lost in the process tends to infinity.

We end Chapter~\ref{chapter:energyestimatescausallyloc} by discussing the particular case that the coefficients $Z^{0}_{\roloc}$ and $\hal_{\roloc}$ converge
at a sufficiently fast polynomial rate along a causal curve. In this case, $d_{A}$ and $\varpi_{A}$ can be calculated in terms of the limiting matrix.

\textbf{Deriving asymptotics.} In Chapter~\ref{chapter:derivingasymptotics}, we turn to the problem of deriving asymptotics, assuming $Z^{0}_{\roloc}$ and
$\hal_{\roloc}$ to converge exponentially. We begin by deriving estimates in the model case of a system of ODE's with an error term; cf.
Lemma~\ref{lemma:asymptoticsmodelODE}. Given the corresponding result and the estimates already derived, we are in a position to prove results such as
Theorem~\ref{thm:leadingorderasymptoticszerointro}. In order to obtain higher order asymptotics, we first need to derive appropriate model equations.
We do so in Section~\ref{section:asymptoticshigherorderderivatives}. Deriving asymptotics for the higher order derivatives is somewhat more complicated
than for the zeroth order derivatives, since we need to proceed inductively; only after we have derived the asymptotics for the lower order derivatives can
we phrase the equation for the higher order derivatives. The associated technical complications necessitates an argument which is substantially longer than
the one concerning the zeroth order derivatives.

\textbf{Specifying asymptotics.} Finally, in Chapter~\ref{chapter:specifyingtheasymptotics}, we turn to the problem of specifying the asymptotics. We do
so by defining an appropriate map from initial data to asymptotic data. Setting up an appropriate finite dimensional class of initial data (such that its
dimension coincides with the dimension of the asymptotic data one wishes to specify), the idea is then to prove that the map from initial data to
asymptotic data is linear and injective (and, thereby, by the choice of class of initial data, bijective). It is important to note that the argument
applies even in situations where the spatial derivatives of the coefficients of the equation diverge along $\g$. 

\subsection{Part IV: Appendices}

In the final part of these notes we discuss technical issues we do not wish to address in the main body of the text. To begin with, we discuss the
existence of a global frame in Section~\ref{section:globalframefinitecover} and define $\ml_{U}\mK$ in Section~\ref{section:timederivativeofmK}.
In Section~\ref{section:syncblowupmc}, we discuss conditions ensuring that the spatial derivatives of $\ln\theta$ do not diverge faster than
polynomially in $\varrho$. This section serves as a motivation for the conditions imposed on $\ln\theta$. 

\textbf{Gagliardo-Nirenberg estimates.} In Appendix~\ref{chapter:gagnir}, we derive Gagliardo-Nirenberg estimates in the case of weighted Sobolev spaces
on manifolds. The weight is allowed to be time dependent, and in order to also allow frameworks which are adapted to the geometry, we consider collections
of vector fields (in the definitions of the Sobolev-type spaces) which are not necessarily a frame, and which are time dependent. Using the
Gagliardo-Nirenberg estimates, we derive Moser type estimates which are then used as a basis for deriving the higher order energy estimates.

\textbf{Examples.} In Appendix~\ref{chapter:examples}, we give examples of classes of spacetimes for which the asymptotic behaviour in the direction
of the singularity is understood. These examples serve the purpose of justifying the assumptions we impose. We begin by discussing spatially homogeneous
solutions. Next, we discuss some classes of solutions constructed by specifying initial data on the singularity. We continue by describing results
concerning stable big bang formation. Finally, we discuss $\tn{3}$-Gowdy symmetric spacetimes. 

\part{Geometry}

\chapter[Properties, frame adapted to the eigenspaces of $\mK$]{Basic properties of the frame adapted to 
the eigenspaces of $\mK$}\label{chapter:propertiesframe}

The assumptions concerning the geometry are expressed using norms associated with the fixed metric $\bge_{\refer}$. However, in many of the 
arguments, the frame $\{X_{A}\}$, its dual $\{Y^{A}\}$ and the eigenvalues $\ell_{A}$, introduced in Definition~\ref{def:XAellA}, appear naturally. 
We therefore need to control these objects and their covariant derivatives in terms of the assumptions. The main purpose of the present chapter is
to take a first step in this direction. We also estimate $E_{\bfI}(P)$, cf. Definition~\ref{def:multiindexnotation}, for a general
product $P$ consisting of factors of several different types (eigenvalues of $\mK$, tensor fields evaluated on the frames $\{X_{A}\}$ and $\{Y^{A}\}$,
Lie derivatives with respect to the shift vector field etc.). This simplifies the derivation of estimates in the chapters to follow. 

\section{Constructing a frame}

Given that $\mK$ is non-degenerate and has a global frame, there is a natural frame on the spacetime; cf. Definition~\ref{def:XAellA}. In the
following lemma, we clarify the properties of this frame.  

\begin{lemma}\label{lemma:framedef}
  Let $(M,g)$ be a time oriented Lorentz manifold. Assume that it has an expanding partial pointed foliation. Assume, moreover, $\mK$ to be
  non-degenerate on $I$ and to have a global frame. Then there is a collection of smooth vector fields $\{X_{A}\}$ and covector
  fields $\{Y^{A}\}$, $A=1,\dots,n$, on $\bM$ such that for each $t\in I$, $\{ X_{A}\}$ and $\{Y^{A}\}$ are frames on $T\bM_{t}$ and $T^{*}\bM_{t}$
  respectively. Moreover, $\mK X_{A}=\ell_{A}X_{A}$, $\mK^{T}Y^{A}=\ell_{A}Y^{A}$ and $\ell_{1}<\cdots<\ell_{n}$ (no summation on $A$). Finally,
  $\bge_{\refer}(X_{A},X_{A})=1$ (no summation on $A$); $\{ X_{A}\}$ is an orthogonal frame with respect to $\chg$; and $Y^{A}(X_{B})=\de^{A}_{B}$. 
\end{lemma}
\begin{remark}
  The map $\mK^{T}$ is defined by the condition that if $\eta\in T^{*}_{p}\bM_{t}$ and $\xi\in T_{p}\bM_{t}$, then $(\mK^{T}\eta)(\xi):=\eta(\mK \xi)$.
\end{remark}
\begin{remark}\label{remark:ellAsum}
  It is of interest to keep in mind that $\textstyle{\sum}_{A}\ell_{A}=1$, since $\tr\mK=1$. 
\end{remark}
\begin{remark}
  The combination of $\{\hU\}$ and $\{X_{A}\}$, $A=1,\dots,n$, is a frame on $\bM\times I$. Moreover, this frame is orthogonal with respect to both $g$
  and $\hg$. 
\end{remark}
\begin{proof}
  The frame $\{X_{A}\}$ is given by Definition~\ref{def:XAellA}. Let $\{Y^{A}\}$ be the dual frame associated with $\{X_{A}\}$. Then
  \[
  (\mK^{T}Y^{A})(X_{B})=Y^{A}(\mK X_{B})=Y^{A}(\ell_{B}X_{B})=\ell_{B}\de^{A}_{B}=\ell_{A}Y^{A}(X_{B})
  \]
  (no summation), so that $\mK^{T}Y^{A}=\ell_{A}Y^{A}$ (no summation). In order to verify the orthogonality of the frame with respect to $\bge$ (and thereby
  with respect to $\chg$), note that
  \begin{equation}\label{eq:XAorthogonal}
    \ell_{A}\bge(X_{A},X_{B})=\bge(\mK X_{A},X_{B})=\theta^{-1}\bk_{ij}X_{A}^{i}X_{B}^{j}=\ell_{B}\bge(X_{A},X_{B}).
  \end{equation}
  The lemma follows. 
\end{proof}

\section{Basic estimates}

In these notes, we use the frame $\{X_{A}\}$, introduced in Definition~\ref{def:XAellA}, and the frame $\{E_{i}\}$, introduced in
Remark~\ref{remark:framenondegenerate}. We also use the terminology introduced in Definition~\ref{def:multiindexnotation}. In the present
section, we estimate the $Y^{A}$ and make elementary observations concerning the relation between $\bD_{\bfI}$ and $\bD^{k}$ applied to tensor fields.

\subsection{Estimating the norm of the elements of the frame $\{Y^{A}\}$}

In order to construct the frame $\{X_{A}\}$, the conditions of Definition~\ref{def:XAellA} are sufficient. However, in order to obtain 
quantitative control of the frame, we need to use the assumption that $\mK$ is bounded with respect to $\bge_{\refer}$.
We begin by estimating the norms of the $Y^{A}$ with respect to $\bge_{\refer}$.

\begin{lemma}\label{lemma:frameinvest}
  Let $(M,g)$ be a time oriented Lorentz manifold. Assume that it has an expanding partial pointed foliation. Assume, moreover, $\mK$ to be
  non-degenerate on $I$, to have a global frame and to be $C^{0}$-uniformly bounded on $I_{-}$; i.e., (\ref{eq:mKsupbasest}) to hold. Then there
  is a constant $C_{Y}$, depending only on $n$, $\mKsup$ and $\e_{\rond}$, such that $|Y^{A}|_{\bge_{\refer}}\leq C_{Y}$ on $\bM_{t}$ for all $A$ and
  $t\in I_{-}$.
\end{lemma}
\begin{proof}
Let $\{E_{i}\}$ and $\{\omega^{i}\}$ be chosen as in Remark~\ref{remark:framenondegenerate}. If $\eta\in T^{*}_{p}\bM$, then $\eta=\eta_{i}\omega^{i}$, 
where $\eta_{i}:=\eta(E_{i})$ and 
\[
|\eta|_{\bge_{\refer}}=\left(\textstyle{\sum}_{i}\eta_{i}^{2}\right)^{1/2}.
\]
By definition,
\begin{equation}\label{eq:YXeqId}
\de^{A}_{B}=Y^{A}(X_{B})=Y^{A}_{i}\omega^{i}(X^{j}_{B}E_{j})=Y^{A}_{i}X^{i}_{B}
\end{equation}
on $\bM\times I$. In other words, if we let $X$ denote the matrix with elements $X^{i}_{B}$ and $Y$ denote the matrix with elements $Y^{A}_{i}$, 
then $YX=\Id$; i.e., $Y$ is the inverse of $X$. Here we consider $X$ and $Y$ to be maps from $\bM\times I$ to $\Mn{n}{\ro}$. Note that 
\[
1=\bge_{\refer}(X_{A},X_{A})=\bge_{\refer}(X_{A}^{i}E_{i},X_{A}^{j}E_{j})=\de_{ij}X_{A}^{i}X_{A}^{j}
\]
(no summation on $A$). Thus the columns of $X$ are unit vectors with respect to the standard Euclidean metric. Let 
$K:\bM\times I\rightarrow\Mn{n}{\ro}$ be the matrix valued function with components $\mK^{i}_{\phantom{i}j}$ (where the components of $\mK$ are 
calculated with respect to the frame $\{E_{i}\}$). Then $\|K\|\leq \mKsup$ on $\bM\times I_{-}$. Moreover, the eigenvalues 
of $K$ are distinct and the minimal distance between two distinct eigenvalues is $\e_{\rond}$. Assume that there
is a sequence $(p_{l},t_{l})$ in $\bM\times I_{-}$ such that $\det X_{l}\rightarrow 0$, where $X_{l}:=X(p_{l},t_{l})$. Then the sequences defined by 
$K_{l}:=K(p_{l},t_{l})$ and $X_{l}$ are contained in a compact set. By choosing subsequences, which we still denote by $\{K_{l}\}$ and $\{X_{l}\}$, 
we can assume $K_{l}$ and $X_{l}$ to converge to, say, $K_{*}$ and $X_{*}$ respectively. 
Clearly, $\|K_{*}\|\leq C_{\mK}$ and the eigenvalues of $K_{*}$ are distinct (due to the continuous dependence of the eigenvalues on the matrix). 
In fact, the minimal distance between two distinct eigenvalues of $K_{*}$ is $\e_{\rond}$. Since the columns of $X_{l}$ converge to eigenvectors of 
$K_{*}$, we obtain a contradiction. In fact, it is clear that there is a positive lower bound $C_{X}>0$, depending only on $n$, $\e_{\rond}$ and 
$\mKsup$, such that $|\det X|\geq C_{X}$ on $\bM\times I_{-}$. In particular, there is a constant $C_{Y}$, with the same dependence, such that
$\|Y\|\leq C_{Y}$ on $\bM\times I_{-}$. Since $|Y^{A}|_{\bge_{\refer}}$ can be bounded in terms of $\|Y\|$, the statement follows. 
\end{proof}

\subsection{Basic conversions}

Next, we make two elementary observations for future reference. 

\begin{lemma}\label{lemma:bDbfAbDmKexp}
  Let $(M,g)$ be a time oriented Lorentz manifold. Assume that it has an expanding partial pointed foliation. Assume, moreover, $\mK$ to be
  non-degenerate on $I$ and to have a global frame. Let $\mt$ be a family of tensor fields on $\bM$ for $t\in I$. For every $1\leq j\leq l\in\zo$
  and every pair of vector field multiindices $\bfI_{i}$, $i=1,2$, with $|\bfI_{1}|=j$ and $|\bfI_{2}|=l-j$,
  \begin{equation}\label{eq:canexpr1}
    (\bD_{\bfI_{1}}\bD^{l-j}\mt)(\bfE_{\bfI_{2}})
  \end{equation}
  is a tensor field of the same type as $\mt$ which can be written as a linear combination of expressions of the form
  \begin{equation}\label{eq:bDbfBlinco}
    (\bD_{\bfJ}\bD^{l-k}\mt)(\bD_{\bfJ_{1}}E_{J_{1}},\dots,\bD_{\bfJ_{l-k}}E_{J_{l-k}}),
  \end{equation}
  where $\bfJ$ and $\bfJ_{i}$ are vector field multiindices and $k$ is an integer satisfying
  \begin{equation}\label{eq:ksplitting}
    |\bfJ|+\textstyle{\sum}_{i=1}^{l-k}|\bfJ_{i}|=k<j.
  \end{equation}
\end{lemma}
\begin{proof}
  We prove a somewhat more general statement by induction on $j$. Assume, inductively, that if $|\bfI|=j$ and $1\leq j\leq l$,
  \begin{equation}\label{eq:canexpr1gen}
    (\bD_{\bfI}\bD^{l-j}\mt)(\bD_{\bfK_{1}}E_{K_{1}},\dots,\bD_{\bfK_{l-j}}E_{K_{l-j}})
  \end{equation}
  can be written as a linear combination of expressions of the form
  \begin{equation*}
    (\bD_{\bfJ}\bD^{l-k}\mt)(\bD_{\bfJ_{1}}E_{J_{1}},\dots,\bD_{\bfJ_{l-k}}E_{J_{l-k}}),
  \end{equation*}
  where $\bfJ$ and $\bfJ_{i}$ are vector field multiindices, $k$ is an integer satisfying satisfying $k<j$ and 
  \begin{equation*}
    |\bfJ|+\textstyle{\sum}_{i=1}^{l-k}|\bfJ_{i}|=k+\textstyle{\sum}_{i=1}^{l-j}|\bfK_{i}|.
  \end{equation*}
  To begin with, the inductive assumption holds for $j=1$:
  \begin{equation}\label{eq:eqz}
    \begin{split}
      & (\bD_{E_{K}}\bD^{l-1}\mt)(\bD_{\bfK_{1}}E_{K_{1}},\dots,\bD_{\bfK_{l-1}}E_{K_{l-1}})\\
      = & (\bD^{l}\mt)(E_{K},\bD_{\bfK_{1}}E_{K_{1}},\dots,\bD_{\bfK_{l-1}}E_{K_{l-1}})
    \end{split}    
  \end{equation}
  Next, assume that the inductive statement holds up to some $1\leq j$ and for all $l\geq j$. Fix an $l$ such that $l\geq j+1$. Then the inductive
  hypothesis holds with $l$ replaced by $l-1$. Applying $\bD_{E_{I_{1}}}$ to the expression (\ref{eq:canexpr1gen}) (with $l$ replaced by $l-1$) yields
  \begin{equation*}
    \begin{split}
      & \bD_{E_{I_{1}}}[(\bD_{E_{I_{2}}}\cdots \bD_{E_{I_{j+1}}}\bD^{l-j-1}\mt)(\bD_{\bfK_{1}}E_{K_{1}},\dots,\bD_{\bfK_{l-j-1}}E_{K_{l-j-1}})]\\
      = & (\bD_{E_{I_{1}}}\cdots \bD_{E_{I_{j+1}}}\bD^{l-j-1}\mt)(\bD_{\bfK_{1}}E_{K_{1}},\dots,\bD_{\bfK_{l-j-1}}E_{K_{l-j-1}})\\
      & +(\bD_{E_{I_{2}}}\cdots \bD_{E_{I_{j+1}}}\bD^{l-j-1}\mt)(\bD_{E_{I_{1}}}\bD_{\bfK_{1}}E_{K_{1}},\dots,\bD_{\bfK_{l-j-1}}E_{K_{l-j-1}})\\
      & +\dots+(\bD_{E_{I_{2}}}\cdots \bD_{E_{I_{j+1}}}\bD^{l-j-1}\mt)(\bD_{\bfK_{1}}E_{K_{1}},\dots,\bD_{E_{I_{1}}}\bD_{\bfK_{l-j-1}}E_{K_{l-j-1}}).
    \end{split}
  \end{equation*}
  Note that the first term on the right hand side is the one we want to calculate. The remaining terms on the right hand side fit into the
  induction hypothesis. Appealing to the inductive hypothesis, $\bD_{E_{I_{1}}}$ applied to the expression (\ref{eq:canexpr1gen}) (with $l$ replaced
  by $l-1$) can be written as a linear combination of terms of the form
  \begin{equation*}
    \begin{split}
      \bD_{E_{I_{1}}}[\bD_{\bfJ}\bD^{l-1-k}\mt(\bD_{\bfJ_{1}}E_{J_{1}},\dots,\bD_{\bfJ_{l-1-k}}E_{J_{l-1-k}})].
    \end{split}
  \end{equation*}
  Expanding this expression leads to the conclusion that all the corresponding terms satisfy the conditions of the induction hypothesis (with
  $j$ replaced by $j+1$). Thus the inductive statement holds. Applying the inductive statement with all the $\bfK_{i}=0$ yields the
  desired conclusion. 
\end{proof}

\begin{lemma}\label{lemma:bDbfAbDkequiv}
  Let $(M,g)$ be a time oriented Lorentz manifold. Assume that it has an expanding partial pointed foliation. Assume, moreover, $\mK$ to be
  non-degenerate on $I$ and to have a global frame. Let $\mt$ be a family of tensor fields on $\bM$ for $t\in I$. Then $\bD_{\bfI}\mt$ can be
  written as a linear combination of terms of the form
  \[
  (\bD^{k}\mt)(\bfE_{\bfI_{1}})\omega^{J_{1}}(\bD_{\bfJ_{1}}E_{K_{1}})\cdots \omega^{J_{l}}(\bD_{\bfJ_{l}}E_{K_{l}}),
  \]
  where $|\bfI|=k+|\bfJ_{1}|+\dots+|\bfJ_{l}|$ and $k\geq 1$ if $|\bfI|\geq 1$. Similarly, if $k=|\bfI|$, then $(\bD^{k}\mt)(\bfE_{\bfI})$ can be
  written as a linear combination of terms of the form
  \[
  (\bD_{\bfJ}\mt)\omega^{I_{1}}(\bD_{\bfJ_{1}}E_{K_{1}})\cdots \omega^{I_{l}}(\bD_{\bfJ_{l}}E_{K_{l}}),
  \]
  where $k=|\bfJ|+|\bfJ_{1}|+\dots+|\bfJ_{l}|$ and $|\bfJ|\geq 1$ if $k\geq 1$.
\end{lemma}
\begin{proof}
  Note that (\ref{eq:eqz}) holds for $l=1$. This demonstrates that the first statement of the lemma holds for $|\bfI|=1$. The general statement follows
  by means of an induction argument. 

  In order to demonstrate the second statement of the lemma, note that
  \begin{equation*}
    \begin{split}
      (\bD^{k}\mt)(E_{I_{1}},\dots,E_{I_{k}}) = & \bD_{E_{I_{1}}}[(\bD^{k-1}\mt)(E_{I_{2}},\dots,E_{I_{k}})]-(\bD^{k-1}\mt)(\bD_{E_{I_{1}}}E_{I_{2}},\dots,E_{I_{k}})\\
      & -\dots-(\bD^{k-1}\mt)(E_{I_{2}},\dots,\bD_{E_{I_{1}}}E_{I_{k}}).
    \end{split}
  \end{equation*}
  Combining this observation with an induction argument yields the second statement and completes the proof of the lemma. 
\end{proof}

\section[Basic formulae and estimates]{Basic formulae and estimates for the covariant derivatives of the eigenvalues and frame}

Next, we express the covariant derivatives of the $\ell_{A}$ and the $X_{A}$ with respect to $\bge_{\refer}$ in terms of covariant derivatives
of $\mK$. 

\begin{lemma}\label{lemma:covderofframe}
  Let $(M,g)$ be a time oriented Lorentz manifold. Assume it to have an expanding partial pointed foliation and $\mK$ to be non-degenerate on
  $I$ and to have a global frame. Let $\xi$ be a vector field on $\bM\times I$ which is tangent to the constant-$t$ hypersurfaces. Then
  \begin{align}
    \bD_{\xi}\ell_{A} = & (\bD_{\xi}\mK)(Y^{A},X_{A}),\label{eq:covderlambdaA}\\
    Y^{A}(\bD_{\xi}X_{A}) = & -\sum_{B\neq A}\frac{1}{\ell_{A}-\ell_{B}}(\bD_{\xi}\mK)(Y^{B},X_{A})\bge_{\refer}(X_{B},X_{A})\label{eq:YAbDEiXA}
  \end{align}
  (no summation on $A$). Moreover, for $A\neq B$,
  \begin{equation}\label{eq:YBbDEiXA}
    Y^{B}(\bD_{\xi}X_{A})=\frac{1}{\ell_{A}-\ell_{B}}(\bD_{\xi}\mK)(Y^{B},X_{A}).
  \end{equation}
\end{lemma}
\begin{proof}
  Applying $\bD_{\xi}$ to
  \[
  \mK(Y^{B},X_{A})=\ell_{A}\de^{B}_{A}
  \]
  (no summation on $A$) yields
  \begin{equation}\label{eq:bDXAchK}
    (\bD_{\xi}\mK)(Y^{B},X_{A})+\mK(\bD_{\xi}Y^{B},X_{A})+\mK(Y^{B},\bD_{\xi}X_{A})=(\bD_{\xi}\ell_{A})\de^{B}_{A}.
  \end{equation}
  On the other hand,
  \begin{equation}\label{eq:bDXCXAYDformulae}
    \bD_{\xi}X_{A}=Y^{D}(\bD_{\xi}X_{A})X_{D},\ \ \
    \bD_{\xi}Y^{B}=-Y^{B}(\bD_{\xi}X_{D})Y^{D}.
  \end{equation}
  Inserting this information into (\ref{eq:bDXAchK}) yields
  \[
  (\bD_{\xi}\mK)(Y^{B},X_{A})+(\ell_{B}-\ell_{A})Y^{B}(\bD_{\xi}X_{A})=(\bD_{\xi}\ell_{A})\de^{B}_{A}
  \]
  (no summation). In particular, (\ref{eq:YBbDEiXA}) holds for $B\neq A$ and (\ref{eq:covderlambdaA}) holds. In order to calculate
  $Y^{A}(\bD_{\xi}X_{A})$ (no summation on $A$), note that
  \begin{equation*}
    \begin{split}
      0 = & \bD_{\xi}[\bge_{\refer}(X_{A},X_{A})]=2\bge_{\refer}(\bD_{\xi}X_{A},X_{A})=2Y^{B}(\bD_{\xi}X_{A})\bge_{\refer}(X_{B},X_{A})\\
      = & 2Y^{A}(\bD_{\xi}X_{A})+2\textstyle{\sum}_{B\neq A}Y^{B}(\bD_{\xi}X_{A})\bge_{\refer}(X_{B},X_{A})
    \end{split}
  \end{equation*}
  (no summation on $A$). Combining this observation with (\ref{eq:YBbDEiXA}) yields (\ref{eq:YAbDEiXA}). The lemma follows. 
\end{proof}

These formulae have the following immediate consequences.

\begin{cor}\label{cor:covderofframe}
  Let $(M,g)$ be a time oriented Lorentz manifold. Assume that it has an expanding partial pointed foliation. Assume, moreover, $\mK$ to be
  non-degenerate on $I$, to have a global frame and to be $C^{0}$-uniformly bounded on $I_{-}$; i.e. (\ref{eq:mKsupbasest}) to hold. Let $\xi$
  be a vector field on $\bM\times I$ which is tangent to the constant-$t$ hypersurfaces. Then there is a constant $C_{1}$, depending only on $n$, $\mKsup$
  and $\e_{\rond}$ such that
  \begin{equation}\label{eq:covderofframe}
    |\bD_{\xi}\ell_{A}|+|\bD_{\xi} Y^{A}|_{\bge_{\refer}}+|\bD_{\xi} X_{A}|_{\bge_{\refer}}\leq C_{1} |\xi|_{\bge_{\refer}}|\bD\mK|_{\bge_{\refer}}
  \end{equation}
  on $\bM_{t}$ for all $A\in \{1,\dots,n\}$ and $t\in I_{-}$. Defining the structure constants, say $\g_{AB}^{C}$, of the $X_{A}$ by
  $[X_{A},X_{B}]=\g_{AB}^{C}X_{C}$, the estimate
  \begin{equation}\label{eq:gaCABbasest}
    |\g^{C}_{AB}|\leq C_{1}|\bD\mK|_{\bge_{\refer}}
  \end{equation}
  also holds on $\bM_{t}$ for all $A,B,C\in \{1,\dots,n\}$ and $t\in I_{-}$.
\end{cor}
\begin{proof}
  Due to (\ref{eq:covderlambdaA}), it is clear that
  \[
  |\bD_{\xi}\ell_{A}|\leq |\bD\mK|_{\bge_{\refer}}|\xi|_{\bge_{\refer}}|X_{A}|_{\bge_{\refer}}|Y^{A}|_{\bge_{\refer}}
  =|\bD\mK|_{\bge_{\refer}}|\xi|_{\bge_{\refer}}|Y^{A}|_{\bge_{\refer}}
  \]
  (no summation on $A$). On the other hand, due to Lemma~\ref{lemma:frameinvest}, the right hand side can be estimated by the right hand side
  of (\ref{eq:covderofframe}) for a $C_{1}$ with the dependence stated in the lemma. The first equality in
  (\ref{eq:bDXCXAYDformulae}), combined with (\ref{eq:YAbDEiXA}), (\ref{eq:YBbDEiXA}), the assumptions and arguments similar to the above yields
  the desired estimate for the third term on the left hand side of (\ref{eq:covderofframe}). Next, the second equality in (\ref{eq:bDXCXAYDformulae}),
  combined with the above, yields the desired estimate of the second term on the left hand side of (\ref{eq:covderofframe}). Finally, note that
  \begin{equation}\label{eq:gABCformula}
    \begin{split}
      \g_{AB}^{C} = & Y^{C}([X_{A},X_{B}])=Y^{C}(\bD_{X_{A}}X_{B}-\bD_{X_{B}}X_{A}).
    \end{split}
  \end{equation}
  Arguments similar to the above yield the desired estimate for the structure constants. 
\end{proof}

\section{Higher order derivatives}

Corollary~\ref{cor:covderofframe} yields bounds on the first order derivatives of $\ell_{A}$, $X_{A}$ and $Y^{A}$. However, it is also of interest
to estimate higher order covariant derivatives. Before doing so, it is convenient to introduce some terminology. 
\begin{definition}\label{def:mfPmKhN}
  Let $(M,g)$ be a time oriented Lorentz manifold. Assume it to have an expanding partial pointed foliation. Given $0\leq m\in\zo$, let
  \begin{align*}
    \mfP_{\mK,m} := & \textstyle{\sum}_{m_{1}+\dots+m_{j}=m,m_{i}\geq 1}|\bD^{m_{1}}\mK|_{\bge_{\refer}}\cdots |\bD^{m_{j}}\mK|_{\bge_{\refer}},\\
    \mfP_{N,m} := & \textstyle{\sum}_{m_{1}+\dots+m_{j}=m,m_{i}\geq 1}|\bD^{m_{1}}\ln\hN|_{\bge_{\refer}}\cdots |\bD^{m_{j}}\ln\hN|_{\bge_{\refer}},\\
    \mfP_{\mK,N,m} := & \textstyle{\sum}_{m_{1}+m_{2}=m}\mfP_{\mK,m_{1}}\mfP_{N,m_{2}},
  \end{align*}
  \index{$\a$Aa@Notation!Functions!$\mfP_{\mK,m}$}%
  \index{$\a$Aa@Notation!Functions!$\mfP_{N,m}$}%
  \index{$\a$Aa@Notation!Functions!$\mfP_{\mK,N,m}$}%
  with the convention that $\mfP_{\mK,0}=1$ and $\mfP_{N,0}=1$.
\end{definition}

Next, we estimate higher order derivatives of $\ell_{A}$, $X_{A}$ and $Y^{A}$. 

\begin{lemma}\label{lemma:bDbfabDlmjchKest}
  Let $(M,g)$ be a time oriented Lorentz manifold. Assume it to have an expanding partial pointed foliation. Assume, moreover, $\mK$ to be
  non-degenerate on $I$, to have a global frame and to be $C^{0}$-uniformly bounded on $I_{-}$; i.e. (\ref{eq:mKsupbasest}) to hold. Then, for
  every pair of integers $j$ and $l$ satisfying $1\leq j\leq l$, and every multiindex $\bfI$ with $|\bfI|=j$, there is a constant $D_{\mK,l}$,
  depending only on $l$, $n$ and $(\bM,\bge_{\refer})$, such that
  \begin{equation}\label{eq:bDbfAbDlmjmKpteststmtEi}
    |\bD_{\bfI}\bD^{l-j}\mK|_{\bge_{\refer}}\leq D_{\mK,l}\textstyle{\sum}_{m=l-j+1}^{l}|\bD^{m}\mK|_{\bge_{\refer}}
  \end{equation}
  on $\bM\times I_{-}$. Similarly, there is a constant $\md_{\mK,j}$ depending only on $\mKsup$, $n$, $l$, $\e_{\rond}$ and $(\bM,\bge_{\refer})$ such that
  \begin{equation}\label{eq:bDbfAellAetcpteststmtEi}
    |\bD_{\bfI}\ell_{A}|+|\bD_{\bfI}X_{A}|_{\bge_{\refer}}+|\bD_{\bfI}Y^{A}|_{\bge_{\refer}}\leq \md_{\mK,j}\textstyle{\sum}_{m=1}^{j}\mfP_{\mK,m}
  \end{equation}
  on $\bM\times I_{-}$.
\end{lemma}
\begin{proof}
  The estimate (\ref{eq:bDbfAbDlmjmKpteststmtEi}) can be demonstrated by means of an induction argument, where the inductive step follows from
  Lemma~\ref{lemma:bDbfAbDmKexp}. In order to prove (\ref{eq:bDbfAellAetcpteststmtEi}), it is sufficient to
  proceed by induction and to appeal to (\ref{eq:covderlambdaA}), (\ref{eq:YAbDEiXA}), (\ref{eq:YBbDEiXA}) and (\ref{eq:bDbfAbDlmjmKpteststmtEi}).
\end{proof}

\section{Composite estimates}

In the chapters to follow, we need to estimate composite expressions. The purpose of the present section is to prove general estimates
to which we can refer in that context.

\begin{lemma}\label{lemma:compositeestEi}
  Let $(M,g)$ be a time oriented Lorentz manifold. Assume that it has an expanding partial pointed foliation. Assume, moreover, $\mK$ to be
  non-degenerate on $I$, to have a global frame and to be $C^{0}$-uniformly bounded on $I_{-}$; i.e. (\ref{eq:mKsupbasest}) to hold. Let $\{E_{i}\}$
  and $\{\omega^{i}\}$ be frames of the type
  introduced in Remark~\ref{remark:framenondegenerate}. Consider a product $P$ consisting of $k_{1}$ factors of type $\mrI$:
  $(\ell_{A}-\ell_{B})^{-1}f(\ell)$, where $f\in C^{\infty}(\rn{n},\ro)$, $A\neq B$ and $\ell=(\ell_{1},\dots,\ell_{n})$; $k_{2}$ factors of type
  $\mrII$: $\mt(Y^{A},X_{B})$ where $\mt$ is a $(1,1)$-tensor field on $\bM$; $k_{3}$ factors of type $\mrIII$: $\bge_{\refer}(X_{A},X_{B})$;
  $k_{4}$ factors of type $\mrIV$: $\hU(\ln\hN)$; $k_{5}$ factors of type $\mrV$: $\omega^{k}(\hN^{-1}\xi)$; $k_{6}$ factors of type $\mrVI$:
  $\omega^{i}(X_{A})$; $k_{7}$ factors of type $\mrVII$: $\hN^{-1}(\ml_{\zeta}\bge_{\refer})(X_{A},X_{B})$; and $k_{8}$ factors of type $\mrVIII$:
  $\hN^{-1}\omega^{k}(\ml_{\eta}E_{j})$. Let $\bfI$ be a vector field multiindex and $l:=|\bfI|$. Then, up to a constant depending only on $l$, $n$, $\e_{\rond}$,
  $\mKsup$, $(\bM,\bge_{\refer})$, the functions $f$ and the $k_{i}$, the expression $|E_{\bfI}(P)|$ can be estimated by a sum of products consisting
  of one factor of the form $\mfP_{\mK,N,m}$; $k_{2}$ factors of the form $|\bD^{p}\mt|_{\bge_{\refer}}$; $k_{4}$ factors of the form
  $|\bD^{q}\hU(\ln\hN)|_{\bge_{\refer}}$; $k_{5}$ factors of the form $\hN^{-1}|\bD_{\bfI}\xi|_{\bge_{\refer}}$; $k_{7}$ factors of the form
  $\hN^{-1}|\bD_{\bfJ}\bD_{\bfK}\zeta|_{\bge_{\refer}}$ (where $|\bfK|=1$);
  $k_{8,1}$ factors of the form $\hN^{-1}|\bD_{\bfL}\bD_{\bfM}\eta|_{\bge_{\refer}}$ (where $|\bfM|=1$) and $k_{8,2}$ factors of the form
  $\hN^{-1}|\bD_{\bfN}\eta|_{\bge_{\refer}}$, where $k_{8,1}+k_{8,2}=k_{8}$, and the sum of $m$, the $p$'s, the $q$'s, 
  the $|\bfI|$'s, the $|\bfJ|$'s, the $|\bfL|$'s and the $|\bfN|$'s is bounded from above by $l$.
\end{lemma}
\begin{remark}
  When we say that there are $k_{2}$ factors of the form $|\bD^{p}\mt|_{\bge_{\refer}}$, what we mean is that if the factors of type $\mrII$ are 
  $\mt_{i}(Y^{A_{i}},X_{B_{i}})$, $i=1,\dots,k_{2}$, then the $k_{2}$ factors of the form $|\bD^{p}\mt|_{\bge_{\refer}}$ are given by 
  $|\bD^{p_{i}}\mt_{i}|_{\bge_{\refer}}$, where the $p_{i}$'s are the $p$'s referred to at the end of the statement. Similar comments apply
  to the other factors. 
\end{remark}
\begin{remark}\label{remark:compositeestEi}
  In case $k_{5}=k_{7}=k_{8}=0$, the statement can be improved as follows: $|E_{\bfI}(P)|$ can, up to a constant depending only on $l$, $n$,
  $\e_{\rond}$, $\mKsup$, $(\bM,\bge_{\refer})$, the functions $f$ and the $k_{i}$, be estimated by a sum of products consisting of
  $\mfP_{\mK,q}$; $k_{2}$ factors of the form $|\bD^{r}\mt|_{\bge_{\refer}}$; and $k_{4}$ factors of the form $|\bD^{s}\hU(\ln\hN)|_{\bge_{\refer}}$, where
  the sum of $q$, the $r$'s and the $s$'s is bounded from above by $l$. Moreover, if, in addition to the above, $k_{6}=0$, then
  the sum of $q$, the $r$'s and the $s$'s is bounded from below by $\min\{1,l\}$.
  In case $k_{8}=0$, the statement can be improved as follows: $|E_{\bfI}(P)|$ can, up to a constant depending only
  on $l$, $n$, $\e_{\rond}$, $\mKsup$, $(\bM,\bge_{\refer})$, the functions $f$ and the $k_{i}$, be estimated by a sum of products consisting
  of $\mfP_{\mK,N,q}$; $k_{2}$ factors of the form $|\bD^{r}\mt|_{\bge_{\refer}}$; $k_{4}$ factors of the form $|\bD^{s}\hU(\ln\hN)|_{\bge_{\refer}}$;
  $k_{5}$ factors of the form $\hN^{-1}|\bD_{\bfI}\xi|_{\bge_{\refer}}$; and $k_{7}$ factors of the form $\hN^{-1}|\bD_{\bfJ}\bD_{\bfK}\zeta|_{\bge_{\refer}}$
  (where $|\bfK|=1$), where the sum of $q$, the $r$'s, the $s$'s, the $|\bfI|$'s and the $|\bfJ|$'s is bounded from above by $l$. 
\end{remark}
\begin{proof}
  In order to estimate $E_{\bfI}(P)$, note that if $E_{\bfI_{1}}$ hits a factor of type $\mrI$, then the result can be estimated by a sum of
  terms of the form $C\mfP_{\mK,l_{a}}$, where $l_{a}\leq l_{1}:=|\bfI_{1}|$ and $C$ only depends on $f$, $\mKsup$, $\e_{\rond}$, $l_{1}$,
  $(\bM,\bge_{\refer})$ and $n$, and we appealed to (\ref{eq:bDbfAellAetcpteststmtEi}). Next, if $E_{\bfI_{2}}$ hits a factor of type $\mrII$,
  then we need to estimate
  \[
  (\bD_{\bfJ}\mt)(\bD_{\bfK}Y^{A},\bD_{\bfL}X_{B}),
  \]
  where $|\bfJ|+|\bfK|+|\bfL|=|\bfI_{2}|$. Due to Lemma~\ref{lemma:bDbfAbDkequiv} and
  (\ref{eq:bDbfAellAetcpteststmtEi}), $E_{\bfI_{2}}$ applied to a factor of type $\mrII$ can be estimated by
  \begin{equation}\label{eq:estoftypeIIfactorEi}
    C\textstyle{\sum}_{l_{a}+l_{b}\leq l_{2}}\mfP_{\mK,l_{a}}|\bD^{l_{b}}\mt|_{\bge_{\refer}},
  \end{equation}
  where $C$ only depends on $\mKsup$, $\e_{\rond}$, $l_{2}:=|\bfI_{2}|$, $(\bM,\bge_{\refer})$ and $n$. If $E_{\bfI_{3}}$ hits a factor of type
  $\mrIII$, then the result can be estimated by a sum of terms of the form $C\mfP_{\mK,l_{b}}$, where $l_{b}\leq l_{3}:=|\bfI_{3}|$ and $C$ only
  depends on $\mKsup$, $\e_{\rond}$, $l_{3}$, $(\bM,\bge_{\refer})$ and $n$, and we appealed to (\ref{eq:bDbfAellAetcpteststmtEi}). Due to
  Lemma~\ref{lemma:bDbfAbDkequiv}, $E_{\bfI_{4}}$ applied to a factor of type $\mrIV$ can be estimated by
  a sum of expressions of the form $C|\bD^{l_{a}}\hU(\ln\hN)|_{\bge_{\refer}}$, where $l_{a}\leq l_{4}:=|\bfI_{4}|$, where $C$ only depends on
  $l_{4}$, $n$ and $(\bM,\bge_{\refer})$. Applying $E_{\bfI_{5}}$ to a factor of type $\mrV$, we need to estimate
  \[
  (\bD_{\bfJ}\omega^{k})(\hN^{-1}\bD_{\bfK}\xi)\cdot [\hN E_{\bfL}(\hN^{-1})]
  \]
  where $|\bfJ|+|\bfK|+|\bfL|=|\bfI_{5}|$. Similarly to the above arguments, when $E_{\bfI_{5}}$ hits a factor of type $\mrV$, the result can
  thus be estimated by
  \begin{equation}\label{eq:typeVfactorestimate}
  C\textstyle{\sum}_{l_{a}+|\bfJ|\leq l_{c}}\mfP_{N,l_{a}}\hN^{-1}|\bD_{\bfJ}\xi|_{\bge_{\refer}},
  \end{equation}
  where $l_{c}\leq l_{5}:=|\bfI_{5}|$ and $C$ only depends on $l_{5}$, $(\bM,\bge_{\refer})$ and $n$. The contribution
  arising when applying $E_{\bfI_{6}}$ to a factor of type $\mrVI$ can be estimated as in the case of factors of type $\mrIII$. Before
  considering terms of type $\mrVII$, note that
  \begin{equation*}
    \begin{split}
      (\ml_{\zeta}\bge_{\refer})(X_{A},X_{B}) = & \bge_{\refer}(\bD_{X_{A}}\zeta,X_{B})+\bge_{\refer}(X_{A},\bD_{X_{B}}\zeta)\\
      = & \omega^{i}(X_{A})\omega_{j}(X_{B})[\bge_{\refer}(\bD_{E_{i}}\zeta,E_{j})+\bge_{\refer}(E_{i},\bD_{E_{j}}\zeta)].
    \end{split}
  \end{equation*}
  Due to this observation, the desired estimate for factors of type $\mrVII$ follows by combining the arguments in the case of factors of
  type $\mrV$ and $\mrVI$. To conclude, if $E_{\bfI_{7}}$ hits a factor of type $\mrVII$, the result can be estimated by
  \[
  C\textstyle{\sum}_{l_{a}+|\bfI|\leq l_{7}, |\bfJ|=1}\mfP_{\mK,N,l_{a}}\hN^{-1}|\bD_{\bfI}\bD_{\bfJ}\xi|_{\bge_{\refer}},
  \]
  where $C$ only depends on $\mKsup$, $\e_{\rond}$, $l_{7}:=|\bfI_{7}|$, $n$ and $(\bM,\bge_{\refer})$. Since
  \begin{equation}\label{eq:hNinvomegaimletaEjform}
  \hN^{-1}\omega^{i}(\ml_{\eta}E_{j})=\hN^{-1}\omega^{i}(\bD_{\eta}E_{j}-\bD_{E_{j}}\eta),
  \end{equation}
  terms of type $\mrVIII$ can be estimated similarly to the above. In fact, if $E_{\bfI_{8}}$ hits a factor of type $\mrVIII$, the result can be
  estimated by
  \begin{equation}\label{eq:estimatefortypeVIIIfactors}
    \begin{split}
      &C\textstyle{\sum}_{l_{a}+|\bfJ|\leq l_{8}}\mfP_{N,l_{a}}\hN^{-1}(|\bD_{\bfJ}\bD_{\bfK}\eta|_{\bge_{\refer}}+|\bD_{\bfJ}\eta|_{\bge_{\refer}}),
    \end{split}
  \end{equation}
  where $|\bfK|=1$, $l_{8}:=|\bfI_{8}|$ and $C$ only depends on $l_{8}$, $(\bM,\bge_{\refer})$ and $n$. Combining the above estimates yields
  the conclusion of the lemma, as well as the statements made in the remarks. 
\end{proof}

\chapter{Lie derivatives of the frame}\label{chapter:liederivativesframe}

The main purpose of the present chapter is to derive formulae for Lie derivatives of the elements of the frame $\{X_{A}\}$ with respect to 
the future pointing unit normal. However, we also wish to relate geometric and non-geometric norms of the normal derivative of the expansion
normalised Weingarten map. The reason for this is that the main assumptions in these notes are expressed using non-geometric norms. It is 
therefore of interest to relate the two perspectives. We end the chapter by considering the commutator of $\hU$ and $E_{i}$. In particular, we
derive expressions and estimates for the corresponding coefficients and their normal derivatives.

\section{Time derivative, geometric perspective}

Define $\bmu_{A}$ by the requirement that (\ref{eq:bgenormXA}) holds; recall that $\{X_{A}\}$ is an orthogonal frame with respect to $\bge$. Introduce
\[
\sfX_{A}:=e^{-\bmu_{A}}X_{A}.
\]
\index{$\a$Aa@Notation!Vector fields!$\sfX_{A}$}%
Then $\{\sfX_{A}\}$ is an orthonormal frame with respect to $\bge$ with dual frame $\{\sfY^{A}\}$.
\index{$\a$Aa@Notation!One form fields!$\sfY^{A}$}%
However, \textit{we extend $\sfY^{A}$ in such a way that $\sfY^{A}(U)=0$}. In what follows, it will also be convenient to use the notation
\index{$\a$Aa@Notation!Operators!$\hml_{U}$}%
\begin{equation}\label{eq:hmlUdef}
\hml_{U}:=\theta^{-1}\ml_{U}. 
\end{equation}

\begin{lemma}\label{lemma:mMstructure}
  Let $(M,g)$ be a time oriented Lorentz manifold. Assume that it has an expanding partial pointed foliation. Assume, moreover, $\mK$ to be
  non-degenerate on $I$ and to have a global frame. Let $\mM$ and $\mL$ be the matrix valued functions on $\bM\times I$ whose components are
  given by
  \begin{equation}\label{eq:mMBCmLBCdef}
    \mM^{B}_{C}:=(\hml_{U}\sfY^{B})(\sfX_{C}),\ \ \
    \mL^{B}_{C}:=\ell_{C}\de^{B}_{C}
  \end{equation}
  (no summation on $C$). Then $\mM=\mL+\mA$, where $\mA:=(\mM-\mM^{T})/2$. In particular, $\mM$ is the sum of a diagonal matrix plus an
  antisymmetric matrix.
\end{lemma}
\begin{proof}
Let $X$ and $Y$ be vector fields on $\bM\times I$ tangent to $\bM$. Then it can be calculated that 
\[
\bk(X,Y)=\frac{1}{2}(\ml_{U}g)(X,Y).
\]

Next, note that 
\[
g=-U^{\flat}\otimes U^{\flat}+\textstyle{\sum}_{A}\sfY^{A}\otimes\sfY^{A}.
\]
In particular, 
\[
\ml_{U}g=-(\ml_{U}U^{\flat})\otimes U^{\flat}-U^{\flat}\otimes (\ml_{U}U^{\flat})+\textstyle{\sum}_{A}(\ml_{U}\sfY^{A})\otimes\sfY^{A}
+\textstyle{\sum}_{A}\sfY^{A}\otimes (\ml_{U}\sfY^{A}).
\]
Thus
\begin{equation}\label{eq:mlUgformula}
(\ml_{U}g)(\sfX_{B},\sfX_{C})=(\ml_{U}\sfY^{C})(\sfX_{B})+(\ml_{U}\sfY^{B})(\sfX_{C}).
\end{equation}
On the other hand, 
\[
(\ml_{U}g)(\sfX_{B},\sfX_{C})=2\bk(\sfX_{B},\sfX_{C})=2\bge(\bK \sfX_{B},\sfX_{C})=2\theta\bge(\mK \sfX_{B},\sfX_{C})=2\theta\ell_{B}\de_{BC}
\]
(no summation on $B$). Let $\mM$ and $\mL$ be defined as in the statement of the lemma. Then the equality (\ref{eq:mlUgformula}) can be written
\[
2\mL=\mM+\mM^{T}. 
\]
The lemma follows. 
\end{proof}

\section{Formulae, geometric and non-geometric perspectives}

Let $\ml_{U}\mK$ be defined by (\ref{eq:mlUchKdefgeometric}). Then 
\begin{equation}\label{eq:mlhUchKYBXAIIUv}
  (\ml_{U}\mK)(Y^{B},X_{A})=U[Y^{B}(\mK X_{A})]-(\ml_{U}Y^{B})(\mK X_{A})-Y^{B}[\mK \overline{\ml_{U}X_{A}}],
\end{equation}
where the overline signifies orthogonal projection. Note also that we here think of $Y^{A}$ as being extended to $\bM\times I$ in such a way that 
$Y^{A}(U)=0$. In what follows, we wish to relate $\hml_{U}\mK$ to $\hml_{U}X_{A}$. Let, to this end, 
\begin{equation}\label{eq:mWdefrel}
  \hml_{U}X_{A}=\mW^{B}_{A}X_{B}+\omW^{0}_{A}U
\end{equation}
\index{$\a$Aa@Notation!Functions!$\mW^{B}_{A}$}%
\index{$\a$Aa@Notation!Functions!$\omW^{0}_{A}$}%
define $\mW^{A}_{B}$ and $\omW^{0}_{B}$, where $\hml_{U}$ is introduced in (\ref{eq:hmlUdef}).
\begin{lemma}
  Let $(M,g)$ be a time oriented Lorentz manifold. Assume that it has an expanding partial pointed foliation. Assume, moreover, $\mK$ to be
  non-degenerate on $I$ and to have a global frame. Then
  \begin{align}
    \hU(\ell_{A}) = & (\hml_{U}\mK)(Y^{A},X_{A}),\label{eq:hUellAnongeo}\\
    \mW^{A}_{A} = & -\textstyle{\sum}_{B\neq A}\mW^{B}_{A}\bge_{\refer}(X_{B},X_{A})+\frac{1}{2\hN}(\ml_{\chi}\bge_{\refer})(X_{A},X_{A}),
    \label{eq:mWAAformulanongeo}\\
    \mW^{B}_{A} = & \textstyle{\frac{1}{\ell_{A}-\ell_{B}}}(\hml_{U}\mK)(Y^{B},X_{A}),\label{eq:mWBAformulanongeo}
  \end{align}
  where there is no summation on $A$ in the first and second equalities and $A\neq B$ in the third equality. Moreover, if $\mM^{0}_{A}$ is defined by
  \[
  \hml_{U}\sfX_{A}=-\mM^{0}_{A}U-\mM^{B}_{A}\sfX_{B},
  \]
  where $\mM$ is the matrix introduced in Lemma~\ref{lemma:mMstructure}, then
  \begin{align}
    \hU(\ell_{A}) = & (\hml_{U}\mK)(\sfY^{A},\sfX_{A}),\label{eq:hUellAgeo}\\
    \mM^{B}_{A} = & \textstyle{\frac{1}{\ell_{B}-\ell_{A}}}(\hml_{U}\mK)(\sfY^{B},\sfX_{A}),\label{eq:mMBAformula}
  \end{align}
  where there is no summation on $A$ in the first equality and $A\neq B$ in the second equality. Note also that $\mM^{A}_{A}$ (no summation on
  $A$) equals $\ell_{A}$ due to Lemma~\ref{lemma:mMstructure}. Finally,
  \begin{align}
    \omW^{0}_{A} = & \theta^{-1}X_{A}(\ln N),\label{eq:mWzAform}\\
    \mM^{0}_{A} = & -\theta^{-1}\sfX_{A}(\ln N),\label{eq:mMzAform}
  \end{align}
\end{lemma}
\begin{proof}
The first term on the right hand side of (\ref{eq:mlhUchKYBXAIIUv}) is given by 
\[
U[Y^{B}(\mK X_{A})]=U(\ell_{A})\de^{B}_{A}
\]
(no summation on $A$). Due to (\ref{eq:mWdefrel}), the relation $\overline{\ml_{U}X_{A}}=\theta\mW^{B}_{A}X_{B}$ holds, so that 
\[
-Y^{B}[\mK \overline{\ml_{U}X_{A}}]=-Y^{B}[\mK \theta\mW^{C}_{A}X_{C}]=-\theta\textstyle{\sum}_{C}\ell_{C}\mW^{C}_{A}Y^{B}(X_{C})=-\theta\ell_{B}\mW^{B}_{A}
\] 
(no summation on $B$). Combining $(\ml_{U}Y^{B})(X_{A})=-Y^{B}(\ml_{U}X_{A})$ with (\ref{eq:mWdefrel}) and the fact that $Y^{B}(U)=0$ yields
\begin{equation}\label{eq:mlhTYBXAUv}
(\ml_{U}Y^{B})(X_{A})=-Y^{B}(\theta\mW_{A}^{C}X_{C})=-\theta\mW^{B}_{A}.
\end{equation}
In particular, 
\[
-(\ml_{U}Y^{B})(\mK X_{A})=-\ell_{A}(\ml_{U}Y^{B})(X_{A})=\ell_{A}\theta\mW^{B}_{A}
\]
(no summation on $A$). Summing up the above observations yields 
\begin{equation}\label{eq:mlhTYBXAformulaUv}
(\ml_{U}\mK)(Y^{B},X_{A})=U(\ell_{A})\de^{B}_{A}+\theta\ell_{A}\mW^{B}_{A}-\theta\ell_{B}\mW^{B}_{A}
\end{equation}
(no summation on $A$ or $B$). In particular, (\ref{eq:hUellAnongeo}) and (\ref{eq:mWBAformulanongeo}) hold. We can also carry through the above
argument with $X_{A}$, $Y^{B}$ and $\mW^{A}_{B}$ replaced by $\sfX_{A}$, $\sfY^{B}$ and $-\mM^{A}_{B}$ respectively. This yields (\ref{eq:hUellAgeo})
and (\ref{eq:mMBAformula}).

Let $\{E_{i}\}$ be an orthonormal basis with respect to $\bge_{\refer}$ as in Remark~\ref{remark:framenondegenerate} and let $X^{i}_{A}$ be the
components of $X_{A}$ with respect to this basis. Then 
\begin{equation}\label{eq:hTbgeXAXAexpUv}
U[\bge_{\refer}(X_{A},X_{A})]=2\de_{ij}U(X^{i}_{A})X^{j}_{A}=2\bge_{\refer}(U(X^{i}_{A})E_{i},X_{A}). 
\end{equation}
On the other hand, 
\[
\ml_{U}X_{A}=U(X^{i}_{A})E_{i}+X^{i}_{A}\ml_{U}E_{i}. 
\]
Moreover, (\ref{eq:mlUEkform}) yields $\overline{\ml_{U}E_{i}}=-N^{-1}\ml_{\chi}E_{i}$, so that 
\begin{equation}\label{eq:overlinemlUitoEietc}
\overline{\ml_{U}X_{A}}=U(X^{i}_{A})E_{i}-\frac{1}{N}X^{i}_{A}\ml_{\chi}E_{i}. 
\end{equation}
Adding up the above yields
\begin{equation*}
\begin{split}
0 = & U[\bge_{\refer}(X_{A},X_{A})]=2\bge_{\refer}(\overline{\ml_{U}X_{A}}+N^{-1}X^{i}_{A}\ml_{\chi}E_{i},X_{A})\\
 = & 2\bge_{\refer}(\overline{\ml_{U}X_{A}},X_{A})-N^{-1}(\ml_{\chi}\bge_{\refer})(X_{A},X_{A}) 
\end{split}
\end{equation*}
(no summation on $A$). On the other hand, 
\[
\bge_{\refer}(\overline{\ml_{U}X_{A}},X_{A})=\bge_{\refer}(\theta\mW^{B}_{A}X_{B},X_{A})=\theta\mW^{A}_{A}
+\textstyle{\sum}_{B\neq A}\theta\mW^{B}_{A}\bge_{\refer}(X_{B},X_{A})
\]
(no summation on $A$). Combining the last two equalities yields (\ref{eq:mWAAformulanongeo}). The derivations of (\ref{eq:mWzAform}) and
(\ref{eq:mMzAform}) are similar to the above. 
\end{proof}

\subsection{Norm equivalences} 
One particular consequence of (\ref{eq:hUellAgeo}) and (\ref{eq:mMBAformula}) is that there is a numerical constant $C$ such that 
\[
\textstyle{\sum}_{A}|\hU(\ell_{A})|+\|\mA\|\leq C\left(1+\textstyle{\sum}_{A\neq B}|\ell_{A}-\ell_{B}|^{-1}\right)|\hml_{U}\mK|_{\bge},
\]
where $\mA$ is the antisymmetric matrix introduced in the statement of Lemma~\ref{lemma:mMstructure}. Moreover, (\ref{eq:hUellAgeo}) and
(\ref{eq:mMBAformula}) also imply that there is a numerical constant $C$ such that 
\[
|\hml_{U}\mK|_{\bge}\leq C\textstyle{\sum}_{A}\left(|\hU(\ell_{A})|+|\ell_{A}|\cdot \|\mA\|\right)
\]
In other words, controlling $|\hml_{U}\mK|_{\bge}$ is equivalent to controlling $|\hU(\ell_{A})|$ and $\|\mA\|$, given that the $\ell_{A}$
and the $|\ell_{A}-\ell_{B}|^{-1}$ ($A\neq B$) are bounded. Considering (\ref{eq:hUellAnongeo}) and (\ref{eq:mWBAformulanongeo}), it is clear that 
there is a similar statement concerning $|\hml_{U}\mK|_{\bge_{\refer}}$. However, in order to obtain such a statement, we need to assume $\mK$ 
to be non-degenerate, to have a global frame and to be $C^{0}$-uniformly bounded. The equivalent objects in this case are $|\hU(\ell_{A})|$ and
$\|\mW_{\rood}\|$; here $\mW_{\rood}$ is the matrix whose off-diagonal components equal those of $\mW$ and whose diagonal components vanish.

\subsection{Relating geometric and non-geometric norms}\label{ssection:relgeoandnongeonorms}
Next, let us estimate $\|\mA\|$ in terms of $\|\mW_{\rood}\|$. Compute, to this end, 
\begin{equation}\label{eq:mMABrel}
\begin{split}
\mM^{A}_{B} = & (\hml_{U}\sfY^{A})(\sfX_{B})=-\sfY^{A}(\hml_{U}\sfX_{B})=\hU(\bmu_{B})\de^{A}_{B}
-\sfY^{A}(e^{-\bmu_{B}}\hml_{U}X_{B})\\
 = & \hU(\bmu_{B})\de^{A}_{B}-\sfY^{A}(e^{\bmu_{C}-\bmu_{B}}\mW_{B}^{C}\sfX_{C})=\hU(\bmu_{B})\de^{A}_{B}-e^{\bmu_{A}-\bmu_{B}}\mW_{B}^{A}
\end{split}
\end{equation}
(no summation). In particular, 
\begin{equation}\label{eq:mMAAhUbmuAmWAA}
\mM^{A}_{A}=\hU(\bmu_{A})-\mW^{A}_{A}
\end{equation}
(no summation on $A$). Moreover, if $A\neq B$, then 
\begin{equation}\label{eq:mAABmWABrel}
-e^{\bmu_{A}-\bmu_{B}}\mW_{B}^{A}=\mA^{A}_{B}
\end{equation}
(no summation). 
At this point, the fact that the right hand side of this equality is antisymmetric has important consequences. In fact, combining (\ref{eq:mAABmWABrel})
with the antisymmetry of $\mA$ yields
\begin{equation}\label{eq:mAbdbymWrood}
|\mA^{A}_{B}|\leq e^{-|\bmu_{A}-\bmu_{B}|}\|\mW_{\rood}\|,
\end{equation}
where $\mW_{\rood}$ is the matrix whose off-diagonal components equal those of $\mW$ and whose diagonal components vanish. In particular, 
in an anisotropic setting, the $\bmu_{A}$ can be expected to grow linearly at different rates. If, in addition, $\|\mW_{\rood}\|$ is bounded,
then $\|\mA\|$ decays exponentially. Finally, note that since $\|\mA\|$ is dominated by $\|\mW_{\rood}\|$ due to (\ref{eq:mAbdbymWrood}), it 
is clear that non-geometric control on $\hml_{U}\mK$ implies geometric control on $\hml_{U}\mK$.

\section{Contribution from the shift vector field}\label{section:contributionfromshift}

Assume now that there is an orthonormal frame $\{E_{i}\}$ on $\bM$ with respect to $\bge_{\refer}$, with dual frame $\{\omega^{i}\}$.
Note that 
\begin{equation}\label{eq:hUEicomm}
  [\hU,E_{i}]=A_{i}^{0}\hU+A_{i}^{k}E_{k},
\end{equation}
where
\index{$\a$Aa@Notation!Functions!$A^{k}_{i}$}%
\index{$\a$Aa@Notation!Functions!$A^{0}_{i}$}%
\begin{equation}\label{eq:Aialphadef}
A_{i}^{0}:=E_{i}(\ln\hN),\ \ \
A^{k}_{i}:=-\hN^{-1}\omega^{k}(\ml_{\chi}E_{i}).
\end{equation}

\begin{lemma}\label{lemma:hUAik}
  Let $(M,g)$ be a time oriented Lorentz manifold. Assume that it has an expanding partial pointed foliation and that there are frames
  $\{E_{i}\}$ and $\{\omega^{i}\}$ as above. Then
  \begin{equation}\label{eq:hUAik}
    \begin{split}
      \hU(A^{k}_{i}) = & \hN^{-1}\omega^{k}(\ml_{E_{i}}\dotchi)+A_{i}^{0}\hN^{-1}\omega^{k}(\dotchi)-\hU(\ln\hN)A_{i}^{k}\\
      & -\hN^{-1}\chi(A_{i}^{k})-\hN^{-1}\chi(\ln\hN)A_{i}^{k},
    \end{split}
  \end{equation}
  where $\dotchi$ is introduced in (\ref{eq:dotchidef}). In particular,
  \begin{equation}\label{eq:EbfIhUAik}
    \begin{split}
      |E_{\bfI}[\hU(A_{i}^{k})]| \leq & C\textstyle{\sum}_{l_{a}+|\bfJ|\leq l+1}\mfP_{N,l_{a}}\hN^{-1}|\bD_{\bfJ}\dotchi|_{\bge_{\refer}}\\
      & +C\textstyle{\sum}_{l_{a}+l_{b}+|\bfJ|\leq l+1;l_{a}+l_{b}\leq l}\mfP_{N,l_{a}}|\bD^{l_{b}}\hU(\ln\hN)|_{\bge_{\refer}}\hN^{-1}|\bD_{\bfJ}\chi|_{\bge_{\refer}}\\
      & +C\textstyle{\sum}_{l_{a}+|\bfJ|+|\bfK|\leq l+2;|\bfJ|\leq l;l_{a}\leq l+1}\mfP_{N,l_{a}}
      \hN^{-1}|\bD_{\bfJ}\chi|_{\bge_{\refer}}\hN^{-1}|\bD_{\bfK}\chi|_{\bge_{\refer}},  
    \end{split}
  \end{equation}
  where $l:=|\bfI|$ and $C$ only depends on $l$, $n$ and $(\bM,\bge_{\refer})$. 
\end{lemma}
\begin{proof}
  Note that (\ref{eq:Aialphadef}) implies
  \[
  -\hN^{-1}\ml_{\chi}E_{i}=A_{i}^{k}E_{k}.
  \]
  Applying $\ml_{\hU}$ to this equality yields
  \begin{equation}\label{eq:hUmSABderfsEi}
    -\hU(\ln\hN)A_{i}^{k}E_{k}-\hN^{-1}\ml_{\hU}\ml_{\chi}E_{i}=\hU(A_{i}^{k})E_{k}+A_{i}^{k}\ml_{\hU}E_{k}. 
  \end{equation}
  In order to proceed, it is of interest to calculate
  \begin{equation*}
    \begin{split}
      \ml_{\hU}\ml_{\chi}E_{i} = & -\ml_{\hU}\ml_{E_{i}}\chi=-[\hU,[E_{i},\chi]=-\hU(E_{i}\chi-\chi E_{i})+(E_{i}\chi-\chi E_{i})\hU\\
        = & -\hU E_{i}\chi+E_{i}\hU\chi-E_{i}\hU\chi+\hU\chi E_{i}+E_{i}\chi \hU-\chi E_{i}\hU+\chi \hU E_{i}-\chi \hU E_{i}\\
        = & -(\ml_{\hU}E_{i})\chi+\chi(\ml_{\hU}E_{i})-E_{i}\ml_{\hU}\chi+\ml_{\hU}\chi E_{i}\\
        = & -A_{i}^{0}\hU\chi-A_{i}^{k}E_{k}\chi+\chi(A_{i}^{0}\hU+A_{i}^{k}E_{k})-\ml_{E_{i}}\ml_{\hU}\chi\\
        = & -A_{i}^{0}\ml_{\hU}\chi-A_{i}^{k}\ml_{E_{k}}\chi+\chi(A_{i}^{0})\hU+\chi(A_{i}^{k})E_{k}-\ml_{E_{i}}\ml_{\hU}\chi.
    \end{split}
  \end{equation*}
  In particular
  \begin{equation}\label{eq:mlhUmlchiEiformula}
    \begin{split}
      \hN^{-1}\overline{\ml_{\hU}\ml_{\chi}E_{i}} = & -A_{i}^{0}\hN^{-1}\overline{\ml_{\hU}\chi}-A_{i}^{k}\hN^{-1}\ml_{E_{k}}\chi
      +\hN^{-1}\chi(A_{i}^{k})E_{k}-\hN^{-1}\overline{\ml_{E_{i}}\ml_{\hU}\chi}\\
      = & -A_{i}^{0}\hN^{-1}\dotchi-A_{i}^{k}A_{k}^{l}E_{l}+\hN^{-1}\chi(A_{i}^{k})E_{k}-\hN^{-1}\overline{\ml_{E_{i}}\ml_{\hU}\chi}.
    \end{split}
  \end{equation}
  In order to simplify the last expression, note that
  \[
  \ml_{\hU}\chi=\hU(\chi^{k})E_{k}+\chi^{k}A_{k}^{0}\hU+\chi^{k}A_{k}^{l}E_{l}.
  \]
  In particular,
  \begin{equation}\label{eq:dotchirelations}
  \dotchi=\hU(\chi^{k})E_{k}+\chi^{k}A_{k}^{l}E_{l},\ \ \
  \ml_{\hU}\chi=\dotchi+\chi^{k}A_{k}^{0}\hU.
  \end{equation}
  Thus
  \[
  \ml_{E_{i}}\ml_{\hU}\chi=\ml_{E_{i}}\dotchi+E_{i}(\chi^{k}A_{k}^{0})\hU+\chi^{k}A_{k}^{0}\ml_{E_{i}}\hU,
  \]
  so that
  \[
  \overline{\ml_{E_{i}}\ml_{\hU}\chi}=\ml_{E_{i}}\dotchi-\chi(\ln \hN)A_{i}^{l}E_{l}.
  \]
  Combining this calculation with (\ref{eq:mlhUmlchiEiformula}) yields
  \begin{equation*}
    \begin{split}
      \hN^{-1}\overline{\ml_{\hU}\ml_{\chi}E_{i}}
      = & -\hN^{-1}\ml_{E_{i}}\dotchi-A_{i}^{0}\hN^{-1}\dotchi-A_{i}^{k}A_{k}^{l}E_{l}\\
      & +\hN^{-1}\chi(A_{i}^{k})E_{k}+\hN^{-1}\chi(\ln \hN)A_{i}^{l}E_{l}. 
    \end{split}
  \end{equation*}
  Combining this observation with (\ref{eq:hUmSABderfsEi}) yields (\ref{eq:hUAik}).

  In order to prove (\ref{eq:EbfIhUAik}), note that the first term on the right hand side of (\ref{eq:hUAik}) yields expressions that can
  be estimated by the first term on the right hand side of (\ref{eq:EbfIhUAik}). This follows from the end of the proof of
  Lemma~\ref{lemma:compositeestEi}, in particular (\ref{eq:estimatefortypeVIIIfactors}). Consider the second term on the right hand side of
  (\ref{eq:hUAik}). It also yields expressions that can be estimated by the first term on the right hand side of (\ref{eq:EbfIhUAik}). This follows
  from the proof of Lemma~\ref{lemma:compositeestEi}, in particular (\ref{eq:typeVfactorestimate}). The third term on the right hand side of
  (\ref{eq:hUAik}) yields expressions that can be estimated by the second term on the right hand side of (\ref{eq:EbfIhUAik}). This follows from
  the proof of Lemma~\ref{lemma:compositeestEi}, in particular the estimates for factors of type $\mrIV$ and $\mrVIII$. Finally, by similar arguments,
  the last two terms on the right hand side of yield expressions that can be estimated by the last term on the right hand side of
  (\ref{eq:EbfIhUAik}).
\end{proof}

\chapter{Estimating the norm of the elements of the frame}\label{chapter:normofelementsofframe}

Recall the notation $\mu_{A}$ and $\bmu_{A}$ introduced in (\ref{eq:muAdef}) and (\ref{eq:bgenormXA}). The asymptotic behaviour of these objects
is of central importance for understanding the causal structure and the asymptotic behaviour of solutions to (\ref{eq:theequation}). 
In particular, we need lower bounds on $\mu_{A}$ on $I_{-}$, where $I_{-}=I\cap (-\infty,t_{0}]$. Deriving such estimates is the main goal of the
present chapter. However, we are also interested in estimating the spatial variation of $\varrho$ and in proving that if $\bx_{0}\in\bM$, then
$\tau(t):=\varrho(\bx_{0},t)$ can be used as a time coordinate. Beyond these main goals, we record additional estimates for, e.g., the weights that
later appear in the energy estimates. 

The lower bound on $\mu_{A}$ is based on considering the evolution of this quantity along the integral curves of $\hU$.
The same is true of $\varrho$. In the course of the estimates, it is necessary to control the divergence of $\chi$ as well as certain Lie
derivatives involving $\chi$. Obtaining such estimates is the purpose of Section~\ref{section:basicestimatesshiftvectorfield}. Needless to say,
we also need to derive formulae for $\hU(\varrho)$ and $\hU(\bmu_{A})$. This is the purpose of Section~\ref{section:geometricidentitites}. Given this
information, we are in a position to derive the main conclusion of the chapter, lower bounds on $\mu_{A}$; cf. Section~\ref{section:mainmuAestimate}.
To achieve this goal, we need to
assume the shift vector field to be small. We also need to assume $\hml_{U}\mK$ to satisfy a weak off-diagonal exponential bound. The proof is based on
a bootstrap argument along the integral curves of $\hU$. The conclusion is that the $\mu_{A}$ grow linearly in $\varrho$ in the direction of the
singularity. This can be interpreted as saying that the conformally rescaled metric $\hg$ exhibits exponential growth in the direction of the
singularity. However, the expansion is not isotropic. 

The next goal is to control the spatial variation of $\varrho$. To this end, we need to commute the evolution equation for $\varrho$ with
$E_{i}$. This leads to the necessity of controlling an additional derivative of $\chi$. Following this estimate, we demonstrate that
$\d_{t}\varrho$ and $\hN$ are comparable; cf. Lemma~\ref{lemma:taurelvaryingbxEi}. This allows us to introduce the time coordinate $\tau$ as
above. We end the chapter by discussing the properties of weight functions that are of importance in the definition of the energies. 

\section{Basic estimates of the shift vector field}\label{section:basicestimatesshiftvectorfield}

Two expressions involving $\chi$ that appear frequently in the analysis are $\rodiv_{\bge_{\refer}}\chi$ and the second term on the right hand side of 
(\ref{eq:mWAAformulanongeo}). We begin by estimating them. 
\begin{lemma}
  Let $(M,g)$ be a time oriented Lorentz manifold. Assume that it has an expanding partial pointed foliation. Assume, moreover, $\mK$ to be
  non-degenerate on $I$ and to have a global frame. Then
  \begin{align}
    (2\hN)^{-1}|(\ml_{\chi}\bge_{\refer})(X_{A},X_{A})| \leq & n^{1/2}e^{-\mu_{\min}}|\bD\chi|_{\rohy},\label{eq:WAAchicontr}\\
    \hN^{-1}|\rodiv_{\bge_{\refer}}\chi| \leq & n^{1/2}e^{-\mu_{\min}}|\bD\chi|_{\rohy}\label{eq:rodivestimateechi}
  \end{align}
  on $I_{-}$, where
  \index{$\a$Aa@Notation!Functions!$\mu_{\min}$}%
  \begin{equation}\label{eq:mumindef}
    \mu_{\min}:=\min_{A}\mu_{A}.
  \end{equation}
\end{lemma}
\begin{proof}
Due to (\ref{eq:bDkchirohydef}) and (\ref{eq:bgenormXA}), it is clear that 
\begin{equation}\label{eq:bDchireformnorm}
|\bD\chi|_{\rohy}^{2}=N^{-2}\textstyle{\sum}_{l}\bge_{ij}(\bD_{E_{l}}\chi)^{i}(\bD_{E_{l}}\chi)^{j}=N^{-2}\textstyle{\sum}_{A,l}e^{2\bmu_{A}}|(\bD_{E_{l}}\chi)^{A}|^{2}. 
\end{equation}
On the other hand
\[
|(\bD_{X_{A}}\chi)^{B}|=\left|\textstyle{\sum}_{i}X^{i}_{A}(\bD_{E_{i}}\chi)^{B}\right|
\leq\left(\textstyle{\sum}_{i}|X^{i}_{A}|^{2}\right)^{1/2}\left(\textstyle{\sum}_{i}|(\bD_{E_{i}}\chi)^{B}|^{2}\right)^{1/2}.
\]
Combining this estimate with (\ref{eq:bDchireformnorm}) yields
\begin{equation}\label{eq:bDXAchiBest}
N^{-2}\textstyle{\sum}_{B}e^{2\bmu_{B}}|(\bD_{X_{A}}\chi)^{B}|^{2}\leq |\bD\chi|_{\rohy}^{2}.
\end{equation}
Next, let us consider 
\begin{equation}\label{eq:shiftvectorcontrtoWAA}
\frac{1}{2\hN}(\ml_{\chi}\bge_{\refer})(X_{A},X_{A})=\frac{1}{\hN}\bge_{\refer}(\bD_{X_{A}}\chi,X_{A})
=\frac{1}{\hN}(\bD_{X_{A}}\chi)^{B}\bge_{\refer}(X_{B},X_{A}). 
\end{equation}
In particular, 
\[
\frac{1}{2\hN}|(\ml_{\chi}\bge_{\refer})(X_{A},X_{A})|\leq \frac{n^{1/2}}{\hN}\left(\textstyle{\sum}_{B}|(\bD_{X_{A}}\chi)^{B}|^{2}\right)^{1/2}
\leq n^{1/2}e^{-\mu_{\min}}|\bD\chi|_{\rohy}, 
\]
where we use the notation introduced in (\ref{eq:mumindef}). Thus (\ref{eq:WAAchicontr}) holds. Next, note that 
$\rodiv_{\bge_{\refer}}\chi=Y^{A}(\bD_{X_{A}}\chi)$. Thus 
\begin{equation}\label{eq:rodivchiestimate}
\begin{split}
\hN^{-1}|\rodiv_{\bge_{\refer}}\chi| \leq & \hN^{-1}\textstyle{\sum}_{A}|Y^{A}(\bD_{X_{A}}\chi)|
 \leq \hN^{-1}\textstyle{\sum}_{A}\sum_{i}|X^{i}_{A}||Y^{A}(\bD_{E_{i}}\chi)| \\
 \leq & \hN^{-1}\textstyle{\sum}_{A}\left(\sum_{i}|(\bD_{E_{i}}\chi)^{A}|^{2}\right)^{1/2}\leq 
n^{1/2}\hN^{-1}\left(\textstyle{\sum}_{A,i}|(\bD_{E_{i}}\chi)^{A}|^{2}\right)^{1/2}\\
 \leq & n^{1/2}e^{-\bmu_{\min}}\hN^{-1}\left(\textstyle{\sum}_{A,i}e^{2\bmu_{A}}|(\bD_{E_{i}}\chi)^{A}|^{2}\right)^{1/2}\leq n^{1/2}e^{-\mu_{\min}}|\bD\chi|_{\rohy},
\end{split}
\end{equation}
where $\bmu_{\min}:=\min_{A}\bmu_{A}$. Thus (\ref{eq:rodivestimateechi}) holds and the lemma follows.
\end{proof}

\section{Geometric identities}\label{section:geometricidentitites}

Before proceeding, we derive some geometric identities.

\begin{lemma}\label{lemma:geometricidentitiesbmuAvarrho}
  Let $(M,g)$ be a time oriented Lorentz manifold. Assume that it has an expanding partial pointed foliation. Assume, moreover, $\mK$ to be
  non-degenerate on $I$ and to have a global frame. Then
  \begin{align}
    \hU(\bmu_{A}) = & \ell_{A}+\mW^{A}_{A},\label{eq:hUbmuAform}\\
    \hU(\varrho) = & 1+\hN^{-1}\rodiv_{\bge_{\refer}}\chi\label{eq:hUvarrhoident}
  \end{align}
  where there is no summation on $A$ in the first equality. 
\end{lemma}
\begin{remark}\label{remark:rodivbgerodivbgeref}
Due to the fact that 
\begin{equation*}
\begin{split}
(\rodiv_{\bge}\chi)\mu_{\bge} = & d(\iota_{\chi}\mu_{\bge})=d[\iota_{\chi}(\varphi\mu_{\bge_{\refer}})]=d(\iota_{\varphi\chi}\mu_{\bge_{\refer}})=
[\rodiv_{\bge_{\refer}}(\varphi\chi)]\mu_{\bge_{\refer}}\\
 = & (\varphi\rodiv_{\bge_{\refer}}\chi+\chi(\varrho)\varphi)\mu_{\bge_{\refer}}
=(\rodiv_{\bge_{\refer}}\chi+\chi(\varrho))\mu_{\bge},
\end{split}
\end{equation*}
the equality (\ref{eq:hUvarrhoident}) can also be written 
\begin{equation}\label{eq:hNinvvarrhothNinvrodivbgechi}
\hN^{-1}\varrho_{t}=1+\hN^{-1}\rodiv_{\bge}\chi.
\end{equation}
\end{remark}
\begin{remark}\label{remark:Cdetronddef}
  If, in addition to the assumptions of the lemma, (\ref{eq:mKsupbasest}) holds, then there is a constant $C_{\det,\rond}$, depending only on $n$, 
  $\mKsup$ and $\e_{\rond}$, such that 
  \begin{equation}\label{eq:sumbmuAvarrhodiff}
    \left|\textstyle{\sum}_{A}\bmu_{A}-\varrho\right|\leq C_{\det,\rond}
  \end{equation}
  on $M_{-}:=\bM\times I_{-}$. This is an immediate consequence of Lemma~\ref{lemma:frameinvest} and (\ref{eq:varphidefXAver}) below.
\end{remark}
\begin{proof}
  Combining Lemma~\ref{lemma:mMstructure} and (\ref{eq:mMAAhUbmuAmWAA}) yields (\ref{eq:hUbmuAform}). Next, consider (\ref{eq:varphidefitobmubgedp}).
  Evaluating this equality with respect to the frame $\{ X_{A}\}$ yields
  \begin{equation}\label{eq:varphidefXAver}
    \exp\left(\textstyle{\sum}_{A}\bmu_{A}\right)=\varphi\cdot (\det\bGe_{\refer})^{1/2},
  \end{equation}
  where $\bGe_{\refer}$ is the matrix with components
  \[
  \bGe_{\refer,AB}=\bge_{\refer}(X_{A},X_{B})=\textstyle{\sum}_{i}X_{A}^{i}X_{B}^{i}
  \]
  and $X_{A}^{i}$ and $Y^{A}_{i}$ are the components of $X_{A}$ and $Y^{A}$ respectively with respect to an orthonormal frame as in 
Remark~\ref{remark:framenondegenerate}. Note also that if $\bGe_{\refer}^{AB}$ denotes the components of the inverse of $\bGe_{\refer}$, then
\[
\bGe_{\refer}^{AB}=\textstyle{\sum}_{i}Y^{A}_{i}Y^{B}_{i}.
\]
Differentiating (\ref{eq:varphidefXAver}) with respect to $\hU$ yields
\[
\exp\left(\textstyle{\sum}_{A}\bmu_{A}\right)\textstyle{\sum_{B}}\hU(\bmu_{B})=\hU(\ln\varphi)\varphi (\det\bGe_{\refer})^{1/2}
+\frac{1}{2}\bGe_{\refer}^{AB}\hU(\bGe_{\refer,AB})\varphi (\det\bGe_{\refer})^{1/2}.
\]
Appealing to (\ref{eq:varphidefXAver}) again yields
\begin{equation}\label{eq:sumAhUbmuA}
\textstyle{\sum_{A}}\hU(\bmu_{A})=\hU(\ln\varphi)+\frac{1}{2}\bGe_{\refer}^{AB}\hU(\bGe_{\refer,AB}).
\end{equation}
On the other hand, Remark~\ref{remark:ellAsum} and (\ref{eq:hUbmuAform}) yield
\begin{equation}\label{eq:hUbmusumsimpl}
\textstyle{\sum_{A}}\hU(\bmu_{A})=1+\textstyle{\sum_{A}}\mW^{A}_{A}.
\end{equation}
Next, let us consider
\begin{equation}\label{eq:bGehUbgeform}
\begin{split}
\bGe_{\refer}^{AB}\hU(\bGe_{\refer,AB}) = & \textstyle{\sum}_{i,j}Y^{A}_{j}Y^{B}_{j}\left[\hU(X_{A}^{i})X_{B}^{i}+X_{A}^{i}\hU(X_{B}^{i})\right]
 =  2Y^{A}_{i}\hU(X^{i}_{A}).
\end{split}
\end{equation}
Due to (\ref{eq:overlinemlUitoEietc}),
\[
\hU(X^{j}_{A})=\omega^{j}(\overline{\hml_{U}X_{A}})+\hN^{-1}X^{i}_{A}\omega^{j}(\ml_{\chi}E_{i}).
\]
Due to (\ref{eq:mWdefrel}), the first term on the right hand side equals $\mW^{B}_{A}X_{B}^{j}$. Thus
\[
Y^{A}_{j}\hU(X^{j}_{A})=\mW^{B}_{A}Y^{A}_{j}X_{B}^{j}+\hN^{-1}Y^{A}_{j}X^{i}_{A}\omega^{j}(\ml_{\chi}E_{i})
=\textstyle{\sum}_{A}\mW^{A}_{A}+\hN^{-1}\omega^{i}(\ml_{\chi}E_{i}).
\]
Combining this equality with (\ref{eq:sumAhUbmuA}), (\ref{eq:hUbmusumsimpl}) and (\ref{eq:bGehUbgeform}) yields
\[
1+\textstyle{\sum_{A}}\mW^{A}_{A}=\hU(\ln\varphi)+\textstyle{\sum}_{A}\mW^{A}_{A}+\hN^{-1}\omega^{i}(\ml_{\chi}E_{i}).
\]
Thus
\[
\hU(\ln\varphi)=1+\hN^{-1}\omega^{i}(\ml_{E_{i}}\chi).
\]
On the other hand
\[
\omega^{i}(\ml_{E_{i}}\chi)=\textstyle{\sum}_{i}\bge_{\refer}(\bD_{E_{i}}\chi-\bD_{\chi}E_{i},E_{i})
=\textstyle{\sum}_{i}\bge_{\refer}(\bD_{E_{i}}\chi,E_{i})=\omega^{i}(\bD_{E_{i}}\chi)=\rodiv_{\bge_{\refer}}\chi,
\]
where we used the fact that $\{E_{i}\}$ is an orthonormal frame with respect to $\bge_{\refer}$. Thus (\ref{eq:hUvarrhoident}) holds and the
lemma follows. 
\end{proof}

\section{Estimating the norm of the elements of the frame}\label{section:mainmuAestimate}

Next, we wish to estimate $\mu_{\min}$, introduced in (\ref{eq:mumindef}). In order to obtain conclusions, we have to assume 
$\chK$ to have a silent upper bound on $I$; cf. Definition~\ref{def:silenceandnondegeneracy}. Moreover, we have to assume $\chi$ to be small enough. 
In fact, the estimate of $\mu_{\min}$ is based on a bootstrap argument which goes through if the shift vector field is small enough. 

\begin{lemma}\label{lemma:lowerbdonmumin}
  Let $(M,g)$ be a time oriented Lorentz manifold with an expanding partial pointed foliation. Assume $\mK$ to be non-degenerate on $I$, to have a
  global frame and to be $C^{0}$-uniformly bounded on $I_{-}$; i.e. (\ref{eq:mKsupbasest}) to hold. Assume $\chK$ to have a silent upper bound on $I$.
  Assume, moreover, that $\hml_{U}\mK$ satisfies a weak off-diagonal exponential bound; cf. Definition~\ref{def:offdiagonalexpdec}. Let $\e_{\chi}$ be
  defined by
  \begin{equation}\label{eq:echibasicassumptionmod}
    \e_{\chi}:=\frac{1}{4}e^{-M_{\mu}}\min\{1,\e_{\Spe}\},
  \end{equation}
  where $M_{\mu}$ is defined by
  \begin{equation}\label{eq:Mmudef}
    M_{\mu}:=(n+1)M_{0}+C_{\det,\rond}+\frac{1}{2};
  \end{equation}
  $C_{\det,\rond}$ is the constant introduced in Remark~\ref{remark:Cdetronddef}; $M_{0}$ is defined by
  \begin{equation}\label{eq:Mzdef}
    M_{0}:=\frac{3(n-1)}{\e_{\rond}\e_{\mK}}(C_{\mK,\mrod}+3M_{\mK,\mrod})+\frac{1}{2};
  \end{equation}
  and $\e_{\rond}$ is the constant appearing in Definition~\ref{def:silenceandnondegeneracy}. Assume, finally, that
  \begin{align}
    n^{1/2}\theta_{0,-}^{-1}|\bD\chi|_{\rohy} \leq & \e_{\chi},\label{eq:chiestfirststepestnoff}
  \end{align}
  for all $t\in I_{-}$, where $\theta_{0,-}$ is defined by (\ref{eq:thetazdef}). Then
  \begin{align}
    \hN^{-1}|\rodiv_{\bge_{\refer}}\chi| \leq & \min\{1,\e_{\Spe}\}e^{\e_{\Spe}\varrho},\label{eq:rodivchiestimpr}\\
    (2\hN)^{-1}|(\ml_{\chi}\bge_{\refer})(X_{A},X_{A})| \leq & \min\{1,\e_{\Spe}\}e^{\e_{\Spe}\varrho},\label{eq:WAAchicontrestimpr}\\
    \mu_{\min} \geq & -\e_{\Spe}\varrho+\ln\theta_{0,-}-M_{\min}\label{eq:muminmainlowerbound}
  \end{align}
  (no summation on $A$ in the second estimate) on $M_{-}$, where $M_{\min}:=M_{\mu}+1$. Moreover, if $\g$ is an integral curve of
  $\hU$ with $\g(0)\in\bM\times \{t_{0}\}$, then
  \begin{align}
    [\hN^{-1}|\rodiv_{\bge_{\refer}}\chi|]\circ\g(s) \leq & \frac{1}{4}\min\{1,\e_{\Spe}\}e^{\e_{\Spe}s},\label{eq:rodivchiroughestimpr}\\
    [(2\hN)^{-1}|(\ml_{\chi}\bge_{\refer})(X_{A},X_{A})|]\circ\g(s)
    \leq & \frac{1}{4}\min\{1,\e_{\Spe}\}e^{\e_{\Spe}s},\label{eq:WAAchicontrroughestimpr}\\
    \mu_{\min}\circ\g(s) \geq & -\e_{\Spe}s+\ln\theta_{0,-}-M_{\mu}\label{eq:mumincircgammalowerbound}
  \end{align}
  for all $s\leq 0$ such that $\g(s)\in M_{-}$. Moreover,
  \begin{equation}\label{eq:varrhosequivalencestmt}
    s-1/2\leq \varrho\circ\g(s)\leq s+1/2
  \end{equation}
  for all $s\leq 0$ such that $\g(s)\in M_{-}$.
\end{lemma}
\begin{remark}\label{remark:Bgtzero}
If one would assume $\hml_{U}\mK$ to satisfy an off-diagonal exponential bound, then the proof could be simplified somewhat. In particular, 
it would not be necessary to carry out a separate argument for $\mu_{1}$. 
\end{remark}
\begin{proof}
The proof is based on a bootstrap argument along integral curves of $\hU$. Let, to this end, $\g$ be a curve such that $\g(0)\in\bM_{t_{0}}$ 
and such that $\dot{\g}(s)=\hU_{\g(s)}$. Let, moreover, $J_{-}:=\g^{-1}(M_{-})$ (which is an interval since the $t$-coordinate of $\g$ is strictly
monotonically increasing due to the fact that $\g$ is future pointing timelike). 

\textbf{Bootstrap assumption:} Assume that $\e_{\chi}$ (appearing in (\ref{eq:chiestfirststepestnoff})) and $\mu_{\min}$ are such that 
\begin{equation}\label{eq:indassumptionmuminechi}
\theta_{0,-}\e_{\chi}e^{-\mu_{\min}\circ\g(s)}\leq \frac{1}{2}\min\{1,\e_{\Spe}\}e^{\e_{\Spe}s}
\end{equation}
on some open subinterval $J_{1}$ of $J_{-}$ containing $0$. Note that, due to (\ref{eq:echibasicassumptionmod}), the bootstrap assumption 
is satisfied with a margin in a neighbourhood of $0$. To obtain this conclusion we used the fact that $\bmu_{A}(\bx,t_{0})=0$; this follows
from the definition of $\bge_{\refer}$ and the normalisation of the $X_{A}$. 

\textbf{Basic conclusions.} Combining the bootstrap assumption with (\ref{eq:WAAchicontr}), (\ref{eq:rodivestimateechi}) and 
(\ref{eq:chiestfirststepestnoff}) yields 
\begin{align}
[\hN^{-1}|\rodiv_{\bge_{\refer}}\chi|]\circ\g(s)\leq & \frac{1}{2}\min\{1,\e_{\Spe}\}e^{\e_{\Spe}s},\label{eq:rodivchiroughest}\\
[(2\hN)^{-1}|(\ml_{\chi}\bge_{\refer})(X_{A},X_{A})|]\circ\g(s)\leq & \frac{1}{2}\min\{1,\e_{\Spe}\}e^{\e_{\Spe}s}\label{eq:WAAchicontrroughest}
\end{align}
on $J_{1}$ (no summation on $A$). 

\textbf{Estimating $\varrho$.} Next, note that (\ref{eq:hUvarrhoident}) yields 
\begin{equation}\label{eq:dvarrhocircgds}
\frac{d}{ds}\varrho\circ\g(s)=\hU(\varrho)|_{\g(s)}=1+(\hN^{-1}\rodiv_{\bge_{\refer}}\chi)[\g(s)]. 
\end{equation}
Combining this equality with (\ref{eq:rodivchiroughest}) yields
\[
\left|\frac{d}{ds}\varrho\circ\g(s)-1\right|\leq\frac{1}{2}\min\{1,\e_{\Spe}\}e^{\e_{\Spe}s}. 
\]
Integrating this estimate from $s\in J_{1}$ to $0$ yields
\begin{equation}\label{eq:varrhosequivalence}
s-1/2\leq \varrho\circ\g(s)\leq s+1/2;
\end{equation}
note that $\varrho\circ\g(0)=0$ due to the definition of $\varrho$ and $\bge_{\refer}$. 
In particular, $\varrho\circ\g(s)$ and $s$ are comparable for $s\in J_{1}$. 

\textbf{Estimating $\mu_{A}$ for $A>1$.} Next, let us turn to $\bmu_{A}$, $\ln\theta$ and $\mu_{A}$ in the case that $A>1$. Recall that 
(\ref{eq:hUnlnthetamomqbas}) holds and that $\mu_{A}=\bmu_{A}+\ln\theta$; cf. the text adjacent to (\ref{eq:bgenormXA}). Thus
\[
\hU(\mu_{A})=\hU(\bmu_{A})+\hU(\ln\theta)=\ell_{A}-n^{-1}(1+q)+\mW^{A}_{A},
\]
where we appealed to (\ref{eq:hUbmuAform}). Next, let $\lambda_{A}$ be the eigenvalues of $\chK$. In other words, $\chK X_{A}=\lambda_{A}X_{A}$ (no summation). Then
\begin{equation}\label{eq:lambdaArelqellA}
\lambda_{A}=\ell_{A}+\hU(\ln\theta)=\ell_{A}-n^{-1}(1+q),
\end{equation}
where we appealed to to (\ref{eq:chKmKthetarelation}). Thus
\begin{equation}\label{eq:hUmuAformula}
\hU(\mu_{A})=\lambda_{A}+\mW^{A}_{A}. 
\end{equation}
On the other hand, due to the assumption that $\chK\leq -\e_{\Spe}$, it follows that $\lambda_{A}\leq -\e_{\Spe}$, so that
\begin{equation}\label{eq:hUmuAfirstroughest}
\hU(\mu_{A})\leq -\e_{\Spe}+\mW^{A}_{A}. 
\end{equation}
In particular, 
\begin{equation}\label{eq:ddsmuAestimatealonggamma}
\frac{d}{ds}\mu_{A}\circ\g(s)\leq -\e_{\Spe}+\mW^{A}_{A}\circ\g(s). 
\end{equation}
Due to this inequality, it is of interest to estimate the integral of $\mW^{A}_{A}\circ\g$ from $s$ to $0$. Note, to this end, that 
for $s\in J_{1}$:
\begin{equation}\label{eq:mWAAcrudeest}
|\mW^{A}_{A}\circ\g(s)|\leq \sum_{B\neq A}|\mW^{B}_{A}\circ\g(s)|+\frac{1}{2}\min\{1,\e_{\Spe}\}e^{\e_{\Spe}s}
\end{equation}
(no summation on $A$), where we appealed to (\ref{eq:mWAAformulanongeo}) and (\ref{eq:WAAchicontrroughest}). 
In particular, 
\begin{equation}\label{eq:intstozmWAAest}
\int_{s}^{0}|\mW^{A}_{A}\circ\g(u)|du \leq \sum_{B\neq A}\int_{s}^{0}|\mW^{B}_{A}\circ\g(u)|du+\frac{1}{2}
\end{equation}
for all $s\in J_{1}$. Clearly, we need to estimate the first term on the right hand side. By assumption, 
(\ref{eq:hmlUhmlUsqmKoffdiagonalexpbd}) and (\ref{eq:expgrowthwithupperbound}) hold for $B>1$ and $A\neq B$. Thus, for $A>1$ and $B\neq A$, 
\begin{equation}\label{eq:mWAAintestalongintcurveexpdec}
\begin{split}
\int_{s}^{0}|\mW^{B}_{A}\circ\g(u)|du \leq & \e_{\rond}^{-1}\int_{s}^{0}(C_{\mK,\mrod}e^{\e_{\mK}\varrho\circ\g(u)}+G_{\mK,\mrod}e^{-\e_{\mK}\varrho\circ\g(u)})du\\
 \leq & 3\e_{\rond}^{-1}\e_{\mK}^{-1}(C_{\mK,\mrod}+G_{\mK,\mrod}e^{-\e_{\mK}s})\\
 \leq & 3\e_{\rond}^{-1}\e_{\mK}^{-1}(C_{\mK,\mrod}+3M_{\mK,\mrod}),
\end{split}
\end{equation}
where we appealed to (\ref{eq:mWBAformulanongeo}), (\ref{eq:varrhosequivalence}), the fact that $\mK$ is non-degenerate and the fact that 
$\e_{\mK}\leq 2$. Combining (\ref{eq:intstozmWAAest}) and (\ref{eq:mWAAintestalongintcurveexpdec}) yields
\begin{equation}\label{eq:intstozmWAAestfinal}
\int_{s}^{0}|\mW^{A}_{A}\circ\g(u)|du \leq M_{0}
\end{equation}
for all $s\in J_{1}$, where $M_{0}$ is given by (\ref{eq:Mzdef}). Combining this estimate with (\ref{eq:ddsmuAestimatealonggamma}) yields 
\begin{equation}\label{eq:muminlowerbdalonggammaprel}
\mu_{A}\circ\g(s)\geq -\e_{\Spe}s+\ln\theta_{0,-}-M_{0}
\end{equation}
for all $s\in J_{1}$ and all $A>1$. 

\textbf{Estimating $\mu_{1}$.} In order to estimate $\mu_{1}$, we have to proceed differently. The reason for this is that we do not assume the estimates
leading to (\ref{eq:intstozmWAAestfinal}) to hold. On the other hand, we know that for $A>1$ and $s\in J_{1}$, 
\[
\left|\int_{s}^{0}(\bmu_{A}\circ\g)'(u)du-\int_{s}^{0}\ell_{A}\circ\g(u)du\right|\leq M_{0},
\]
where we appealed to (\ref{eq:hUbmuAform}) and (\ref{eq:intstozmWAAestfinal}). Thus
\begin{equation}\label{eq:bmuAeqintellA}
\left|\bmu_{A}\circ\g(s)+\int_{s}^{0}\ell_{A}\circ\g(u)du\right|\leq M_{0}
\end{equation}
for all $A>1$ and $s\in J_{1}$. In particular, 
\[
\left|\int_{s}^{0}\textstyle{\sum}_{A>1}\ell_{A}\circ\g(u)du+\textstyle{\sum}_{A>1}\bmu_{A}\circ\g(s)\right|\leq (n-1)M_{0}
\]
for all $s\in J_{1}$. Due to the fact that the sum of the $\ell_{A}$ equals $1$ and the fact that (\ref{eq:sumbmuAvarrhodiff}) holds, 
this estimate yields
\[
\left|\int_{s}^{0}[1-\ell_{1}\circ\g(u)]du-\bmu_{1}\circ\g(s)+\varrho\circ\g(s)\right|\leq (n-1)M_{0}+C_{\det,\rond}
\]
for all $s\in J_{1}$. Combining this estimate with (\ref{eq:varrhosequivalence}) yields the conclusion that 
\begin{equation}\label{eq:bmuoneeqintellone}
\left|\int_{s}^{0}\ell_{1}\circ\g(u)du+\bmu_{1}\circ\g(s)\right|\leq (n-1)M_{0}+C_{\det,\rond}+\frac{1}{2}
\end{equation}
for all $s\in J_{1}$. In particular, since $\ell_{1}<\ell_{2}$, 
\begin{equation}\label{eq:muonelowerbd}
\begin{split}
\mu_{1}\circ\g(s) \geq & -\int_{s}^{0}\ell_{1}\circ\g(u)du+\ln\theta\circ\g(s)-(n-1)M_{0}-C_{\det,\rond}-\frac{1}{2}\\
 \geq & -\int_{s}^{0}\ell_{2}\circ\g(u)du+\ln\theta\circ\g(s)-(n-1)M_{0}-C_{\det,\rond}-\frac{1}{2}\\
 \geq & \mu_{2}\circ\g(s)-nM_{0}-C_{\det,\rond}-\frac{1}{2}\\
 \geq & -\e_{\Spe}s+\ln\theta_{0,-}-(n+1)M_{0}-C_{\det,\rond}-\frac{1}{2}
\end{split}
\end{equation}
for all $s\in J_{1}$, where we appealed to (\ref{eq:muminlowerbdalonggammaprel}) and (\ref{eq:bmuAeqintellA}). In particular, 
\begin{equation}\label{eq:muminlowerbdalonggamma}
\mu_{\min}\circ\g(s)\geq -\e_{\Spe}s+\ln\theta_{0,-}-M_{\mu}
\end{equation}
for all $s\in J_{1}$, where $M_{\mu}$ is given by (\ref{eq:Mmudef}). 

\textbf{Improving the bootstrap assumptions.} One particular consequence of (\ref{eq:muminlowerbdalonggamma}) is that  
\[
\theta_{0,-}\e_{\chi}e^{-\mu_{\min}\circ\g(s)}\leq e^{M_{\mu}}\e_{\chi}e^{\e_{\Spe}s}\leq \frac{1}{4}\min\{1,\e_{\Spe}\}e^{\e_{\Spe}s}
\]
for all $s\in J_{1}$. Thus the bootstrap assumption is satisfied with a margin, and can be extended beyond the lower bound on $J_{1}$. Thus the bootstrap 
assumption holds on all of $J_{-}$. In fact, (\ref{eq:rodivchiroughest}) and (\ref{eq:WAAchicontrroughest}) can be improved to 
(\ref{eq:rodivchiroughestimpr}) and (\ref{eq:WAAchicontrroughestimpr}) respectively. Note also that (\ref{eq:muminlowerbdalonggamma}) yields
(\ref{eq:mumincircgammalowerbound}) and that (\ref{eq:varrhosequivalence}) yields (\ref{eq:varrhosequivalencestmt}). Combining these improved estimates 
with (\ref{eq:varrhosequivalence}), (\ref{eq:muminlowerbdalonggamma}) and the fact that $\e_{\Spe}\leq 2$ yields
\begin{align*}
\left[\hN^{-1}|\rodiv_{\bge_{\refer}}\chi|\right]\circ\g(s) \leq & \min\{1,\e_{\Spe}\}e^{\e_{\Spe}\varrho\circ\g(s)},\\
\left[(2\hN)^{-1}|(\ml_{\chi}\bge_{\refer})(X_{A},X_{A})|\right]\circ\g(s)
\leq & \min\{1,\e_{\Spe}\}e^{\e_{\Spe}\varrho\circ\g(s)},\\
\mu_{\min}\circ\g(s) \geq & -\e_{\Spe}\varrho\circ\g(s)+\ln\theta_{0,-}-M_{\mu}-1.
\end{align*}
Since these estimates hold along all integral curves of $\hU$ to the past of $\bM_{t_{0}}$, we conclude that (\ref{eq:rodivchiestimpr}), (\ref{eq:WAAchicontrestimpr})
and (\ref{eq:muminmainlowerbound}) hold. The lemma follows. 
\end{proof}

Due to this lemma, we can estimate $\mW^{A}_{A}$. In fact, we have the following corollary. 

\begin{cor}\label{cor:expdecmWABmWAAetc}
Given that the assumptions of Lemma~\ref{lemma:lowerbdonmumin} hold, the estimate
\begin{align}
|\mW^{A}_{B}| \leq & \e_{\rond}^{-1}C_{\mK,\mrod}e^{\e_{\mK}\varrho}+\e_{\rond}^{-1}G_{\mK,\mrod}e^{-\e_{\mK}\varrho}\label{eq:mWAneqBbasicestimate}
\end{align}
holds on $M_{-}$ for all $A\neq B$ and $B>1$. Moreover, (\ref{eq:expgrowthwithupperbound}) and
\begin{align}
|\mW^{A}_{A}| \leq & \textstyle{\sum}_{B\neq A}|\mW^{B}_{A}|+\min\{1,\e_{\Spe}\}e^{\e_{\Spe}\varrho}\label{eq:mWAAexpdecest}
\end{align}
(no summation on $A$) hold on $M_{-}$. Next, let $\g$ be a curve with the properties stated in Lemma~\ref{lemma:lowerbdonmumin} and
$J_{-}:=\g^{-1}(M_{-})$. Then, assuming $A\neq B$ and $B>1$, 
\begin{align}
|\mW^{A}_{B}\circ\g(s)| \leq & 3\e_{\rond}^{-1}C_{\mK,\mrod}e^{\e_{\mK}s}+3\e_{\rond}^{-1}G_{\mK,\mrod}e^{-\e_{\mK}s}\label{eq:mWAneqBbasicestimatealongg}
\end{align}
on $J_{-}$. Moreover, 
\begin{equation}\label{eq:expgrowthwithupperboundalonggamma}
G_{\mK,\mrod}e^{-\e_{\mK}s}\leq 3M_{\mK,\mrod}
\end{equation}
for all $s\in J_{-}$ and
\begin{align}
|\mW^{A}_{A}\circ\g(s)| \leq & \textstyle{\sum}_{B\neq A}|\mW^{B}_{A}\circ\g(s)|+\frac{1}{4}\min\{1,\e_{\Spe}\}e^{\e_{\Spe}s}\label{eq:mWAAexpdecestalongg}
\end{align}
(no summation on $A$) for all $s\in J_{-}$. Finally, there is a constant $M_{\rodiff}$, given by (\ref{eq:Mdiffdef}) below, such that,
assuming $A>B$, 
\begin{align}
\bmu_{A}-\bmu_{B} \leq & (A-B)\e_{\rond}\varrho+M_{\rodiff},\label{eq:bmuAmbmuBlowbd}\\
\ln\theta \geq & -(n^{-1}+\e_{\Spe})\varrho+\ln\theta_{0,-}-2\label{eq:lnthetalowbd}
\end{align}
on $M_{-}$. 
\end{cor}
\begin{remark}\label{remark:lnthetaupperbd}
Assuming, in addition to the conditions of the lemma, $q$ to be $C^{0}$-uniformly bounded on $I_{-}$ with constant $C_{q}$ yields 
\begin{equation}\label{eq:lnthetaupperbd}
\ln\theta\leq -\frac{1}{n}(1+C_{q})\varrho+\ln\theta_{0,+}+\frac{1}{2n}(1+C_{q})
\end{equation}
on $M_{-}$,
where $\theta_{0,+}$ is given by (\ref{eq:thetazdef}). Combining (\ref{eq:lnthetalowbd}) and (\ref{eq:lnthetaupperbd}) yields the conclusion that
$\varrho\rightarrow -\infty$ if and only if $\theta\rightarrow\infty$. In fact, $\ldr{\ln\theta}$ and $\ldr{\varrho}$ are equivalent. 
\end{remark}
\begin{remark}
Assume, in addition to the conditions of the lemma, that, for some $A>1$, there is a constant $L_{A}$ such that $\ell_{A}\geq L_{A}$ on $M_{-}$. Then 
\begin{equation}\label{eq:bmuAupperbditoLA}
\bmu_{A}\leq L_{A}\varrho+M_{0}+\frac{1}{2}|L_{A}|
\end{equation}
on $M_{-}$, where we appealed to (\ref{eq:varrhosequivalencestmt}) and (\ref{eq:bmuAeqintellA}), and $M_{0}$ is given by (\ref{eq:Mzdef}). Similarly, if 
there is a constant $L_{1}$ such that $\ell_{1}\geq L_{1}$ on $M_{-}$, then 
\begin{equation}\label{eq:bmuoneupperbditoLone}
\bmu_{1}\leq L_{1}\varrho+(n-1)M_{0}+C_{\det,\rond}+\frac{1}{2}(|L_{1}|+1)
\end{equation}
on $M_{-}$, where we appealed to (\ref{eq:varrhosequivalencestmt}) and (\ref{eq:bmuoneeqintellone}). 
\end{remark}
\begin{proof}
  By assumption, (\ref{eq:hmlUhmlUsqmKoffdiagonalexpbd}) and (\ref{eq:expgrowthwithupperbound}) hold for $A\neq B$ and $B>1$. Combining this assumption
  with (\ref{eq:mWBAformulanongeo}) and the assumed non-degeneracy yields (\ref{eq:mWAneqBbasicestimate}). The estimate (\ref{eq:mWAAexpdecest}) is an
  immediate consequence of (\ref{eq:mWAAformulanongeo}) and (\ref{eq:WAAchicontrestimpr}). The estimate (\ref{eq:mWAneqBbasicestimatealongg}) follows
  from (\ref{eq:mWAneqBbasicestimate}), (\ref{eq:varrhosequivalencestmt}) and the fact that $\e_{\mK}\leq 2$. In addition,
  (\ref{eq:expgrowthwithupperboundalonggamma}) follows from (\ref{eq:expgrowthwithupperbound}), (\ref{eq:varrhosequivalencestmt}) and
  $\e_{\mK}\leq 2$. Next, (\ref{eq:mWAAexpdecestalongg}) follows from (\ref{eq:mWAAformulanongeo}) and (\ref{eq:WAAchicontrroughestimpr}). 

  In order to prove (\ref{eq:bmuAmbmuBlowbd}), it is convenient to divide the analysis into two cases. If $1<B<A$, then
  (\ref{eq:varrhosequivalencestmt}) and (\ref{eq:bmuAeqintellA}) imply that 
\begin{equation}\label{eq:bmuABdiffhigherA}
\begin{split}
\bmu_{A}\circ\g(s)-\bmu_{B}\circ\g(s) \leq & \int_{s}^{0}(\ell_{B}-\ell_{A})\circ\g(u)du+2M_{0}\leq (A-B)\e_{\rond}s+2M_{0}\\
 \leq & (A-B)\e_{\rond}\varrho\circ\g(s)+\frac{1}{2}(n-2)\e_{\rond}+2M_{0}
\end{split}
\end{equation}
for all $s\in J_{-}$. If $B=1$ and $A>1$, then (\ref{eq:varrhosequivalencestmt}), (\ref{eq:bmuAeqintellA}), (\ref{eq:bmuoneeqintellone})
and the fact that $\ell_{A}-\ell_{1}>(A-1)\e_{\rond}$ yield 
\begin{equation}\label{eq:bmuABdiffAeqone}
\begin{split}
\bmu_{A}\circ\g(s)-\bmu_{1}\circ\g(s) \leq & (A-1)\e_{\rond}s+nM_{0}+C_{\det,\rond}+\frac{1}{2}\\
 \leq & (A-1)\e_{\rond}\varrho\circ\g(s)+\frac{1}{2}(n-1)\e_{\rond}+nM_{0}+C_{\det,\rond}+\frac{1}{2}
\end{split}
\end{equation}
for all $s\in J_{-}$. Defining $M_{\rodiff}$ by 
\begin{equation}\label{eq:Mdiffdef}
M_{\rodiff}:=\frac{1}{2}(n-1)\e_{\rond}+nM_{0}+C_{\det,\rond}+\frac{1}{2},
\end{equation}
where $M_{0}$ is given by (\ref{eq:Mzdef}), the estimates (\ref{eq:bmuABdiffhigherA}) and (\ref{eq:bmuABdiffAeqone}) yield the conclusion that 
(\ref{eq:bmuAmbmuBlowbd}) holds. Turning to $\theta$, note that (\ref{eq:hUnlnthetamomqbas}) and Remark~\ref{remark:qlwbd} yields
\[
\hU(\ln\theta)\leq -n^{-1}-\e_{\Spe}, 
\]
so that, by arguments similar to the above, (\ref{eq:lnthetalowbd}) holds. The proofs of (\ref{eq:lnthetaupperbd}) and (\ref{eq:bmuAupperbditoLA}) are similar
to the above.  The lemma follows. 
\end{proof}

\subsection{Rough estimate of $\bmu_{A}$}

In what follows, it will be of interest to have a rough estimate of $\bmu_{A}$. 

\begin{cor}\label{cor:roughestbmuA}
Given that the assumptions of Lemma~\ref{lemma:lowerbdonmumin} hold, the estimate
\begin{equation}\label{eq:bmuAroughtest}
|\bmu_{A}|\leq L_{\max}|\varrho|+M_{\max}
\end{equation}
holds on $M_{-}$ for all $A$, where 
\[
L_{\max}:=\sup_{x\in M_{-}}\sup_{A}|\ell_{A}(x)|,\ \ \
M_{\max}:=(n-1)M_{0}+C_{\det,\rond}+\frac{1}{2}(L_{\max}+1)
\]
and $M_{0}$ is given by (\ref{eq:Mzdef}).
\end{cor}
\begin{proof}
The statement is an immediate consequence of (\ref{eq:varrhosequivalencestmt}), (\ref{eq:bmuAeqintellA}) and (\ref{eq:bmuoneeqintellone}). 
\end{proof}

\subsection{Revisiting the assumptions}

At this stage, we are in a position to revisit the assumptions and to strengthen some of them. Recall, to this end, that 
(\ref{eq:mAABmWABrel}) holds and that the right hand side of this equality is antisymmetric. This yields the following conclusion. 

\begin{prop}\label{prop:strengthenassumpAltB}
Given that the assumptions of Lemma~\ref{lemma:lowerbdonmumin} hold and that there is a $(\weight_{a},\weight_{b})=\weight\in\Weight$ and
a constant $D_{\mK,\weight}$ such that 
\[
\|\hml_{U}\mK\|_{C^{0}_{\weight}(\bM)}\leq D_{\mK,\weight}
\]
on $I_{-}$, there is a constant $C$ such that for $A<B$,
\[
|(\hml_{U}\mK)(Y^{A},X_{B})|\leq C\ldr{\varrho}^{\weight_{a}}e^{2(B-A)\e_{\rond}\varrho}
\]
on $I_{-}$, where $C$ only depends on $n$, $D_{\mK,\weight}$, $\mKsup$, $\e_{\rond}$ and the constant $M_{\rodiff}$ introduced in (\ref{eq:Mdiffdef}). 
\end{prop}
\begin{proof}
Due to (\ref{eq:mAABmWABrel}) and the fact that the right hand side of this equality is antisymmetric, it is clear that 
\begin{equation*}
\begin{split}
|(\hml_{U}\mK)(Y^{A},X_{B})| = & |\ell_{A}-\ell_{B}|\cdot|\mW^{A}_{B}|=e^{2(\bmu_{B}-\bmu_{A})}|\ell_{A}-\ell_{B}||\mW^{B}_{A}|\\
= & e^{2(\bmu_{B}-\bmu_{A})}|(\hml_{U}\mK)(Y^{B},X_{A})|\\
\leq & C_{Y}D_{\mK,\weight}e^{2M_{\rodiff}}\ldr{\varrho}^{\weight_{a}}e^{2(B-A)\e_{\rond}\varrho}
\end{split}
\end{equation*}
where we appealed to (\ref{eq:mWBAformulanongeo}), (\ref{eq:bmuAmbmuBlowbd}) and the non-degeneracy of $\mK$. The proposition follows. 
\end{proof}

\section{Estimating the relative spatial variation of $\varrho$}\label{section:estimatingrelativespatialvariation}

Next, we estimate the spatial variation of $\varrho$. 

\begin{lemma}\label{lemma:respvarvarrhoEi}
  Given that the conditions of Lemma~\ref{lemma:lowerbdonmumin} are fulfilled, assume (\ref{eq:bDlnNbDlnthetabd}) to hold. Let, moreover,
  $(0,\cweight)=\weight_{0}\in \Weight$ and assume that there is a constant $c_{\chi,2}$ such that 
  \begin{equation}\label{eq:cchitwoestimate}
    \theta_{0,-}^{-1}\|\chi\|_{C^{2,\weight_{0}}_{\rohy}(\bM)}\leq c_{\chi,2}
  \end{equation}
  on $I_{-}$. Then there is a constant $C_{\varrho}$, depending only on $\cweight$, $c_{\chi,2}$,
  $\bDlnhNsup$, $\mKsup$, $C_{\mK,\mrod}$, $M_{\mK,\mrod}$, $\e_{\Spe}$, $\e_{\rond}$, $\e_{\mK}$, $n$ and $(\bM,\bge_{\refer})$, such that 
  \begin{equation}\label{eq:bDvarrhobdEi}
    |\bD\varrho|_{\bge_{\refer}}\leq C_{\varrho}\ldr{\varrho}
  \end{equation}
  on $M_{-}$. In particular, there is a constant $C_{\rovar}\geq 1$ such that
  \begin{equation}\label{eq:varrhorelvarestEi}
    C_{\rovar}^{-1}\leq \frac{1-\varrho(\bx_{1},t)}{1-\varrho(\bx_{2},t)}\leq C_{\rovar}
  \end{equation}
  for all $t\in I_{-}$ and $\bx_{i}\in\bM$, $i=1,2$; recall that $\varrho\leq 0$ on $M_{-}$. Here $C_{\rovar}$ is of the form
  $C_{\rovar}=\exp\left(K_{\varrho}d_{\bM}\right)$, where $d_{\bM}$ is the diameter of $\bM$ with respect to $\bge_{\refer}$ and $K_{\varrho}$ has
  the same dependence as $C_{\varrho}$.
\end{lemma}
\begin{proof}
  The starting point is (\ref{eq:hUvarrhoident}). Commuting the right hand side with $E_{i}$, chosen as in Remark~\ref{remark:framenondegenerate},
  yields, cf. (\ref{eq:hUEicomm}) and (\ref{eq:Aialphadef}),
  \begin{equation}\label{eq:Eivarrhoevolution}
    \hU[E_{i}(\varrho)]=E_{i}(\ln\hN)+\hN^{-1}E_{i}(\rodiv_{\bge_{\refer}}\chi)-\hN^{-1}(\ml_{\chi}E_{i})(\varrho).
  \end{equation}
  We assume the first term on the right hand side to be bounded. However, we need to estimate the second and third terms. Note, to this end, that
  \begin{equation}\label{eq:Eirodivchi}
    \begin{split}
      E_{i}\left(\rodiv_{\bge_{\refer}}\chi\right) = & E_{i}\left[\textstyle{\sum}_{j}(\bD\chi)(\omega^{j},E_{j})\right]\\
      = & \textstyle{\sum}_{j}(\bD^{2}\chi)(\omega^{j},E_{i},E_{j})+\textstyle{\sum}_{j}(\bD\chi)(\bD_{E_{i}}\omega^{j},E_{j})\\
      & +\textstyle{\sum}_{j}(\bD\chi)(\omega^{j},\bD_{E_{i}}E_{j}).
    \end{split}
  \end{equation}
  On the other hand,
  \begin{equation*}
    \begin{split}
      |(\bD^{2}\chi)(\omega^{j},E_{i},E_{j})| \leq & \textstyle{\sum}_{A}e^{-\bmu_{A}}e^{\bmu_{A}}|(\bD^{2}\chi)(Y^{A},E_{i},E_{j})|\cdot |\omega^{j}(X_{A})|\\
      \leq & C\hN e^{-\mu_{\min}}|\bD^{2}\chi|_{\rohy},
    \end{split}
  \end{equation*}
  where $C$ only depends on $n$. The second and third terms on the right hand side of (\ref{eq:Eirodivchi}) can be estimated similarly. To conclude,
  \[
  \hN^{-1}\left|E_{i}\left(\rodiv_{\bge_{\refer}}\chi\right)\right|\leq C_{a} e^{-\mu_{\min}}|\bD^{2}\chi|_{\rohy}+C_{b} e^{-\mu_{\min}}|\bD\chi|_{\rohy},
  \]
  where $C_{a}$ only depends on $n$ and $C_{b}$ only depends on $n$ and $(\bM,\bge_{\refer})$. Combining this estimate with the assumptions and
  (\ref{eq:muminmainlowerbound}) yields
  \begin{equation}\label{eq:Eirodivchiestimate}
    \hN^{-1}\left|E_{i}\left(\rodiv_{\bge_{\refer}}\chi\right)\right|\leq C\ldr{\varrho}^{2\cweight}e^{\e_{\Spe}\varrho},
  \end{equation}
  where $C$ only depends on $c_{\chi,2}$, $n$, $(\bM,\bge_{\refer})$ and the constant $M_{\min}$ appearing in (\ref{eq:muminmainlowerbound}). Next, we need
  to estimate
  \begin{equation}\label{eq:omegakhNinvmlchiEi}
    -\omega^{k}(\hN^{-1}\ml_{\chi}E_{i})=-\hN^{-1}\chi^{A}\omega^{k}(\bD_{X_{A}}E_{i})+\omega^{k}(X_{A})\hN^{-1}Y^{A}(\bD_{E_{i}}\chi).
  \end{equation}
  This expression can be estimated by arguments similar to the above. This yields
  \[
  |\omega^{k}(\hN^{-1}\ml_{\chi}E_{i})|\leq Ce^{-\mu_{\min}}(|\bD\chi|_{\rohy}+|\chi|_{\rohy}),
  \]
  where $C$ only depends on $n$ and $(\bM,\bge_{\refer})$. Combining this estimate with the assumptions and
  (\ref{eq:muminmainlowerbound}) yields
  \begin{equation}\label{eq:omegakmlchiestimate}
    |\omega^{k}(\hN^{-1}\ml_{\chi}E_{i})|\leq C\ldr{\varrho}^{\cweight}e^{\e_{\Spe}\varrho},
  \end{equation}
  where $C$ only depends on $c_{\chi,2}$, $n$, $(\bM,\bge_{\refer})$ and the constant $M_{\min}$ appearing in (\ref{eq:muminmainlowerbound}).

  \textbf{Estimating the evolution along an integral curve.} Let $\g$ be an integral curve of $\hU$ such that $\g(0)\in\bM\times\{t_{0}\}$. Let,
  moreover, $\xi$ be the $\rn{n}$-valued function whose components are $[E_{i}(\varrho)]\circ\g$; let $A$ be the matrix whose components are given by
  the left hand side of (\ref{eq:omegakhNinvmlchiEi}), evaluated along $\g$ and where the order of the components is such that (\ref{eq:xieqEi}) below
  holds; and let $f$ be the $\rn{n}$-valued function whose components are the sum of the first and the second term on the right hand side of
  (\ref{eq:Eivarrhoevolution}), evaluated along $\g$. Then (\ref{eq:Eivarrhoevolution}) implies that
  \begin{equation}\label{eq:xieqEi}
    \frac{d\xi}{ds}-A\xi=f.
  \end{equation}
  In particular,
  \[
  \frac{d}{ds}\ldr{\xi}=\ldr{\xi}^{-1}\xi\cdot\frac{d\xi}{ds}\geq -\|A\|\ldr{\xi}-|f|.
  \]
  Integrating from $s$ to $0$ yields
  \[
  1-\ldr{\xi(s)}\geq -\int_{s}^{0}\|A(s')\|\ldr{\xi(s')}ds'-\int_{s}^{0}|f(s')|ds'
  \]
  recall that $\varrho(\bx,t_{0})=0$. In particular, if $s_{0}\leq s\leq 0$, then
  \[
  \ldr{\xi(s)}\leq 1+\int_{s_{0}}^{0}|f(s')|ds'+\int_{s}^{0}\|A(s')\|\ldr{\xi(s')}ds'.
  \]
  Gr\"{o}nwall's lemma then yields
  \begin{equation}\label{eq:resultofgronwallslemma}
    \ldr{\xi(s_{0})}\leq \left(1+\int_{s_{0}}^{0}|f(s')|ds'\right)\exp\left(\int_{s_{0}}^{0}\|A(s)\|ds\right)
  \end{equation}
  for all $s_{0}\leq 0$. In order to estimate the right hand side, note that (\ref{eq:varrhosequivalencestmt}), (\ref{eq:Eirodivchiestimate}) and
  the assumptions yield
  \begin{equation}\label{eq:fofsestimate}
    |f(s)|\leq \bDlnhNsup+C_{b}\ldr{s}^{2\cweight}e^{\e_{\Spe}s},
  \end{equation}
  where $C_{b}$ only depends on $c_{\chi,2}$, $n$, $(\bM,\bge_{\refer})$, $\cweight$ and the constant $M_{\min}$ appearing in
  (\ref{eq:muminmainlowerbound}). Next, note that (\ref{eq:varrhosequivalencestmt}) and (\ref{eq:omegakmlchiestimate}) yield
  \begin{equation}\label{eq:normAofsestimate}
    \|A(s)\|\leq C\ldr{s}^{\cweight}e^{\e_{\Spe}s},
  \end{equation}
  where $C$ only depends on $c_{\chi,2}$, $n$, $(\bM,\bge_{\refer})$, $\cweight$ and the constant $M_{\min}$ appearing in
  (\ref{eq:muminmainlowerbound}). Integrating the estimates (\ref{eq:fofsestimate}) and (\ref{eq:normAofsestimate}) and combining the result with
  (\ref{eq:varrhosequivalencestmt}) and (\ref{eq:resultofgronwallslemma}) yields (\ref{eq:bDvarrhobdEi}). 

  Next, let $t\in I_{-}$ and $\xi$ be a curve in $\bM\times \{t\}$ such that $\xi(0)=(\bx_{1},t)$ and $\xi(1)=(\bx_{2},t)$. Then
  \[
  \frac{d}{ds}\ln[1-\varrho\circ\xi]=-\frac{1}{1-\varrho\circ\xi}\dot{\xi}(\varrho)
  =-\frac{1}{1-\varrho\circ\xi}\dot{\xi}^{i}E_{i}|_{\xi}(\varrho).
  \]
  Thus
  \[
  \left|\frac{d}{ds}\ln[1-\varrho\circ\xi]\right|\leq C_{\varrho,2}|\dot{\xi}|_{\bge_{\refer}},
  \]
  where $C_{\varrho,2}$ has the same dependence as $C_{\varrho}$. Integrating this estimate and taking the infimum over the curves connecting
  $(\bx_{i},t)$ yields (\ref{eq:varrhorelvarestEi}). The lemma follows.  
\end{proof}

In what follows, it is also convenient to know that the following estimate holds. 

\begin{lemma}\label{lemma:taurelvaryingbxEi}
  Given that the assumptions of Lemma~\ref{lemma:respvarvarrhoEi} hold, assume
  \[
  n^{1/2}\theta_{0,-}^{-1}|\chi|_{\rohy} \leq \de_{\chi}
  \]
  to hold on $M_{-}$. Assuming $\de_{\chi}\leq 1$ to be small enough, the bound depending only on $\cweight$, $c_{\chi,2}$, $\bDlnhNsup$, $\mKsup$,
  $C_{\mK,\mrod}$, $M_{\mK,\mrod}$, $\e_{\Spe}$, $\e_{\rond}$, $\e_{\mK}$, $n$ and $(\bM,\bge_{\refer})$, the estimate
  \begin{equation}\label{eq:hNinvdtvarrhoestEi}
    \frac{1}{2}\leq \hN^{-1}\d_{t}\varrho\leq \frac{3}{2}
  \end{equation}
  holds on $M_{-}$. Fix $\bx_{1},\bx_{2}\in\bM$ and $t_{1},t_{2}\in I_{-}$ such that $t_{1}<t_{2}$. Then
  \begin{equation}\label{eq:DeltavarrhorelvariationEi}
    \frac{1}{3K_{\rovar}}\leq \frac{\varrho(\bx_{2},t_{2})-\varrho(\bx_{2},t_{1})}{\varrho(\bx_{1},t_{2})-\varrho(\bx_{1},t_{1})}\leq 3K_{\rovar},
  \end{equation}
  where
  \begin{equation}\label{eq:KrovarEi}
    K_{\rovar}:=\exp(\bDlnhNsup d_{\bM})
  \end{equation}
  \index{$\a$Aa@Notation!Constants!$K_{\rovar}$}%
  and $d_{\bM}$ is the diameter of $\bM$ with respect to $\bge_{\refer}$.
\end{lemma}
\begin{remark}
  If the standard assumptions are satisfied, then the conditions of the lemma are satisfied; cf. Lemma~\ref{lemma:smallnessshiftconsequences} and
  Definition~\ref{def:standardassumptions}.
\end{remark}
\begin{proof}
  Due to (\ref{eq:hUvarrhoident}),
  \begin{equation}\label{eq:tNinverseeqonepluserror}
    \hN^{-1}\d_{t}\varrho=1+\hN^{-1}\chi(\varrho)+\hN^{-1}\rodiv_{\bge_{\refer}}\chi.
  \end{equation}
  Due to (\ref{eq:rodivchiroughestimpr}), it is clear that the third term on the right hand side is bounded from above by $1/4$ in absolute
  value on $M_{-}$. Next, note that
  \begin{equation}\label{eq:firststepvarrhospatialvar}
    \begin{split}
      \hN^{-1}|\chi(\varrho)|
      \leq & n^{1/2}\hN^{-1}\left(\textstyle{\sum}_{A}|\chi^{A}|^{2}\right)^{1/2}|\bD\varrho|_{\bge_{\refer}}
      \leq n^{1/2}e^{-\mu_{\min}}|\chi|_{\rohy}|\bD\varrho|_{\bge_{\refer}}\\
      \leq & n^{1/2}e^{M_{\min}}C_{\varrho}\ldr{\varrho}e^{\e_{\Spe}\varrho}\theta_{0,-}^{-1}|\chi|_{\rohy}
      \leq n^{1/2}e^{M_{\min}}C_{\varrho}(1+\e_{\Spe}^{-1})\theta_{0,-}^{-1}|\chi|_{\rohy},
    \end{split}
  \end{equation}
  where $M_{\min}$ is introduced in connection with (\ref{eq:muminmainlowerbound}). Assuming $\de_{\chi}$ to be small enough, the bound depending
  only on the quantities listed in the statement of the lemma, it is clear that the right hand side is bounded by $1/4$ on $M_{-}$. Combining the
  above observations yields the conclusion that (\ref{eq:hNinvdtvarrhoestEi}) holds. Fix $\bx_{1},\bx_{2}\in\bM$ and $t_{1},t_{2}\in I_{-}$ such that
  $t_{1}<t_{2}$. Then
  \begin{align}
    \frac{1}{2}\hN(\bx_{1},t)\leq & \d_{t}\varrho(\bx_{1},t)\leq \frac{3}{2}\hN(\bx_{1},t),\label{eq:tNestimatepreliminary}\\
    \frac{1}{2K_{\rovar}}\hN(\bx_{1},t)\leq & \d_{t}\varrho(\bx_{2},t)\leq \frac{3}{2}K_{\rovar}\hN(\bx_{1},t),\label{eq:dtvarrhoreltohNatotherpointEi}
  \end{align}
  where $K_{\rovar}$ is given by (\ref{eq:KrovarEi}). Integrating these estimates from $t_{1}$ to $t_{2}$ and carrying out appropriate divisions
  yields (\ref{eq:DeltavarrhorelvariationEi}). The lemma follows.
\end{proof}

\section{Relating the mean curvature and the logarithmic volume density}

Many solutions to Einstein's equations are such that the deceleration parameter converges to $n-1$. It is of interest to relate $\ln\theta$ and
$\varrho$ under these circumstances.

\begin{lemma}\label{lemma:thetavarrhorelqconvtonmo}
  Assume that the conditions of Lemma~\ref{lemma:taurelvaryingbxEi} are fulfilled. Assume, moreoever, that there is a constant $d_{q}$ such that 
  \begin{equation}\label{eq:qconvergence}
    \|\ldr{\varrho(\cdot,t)}^{3/2}[q(\cdot,t)-(n-1)]\|_{C^{0}(\bM)} \leq d_{q}
  \end{equation}
  for all $t\in I_{-}$. Then there is a constant $R_{q}$, depending only on $d_{q}$, such that
  \begin{equation}\label{eq:varrhoplnthetaglobbound}
    \|\varrho+\ln\theta-\ln\theta_{0,-}\|_{C^{0}(M_{-})}\leq R_{q}+\Theta_{+},
  \end{equation}
  where $\theta_{0,\pm}$ is defined in (\ref{eq:thetazdef}) and 
  \begin{equation}\label{eq:Thetaplusdef}
    \Theta_{+}:=\ln\frac{\theta_{0,+}}{\theta_{0,-}}.
  \end{equation}  
\end{lemma}
\begin{remark}\label{remark:Thetaplusestimate}
  In most of these notes, we assume an estimate of the form
  \begin{equation}\label{eq:Kthetaoneestimate}
    \|\ln\theta\|_{C^{\bfl_{0}}_{\weight_{0}}(\bM)}\leq c_{\theta,1}
  \end{equation}
  to be satisfied for all $t\in I_{-}$, where $\bfl_{0}:=(1,1)$. If such an estimate holds, then $\Theta_{+}$ is bounded by a constant depending only
  on $c_{\theta,1}$ and $(\bM,\bge_{\refer})$. 
\end{remark}
\begin{proof}
  Note that combining (\ref{eq:hUnlnthetamomqbas}) and (\ref{eq:hUvarrhoident}) yields
  \begin{equation}\label{eq:varrhoplnthetaevo}
    \hU(\varrho+\ln\theta)=\hN^{-1}\rodiv_{\bge_{\refer}}\chi-\frac{1}{n}[q-(n-1)].
  \end{equation}
  Let $\g$ be an integral curve of $\hU$ with the properties stated in Lemma~\ref{lemma:lowerbdonmumin}. Combining (\ref{eq:rodivchiroughestimpr}),
  (\ref{eq:varrhosequivalencestmt}), (\ref{eq:qconvergence}) and (\ref{eq:varrhoplnthetaevo}) yields
  \[
  \left|\frac{d}{ds}[(\varrho+\ln\theta)\circ\g](s)\right|\leq \frac{1}{4}\min\{1,\e_{\Spe}\}e^{\e_{\Spe}s}+\frac{1}{n}d_{q}\ldr{s+1/2}^{-3/2}
  \]
  for all $s\leq 0$. Integrating this estimate yields a bound on $\varrho+\ln\theta-\ln\theta_{0,-}$ for $s\leq 0$. Since this estimate holds
  regardless of the choice of integral curve of $\hU$, the conclusion of the lemma follows. 
\end{proof}

\section{Changing the time coordinate}

In the arguments to follow, it is convenient to change the time coordinate. Fix, to this end, $\bx_{0}\in\bM$ and let
\begin{equation}\label{eq:taudefinitionEi}
  \tau(t):=\varrho(\bx_{0},t).
\end{equation}
\index{$\a$Aa@Notation!Functions!$\tau$}%
To begin with, it is of interest to note that we can use $\tau$ instead of $\varrho$ in many of the estimates stated earlier. 

\begin{lemma}\label{lemma:epsilonlowdefEi}
  Given that the assumptions of Lemma~\ref{lemma:taurelvaryingbxEi} hold, let $\tau$ be defined by (\ref{eq:taudefinitionEi}).
  Then
  \begin{equation}\label{eq:eSpevarrhoeelowtaurelEi}
    e^{\e_{\Spe}\varrho(\bx,t)}\leq e^{\eSpe\tau(t)}
  \end{equation}
  for all $(\bx,t)\in M_{-}$, where $\eSpe:=\e_{\Spe}/(3K_{\rovar})$
  \index{$\a$Aa@Notation!Constants!$\eSpe$}%
  and $K_{\rovar}$ is given by (\ref{eq:KrovarEi}). Similarly,
  if $t_{1}\leq t_{2}\leq t_{0}$ and $\bx\in\bM$,
  \begin{equation}\label{eq:emKbxtottEi}
    e^{\e_{\mK}[\varrho(\bx,t_{1})-\varrho(\bx,t_{2})]}\leq e^{\emK[\tau(t_{1})-\tau(t_{2})]}
  \end{equation}
  where $\emK:=\e_{\mK}/(3K_{\rovar})$.
  \index{$\a$Aa@Notation!Constants!$\emK$}%
  Finally,
  \begin{equation}\label{eq:hNtaudotequivEi}
    (2K_{\rovar})^{-1}\leq [\hN(\bx,t)]^{-1}\d_{t}\tau(t)\leq 2K_{\rovar}
  \end{equation}
  for all $t\in I_{-}$ and $\bx\in \bM$.
\end{lemma}
\begin{proof}
Due to the assumptions, (\ref{eq:DeltavarrhorelvariationEi}) holds. Applying this estimate with $t_{1}=t$, $t_{2}=t_{0}$, $\bx_{2}=\bx$
and $\bx_{1}=\bx_{0}$ yields (\ref{eq:eSpevarrhoeelowtaurelEi}). The proof of (\ref{eq:emKbxtottEi}) is similar. Finally, 
(\ref{eq:hNtaudotequivEi}) is an immediate consequence of (\ref{eq:dtvarrhoreltohNatotherpointEi}). 
\end{proof}

At this stage, it is of interest to rephrase the conditions (\ref{eq:hmlUhmlUsqmKoffdiagonalexpbd}) and (\ref{eq:expgrowthwithupperbound})
in terms of $\tau$. 

\begin{lemma}\label{lemma:hmlUhmlUsqmKoffdiagonalexpbdtauEi}
Given that the conditions of Lemma~\ref{lemma:taurelvaryingbxEi} are fulfilled, assume that (\ref{eq:hmlUhmlUsqmKoffdiagonalexpbd}) and 
(\ref{eq:expgrowthwithupperbound}) are satisfied for some $A\neq B$. Then 
\begin{equation}\label{eq:hmlUhmlUsqmKoffdiagonalexpbdtauEi}
  |(\hml_{U}\mK)(Y^{A},X_{B})|\leq C_{\mK,\mrod}e^{\emK\tau}+M_{\mK,\mrod}e^{\emK(\tau_{-}-\tau)}
\end{equation}
on $M_{-}$. Here $\tau_{-}$ is the limit of $\tau(t)$ as $t\rightarrow t_{-}$. 
\end{lemma}
\begin{proof}
Appealing to (\ref{eq:emKbxtottEi}) with $t_{1}=t$ and $t_{2}=t_{0}$ yields $e^{\e_{\mK}\varrho}\leq e^{\emK\tau}$. Assuming that $t_{1}\leq t\leq t_{0}$, the 
estimate (\ref{eq:expgrowthwithupperbound}) yields
\[
G_{\mK,\mrod}\leq M_{\mK,\mrod}e^{\e_{\mK}\varrho(\bx,t_{1})},
\]
so that 
\[
G_{\mK,\mrod}e^{-\e_{\mK}\varrho(\bx,t)}\leq M_{\mK,\mrod}e^{\e_{\mK}[\varrho(\bx,t_{1})-\varrho(\bx,t)]}
\leq M_{\mK,\mrod}e^{\emK[\tau(t_{1})-\tau(t)]},
\]
where we appealed to (\ref{eq:emKbxtottEi}) in the last step. In the right hand side, we can let $t_{1}$ tend to $t_{-}$. Denoting the corresponding
limit of $\tau(t_{1})$ by $\tau_{-}$, we obtain
\[
G_{\mK,\mrod}e^{-\e_{\mK}\varrho(\bx,t)}\leq M_{\mK,\mrod}e^{\emK[\tau_{-}-\tau(t)]}.
\]
Combining the above estimates with (\ref{eq:hmlUhmlUsqmKoffdiagonalexpbd}) and (\ref{eq:expgrowthwithupperbound}) yields the conclusion of the lemma. 
\end{proof}

\section{Relating the mean curvature and the logarithmic volume density II}

The following observation will be of importance in the discussion of the energies. 

\begin{lemma}\label{lemma:thetavarrhorelqconvtonmotwo}
  Assume the conditions of Definition~\ref{def:basicassumptions} and of Lemma~\ref{lemma:taurelvaryingbxEi} to be satisfied. Assume, moreover,
  (\ref{eq:Kthetaoneestimate}) to be fulfilled. Let $t_{c}\in I_{-}$ and $\tvarphi:=\theta\varphi$, where $\varphi$ is defined by
  (\ref{eq:varphidefitobmubgedp}). Define $\tvarphi_{c}$ by $\tvarphi_{c}(\bx,t):=\tvarphi(\bx,t_{c})$. Finally, let
  \begin{equation}\label{eq:tetaonedefprel}
    \teta_{1} := \textstyle{\frac{1}{n}}|q-(n-1)|.
  \end{equation}
  Then
  \begin{equation}\label{eq:lntvarphimlntvarphic}
    |\ln\tvarphi-\ln\tvarphi_{c}|\leq C_{a}\ldr{\tau_{c}}^{\bcweight}e^{\eSpe\tau_{c}}+2K_{\rovar}\int_{\tau}^{\tau_{c}}\teta_{1}(\cdot,s)ds
  \end{equation}
  on $M_{c}:=\{(\bx,t)\in\bM\times I:t\leq t_{c}\}$, where $\tau_{c}:=\tau(t_{c})$, $\bcweight:=\max\{1,\cweight\}$ and $C_{a}$ only depends on
  $c_{\robas}$, $c_{\chi,2}$, $c_{\theta,1}$ and $(\bM,\bge_{\refer})$; here $c_{\robas}$ is given by (\ref{eq:crobasdef}). Assuming, in addition to the
  above, that (\ref{eq:qconvergence}) holds,
  \begin{equation}\label{eq:lntvarphimlntvarphicimp}
    |\ln\tvarphi-\ln\tvarphi_{c}|\leq C_{a}\ldr{\tau_{c}}^{\bcweight}e^{\eSpe\tau_{c}}+C_{b}\ldr{\tau_{c}}^{-1/2}
  \end{equation}
  on $M_{c}$, where $C_{a}$ has the same dependence as in the case of (\ref{eq:lntvarphimlntvarphic}) and $C_{b}$ only depends on $K_{\rovar}$
  and $d_{q}$. 
\end{lemma}
\begin{remark}
  In many convergent settings of interest in general relativity, $q-(n-1)$ converges to zero exponentially, so that (\ref{eq:qconvergence}) holds.
  However, even in oscillatory cases, the average of $\teta_{1}$ over large time intervals tends to zero. To be more precise, it is not unreasonable
  to assume that for every $\e>0$, there is a $T\leq \tau_{c}$ such that for all $\tau\leq T$,
  \[
  \int_{\tau}^{\tau_{c}}\teta_{1}ds\leq \e (\tau_{c}-\tau).
  \]
\end{remark}
\begin{proof}
  Note, to begin with, that
  \begin{equation}\label{eq:dtaulnvarphi}
    \d_{\tau}\ln\tvarphi=\tN\hN^{-1}\d_{t}\ln\tvarphi=\tN(\hU+\hN^{-1}\chi)\ln\tvarphi.
  \end{equation}
  Here $\tN:=\hN/\d_{t}\tau$. Note that $\tN$ is bounded due to (\ref{eq:hNtaudotequivEi}). On the other hand, combining (\ref{eq:rodivchiestimpr}),
  (\ref{eq:varrhoplnthetaevo}), (\ref{eq:eSpevarrhoeelowtaurelEi}) and (\ref{eq:hNtaudotequivEi}) yields
  \[
  |\tN\hU\ln\tvarphi|=|\tN\hU(\varrho+\ln\theta)|\leq 2K_{\rovar}e^{\eSpe\tau}+2K_{\rovar}|q-(n-1)|/n
  \]
  on $M_{-}$. Note that the second term on the far right hand side is bounded by $2K_{\rovar}\teta_{1}$. Next, we wish to estimate $\hN^{-1}\chi(\tvarphi)$.
  Note, to this end, that
  \[
  \hN^{-1}|\chi(\ln\tvarphi)|\leq \hN^{-1}|\chi|_{\bge_{\refer}}|\bD\ln\tvarphi|_{\bge_{\refer}}.  
  \]
  However,
  \begin{equation*}
    \begin{split}
      \hN^{-1}|\chi|_{\bge_{\refer}} \leq & \hN^{-1}|\chi^{A}X_{A}|_{\bge_{\refer}}\leq \hN^{-1}\left(\textstyle{\sum}_{A}(\chi^{A})^{2}\right)^{1/2}\sqrt{n}\\
      \leq & \hN^{-1}e^{-\bmu_{\min}}\left(\textstyle{\sum}_{A}e^{2\bmu_{A}}(\chi^{A})^{2}\right)^{1/2}\sqrt{n}=\sqrt{n}e^{-\mu_{\min}}N^{-1}|\chi|_{\bge}.
    \end{split}
  \end{equation*}
  Combining this estimate with (\ref{eq:muminmainlowerbound}), (\ref{eq:cchitwoestimate}), (\ref{eq:eSpevarrhoeelowtaurelEi}) and the fact
  that $|\chi|_{\rohy}=N^{-1}|\chi|_{\bge}$ yields
  \begin{equation}\label{eq:hNinvchibgereferestimate}
    \hN^{-1}|\chi|_{\bge_{\refer}}\leq Ce^{\eSpe\tau}
  \end{equation}
  on $M_{-}$, where $C$ only depends on $c_{\chi,2}$ and $c_{\robas}$. Next, note that
  \begin{equation}\label{eq:bDlntvarphicestpre}
    \begin{split}
      |\bD\ln\tvarphi|_{\bge_{\refer}} \leq & |\bD\ln\theta|_{\bge_{\refer}}+|\bD\varrho|_{\bge_{\refer}}\\
      \leq & c_{\theta,1}\ldr{\varrho}^{\cweight}+C_{\varrho}\ldr{\varrho}\leq C_{a}\ldr{\tau}^{\bcweight},
    \end{split}
  \end{equation}
  where we appealed to (\ref{eq:bDvarrhobdEi}) and (\ref{eq:Kthetaoneestimate}) in the second step and to
  (\ref{eq:DeltavarrhorelvariationEi}) in the last step. Here $C_{a}$ only depends on $c_{\robas}$, $c_{\chi,2}$, $c_{\theta,1}$ and
  $(\bM,\bge_{\refer})$; and $\bcweight:=\max\{\cweight,1\}$. To conclude, 
  \[
  \hN^{-1}|\chi(\ln\tvarphi)|\leq C_{a}\ldr{\tau}^{\bcweight}e^{\eSpe\tau}
  \]
  on $M_{-}$, where $C_{a}$ only depends on $c_{\robas}$, $c_{\chi,2}$, $c_{\theta,1}$ and $(\bM,\bge_{\refer})$. Combining the above estimates
  yields the conclusion that
  \[
  |\d_{\tau}\ln\tvarphi|\leq C_{a}\ldr{\tau}^{\bcweight}e^{\eSpe\tau}+2K_{\rovar}\teta_{1}
  \]
  on $M_{-}$, where $C_{a}$ only depends on $c_{\robas}$, $c_{\chi,2}$, $c_{\theta,1}$ and $(\bM,\bge_{\refer})$.
  Thus (\ref{eq:lntvarphimlntvarphic}) holds. Assuming, in addition, that (\ref{eq:qconvergence}) holds,
  \[
  2K_{\rovar}\teta_{1}\leq C_{b}\ldr{\tau}^{-3/2}
  \]
  on $M_{-}$, where we appealed to (\ref{eq:DeltavarrhorelvariationEi}), and $C_{b}$ only depends on $K_{\rovar}$ and $d_{q}$. Combining this
  estimate with (\ref{eq:lntvarphimlntvarphic}) yields (\ref{eq:lntvarphimlntvarphicimp}). 
\end{proof}

\chapter{Function spaces and estimates}\label{chapter:functionspacesandestimates}

In the present chapter, we introduce weighted spaces and derive some basic estimates. In (\ref{eq:mtClbS}) and (\ref{eq:mtHlbS}), we introduced weighted 
spaces using the Riemannian metric $\bge_{\refer}$. However, in many applications, it is more convenient to use the frame $\{E_{i}\}$ in combination with
$\bge_{\refer}$. We begin by defining the corresponding spaces. We then prove relations and equivalences between different norms. Moser estimates are of
particular importance, and appealing to Appendix~\ref{chapter:gagnir}, we derive such estimates in Section~\ref{section:MoserestimatesbbEversion}.
We end the chapter by recording weighted Sobolev estimates for $\ell_{A}$, $X_{A}$ and $Y^{A}$. 

\section{Function spaces}

Using the notation introduced in Definition~\ref{def:multiindexnotation}, the following spaces will be of interest.

\begin{definition}
  Let $(M,g)$ be a time oriented Lorentz manifold. Assume it to have an expanding partial pointed foliation and $\mK$ to be non-degenerate on $I$ and to
  have a global frame. Let $\{E_{i}\}$ be the frame introduced in Remark~\ref{remark:framenondegenerate}. Let $(\weight_{a},\weight_{b})=\weight\in\Weight$
  and $(l_{0},l_{1})=\bfl\in\Index$. Define, using the notation introduced in Definition~\ref{def:multiindexnotation},
  \begin{align}
    \|\mt(\cdot,t)\|_{\mc^{\bfl}_{\bbE,\weight}(\bM)} := & \textstyle{\sup}_{\bx\in\bM}\left(
    \textstyle{\sum}_{j=l_{0}}^{l_{1}}\sum_{|\bfI|=j}
    \ldr{\varrho(\bx,t)}^{-2\weight_{a}-2j\weight_{b}}|\bD_{\bfI}\mt(\bx,t)|_{\bge_{\refer}}^{2}\right)^{1/2},\label{eq:mtmClbbEbS}\\
    \|\mt(\cdot,t)\|_{\mH^{\bfl}_{\bbE,\weight}(\bM)} := & \left(\int_{\bM}\textstyle{\sum}_{j=l_{0}}^{l_{1}}\sum_{|\bfI|=j}
    \ldr{\varrho(\cdot,t)}^{-2\weight_{a}-2j\weight_{b}}|\bD_{\bfI}\mt(\cdot,t)|_{\bge_{\refer}}^{2}\mu_{\bge_{\refer}}\right)^{1/2}.\label{eq:mtmHlbbEbS}
  \end{align}
  \index{$\a$Aa@Notation!Norms!$\mc^{\bfl}_{\bbE,\weight}(\bM)$}%
  \index{$\a$Aa@Notation!Norms!$\mH^{\bfl}_{\bbE,\weight}(\bM)$}%
  If $l_{0}=0$, then we replace $\bfl$ in (\ref{eq:mtmClbbEbS})--(\ref{eq:mtmHlbbEbS}) with $l:=l_{1}$. Next define, in analogy with the $C^{l,\weight}_{\rohy}$-
  and $H^{l,\weight}_{\rohy}$-norms introduced in (\ref{eq:Hlrohydef}) and (\ref{eq:Clrohydef}),
  \index{$\a$Aa@Notation!Norms!$\mH^{l,\weight}_{\bbE,\rohy}(\bM)$}%
  \index{$\a$Aa@Notation!Norms!$\mc^{l,\weight}_{\bbE,\rohy}(\bM)$}%
  \begin{align}
    \|\chi(\cdot,t)\|_{\mH^{l,\weight}_{\bbE,\rohy}(\bM)} := & \left(\int_{\bM}\textstyle{\sum}_{|\bfI|\leq l}
    \ldr{\varrho(\cdot,t)}^{-2\weight_{a}-2|\bfI|\weight_{b}}N^{-2}|\bD_{\bfI}\chi(\cdot,t)|_{\bge}^{2}\mu_{\bge_{\refer}}\right)^{1/2},\label{eq:mHlbErohydef}\\
    \|\chi(\cdot,t)\|_{\mc^{l,\weight}_{\bbE,\rohy}(\bM)} := & \sup_{\bx\in\bM}\left(\textstyle{\sum}_{|\bfI|\leq l}
    \ldr{\varrho(\cdot,t)}^{-2\weight_{a}-2|\bfI|\weight_{b}}N^{-2}|\bD_{\bfI}\chi(\bx,t)|_{\bge}^{2}\right)^{1/2}. \label{eq:mClbErohydef}
  \end{align}
\end{definition}

\subsection{Basic equivalences and estimates}

In what follows, it is of interest to compare the different norms. Some of the comparisons are straightforward, and we record them
in the present subsection. Others require more of an effort and will only be carried out later on.

\begin{lemma}\label{lemma:basequiv}
  Let $(M,g)$ be a time oriented Lorentz manifold. Assume it to have an expanding partial pointed foliation and $\mK$ to be non-degenerate on $I$ and to
  have a global frame. Let $(l_{0},l_{1})=\bfl\in\Index$ and $(\weight_{a},\weight_{b})=\weight\in \Index$. Then, assuming $l_{0}\leq 1$, there are constants
  $C_{\sup,\bfl},C_{\rosob,\bfl}\geq 1$, depending only on $\bfl$, $n$, $(\bM,\bge_{\refer})$ and the type of the tensor field, such that
  \begin{align}
    C_{\sup,\bfl}^{-1}\|\mt(\cdot,t)\|_{C^{\bfl}_{\weight}(\bM)} \leq & \|\mt(\cdot,t)\|_{\mc^{\bfl}_{\bbE,\weight}(\bM)}
    \leq C_{\sup,\bfl}\|\mt(\cdot,t)\|_{C^{\bfl}_{\weight}(\bM)},\label{eq:Clmclequiv}\\
    C_{\rosob,\bfl}^{-1}\|\mt(\cdot,t)\|_{H^{\bfl}_{\weight}(\bM)} \leq & \|\mt(\cdot,t)\|_{\mH^{\bfl}_{\bbE,\weight}(\bM)}
    \leq C_{\rosob,\bfl}\|\mt(\cdot,t)\|_{H^{\bfl}_{\weight}(\bM)}.\label{eq:HlmHlequiv}
  \end{align}
  Similarly, given $0\leq l\in\zo$ and $\weight$ as above, there are constants $C_{\rohc,l},C_{\rohs,l}\geq 1$, depending only on $0\leq l\in\zo$ and
  $(\bM,\bge_{\refer})$, such that
  \begin{align}
    C_{\rohc,l}^{-1}\|\chi(\cdot,t)\|_{C^{l,\weight}_{\rohy}(\bM)} \leq & \|\chi(\cdot,t)\|_{\mc^{l,\weight}_{\bbE,\rohy}(\bM)}
    \leq C_{\rohc,l}\|\chi(\cdot,t)\|_{C^{l,\weight}_{\rohy}(\bM)},\label{eq:hyClmclequiv}\\
    C_{\rohs,l}^{-1}\|\chi(\cdot,t)\|_{H^{l,\weight}_{\rohy}(\bM)} \leq & \|\chi(\cdot,t)\|_{\mH^{l,\weight}_{\bbE,\rohy}(\bM)}
    \leq C_{\rohs,l}\|\chi(\cdot,t)\|_{H^{l,\weight}_{\rohy}(\bM)}.\label{eq:hyHlmHlequiv}
  \end{align}
\end{lemma}
\begin{proof}
Due to Lemma~\ref{lemma:bDbfAbDkequiv}, the fact that $\weight_{a},\weight_{b}\geq 0$ and the fact that $l_{0}\leq 1$, it is clear that 
(\ref{eq:Clmclequiv}) and (\ref{eq:HlmHlequiv}) hold.

Next, let $\{\Omega^{i}\}$ be a frame of one-form fields which are orthonormal with respect to $\bge$. Then estimating $|\bD^{k}\chi|_{\rohy}$ is 
equivalent to estimating a sum of expressions of the form $N^{-1}|\Omega^{i}[(\bD^{k}\chi)(\bfE_{\bfI})]|$. Combining this fact with 
Lemma~\ref{lemma:bDbfAbDkequiv} yields the conclusion that 
\[
|\bD^{k}\chi|_{\rohy}\leq C\textstyle{\sum}_{|\bfI|\leq k}N^{-1}|\bD_{\bfI}\chi|_{\bge},
\]
where $C$ only depends on $n$, $k$ and $(\bM,\bge_{\refer})$. Thus the left hand side estimates in (\ref{eq:hyClmclequiv}) and (\ref{eq:hyHlmHlequiv}) hold. 
Next, note that $|\bD_{\bfI}\chi|_{\bge}$ can be estimated by a sum of terms of the form $|\Omega^{i}(\bD_{\bfI}\chi)|$. Combining this observation with 
Lemma~\ref{lemma:bDbfAbDkequiv} yields the right hand side estimates in (\ref{eq:hyClmclequiv}) and 
(\ref{eq:hyHlmHlequiv}). The lemma follows. 
\end{proof}

For future reference, it is of interest to record a relation between $C^{k}$- and $\mc^{k}$-norms. Introduce, to this end, the following
notation. 
\begin{definition}\label{def:msfPmKhN}
Let $(M,g)$ be a time oriented Lorentz manifold. Assume it to have an expanding partial pointed foliation. Let $0\leq m\in\zo$ and $0\leq \cweight\in \ro$. 
Then 
\begin{align*}
\msfP_{\mK,m,\cweight} := & \textstyle{\sum}_{m_{1}+\dots+m_{j}=m,m_{i}\geq 1}\|\ldr{\varrho}^{-m_{1}\cweight}\bD^{m_{1}}\mK\|_{C^{0}(\bM)}\cdots 
\|\ldr{\varrho}^{-m_{j}\cweight}\bD^{m_{j}}\mK\|_{C^{0}(\bM)},\\
\msfP_{N,m,\cweight} := & \textstyle{\sum}_{m_{1}+\dots+m_{j}=m,m_{i}\geq 1}\|\ldr{\varrho}^{-m_{1}\cweight}\bD^{m_{1}}\ln\hN\|_{C^{0}(\bM)}\cdots 
\|\ldr{\varrho}^{-m_{j}\cweight}\bD^{m_{j}}\ln\hN\|_{C^{0}(\bM)},\\
\msfP_{\mK,N,m,\cweight} := & \textstyle{\sum}_{m_{1}+m_{2}=m}\msfP_{\mK,m_{1},\cweight}\msfP_{N,m_{2},\cweight},
\end{align*}
\index{$\a$Aa@Notation!Functions!$\msfP_{\mK,m,\cweight}$}%
\index{$\a$Aa@Notation!Functions!$\msfP_{N,m,\cweight}$}%
\index{$\a$Aa@Notation!Functions!$\msfP_{\mK,N,m,\cweight}$}%
with the convention that $\msfP_{\mK,0,\cweight}=1$ and $\msfP_{N,0,\cweight}=1$. 
\end{definition}

\section{Estimating the shift vector field}\label{section:estshiftvect}

In Subsection~\ref{ssection:asslash}, we introduce weighted supremum and Sobolev norms for the shift vector field. It is of interest to compare them
with the following norms:
\begin{align}
\|\mt(\cdot,t)\|_{\mH^{\bfl,\weight}_{\bbE,\rocon}(\bM)} := & 
\left(\int_{\bM}\textstyle{\sum}_{l_{0}\leq|\bfI|\leq l_{1}}\hN^{-2}(\cdot,t)
\ldr{\varrho(\cdot,t)}^{-2\weight_{a}-2|\bfI|\weight_{b}}|\bD_{\bfI}\mt(\cdot,t)|_{\bge_{\refer}}^{2}\bmu_{\bge_{\refer}}\right)^{1/2},\label{eq:mHlbbeWnorm}\\
\|\mt(\cdot,t)\|_{\mc^{\bfl,\weight}_{\bbE,\rocon}(\bM)} := & 
\sup_{\bx\in\bM}\left(\textstyle{\sum}_{l_{0}\leq |\bfI|\leq l_{1}}\hN^{-2}(\bx,t)
\ldr{\varrho(\bx,t)}^{-2\weight_{a}-2|\bfI|\weight_{b}}|\bD_{\bfI}\mt(\bx,t)|_{\bge_{\refer}}^{2}\right)^{1/2}.\label{eq:mClbbeWnorm}
\end{align}
\index{$\a$Aa@Notation!Norms!$\mH^{\bfl,\weight}_{\bbE,\rocon}(\bM)$}%
\index{$\a$Aa@Notation!Norms!$\mc^{\bfl,\weight}_{\bbE,\rocon}(\bM)$}%
Here $(l_{0},l_{1})=\bfl\in\Index$, $(\weight_{a},\weight_{b})=\weight\in\Weight$ and we use the notation introduced in
Definition~\ref{def:multiindexnotation}. 

\begin{lemma}\label{lemma:chimclbbeWClhy}
  Given that the assumptions of Lemma~\ref{lemma:taurelvaryingbxEi} hold, let $\tau$ be defined by (\ref{eq:taudefinitionEi}). Let $\xi$ be
  a vector field on $\bM$, $(l_{0},l_{1})=\bfl\in\Index$ and $(\weight_{a},\weight_{b})=\weight\in\Weight$. Then, assuming $l_{0}\leq 1$,
  \begin{align}
    \|\xi(\cdot,t)\|_{\mH^{\bfl,\weight}_{\bbE,\rocon}(\bM)} \leq & Ce^{M_{\min}}e^{\eSpe\tau}\theta_{0,-}^{-1}
    \|\xi(\cdot,t)\|_{H^{\bfl,\weight}_{\rohy}(\bM)},\label{eq:chimHlHlest}\\
    \|\xi(\cdot,t)\|_{\mc^{\bfl,\weight}_{\bbE,\rocon}(\bM)} \leq & Ce^{M_{\min}}e^{\eSpe\tau}\theta_{0,-}^{-1}
    \|\xi(\cdot,t)\|_{C^{\bfl,\weight}_{\rohy}(\bM)},\label{eq:chimClClest}
  \end{align}
  where $C$ only depends on $n$, $\bfl$, $\weight$ and $(\bM,\bge_{\refer})$; $M_{\min}$ is defined in the text adjacent to
  (\ref{eq:muminmainlowerbound}); and $\eSpe$ is defined in the text adjacent to (\ref{eq:eSpevarrhoeelowtaurelEi}).
  Similarly, assuming $l_{0}\leq 1$,
  \begin{align}
    \|\xi(\cdot,t)\|_{\mH^{\bfl}_{\bbE,\weight}(\bM)} \leq & Ce^{M_{\min}}e^{\eSpe\tau}\theta_{0,-}^{-1}
    \|\xi(\cdot,t)\|_{H^{\bfl,\weight}_{\rohc}(\bM)},\label{eq:chimHlHlesthc}\\
    \|\xi(\cdot,t)\|_{\mc^{\bfl}_{\bbE,\weight}(\bM)} \leq & Ce^{M_{\min}}e^{\eSpe\tau}\theta_{0,-}^{-1}
    \|\xi(\cdot,t)\|_{C^{\bfl,\weight}_{\rohc}(\bM)},\label{eq:chimClClesthc}
  \end{align}
  where $C$ only depends on $n$, $\bfl$, $\weight$ and $(\bM,\bge_{\refer})$
\end{lemma}
\begin{remark}\label{remark:chiclvarrhodecay}
  Arguments similar to the proof give the following conclusion. Given that the conditions of Lemma~\ref{lemma:lowerbdonmumin} are fulfilled
  and that $\bfl$ and $\weight$ are as in the statement of the lemma, 
  \[
  \ldr{\varrho}^{-\weight_{a}-|\bfI|\weight_{b}}\hN^{-1}|\bD_{\bfI}\xi|_{\bge_{\refer}}\leq Ce^{M_{\min}}e^{\e_{\Spe}\varrho}\theta_{0,-}^{-1}
  \|\xi(\cdot,t)\|_{C^{\bfl,\weight}_{\rohy}(\bM)}
  \]
  for all $(\bx,t)\in M_{-}$ and $l_{0}\leq |\bfI|\leq l_{1}$, where $C$ only depends on $n$, $\bfl$, $\weight$ and $(\bM,\bge_{\refer})$. Moreover,
  \begin{equation}\label{eq:ClbbEweightestimatepointwise}
    \ldr{\varrho}^{-\weight_{a}-|\bfI|\weight_{b}}|\bD_{\bfI}\xi|_{\bge_{\refer}}\leq Ce^{M_{\min}}e^{\e_{\Spe}\varrho}\theta_{0,-}^{-1}
    \|\xi(\cdot,t)\|_{C^{\bfl,\weight}_{\rohc}(\bM)}
  \end{equation}
  for all $(\bx,t)\in M_{-}$ and $l_{0}\leq |\bfI|\leq l_{1}$, where $C$ only depends on $n$, $\bfl$, $\weight$ and $(\bM,\bge_{\refer})$.
\end{remark}
\begin{proof}
  Note that $\bD_{\bfI}\xi$ can be written as a linear combination of terms of the form
  \[
  (\bD^{j}\xi)(E_{k_{1}},\dots,E_{k_{j}})\omega^{I_{1}}(\bD_{\bfI_{1}}E_{K_{1}})\cdots\omega^{I_{l}}(\bD_{\bfI_{l}}E_{K_{l}})
  \]
  where $j_{0}\leq j\leq |\bfI|$ and $j_{0}:=\min\{1,|\bfI|\}$; cf. Lemma~\ref{lemma:bDbfAbDkequiv}. The last $l$ factors can
  be estimated in absolute value by a constant depending only on $(\bM,\bge_{\refer})$ and $|\bfI|$. Thus $\hN^{-1}|\bD_{\bfI}\xi|_{\bge_{\refer}}$ can
  be estimated by a linear combination of terms of the form
  \begin{equation*}
    \begin{split}
      \hN^{-1}|\omega^{i}[(\bD^{j}\xi)(E_{k_{1}},\dots,E_{k_{j}})]| \leq & \hN^{-1}|\omega^{i}(X_{A})Y^{A}[(\bD^{j}\xi)(E_{k_{1}},\dots,E_{k_{j}})]|\\
      \leq & \textstyle{\sum}_{A}e^{-\mu_{A}}N^{-1}e^{\bmu_{A}}|Y^{A}[(\bD^{j}\xi)(E_{k_{1}},\dots,E_{k_{j}})]|,
    \end{split}
  \end{equation*}
  where $j_{0}\leq j\leq |\bfI|$. Summing up,
  \begin{equation}\label{eq:hNinvbDbfIbgerefer}
    \begin{split}
      \hN^{-1}|\bD_{\bfI}\xi|_{\bge_{\refer}} \leq & C\textstyle{\sum}_{j_{0}\leq j\leq |\bfI|}e^{-\mu_{\min}}|\bD^{j}\xi|_{\rohy}\leq Ce^{M_{\min}}e^{\eSpe\tau}
      \theta_{0,-}^{-1}\textstyle{\sum}_{j_{0}\leq j\leq |\bfI|}|\bD^{j}\xi|_{\rohy},
    \end{split}
  \end{equation}
  where $C$ only depends on $n$, $|\bfI|$ and $(\bM,\bge_{\refer})$; and we appealed to (\ref{eq:muminmainlowerbound}) and
  (\ref{eq:eSpevarrhoeelowtaurelEi}). The estimates (\ref{eq:chimHlHlest}) and (\ref{eq:chimClClest}) follow. The proof of
  (\ref{eq:chimHlHlesthc}) and (\ref{eq:chimClClesthc}) is similar. The lemma follows. 
\end{proof}

\section{Moser estimates}\label{section:MoserestimatesbbEversion}

In Appendix~\ref{chapter:gagnir}, we derive Gagliardo-Nirenberg as well as Moser estimates with respect to different frames on $\bM$. Here we combine
these results with estimates of the spatial variation of $\varrho$ in order to derive weighted versions of the Moser estimates. Before stating
the estimates, it is convenient to introduce the following terminology. If $\mt$ is a family of smooth tensor fields on $\bM$ for
$t\in I$ and $0\leq l\in\zo$, then 
\begin{align}
|(\bD^{l}_{\bbE}\mt)(\bx,t)|_{\bge_{\refer}} := & \left(\textstyle{\sum}_{|\bfI|=l}|\bD_{\bfI}\mt(\bx,t)|_{\bge_{\refer}}^{2}\right)^{1/2},\label{eq:bDlbbEmtdef} 
\end{align}
where we use the notation introduced in Definition~\ref{def:multiindexnotation}.

\begin{prop}\label{prop:moserestwightEi}
  Assume that the conditions of Lemma~\ref{lemma:taurelvaryingbxEi} hold. Let $0\leq l_{i}\in\mathbb{Z}$ and $l=l_{1}+\dots+l_{j}$. Then there is a constant
  $C$ such that if $\mt_{1},\dots,\mt_{j}$ are families of smooth tensor fields on $\bM$ for $t\in I$; and
  $(\weight_{m,a},\weight_{m,b})=\weight_{m}\in\Weight$, $m=1,\dots j$; then
  \begin{equation}\label{eq:mosergagliardonirenbergwightEi}
    \begin{split}
      & \left\| \textstyle{\prod}_{m=1}^{j}\ldr{\varrho(\cdot,t)}^{-\weight_{m,a}-l_{m}\weight_{m,b}}|(\bD^{l_{m}}_{\bbE}\mt_{m})(\cdot,t)|_{\bge_{\refer}}\right\|_{2}\\
      \leq & C\textstyle{\sum}_{i}\|\mt_{i}(\cdot,t)\|_{\mH^{l}_{\bbE,\weight_{i}}}
      \textstyle{\prod}_{m\neq i}\|\mt_{m}(\cdot,t)\|_{C^{0}_{\weight_{m}}(\bM)};
    \end{split}
  \end{equation}
  cf. the notation introduced in (\ref{eq:bDlbbEmtdef}) and (\ref{eq:mtmHlbbEbS}). Moreover, the constant $C$ only depends on $\bDlnhNsup$,
  $\weight_{m}$ ($m=1,\dots,j$), $n$, $l$ and $(\bM,\bge_{\refer})$.
\end{prop}
\begin{proof}
  First note that we can apply Corollary~\ref{cor:mixedmoserestweight} with $q=r=0$; $v_{m}=\ldr{\tau}^{-\weight_{m,a}}$; and $h_{m}=\ldr{\tau}^{-\weight_{m,b}}$.
  This yields
  \begin{equation*}
    \begin{split}
      & \left\|\textstyle{\prod}_{m=1}^{j}\ldr{\tau}^{-\weight_{m,a}-l_{m}\weight_{m,b}}|(\bD^{l_{m}}_{\bbE}\mt_{m})(\cdot,t)|_{\bge_{\refer}}\right\|_{2}\\
      \leq & C_{a}\textstyle{\sum}_{i}\textstyle{\sum}_{k\leq l}\|\ldr{\tau}^{-\weight_{i,a}-k\weight_{i,b}}|(\bD^{k}_{\bbE}\mt_{i})(\cdot,t)|_{\bge_{\refer}}\|_{2}
      \textstyle{\prod}_{o\neq i}\|\ldr{\tau}^{-\weight_{o,a}}\mt_{o}(\cdot,t)\|_{\infty},
    \end{split}
  \end{equation*}
  where the constant $C_{a}$ only depends on $l$, $n$ and $(\bM,\bge_{\refer})$. At this stage, we can appeal to (\ref{eq:DeltavarrhorelvariationEi}) in
  order to deduce the conclusion of the proposition.
\end{proof}
Next, we formulate a version without a frame. 

\begin{prop}\label{prop:moserestwightnoframe}
  Assume that the conditions of Lemma~\ref{lemma:taurelvaryingbxEi} hold. Let $0\leq l_{i}\in\mathbb{Z}$ and $l=l_{1}+\dots+l_{j}$. Then there is a constant
  $C$ such that if $\mt_{1},\dots,\mt_{j}$ are families of smooth tensor fields on $\bM$ for $t\in I$; and
  $(\weight_{m,a},\weight_{m,b})=\weight_{m}\in\Weight$, $m=1,\dots j$; then
  \begin{equation}\label{eq:mosergagliardonirenbergwightEibD}
    \begin{split}
      & \left\| \textstyle{\prod}_{m=1}^{j}\ldr{\varrho(\cdot,t)}^{-\weight_{m,a}-l_{m}\weight_{m,b}}|(\bD^{l_{m}}\mt_{m})(\cdot,t)|_{\bge_{\refer}}\right\|_{2}\\
      \leq & C\textstyle{\sum}_{i}\|\mt_{i}(\cdot,t)\|_{H^{l}_{\weight_{i}}}
      \textstyle{\prod}_{m\neq i}\|\mt_{m}(\cdot,t)\|_{C^{0}_{\weight_{m}}(\bM)}.
    \end{split}
  \end{equation}
  Moreover, the constant $C$ only depends on $\bDlnhNsup$, $\weight_{m}$ ($m=1,\dots,j$), $n$, $l$ and $(\bM,\bge_{\refer})$. 
\end{prop}
\begin{proof}
  As in the proof of Proposition~\ref{prop:moserestwightEi}, the statement follows by an application of Corollary~\ref{cor:mixedmoserestweight},
  keeping (\ref{eq:DeltavarrhorelvariationEi}) in mind. 
\end{proof}

\section{Estimating derivatives of the frame and the eigenvalues in $L^{2}$}

\begin{lemma}\label{lemma:bDbfabDlmjchKestSobEi}
  Assume that (\ref{eq:mKComfwbd}) and the conditions of Lemma~\ref{lemma:taurelvaryingbxEi} hold. Let $1\leq l\in\zo$ and
  $(0,\cweight)=\weight_{0}\in\Weight$. Then there is a constant $\mc_{\mK,l}$ depending only on $\mKsup$, $K_{\cweight}$, $\bDlnhNsup$,
  $\e_{\rond}$, $l$, $n$, $\cweight$ and $(\bM,\bge_{\refer})$,
  such that the following holds. For every $1\leq j\leq l\in\zo$ and every choice of vector field multiindex $\bfI$ with $|\bfI|=j$,
  \begin{equation}\label{eq:bDindasslambdaAXASobEi}
    \begin{split}
      & \|\ldr{\varrho(\cdot,t)}^{-j\cweight}\bD_{\bfI}\ell_{A}(\cdot,t)\|_{L^{2}(\bM)}
      +\|\ldr{\varrho(\cdot,t)}^{-j\cweight}\bD_{\bfI}X_{A}(\cdot,t)\|_{L^{2}(\bM)}\\
      + & \|\ldr{\varrho(\cdot,t)}^{-j\cweight}\bD_{\bfI}Y^{A}(\cdot,t)\|_{L^{2}(\bM)} \leq \mc_{\mK,l}\|\mK(\cdot,t)\|_{H^{\bfl}_{\weight_{0}}(\bM)}
    \end{split}
  \end{equation}
  for all $t\in I_{-}$ and all $A\in\{1,\dots,n\}$, where $\bfl:=(1,l)$. Finally, if $m=|\bfI|\leq l$, then
  \begin{equation}\label{eq:bDbfAgABCestSobEi}
    \|\ldr{\varrho(\cdot,t)}^{-(m+1)\cweight}\bD_{\bfI}\g^{A}_{BC}(\cdot,t)\|_{L^{2}(\bM)}\leq \mc_{\mK,l+1}\|\mK(\cdot,t)\|_{H^{\bfl_{1}}_{\weight_{0}}(\bM)}
  \end{equation}
  for all $t\in I_{-}$ and all $A,B,C\in\{1,\dots,n\}$, where $\bfl_{1}=(1,l+1)$. 
\end{lemma}
\begin{remark}
  Similar estimates can be derived without assuming (\ref{eq:mKComfwbd}) to hold. However, then $\bfl$ has to be replaced by $l$ on the right
  hand side of (\ref{eq:bDindasslambdaAXASobEi}), and $\bfl_{1}$ has to be replaced by $l+1$ on the right hand side of (\ref{eq:bDbfAgABCestSobEi}).
  Moreover, the corresponding constants are independent of $K_{\cweight}$. 
\end{remark}
\begin{proof}
  Consider (\ref{eq:bDindasslambdaAXASobEi}). Due to (\ref{eq:bDbfAellAetcpteststmtEi}), it is sufficient to estimate $\ldr{\varrho}^{-l\cweight}\mfP_{\mK,p}$
  in $L^{2}$ for $1\leq p\leq l$. Apply Proposition~\ref{prop:moserestwightnoframe} to this expression with the $\mt_{m}$ replaced by $\bD\mK$;
  $\weight_{m,a}=\cweight$; and $\weight_{m,b}=\cweight$. This yields (\ref{eq:bDindasslambdaAXASobEi}). The proof of (\ref{eq:bDbfAgABCestSobEi}) is
  similar. 
\end{proof}

\chapter[Higher order estimates, Lie derivatives]{Higher order estimates of the Lie derivatives 
of the elements of the frame}\label{chapter:higherorderestimatesnormsLieframe}

Consider $\mW^{A}_{B}$ and $\overline{\mW}^{0}_{A}$ defined by (\ref{eq:mWdefrel}). When deriving energy estimates, we need to estimate these quantities
in weighted $C^{k}$- and $H^{k}$-spaces. This is the main purpose of the present chapter. However, we also need to estimate $A_{i}^{k}$ introduced in
(\ref{eq:Aialphadef}) as well as its first normal derivative. We end the chapter by recording the consequences of combining these
estimates with the higher order $C^{k}$- and Sobolev assumptions. 

\section{Estimating $\mW^{A}_{B}$}

The main estimate of the present chapter is the following: 

\begin{lemma}\label{lemma:hoemWEi}
  Let $(M,g)$ be a time oriented Lorentz manifold. Assume that it has an expanding partial pointed foliation. Assume, moreover, $\mK$ to be
  non-degenerate on $I$, to have a global frame and to be $C^{0}$-uniformly bounded on $I_{-}$; i.e., (\ref{eq:mKsupbasest}) to hold. Then, if
  $B\neq C$,
  \begin{equation}\label{eq:XbfBWCBestCkEi}
    |E_{\bfI}(\mW^{C}_{B})|\leq C_{a}\textstyle{\sum}_{l_{\min}\leq l_{a}+l_{b}\leq |\bfI|}\mfP_{\mK,l_{a}}|\bD^{l_{b}}\hml_{U}\mK|_{\bge_{\refer}}
  \end{equation}
  on $\bM\times I_{-}$, where $l_{\min}:=\min\{1,|\bfI|\}$, and $C_{a}$ only depends on $n$, $\e_{\rond}$, $\mKsup$, $|\bfI|$ and $(\bM,\bge_{\refer})$.
  In particular, if $(\cweight,\cweight)=\weight\in\Weight$, $(l_{0},l_{1})=\bfl\in\Index$ and $B\neq C$, then
  \begin{equation}\label{eq:WCneqBmckestEi}
    \|\mW^{C}_{B}\|_{\mc^{\bfl}_{\bbE,\weight}(\bM)}\leq C_{a}
    \textstyle{\sum}_{k_{\min}\leq l_{a}+l_{b}\leq l_{1}}\msfP_{\mK,l_{a},\cweight}\|\ldr{\varrho}^{-(l_{b}+1)\cweight}\bD^{l_{b}}\hml_{U}\mK\|_{C^{0}(\bM)}
  \end{equation}
  on $I_{-}$, where $k_{\min}:=\min\{l_{0},1\}$ and the constant $C_{a}$ only depends on $n$, $\e_{\rond}$, $\mKsup$, $\bfl$ and $(\bM,\bge_{\refer})$.

  Moreover,
  \begin{equation}\label{eq:finalproducttobeestimatedCkstmtEi}
    \begin{split}
      |E_{\bfI}(\mW^{A}_{A})| \leq & C_{a}\textstyle{\sum}_{l_{\min}\leq l_{a}+l_{b}\leq|\bfI|}\mfP_{\mK,l_{a}}|\bD^{l_{b}}\hml_{U}\mK|_{\bge_{\refer}}\\
      & +C_{a}\textstyle{\sum}_{l_{a}+|\bfJ|\leq |\bfI|,|\bfK|=1}\mfP_{\mK,N,l_{a}}\hN^{-1}|\bD_{\bfJ}\bD_{\bfK}\chi|_{\bge_{\refer}}
    \end{split}
  \end{equation}
  (no summation on $A$), where $l_{\min}$ is defined as above and $C_{a}$ only depends on $n$, $\e_{\rond}$, $\mKsup$, $(\bM,\bge_{\refer})$ and $|\bfI|$.
  In particular, if $(0,\cweight)=\weight_{0}$, $\bfl_{b}=(1,l_{b})$ and $(l_{0},l_{1})=\bfl\in\Index$, then
  \begin{equation}\label{eq:mWAAmckestEi}
    \begin{split}
      \|\mW^{A}_{A}\|_{\mc^{\bfl}_{\bbE,\weight}(\bM)} \leq & C_{a}\textstyle{\sum}_{k_{\min}\leq l_{a}+l_{b}\leq l_{1}}\msfP_{\mK,l_{a},\cweight}
      \|\ldr{\varrho}^{-(l_{b}+1)\cweight}\bD^{l_{b}}\hml_{U}\mK\|_{C^{0}(\bM)}\\
      & +C_{a}\textstyle{\sum}_{l_{a}+l_{b}\leq l_{1}+1,l_{b}\geq 1}\msfP_{\mK,N,l_{a},\cweight}\|\chi\|_{\mc^{\bfl_{b},\weight_{0}}_{\bbE,\rocon}(\bM)}
    \end{split}
  \end{equation}
  (no summation on $A$), where $k_{\min}$ is defined as above and $C_{a}$ only depends on $n$, $\e_{\rond}$, $\mKsup$, $(\bM,\bge_{\refer})$ and $l$; and the
  $\mc^{\bfl,\weight}_{\bbE,\rocon}(\bM)$-norm is introduced in (\ref{eq:mClbbeWnorm}).
\end{lemma}
\begin{remark}
  Considering (\ref{eq:mWAAmckestEi}), it is clear that estimates of the form (\ref{eq:chimClClest}) are of interest. 
\end{remark}
\begin{proof}
  When $B\neq C$, Lemma~\ref{lemma:compositeestEi}, Remark~\ref{remark:compositeestEi} and (\ref{eq:mWBAformulanongeo}) yield
  (\ref{eq:XbfBWCBestCkEi}), an estimate which implies (\ref{eq:WCneqBmckestEi}); cf. Definitions~\ref{def:mfPmKhN} and \ref{def:msfPmKhN}.
  In order to estimate $\mW^{A}_{A}$ (no summation), it is sufficient to appeal to Lemma~\ref{lemma:compositeestEi}, Remark~\ref{remark:compositeestEi}
  and (\ref{eq:mWAAformulanongeo}). This yields (\ref{eq:finalproducttobeestimatedCkstmtEi}), an estimate which immediately implies
  (\ref{eq:mWAAmckestEi}).
\end{proof}

Next we turn to Sobolev estimates.

\begin{lemma}\label{lemma:mWmHlestEi}
  Given that (\ref{eq:mKComfwbd}) and the assumptions of Lemma~\ref{lemma:taurelvaryingbxEi} are satisfied, let $1\leq l\in\zo$,
  $(\cweight,\cweight)=\weight\in \Weight$ and $\weight_{0}=(0,\cweight)$. Then there is a constant $C_{a}$ such that, for $A\neq B$,
  \begin{equation}\label{eq:WCBmHlestEi}
    \|\mW^{A}_{B}\|_{\mH^{l}_{\bbE,\weight}(\bM)} \leq C_{a}(\|\hml_{U}\mK\|_{C^{0}_{\weight}(\bM)}\|\mK\|_{H^{l}_{\weight_{0}}(\bM)}+\|\hml_{U}\mK\|_{H^{l}_{\weight}(\bM)})
  \end{equation}
  on $I_{-}$, where $C_{a}$ only depends on $C_{\mK}$, $\e_{\rond}$, $\bDlnhNsup$, $\cweight$, $n$, $(\bM,\bge_{\refer})$ and an upper
  bound on $l$. Moreover, 
  \begin{equation}\label{eq:WAAmHlestEi}
    \begin{split}
      \|\mW^{A}_{A}\|_{\mH^{l}_{\bbE,\weight}(\bM)} \leq & C_{a}(\|\hml_{U}\mK\|_{C^{0}_{\weight}(\bM)}\|\mK\|_{H^{l}_{\weight_{0}}(\bM)}+\|\hml_{U}\mK\|_{H^{l}_{\weight}(\bM)})\\
      & +C_{b}e^{M_{\min}}e^{\eSpe\tau}\left(\|\mK\|_{H^{\bfl}_{\weight_{0}}(\bM)}+\theta_{0,-}^{-1}\|\chi\|_{H^{\bfl_{1},\weight_{0}}_{\rohy}(\bM)}
      +\|\ln\hN\|_{H^{\bfl}_{\weight_{0}}(\bM)}\right)
    \end{split}
  \end{equation}
  on $I_{-}$ (no summation on $A$), where $\bfl:=(1,l)$, $\bfl_{1}:=(1,l+1)$ and $C_{b}$ only depends on $\mKsup$, $\bDlnhNsup$, $K_{\cweight}$,
  $c_{\chi,2}$, $\cweight$, $\e_{\rond}$, $n$, $l$ and $(\bM,\bge_{\refer})$.
\end{lemma}
\begin{remark}
  The estimate (\ref{eq:WCBmHlestEi}) holds without assuming (\ref{eq:mKComfwbd}) to hold. Moreover an estimate similar to (\ref{eq:WAAmHlestEi})
  holds without assuming (\ref{eq:mKComfwbd}) to hold. However, we then have to replace $\bfl$ by $l$ in the norm of $\mK$ in the second term on the
  right hand side of (\ref{eq:WAAmHlestEi}). Moreover, the constants are then independent of $K_{\cweight}$. 
\end{remark}
\begin{proof}
  The estimate (\ref{eq:WCBmHlestEi}) follows by applying Proposition~\ref{prop:moserestwightnoframe} to (\ref{eq:XbfBWCBestCkEi}).

  Next, let us turn to $\mW^{A}_{A}$ (no summation). Consider (\ref{eq:finalproducttobeestimatedCkstmtEi}). The first term on the right hand side gives
  rise to the first term on the right hand side of (\ref{eq:WAAmHlestEi}). The argument to prove this is identical to the proof of (\ref{eq:WCBmHlestEi}).
  Turning to the second term on the right hand side of (\ref{eq:finalproducttobeestimatedCkstmtEi}), we, up to constants depending only on $n$,
  $\e_{\rond}$, $\mKsup$, $(\bM,\bge_{\refer})$ and $l$, need to estimate expressions of the form
  \[
  \ldr{\varrho}^{-(l+1)\cweight}\textstyle{\prod}_{i=1}^{j}|\bD^{m_{i}+1}\mK|_{\bge_{\refer}}
  \textstyle{\prod}_{k=1}^{p}|\bD^{l_{k}+1}\ln\hN|_{\bge_{\refer}}
  \cdot\hN^{-1}|\bD_{\bfJ}\bD_{\bfK}\chi|_{\bge_{\refer}}
  \]
  in $L^{2}$, where the sum of the $m_{i}$, the $l_{i}$, $j$, $p$ and $|\bfJ|$ is less than or equal to $l$; and $|\bfK|=1$. At this stage, we can
  appeal to (\ref{eq:DeltavarrhorelvariationEi}) and (\ref{eq:hNtaudotequivEi}) in order to exchange $\varrho$ with $\tau$ and $\hN$ with
  $\d_{t}\tau$. Appealing to Corollary~\ref{cor:mixedmoserestweight} with appropriate choices of weights etc., as well as (\ref{eq:bDlnNbDlnthetabd})
  and (\ref{eq:mKComfwbd}), it is thus clear that it is sufficient to estimate 
  \begin{equation*}
    \begin{split}
      & C(\|\mK\|_{H^{\bfl}_{\weight_{0}}(\bM)}+\|\ln\hN\|_{H^{\bfl}_{\weight_{0}}(\bM)})\|\chi\|_{\mc^{\bfl_{0},\weight_{0}}_{\bbE,\rocon}(\bM)}
      +C\|\chi\|_{\mH^{\bfl_{1},\weight_{0}}_{\bbE,\rocon}(\bM)},
    \end{split}
  \end{equation*}
  where $\bfl_{0}=(1,1)$, $\bfl_{1}=(1,l+1)$ and $C$ only depends on $\bDlnhNsup$, $\mKsup$, $K_{\cweight}$, $\cweight$, $\e_{\rond}$,
  $n$, $l$ and $(\bM,\bge_{\refer})$. We also obtain a similar estimate with $\bfl$ replaced by $l$ in the norm of $\mK$. The constant in the
  corresponding estimate is independent of $K_{\cweight}$, and the estimate holds without demanding that (\ref{eq:mKComfwbd}) hold. Next,
  \begin{align*}    
    \|\chi\|_{\mc^{\bfl_{0},\weight_{0}}_{\bbE,\rocon}(\bM)}
    \leq Ce^{M_{\min}}e^{\vare_{\Spe}\tau}\theta_{0,-}^{-1}\|\chi\|_{C^{\bfl_{0},\weight_{0}}_{\rohy}(\bM)}
    \leq & Ce^{M_{\min}}e^{\vare_{\Spe}\tau}c_{\chi,2}
  \end{align*}
  where $C$ only depends $n$, $\cweight$ and $(\bM,\bge_{\refer})$ and we appealed to (\ref{eq:chimClClest}) and the assumptions. Finally,
  \begin{equation*}
    \begin{split}
      \|\chi\|_{\mH^{\bfl_{1},\weight_{0}}_{\bbE,\rocon}(\bM)}\leq Ce^{M_{\min}}e^{\vare_{\Spe}\tau}\theta_{0,-}^{-1}\|\chi(\cdot,t)\|_{H^{\bfl_{1},\weight_{0}}_{\rohy}(\bM)},
    \end{split}
  \end{equation*}
  where $C$ only depends on $n$, $l$, $\cweight$ and $(\bM,\bge_{\refer})$, and we appealed to (\ref{eq:chimHlHlest}). Combining the above estimates
  yields the conclusion of the lemma. 
\end{proof}

\section{Estimating $A_{i}^{k}$ and $\hU(A_{i}^{k})$}

Returning to Section~\ref{section:contributionfromshift}, we next wish to estimate $A_{i}^{k}$ and $\hU(A_{i}^{k})$.

\begin{lemma}\label{lemma:Aiksupest}
  Given that the assumptions of Lemma~\ref{lemma:taurelvaryingbxEi} hold, let $\tau$ be defined by (\ref{eq:taudefinitionEi}). Let
  $(\cweight,\cweight)=\weight\in\Weight$ and $(0,\cweight)=\weight_{0}$. Then 
  \begin{equation}\label{eq:AikCzest}
    \|A_{i}^{k}(\cdot,t)\|_{C^{0}_{\weight}(\bM)}\leq Ce^{M_{\min}}e^{\eSpe\tau(t)}\theta_{0,-}^{-1}\|\chi(\cdot,t)\|_{C^{1,\weight_{0}}_{\rohy}(\bM)}
  \end{equation}
  for $t\in I_{-}$, where $C$ only depends on $n$, $\cweight$ and $(\bM,\bge_{\refer})$; $M_{\min}$ is defined in the text adjacent to
  (\ref{eq:muminmainlowerbound}); and $\eSpe$ is defined in the text adjacent to (\ref{eq:eSpevarrhoeelowtaurelEi}). Let $1\leq l\in\zo$
  and assume, in addition to the above, that
  \begin{equation}\label{eq:bDlnfckmoest}
    \|\ln\hN\|_{C^{\bfl}_{\weight_{0}}(\bM)}\leq C_{\rorel,\bfl}
  \end{equation}
  on $I_{-}$, where $\bfl=(1,l)$. Then
  \[
  \|A_{i}^{k}(\cdot,t)\|_{C^{l}_{\weight}(\bM)}\leq Ce^{M_{\min}}e^{\eSpe\tau(t)}\theta_{0,-}^{-1}\|\chi(\cdot,t)\|_{C^{l+1,\weight_{0}}_{\rohy}(\bM)}
  \]
  for $t\in I_{-}$, where $C$ only depends on $C_{\rorel,\bfl}$, $l$, $n$, $\cweight$, and $(\bM,\bge_{\refer})$.
\end{lemma}
\begin{remark}
  Given that the conditions of Lemma~\ref{lemma:lowerbdonmumin} are fulfilled, an argument similar to the proof, combined with
  Remark~\ref{remark:chiclvarrhodecay}, yields
  \begin{equation}\label{eq:AikCzestvarrho}
    \ldr{\varrho}^{-\cweight}|A_{i}^{k}|\leq Ce^{M_{\min}}e^{\e_{\Spe}\varrho}\theta_{0,-}^{-1}\|\chi(\cdot,t)\|_{C^{1,\weight_{0}}_{\rohy}(\bM)}
  \end{equation}
  on $M_{-}$, where $C$ only depends on $n$, $\cweight$ and $(\bM,\bge_{\refer})$; and $M_{\min}$ is defined in the text adjacent to
  (\ref{eq:muminmainlowerbound}). Assume, in addition, that the estimate (\ref{eq:bDlnfckmoest}) holds. Then an argument similar to the proof,
  combined with Remark~\ref{remark:chiclvarrhodecay}, yields
  \begin{equation}\label{eq:EbfIAikvarrhoestimate}
  \ldr{\varrho}^{-(|\bfI|+1)\cweight}|E_{\bfI}A_{i}^{k}|
  \leq Ce^{M_{\min}}e^{\e_{\Spe}\varrho}\theta_{0,-}^{-1}\|\chi(\cdot,t)\|_{C^{l+1,\weight_{0}}_{\rohy}(\bM)}
  \end{equation}
  on $M_{-}$ for all $|\bfI|\leq l$, where $C$ only depends on $C_{\rorel,\bfl}$, $l$, $n$, $\cweight$ and $(\bM,\bge_{\refer})$.
\end{remark}
\begin{proof}
  Combining the end of the proof of Lemma~\ref{lemma:compositeestEi} with (\ref{eq:Aialphadef}) yields
  \[
  |E_{\bfI}(A_{i}^{k})|\leq C\textstyle{\sum}_{l_{a}+|\bfJ|\leq |\bfI|+1,l_{a}\leq |\bfI|}\mfP_{N,l_{a}}\hN^{-1}|\bD_{\bfJ}\chi|_{\bge_{\refer}},
  \]
  where $C$ only depends on $n$, $|\bfI|$ and $(\bM,\bge_{\refer})$. In particular,
  \[
  \|A_{i}^{k}(\cdot,t)\|_{C^{0}_{\weight}(\bM)}\leq C\|\chi\|_{C^{1,\weight_{0}}_{\bbE,\rocon}}
  \leq Ce^{M_{\min}}e^{\eSpe\tau(t)}\theta_{0,-}^{-1}\|\chi(\cdot,t)\|_{C^{1,\weight_{0}}_{\rohy}(\bM)},
  \]
  where we appealed to Lemmas~\ref{lemma:basequiv} and \ref{lemma:chimclbbeWClhy}. This yields (\ref{eq:AikCzest}). Assuming, in addition, the stated
  bound on $\ln\hN$,
  \[
  \|A_{i}^{k}(\cdot,t)\|_{C^{l}_{\weight}(\bM)}\leq C\|\chi\|_{C^{l+1,\weight_{0}}_{\bbE,\rocon}}
  \leq Ce^{M_{\min}}e^{\eSpe\tau(t)}\theta_{0,-}^{-1}\|\chi(\cdot,t)\|_{C^{l+1,\weight_{0}}_{\rohy}(\bM)},
  \]
  where we appealed to Lemmas~\ref{lemma:basequiv} and \ref{lemma:chimclbbeWClhy}. The lemma follows. 
\end{proof}

\begin{lemma}\label{lemma:hUAiksupest}
  Given that the assumptions of Lemma~\ref{lemma:taurelvaryingbxEi} hold, let $\tau$ be defined by (\ref{eq:taudefinitionEi}). Let
  $(\cweight,\cweight)=\weight\in\Weight$, $\weight_{1}:=(2\cweight,\cweight)$ and $(0,\cweight)=\weight_{0}$. Let $0\leq l\in\zo$
  and assume, in addition, that the estimate (\ref{eq:bDlnfckmoest}) holds with $\bfl$ replaced by $\bfl_{1}:=(1,l+1)$. Then
  \begin{equation*}
    \begin{split}
      & \|\hU(A_{i}^{k})(\cdot,t)\|_{C^{l}_{\weight_{1}}(\bM)}\\
      \leq & Ce^{M_{\min}}e^{\eSpe\tau(t)}\theta_{0,-}^{-1}\|\dotchi(\cdot,t)\|_{C^{l+1,\weight}_{\rohy}(\bM)}\\
      & +Ce^{M_{\min}}e^{\eSpe\tau(t)}\textstyle{\sum}_{l_{a}+l_{b}\leq l}\|\hU(\ln\hN)(\cdot,t)\|_{C^{l_{a}}_{\weight}(\bM)}
      \theta_{0,-}^{-1}\|\chi(\cdot,t)\|_{C^{l_{b}+1,\weight_{0}}_{\rohy}(\bM)}\\
      & +Ce^{2M_{\min}}e^{2\eSpe\tau(t)}\textstyle{\sum}_{l_{a}+l_{b}\leq l+2;l_{a}\leq l}
      \theta_{0,-}^{-2}\|\chi(\cdot,t)\|_{C^{l_{a},\weight_{0}}_{\rohy}(\bM)}\|\chi(\cdot,t)\|_{C^{l_{b},\weight_{0}}_{\rohy}(\bM)}
    \end{split}
  \end{equation*}
  for $t\in I_{-}$, where $C$ only depends on $C_{\rorel,\bfl_{1}}$, $l$, $n$, $\cweight$ and $(\bM,\bge_{\refer})$.
\end{lemma}
\begin{remark}
  Given that the conditions of Lemma~\ref{lemma:lowerbdonmumin} are fulfilled, let $\weight$, $\weight_{1}$, $\weight_{0}$ be as in the statement
  of the lemma. Let $0\leq l\in\zo$ and assume that the estimate (\ref{eq:bDlnfckmoest}) holds with $\bfl$ replaced by $\bfl_{1}:=(1,l+1)$. Then an
  argument similar to the proof, combined with Remark~\ref{remark:chiclvarrhodecay}, yields
  \begin{equation}\label{eq:hUAikweightedvarrhoestimate}
    \begin{split}
      & \ldr{\varrho}^{-(l+2)\cweight}|E_{\bfI}\hU(A_{i}^{k})|\\
      \leq & Ce^{M_{\min}}e^{\e_{\Spe}\varrho}\theta_{0,-}^{-1}\|\dotchi(\cdot,t)\|_{C^{l+1,\weight}_{\rohy}(\bM)}\\
      & +Ce^{M_{\min}}e^{\e_{\Spe}\varrho}\textstyle{\sum}_{l_{a}+l_{b}\leq l}\|\hU(\ln\hN)(\cdot,t)\|_{C^{l_{a}}_{\weight}(\bM)}
      \theta_{0,-}^{-1}\|\chi(\cdot,t)\|_{C^{l_{b}+1,\weight_{0}}_{\rohy}(\bM)}\\
      & +Ce^{2M_{\min}}e^{2\e_{\Spe}\varrho}\textstyle{\sum}_{l_{a}+l_{b}\leq l+2;l_{a}\leq l}
      \theta_{0,-}^{-2}\|\chi(\cdot,t)\|_{C^{l_{a},\weight_{0}}_{\rohy}(\bM)}\|\chi(\cdot,t)\|_{C^{l_{b},\weight_{0}}_{\rohy}(\bM)}
    \end{split}
  \end{equation}
  on $M_{-}$ for $|\bfI|\leq l$, where $C$ only depends on $C_{\rorel,\bfl_{1}}$, $l$, $n$, $\cweight$ and $(\bM,\bge_{\refer})$.  
\end{remark}
\begin{proof}
  The statement is an immediate consequence of (\ref{eq:EbfIhUAik}) and arguments similar to the proof of the previous lemma.
\end{proof}

We also need to estimate $A_{i}^{k}$ and $\hU(A_{i}^{k})$ with respect to weighted Sobolev norms.

\begin{lemma}\label{lemma:AikmHlestEi}
  Given that the assumptions of Lemma~\ref{lemma:taurelvaryingbxEi} are satisfied, let $1\leq l\in\zo$, $(\cweight,\cweight)=\weight\in \Weight$
  and $\weight_{0}=(0,\cweight)$. Then
  \begin{equation}\label{eq:AikmHlestEi}
    \begin{split}
      \|A^{k}_{i}(\cdot,t)\|_{\mH^{l}_{\bbE,\weight}(\bM)} \leq & C_{a}e^{M_{\min}}e^{\eSpe\tau(t)}\left(\theta_{0,-}^{-1}\|\chi(\cdot,t)\|_{H^{l+1,\weight_{0}}_{\rohy}(\bM)}
      +\|\ln\hN(\cdot,t)\|_{H^{\bfl}_{\weight_{0}}(\bM)}\right)
    \end{split}
  \end{equation}
  on $I_{-}$, where $\bfl:=(1,l)$ and $C_{a}$ only depends on $\bDlnhNsup$, $c_{\chi,2}$, $\cweight$, $n$, $l$ and $(\bM,\bge_{\refer})$.  
\end{lemma}
\begin{proof}
  Due to Lemma~\ref{lemma:compositeestEi} and its proof, it is clear that when applying $\bD_{\bfI}$ to $A_{i}^{k}$, the resulting
  expression can be estimated by
  \begin{equation*}
    \begin{split}
      &C\textstyle{\sum}_{l_{a}+|\bfJ|\leq l}\mfP_{N,l_{a}}\hN^{-1}(|\bD_{\bfJ}\bD_{\bfK}\chi|_{\bge_{\refer}}+|\bD_{\bfJ}\chi|_{\bge_{\refer}}),
    \end{split}
  \end{equation*}
  where $|\bfK|=1$, $l:=|\bfI|$ and $C$ only depends on $l$, $(\bM,\bge_{\refer})$ and $n$. In order to estimate this expression in the
  appropriate weighted $L^{2}$-spaces, we can proceed as in the proof of Lemma~\ref{lemma:mWmHlestEi}. The lemma follows. 
\end{proof}

Finally, we have the following estimate.

\begin{lemma}\label{lemma:hUAikmHlestEi}
  Given that the assumptions of Lemma~\ref{lemma:taurelvaryingbxEi} are satisfied, let $1\leq l\in\zo$, $(\cweight,\cweight)=\weight\in \Weight$,
  $\weight_{0}:=(0,\cweight)$ and $\weight_{1}:=(2\cweight,\cweight)$. Assume that there is a constant $C_{\chi}$ such that 
  \[
  \theta_{0,-}^{-1}\|\chi(\cdot,t)\|_{C^{1,\weight_{0}}_{\rohy}(\bM)}+\theta_{0,-}^{-1}\|\dotchi(\cdot,t)\|_{C^{0,\weight}_{\rohy}(\bM)}\leq C_{\chi}
  \]
  on $I_{-}$. Then
  \begin{equation}\label{eq:hUAikmHlestEi}
    \begin{split}
      & \|\hU(A^{k}_{i})\|_{\mH^{l}_{\bbE,\weight_{1}}(\bM)}\\
      \leq & C_{a}e^{M_{\min}}e^{\eSpe\tau}\left(\theta_{0,-}^{-1}\|\dotchi\|_{H^{l+1,\weight}_{\rohy}(\bM)}
      +\|\ln\hN\|_{H^{\bfl_{1}}_{\weight_{0}}(\bM)}\right)\\
      & +C_{a}e^{M_{\min}}e^{\eSpe\tau}\|\hU(\ln\hN)\|_{C^{0}_{\weight}(\bM)}\left(\theta_{0,-}^{-1}\|\chi\|_{H^{l+1,\weight_{0}}_{\rohy}(\bM)}
      +\|\ln\hN\|_{H^{\bfl}_{\weight_{0}}(\bM)}\right)\\
      &+C_{a}e^{M_{\min}}e^{\eSpe\tau}\|\hU(\ln\hN)\|_{H^{l}_{\weight}(\bM)}\\
      &+C_{a}e^{2M_{\min}}e^{2\eSpe\tau}\left(\theta_{0,-}^{-1}\|\chi\|_{H^{l+2,\weight_{0}}_{\rohy}(\bM)}
      +\|\ln\hN\|_{H^{\bfl_{1}}_{\weight_{0}}(\bM)}\right)
    \end{split}
  \end{equation}
  on $I_{-}$, where $\bfl:=(1,l)$, $\bfl_{1}:=(1,l+1)$ and $C_{a}$ only depends on $\bDlnhNsup$, $C_{\chi}$, $\cweight$, $n$, $l$ and $(\bM,\bge_{\refer})$.
\end{lemma}
\begin{proof}
  Consider (\ref{eq:EbfIhUAik}). We need to estimate weighted versions of the terms on the right hand side in $L^{2}$. Due to an argument similar
  to the proof of Lemma~\ref{lemma:mWmHlestEi}, we conclude that the first term on the right hand side of (\ref{eq:EbfIhUAik}) gives rise to
  expressions that can be estimated by the first term on the right hand side of (\ref{eq:hUAikmHlestEi}). By a similar argument, the second term
  on the right hand side of (\ref{eq:EbfIhUAik}) gives rise to expressions that can be estimated by the sum of the second and third terms on the
  right hand side of (\ref{eq:hUAikmHlestEi}). Finally, the last term on the right hand side of (\ref{eq:EbfIhUAik}) gives rise to expressions that
  can be estimated by the last term on the right hand side of (\ref{eq:hUAikmHlestEi}). 
\end{proof}

\section{Consequences of the higher order Sobolev assumptions}

Given that the higher order Sobolev assumptions hold, cf. Definition~\ref{def:sobklassumptions}, we obtain the following conclusions. 

\begin{lemma}\label{lemma:mWAhUASobestimates}
  Fix $l$, $\bfl_{0}$, $\bfl$, $\bfl_{1}$, $\cweight$, $\weight_{0}$ and $\weight$ as in Definition~\ref{def:sobklassumptions}.
  Let $\weight_{1}:=(2\cweight,\cweight)$. Then, given that the assumptions of Lemma~\ref{lemma:taurelvaryingbxEi} as well as the
  $(\cweight,l)$-Sobolev assumptions are satisfied,
  \begin{align}
    \|\mW^{A}_{B}(\cdot,t)\|_{H^{l+1}_{\weight}(\bM)} \leq & C_{a},\label{eq:mWABmHlmfwbdbconstant}\\
    \|A^{k}_{i}(\cdot,t)\|_{H^{l+1}_{\weight}(\bM)} \leq & C_{a}e^{\eSpe\tau(t)},\label{eq:AikmHlestsobassumpEi}\\
    \|\hU(A^{k}_{i})(\cdot,t)\|_{H^{l-1}_{\weight_{1}}(\bM)} \leq & C_{a}e^{\eSpe\tau(t)}\label{eq:hUAikSoblestulsobass}
  \end{align}
  for all $t\in I_{-}$, all $A,B$ and all $i,k$, where $C_{a}$ only depends on $s_{\cweight,l}$ and $(\bM,\bge_{\refer})$. Moreover,
  \begin{align}
    \|\mW^{A}_{B}(\cdot,t)\|_{C^{0}_{\weight}(\bM)} \leq & C_{a},\label{eq:mWABmHCzeromfwbdbconstant}\\
    \|A^{k}_{i}(\cdot,t)\|_{C^{0}_{\weight}(\bM)} \leq & C_{a}e^{\eSpe\tau(t)},\label{eq:AikmCzeroestsobassumpEi}\\
    \|\hU(A^{k}_{i})(\cdot,t)\|_{C^{0}_{\weight_{1}}(\bM)} \leq & C_{a}e^{\eSpe\tau(t)}\label{eq:hUAikCzeroestulsobass}
  \end{align}
  for all $t\in I_{-}$, all $A,B$ and all $i,k$, where $C_{a}$ only depends on $s_{\cweight,l}$ and $(\bM,\bge_{\refer})$.  
\end{lemma}
\begin{proof}
  The estimate (\ref{eq:mWABmHlmfwbdbconstant}) follows immediately from (\ref{eq:HlmHlequiv}), (\ref{eq:WCBmHlestEi}), (\ref{eq:WAAmHlestEi})
  and the assumptions. The estimate (\ref{eq:AikmHlestsobassumpEi}) follows immediately from (\ref{eq:HlmHlequiv}), (\ref{eq:AikmHlestEi})
  and the assumptions. Moreover, the estimate (\ref{eq:hUAikSoblestulsobass}) follows immediately from (\ref{eq:HlmHlequiv}), (\ref{eq:hUAikmHlestEi})
  and the assumptions. Finally, (\ref{eq:mWABmHCzeromfwbdbconstant}) follows from (\ref{eq:Clmclequiv}), (\ref{eq:chimClClest}),
  (\ref{eq:WCneqBmckestEi}), (\ref{eq:mWAAmckestEi}) and the assumptions; (\ref{eq:AikmCzeroestsobassumpEi}) follows from (\ref{eq:AikCzest}) and
  the assumptions; and (\ref{eq:hUAikCzeroestulsobass}) follows from Lemma~\ref{lemma:hUAiksupest} and the assumptions. 
\end{proof}

\section{Consequences of the higher order $C^{k}$-assumptions}

The following consequences of the higher order $C^{k}$-assumptions will be of interest in what follows.

\begin{lemma}\label{lemma:mWAhUAClestimates}
  Fix $l$, $\cweight$ and $\weight$ as in Definition~\ref{def:supmfulassumptions} and let $\weight_{1}:=(2\cweight,\cweight)$. Then, given that the
  assumptions of Lemma~\ref{lemma:taurelvaryingbxEi} as well as the $(\cweight,l)$-supremum assumptions are satisfied,
  \begin{align}
    \|\mW^{A}_{B}(\cdot,t)\|_{C^{l+1}_{\weight}(\bM)} \leq & C_{a},\label{eq:mWABClmfwbdbconstant}\\
    \|A^{k}_{i}(\cdot,t)\|_{C^{l+1}_{\weight}(\bM)} \leq & C_{a}e^{\eSpe\tau(t)},\label{eq:AikClestsobassumpEi}\\
    \|\hU(A^{k}_{i})(\cdot,t)\|_{C^{l-1}_{\weight_{1}}(\bM)} \leq & C_{a}e^{\eSpe\tau(t)}\label{eq:hUAikClestulsobass}
  \end{align}
  for all $t\in I_{-}$, all $A,B$ and all $i,k$, where $C_{a}$ only depends on $c_{\cweight,l}$ and $(\bM,\bge_{\refer})$. Here $l$ is required
  to satisfy $l\geq 1$ in order for the last estimate to hold. 
\end{lemma}
\begin{remark}\label{remark:mWAhUAClestimates}
  In certain situations, it is of interest to keep in mind that the estimates
  (\ref{eq:AikClestsobassumpEi}) and (\ref{eq:hUAikClestulsobass}) can be improved to
  \begin{align}
    \ldr{\varrho}^{-(|\bfI|+1)\cweight}|E_{\bfI}A_{i}^{k}|\leq & C_{a}e^{\e_{\Spe}\varrho},\label{eq:AikClestsobassumpEivarrho}\\
    \ldr{\varrho}^{-(|\bfJ|+2)\cweight}|E_{\bfJ}\hU(A_{i}^{k})|\leq & C_{a}e^{\e_{\Spe}\varrho}\label{eq:hUAikClestulsobassvarrho}
  \end{align}
  on $M_{-}$, for all $i,k$ and all $|\bfI|\leq l+1$ and $|\bfJ|\leq l-1$, where $C_{a}$ only depends on $c_{\cweight,l}$ and $(\bM,\bge_{\refer})$. Here
  (\ref{eq:AikClestsobassumpEivarrho}) follows from (\ref{eq:EbfIAikvarrhoestimate}) and the assumptions. Moreover,
  (\ref{eq:hUAikClestulsobassvarrho}) follows from (\ref{eq:hUAikweightedvarrhoestimate}) and the assumptions. 
\end{remark}
\begin{proof}
  The estimate (\ref{eq:mWABClmfwbdbconstant}) is an immediate consequence of (\ref{eq:Clmclequiv}), (\ref{eq:chimClClest}), 
  (\ref{eq:WCneqBmckestEi}), (\ref{eq:mWAAmckestEi}) and the assumptions. The estimate (\ref{eq:AikClestsobassumpEi}) is an immediate 
  consequence of Lemma~\ref{lemma:Aiksupest} and the assumptions. Finally, estimate (\ref{eq:hUAikClestulsobass}) is an immediate 
  consequence of Lemma~\ref{lemma:hUAiksupest} and the assumptions. 
\end{proof}

\chapter{Estimates of the components of the metric}\label{chapter:estimatingthecomponentsofmetric}

When deriving energy estimates, we need to control weighted Sobolev and $C^{k}$-norms of $\mu_{A}$. Due to the assumptions concerning $\theta$,
it is sufficient to derive such estimates for $\bmu_{A}$. This is the main purpose of the present chapter. We begin, in
Section~\ref{section:sobestofbmuAEi}, by deriving expressions for $\hU[E_{\bfI}(\bmu_{A})]$. Combining these expressions with the assumptions; energy
type estimates; the previously derived Moser estimates; and the weighted Sobolev estimates for $A_{i}^{k}$, we obtain weighted Sobolev estimates for
$\bmu_{A}$ in Section~\ref{section:energyestimatesbmuA}. In order to obtain weighted $C^{k}$-estimates, we carry out energy estimates for
$E_{\bfI}(\bmu_{A})$ along integral curves of $\hU$. We end the chapter by deriving weighted $C^{k}$-estimates for $\varrho$. 

\section{An equation for higher order derivatives of $\bmu_{A}$}\label{section:sobestofbmuAEi}

Our next goal is to derive $L^{2}$-based energy estimates for $\bmu_{A}$. As a preliminary step, it is of interest to commute the equation
(\ref{eq:hUbmuAform}) with $E_{\bfI}$. Note, to this end, that (\ref{eq:hUEicomm}) and (\ref{eq:Aialphadef}) hold. Combining (\ref{eq:hUbmuAform})
with (\ref{eq:hUEicomm}) yields
\begin{equation}\label{eq:hUEibmuAbasformula}
  \begin{split}
    \hU[E_{i}(\bmu_{A})] = & A_{i}^{k}E_{k}(\bmu_{A})+E_{i}(\ell_{A}+\mW^{A}_{A})+A_{i}^{0}(\ell_{A}+\mW^{A}_{A}). 
  \end{split}
\end{equation}

\begin{lemma}\label{lemma:hUXbfAbmuAtermsEi}
  Let $(M,g)$ be a time oriented Lorentz manifold. Assume that it has an expanding partial pointed foliation. Assume, moreover, $\mK$ to be
  non-degenerate on $I$ and to have a global frame. Let $\bfI$ be a vector field multiindex. Then $\hU[E_{\bfI}(\bmu_{A})]$ is a linear combination of
  terms of the form
  \begin{equation}\label{eq:typeonetermEi}
    E_{\bfI_{1}}(\ln\hN)\cdots E_{\bfI_{k}}(\ln\hN)E_{\bfJ}(A_{i}^{j})E_{\bfK}(\bmu_{A}),
  \end{equation}
  where $|\bfI_{1}|+\dots+|\bfI_{k}|+|\bfJ|+|\bfK|=|\bfI|$, $|\bfI_{i}|\neq 0$; and terms of the form
  \begin{equation}\label{eq:typetwotermEi}
    E_{\bfI_{1}}(\ln\hN)\cdots E_{\bfI_{k}}(\ln\hN)E_{\bfJ}(\ell_{A}+\mW^{A}_{A}),
  \end{equation}
  where $|\bfI_{1}|+\dots+|\bfI_{k}|+|\bfJ|=|\bfI|$, $|\bfI_{i}|\neq 0$.
\end{lemma}
\begin{remark}
  In case $k=0$, there are no terms of the form $E_{\bfI_{i}}(\ln\hN)$ in the expressions (\ref{eq:typeonetermEi}) and (\ref{eq:typetwotermEi}). 
\end{remark}
\begin{proof}
  Due to (\ref{eq:hUEibmuAbasformula}), the statement holds for $|\bfI|=1$. Let us therefore assume that it holds for all $|\bfI|\leq l$ and some
  $1\leq l\in\zo$. Given such an $\bfI$, compute
  \[
  \hU[E_{m}E_{\bfI}(\bmu_{A})]=A_{m}^{0}\hU[E_{\bfI}(\bmu_{A})]+A_{m}^{k}E_{k}E_{\bfI}(\bmu_{A})+E_{m}\hU[E_{\bfI}(\bmu_{A})],
  \]
  where we appealed to (\ref{eq:hUEicomm}). Combining this equality with the inductive assumption yields the conclusion of the lemma. 
\end{proof}

\section{Energy estimates}\label{section:energyestimatesbmuA}

In the present section, we use Lemma~\ref{lemma:hUXbfAbmuAtermsEi} to derive weighted Sobolev estimates of $\bmu_{A}$. Let $1\leq l\in\zo$, 
$(\weight_{a},\weight_{b})=\weight\in\Weight$ and consider the following energy:
\[
\me_{\bmu,\weight,l}(\tau):=\frac{1}{2}\int_{\bM}\textstyle{\sum}_{A}\textstyle{\sum}_{|\bfI|\leq l+1}
\ldr{\tau}^{-2\weight_{a}-2|\bfI|\weight_{b}}|(E_{\bfI}\bmu_{A})(\cdot,t(\tau))|^{2}\mu_{\bge_{\refer}}.
\]
In what follows, we also use the notation $\me_{\bmu,\weight}:=\me_{\bmu,\weight,0}$. 

\begin{lemma}\label{lemma:energyestimatesbmuA}
  Fix $l$, $\bfl_{0}$, $\bfl$, $\bfl_{1}$, $\cweight$, $\weight_{0}$ and $\weight$ as in Definition~\ref{def:sobklassumptions}.
  Given that the the assumptions of Lemma~\ref{lemma:taurelvaryingbxEi} as well as the $(\cweight,l)$-Sobolev assumptions are satisfied, there 
  is a constant $C_{\bmu,l}$ such that
  \begin{equation}\label{eq:bmuAmHlestimate}
    \|\bmu_{A}(\cdot,\tau)\|_{\mH^{l+1}_{\bbE,\weight}(\bM)}\leq C_{\bmu,l}\ldr{\tau}
  \end{equation}
  on $I_{-}$ for all $A$, where $C_{\bmu,l}$ only depends on $s_{\cweight,l}$ and $(\bM,\bge_{\refer})$.
\end{lemma}
\begin{remark}\label{remark:muAmhlestimate}
  Combining (\ref{eq:bmuAmHlestimate}) with the assumptions and the fact that $\mu_{A}=\bmu_{A}+\ln\theta$ yields the conclusion that 
  \begin{equation}\label{eq:muAmHlestimatermk}
    \|\mu_{A}(\cdot,\tau)\|_{\mH^{\bfl_{1}}_{\bbE,\weight}(\bM)}\leq C_{\mu,l}\ldr{\tau}
  \end{equation}
  on $I_{-}$ for all $A$, where $C_{\mu,l}$ only depends on $s_{\cweight,l}$ and $(\bM,\bge_{\refer})$. 
\end{remark}
\begin{proof}
  Let $\weight_{a}=\weight_{b}=\cweight$, and estimate
  \begin{equation}\label{eq:dtaumemuk}
    \begin{split}
      \d_{\tau}\me_{\bmu,\weight,l} \geq & \int_{\bM}\textstyle{\sum}_{A}\textstyle{\sum}_{|\bfI|\leq l+1}
      \ldr{\tau}^{-2\weight_{a}-2|\bfI|\weight_{b}}E_{\bfI}\bmu_{A}\cdot\d_{\tau}(E_{\bfI}\bmu_{A})\mu_{\bge_{\refer}}
    \end{split}
  \end{equation}
  for all $\tau\leq 0$. In order to estimate the right hand side, note that
  \begin{equation}\label{eq:dtauXbfAbmuA}
    \d_{\tau}(E_{\bfI}\bmu_{A})=\frac{\hN}{\dot{\tau}}\hU(E_{\bfI}\bmu_{A})+\frac{1}{\dot{\tau}}\chi(E_{\bfI}\bmu_{A}),
  \end{equation}
  where we appealed to (\ref{eq:futurenormalNchiexpr}). Combining this observation with (\ref{eq:dtaumemuk}), we need to estimate
  \[
  \int_{\bM}E_{\bfI}\bmu_{A}\frac{1}{\dot{\tau}}\chi(E_{\bfI}\bmu_{A})\mu_{\bge_{\refer}}
  =\frac{1}{2\dot{\tau}}\int_{\bM}\chi(|E_{\bfI}\bmu_{A}|^{2})\mu_{\bge_{\refer}}
  =-\frac{1}{2\dot{\tau}}\int_{\bM}|E_{\bfI}\bmu_{A}|^{2}(\rodiv_{\bge_{\refer}}\chi)\mu_{\bge_{\refer}}.
  \]
  In particular,
  \begin{equation*}
    \begin{split}
      \left|\int_{\bM}E_{\bfI}\bmu_{A}\frac{1}{\dot{\tau}}\chi(E_{\bfI}\bmu_{A})\mu_{\bge_{\refer}}\right|
      \leq & K_{\rovar}\int_{\bM}|E_{\bfI}\bmu_{A}|^{2}\hN^{-1}|\rodiv_{\bge_{\refer}}\chi|\mu_{\bge_{\refer}}\\
      \leq & K_{\rovar}\int_{\bM}|E_{\bfI}\bmu_{A}|^{2}e^{\eSpe\tau}\mu_{\bge_{\refer}}
    \end{split}
  \end{equation*}
  where we appealed to (\ref{eq:rodivchiestimpr}), (\ref{eq:eSpevarrhoeelowtaurelEi}) and (\ref{eq:hNtaudotequivEi}). Combining this observation with
  (\ref{eq:dtaumemuk}) and (\ref{eq:dtauXbfAbmuA}) yields
  \begin{equation}\label{eq:memukderest}
    \begin{split}
      \d_{\tau}\me_{\bmu,\weight,l}\geq & -2K_{\rovar}e^{\eSpe\tau}\me_{\bmu,\weight,l}\\
      & -2K_{\rovar}\int_{\bM}\textstyle{\sum}_{A}\textstyle{\sum}_{|\bfI|\leq l+1}\ldr{\tau}^{-2\weight_{a}-2|\bfI|\weight_{b}}|E_{\bfI}\bmu_{A}|
      \cdot|\hU(E_{\bfI}\bmu_{A})|\mu_{\bge_{\refer}},
    \end{split}
  \end{equation}
  where we appealed to (\ref{eq:hNtaudotequivEi}). In particular, it is thus clear that we need to estimate $\hU(E_{\bfI}\bmu_{A})$ in $L^{2}$. In other words,
  we need to estimate terms of the form (\ref{eq:typeonetermEi}) and (\ref{eq:typetwotermEi}) in $L^{2}$. 

  \textbf{Estimating expressions of the form (\ref{eq:typeonetermEi}).} Before estimating the expression appearing in (\ref{eq:typeonetermEi})
  in $L^{2}$, we write $E_{\bfI_{i}}=E_{\bfL_{i}}E_{I_{i}}$ for some $I_{i}$. Next, we appeal to Corollary~\ref{cor:mixedmoserestweight}. When we do so,
  all the $\mU_{i}$ are functions: $E_{I_{j}}(\ln\hN)$, $A^{q}_{m}$ and $\bmu_{A}$. This yields
  \begin{equation}\label{eq:typeonetermsroughestimate}
    \begin{split}
      & \|\ldr{\tau}^{-\weight_{a}-|\bfI|\weight_{b}}E_{\bfI_{1}}(\ln\hN)\cdots E_{\bfI_{k}}(\ln\hN)E_{\bfJ}(A_{m}^{q})E_{\bfK}(\bmu_{A})\|_{2}\\
      \leq & C\left(\|\bmu_{A}\|_{\infty}\|A^{q}_{m}\|_{\mfH^{l_{1}}_{\weight}(\bM)}+\|A^{q}_{m}\|_{\infty}
      \|\bmu_{A}\|_{\mfH^{l_{1}}_{\weight}(\bM)}\right.\\
      & \left. \phantom{C(}+\|A^{q}_{m}\|_{\infty}\|\bmu_{A}\|_{\infty}\textstyle{\sum}_{p}\|\ldr{\tau}^{-\weight_{b}}E_{I_{p}}\ln\hN\|_{\mfH^{l_{1}}_{\weight}(\bM)}\right),
    \end{split}
  \end{equation}
  where $l_{1}=|\bfI|-k$ and $C$ only depends on $n$, $l$, $\bDlnhNsup$ and $(\bM,\bge_{\refer})$. Here the $\mfH^{l}_{\weight}(\bM)$-norm is defined as
  follows: 
  \[
  \|\mt(\cdot,t)\|_{\mfH^{l}_{\weight}(\bM)} := \left(\int_{\bM}\textstyle{\sum}_{j=0}^{l}\sum_{|\bfI|=j}\ldr{\tau(t)}^{-2\weight_{a}-2j\weight_{b}}
  |\bD_{\bfI}\mt(\cdot,t)|_{\bge_{\refer}}^{2}\mu_{\bge_{\refer}}\right)^{1/2}.
  \]
  Combining Corollary~\ref{cor:roughestbmuA} and Lemma~\ref{lemma:taurelvaryingbxEi}, it is clear that
  \begin{equation}\label{eq:bmuALinfty}
    \|\bmu_{A}(\cdot,t)\|_{C^{0}(\bM)}\leq \mc_{\bmu}\ldr{\tau(t)}
  \end{equation}
  for all $t\in I_{-}$, where $\mc_{\bmu}$ only depends on $n$, $\e_{\rond}$, $\e_{\mK}$, $\mKsup$, $C_{\mK,\mrod}$, $M_{\mK,\mrod}$, $\bDlnhNsup$ and
  $(\bM,\bge_{\refer})$. Moreover,
  \[
  \|\bmu_{A}(\cdot,t)\|_{\mfH^{l+1}_{\weight}(\bM)}^{2}\leq 2\me_{\bmu,\weight,l}(\tau(t)).
  \]
  Next, note that the conclusions of Lemma~\ref{lemma:mWAhUASobestimates} hold. Moreover, due to Lemma~\ref{lemma:taurelvaryingbxEi},
  \begin{equation}\label{eq:mfHlmfvfrommHlmfvnorm}
    \|A_{i}^{j}(\cdot,t)\|_{\mfH^{m}_{\weight}(\bM)}\leq C\|A_{i}^{j}(\cdot,t)\|_{\mH^{m}_{\bbE,\weight}(\bM)}
  \end{equation}
  for all $t\in I_{-}$, where $C$ only depends on $n$, $m$, $\cweight$ and $K_{\rovar}$. Moreover, the right hand side of (\ref{eq:mfHlmfvfrommHlmfvnorm})
  is bounded by the right hand side of (\ref{eq:AikmHlestsobassumpEi}) for $m\leq l+1$. Next, note that
  \[
  \|\ldr{\tau}^{-\weight_{b}}E_{p}\ln\hN\|_{\mfH^{l_{1}}_{\weight}(\bM)}\leq C\|\ln\hN\|_{\mH^{\bfl_{1}}_{\bbE,\weight}(\bM)}
  \]
  on $I_{-}$, where $\bfl_{1}=(1,l_{1}+1)$, and $C$ only depends on $n$, $l_{1}$, $K_{\rovar}$ and $\cweight$. Combining this estimate with
  the assumptions yields the conclusion that for $l_{1}\leq l$, the right hand side is bounded by a constant depending only on $s_{\cweight,l}$
  and $(\bM,\bge_{\refer})$. Summing up the above observations yields
  \begin{equation*}
    \begin{split}
      \|\ldr{\tau}^{-\weight_{a}-|\bfI|\weight_{b}}E_{\bfI_{1}}(\ln\hN)\cdots E_{\bfI_{k}}(\ln\hN)E_{\bfJ}(A_{i}^{j})E_{\bfK}(\bmu_{A})\|_{2}
      \leq & C\ldr{\tau}^{\cweight+1}e^{\eSpe\tau}+C\ldr{\tau}^{\cweight}e^{\eSpe\tau}\me_{\bmu,\weight,l}^{1/2}
    \end{split}
  \end{equation*}
  on $I_{-}$, where $C$ only depends on $s_{\cweight,l}$ and $(\bM,\bge_{\refer})$.

  \textbf{Estimating expressions of the form (\ref{eq:typetwotermEi}).} Expressions of the form (\ref{eq:typetwotermEi}) can be estimated similarly
  to the above. In fact, an estimate analogous to (\ref{eq:typeonetermsroughestimate}) combined with the equivalence of $\ldr{\tau}$ and
  $\ldr{\varrho}$ yields
  \begin{equation}\label{eq:typetwotermmfHmlprelest}
    \begin{split}
      & \|\ldr{\tau}^{-\weight_{a}-|\bfI|\weight_{b}}E_{\bfI_{1}}(\ln\hN)\cdots E_{\bfI_{k}}(\ln\hN)E_{\bfJ}(\ell_{A}+\mW^{A}_{A})\|_{2}\\
      \leq & C\left(\|\ell_{A}+\mW^{A}_{A}\|_{\mH^{l_{1}}_{\bbE,\weight}(\bM)}+\|\ell_{A}+\mW^{A}_{A}\|_{C^{0}_{\weight}(\bM)}
      \|\ln\hN\|_{\mH^{\bfl_{1}}_{\bbE,\weight_{0}}(\bM)}\right)
    \end{split}
  \end{equation}
  where $l_{1}=|\bfI|-k$, $\bfl_{1}=(1,l_{1}+1)$ and $C$ only depends on $s_{\cweight,l}$ and $(\bM,\bge_{\refer})$. Next, note that $\ell_{A}=\mK(Y^{A},X_{A})$
  (no summation on $A$), so that $\ell_{A}$ is bounded. Combining this observation with (\ref{eq:mWABmHCzeromfwbdbconstant}) yields the conclusion that
  $\|\ell_{A}+\mW^{A}_{A}\|_{C^{0}_{\weight}(\bM)}$ is bounded by a constant depending only on $s_{\cweight,l}$. Due to (\ref{eq:mWABmHlmfwbdbconstant}) and the
  assumptions, the only thing that remains to be estimated is the weighted Sobolev norm of $\ell_{A}$. However, such an estimate follows from
  (\ref{eq:bDindasslambdaAXASobEi}). To conclude, the right hand side of (\ref{eq:typetwotermmfHmlprelest}) can be estimated by a constant depending only
  on $s_{\cweight,l}$ and $(\bM,\bge_{\refer})$.

  \textbf{Estimating $\hU(E_{\bfI}\bmu_{A})$ in $L^{2}$.} Summing up the above estimates yields
  \begin{equation}\label{eq:hUXbfAbmuAest}
    \left(\textstyle{\sum}_{A}\textstyle{\sum}_{|\bfI|\leq l+1}\ldr{\tau}^{-2\weight_{a}-2|\bfI|\weight_{b}}\|\hU(E_{\bfI}\bmu_{A})\|_{2}^{2}\right)^{1/2}
    \leq C_{a}+C_{b}\ldr{\tau}^{\cweight}e^{\eSpe\tau}\me_{\bmu,\weight,l}^{1/2},
  \end{equation}
  where $C_{a}$ and $C_{b}$ only depend on $s_{\cweight,l}$ and $(\bM,\bge_{\refer})$. 

  \textbf{Estimating $\bmu_{A}$ in $\mH^{l}$.} Combining (\ref{eq:memukderest}) and (\ref{eq:hUXbfAbmuAest}) yields
  \begin{equation}\label{eq:memukderestfinal}
    \d_{\tau}\me_{\bmu,\weight,l}\geq -C_{c}\me_{\bmu,\weight,l}^{1/2}-C_{d}\ldr{\tau}^{\cweight}e^{\eSpe\tau}\me_{\bmu,\weight,l}
  \end{equation}
  on $I_{-}$, where $C_{c}$ and $C_{d}$ only depend on $s_{\cweight,l}$ and $(\bM,\bge_{\refer})$. Thus
  \[
  \d_{\tau}E_{\bmu,\weight,l}^{1/2}\geq-\frac{1}{2}C_{c}-\frac{1}{2}C_{d}\ldr{\tau}^{\cweight}e^{\eSpe\tau}E_{\bmu,\weight,l}^{1/2}
  \]
  on $I_{-}$, where $E_{\bmu,\weight,l}:=\me_{\bmu,\weight,l}+1$. This estimate implies that
  \[
  E_{\bmu,\weight,l}^{1/2}(\tau)\leq E_{\bmu,\weight,l}^{1/2}(0)+C_{c}\ldr{\tau}+\int_{\tau}^{0}C_{d}\ldr{s}^{\cweight}e^{\eSpe s}E_{\bmu,\weight,l}^{1/2}(s)ds
  \]
  on $I_{-}$. Combining this estimate with an argument similar to the proof of Gr\"{o}nwall's lemma yields
  \[
  E_{\bmu,\weight,l}^{1/2}(\tau)\leq C\ldr{\tau}
  \]
  on $I_{-}$, where $C$ only depends on $s_{\cweight,l}$ and $(\bM,\bge_{\refer})$. 
\end{proof}

\section{$C^{k}$-estimates of $\bmu_{A}$}\label{section:CkestofbmuAEi}

The purpose of the present section is to derive weighted $C^{k}$-estimates of $\bmu_{A}$.

\begin{lemma}\label{lemma:CkestofbmuAEi}
  Fix $l$, $\bfl_{1}$, $\cweight$, $\weight_{0}$ and $\weight$ as in Definition~\ref{def:supmfulassumptions}.
  Then, given that the assumptions of Lemma~\ref{lemma:taurelvaryingbxEi} as well as the $(\cweight,l)$-supremum assumptions are satisfied,
  there is a constant $C_{\bmu,l}$ such that
  \begin{equation}\label{eq:bmuAmclmfwestEi}
    \|\bmu_{A}(\cdot,t)\|_{\mc^{l+1}_{\bbE,\weight}(\bM)}\leq C_{\bmu,l}\ldr{\tau}
  \end{equation}
  for all $t\in I_{-}$, where $C_{\bmu,l}$ only depends on $c_{\cweight,l}$ and $(\bM,\bge_{\refer})$. 
\end{lemma}
\begin{remark}\label{remark:CkestofmuAEi}
  Similarly to Remark~\ref{remark:muAmhlestimate}, combining (\ref{eq:bmuAmclmfwestEi}) with the assumptions and the fact that 
  $\mu_{A}=\bmu_{A}+\ln\theta$ yields the conclusion that 
  \begin{equation}\label{eq:muAClestimatermk}
    \|\mu_{A}(\cdot,\tau)\|_{\mc^{\bfl_{1}}_{\bbE,\weight}(\bM)}\leq C_{\mu,l}\ldr{\tau}
  \end{equation}
  on $I_{-}$ for all $A$, where $C_{\mu,l}$ only depends on $c_{\cweight,l}$ and $(\bM,\bge_{\refer})$. 
\end{remark}
\begin{proof}
  Fix an integral curve $\g$ of $\hU$ such that $\g(0)\in\bM_{t_{0}}$, let $\weight_{a}=\weight_{b}=\cweight$ and define
  \[
  \mfe_{\weight,k}(s)=\textstyle{\sum}_{|\bfI|\leq k}\sum_{A}\ldr{s}^{-2\weight_{a}-2|\bfI|\weight_{b}}[(E_{\bfI}\bmu_{A})\circ\g(s)]^{2}.
  \]
  Note that, by definition, $\mfe_{\weight,k}(0)=0$. Differentiating $\mfe_{\weight,k}$ yields
  \begin{equation}\label{eq:mfekprEi}
    \mfe_{\weight,k}'(s)\geq
    2\textstyle{\sum}_{|\bfI|\leq k}\sum_{A}\ldr{s}^{-2\weight_{a}-2|\bfI|\weight_{b}}[\hU(E_{\bfI}\bmu_{A})]\circ\g(s)\cdot (E_{\bfI}\bmu_{A})\circ\g(s)
  \end{equation}
  for all $s\leq 0$. Thus it is clearly of interest to estimate $\hU(E_{\bfI}\bmu_{A})$ along $\g$. To this end, we appeal to
  Lemma~\ref{lemma:hUXbfAbmuAtermsEi}. We thus need to estimate the contribution from terms of the form (\ref{eq:typeonetermEi}) and terms of the
  form (\ref{eq:typetwotermEi}). We begin with some preliminary observations. 

  \textbf{Preliminary estimates.} Before proceeding, it is of interest to note that
  \begin{equation}\label{eq:svarrhotauequivalenceEi}
    \ldr{s}\leq 2\ldr{\varrho\circ\g(s)}\leq C_{1}\ldr{\tau\circ\g^{0}(s)},\ \ \
    \ldr{\tau\circ\g^{0}(s)}\leq C_{2}\ldr{\varrho\circ\g(s)}\leq 2C_{2}\ldr{s}
  \end{equation}
  for all $s\leq 0$, where $C_{1}$ and $C_{2}$ only depend on $K_{\rovar}$ and we appealed to (\ref{eq:varrhosequivalencestmt}) and
  (\ref{eq:DeltavarrhorelvariationEi}). Next, note that Lemma~\ref{lemma:bDbfAbDkequiv} yields
  \begin{equation}\label{eq:weEbfIilnhNalchar}
    \begin{split}
      & \ldr{s}^{-|\bfI_{i}|\weight_{b}}|(E_{\bfI_{i}}\ln\hN)\circ\g(s)|\leq C
      \textstyle{\sum}_{m=1}^{|\bfI_{i}|}\ldr{\varrho\circ\g(s)}^{-m\weight_{b}}|(\bD^{m}\ln\hN)\circ\g(s)|_{\bge_{\refer}}
      \leq C
    \end{split}
  \end{equation}
  for all $s\leq 0$ and all $\bfI_{i}$ such that $1\leq |\bfI_{i}|\leq l+1$, where $C$ only depends on $c_{\cweight,l}$ and
  $(\bM,\bge_{\refer})$. Next, combining (\ref{eq:Clmclequiv}), (\ref{eq:mWABClmfwbdbconstant}), (\ref{eq:svarrhotauequivalenceEi})
  and the assumptions yields
  \[
  \ldr{s}^{-\weight_{a}-|\bfI|\weight_{b}}|[E_{\bfI}(\mW^{A}_{B})]\circ\g(s)|\leq C
  \]
  for all $s\leq 0$ and all $\bfI$ such that $|\bfI|\leq l+1$, where $C$ only depends on $c_{\cweight,l}$ and $(\bM,\bge_{\refer})$.
  Moreover, due to (\ref{eq:bDbfAellAetcpteststmtEi}), (\ref{eq:svarrhotauequivalenceEi}) and the assumptions, it is clear that
  \[
  \ldr{s}^{-|\bfJ|\weight_{b}}|[E_{\bfJ}(\ell_{A})]\circ\g(s)|\leq C
  \]
  for all $s\leq 0$ and all $\bfJ$ such that $|\bfJ|\leq l+1$, where $C$ only depends on $c_{\cweight,l}$ and $(\bM,\bge_{\refer})$. Finally, note that
  combining (\ref{eq:AikClestsobassumpEivarrho}) with (\ref{eq:svarrhotauequivalenceEi}) and the assumptions yields 
  \begin{equation}\label{eq:weEbfJAijestalchar}
  \ldr{s}^{-|\bfJ|\weight_{b}}|[E_{\bfJ}(A_{i}^{j})]\circ\g(s)|\leq C\ldr{s}^{\cweight}e^{\e_{\Spe}s}
  \end{equation}
  for all $s\leq 0$ and all $\bfJ$ such that $|\bfJ|\leq l+1$, where $C$ only depends on $c_{\cweight,l}$ and $(\bM,\bge_{\refer})$. Next, we consider the 
  contributions from terms of the form (\ref{eq:typeonetermEi}) and terms of the form (\ref{eq:typetwotermEi}) separately. 
    
  \textbf{Estimating the contribution from terms of the form (\ref{eq:typeonetermEi}).} The contribution from terms of the form (\ref{eq:typeonetermEi})
  can be estimated by
  \begin{equation*}
    \begin{split}
      \ldr{s}^{-\weight_{a}-|\bfI|\weight_{b}}|[E_{\bfI_{1}}(\ln\hN)\cdots E_{\bfI_{k}}(\ln\hN)E_{\bfJ}(A_{i}^{j})E_{\bfK}(\bmu_{A})]\circ\g(s)|
      \leq & C\ldr{s}^{\cweight}e^{\e_{\Spe}s}\mfe_{\weight,|\bfK|}^{1/2}(s)
    \end{split}
  \end{equation*}
  for all $s\leq 0$, where we appealed to (\ref{eq:weEbfIilnhNalchar}) and (\ref{eq:weEbfJAijestalchar}) and the constant only depends on
  $c_{\cweight,l}$ and $(\bM,\bge_{\refer})$. 

  \textbf{Estimating the contribution from terms of the form (\ref{eq:typetwotermEi}).} Due to the preliminary estimates, the contribution
  from terms of the form (\ref{eq:typetwotermEi}) can be estimated by $C_{b}$ for all $s\leq 0$, where the constant $C_{b}$ only depends on
  $c_{\cweight,l}$ and $(\bM,\bge_{\refer})$.

  \textbf{Summing up.} Combining the above estimates yields the conclusion that
  \begin{equation}\label{eq:mfeweightkprimelb}
  \mfe_{\weight,k}'(s)\geq -C_{a}\ldr{s}^{\cweight}e^{\e_{\Spe}s}\mfe_{\weight,k}(s)-C_{b}\mfe_{\weight,k}^{1/2}(s)
  \end{equation}
  for all $s\leq 0$ and $k\leq l+1$, where $C_{a}$ and $C_{b}$ only depend on $c_{\cweight,l}$ and $(\bM,\bge_{\refer})$. This estimate can be integrated
  in order to yield the conclusion that $\mfe_{\weight,l+1}$ does not grow faster than $\ldr{s}^{2}$. Since the relevant constants only depend on $c_{\cweight,l}$
  and $(\bM,\bge_{\refer})$ and not on the integral curve, the desired conclusion follows by appealing to (\ref{eq:svarrhotauequivalenceEi}). 
\end{proof}

\section{$C^{k}$-estimates of $\varrho$}

In various contexts, it is of interest to estimate $\varrho$ separately. Note that the relation (\ref{eq:varphidefXAver}), combined with
Lemma~\ref{lemma:CkestofbmuAEi}, yields estimates for $\varrho$. However, the corresponding arguments are based on stronger assumptions than
necessary. Here, we therefore use the arguments of Lemma~\ref{lemma:respvarvarrhoEi} as a starting point.

\begin{lemma}\label{lemma:CkestofvarrhoEi}
  Let $1\leq l\in\zo$ and $(0,\cweight)=\weight_{0}\in\Weight$. Given that the conditions of Lemma~\ref{lemma:taurelvaryingbxEi} are fulfilled, assume
  that the basic assumptions, cf. Definition~\ref{def:basicassumptions}, are satisfied. Assume that there is a constant $c_{\chi,l+1}$ such that 
  \[
  \theta_{0,-}^{-1}\|\chi\|_{C^{l+1,\weight_{0}}_{\rohy}(\bM)}\leq c_{\chi,l+1}
  \]
  on $I_{-}$. Assume, moreover, that there is a constant $C_{\rorel,\bfl}$ such that (\ref{eq:bDlnfckmoest}) holds with $\bfl=(1,l)$. Then there
  is a constant $C_{\varrho,\weight_{0},l}$ such that
  \begin{equation}\label{eq:varrhomclmfwestEi}
    \|\varrho(\cdot,t)\|_{\mc^{l}_{\bbE,\weight_{0}}(\bM)}\leq C_{\varrho,\weight_{0},l}\ldr{\tau}
  \end{equation}
  for all $t\in I_{-}$, where $C_{\varrho,\weight_{0},l}$ only depends on $c_{\robas}$, $c_{\chi,l+1}$, $C_{\rorel,\bfl}$, $l$ and $(\bM,\bge_{\refer})$.
\end{lemma}
\begin{proof}
  Note, first of all, that (\ref{eq:Eivarrhoevolution}) can be written
  \begin{equation}\label{eq:hUEivarrhobasiformul}
    \hU[E_{i}(\varrho)]=E_{i}(\ln\hN)+\hN^{-1}E_{i}(\rodiv_{\bge_{\refer}}\chi)+A_{i}^{k}E_{k}(\varrho),
  \end{equation}
  where we used the notation introduced in (\ref{eq:Aialphadef}). Appealing to (\ref{eq:hUEicomm}), (\ref{eq:hUEivarrhobasiformul}) and an inductive
  argument, it can be demonstrated that
  \[
  \hU[E_{\bfI}(\varrho)]=A_{\bfI}+B_{\bfI}+\textstyle{\sum}_{1\leq |\bfJ|\leq |\bfI|}C_{\bfI,\bfJ}E_{\bfJ}(\varrho),
  \]
  where $A_{\bfI}$ is a linear combination of terms of the form
  \[
  E_{\bfI_{1}}(\ln\hN)\cdots E_{\bfI_{k}}(\ln\hN),
  \]
  where $\bfI_{j}\neq 0$ and $|\bfI_{1}|+\dots+|\bfI_{k}|=|\bfI|$; $B_{\bfI}$ is a linear combination of terms of the form
  \[
  E_{\bfI_{1}}(\ln\hN)\cdots E_{\bfI_{k}}(\ln\hN)\hN^{-1}E_{\bfJ}(\rodiv_{\bge_{\refer}}\chi),
  \]
  where $\bfI_{j}\neq 0$, $\bfJ\neq 0$ and $|\bfI_{1}|+\dots+|\bfI_{k}|+|\bfJ|=|\bfI|$; and $C_{\bfI,\bfJ}$ is a linear combination of terms of
  the form
  \[
  E_{\bfI_{1}}(\ln\hN)\cdots E_{\bfI_{k}}(\ln\hN)E_{\bfK}(A_{i}^{k})
  \]
  where $\bfI_{j}\neq 0$ and $|\bfI_{1}|+\dots+|\bfI_{k}|+|\bfK|=|\bfI|-|\bfJ|$. At this stage, we can proceed as in the proof of
  Lemma~\ref{lemma:CkestofbmuAEi}. In fact, fix a curve $\g$ as in the proof of Lemma~\ref{lemma:CkestofbmuAEi} and define
  \[
  \mfF_{\weight,k}(s)=\textstyle{\sum}_{|\bfI|\leq k}\ldr{s}^{-2|\bfI|\cweight}[(E_{\bfI}\varrho)\circ\g(s)]^{2}.
  \]
  Note that, by definition, $\mfF_{\weight,k}(0)=0$. Moreover (\ref{eq:weEbfIilnhNalchar}) holds for $1\leq |\bfI_{i}|\leq l$, with a constant
  depending only on $n$, $l$, $\cweight$, $C_{\rorel,\bfl}$ and $(\bM,\bge_{\refer})$; and (\ref{eq:weEbfJAijestalchar}) holds for $|\bfJ|\leq l$
  due to (\ref{eq:EbfIAikvarrhoestimate}) and the assumptions, where the constant $C$ only depends on $c_{\robas}$, $C_{\rorel,\bfl}$, $c_{\chi,l+1}$,
  $l$ and $(\bM,\bge_{\refer})$. Finally, we need to estimate
  \[
  \ldr{s}^{-|\bfJ|\cweight}|[\hN^{-1}E_{\bfJ}(\rodiv_{\bge_{\refer}}\chi)]\circ\g(s)|\leq C_{a}\ldr{s}^{\cweight}e^{\e_{\Spe} s},
  \]
  where we used the fact that $\rodiv_{\bge_{\refer}}\chi=\omega^{i}(\bD_{E_{i}}\chi)$. Moreover, we appealed to Remark~\ref{remark:chiclvarrhodecay} and 
  the assumptions. Finally, $C_{a}$ only depends on $c_{\robas}$, $c_{\chi,l+1}$, $l$ and $(\bM,\bge_{\refer})$. 
  Combining the above estimates yields the conclusion that
  \[
  \ldr{s}^{-|\bfI|\cweight}|[\hU E_{\bfI}(\varrho)]\circ\g(s)|\leq C_{a}+\textstyle{\sum}_{m=1}^{|\bfI|}C_{b}\ldr{s}^{\cweight}e^{\e_{\Spe} s}\mfF_{\weight,m}^{1/2}(s)
  \]
  for all $s\leq 0$, where $C_{a}$ and $C_{b}$ only depend on $c_{\robas}$, $c_{\chi,l+1}$, $C_{\rorel,\bfl}$, $l$ and
  $(\bM,\bge_{\refer})$. At this stage, we can proceed as in the proof of Lemma~\ref{lemma:CkestofbmuAEi} in order to deduce the conclusion of the
  lemma. 
\end{proof}

\part{Wave equations}

\chapter{Systems of wave equations, basic energy estimate}\label{chapter:systemsofwaveequations}

The main purpose of these notes is to analyse the asymptotic behaviour of solutions to (\ref{eq:theequation}). It is natural to begin by obtaining energy
estimates. In the present chapter we take a first step in this direction by deriving a zeroth order energy estimate. This estimate is
based on an energy identity we derive in Section~\ref{section:conformaleqabaseneestimates}. In order to take the step from the energy identity to an energy
estimate, we need to impose conditions on the coefficients of the equation. We discuss this topic in Section~\ref{section:assumptionsconcerningthecoeff}
below. Given these preliminaries, we obtain the basic energy estimate in Section~\ref{section:basicenergyestimates}. We end the chapter by expressing the
wave operator associated with $\hg$ with respect to the frame given by $\hU$ and the $X_{A}$. This also leads to a reformulation of (\ref{eq:theequation})
as (\ref{eq:equationintermsofcanonicalframeintro}). Note that this reformulation is important in the derivation of a model equation for the
asymptotic behaviour; cf. the heuristic discussions in Sections~\ref{section:resultsintrointo} and \ref{section:energyestimatesincausallylocalisedregions}.

\section{Conformal equation and basic energy estimates}\label{section:conformaleqabaseneestimates}

In the present paper, we are interested in equations of the form (\ref{eq:theequation}). However, it is convenient to rewrite this equation in terms of 
the conformal metric $\hg$. We do so in Subsection~\ref{ssection:confeq}. There, we also introduce a stress energy tensor which gives rise to the basic 
energy. Using this information, we derive the basic energy identity in Subsection~\ref{ssection:setbasen}. Throughout this section, we assume $(M,g)$ 
to be a time oriented Lorentz manifold. Moreover, we assume $(M,g)$ to have an expanding partial pointed foliation and $\mK$ to be non-degenerate on $I$ and
to have a global frame.

\subsection{Expressing the equation with respect to the conformal metric}\label{ssection:confeq}

\textbf{The wave operator.} To begin with, note that the wave operator is given by 
\begin{equation}\label{eq:waveoperator}
\Box_{g}u:=\frac{1}{\sqrt{-\det g}}\d_{\a}(\sqrt{-\det g}\ g^{\a\b}\d_{\b}u).
\end{equation}
\index{$\a$Aa@Notation!Operators!$\Box_{g}$}%
If $\hg$ is given by Definition~\ref{def:basicnotions}, then 
\[
\Box_{\hg}u=\frac{1}{\theta^{n+1}\sqrt{-\det g}}\d_{\a}(\theta^{n-1}\sqrt{-\det g}\ g^{\a\b}\d_{\b}u)
=\theta^{-2}\Box_{g}u+(n-1)\theta^{-3}g^{\a\b}\d_{\a}\theta\d_{\b}u,
\]
where $n=\dim\bM$. Thus
\begin{equation}\label{eq:Boxgconfresc}
\Box_{g}u=\theta^{2}\Box_{\hg}u-(n-1)\theta\hg(\grad_{\hg}\theta,\grad_{\hg}u). 
\end{equation}
It is convenient to split the first order expressions into time and space derivatives. Note, to this end, that 
\begin{align*}
  \hg(\grad_{\hg}\phi,\grad_{\hg}\psi) = & -\hU(\phi)\hU(\psi)+\textstyle{\sum}_{A}e^{-2\mu_{A}}X_{A}(\phi)X_{A}(\psi)
\end{align*}
Combining these observations yields
\begin{equation}\label{eq:waveopconfwaveop}
\theta^{-2}\Box_{g}u=\Box_{\hg}u+(n-1)\hU(\ln\theta)\hU(u)-(n-1)\textstyle{\sum}_{A}e^{-2\mu_{A}}X_{A}(\ln\theta)X_{A}(u).
\end{equation}

\textbf{The equation.} Combining (\ref{eq:waveopconfwaveop}) with (\ref{eq:theequation}) yields
\begin{equation}\label{eq:theequationwrthg}
\Box_{\hg}u+(n-1)\hU(\ln\theta)\hU(u)-(n-1)\textstyle{\sum}_{B}e^{-2\mu_{B}}X_{B}(\ln\theta)X_{B}(u)+\hmcX(u)+\hal u=\hf,
\end{equation}
where $\hmcX:=\theta^{-2}\mcX$,
\index{$\a$Aa@Notation!Coefficients of the equation!$\hmcX$}%
$\hal:=\theta^{-2}\a$
\index{$\a$Aa@Notation!Coefficients of the equation!$\hal$}%
and $\hf:=\theta^{-2}f$.
\index{$\a$Aa@Notation!Functions!$\hf$}%
It is convenient to decompose $\hmcX$ according to 
\begin{equation}\label{eq:hmcXalphadef}
\hmcX=\hmcX^{0}\hU+\hmcX^{A}X_{A},
\end{equation}
\index{$\a$Aa@Notation!Coefficients of the equation!$\hmcX^{0}$}%
where $\hmcX^{0}$ and $\hmcX^{A}$ are matrix valued functions on $M$. Appealing, additionally, to (\ref{eq:chKmKthetarelation}), 
the equation can be written
\begin{equation}\label{eq:theconfresceq}
\begin{split}
 & \Box_{\hg}u+\frac{n-1}{n}(\chth-1)\hU(u)-(n-1)\textstyle{\sum}_{B}e^{-2\mu_{B}}X_{B}(\ln\theta)X_{B}(u)\\
 & +\hmcX^{0}\hU(u)+\hmcX^{B}X_{B}(u)+\hal u=\hf.
\end{split}
\end{equation}

\subsection{The basic energy identity}\label{ssection:setbasen}

In order to estimate the evolution of $u$, it is convenient to let $\tau_{c}\leq 0$ and to introduce the stress energy tensor
\[
T_{\a\b}=\hna_{\a}u\cdot\hna_{\b}u-\frac{1}{2}\left(\hna^{\g}u\cdot\hna_{\g}u+\iota_{a}|u|^{2}+\iota_{b}\ldr{\tau-\tau_{c}}^{-3}|u|^{2}\right)\hg_{\a\b},
\]
where $\iota_{a}$ and $\iota_{b}$ are constants and $\hna$ is the Levi-Civita connection associated with $\hg$. We choose these constants as follows.
If there is a constant $d_{\a}$ such that
\begin{equation}\label{eq:haldecay}
  \sup_{\bx\in\bM}\|\hal(\bx,t)\|\leq d_{\a}\ldr{\tau(t)-\tau_{c}}^{-3}
\end{equation}
for all $t\leq t_{c}$, where $\tau(t_{c})=\tau_{c}$, we choose $\iota_{a}=0$ and $\iota_{b}=1$. Otherwise, we choose $\iota_{a}=1$ and $\iota_{b}=0$.
The reason for choosing
$\iota_{a}=0$, $\iota_{b}=1$ and the factor $\ldr{\tau-\tau_{c}}^{-3}$ in case $\hal$ satisfies the estimate (\ref{eq:haldecay}) is that, first of all,
this choice ensures that the zeroth order term does not contribute to the growth of the energy; and, second, controlling the energy gives
control of the $L^{2}$-norm of $u$ up to a polynomial weight in $\tau$ (and most of the estimates derived below will be up to polynomial
weights). In particular, 
\[
T(\hU,\hU)=\textstyle{\frac{1}{2}}|\hU(u)|^{2}+\frac{1}{2}\sum_{A}e^{-2\mu_{A}}|X_{A}(u)|^{2}
+\frac{1}{2}\iota_{a}|u|^{2}+\frac{1}{2}\iota_{b}\ldr{\tau-\tau_{c}}^{-3}|u|^{2},
\]
where $|\cdot|$ denotes the ordinary Euclidean norm of a vector in $\rn{m_{\ros}}$. It is thus natural to define an energy
\begin{equation}\label{eq:hFtaudef}
  \begin{split}
    \msE[u](\tau) := & \frac{1}{2}\int_{\bM_{\tau}}\left(|\hU(u)|^{2}+\textstyle{\sum}_{A}e^{-2\mu_{A}}|X_{A}(u)|^{2}
    +\iota_{a}|u|^{2}+\iota_{b}\ldr{\tau-\tau_{c}}^{-3}|u|^{2}\right)\mu_{\chg},
  \end{split}  
\end{equation}
where we abuse notation in that if $\tau_{a}=\tau(t_{a})$, then $\bM_{\tau_{a}}$ is understood to equal $\bM_{t_{a}}$ etc. With this definition, the 
following basic energy identity holds. 

\begin{lemma}\label{lemma:basicenergyestimate}
  Let $(M,g)$ be a time oriented Lorentz manifold. Assume it to have an expanding partial pointed foliation and $\mK$ to be non-degenerate on $I$ and to
  have a global frame. Then
  \begin{equation}\label{eq:baenidprel}
    \msE(\tau_{b})=\msE(\tau_{a})-\int_{\tau_{a}}^{\tau_{b}}\left(\int_{\bM_{\tau}}\tN\mP\mu_{\chg}\right)d\tau,
  \end{equation}
  where $\tau_{a}\leq \tau_{b}\leq\tau_{c}\leq 0$, $\tN:=\hN/\d_{t}\tau$, $\tau$ is introduced in (\ref{eq:taudefinitionEi}) and
  \begin{equation}\label{eq:mPdef}
    \begin{split}
      \mP := & \left(\textstyle{\frac{n-1}{n}}-\textstyle{\frac{n-2}{2n}}\chth\right)|\hU(u)|^{2}
      +\textstyle{\sum}_{A}e^{-2\mu_{A}}X_{A}\left(\ln\frac{\theta^{n-1}}{\hN}\right)X_{A}(u)\cdot\hU(u)\\
      & +\textstyle{\sum}_{A}\left(\lambda_{A}-\frac{1}{2}\chth\right)e^{-2\mu_{A}}|X_{A}(u)|^{2}
      -\textstyle{\frac{1}{2}}\chth(\iota_{a}+\iota_{b}\ldr{\tau-\tau_{c}}^{-3})|u|^{2}\\
      & +\textstyle{\frac{3}{2}}\iota_{b}\tN^{-1}\ldr{\tau-\tau_{c}}^{-5}(\tau-\tau_{c}) |u|^{2}-[\hmcX^{0}\hU(u)]\cdot\hU(u)-[\hmcX^{A}X_{A}(u)]\cdot\hU(u)\\
      & -(\hal u)\cdot\hU(u)-(\iota_{a}+\iota_{b}\ldr{\tau-\tau_{c}}^{-3}) u\cdot\hU(u)+\hf\cdot\hU(u).
    \end{split}
  \end{equation}
  Here $\hmcX^{0}$ and $\hmcX^{A}$ are defined by (\ref{eq:hmcXalphadef}) and $\lambda_{A}$ is the eigenvalue of $\chK$ corresponding to $X_{A}$; i.e.,
  $\chK X_{A}=\lambda_{A}X_{A}$ (no summation). 
\end{lemma}
\begin{remark}
  For many solutions to Einstein's equations, $q$ converges exponentially to $n-1$. For this reason, it is of interest to note that $\mP$ can
  be rewritten
  \begin{equation}\label{eq:mPdefvtwo}
    \begin{split}
      \mP := & -\chth T(\hU,\hU)+\textstyle{\frac{1}{n}}[(n-1)-q]|\hU(u)|^{2}
      +\textstyle{\sum}_{A}e^{-2\mu_{A}}X_{A}\left(\ln\frac{\theta^{n-1}}{\hN}\right)X_{A}(u)\cdot\hU(u)\\
      & +\textstyle{\sum}_{A}\lambda_{A}e^{-2\mu_{A}}|X_{A}(u)|^{2}+\textstyle{\frac{3}{2}}\iota_{b}\tN^{-1}\ldr{\tau-\tau_{c}}^{-5}(\tau-\tau_{c})|u|^{2}
      -[\hmcX^{0}\hU(u)]\cdot\hU(u)\\
      & -[\hmcX^{A}X_{A}(u)]\cdot\hU(u)-(\hal u)\cdot\hU(u)-(\iota_{a}+\iota_{b}\ldr{\tau-\tau_{c}}^{-3}) u\cdot\hU(u)+\hf\cdot\hU(u).
    \end{split}
  \end{equation}  
\end{remark}
\begin{proof}
  Compute
  \begin{equation*}
    \begin{split}
      \hna^{\a}T_{\a\b} 
      = & (\Box_{\hg}u-\iota_{a}u-\iota_{b}\ldr{\tau-\tau_{c}}^{-3}u)\cdot\hna_{\b}u
      +\textstyle{\frac{3}{2}}\iota_{b}\ldr{\tau-\tau_{c}}^{-5}(\tau-\tau_{c}) (\hna_{\b}\tau)|u|^{2}.
    \end{split}
  \end{equation*}
  In particular,
  \begin{equation}\label{eq:divmcV}
    \begin{split}
      \hna^{\a}(T_{\a\b}\hU^{\b}) = & (\hna^{\a}T_{\a\b})\hU^{\b}+T_{\a\b}\hna^{\a}\hU^{\b}\\
      = & (\Box_{\hg}u-\iota_{a}u-\iota_{b}\ldr{\tau-\tau_{c}}^{-3}u)\cdot \hU(u)\\
      & +\textstyle{\frac{3}{2}}\iota_{b}\tN^{-1}\ldr{\tau-\tau_{c}}^{-5}(\tau-\tau_{c})|u|^{2}+T^{\a\b}\hpi_{\a\b},
    \end{split}
  \end{equation}
  where $\tN$ is defined in the statement of the lemma and the deformation tensor $\hpi$ is defined by
  \[
  \hpi:=\textstyle{\frac{1}{2}}\ml_{\hU}\hg.
  \]
  Let $t_{a}<t_{b}$, where $t_{a},t_{b}\in I$, and
  \begin{equation}\label{eq:Mabdef}
    M_{ab}:=\bM\times [t_{a},t_{b}].
  \end{equation}
  Let, moreover, $\mcV$ be the vector field defined by
  \begin{equation}\label{eq:mcVdef}
    \mcV^{\a}:=T^{\a}_{\phantom{\a}\b}\hU^{\b}.
  \end{equation}
  Then \cite[Lemma~10.8, p.~100]{minbok} yields
  \begin{equation}\label{eq:basicdivergenceidentity}
    \int_{M_{ab}}\mathrm{div}_{\hg}\mcV\mu_{\hg}=-\int_{\bM_{t_{b}}}T(\hU,\hU)\mu_{\chg}
    +\int_{\bM_{t_{a}}}T(\hU,\hU)\mu_{\chg};
  \end{equation}
  here we assume $u$ to be such that the integration makes sense. In particular, letting $\msE$ be defined by (\ref{eq:hFtaudef}),
  it follows that
  \begin{equation*}
    \begin{split}
      \msE(t_{b}) = & \msE(t_{a})-\int_{M_{ab}}\left(\Box_{\hg}u\cdot \hU(u)+T^{\a\b}\hpi_{\a\b}\right)\mu_{\hg}\\
      & -\int_{M_{ab}}\left[-(\iota_{a}+\iota_{b}\ldr{\tau-\tau_{c}}^{-3})u\cdot \hU(u)
        +\textstyle{\frac{3}{2}}\iota_{b}\tN^{-1}\ldr{\tau-\tau_{c}}^{-5}(\tau-\tau_{c})|u|^{2}\right]\mu_{\hg},
    \end{split}
  \end{equation*}  
  where we appealed to (\ref{eq:divmcV}). Let us consider the second term on the right hand side. Since $\det \hg=-\hN^{2}\det\chg$ (with respect to
  standard coordinates; cf. \cite[Remark~25.3, p. 469]{finallinsys}), it can, ignoring the sign, be written
  \begin{equation*}
    \begin{split}
      & \int_{M_{ab}}\left(\Box_{\hg}u\cdot \hU(u)+T^{\a\b}\hpi_{\a\b}\right)\mu_{\hg}\\
      = & \int_{\tau_{a}}^{\tau_{b}}\left(\int_{\bM_{\tau}}\tN\left(\Box_{\hg}u\cdot \hU(u)+T^{\a\b}\hpi_{\a\b}\right)\mu_{\chg}\right)d\tau,
    \end{split}
  \end{equation*}
  where $\tN$ is defined in the statement of the lemma. Here we abuse notation in that if $\tau_{a}=\tau(t_{a})$, then $\bM_{\tau_{a}}$ is
  understood to equal $\bM_{t_{a}}$ etc. In order to simplify the expression involving $\hpi$, note that
  \[
  (\ml_{\hU}\hg)(X,Y)=\ldr{\hna_{X}\hU,Y}+\ldr{\hna_{Y}\hU,X},
  \]
  where $\ldr{\cdot,\cdot}:=\hg$. In particular, $(\ml_{\hU}\hg)(\hU,\hU)=0$ and
  \[
  (\ml_{\hU}\hg)(X_{A},X_{B})=2\chk(X_{A},X_{B})=2\chg(\chK X_{A},X_{B})=2\lambda_{A}e^{2\mu_{A}}\de_{AB}
  \]
  (no summation on $A$). Next, note that $\ldr{\hna_{X_{A}}\hU,\hU}=0$ and that
  \begin{equation}\label{eq:ldrnablaUUdj}
    \ldr{\hna_{\hU}\hU,X_{A}}=-\ldr{\hU,\hna_{\hU}X_{A}}=-\ldr{\hU,[\hU,X_{A}]+\hna_{X_{A}}\hU}=X_{A}\ln \hN.
  \end{equation}
  Thus
  \[
  (\ml_{\hU}\hg)(\hU,X_{A})=\ldr{\hna_{\hU}\hU,X_{A}}=X_{A}\ln \hN,
  \]
  where we appealed to (\ref{eq:ldrnablaUUdj}). Thus
  \begin{equation*}
    \begin{split}
      T^{\a\b}\hpi_{\a\b} = & -\textstyle{\sum}_{A}e^{-2\mu_{A}}X_{A}(\ln\hN)X_{A}(u)\cdot\hU(u)
      +\textstyle{\sum}_{A}\lambda_{A}e^{-2\mu_{A}}|X_{A}(u)|^{2}\\
      & -\textstyle{\frac{1}{2}}\chth\left(-|\hU(u)|^{2}+\textstyle{\sum}_{A}e^{-2\mu_{A}}|X_{A}(u)|^{2}+\iota_{a}|u|^{2}
      +\iota_{b}\ldr{\tau-\tau_{c}}^{-3}|u|^{2}\right).
    \end{split}
  \end{equation*}
  Next, appealing to (\ref{eq:theconfresceq}) yields
  \begin{equation*}
    \begin{split}
      \Box_{\hg}u\cdot \hU(u) = & -\textstyle{\frac{n-1}{n}}(\chth-1)|\hU(u)|^{2}+(n-1)\textstyle{\sum}_{A}e^{-2\mu_{A}}X_{A}(\ln\theta)X_{A}(u)\cdot\hU(u)\\
      & -[\hmcX^{0}\hU(u)]\cdot\hU(u)-[\hmcX^{A}X_{A}(u)]\cdot\hU(u)-(\hal u)\cdot\hU(u)+\hf\cdot\hU(u).
    \end{split}
  \end{equation*}
  Summing up the above computations yields the conclusion of the lemma. 
\end{proof}

In some settings, it is convenient to rescale the stress energy tensor as follows. First, let
\begin{equation}\label{eq:tvarphidefinition}
  \tvarphi:=\theta\varphi,
\end{equation}
\index{$\a$Aa@Notation!Functions!$\tvarphi$}%
where $\varphi$ is defined by (\ref{eq:varphidefitobmubgedp}). Second, fix a $t_{c}\leq t_{0}$ and define $\tvarphi_{c}$ by
\begin{equation}\label{eq:tvarphicdef}
  \tvarphi_{c}(\bx,t):=\tvarphi(\bx,t_{c}).
\end{equation}
\index{$\a$Aa@Notation!Functions!$\tvarphi_{c}$}%
Finally, rescale the stress energy tensor according to
\begin{equation}\label{eq:hTalbec}
  \hT_{\a\b}:=\tvarphi_{c}^{-1}\theta^{-(n-1)}T_{\a\b}.
\end{equation}
This leads to an energy analogous to (\ref{eq:hFtaudef}). If $\tau_{c}=\tau(t_{c})$, it can be written 
\begin{equation}\label{eq:hEtaudef}
  \begin{split}
    \hE[u](\tau;\tau_{c}) := & \int_{\bM_{\tau}}T(\hU,\hU)\tvarphi_{c}^{-1}\theta^{-(n-1)}\mu_{\chg}.
  \end{split}  
\end{equation}
\index{$\a$Aa@Notation!Energies!$\hE[u](\tau;\tau_{c})$}%
Note that the rescaling given by (\ref{eq:hTalbec}) is such that
\begin{equation}\label{eq:hEutauctauc}
  \hE[u](\tau_{c};\tau_{c})=\int_{\bM_{\tau_{c}}}T(\hU,\hU)\mu_{\bge_{\refer}}.
\end{equation}
\begin{cor}
  Let $(M,g)$ be a time oriented Lorentz manifold. Assume it to have an expanding partial pointed foliation and $\mK$ to be non-degenerate on $I$ and to
  have a global frame. Then, if $\tau_{a}\leq\tau_{b}\leq \tau_{c}\leq 0$, 
  \begin{equation}\label{eq:baenid}
    \hE(\tau_{b};\tau_{c})=\hE(\tau_{a};\tau_{c})-\int_{\tau_{a}}^{\tau_{b}}\left(\int_{\bM_{\tau}}\tN\mQ\tvarphi_{c}^{-1}\theta^{-(n-1)}\mu_{\chg}\right)d\tau,
  \end{equation}
  where $\tN:=\hN/\d_{t}\tau$, $\tau$ is introduced in (\ref{eq:taudefinitionEi}) and
  \begin{equation}\label{eq:mQdefvtwo}
    \begin{split}
      \mQ := & \textstyle{\frac{1}{n}}[q-(n-1)]T(\hU,\hU)+\textstyle{\frac{1}{n}}[(n-1)-q]|\hU(u)|^{2}-\hN^{-1}\chi(\ln\tvarphi_{c})T(\hU,\hU)\\
      & -\textstyle{\sum}_{A}e^{-2\mu_{A}}X_{A}[\ln(\tvarphi_{c}\hN)]X_{A}(u)\cdot\hU(u)
      +\textstyle{\sum}_{A}\lambda_{A}e^{-2\mu_{A}}|X_{A}(u)|^{2}\\
      & +\textstyle{\frac{3}{2}}\iota_{b}\tN^{-1}\ldr{\tau-\tau_{c}}^{-5}(\tau-\tau_{c}) |u|^{2}
      -[\hmcX^{0}\hU(u)]\cdot\hU(u)-[\hmcX^{A}X_{A}(u)]\cdot\hU(u)\\
      & -(\hal u)\cdot\hU(u)-(\iota_{a}+\iota_{b}\ldr{\tau-\tau_{c}}^{-3}) u\cdot\hU(u)+\hf\cdot\hU(u).
    \end{split}
  \end{equation}  
\end{cor}
\begin{proof}
  The proof is essentially identical to the proof of Lemma~\ref{lemma:basicenergyestimate}; we only need to calculate the changes caused by
  the rescaling of the stress energy tensor. Note, to this end, that
  \[
  \hna^{\a}\hT_{\a\b}=\hna^{\a}[\ln(\tvarphi_{c}^{-1}\theta^{-(n-1)})]\hT_{\a\b}+\tvarphi_{c}^{-1}\theta^{-(n-1)}\hna^{\a}T_{\a\b}. 
  \]
  Define $\hmcV$ in analogy with (\ref{eq:mcVdef}); we simply replace $T$ with $\hT$. Then
  \begin{equation}\label{eq:rodivhghmcV}
    \rodiv_{\hg}\hmcV=\hna^{\a}[\ln(\tvarphi_{c}^{-1}\theta^{-(n-1)})]\hT_{\a\b}\hU^{\b}+\tvarphi_{c}^{-1}\theta^{-(n-1)}\rodiv_{\hg}\mcV.
  \end{equation}
  Beyond the rescaling, the only correction to the previous calculations thus consists in the first term on the right hand side of
  (\ref{eq:rodivhghmcV}). However, 
  \begin{equation*}
    \begin{split}
      \hna^{\a}[\ln(\tvarphi_{c}^{-1}\theta^{-(n-1)})]\hT_{\a\b}\hU^{\b} = & -\frac{n-1}{n}(q+1)\hT(\hU,\hU)-\hN^{-1}\chi(\ln\tvarphi_{c})\hT(\hU,\hU)\\
      & +\tvarphi_{c}^{-1}\theta^{-(n-1)}\textstyle{\sum}_{A}e^{-2\mu_{A}}X_{A}[\ln(\tvarphi_{c}^{-1}\theta^{-(n-1)})]X_{A}(u)\cdot \hU(u). 
    \end{split}
  \end{equation*}
  Adding this correction to the previous calculations yields the conclusion of the corollary. 
\end{proof}

\section{Assumptions concerning the coefficients}\label{section:assumptionsconcerningthecoeff}

In order to derive estimates for the energy using (\ref{eq:baenid}), it is necessary to impose conditions on $\hmcX$ and $\hal$. 

\begin{definition}\label{def:Czerobalance}
Let $(M,g)$ be a time oriented Lorentz manifold. Assume it to have an expanding partial pointed foliation. Consider the equation
(\ref{eq:theequation}) and define $\mcX^{\perp}$ by the conditions that its components are vector fields which are perpendicular to 
$\hU$ and that it is such that there is a matrix valued function $\mcX^{0}$ with the property that $\mcX=\mcX^{0}U+\mcX^{\perp}$. Then
(\ref{eq:theequation}) is said to be $C^{0}$-\textit{balanced}
\index{C@$C^{0}$-balanced equation}%
\index{Equation!$C^{0}$-balanced}%
on $I_{-}$ if there is a constant $C_{\robal,0}>0$ such that 
\begin{equation}\label{eq:Czbalanced}
\theta^{-1}\|\mcX^{0}\|+\textstyle{\sum}_{i,j=1}^{m_{\ros}}\theta^{-1}|\mcX^{\perp}_{ij}|_{\bge}+\theta^{-2}\|\a\|\leq C_{\robal,0}
\end{equation}
on $M_{-}$. 
\end{definition}
\begin{remark}
Note that $\mcX^{\perp}$ is a family of matrices of vector fields on $\bM$. In particular, $\mcX^{\perp}_{ij}$ is a family of vector fields 
on $\bM$. 
\end{remark}
\begin{remark}\label{remark:Czbalancediffcond}
  Dividing $\hmcX$ according to $\hmcX=\hmcX^{0}\hU+\hmcX^{\perp}$, where the components of $\hmcX^{\perp}$ are perpendicular to $\hU$, the estimate
  (\ref{eq:Czbalanced}) can be written
  \begin{equation}\label{eq:Czbalancedhat}
    \|\hmcX^{0}\|+\textstyle{\sum}_{i,j=1}^{m_{\ros}}|\hmcX^{\perp}_{ij}|_{\chg}+\|\hal\|\leq C_{\robal,0},
  \end{equation}
  where $\hal$ is defined below (\ref{eq:theequationwrthg}). In particular, if (\ref{eq:coefflassumptions}) holds for $l=0$, then (\ref{eq:theequation}) is
  $C^{0}$-balanced on $I_{-}$.
\end{remark}
Next, we derive some basic consequences of the assumption of $C^{0}$-balance. 

\begin{lemma}\label{lemma:Czcoefficientbounds}
  Let $(M,g)$ be a time oriented Lorentz manifold. Assume it to have an expanding partial pointed foliation and $\mK$ to be non-degenerate on $I$ and to have
  a global frame. If (\ref{eq:theequation}) is $C^{0}$-balanced on $I_{-}$, there is a constant $K_{\robal,0}>0$, depending only on $C_{\robal,0}$, $m_{\ros}$
  and $n$, such that if $\hmcX^{0}$ and $\hmcX^{A}$ are defined by (\ref{eq:hmcXalphadef}) and $\hal:=\theta^{-2}\a$, then
  \begin{equation}\label{eq:Czcoefficientbounds}
    \|\hal\|+\left(\textstyle{\sum}_{A}e^{2\mu_{A}}\|\hmcX^{A}\|^{2}\right)^{1/2}+\|\hmcX^{0}\|\leq K_{\robal,0}
  \end{equation}
  on $M_{-}$. 
\end{lemma}
\begin{proof}
The bound on $\|\hal\|$ follows immediately from (\ref{eq:Czbalanced}). Since $\hmcX^{0}=\theta^{-1}\mcX^{0}$, the same is true of the estimate 
for $\hmcX^{0}$. In order to estimate $\hmcX^{A}$, note that $\theta^{-2}\mcX^{\perp}=\hmcX^{A}X_{A}$. Thus
\[
\theta^{-2}|\mcX^{\perp}_{ij}|_{\bge}^{2}=\chg(\hmcX^{A}_{ij}X_{A},\hmcX^{B}_{ij}X_{B})=\textstyle{\sum}_{A}e^{2\mu_{A}}|\hmcX^{A}_{ij}|^{2}.
\]
Combining this equality with (\ref{eq:Czbalanced}) yields the desired bound on $e^{\mu_{A}}\|\hmcX^{A}\|$.
\end{proof}

In the estimates to follow, it is convenient to use the following notation:
\begin{equation}\label{eq:hmcXperpchgnorm}
  \|\hmcX^{\perp}\|_{\chg}:=\left(\textstyle{\sum}_{A}e^{2\mu_{A}}\|\hmcX^{A}\|^{2}\right)^{1/2}.
\end{equation}

\section{Basic energy estimate}\label{section:basicenergyestimates}

Given that (\ref{eq:theequation}) is $C^{0}$-balanced on $I_{-}$, we obtain a basic energy estimate. In the derivation, it is convenient to use the notation
\begin{equation}\label{eq:medef}
  \me[u]:=\frac{1}{2}\left(|\hU(u)|^{2}+\textstyle{\sum}_{A}e^{-2\mu_{A}}|X_{A}(u)|^{2}+\iota_{a}|u|^{2}+\iota_{b}\ldr{\tau-\tau_{c}}^{-3}|u|^{2}\right), 
\end{equation}
\index{$\a$Aa@Notation!Energy densities!$\me[u]$}%
where the constants $\iota_{a}$ and $\iota_{b}$ are chosen as at the beginning of Subsection~\ref{ssection:setbasen}.

\begin{lemma}\label{lemma:basicenergyestimatequant}
  Assume the conditions of Definition~\ref{def:basicassumptions} and of Lemma~\ref{lemma:taurelvaryingbxEi} to be fulfilled. Assume, moreover, that
  there is a constant $c_{\theta,1}$ such that
  \begin{equation}\label{eq:cthetaoneestimate}
    \|(\ln\theta)(\cdot,t)\|_{C^{\bfl_{0}}_{\weight_{0}}(\bM)}\leq c_{\theta,1}
  \end{equation}
  for all $t\leq t_{0}$, where $\bfl_{0}:=(1,1)$. Then 
  \begin{equation}\label{eq:basicenergyestimateprel}
    \begin{split}
      \hE(\tau_{a};\tau_{c}) \leq &  \hE(\tau_{b};\tau_{c})+\int_{\tau_{a}}^{\tau_{b}}\zeta(\tau)\hE(\tau;\tau_{c})d\tau
      +\int_{\tau_{a}}^{\tau_{b}}\int_{\bM_{\tau}}\tN|\hf|\cdot |\hU(u)|\tvarphi_{c}^{-1}\theta^{-(n-1)}\mu_{\chg}d\tau
    \end{split}
  \end{equation}
  for all $\tau_{a}\leq\tau_{b}\leq \tau_{c}\leq 0$. Here $\hE$ is defined by (\ref{eq:hEtaudef}), $\tvarphi_{c}$ is defined by
  (\ref{eq:tvarphicdef}), 
  \[
  \zeta=2K_{\rovar}(\zeta_{1}+\zeta_{2}+\iota_{a}\zeta_{3,a}+\iota_{b}\zeta_{3,b}),
  \]
  $K_{\rovar}$ is defined in (\ref{eq:KrovarEi}) and
  \begin{align}
    \zeta_{1}(\tau) := & \sup_{\bx\in\bM}\textstyle{\frac{1}{n}}|q(\bx,\tau)-(n-1)|,\label{eq:etaonedef}\\
    \zeta_{2}(\tau) := & C_{b}\ldr{\tau}^{\bcweight}e^{\eSpe\tau},\label{eq:etatwodef}\\
    \zeta_{3,a}(\tau) := & \sup_{\bx\in\bM}\left(2\|\hmcX^{0}(\bx,\tau)\|+\|\hmcX^{\perp}(\bx,\tau)\|_{\chg}+\|\hal(\bx,\tau)\|+1\right),\label{eq:etathreeadef}\\
    \zeta_{3,b}(\tau) := & \sup_{\bx\in\bM}\left(2\|\hmcX^{0}(\bx,\tau)\|+\|\hmcX^{\perp}(\bx,\tau)\|_{\chg}\right)
    +(d_{\a}+1)\ldr{\tau-\tau_{c}}^{-3/2},\label{eq:etathreebdef}
  \end{align}
  where $\bcweight:=\max\{\cweight,1\}$. Here $C_{b}$ only depends on $c_{\robas}$, $c_{\chi,2}$, $c_{\theta,1}$, $(\bM,\bge_{\refer})$ and a lower bound on
  $\theta_{0,-}$. Note also that $\zeta_{3,b}$ only enters the definition of $\zeta$ in case (\ref{eq:haldecay}) holds. 
\end{lemma}
\begin{proof}
  Recall the notation (\ref{eq:medef}) and consider (\ref{eq:baenid}). We already know $\tN$ to be bounded; cf. (\ref{eq:hNtaudotequivEi}). We therefore
  need to estimate $\mQ$, defined by (\ref{eq:mQdefvtwo}), from above. Consider the first two terms appearing on the right hand side of
  (\ref{eq:mQdefvtwo}). If the first one is negative, the second one is non-negative and vice versa. This means that we only have to include one of the
  terms. In fact, the sum of the first two terms can be estimated from above by $\zeta_{1}\me$, where $\zeta_{1}$ is defined by (\ref{eq:etaonedef}).
  Turning to the third term, note that 
  \[
  \hN^{-1}|\chi(\tvarphi_{c})|\leq \hN^{-1}|\chi|_{\bge_{\refer}}|\bD\ln\tvarphi_{c}|_{\bge_{\refer}}.  
  \]
  However, the first two factors can be estimated by appealing to (\ref{eq:hNinvchibgereferestimate}). Moreover, the last factor can be estimated
  by appealing to (\ref{eq:bDlntvarphicestpre}) with $\tau$ replaced by $\tau_{c}$. To conclude, 
  \[
  \hN^{-1}|\chi(\tvarphi_{c})|\leq C_{a}\ldr{\tau_{c}}^{\bcweight}e^{\eSpe\tau}
  \]
  for all $\tau\leq \tau_{c}$, where $C_{a}$ only depends on $c_{\robas}$, $c_{\chi,2}$, $c_{\theta,1}$ and $(\bM,\bge_{\refer})$. In particular, the
  third term on the right hand side of (\ref{eq:mQdefvtwo}) gives rise to an expression that can be estimated by a contribution to $\zeta$ of
  the form (\ref{eq:etatwodef}). Turning to the fourth term on the right hand side, appealing to (\ref{eq:bDlnNbDlnthetabd}),
  (\ref{eq:muminmainlowerbound}), (\ref{eq:eSpevarrhoeelowtaurelEi}) and (\ref{eq:bDlntvarphicestpre}) with $\tau$ replaced by $\tau_{c}$
  yields the conclusion that it can be estimated in the same way. The fifth and sixth terms on the right hand side of (\ref{eq:mQdefvtwo}) are
  both negative and can therefore be ignored. In case $\iota_{a}=1$ and $\iota_{b}=0$, the sum of terms seven to ten can be estimated by $\zeta_{3,a}\me$,
  where $\zeta_{3,a}$ is defined by (\ref{eq:etathreeadef}). In case $\iota_{a}=0$ and $\iota_{b}=1$, the sum of terms seven to ten can be estimated by
  $\zeta_{3,b}\me$, where $\zeta_{3,b}$ is defined by (\ref{eq:etathreebdef}). Combining the above estimates with (\ref{eq:hNtaudotequivEi}) and
  (\ref{eq:baenid}) yields the conclusion of the lemma. 
\end{proof}

\begin{cor}\label{cor:basicenergyestimate}
  Assume the conditions of Definition~\ref{def:basicassumptions} and of Lemma~\ref{lemma:taurelvaryingbxEi} to be fulfilled. Assume, moreover,
  (\ref{eq:theequation}) to be $C^{0}$-balanced on $I_{-}$, (\ref{eq:cthetaoneestimate}) to hold and $q$ to be bounded on $M_{-}$.
  Then, if $u$ is a solution to (\ref{eq:theequation}),
  \begin{equation}\label{eq:basicenergyestimate}
    \begin{split}
      \hE(\tau_{a};\tau_{c}) \leq &  \hE(\tau_{b};\tau_{c})+\int_{\tau_{a}}^{\tau_{b}}\kappa(\tau)\hE(\tau;\tau_{c})d\tau
      +\int_{\tau_{a}}^{\tau_{b}}\int_{\bM_{\tau}}\tN|\hf|\cdot |\hU(u)|\tvarphi_{c}^{-1}\theta^{-(n-1)}\mu_{\chg}d\tau
    \end{split}
  \end{equation}
  for all $\tau_{a}\leq\tau_{b}\leq \tau_{c}\leq 0$, where
  \begin{align}
    \kappa(\tau) := & c_{0}+\kappa_{\rem}(\tau),\label{eq:kappadef}\\
    c_{0} := & 2K_{\rovar}\sup_{M_{-}}\left(\textstyle{\frac{1}{n}}|q-(n-1)|+2\|\hmcX^{0}\|
    +\|\hmcX^{\perp}\|_{\chg}+\iota_{a}\|\hal\|+\iota_{a}\right)\label{eq:czCbdef}
  \end{align}
  \index{$\a$Aa@Notation!Functions!$\kappa$}%
  \index{$\a$Aa@Notation!Constants!$c_{0}$}%
  and $\kappa_{\rem}\in L^{1}(-\infty,\tau_{c}]$. Moreover, the $L^{1}$-norm of $\kappa_{\rem}$ only depends on $c_{\robas}$, $c_{\chi,2}$, $c_{\theta,1}$,
  $(\bM,\bge_{\refer})$, $d_{\a}$ (in case $\iota_{b}=1$) and a lower bound on $\theta_{0,-}$.

  Assume, in addition to the above, that (\ref{eq:haldecay}) holds and that there are constants $d_{q}$ and $d_{\coeff}$ such that (\ref{eq:qconvergence})
  and
  \begin{align}
    \sup_{\bx\in\bM}[\|\hmcX^{0}(\bx,t)\|+\|\hmcX^{\perp}(\bx,t)\|_{\chg}] \leq & d_{\coeff}\ldr{\tau(t)-\tau_{c}}^{-3/2}\label{eq:coeffconvergence}
  \end{align}
  hold for all $t\leq t_{c}$. Then (\ref{eq:basicenergyestimate}) holds with $\kappa\in L^{1}(-\infty,\tau_{c}]$. Moreover, the $L^{1}$-norm of $\kappa$
  is bounded by a constant depending only on $c_{\robas}$, $c_{\chi,2}$, $c_{\theta,1}$, $(\bM,\bge_{\refer})$, $d_{\a}$, $d_{q}$, $d_{\coeff}$ and a lower bound on
  $\theta_{0,-}$
\end{cor}
\begin{remark}
  One consequence of (\ref{eq:basicenergyestimate}) is that if $f=0$, then $\hE$ does not grow faster than exponentially. It is important to note that
  if the equation is not $C^{0}$-balanced, then the energy could grow superexponentially. For a justification of this statement, see
  \cite[Chapters~2 and 16]{finallinsys}. 
\end{remark}
\begin{remark}\label{remark:tgbgerefequiv}
  If all the conditions of the corollary are satisfied and $f=0$, then $\hE(\tau;\tau_{c})$ is bounded for all $\tau\leq \tau_{c}\leq 0$. Moreover,
  all the conditions of Lemma~\ref{lemma:thetavarrhorelqconvtonmotwo} are satisfied, so that (\ref{eq:lntvarphimlntvarphicimp}) holds. Since 
  \begin{equation}\label{eq:mutgreformulation}
    \begin{split}
      \tvarphi_{c}^{-1}\theta^{-(n-1)}\mu_{\chg} = & \tvarphi_{c}^{-1}\theta^{-(n-1)}\theta^{n}\mu_{\bge}=\tvarphi_{c}^{-1}\theta \varphi\mu_{\bge_{\refer}}\\
      = & \tvarphi_{c}^{-1}\tvarphi\mu_{\bge_{\refer}}=\exp[\ln\tvarphi-\ln\tvarphi_{c}]\mu_{\bge_{\refer}},
    \end{split}    
  \end{equation}
  where we use the notation introduced in (\ref{eq:tvarphidefinition}) and (\ref{eq:tvarphicdef}), this means, in particular, that it does not
  matter if the $L^{2}$ norm is calculated with respect to the measure $\tvarphi_{c}^{-1}\theta^{-(n-1)}\mu_{\chg}$ or with respect to the measure
  $\mu_{\bge_{\refer}}$. Thus
  \[
  \int_{\bM_{\tau}}\left(|\hU(u)|^{2}+\textstyle{\sum}_{A}e^{-2\mu_{A}}|X_{A}(u)|^{2}+\ldr{\tau-\tau_{c}}^{-3}|u|^{2}\right)\mu_{\bge_{\refer}}
  \]
  is bounded. 
\end{remark}
\begin{remark}\label{remark:KGboundedbasicenergy}
  Assuming that (\ref{eq:qconvergence}) holds, the conclusions of Remark~\ref{remark:tgbgerefequiv} apply to the Klein-Gordon equation.
  The reason for this is that in the case of the Klein-Gordon equation, $\hmcX=0$ and $\hal=-\theta^{-2}m^{2}$, where $m$ is a constant. Moreover,
  due to (\ref{eq:hUnlnthetamomqbas}) and the fact that $q\geq n\e_{\Spe}$ (cf. Remark~\ref{remark:qlwbd}), it can be demonstrated that
  $\theta$ tends to infinity exponentially as $\tau\rightarrow-\infty$. 
\end{remark}
\begin{proof}
  Up to arguments that are similar to those of the proof of Lemma~\ref{lemma:basicenergyestimatequant}, the statement follows from
  Lemma~\ref{lemma:basicenergyestimatequant}. 
\end{proof}

\section{Wave operator, conformal rescaling}\label{section:conformalwaveoperator}

Our next goal is to derive energy estimates for higher order energies. However, we then need to commute the wave operator with the 
vector fields $E_{i}$. As a preliminary step, it is of interest to express the wave operator with respect to the frame given by $X_{0}:=\hU$ 
and the $X_{A}$. When doing so, it is convenient to use the following notation. The Christoffel symbols and contracted Christoffel symbols, denoted by
$\hG_{\a\b}^{\g}$ and $\hG^{\g}$ respectively, are defined by 
\begin{equation}\label{eq:ChrConChr}
\hna_{X_{\a}}X_{\b}=\hG_{\a\b}^{\g}X_{\g},\ \ \
\hG^{\g}:=\hg^{\a\b}\hG_{\a\b}^{\g}.
\end{equation}
Next, if the structure constants $\g^{A}_{BC}$ are defined as in Corollary~\ref{cor:covderofframe}, then
\begin{equation}\label{eq:aAdef}
a_{A}:=\frac{1}{2}\g^{B}_{AB}.
\end{equation}
\begin{lemma}\label{lemma:Boxhguocalcoord}
  Let $(M,g)$ be a time oriented Lorentz manifold. Assume it to have an expanding partial pointed foliation and $\mK$ to be non-degenerate on $I$ and to
  have a global frame. Then
  \begin{equation}\label{eq:Boxhguformula}
    \Box_{\hg}u=-\hU^{2}(u)+\textstyle{\sum}_{A}e^{-2\mu_{A}}X_{A}^{2}(u)-\chth\hU(u)-\hG^{A}X_{A}(u),
  \end{equation}
  where
  \begin{equation}\label{eq:hGAdef}
    \hG^{A}=-e^{-2\mu_{A}}X_{A}(\ln \hN)+2e^{-2\mu_{A}}X_{A}(\mu_{A})-e^{-2\mu_{A}}X_{A}(\mu_{\rotot})+2e^{-2\mu_{A}}a_{A}
  \end{equation}
  (no summation), $\mu_{\rotot}:=\textstyle{\sum}_{A}\mu_{A}$ and $a_{A}$ is defined by (\ref{eq:aAdef}). 
\end{lemma}
\begin{remark}
For future reference, it is of interest to note that the conclusion can also be written
\begin{equation}\label{eq:Boxguwrtframe}
\begin{split}
\Box_{\hg}u = & -\hU^{2}(u)+\textstyle{\sum}_{A}e^{-2\mu_{A}}X_{A}^{2}(u)-\chth \hU(u)\\
 & +\textstyle{\sum}_{C}e^{-2\mu_{C}}X_{C}(\ln \hN)X_{C}(u)-2\textstyle{\sum}_{C}e^{-2\mu_{C}}X_{C}(\mu_{C})X_{C}(u)\\
 & +\textstyle{\sum}_{C}e^{-2\mu_{C}}X_{C}(\mu_{\rotot})X_{C}(u)-2\textstyle{\sum}_{C}e^{-2\mu_{C}}a_{C}X_{C}(u).
\end{split}
\end{equation}
\end{remark}
\begin{proof}
Note, to begin with, that if $\hg_{\a\b}=\hg(X_{\a},X_{\b})$, then 
\begin{equation*}
\begin{split}
\Box_{\hg}u = & \hg^{\a\b}(\hna^{2}u)(X_{\a},X_{\b})=\hg^{\a\b}[\hna_{X_{\a}}(\hna u)(X_{\b})]\\
 = & \hg^{\a\b}[X_{\a}X_{\b}(u)-\hna_{\hna_{X_{\a}}X_{\b}}u]=\hg^{\a\b}X_{\a}X_{\b}(u)-\hg^{\a\b}\hG_{\a\b}^{\g}X_{\g}(u),
\end{split}
\end{equation*}
where we use the notation (\ref{eq:ChrConChr}). Thus, again using the notation introduced in (\ref{eq:ChrConChr}), 
\[
\Box_{\hg}u=-\hU^{2}(u)+\textstyle{\sum}_{A}e^{-2\mu_{A}}X_{A}^{2}(u)-\hG^{\g}X_{\g}(u).
\]
In order to proceed, it is of interest to note that if $\ldr{\cdot,\cdot}:=\hg$, then 
\[
\hG_{\a\b}^{0}=-\ldr{\hna_{X_{\a}}X_{\b},X_{0}},\ \ \
\hG_{\a\b}^{A}=e^{-2\mu_{A}}\ldr{\hna_{X_{\a}}X_{\b},X_{A}}
\]
(no summation on $A$). In particular, $\hG^{0}_{00}=0$ and 
\[
\hG_{AB}^{0}=-\ldr{\hna_{X_{A}}X_{B},X_{0}}=\ldr{X_{B},\hna_{X_{A}}X_{0}}=\chk_{AB},
\]
so that $\hG^{0}=\tr_{\chg}\chk=\chth$. Next, note that (\ref{eq:ldrnablaUUdj}) yields 
\[
\ldr{\hna_{X_{0}}X_{0},X_{A}}=X_{A}(\ln \hN).
\]
Moreover, the Koszul formula yields
\begin{equation*}
\begin{split}
\ldr{\hna_{X_{A}}X_{B},X_{C}} = & e^{2\mu_{C}}X_{A}(\mu_{C})\de_{BC}+e^{2\mu_{C}}X_{B}(\mu_{C})\de_{AC}-e^{2\mu_{A}}X_{C}(\mu_{A})\de_{AB}\\
  & -\frac{1}{2}e^{2\mu_{A}}\g^{A}_{BC}+\frac{1}{2}e^{2\mu_{B}}\g^{B}_{CA}+\frac{1}{2}e^{2\mu_{C}}\g^{C}_{AB}
\end{split}
\end{equation*}
(no summation). Combining the above observations yields
\begin{equation*}
\begin{split}
\hG^{C} = & \hg^{\a\b}\hG^{C}_{\a\b}=-\hG^{C}_{00}+\textstyle{\sum}_{A}e^{-2\mu_{A}}\hG^{C}_{AA}\\
 = & -e^{-2\mu_{C}}\ldr{\hna_{X_{0}}X_{0},X_{C}}+\textstyle{\sum}_{A}e^{-2\mu_{A}-2\mu_{C}}\ldr{\hna_{X_{A}}X_{A},X_{C}}\\
 = & -e^{-2\mu_{C}}X_{C}(\ln \hN)+2e^{-2\mu_{C}}X_{C}(\mu_{C})-e^{-2\mu_{C}}X_{C}(\mu_{\rotot})+2e^{-2\mu_{C}}a_{C}.
\end{split}
\end{equation*}
Summing up yields the conclusion of the lemma. 
\end{proof}

\chapter{Commutators}\label{chapter:commutators}

In the previous chapter, we derived zeroth order energy estimates. To obtain higher order energy estimates, we need to commute the differential operator
$L$ (corresponding to the left hand side in (\ref{eq:equationintermsofcanonicalframeintro})) with the spatial frame $\{E_{i}\}$. The purpose of the present
chapter is to derive formulae for the commutators of $E_{\bfI}$ with the individual terms in $L$. We also state estimates for the corresponding coefficients.
In the applications, we either extract the coefficients in $C^{0}$ (in case we make $(\cweight,l)$-supremum assumptions) or apply Moser estimates (in case
we make $(\cweight,l)$-Sobolev assumptions). The exact form of the commutator formulae and estimates that are most convenient depends on which of these
methods we use. For that reason, most of the commutator formulae and estimates come in two forms. 

\section{Commuting spatial derivatives with the wave operator, step I}

As a first step, we need to control the commutator of $E_{i}$ with the second order derivative operators appearing on the right hand side of
(\ref{eq:Boxguwrtframe}). We begin by calculating the commutator with $e^{-2\mu_{A}}X_{A}^{2}$. In the statement of the result, the following notation
will be useful.

\begin{definition}\label{def:mfPmKmuhN}
  Let $(M,g)$ be a time oriented Lorentz manifold. Assume it to have an expanding partial pointed foliation and $\mK$ to be non-degenerate on $I$ and to have
  a global frame. Given $0\leq m,k\in\zo$, let
  \begin{align*}
    \mfP_{\mu,m} := & \textstyle{\sum}_{m_{1}+\dots+m_{j}=m,m_{i}\geq 1}\sum_{A_{1},\dots,A_{j}}|\bD^{m_{1}}\mu_{A_{1}}|_{\bge_{\refer}}
    \cdots |\bD^{m_{j}}\mu_{A_{j}}|_{\bge_{\refer}},\\
    \mfP_{\mu,m,k} := & \textstyle{\sum}_{m_{1}+\dots+m_{j}=m,1\leq m_{i}\leq k}\sum_{A_{1},\dots,A_{j}}|\bD^{m_{1}}\mu_{A_{1}}|_{\bge_{\refer}}
    \cdots |\bD^{m_{j}}\mu_{A_{j}}|_{\bge_{\refer}},\\
    \mfP_{\mK,\mu,m} := & \textstyle{\sum}_{m_{1}+m_{2}=m}\mfP_{\mK,m_{1}}\mfP_{\mu,m_{2}},\\
    \mfP_{\mK,\mu,N,m} := & \textstyle{\sum}_{m_{1}+m_{2}+m_{3}=m}\mfP_{\mK,m_{1}}\mfP_{\mu,m_{2}}\mfP_{N,m_{3}},
  \end{align*}
  \index{$\a$Aa@Notation!Functions!$\mfP_{\mu,m}$}%
  \index{$\a$Aa@Notation!Functions!$\mfP_{\mu,m,k}$}%
  \index{$\a$Aa@Notation!Functions!$\mfP_{\mK,\mu,m}$}%
  \index{$\a$Aa@Notation!Functions!$\mfP_{\mK,\mu,N,m}$}%
  with the convention that $\mfP_{\mu,0}=\mfP_{\mu,0,k}=1$. 
\end{definition}

In situations where we make $(\cweight,l)$-supremum assumptions, the following form of the commutators and estimates are convenient. 

\begin{lemma}\label{lemma:EbfIemmuAXAsqcomm}
  Let $(M,g)$ be a time oriented Lorentz manifold. Assume it to have an expanding partial pointed foliation and $\mK$ to be non-degenerate on $I$, to have
  a global frame and to be $C^{0}$-uniformly bounded on $I_{-}$; i.e. (\ref{eq:mKsupbasest}) to hold. Then
  \begin{equation}\label{eq:EbfIemmuAXAsqcomm}
  [E_{\bfI},e^{-2\mu_{A}}X_{A}^{2}]\psi=\textstyle{\sum}_{1\leq |\bfJ|\leq |\bfI|}D_{\bfI,\bfJ}^{A}e^{-2\mu_{A}}X_{A}E_{\bfJ}\psi
  +\textstyle{\sum}_{1\leq |\bfJ|\leq |\bfI|}F_{\bfI,\bfJ}^{A}e^{-2\mu_{A}}E_{\bfJ}\psi
  \end{equation}
  (no summeation on $A$), where
  \begin{align}
    |D_{\bfI,\bfJ}^{A}| \leq & C\textstyle{\sum}_{m=0}^{l_{a}}\mfP_{\mK,\mu,m},\label{eq:DIJAest}\\
    |F_{\bfI,\bfJ}^{A}| \leq & C\textstyle{\sum}_{m=0}^{l_{b}}\textstyle{\sum}_{m_{1}+m_{2}=m}\mfP_{\mK,m_{1}}\mfP_{\mu,m_{2},l_{a}}\label{eq:FIJAest}
  \end{align}
  on $I_{-}$, $l_{a}:=|\bfI|+1-|\bfJ|$, $l_{b}:=|\bfI|+2-|\bfJ|$, and $C$ only depends on $|\bfI|$, $|\bfJ|$, $n$, $\mKsup$, $\e_{\rond}$ and
  $(\bM,\bge_{\refer})$. 
\end{lemma}
\begin{proof}
  Note that
  \begin{equation}\label{eq:EiXAcommutator}
  [E_{i},X_{A}]=B_{iA}^{k}E_{k},
  \end{equation}
  where
  \begin{equation}\label{eq:EiXAcommutatorterminology}
  B_{iA}^{k}:=E_{i}(X_{A}^{k})+X_{A}^{j}\eta_{ij}^{k},\ \ \
  \eta_{ij}^{k}:=\omega^{k}([E_{i},E_{j}]).
  \end{equation}
  Using this notation, it can be calculated that
  \begin{equation}\label{eq:EiemmuAXAsqcomm}
    \begin{split}
      [E_{i},e^{-2\mu_{A}}X_{A}^{2}] = & 2[B_{iA}^{k}-E_{i}(\mu_{A})X_{A}^{k}]e^{-2\mu_{A}}X_{A}E_{k}\\
      & +e^{-2\mu_{A}}[B_{iA}^{l}B_{lA}^{k}+X_{A}(B_{iA}^{k})-2E_{i}(\mu_{A})X_{A}(X_{A}^{k})]E_{k}.
    \end{split}
  \end{equation}
  Note also that
  \begin{equation}\label{eq:inductivestepcomm}
    [E_{i}E_{\bfI},e^{-2\mu_{A}}X_{A}^{2}]=E_{i}[E_{\bfI},e^{-2\mu_{A}}X_{A}^{2}]+[E_{i},e^{-2\mu_{A}}X_{A}^{2}]E_{\bfI}.
  \end{equation}
  Let $\bfI$ be a frame index with $|\bfI|\geq 1$. Next we prove, by induction, that (\ref{eq:EbfIemmuAXAsqcomm}) holds, where $D_{\bfI,\bfJ}^{A}$
  is a sum of terms of the form
  \[
  E_{\bfI_{1}}(\mu_{A})\cdots E_{\bfI_{m}}(\mu_{A})E_{\bfK}(X_{A}^{l})f,
  \]
  and $f$ is a function all of whose derivatives with respect to the frame $\{E_{i}\}$ can be bounded by constants depending only on
  $(\bM,\bge_{\refer})$ and the order of the derivative. Here $|\bfI_{1}|+\dots+|\bfI_{m}|+|\bfK|\leq |\bfI|+1-|\bfJ|$ and $\bfI_{l}\neq 0$.
  Similarly, $F_{\bfI,\bfJ}^{A}$ is a sum of terms of the form
  \[
  E_{\bfI_{1}}(\mu_{A})\cdots E_{\bfI_{m}}(\mu_{A})E_{\bfK_{1}}(X_{A}^{l_{1}})\cdots E_{\bfK_{p}}(X_{A}^{l_{p}})f,
  \]
  where $f$ is as before. Here $|\bfI_{1}|+\dots+|\bfI_{m}|+|\bfK_{1}|+\dots+|\bfK_{p}|\leq |\bfI|+2-|\bfJ|$ and $1\leq |\bfI_{j}|\leq |\bfI|+1-|\bfJ|$.
  Due to (\ref{eq:EiemmuAXAsqcomm}), the desired statement holds for $|\bfI|=1$. Assuming, inductively, that the desired statement holds
  and keeping (\ref{eq:inductivestepcomm}) in mind, it follows that the desired statement holds for all $\bfI$ such that $|\bfI|\geq 1$.
  Combining the above observation with Lemma~\ref{lemma:bDbfAbDkequiv} and (\ref{eq:bDbfAellAetcpteststmtEi}) yields the statement of the lemma. 
\end{proof}

In situations where we make $(\cweight,l)$-Sobolev assumptions, the following form of the commutators and estimates are convenient. 

\begin{lemma}\label{lemma:EbfIemmuAXAsqcommoo}
  Let $(M,g)$ be a time oriented Lorentz manifold. Assume it to have an expanding partial pointed foliation and $\mK$ to be non-degenerate on $I$, to have a
  global frame and to be $C^{0}$-uniformly bounded on $I_{-}$; i.e., (\ref{eq:mKsupbasest}) to hold. Then
  \begin{equation}\label{eq:EbfIemmuAXAsqcommoo}
  [E_{\bfI},e^{-2\mu_{A}}X_{A}^{2}]\psi=\textstyle{\sum}_{1\leq |\bfJ|\leq |\bfI|}\bD_{\bfI,\bfJ}^{A}e^{-\mu_{A}}E_{\bfJ}(e^{-\mu_{A}}X_{A}\psi)
  +\textstyle{\sum}_{1\leq |\bfJ|\leq |\bfI|}\bF_{\bfI,\bfJ}^{A}e^{-2\mu_{A}}E_{\bfJ}\psi
  \end{equation}
  (no summation on $A$), where
  \begin{align}
    |\bD_{\bfI,\bfJ}^{A}| \leq & C\textstyle{\sum}_{m=0}^{l_{a}}\mfP_{\mK,\mu,m},\label{eq:DIJAestoo}\\
    |\bF_{\bfI,\bfJ}^{A}| \leq & C\textstyle{\sum}_{m=0}^{l_{b}}\textstyle{\sum}_{m_{1}+m_{2}=m}\mfP_{\mK,m_{1}}\mfP_{\mu,m_{2},l_{a}}\label{eq:FIJAestoo}
  \end{align}
  on $I_{-}$,
  $l_{a}:=|\bfI|+1-|\bfJ|$, $l_{b}:=|\bfI|+2-|\bfJ|$, and $C$ only depends on $|\bfI|$, $|\bfJ|$, $n$, $\mKsup$, $\e_{\rond}$ and $(\bM,\bge_{\refer})$. 
\end{lemma}
\begin{proof}
  Note that (\ref{eq:EiXAcommutator}), (\ref{eq:EiXAcommutatorterminology}) and (\ref{eq:EiemmuAXAsqcomm}) hold. On the other hand, 
  \[
  e^{-2\mu_{A}}X_{A}E_{k}\psi=e^{-\mu_{A}}E_{k}(e^{-\mu_{A}}X_{A}\psi)+e^{-2\mu_{A}}E_{k}(\mu_{A})X_{A}\psi-e^{-2\mu_{A}}B_{kA}^{l}E_{l}\psi. 
  \]
  Combining this equality with (\ref{eq:EiemmuAXAsqcomm}) yields 
  \begin{equation}\label{eq:EiemmuAXAsqcommoo}
    \begin{split}
      & [E_{i},e^{-2\mu_{A}}X_{A}^{2}]\psi\\
      = & 2e^{-\mu_{A}}[B_{iA}^{k}-E_{i}(\mu_{A})X_{A}^{k}]E_{k}(e^{-\mu_{A}}X_{A}\psi)\\
      & +e^{-2\mu_{A}}[-B_{iA}^{l}B_{lA}^{k}+X_{A}(B_{iA}^{k})+2B_{iA}^{l}E_{l}(\mu_{A})X_{A}^{k}-2E_{i}(\mu_{A})X_{A}(\mu_{A})X_{A}^{k}]E_{k}\psi.
    \end{split}
  \end{equation}
  Note also that
  \begin{equation}\label{eq:inductivestepcommoo}
    [E_{\bfI}E_{i},e^{-2\mu_{A}}X_{A}^{2}]=E_{\bfI}[E_{i},e^{-2\mu_{A}}X_{A}^{2}]+[E_{\bfI},e^{-2\mu_{A}}X_{A}^{2}]E_{i}.
  \end{equation}
  Let $\bfI$ be a frame index with $|\bfI|\geq 1$. Next we prove, by induction, that (\ref{eq:EbfIemmuAXAsqcommoo}) holds, where $\bD_{\bfI,\bfJ}^{A}$
  is a sum of terms of the form
  \[
  E_{\bfI_{1}}(\mu_{A})\cdots E_{\bfI_{m}}(\mu_{A})E_{\bfK}(X_{A}^{l})f,
  \]
  and $f$ is a function all of whose derivatives with respect to the frame $\{E_{i}\}$ can be bounded by constants depending only on
  $(\bM,\bge_{\refer})$ and the order of the derivative. Here $|\bfI_{1}|+\dots+|\bfI_{m}|+|\bfK|\leq |\bfI|+1-|\bfJ|$ and $\bfI_{l}\neq 0$.
  Similarly, $\bF_{\bfI,\bfJ}^{A}$ is a sum of terms of the form
  \[
  E_{\bfI_{1}}(\mu_{A})\cdots E_{\bfI_{m}}(\mu_{A})E_{\bfK_{1}}(X_{A}^{l_{1}})\cdots E_{\bfK_{p}}(X_{A}^{l_{p}})f,
  \]
  where $f$ is as before. Here $|\bfI_{1}|+\dots+|\bfI_{m}|+|\bfK_{1}|+\dots+|\bfK_{p}|\leq |\bfI|+2-|\bfJ|$ and $1\leq |\bfI_{j}|\leq |\bfI|+1-|\bfJ|$.
  Due to (\ref{eq:EiemmuAXAsqcommoo}), the desired statement holds for $|\bfI|=1$. Assuming, inductively, that the desired statement holds
  and keeping (\ref{eq:inductivestepcommoo}) in mind, it can be demonstrated that the desired statement holds for all $\bfI$ such that $|\bfI|\geq 1$.
  The only nontrivial step consists in rewriting
  \begin{equation*}
    \begin{split}
      & \bD_{\bfI,\bfJ}^{A}e^{-\mu_{A}}E_{\bfJ}(e^{-\mu_{A}}X_{A}E_{i}\psi)\\
      = & \bD_{\bfI,\bfJ}^{A}e^{-\mu_{A}}E_{\bfJ}E_{i}(e^{-\mu_{A}}X_{A}\psi)
      +\bD_{\bfI,\bfJ}^{A}e^{-\mu_{A}}E_{\bfJ}[e^{-\mu_{A}}(E_{i}(\mu_{A})X_{A}^{k}E_{k}\psi-B_{iA}^{k}E_{k}\psi)].
    \end{split}
  \end{equation*}
  The first term on the right hand side is already of the desired form. Moreover, it can be demonstrated that the second term on the right hand side
  is of the form of the second sum on the right hand side of (\ref{eq:EbfIemmuAXAsqcommoo}). In addition, the corresponding contribution to
  $\bF_{\bfI_{a},\bfJ_{a}}$ is such that it satisfies the inductive hypothesis. Combining the above observation with Lemma~\ref{lemma:bDbfAbDkequiv}
  and (\ref{eq:bDbfAellAetcpteststmtEi}) yields the statement of the lemma. 
\end{proof}

\section{Commuting spatial derivatives with the wave operator, step II}

Next, we turn to the commutator with $\hU^{2}$, and we begin by deriving the form of the commutators and estimates that are convenient in the context
of the $(\cweight,l)$-supremum assumptions.

\begin{lemma}\label{lemma:bdbfAhUsqcommformEi}
  Let $(M,g)$ be a time oriented Lorentz manifold. Assume it to have an expanding partial pointed foliation and $\mK$ to be non-degenerate on $I$ and
  to have a global frame. Then
  \begin{equation}\label{eq:bdbfAhUsqcommformEi}
    [\hU^{2},E_{\bfI}]\psi = \textstyle{\sum}_{|\bfJ|\leq |\bfI|}\sum_{k=0}^{1}C_{\bfI,\bfJ}^{k}\hU^{k}E_{\bfJ}\psi
    +\textstyle{\sum}_{|\bfJ|\leq |\bfI|-1}C_{\bfI,\bfJ}^{2}\hU^{2}E_{\bfJ}\psi,
  \end{equation}
  where
  \begin{align}
    |C_{\bfI,\bfJ}^{2}| \leq & C\textstyle{\sum}_{m=1}^{l_{a}}\mfP_{N,m},\label{eq:CIJtwoest}\\
    |C_{\bfI,\bfJ}^{1}| \leq & C\textstyle{\sum}_{m+|\bfK|\leq l_{a}}\textstyle{\sum}_{i,k}\mfP_{N,m}|E_{\bfK}(A_{i}^{k})|\label{eq:CIJoneest}\\
    & +C\textstyle{\sum}_{1\leq m+|\bfK|\leq l_{a}}\mfP_{N,m}|E_{\bfK}\hU(\ln\hN)|\nonumber\\
    |C_{\bfI,\bfJ}^{0}| \leq & C\textstyle{\sum}_{m+|\bfK|\leq l_{a}}\textstyle{\sum}_{i,k}\mfP_{N,m}|E_{\bfK}\hU(A_{i}^{k})|\label{eq:CIJzeroest}\\
    & +C\textstyle{\sum}_{m+|\bfJ_{1}|+|\bfJ_{2}|\leq l_{a}}\textstyle{\sum}_{i,k}\mfP_{N,m}|E_{\bfJ_{1}}(A_{i}^{k})|\cdot|E_{\bfJ_{2}}\hU(\ln\hN)|\nonumber\\
    & +C\textstyle{\sum}_{m+|\bfJ_{1}|+|\bfJ_{2}|\leq l_{a}}\textstyle{\sum}_{i,k,p,q}\mfP_{N,m}|E_{\bfJ_{1}}(A_{i}^{k})|\cdot|E_{\bfJ_{2}}(A_{p}^{q})|,\nonumber
  \end{align}
  $l_{a}:=|\bfI|-|\bfJ|$ and $C$ only depends on $|\bfI|$, $|\bfJ|$, $n$ and $(\bM,\bge_{\refer})$. Finally, if $\bfJ=0$, then
  $C^{0}_{\bfI,\bfJ}=0$. 
  \end{lemma}
\begin{proof}
  Due to (\ref{eq:hUEicomm}),
  \begin{equation}\label{eq:hUsqEicommutator}
  [\hU^{2},E_{i}]=2A_{i}^{0}\hU^{2}+2A_{i}^{k}\hU E_{k}+[\hU(A_{i}^{0})-A_{i}^{k}A_{k}^{0}]\hU+[\hU(A_{i}^{k})-A_{i}^{l}A_{l}^{k}]E_{k}.
  \end{equation}
  Note also that
  \begin{equation}\label{eq:commutatorhUsqindstep}
  [\hU^{2},E_{i}E_{\bfI}]=E_{i}[\hU^{2},E_{\bfI}]+[\hU^{2},E_{i}]E_{\bfI}.
  \end{equation}
  Next, we wish to prove, using an inductive argument, that (\ref{eq:bdbfAhUsqcommformEi}) holds, where $C_{\bfI,\bfJ}^{2}$ is a linear combination
  of expressions of the form
  \begin{equation}\label{eq:CbfIbfJtwoterms}
    E_{\bfI_{1}}\ln\hN\cdots E_{\bfI_{k}}\ln\hN,
  \end{equation}
  where $|\bfI_{1}|+\dots+|\bfI_{k}|=|\bfI|-|\bfJ|$, $k\geq 1$ and $\bfI_{j}\neq 0$. Moreover, $C_{\bfI,\bfJ}^{1}$ is a linear combination of
  expressions of the form
  \begin{align}
    & E_{\bfI_{1}}\ln\hN\cdots E_{\bfI_{k}}\ln\hN\cdot E_{\bfK}(A_{i}^{r}),\label{eq:mfConeterms}\\
    & E_{\bfI_{1}}\ln\hN\cdots E_{\bfI_{k}}\ln\hN\cdot E_{\bfK}\hU\ln\hN,\label{eq:hConeterms}
  \end{align}
  where $|\bfI_{1}|+\dots+|\bfI_{k}|+|\bfK|=|\bfI|-|\bfJ|$, $\bfI_{j}\neq 0$ and $|\bfK|+k\geq 1$ in the second expression. Finally, $C_{\bfI,\bfJ}^{0}$ is a
  linear combination of expressions of the form
  \begin{align}
    & E_{\bfI_{1}}\ln\hN\cdots E_{\bfI_{k}}\ln\hN\cdot E_{\bfK}\hU(A_{i}^{r}),\label{eq:CzerotermsA}\\
    & E_{\bfI_{1}}\ln\hN\cdots E_{\bfI_{k}}\ln\hN\cdot E_{\bfJ_{1}}(A_{r}^{l})\cdot E_{\bfJ_{2}}(A_{p}^{q}),\label{eq:CzerotermsB}\\
    & E_{\bfI_{1}}\ln\hN\cdots E_{\bfI_{k}}\ln\hN\cdot E_{\bfJ_{1}}(A_{i}^{r})\cdot E_{\bfJ_{2}}\hU\ln\hN,\label{eq:CzerotermsC}
  \end{align}
  where $|\bfI_{1}|+\dots+|\bfI_{k}|+|\bfK|=|\bfI|-|\bfJ|$; $|\bfI_{1}|+\dots+|\bfI_{k}|+|\bfJ_{1}|+|\bfJ_{2}|=|\bfI|-|\bfJ|$; $\bfI_{j}\neq 0$; and
  $k+|\bfJ_{2}|\geq 1$ in the last expression. Moreover, if $\bfJ=0$, then $C^{0}_{\bfI,\bfJ}=0$. 

  In order to prove the above statement, note that it holds for $|\bfI|=1$. This follows from (\ref{eq:hUsqEicommutator}), keeping in mind that
  $A_{i}^{0}=E_{i}(\ln\hN)$ and that
  \begin{equation}\label{eq:hUAizrewritten}
    \begin{split}
      \hU(A_{i}^{0}) = & \hU E_{i}(\ln\hN)=[\hU,E_{i}](\ln\hN)+E_{i}[\hU(\ln\hN)]\\
      = & A_{i}^{0}\hU(\ln\hN)+A_{i}^{k}E_{k}(\ln\hN)+E_{i}[\hU(\ln\hN)].
    \end{split}
  \end{equation}
  In order to prove the statement in general, assume that it holds for frame indices $\bfI$ such that $1\leq |\bfI|\leq m$ and let $\bfI$ be
  a frame index such that $|\bfI|=m$. Given $i\in \{1,\dots,n\}$, we wish to prove that the left hand side of (\ref{eq:commutatorhUsqindstep}),
  applied to a function $\psi$, satisfies the desired statement. In the case of the second term on the right hand side of
  (\ref{eq:commutatorhUsqindstep}), this follows from the fact that the inductive assumption holds for $|\bfI|=1$. Concerning the first term on
  the right hand side of (\ref{eq:commutatorhUsqindstep}), combining this term with the inductive assumptions, it can immediately be verified that
  most of the resulting terms are of the desired form. However, special attention needs to be devoted to
  \[
  \textstyle{\sum}_{|\bfJ|\leq |\bfI|}C_{\bfI,\bfJ}^{1}[E_{i},\hU] E_{\bfJ}\psi
  +\textstyle{\sum}_{|\bfJ|\leq |\bfI|-1}C_{\bfI,\bfJ}^{2}[E_{i},\hU^{2}]E_{\bfJ}\psi.
  \]
  However, keeping (\ref{eq:hUEicomm}) and (\ref{eq:hUsqEicommutator}) in mind, the resulting terms also fit into the inductive hypothesis.

  In order to deduce the conclusion of the lemma, it is sufficient to note that the products of the $E_{\bfI_{j}}\ln\hN$ can be estimated by
  sums of $\mfP_{N,m}$.
\end{proof}

When we make $(\cweight,l)$-Sobolev assumptions, the following forms of the commutators and estimates are convenient. 

\begin{lemma}\label{lemma:bdbfAhUsqcommformEioo}
  Let $(M,g)$ be a time oriented Lorentz manifold. Assume it to have an expanding partial pointed foliation and $\mK$ to be non-degenerate on $I$ and to
  have a global frame. Then
  \begin{equation}\label{eq:bdbfAhUsqcommformEioo}
    [\hU^{2},E_{\bfI}]\psi = \textstyle{\sum}_{|\bfJ|\leq |\bfI|}\sum_{k=0}^{1}\bC_{\bfI,\bfJ}^{k}E_{\bfJ}\hU^{k}\psi
    +\textstyle{\sum}_{|\bfJ|\leq |\bfI|-1}\bC_{\bfI,\bfJ}^{2}E_{\bfJ}\hU^{2}\psi,
  \end{equation}
  where
  \begin{align}
    |\bC_{\bfI,\bfJ}^{2}| \leq & C\textstyle{\sum}_{m=1}^{l_{a}}\mfP_{N,m},\label{eq:CIJtwoestoo}\\
    |\bC_{\bfI,\bfJ}^{1}| \leq & C\textstyle{\sum}_{m+|\bfK|\leq l_{a}}\textstyle{\sum}_{i,k}\mfP_{N,m}|E_{\bfK}(A_{i}^{k})|\label{eq:CIJoneestoo}\\
    & +C\textstyle{\sum}_{1\leq m+|\bfK|\leq l_{a}}\mfP_{N,m}|E_{\bfK}\hU(\ln\hN)|\nonumber\\
    |\bC_{\bfI,\bfJ}^{0}| \leq & C\textstyle{\sum}_{m+|\bfK|\leq l_{a}}\textstyle{\sum}_{i,k}\mfP_{N,m}|E_{\bfK}\hU(A_{i}^{k})|\label{eq:CIJzeroestoo}\\
    & +C\textstyle{\sum}_{m+|\bfJ_{1}|+|\bfJ_{2}|\leq l_{a}}\textstyle{\sum}_{i,k}\mfP_{N,m}|E_{\bfJ_{1}}(A_{i}^{k})|\cdot|E_{\bfJ_{2}}\hU(\ln\hN)|\nonumber\\
    & +C\textstyle{\sum}_{m+|\bfJ_{1}|+|\bfJ_{2}|\leq l_{a}}\textstyle{\sum}_{i,k,p,q}\mfP_{N,m}|E_{\bfJ_{1}}(A_{i}^{k})|\cdot|E_{\bfJ_{2}}(A_{p}^{q})|,\nonumber
  \end{align}
  $l_{a}:=|\bfI|-|\bfJ|$ and $C$ only depends on $|\bfI|$, $|\bfJ|$, $n$ and $(\bM,\bge_{\refer})$. Finally, if $\bfJ=0$, then
  $\bC^{0}_{\bfI,\bfJ}=0$. 
  \end{lemma}
\begin{proof}
  Due to (\ref{eq:hUEicomm}),
  \begin{equation}\label{eq:hUsqEicommutatoroo}
    [\hU^{2},E_{i}]=2A_{i}^{0}\hU^{2}+2A_{i}^{k}E_{k}\hU+[\hU(A_{i}^{0})+A_{i}^{k}A_{k}^{0}]\hU+[\hU(A_{i}^{k})+A_{i}^{l}A_{l}^{k}]E_{k}.
  \end{equation}
  Note also that
  \begin{equation}\label{eq:commutatorhUsqindstepoo}
  [\hU^{2},E_{\bfI}E_{i}]=E_{\bfI}[\hU^{2},E_{i}]+[\hU^{2},E_{\bfI}]E_{i}.
  \end{equation}
  Next, we wish to prove, using an inductive argument, that (\ref{eq:bdbfAhUsqcommformEioo}) holds, where $\bC_{\bfI,\bfJ}^{2}$ is a linear combination
  of expressions of the form
  \[
  E_{\bfI_{1}}\ln\hN\cdots E_{\bfI_{k}}\ln\hN,
  \]
  where $|\bfI_{1}|+\dots+|\bfI_{k}|=|\bfI|-|\bfJ|$, $k\geq 1$ and $\bfI_{j}\neq 0$. Moreover, $\bC_{\bfI,\bfJ}^{1}$ is a linear combination of
  expressions of the form
  \begin{align*}
    & E_{\bfI_{1}}\ln\hN\cdots E_{\bfI_{k}}\ln\hN\cdot E_{\bfK}(A_{i}^{k}),\\
    & E_{\bfI_{1}}\ln\hN\cdots E_{\bfI_{k}}\ln\hN\cdot E_{\bfK}\hU\ln\hN,
  \end{align*}
  where $|\bfI_{1}|+\dots+|\bfI_{k}|+|\bfK|=|\bfI|-|\bfJ|$, $\bfI_{j}\neq 0$ and $|\bfK|+k\geq 1$ in the second expression. Finally, 
  $\bC_{\bfI,\bfJ}^{0}$ is a linear combination of expressions of the form
  \begin{align*}
    & E_{\bfI_{1}}\ln\hN\cdots E_{\bfI_{k}}\ln\hN\cdot E_{\bfK}\hU(A_{i}^{k}),\\
    & E_{\bfI_{1}}\ln\hN\cdots E_{\bfI_{k}}\ln\hN\cdot E_{\bfJ_{1}}(A_{k}^{l})\cdot E_{\bfJ_{2}}(A_{p}^{q}),\\
    & E_{\bfI_{1}}\ln\hN\cdots E_{\bfI_{k}}\ln\hN\cdot E_{\bfJ_{1}}(A_{i}^{k})\cdot E_{\bfJ_{2}}\hU\ln\hN,
  \end{align*}
  where $|\bfI_{1}|+\dots+|\bfI_{k}|+|\bfK|=|\bfI|-|\bfJ|$; $|\bfI_{1}|+\dots+|\bfI_{k}|+|\bfJ_{1}|+|\bfJ_{2}|=|\bfI|-|\bfJ|$; $\bfI_{j}\neq 0$; and
  $k+|\bfJ_{2}|\geq 1$ in the last expression. Moreover, if $\bfJ=0$, then $\bC^{0}_{\bfI,\bfJ}=0$. 

  In order to prove the above statement, note that it holds for $|\bfI|=1$. This follows from (\ref{eq:hUsqEicommutatoroo}), keeping in mind that
  (\ref{eq:hUAizrewritten}) and $A_{i}^{0}=E_{i}(\ln\hN)$ hold. In order to prove the statement in general, assume that it holds for frame indices 
  $\bfI$ such that $1\leq |\bfI|\leq m$ and let $\bfI$ be a frame index such that $|\bfI|=m$. Given $i\in \{1,\dots,n\}$, we wish to prove that 
  the left hand side of (\ref{eq:commutatorhUsqindstepoo}), applied to a function $\psi$, satisfies the desired statement. In the case of the first 
  term on the right hand side of (\ref{eq:commutatorhUsqindstepoo}), this follows from the fact that the inductive assumption holds for $|\bfI|=1$.   
  Concerning the second term on the right hand side of (\ref{eq:commutatorhUsqindstepoo}), combining this term with the inductive assumptions, it can 
  immediately be verified that some of the resulting terms are of the desired form. However, special attention needs to be devoted to
  \[
  \textstyle{\sum}_{|\bfJ|\leq |\bfI|}\bC_{\bfI,\bfJ}^{1}E_{\bfJ}[\hU,E_{i}] \psi
  +\textstyle{\sum}_{|\bfJ|\leq |\bfI|-1}\bC_{\bfI,\bfJ}^{2}E_{\bfJ}[\hU^{2},E_{i}]\psi.
  \]
  However, keeping (\ref{eq:hUEicomm}) and (\ref{eq:hUsqEicommutatoroo}) in mind, the resulting terms also fit into the inductive hypothesis.

  In order to deduce the conclusion of the lemma, it is sufficient to note that the products of the $E_{\bfI_{j}}\ln\hN$ can be estimated by
  sums of $\mfP_{N,m}$.
\end{proof}

\section{Commuting the equation with spatial derivatives}

Combining (\ref{eq:theconfresceq}) with  (\ref{eq:Boxhguformula}) yields the conclusion that (\ref{eq:theconfresceq}) can be written
\begin{equation}\label{eq:theeqreformEi}
  \begin{split}
    Lu=\hf,
  \end{split}
\end{equation}
where 
\begin{align}
  L := & -\hU^{2}+\textstyle{\sum}_{A}e^{-2\mu_{A}}X_{A}^{2}+\hmcY^{0}\hU+\hmcY^{B}X_{B}
  +\hmcX^{0}\hU+\hmcX^{B}X_{B}+\hal ,\label{eq:LuformulaEi}\\
  \hmcY^{0} := & -\frac{1}{n}\chth-\frac{n-1}{n},\label{eq:hmcYzdefEi}\\
  \hmcY^{A} = & -\hG^{A}-(n-1)e^{-2\mu_{A}}X_{A}(\ln\theta).\label{eq:hmcYAdefEi}
\end{align}
\index{$\a$Aa@Notation!Operators!$L$}%
\index{$\a$Aa@Notation!Functions!$\hmcY^{0}$}%
\index{$\a$Aa@Notation!Functions!$\hmcY^{A}$}%
Due to the above formulae, it is of interest to calculate the commutator of $E_{\bfI}$ with $Z^{0}\hU$ and $Z^{A}X_{A}$ for matrix valued functions
$Z^{0}$ and $Z^{A}$. 

\begin{lemma}\label{lemma:bdbfAchthhUcommformEi}
  Let $(M,g)$ be a time oriented Lorentz manifold. Assume it to have an expanding partial pointed foliation and $\mK$ to be non-degenerate on $I$ and to
  have a global frame. Let $Z^{0}$ be a smooth matrix valued function on $\bM\times I$. Then
  \begin{equation}\label{eq:bDbfAchthhUcommEi}
    [E_{\bfI},Z^{0}\hU]=\textstyle{\sum}_{|\bfJ|\leq |\bfI|-1}G_{\bfI,\bfJ}^{1}\hU E_{\bfJ}
    +\textstyle{\sum}_{1\leq |\bfJ|\leq |\bfI|}G_{\bfI,\bfJ}^{0}E_{\bfJ},
  \end{equation}
  where
  \begin{align*}
    \|G_{\bfI,\bfJ}^{1}\| \leq & C_{a}\textstyle{\sum}_{k_{a}+|\bfK|\leq l_{a}}\mfP_{N,k_{a}}\|E_{\bfK}(Z^{0})\|,\\
    \|G_{\bfI,\bfJ}^{0}\| \leq & C_{a}\textstyle{\sum}_{k_{a}+|\bfJ_{1}|+|\bfJ_{2}|\leq l_{a}}\textstyle{\sum}_{i,k}\mfP_{N,k_{a}}
    | E_{\bfJ_{1}}(A_{i}^{k})|\cdot\|E_{\bfJ_{2}}(Z^{0})\|,
  \end{align*}
  $l_{a}=|\bfI|-|\bfJ|$ and $C_{a}$ only depends on $|\bfI|$, $n$ and $(\bM,\bge_{\refer})$.
\end{lemma}
\begin{proof}
  We begin by proving the following statement inductively: (\ref{eq:bDbfAchthhUcommEi}) holds, where $G_{\bfI,\bfJ}^{1}$ is a linear combination
  of terms of the form
  \begin{equation}\label{eq:commZzeroEbfIfirsttype}
    E_{\bfI_{1}}(\ln\hN)\cdots E_{\bfI_{k}}(\ln\hN)E_{\bfK}(Z^{0}),
  \end{equation}
  $\bfI_{j}\neq 0$ and $|\bfI_{1}|+\dots+|\bfI_{k}|+|\bfK|=|\bfI|-|\bfJ|$. Moreover, $G_{\bfI,\bfJ}^{0}$ is a linear combination of terms of the form
  \begin{equation}\label{eq:commZzeroEbfIsecondtype}
    E_{\bfI_{1}}(\ln\hN)\cdots E_{\bfI_{k}}(\ln\hN)E_{\bfJ_{1}}(A_{i}^{k})E_{\bfJ_{2}}(Z^{0}),
  \end{equation}
  $\bfI_{j}\neq 0$ and $|\bfI_{1}|+\dots+|\bfI_{k}|+|\bfJ_{1}|+|\bfJ_{2}|=|\bfI|-|\bfJ|$. In order to prove the statement, compute
  \[
  [E_{i},Z^{0}\hU]=E_{i}(Z^{0})\hU+Z^{0} [E_{i},\hU]=E_{i}(Z^{0})\hU-A_{i}^{0}Z^{0}\hU-A_{i}^{k}Z^{0}E_{k}.
  \]
  This equality demonstrates that the statement holds in case $|\bfI|=1$. Next, note that 
  \begin{equation}\label{eq:bDbfAchthhUcommdecompEi}
    [E_{i}E_{\bfI},Z^{0}\hU]=E_{i}[E_{\bfI},Z^{0}\hU]+[E_{i},Z^{0}\hU] E_{\bfI}.
  \end{equation}
  We consider the terms on the right hand side of (\ref{eq:bDbfAchthhUcommdecompEi}) separately. Appealing to the
  inductive assumption, the first term on the right hand side can be written
  \[
  E_{i}\left(\textstyle{\sum}_{|\bfJ|\leq |\bfI|-1}G_{\bfI,\bfJ}^{1}\hU E_{\bfJ}
  +\textstyle{\sum}_{1\leq |\bfJ|\leq |\bfI|}G_{\bfI,\bfJ}^{0}E_{\bfJ}\right).
  \]
  Most of the terms that result when expanding this expression fit into the induction hypothesis. However, we need to consider
  \[
  \textstyle{\sum}_{|\bfJ|\leq |\bfI|-1}G_{\bfI,\bfJ}^{1}[E_{i},\hU]E_{\bfJ}
  \]
  more carefully. However, appealing to (\ref{eq:hUEicomm}), it is clear that this expression also fits into the induction hypothesis.
  Finally, the second term on the right hand side of (\ref{eq:bDbfAchthhUcommdecompEi}) can be rewritten in the desired form by appealing to the induction
  hypothesis for $|\bfI|=1$. Thus the desired statement holds. 

  Given the above statement, the conclusions of the lemma follow by arguments similar to the ones used in the proofs of the previous lemmas. 
\end{proof}

It will also be of interest to know that the following, related, result holds. 

\begin{lemma}\label{lemma:bdbfAchthhUcommformEireverse}
  Let $(M,g)$ be a time oriented Lorentz manifold. Assume it to have an expanding partial pointed foliation and $\mK$ to be non-degenerate on $I$ and to
  have a global frame. Let $Z^{0}$ be a smooth matrix valued function on $\bM\times I$. Then
  \begin{equation}\label{eq:bDbfAchthhUcommEireverse}
    [E_{\bfI},Z^{0}\hU]=\textstyle{\sum}_{|\bfJ|\leq |\bfI|-1}\bGe_{\bfI,\bfJ}^{1}E_{\bfJ}\hU 
    +\textstyle{\sum}_{1\leq |\bfJ|\leq |\bfI|}\bGe_{\bfI,\bfJ}^{0}E_{\bfJ},
  \end{equation}
  where
  \begin{align*}
    \|\bGe_{\bfI,\bfJ}^{1}\| \leq & C_{a}\textstyle{\sum}_{k_{a}+|\bfK|\leq l_{a}}\mfP_{N,k_{a}}\|E_{\bfK}(Z^{0})\|,\\
    \|\bGe_{\bfI,\bfJ}^{0}\| \leq & C_{a}\textstyle{\sum}_{k_{a}+|\bfJ_{1}|+|\bfJ_{2}|\leq l_{a}}\textstyle{\sum}_{i,k}\mfP_{N,k_{a}}
    | E_{\bfJ_{1}}(A_{i}^{k})|\cdot\|E_{\bfJ_{2}}(Z^{0})\|,
  \end{align*}
  $l_{a}:=|\bfI|-|\bfJ|$ and $C_{a}$ only depends on $|\bfI|$, $n$ and $(\bM,\bge_{\refer})$.
\end{lemma}
\begin{proof}
  The proof is similar to that of Lemma~\ref{lemma:bdbfAchthhUcommformEi}.
\end{proof}

Finally, we need to calculate the commutator of $ E_{\bfI}$ and $Z^{A}X_{A}$. 

\begin{lemma}\label{lemma:bdbfAGAXGcommformEi}
  Let $(M,g)$ be a time oriented Lorentz manifold. Assume it to have an expanding partial pointed foliation and $\mK$ to be non-degenerate on $I$, to
  have a global frame and to be $C^{0}$-uniformly bounded on $I_{-}$; i.e., (\ref{eq:mKsupbasest}) to hold. Let $Z^{A}$, $A=1,\dots,n$, be smooth matrix
  valued functions on $\bM\times I$. Then
  \begin{equation}\label{eq:bDbfAGAXAcommEi}
    [E_{\bfI},Z^{A}X_{A}]=\textstyle{\sum}_{1\leq |\bfJ|\leq |\bfI|}H_{\bfI,\bfJ} E_{\bfJ},
  \end{equation}
  where
  \[
  \|H_{\bfI,\bfJ}\|\leq C_{a}\textstyle{\sum}_{k_{a}+|\bfK|\leq l_{b}}\textstyle{\sum}_{A}\mfP_{\mK,k_{a}}\|E_{\bfK}(Z^{A})\|
  \]
  on $I_{-}$, $l_{b}:=|\bfI|-|\bfJ|+1$ and $C_{a}$ only depends on $\mKsup$, $\e_{\rond}$, $|\bfI|$, $n$ and $(\bM,\bge_{\refer})$.
\end{lemma}
\begin{proof}
  We begin by proving the following statement inductively: (\ref{eq:bDbfAGAXAcommEi}) holds, where $H_{\bfI,\bfJ}$ is a sum of
  expressions of the form
  \begin{equation}\label{eq:HABlincombofEi}
     f E_{\bfJ_{1}}(X_{A}^{i})E_{\bfJ_{2}}(Z^{A})
  \end{equation}
  where $|\bfJ_{1}|+|\bfJ_{2}|\leq |\bfI|+1-|\bfJ|$ and $f$ is a function all of whose derivatives with respect to the frame $\{E_{i}\}$ can be
  bounded by constants depending only on $(\bM,\bge_{\refer})$ and the order of the derivative. Compute, to this end,
  \[
  [E_{i},Z^{A}X_{A}]=E_{i}(Z^{A})X_{A}+Z^{A}[E_{i},X_{A}]=E_{i}(Z^{A})X_{A}+Z^{A}B_{iA}^{k}E_{k},
  \]
  where we appealed to (\ref{eq:EiXAcommutator}). This equality demonstrates that (\ref{eq:bDbfAGAXAcommEi}) holds for $|\bfI|=1$. Next, note
  that
  \begin{equation}\label{eq:bDbfAGAXAcommdecompEi}
    [E_{i}E_{\bfI},Z^{A}X_{A}]=E_{i}[E_{\bfI},Z^{A}X_{A}]+[E_{i},Z^{A}X_{A}] E_{\bfI}.
  \end{equation}
  We consider the terms on the right hand side of (\ref{eq:bDbfAGAXAcommdecompEi}) separately. Appealing to the inductive assumption, the first
  term on the right hand side can be written
  \[
  E_{i}\left(\textstyle{\sum}_{1\leq |\bfJ|\leq |\bfI|}H_{\bfI,\bfJ} E_{\bfJ}\right).
  \]
  The terms that result when expanding this expression fit into the induction hypothesis. Finally, the second term on the right hand side of
  (\ref{eq:bDbfAGAXAcommdecompEi}) can be rewritten in the desired form by appealing to the induction hypothesis for $|\bfI|=1$.

  Keeping (\ref{eq:bDbfAellAetcpteststmtEi}) in mind, the conclusions of the lemma follow by arguments similar to the ones used in the proofs
  of the previous lemmas. 
\end{proof}

\chapter{Higher order energy estimates, part I}\label{chapter:higherorderenergyestimatespartI}

Given the material of the previous two chapters, we are in a position to derive higher order energy estimates. Due to the zeroth order
energy estimate stated in Chapter~\ref{chapter:systemsofwaveequations}, it is sufficient to estimate $[L,E_{\bfI}]u$ in $L^{2}$. To obtain such an
estimate, we, in the present chapter, make $(\cweight,l)$-supremum assumptions. This allows us to extract the coefficients of the derivatives of $u$
appearing in $[L,E_{\bfI}]u$ in $C^{0}$ when estimating the commutator. Moreover, the $C^{0}$-estimates of the coefficients follow by combining
the commutator estimates of the previous chapter with the $(\cweight,l)$-supremum assumptions.

In Section~\ref{section:higherorderenergyestimatesulsuprass}, we record the conclusions concerning the higher order energies that can immediately
be obtained from the zeroth order energy estimates. We also isolate the quantities that remain to be estimated. Next, we devote
Sections~\ref{section:commwithhUsqEi}-\ref{section:summingupulsupass} to estimating $[L,E_{\bfI}]u$. The desired conclusions mainly follow from
the commutator estimates of the previous chapter and the $(\cweight,l)$-supremum assumptions. However, it is also necessary to estimate
expressions such as $\hU^{2}E_{\bfI}u$, and to this end, it is necessary to use the fact that (\ref{eq:theequation}) is satisfied. Combining the above
results yields a higher order energy estimate; cf. Section~\ref{section:firstenergyestimateulsupass}. In order to obtain the desired conclusion, we
use induction on the order of the energy.
It is also of interest to obtain weighted $C^{k}$ estimates of the unknown. To this end, we derive weighted Sobolev embedding estimates in
Section~\ref{section:weightedSobembulsupass}. Combining these estimates with the higher order energy estimates yields weighted $C^{k}$-control
of the unknown in Section~\ref{section:Ckestimatesweightedenergydensity}. 

\section{Higher order energies}\label{section:higherorderenergyestimatesulsuprass}

Prior to carrying out estimates, it is convenient to fix $\tau_{c}\leq 0$ and to introduce the notation
\begin{align}
  \me_{k}[u] := & \sum_{|\bfI|\leq k}\me[E_{\bfI}u]\label{eq:mektaudefEi}\\
  = & \frac{1}{2}\sum_{|\bfI|\leq k}\left(|\hU(E_{\bfI}u)|^{2}+\textstyle{\sum}_{A}e^{-2\mu_{A}}|X_{A}(E_{\bfI}u)|^{2}
  +\iota_{a}|E_{\bfI}u|^{2}+\iota_{b}\ldr{\tau-\tau_{c}}^{-3}|E_{\bfI}u|^{2}\right),\nonumber\\
  \hE_{k}[u](\tau;\tau_{c}) := & \sum_{|\bfI|\leq k}\hE[E_{\bfI}u](\tau;\tau_{c})=\int_{\bM_{\tau}}\me_{k}[u]\mutgc\label{eq:hEktaudefEi}
\end{align}
\index{$\a$Aa@Notation!Energy densities!$\me_{k}[u]$}%
\index{$\a$Aa@Notation!Energies!$\hE_{k}[u](\tau;\tau_{c})$}%
for all $\tau\leq\tau_{c}$, where we use the notation introduced in (\ref{eq:hEtaudef}) and (\ref{eq:medef}) as well as 
\begin{equation}\label{eq:mutgdef}
  \mutgc:=\tvarphi_{c}^{-1}\theta^{-(n-1)}\mu_{\chg}.
\end{equation}
\index{$\a$Aa@Notation!Volume forms!$\mutgc$}%
Commuting (\ref{eq:theeqreformEi}) with $E_{\bfI}$ yields
\begin{equation}\label{eq:LbDbfAuexprhfBfAdefEi}
  L(E_{\bfI}u)=E_{\bfI}\hf+[L,E_{\bfI}]u=:\hf_{\bfI}. 
\end{equation}
Assuming the conditions of Definition~\ref{def:basicassumptions} and Lemma~\ref{lemma:taurelvaryingbxEi} to be fulfilled; (\ref{eq:theequation}) to be
$C^{0}$-balanced on $I_{-}$; (\ref{eq:cthetaoneestimate}) to hold; and $q$ to be bounded on $M$, (\ref{eq:basicenergyestimate}) implies that for all
$\tau_{a}\leq \tau_{b}\leq\tau_{c}\leq 0$, 
\begin{equation}\label{eq:elbasenergyestimateEi}
  \begin{split}
    \hE_{k}(\tau_{a};\tau_{c}) \leq &  \hE_{k}(\tau_{b};\tau_{c})+\int_{\tau_{a}}^{\tau_{b}}\kappa(\tau)\hE_{k}(\tau;\tau_{c})d\tau\\
    & +\int_{\tau_{a}}^{\tau_{b}}\int_{\bM_{\tau}}\textstyle{\sum}_{|\bfI|\leq k}\tN|\hf_{\bfI}|\cdot |\hU(E_{\bfI}u)|\mutgc d\tau,
  \end{split}
\end{equation}
where $\kappa$ has the properties stated in Corollary~\ref{cor:basicenergyestimate}. We wish to estimate the last term on the right hand side.
Keeping in mind that $\tN=\hN/\d_{t}\tau$ is globally bounded, cf. (\ref{eq:hNtaudotequivEi}), it is clear that it is bounded by 
\[
C\int_{\tau_{a}}^{\tau_{b}}\left(\int_{\bM_{\tau}}\textstyle{\sum}_{|\bfI|\leq k}|\hf_{\bfI}|^{2}\mutgc\right)^{1/2}\hE_{k}^{1/2}[u]d\tau.
\]
Due to this observation and (\ref{eq:LbDbfAuexprhfBfAdefEi}) it is natural to focus on estimating 
\begin{equation}\label{eq:thecommutatorEi}
  \int_{\bM_{\tau}}\textstyle{\sum}_{|\bfI|\leq k}|[L,E_{\bfI}]u|^{2}\mutgc.
\end{equation}
Keeping (\ref{eq:LuformulaEi}) in mind, the estimate naturally breaks into the following parts. 

\section{Commutator with $\hU^{2}$}\label{section:commwithhUsqEi}

In order to estimate the contribution from $[\hU^{2},E_{\bfI}]u$, we appeal to Lemma~\ref{lemma:bdbfAhUsqcommformEi}. Due to 
(\ref{eq:bdbfAhUsqcommformEi}), we begin by considering 
\[
\textstyle{\sum}_{k=0}^{1}|C_{\bfI,\bfJ}^{k}\hU^{k}E_{\bfJ}u|^{2}.
\]
We need two different types of estimates. Up to a certain degree of regularity, we need to estimate $C_{\bfI,\bfJ}^{k}$ in $L^{\infty}$. The purpose of 
the corresponding energy estimates is to obtain $L^{\infty}$-estimates of $u$, its first derivatives etc. Once these estimates have been obtained, we use 
Moser estimates to control $\bC_{\bfI,\bfJ}^{k}E_{\bfJ}\hU^{k}u$ in $L^{2}$; cf. Chapter~\ref{chapter:hoeepII} below. 

\begin{lemma}\label{lemma:CkbfBbfAClaestEi}
  Fix $l$, $\bfl_{1}$, $\cweight$, $\weight_{0}$ and $\weight$ as in Definition~\ref{def:supmfulassumptions}. Assume that the conditions of
  Lemma~\ref{lemma:taurelvaryingbxEi} and the $(\cweight,l)$-supremum assumptions are satisfied. Let $\bfI$ and $\bfJ$ be frame indices such that
  $l_{a}:=|\bfI|-|\bfJ|$ satisfies $0\leq l_{a}\leq l$. Then, 
  \begin{align}
    \ldr{\varrho}^{-l_{a}\cweight}|C^{2}_{\bfI,\bfJ}| \leq & C_{a},\label{eq:CtwobfIbfJaE}\\
    \ldr{\varrho}^{-(l_{a}+1)\cweight}|C^{1}_{\bfI,\bfJ}| \leq & C_{a}e^{\e_{\Spe}\varrho}+\iota_{l_{a}}C_{a}\label{eq:ConebfIbfJaE}
  \end{align}
  on $M_{-}$, where $\iota_{k}=0$ if $k=0$ and $\iota_{k}=1$ if $k\geq 1$. Moreover, $C_{a}$ only depends on $c_{\cweight,l}$ and $(\bM,\bge_{\refer})$.
  Next, assume, in addition to the above, that $|\bfI|\leq l$. Then 
  \begin{equation}\label{eq:CzerobfIbfJaE}
    \ldr{\varrho}^{-(l_{a}+2)\cweight}|C^{0}_{\bfI,\bfJ}|\leq C_{a}e^{\e_{\Spe}\varrho}
  \end{equation}
  on $M_{-}$, where $C_{a}$ only depends on $c_{\cweight,l}$ and $(\bM,\bge_{\refer})$. 
\end{lemma}
\begin{proof}
  Note, to begin with, that combining (\ref{eq:CIJtwoest}) with the assumptions yields (\ref{eq:CtwobfIbfJaE}). Next, consider (\ref{eq:CIJoneest}).
  In order to estimate weighted versions of the first term on the right hand side, we appeal to (\ref{eq:AikClestsobassumpEivarrho}). The second term
  on the right hand side of (\ref{eq:CIJoneest}) can simply be estimated by appealing to the assumptions; cf. Definition~\ref{def:supmfulassumptions}.
  Note, however, that the second term on the right hand side of (\ref{eq:CIJoneest}) vanishes if $l_{a}=0$. This yields (\ref{eq:ConebfIbfJaE}). Finally,
  consider (\ref{eq:CIJzeroest}). Note that if $|\bfJ|=0$, then $C^{0}_{\bfI,\bfJ}=0$. Only in the case that $|\bfJ|\geq 1$ is there thus something to
  estimate. In particular, we can assume that $l_{a}\leq l-1$, since $|\bfI|\leq l$. In order to estimate weighted versions of the first term on the
  right hand side of (\ref{eq:CIJzeroest}), we appeal to (\ref{eq:hUAikClestulsobassvarrho}). The remaining two terms on the right hand side of
  (\ref{eq:CIJzeroest}) can be estimated similarly to the above. The result is (\ref{eq:CzerobfIbfJaE}).
\end{proof}

This lemma has the following consequences in the context of energy estimates. 

\begin{cor}\label{cor:auxcommhUsqEi}
  Given that all the assumptions of Lemma~\ref{lemma:CkbfBbfAClaestEi} are satisfied and $|\bfI|\leq l$,
  \begin{equation*}
    \begin{split}
      \textstyle{\sum}_{|\bfJ|\leq |\bfI|}\textstyle{\sum}_{k=0}^{1}|C_{\bfI,\bfJ}^{k}\hU^{k}E_{\bfJ}u|^{2}
      \leq & C_{a}\ldr{\varrho}^{4\cweight}\ldr{\tau-\tau_{c}}^{3\iota_{b}}e^{2\e_{\Spe}\varrho}\me_{l}\\
      & +C_{a}\textstyle{\sum}_{m=0}^{l-1}\ldr{\varrho}^{2(l-m+1)\cweight}\me_{m}
    \end{split}
  \end{equation*}  
  for all $\tau\leq\tau_{c}$, where $C_{a}$ only depends on $c_{\cweight,l}$ and $(\bM,\bge_{\refer})$. 
\end{cor}
\begin{remark}
  We only estimate the last term on the right hand side of (\ref{eq:bdbfAhUsqcommformEi}) in terms of the energies later. However, summarising, for
  $|\bfI|\leq l$,
  \begin{equation}\label{eq:bDbfAhUsqcommutatorestimateEi}
    \begin{split}
      |[E_{\bfI},\hU^{2}]u|^{2} \leq & C_{a}\ldr{\varrho}^{4\cweight}\ldr{\tau-\tau_{c}}^{3\iota_{b}}e^{2\e_{\Spe}\varrho}\me_{l}
      +C_{a}\textstyle{\sum}_{m=0}^{l-1}\ldr{\varrho}^{2(l-m+1)\cweight}\me_{m}\\
      & +C_{a}\textstyle{\sum}_{|\bfJ|\leq l-1}\ldr{\varrho}^{2(|\bfI|-|\bfJ|)\cweight}|\hU^{2}E_{\bfJ}u|^{2}
    \end{split}
  \end{equation}
  for all $\tau\leq\tau_{c}$, where $C_{a}$ only depends on $c_{\cweight,l}$ and $(\bM,\bge_{\refer})$. 
\end{remark}
\begin{proof}
  The estimate is an immediate consequence of Lemma~\ref{lemma:CkbfBbfAClaestEi}.
\end{proof}

\section{Commutator with $e^{-2\mu_{A}}X_{A}^{2}$}

In order to estimate the commutator with $e^{-2\mu_{A}}X_{A}^{2}$, let us return to Lemma~\ref{lemma:EbfIemmuAXAsqcomm}. 

\begin{lemma}\label{lemma:commemmuAXAsqEi}
  Fix $l$, $\bfl$, $\bfl_{1}$, $\cweight$, $\weight_{0}$ and $\weight$ as in Definition~\ref{def:supmfulassumptions}.
  Then, given that the assumptions of Lemma~\ref{lemma:taurelvaryingbxEi} as well as the $(\cweight,l)$-supremum assumptions are satisfied,   
  \begin{align*}
    \ldr{\varrho}^{-(l_{a}+1)(2\cweight+1)}|D_{\bfI,\bfJ}^{A}| \leq & C_{a},\\
    \ldr{\varrho}^{-(l_{a}+2)(2\cweight+1)}|F_{\bfI,\bfJ}^{A}| \leq & C_{a}
  \end{align*}
  on $I_{-}$ for all $1\leq |\bfJ|\leq |\bfI|\leq l$, where $l_{a}:=|\bfI|-|\bfJ|$ and $C_{a}$ only depends on $c_{\cweight,l}$ and $(\bM,\bge_{\refer})$. 
\end{lemma}
\begin{proof}
  Combining Remark~\ref{remark:CkestofmuAEi} with Lemma~\ref{lemma:EbfIemmuAXAsqcomm} and the assumptions yields the conclusions of the lemma. 
\end{proof}
This observation has the following corollary. 

\begin{cor}
  Given that the assumptions of Lemma~\ref{lemma:commemmuAXAsqEi} hold,
  \begin{equation}\label{eq:bDbfAemtmuAXAsqcommEi}
    \begin{split}
      |[E_{\bfI},e^{-2\mu_{A}}X_{A}^{2}]u|^{2} \leq & 
      C_{a}\theta_{0,-}^{-2}\textstyle{\sum}_{m=1}^{l}\ldr{\varrho}^{2(l-m+1)(2\cweight+1)}e^{2\e_{\Spe}\varrho}\me_{m}\\
      & +C_{a}\theta_{0,-}^{-4}\textstyle{\sum}_{m=1}^{l}\ldr{\varrho}^{2(l-m+2)(2\cweight+1)}\ldr{\tau-\tau_{c}}^{3\iota_{b}}e^{4\e_{\Spe}\varrho}\me_{m}
    \end{split}
  \end{equation}
  for all $\tau\leq\tau_{c}$ and $|\bfI|\leq l$, where $C_{a}$ only depends on $c_{\cweight,l}$ and $(\bM,\bge_{\refer})$.  
\end{cor}
\begin{proof}
  The corollary is an immediate consequence of (\ref{eq:muminmainlowerbound}) and Lemmas~\ref{lemma:EbfIemmuAXAsqcomm}
  and \ref{lemma:commemmuAXAsqEi}. 
\end{proof}

\section{Commutator with $Z^{0}\hU$}

Considering (\ref{eq:LuformulaEi}), we are next interested in estimating the commutator with $Z^{0}\hU$, where 
\begin{equation}\label{eq:ZzdefEi}
Z^{0}:=\hmcY^{0}\Id+\hmcX^{0}
\end{equation}
\index{$\a$Aa@Notation!Functions!$Z^{0}$}%
and $\hmcY^{0}$ is given by (\ref{eq:hmcYzdefEi}). Before doing so, we need to impose conditions on the coefficients of the equation. Here we demand
the existence of a constant $c_{\rocoeff,l}$ such that (\ref{eq:coefflassumptions}) holds for all $t\in I_{-}$, where $l$ and $\weight_{0}$ 
have the properties stated in Definition~\ref{def:supmfulassumptions}. 

\begin{lemma}\label{lemma:ZzcommEi}
  Fix $l$, $\bfl_{1}$, $\cweight$, $\weight_{0}$ and $\weight$ as in Definition~\ref{def:supmfulassumptions}. Assume the conditions of
  Lemma~\ref{lemma:taurelvaryingbxEi}; the $(\cweight,l)$-supremum assumptions; and (\ref{eq:coefflassumptions}) to hold. Let $G_{\bfI,\bfJ}^{i}$,
  $i=0,1$, be the functions such that (\ref{eq:bDbfAchthhUcommEi}) holds, where $Z^{0}$ is given by (\ref{eq:ZzdefEi}). Then
  \begin{align}
    \ldr{\varrho}^{-l_{a}\cweight}\|G^{1}_{\bfI,\bfJ}\| \leq & C_{a},\label{eq:wGzbfIbfJest}\\
    \ldr{\varrho}^{-(l_{a}+1)\cweight}\|G^{0}_{\bfI,\bfJ}\| \leq & C_{a}e^{\e_{\Spe}\varrho}\label{eq:wGonebfIbfJest}
  \end{align}
  on $M_{-}$, where $l_{a}:=|\bfI|-|\bfJ|$; $|\bfI|\leq l$; $|\bfJ|\leq |\bfI|-1$ in the first estimate; $|\bfJ|\leq |\bfI|$ in the second estimate;
  and $C_{a}$ only depends on $c_{\cweight,l}$, $c_{\rocoeff,l}$, $m_{\ros}$ and $(\bM,\bge_{\refer})$.   
\end{lemma}
\begin{remark}
  The same conclusion holds when $Z^{0}=\mathrm{Id}$, in which case the dependence of the constants on $c_{\rocoeff,l}$ can be omitted. 
\end{remark}
\begin{proof}
  Note that $\chth=-q$ due to (\ref{eq:chthexpressionhUlntheta}). Combining this observation with Lemma~\ref{lemma:bdbfAchthhUcommformEi},
  (\ref{eq:ZzdefEi}) and the assumptions yields (\ref{eq:wGzbfIbfJest}). Similarly, appealing to (\ref{eq:AikClestsobassumpEivarrho}), 
  Lemma~\ref{lemma:bdbfAchthhUcommformEi} as well as the assumptions yields (\ref{eq:wGonebfIbfJest}). 
\end{proof}

\begin{cor}\label{cor:commZzhUEi}
  Assume that the conditions of Lemma~\ref{lemma:ZzcommEi} hold and let $1\leq l\in\zo$. Then, for $|\bfI|\leq l$,
  \begin{equation}\label{eq:ZzhUcommestEi}
    \begin{split}
      |[E_{\bfI},Z^{0}\hU]u|^{2} \leq & C_{a}\textstyle{\sum}_{m=0}^{l-1}\ldr{\varrho}^{2(l-m)\cweight}\me_{m}\\
      &+C_{a}\textstyle{\sum}_{m=1}^{l}\ldr{\varrho}^{2(l-m+1)\cweight}\ldr{\tau-\tau_{c}}^{3\iota_{b}}e^{2\e_{\Spe}\varrho}\me_{m}
    \end{split}    
  \end{equation}
  on $M_{-}$, where $Z^{0}$ is given by (\ref{eq:ZzdefEi}) and $C_{a}$ only depends on $c_{\cweight,l}$, $c_{\rocoeff,l}$, $m_{\ros}$ and $(\bM,\bge_{\refer})$.
\end{cor}
\begin{remark}\label{remark:ZzeqtoIdEi}
The same conclusion holds when $Z^{0}=\mathrm{Id}$, in which case the dependence of the constant on $c_{\rocoeff,l}$ can be omitted. 
\end{remark}
\begin{proof}
  The statement is an immediate consequence of Lemmas~\ref{lemma:bdbfAchthhUcommformEi} and \ref{lemma:ZzcommEi}. 
\end{proof}

\section{Commutator with $Z^{A}X_{A}$}

Next, we wish to estimate the commutator with $Z^{A}X_{A}$, where
\index{$\a$Aa@Notation!Functions!$Z^{A}$}%
\begin{equation}\label{eq:ZAdefEi}
  Z^{A}:=\hmcY^{A}\Id+\hmcX^{A}. 
\end{equation} 

\begin{lemma}\label{lemma:HABestEi}
  Fix $l$, $\bfl_{1}$, $\cweight$, $\weight_{0}$ and $\weight$ as in Definition~\ref{def:supmfulassumptions}. Assume the conditions of
  Lemma~\ref{lemma:taurelvaryingbxEi}; the $(\cweight,l)$-supremum assumptions; and (\ref{eq:coefflassumptions}) to hold. Let $H_{\bfI,\bfJ}$ be
  such that (\ref{eq:bDbfAGAXAcommEi}) holds, where $Z^{A}$ is given by (\ref{eq:ZAdefEi}). Then, if
  $1\leq |\bfJ|\leq |\bfI|\leq l$,
  \begin{equation}\label{eq:bDbfAZAestEi}
    \ldr{\varrho}^{-|\bfI|\cweight}\|E_{\bfI}Z^{A}\|+\ldr{\varrho}^{-(l_{a}+1)\cweight}\|H_{\bfI,\bfJ}\|
    \leq C_{a}\theta_{0,-}^{-1}e^{\e_{\Spe}\varrho}
  \end{equation}
  on $M_{-}$, where $l_{a}:=|\bfI|-|\bfJ|$, and $C_{a}$ only depends on $c_{\cweight,l}$, $c_{\rocoeff,l}$, $m_{\ros}$, $(\bM,\bge_{\refer})$ and a lower bound on 
  $\theta_{0,-}$. 
\end{lemma}
\begin{remark}\label{remark:emuAhmcYAest}
  Due to the proof, it also follows that
  \[
  e^{\mu_{A}}|\hmcY^{A}|\leq C_{a}\theta_{0,-}^{-1}\ldr{\varrho}^{2\cweight+1}e^{\e_{\Spe}\varrho}
  \]
  on $M_{-}$, where $C_{a}$ only depends on $c_{\cweight,0}$ and $(\bM,\bge_{\refer})$. Moreover, 
  \begin{equation}\label{eq:ZACznormestimate}
    \|Z^{A}\|\leq C_{b}\theta_{0,-}^{-1}e^{\e_{\Spe}\varrho}
  \end{equation}
  on $M_{-}$, where $C_{b}$ only depends on $c_{\cweight,0}$, $c_{\rocoeff,0}$, $m_{\ros}$, $(\bM,\bge_{\refer})$ and a lower bound on $\theta_{0,-}$.
\end{remark}
\begin{proof}
  Keeping (\ref{eq:hGAdef}) and (\ref{eq:hmcYAdefEi}) in mind, it follows that
  \begin{equation}\label{eq:EbfKhmcYAest}
    \begin{split}
      |E_{\bfK}(\hmcY^{A})| \leq & 
      C_{a}\textstyle{\sum}_{m=0}^{k}\textstyle{\sum}_{m_{a}+m_{b}=m}e^{-2\mu_{A}}\mfP_{\mK,\mu,m_{a}}|\bD^{m_{b}+1}\ln\theta|_{\bge_{\refer}}\\
      & +C_{a}\textstyle{\sum}_{m=1}^{k+1}e^{-2\mu_{A}}\mfP_{\mK,\mu,N,m}
    \end{split}
  \end{equation}
  on $M_{-}$, where $k:=|\bfK|$ and $C_{a}$ only depends on $\mKsup$, $\e_{\rond}$, $n$, $k$ and $(\bM,\bge_{\refer})$. Combining this observation
  with Lemma~\ref{lemma:bdbfAGAXGcommformEi}, the contribution of $\hmcY^{A}$ to $H_{\bfI,\bfJ}$ can be estimated by the right hand side of
  (\ref{eq:EbfKhmcYAest}) but with $k$ replaced by $l_{b}:=|\bfI|-|\bfJ|+1$. In either case, the contribution to the terms on the left hand side of
  (\ref{eq:bDbfAZAestEi}) can be estimated by the right hand side of (\ref{eq:bDbfAZAestEi}). In order to obtain this conclusion, we appealed to
  Remark~\ref{remark:CkestofmuAEi}, (\ref{eq:muminmainlowerbound}) and the assumptions. 

  Next, note that $E_{\bfI}[\hmcX^{A}_{ij}]$ can be written as a linear combination of terms of the form
  \begin{equation}\label{eq:EbfIhmcXAttbeest}
    (\bD_{\bfI_{1}}Y^{A})(\bD_{\bfI_{2}}\hmcX_{ij}^{\perp}),
  \end{equation}
  where $|\bfI_{1}|+|\bfI_{2}|=|\bfI|$. Appealing to (\ref{eq:bDbfAellAetcpteststmtEi}), (\ref{eq:ClbbEweightestimatepointwise}) and the assumptions
  yields
  \[
  \ldr{\varrho}^{-|\bfI|\cweight}|E_{\bfI}[\hmcX^{A}_{ij}]|\leq C\theta_{0,-}^{-1}e^{\e_{\Spe}\varrho}
  \]
  on $M_{-}$ for $|\bfI|\leq l$, where $C$ only depends on $c_{\cweight,l}$, $c_{\rocoeff,l}$ and $(\bM,\bge_{\refer})$. Again, the contribution to the terms 
  on the left hand side of (\ref{eq:bDbfAZAestEi}) can be estimated by the right hand side of (\ref{eq:bDbfAZAestEi}).  
\end{proof}

\begin{cor}
  Given that the assumptions of Lemma~\ref{lemma:HABestEi} are satisfied and $1\leq |\bfI|=l$,
  \begin{equation}\label{eq:ZAXAcommestEi}
    |[E_{\bfI},Z^{A}X_{A}]u|^{2}\leq C_{a}\theta_{0,-}^{-2}\textstyle{\sum}_{m=1}^{l}\ldr{\varrho}^{2(l-m+1)\cweight}
    \ldr{\tau-\tau_{c}}^{3\iota_{b}}e^{2\e_{\Spe}\varrho}\me_{m}
  \end{equation}
  on $M_{-}$, where $C_{a}$ only depends on $c_{\cweight,l}$, $c_{\rocoeff,l}$, $m_{\ros}$, $(\bM,\bge_{\refer})$ and a lower bound on 
  $\theta_{0,-}$. 
\end{cor}

It is of interest to record a related result. 

\begin{lemma}\label{lemma:emmuAXAEbfIcomm}
  Fix $l$, $\bfl_{1}$, $\cweight$, $\weight_{0}$ and $\weight$ as in Definition~\ref{def:supmfulassumptions}. Assume the conditions of
  Lemma~\ref{lemma:taurelvaryingbxEi} and the $(\cweight,l)$-supremum assumptions to hold. Then, if $\psi$ is a
  smooth function on $\bM\times I$ and $|\bfI|\leq l$,
  \begin{equation}\label{eq:emmuAXAEbfIcomm}
    \left|\left[E_{\bfI},e^{-\mu_{A}}X_{A}\right]\psi\right|
    \leq C_{a}\theta_{0,-}^{-1}\ldr{\varrho}^{l(2\cweight+1)}e^{\e_{\Spe}\varrho}\textstyle{\sum}_{1\leq |\bfJ|\leq l}|E_{\bfJ}\psi|
  \end{equation}
  on $M_{-}$ (no summation on $A$), where $C_{a}$ only depends on $c_{\cweight,l}$ and $(\bM,\bge_{\refer})$.
\end{lemma}
\begin{proof}
  Due to Lemma~\ref{lemma:bdbfAGAXGcommformEi}, we know that 
  \[
  \left|\left[E_{\bfI},e^{-\mu_{A}}X_{A}\right]\psi\right|
  \leq C_{a}\textstyle{\sum}_{1\leq |\bfJ|\leq |\bfI|}\textstyle{\sum}_{k_{a}+|\bfK|\leq |\bfI|-|\bfJ|+1}
  \mfP_{\mK,k_{a}}|E_{\bfK}(e^{-\mu_{A}})|\cdot |E_{\bfJ}(\psi)| 
  \]
  where $C_{a}$ only depends on $\mKsup$, $\e_{\rond}$, $|\bfI|$, $n$ and $(\bM,\bge_{\refer})$. Combining this estimate with (\ref{eq:muminmainlowerbound}),
  Remark~\ref{remark:CkestofmuAEi} and the assumptions yields the conclusion. 
\end{proof}

\section{Commutator with $\hal$} 

\begin{lemma}\label{lemma:halcommutator}
  Let $(M,g)$ be a time oriented Lorentz manifold and $1\leq l\in\zo$. Assume it to have an expanding partial pointed foliation and $\mK$ to be
  non-degenerate on $I$ and to have a global frame. Assume, moreover, (\ref{eq:coefflassumptions}) to hold. Then, if $1\leq |\bfI|\leq l$,
  \begin{equation}\label{eq:halcommestEi}
    |[E_{\bfI},\hal]u|^{2}\leq C_{a}\textstyle{\sum}_{m=0}^{l-1}\ldr{\varrho}^{2(l-m)\cweight}\ldr{\tau-\tau_{c}}^{3\iota_{b}}\me_{m}
  \end{equation}
  on $M_{-}$, where $C_{a}$ only depends on $c_{\rocoeff,l}$, $m_{\ros}$, $n$, $l$ and $(\bM,\bge_{\refer})$.
\end{lemma}
\begin{proof}
  Note that $[E_{\bfI},\hal]$ can be written as a linear combination of terms of the form $(E_{\bfJ}\hal)E_{\bfK}$, where $|\bfJ|\geq 1$ and
  $|\bfJ|+|\bfK|=|\bfI|$. The statement of the lemma is thus an immediate consequence of the assumptions.  
\end{proof}

\section{Estimating $\hU^{2}E_{\bfI}u$}\label{ssection:hUsqbDbfAestEi}

\begin{lemma}\label{lemma:hUsquestEi}
  Let $l=0$. Given this $l$, fix $\bfl_{1}$, $\cweight$, $\weight_{0}$ and $\weight$ as in 
  Definition~\ref{def:supmfulassumptions}. Assume the conditions of Lemma~\ref{lemma:taurelvaryingbxEi}; the $(\cweight,l)$-supremum
  assumptions; and (\ref{eq:coefflassumptions}) to hold. Then, if $u$ is a solution to (\ref{eq:theequation}), 
  \begin{equation}\label{eq:hUsquestimateprel}
    |\hU^{2}u|\leq C_{a}\theta_{0,-}^{-1}e^{\e_{\Spe}\varrho}\me_{1}^{1/2}+\sqrt{2}\eta\me^{1/2}+|\hf|
  \end{equation}
  on $M_{c}$, where $M_{c}$ is the subset of $M_{-}$ corresponding to $\tau\leq\tau_{c}$; $C_{a}$ only depends on $c_{\robas}$, $(\bM,\bge_{\refer})$
  and a lower bound on $\theta_{0,-}$;
  \begin{equation}\label{eq:etadef}
    \begin{split}
      \eta := & \frac{1}{n}|q-(n-1)|+\|\hmcX^{0}\|+\|\hmcX^{\perp}\|_{\chg}+\iota_{a}\|\hal\|+\iota_{b}\ldr{\tau-\tau_{c}}^{3/2}\|\hal\|\\
      & +C_{b}\theta_{0,-}^{-1}\ldr{\varrho}^{2\cweight+1}e^{\e_{\Spe}\varrho};
    \end{split}
  \end{equation}  
  and $C_{b}$ only depends on $c_{\cweight,0}$ and $(\bM,\bge_{\refer})$. In particular,
  \begin{equation}\label{eq:hUsquestimate}
    |\hU^{2}u|\leq C_{a}\theta_{0,-}^{-1}e^{\e_{\Spe}\varrho}\me_{1}^{1/2}+\bc_{0}\me^{1/2}
    +C_{c}\theta_{0,-}^{-1}\ldr{\varrho}^{2\cweight+1}e^{\e_{\Spe}\varrho}\me^{1/2}+|\hf|
  \end{equation}
  on $M_{c}$, where $C_{c}$ only depends on $c_{\cweight,0}$ and $(\bM,\bge_{\refer})$; and 
  \[
  \bc_{0}:=\sqrt{2}\sup_{M_{c}}\left(\frac{1}{n}|q-(n-1)|+\|\hmcX^{0}\|+\|\hmcX^{\perp}\|_{\chg}+\iota_{a}\|\hal\|
  +\iota_{b}\ldr{\tau-\tau_{c}}^{3/2}\|\hal\|\right).
  \]
\end{lemma}
\begin{remark}
  If $\iota_{b}\neq 0$, then $\ldr{\tau-\tau_{c}}^{3/2}\|\hal\|$ is bounded on $M_{c}$; cf. Subsection~\ref{ssection:setbasen}. 
\end{remark}
\begin{remark}
  Note that if the all the conditions of Corollary~\ref{cor:basicenergyestimate} are satisfied, then $\bareta\in L^{1}(-\infty,\tau_{c}]$, where
  \[
  \bareta(\tau):=\sup_{\bx\in\bM}\eta(\bx,\tau)
  \]
  and $\eta$ is defined by (\ref{eq:etadef}).
\end{remark}
\begin{proof}
  Due to (\ref{eq:theeqreformEi}) and the definitions (\ref{eq:ZzdefEi}) and (\ref{eq:ZAdefEi}),
  \[
  |\hU^{2}u|\leq \textstyle{\sum}_{A}e^{-2\mu_{A}}|X_{A}^{2}u|+|Z^{0}\hU u|+|Z^{A}X_{A}u|+|\hal u|+|\hf|.
  \]
  However,
  \begin{equation}\label{eq:emtwomuAXAsquest}
    \begin{split}
      |e^{-2\mu_{A}}X_{A}^{2}u| \leq & \textstyle{\sum}_{i}e^{-2\mu_{A}}|X_{A}(X_{A}^{i})|\cdot |E_{i}u|
      +e^{-\mu_{A}}\left(\textstyle{\sum}_{i}e^{-2\mu_{A}}|X_{A}E_{i}u|^{2}\right)^{1/2}\\
      \leq & C_{a}\theta_{0,-}^{-2}\ldr{\varrho}^{\cweight}e^{2\e_{\Spe}\varrho}\left(\textstyle{\sum}_{i}|E_{i}u|^{2}\right)^{1/2}
      +C_{a}\theta_{0,-}^{-1}e^{\e_{\Spe}\varrho}\left(\textstyle{\sum}_{i}e^{-2\mu_{A}}|X_{A}E_{i}u|^{2}\right)^{1/2}\\
      \leq & C_{b}\theta_{0,-}^{-1}e^{\e_{\Spe}\varrho}\me_{1}^{1/2}
    \end{split}
  \end{equation}
  on $M_{-}$, where $C_{a}$ only depends on $c_{\robas}$ and $(\bM,\bge_{\refer})$; and $C_{b}$ only depends on $c_{\robas}$, $(\bM,\bge_{\refer})$ and a
  lower bound on $\theta_{0,-}$. Next, note that one consequence of (\ref{eq:coefflassumptions}) is that (\ref{eq:Czbalancedhat}) holds. In other
  words, (\ref{eq:theequation}) is $C^{0}$-balanced on $I_{-}$ and (\ref{eq:Czcoefficientbounds}) holds. On the other hand,
  \begin{align*}
    |Z^{0}\hU u| \leq & \left(\frac{1}{n}|q-(n-1)|+\|\hmcX^{0}\|\right)|\hU u|,\\
    |Z^{A}X_{A}u| \leq & \left[\|\hmcX^{\perp}\|_{\chg}
      +\left(\textstyle{\sum}_{A}e^{2\mu_{A}}|\hmcY^{A}|^{2}\right)^{1/2}\right]\left(\textstyle{\sum}_{A}e^{-2\mu_{A}}|X_{A}u|^{2}\right)^{1/2},
  \end{align*}
  where we use the notation introduced in (\ref{eq:hmcXperpchgnorm}). In order to obtain these estimates, we appealed to (\ref{eq:hmcYzdefEi}),
  (\ref{eq:ZzdefEi}) and (\ref{eq:ZAdefEi}). Combining these estimates with Remark~\ref{remark:emuAhmcYAest} yields the conclusion of the lemma. 
\end{proof}

Next, we consider higher order derivatives. 

\begin{lemma}\label{lemma:hUsquncondestEi}
  Fix $l$, $\bfl_{1}$, $\cweight$, $\weight_{0}$ and $\weight$ as in Definition~\ref{def:supmfulassumptions}. Assume the conditions of
  Lemma~\ref{lemma:taurelvaryingbxEi}; the $(\cweight,l)$-supremum assumptions; and (\ref{eq:coefflassumptions}) to hold. Then, if $u$ is a solution
  to (\ref{eq:theequation}),
  \begin{equation}\label{eq:hUsqbDbfAusqindassumpEi}
    \begin{split}
      |\hU^{2}E_{\bfI}u| \leq & C_{a}e^{\e_{\Spe}\varrho}\me_{l+1}^{1/2}+C_{b}\ldr{\varrho}^{\a_{l}\cweight+l\cweight}\ldr{\tau-\tau_{c}}^{3\iota_{b}/2}\me_{l}^{1/2}\\
      & +C_{f}\textstyle{\sum}_{m=0}^{l}\textstyle{\sum}_{|\bfJ|=m}\ldr{\varrho}^{(l-m)\cweight}|E_{\bfJ}\hf|
    \end{split}
  \end{equation}
  on $M_{c}$ for all $|\bfI|\leq l$, where $\a_{0}=0$ and $\a_{j}=1$ for $j\geq 1$; $C_{a}$ only depends on $c_{\cweight,l}$, $(\bM,\bge_{\refer})$
  and a lower bound on $\theta_{0,-}$; and $C_{b}$ only depends on $c_{\cweight,l}$, $c_{\rocoeff,l}$, $m_{\ros}$, $d_{\a}$ (in case $\iota_{b}\neq 0$),
  $(\bM,\bge_{\refer})$ and a lower bound on $\theta_{0,-}$. Finally, $C_{f}$ only depends on $c_{\cweight,l}$ and $(\bM,\bge_{\refer})$.
\end{lemma}
\begin{proof}
  Assume, inductively, that if $j:=|\bfI|\leq k$, then 
  \begin{equation}\label{eq:indasshUsq}
    \begin{split}
      |\hU^{2}E_{\bfI}u| \leq & C_{a}e^{\e_{\Spe}\varrho}\me_{j+1}^{1/2}+C_{b}\ldr{\varrho}^{\a_{j}\cweight+j\cweight}\ldr{\tau-\tau_{c}}^{3\iota_{b}/2}\me_{j}^{1/2}\\
      & +C_{f}\textstyle{\sum}_{m=0}^{j}\textstyle{\sum}_{|\bfJ|=m}\ldr{\varrho}^{(j-m)\cweight}|E_{\bfJ}\hf|
    \end{split}
  \end{equation}
  on $M_{c}$, where $C_{a}$, $C_{b}$ and $C_{f}$ have the dependence stated in the lemma. Moreover, $\a_{0}=0$ and 
  $\a_{j}=1$ for $j\geq 1$. Due to Lemma~\ref{lemma:hUsquestEi}, we know this estimate to hold if $k=0$. Moreover, for $k=0$, $C_{a}$ only depends
  on $c_{\robas}$, $(\bM,\bge_{\refer})$ and a lower bound on $\theta_{0,-}$; $C_{b}$ only depends on $c_{\cweight,0}$, $c_{\rocoeff,0}$, $m_{\ros}$, $d_{\a}$ (in case
  $\iota_{b}\neq 0$), $(\bM,\bge_{\refer})$ and a lower bound on $\theta_{0,-}$; and $C_{f}=1$. Assume that (\ref{eq:indasshUsq}) holds for $k\geq 0$
  and let $|\bfI|=k+1$. Due to the equation,
  \begin{equation}\label{eq:lEbfIuid}
    LE_{\bfI}u=[L,E_{\bfI}]u+E_{\bfI}\hf. 
  \end{equation}
  Combining this equality with Lemma~\ref{lemma:hUsquestEi} with $u$ replaced by $E_{\bfI}u$ and $\hf$ replaced by the right hand side of 
  (\ref{eq:lEbfIuid}) yields
  \begin{equation}\label{eq:hUsqbDbfAfirststepEi}
    |\hU^{2}E_{\bfI}u|\leq C_{a}e^{\e_{\Spe}\varrho}\me_{k+2}^{1/2}+C_{b}\me^{1/2}_{k+1}+|E_{\bfI}\hf|+|[L,E_{\bfI}]u|.
  \end{equation}
  For this reason, it is clearly of interest to estimate $|[L,E_{\bfI}]u|$. Since
  \[
  L=-\hU^{2}+\textstyle{\sum}_{A}e^{-2\mu_{A}}X_{A}^{2}+Z^{0}\hU+Z^{A}X_{A}+\hal,
  \]
  it is sufficient to appeal to (\ref{eq:bDbfAhUsqcommutatorestimateEi}), (\ref{eq:bDbfAemtmuAXAsqcommEi}), (\ref{eq:ZzhUcommestEi}),
  (\ref{eq:ZAXAcommestEi}), (\ref{eq:halcommestEi}) and the inductive hypothesis. This yields 
  \[
  |[L,E_{\bfI}]u|\leq C_{b}\ldr{\varrho}^{(k+2)\cweight}\ldr{\tau-\tau_{c}}^{3\iota_{b}/2}\me_{k+1}^{1/2}
  +C_{f}\textstyle{\sum}_{p=0}^{k}\textstyle{\sum}_{|\bfK|=p}\ldr{\varrho}^{(k+1-p)\cweight}|E_{\bfK}\hf|.
  \]
  Moreover, given that $k+1\leq l$, the constants have the desired dependence. Combining this estimate with (\ref{eq:hUsqbDbfAfirststepEi})  
  yields the conclusion that the inductive assumption holds with $k$ replaced by $k+1$. The lemma follows. 
\end{proof}

\section{Summing up}\label{section:summingupulsupass}

Finally, we are in a position to estimate the expression (\ref{eq:thecommutatorEi}). 

\begin{lemma}\label{lemma:LbDbfAunconditionalestimateEi}
  Fix $l$, $\bfl_{1}$, $\cweight$, $\weight_{0}$ and $\weight$ as in Definition~\ref{def:supmfulassumptions}. Assume the conditions of
  Lemma~\ref{lemma:taurelvaryingbxEi}; the $(\cweight,l)$-supremum assumptions; and (\ref{eq:coefflassumptions}) to hold. Then, if $u$ is a
  solution to (\ref{eq:theequation}),
  \begin{equation}\label{eq:LbDbfAunconditionalestimateEi}
    \begin{split}
      |[L,E_{\bfI}]u| \leq & C_{a}\ldr{\varrho}^{2\cweight+1}\ldr{\tau-\tau_{c}}^{3\iota_{b}/2}e^{\e_{\Spe}\varrho}\me_{l}^{1/2}
      +C_{b}\ldr{\varrho}^{(l+1)\cweight}\ldr{\tau-\tau_{c}}^{3\iota_{b}/2}\me_{l-1}^{1/2}\\
      &+C_{f}\textstyle{\sum}_{m=0}^{l-1}\textstyle{\sum}_{|\bfJ|=m}\ldr{\varrho}^{(l-m)\cweight}|E_{\bfJ}\hf|
    \end{split}    
  \end{equation}
  on $M_{c}$ for all $|\bfI|\leq l$, where $C_{a}$ and $C_{b}$ only depend on $c_{\cweight,l}$, $c_{\rocoeff,l}$, $m_{\ros}$, $d_{\a}$ (in case $\iota_{b}\neq 0$),
  $(\bM,\bge_{\refer})$ and a lower bound on $\theta_{0,-}$. Moreover, $C_{f}$ only depends on $c_{\cweight,l}$ and $(\bM,\bge_{\refer})$.
\end{lemma}
\begin{remark}
  Combining (\ref{eq:LbDbfAunconditionalestimateEi}) with (\ref{eq:DeltavarrhorelvariationEi}) and (\ref{eq:eSpevarrhoeelowtaurelEi}) yields the
  conclusion that 
  \begin{equation}\label{eq:LbDbfAunconditionalestimateEitau}
    \begin{split}
      |[L,E_{\bfI}]u| \leq & C_{a}\ldr{\tau}^{2\cweight+1}\ldr{\tau-\tau_{c}}^{3\iota_{b}/2}e^{\eSpe\tau}\me_{l}^{1/2}
      +C_{b}\ldr{\tau}^{(l+1)\cweight}\ldr{\tau-\tau_{c}}^{3\iota_{b}/2}\me_{l-1}^{1/2}\\
      &+C_{f}\textstyle{\sum}_{m=0}^{l-1}\textstyle{\sum}_{|\bfJ|=m}\ldr{\tau}^{(l-m)\cweight}|E_{\bfJ}\hf|
    \end{split}    
  \end{equation}
  on $M_{c}$ for all $|\bfI|\leq l$, where $C_{a}$, $C_{b}$ and $C_{f}$ have the same dependence as in the case of 
  (\ref{eq:LbDbfAunconditionalestimateEi}).
\end{remark}
\begin{proof}
  The estimate follows from an argument which is similar to the proof of Lemma~\ref{lemma:hUsquncondestEi}.
\end{proof}

\section{First energy estimate}\label{section:firstenergyestimateulsupass}

Fix $\tau_{c}\leq 0$. Then, due to (\ref{eq:LbDbfAunconditionalestimateEitau}),
\begin{equation*}
  \begin{split}
    \int_{\bM_{\tau}}\textstyle{\sum}_{|\bfI|\leq k}|[L,E_{\bfI}]u|^{2}\mutgc \leq &
    C_{a}\ldr{\tau}^{4\cweight+2}\ldr{\tau-\tau_{c}}^{3\iota_{b}}e^{2\eSpe\tau}\hE_{k}(\tau;\tau_{c})\\
    & +C_{b}\ldr{\tau}^{2(k+1)\cweight}\ldr{\tau-\tau_{c}}^{3\iota_{b}}\hE_{k-1}(\tau;\tau_{c})\\
    & +C_{f}\int_{\bM_{\tau}}\textstyle{\sum}_{m=0}^{k-1}\textstyle{\sum}_{|\bfJ|=m}\ldr{\tau}^{2(k-m)\cweight}|E_{\bfJ}\hf|^{2}\mutgc
  \end{split}
\end{equation*}
for all $\tau\leq \tau_{c}$, 
where the constants have the same dependence as in (\ref{eq:LbDbfAunconditionalestimateEi}). Combining this estimate with
(\ref{eq:elbasenergyestimateEi}) yields the conclusion that for all $\tau_{a}\leq\tau_{b}\leq\tau_{c}\leq 0$,
\begin{equation}\label{eq:elbasenergyestimatesvEi}
  \begin{split}
    \hE_{k}(\tau_{a};\tau_{c}) \leq &  \hE_{k}(\tau_{b};\tau_{c})+\int_{\tau_{a}}^{\tau_{b}}\kappa(\tau)\hE_{k}(\tau;\tau_{c})d\tau\\
    & +C_{a}\int_{\tau_{a}}^{\tau_{b}}\ldr{\tau}^{2\cweight+1}\ldr{\tau-\tau_{c}}^{3\iota_{b}/2}e^{\eSpe\tau}\hE_{k}(\tau;\tau_{c})d\tau\\
    & +C_{b}\int_{\tau_{a}}^{\tau_{b}}\ldr{\tau}^{(k+1)\cweight}\ldr{\tau-\tau_{c}}^{3\iota_{b}/2}\hE_{k-1}^{1/2}(\tau;\tau_{c})\hE_{k}^{1/2}(\tau;\tau_{c})d\tau\\
    & +C_{f}\int_{\tau_{a}}^{\tau_{b}}\hF_{k}(\tau)\hE_{k}^{1/2}(\tau;\tau_{c})d\tau,
  \end{split}
\end{equation}
where
\[
\hF_{l}(\tau):=\left(\int_{\bM_{\tau}}\textstyle{\sum}_{m=0}^{l}\textstyle{\sum}_{|\bfJ|=m}\ldr{\tau}^{2(l-m)\cweight}|E_{\bfJ}\hf|^{2}\mutgc\right)^{1/2}.
\]
Here $\kappa$ is the function introduced in (\ref{eq:kappadef}) and the constants $C_{a}$ and $C_{b}$ have the dependence stated in connection
with (\ref{eq:LbDbfAunconditionalestimateEi}). Let us derive energy estimates in the case that $\hf=0$.
\begin{lemma}\label{lemma:hEkenergyestimate}
  Fix $l$, $\bfl_{1}$, $\cweight$, $\weight_{0}$ and $\weight$ as in Definition~\ref{def:supmfulassumptions}. Assume the conditions of
  Lemma~\ref{lemma:taurelvaryingbxEi}; the $(\cweight,l)$-supremum assumptions; and (\ref{eq:coefflassumptions}) to hold. Then, if $u$ is a
  solution to (\ref{eq:theequation}) and $\hf=0$,
  \begin{equation}\label{eq:basichigherorderenergyestimate}
    \hE_{k}(\tau_{a};\tau_{c}) \leq
    C_{k}\textstyle{\sum}_{m=0}^{k}\ldr{\tau_{a}}^{2a_{k,m}\cweight}\ldr{\tau_{c}-\tau_{a}}^{2b_{k,m}}
    \ldr{\tau_{b}-\tau_{a}}^{2c_{k,m}}e^{c_{0}(\tau_{b}-\tau_{a})}\hE_{m}(\tau_{b};\tau_{c})
  \end{equation}
  for all $\tau_{a}\leq\tau_{b}\leq \tau_{c}\leq 0$ and $0\leq k\leq l$, where
  \begin{align*}
    a_{k,m} = & (m+k+3)(k-m)/2,\\
    b_{k,m} = & 3(k-m)\iota_{b}/2,\\
    c_{k,m} = & k-m
  \end{align*}
  for all $0\leq m\leq k$. Moreover, $C_{k}$ only depends on $c_{\cweight,l}$, $c_{\rocoeff,l}$, $m_{\ros}$, $d_{\a}$ (in case $\iota_{b}\neq 0$), $(\bM,\bge_{\refer})$
  and a lower bound on $\theta_{0,-}$. Here $c_{0}$ is defined by (\ref{eq:czCbdef}).
\end{lemma}
\begin{remark}\label{remark:basichigherorderenergyestimateimp}
  If, in addition to the assumptions of the lemma, all the conditions of Corollary~\ref{cor:basicenergyestimate} are satisfied, the estimate
  (\ref{eq:basichigherorderenergyestimate}) can be improved to
  \begin{equation}\label{eq:basichigherorderenergyestimateimp}
    \hE_{k}(\tau_{a};\tau_{c}) \leq
    C_{k}\textstyle{\sum}_{m=0}^{k}\ldr{\tau_{a}}^{2a_{k,m}\cweight}\ldr{\tau_{c}-\tau_{a}}^{2b_{k,m}}\ldr{\tau_{b}-\tau_{a}}^{2c_{k,m}}\hE_{m}(\tau_{b};\tau_{c})
  \end{equation}
  for all $\tau_{a}\leq\tau_{b}\leq\tau_{c}\leq 0$ and $0\leq k\leq l$, where $a_{k,m}$, $b_{k,m}$ and $c_{k,m}$ are as in the statement of the lemma
  and $C_{k}$ only depends on $c_{\cweight,l}$, $d_{q}$, $c_{\rocoeff,l}$, $d_{\coeff}$, $d_{\a}$, $m_{\ros}$, $(\bM,\bge_{\refer})$ and a lower bound on $\theta_{0,-}$.
  Here $d_{q}$ and $d_{\coeff}$ are the constants appearing in (\ref{eq:qconvergence}) and (\ref{eq:coeffconvergence}) respectively. Combining this
  estimate, with $\tau_{b}=\tau_{c}=0$ and $\tau_{a}=\tau\leq 0$, with (\ref{eq:hEutauctauc}) and the observations made in Remark~\ref{remark:tgbgerefequiv}
  yields the conclusion that for $|\bfI|\leq l$, 
  \begin{equation*}
    \begin{split}
      & \int_{\bM_{\tau}}\left(|\hU E_{\bfI}u|^{2}+\textstyle{\sum}_{A}e^{-2\mu_{A}}|X_{A}E_{\bfI}u|^{2}+\ldr{\tau}^{-3}|E_{\bfI}u|^{2}\right)\mu_{\bge_{\refer}}\\
      \leq & C_{l}\ldr{\tau}^{\g_{l}\cweight+\de_{l}}\sum_{|\bfJ|\leq l}
      \int_{\bM_{t_{0}}}\left(|\hU E_{\bfJ}u|^{2}+\textstyle{\sum}_{A}e^{-2\mu_{A}}|X_{A}E_{\bfJ}u|^{2}+|E_{\bfJ}u|^{2}\right)\mu_{\bge_{\refer}}      
    \end{split}
  \end{equation*}  
  for all $\tau\leq 0$, where $C_{l}$ only depends on $c_{\cweight,l}$, $d_{q}$, $c_{\rocoeff,l}$, $d_{\coeff}$, $d_{\a}$, $m_{\ros}$, $(\bM,\bge_{\refer})$ and a
  lower bound on $\theta_{0,-}$. Moreover, $\g_{l}$, $\de_{l}$ are constants depending only on $l$. 
\end{remark}
\begin{proof}
  In case $\hf=0$, (\ref{eq:basicenergyestimate}) takes the form
  \begin{equation}\label{eq:basicenergyestimatewohfEi}
    \begin{split}
      \hE(\tau_{a};\tau_{c}) \leq &  \hE(\tau_{b};\tau_{c})+\int_{\tau_{a}}^{\tau_{b}}\kappa(\tau)\hE(\tau;\tau_{c})d\tau
    \end{split}
  \end{equation}
  for all $\tau_{a}\leq\tau_{b}\leq\tau_{c}\leq 0$. Combining this estimate with a Gr\"{o}nwall's lemma type argument and the properties of $\kappa$,
  stated in Corollary~\ref{cor:basicenergyestimate}, yields
  \begin{equation}\label{eq:hEtauataubest}
    \hE(\tau_{a};\tau_{c})\leq C_{a}e^{c_{0}(\tau_{b}-\tau_{a})}\hE(\tau_{b};\tau_{c})
  \end{equation}
  for all $\tau_{a}\leq\tau_{b}\leq \tau_{c}\leq 0$, where $C_{a}$ only depends $c_{\cweight,0}$, $d_{\a}$ (in case $\iota_{b}\neq 0$),
  $(\bM,\bge_{\refer})$ and a
  lower bound on $\theta_{0,-}$. Here $c_{0}$ is defined by (\ref{eq:czCbdef}). If the conditions of Remark~\ref{remark:basichigherorderenergyestimateimp}
  are satisfied, the estimate (\ref{eq:hEtauataubest}) holds with $c_{0}$ set to zero. However, the constant $C_{a}$ then depends on $c_{\cweight,0}$,
  $d_{q}$, $d_{\a}$, $d_{\coeff}$, $(\bM,\bge_{\refer})$ and a lower bound on $\theta_{0,-}$.
  
  \textbf{Inductive assumption.} Let us make the inductive assumption that 
  \[
  \hE_{k}(\tau_{a};\tau_{c}) \leq C_{k}e^{c_{0}(\tau_{b}-\tau_{a})}
  \textstyle{\sum}_{m=0}^{k}\ldr{\tau_{a}}^{2a_{k,m}\cweight}\ldr{\tau_{c}-\tau_{a}}^{2b_{k,m}}\ldr{\tau_{b}-\tau_{a}}^{2c_{k,m}}\hE_{m}(\tau_{b};\tau_{c})
  \]
  for all $\tau_{a}\leq\tau_{b}\leq \tau_{c}\leq 0$, where $a_{k,m}$, $b_{k,m}$ and $c_{k,m}$ remain to be determined, and $C_{k}$ only depends on
  $c_{\cweight,l}$, $c_{\rocoeff,l}$, $d_{\a}$ (in case $\iota_{b}\neq 0$), $m_{\ros}$, $(\bM,\bge_{\refer})$ and a lower bound on $\theta_{0,-}$. We know this statement
  to be true for $k=0$ with $a_{0,0}=b_{0,0}=c_{0,0}=0$. Again, if the conditions of Remark~\ref{remark:basichigherorderenergyestimateimp} are satisfied,
  the estimate (\ref{eq:hEtauataubest}) holds with $c_{0}$ set to zero, at the expense of demanding that the constant
  $C_{k}$, additionally, depend on $d_{q}$ and $d_{\coeff}$. 

  \textbf{Inductive argument.} Given that the inductive assumption holds for $k-1$, we prove that it holds for $k$. Denote, to this end, the
  right hand side of (\ref{eq:elbasenergyestimatesvEi}) by $\xi(\tau_{a})$. Then, appealing to (\ref{eq:elbasenergyestimatesvEi}) and the definition
  of $\xi$, 
  \[
  \xi'\geq -H'\xi-g\xi^{1/2},
  \]
  where 
  \begin{align*}
    H'(\tau) := & \kappa(\tau)+C_{a}\ldr{\tau}^{2\cweight+1}\ldr{\tau-\tau_{c}}^{3\iota_{b}/2}e^{\eSpe\tau},\\
    g(\tau) := & C_{b}\ldr{\tau}^{(k+1)\cweight}\ldr{\tau-\tau_{c}}^{3\iota_{b}/2}\hE_{k-1}^{1/2}(\tau;\tau_{c}),
  \end{align*}
  and the constants $C_{a}$ and $C_{b}$ are the ones appearing in (\ref{eq:elbasenergyestimatesvEi}). Using this estimate, it can be verified that for
  $\tau_{a}\leq \tau_{b}$, 
  \begin{equation}\label{eq:firstenestxihalfest}
    \xi^{1/2}(\tau_{a})\leq e^{[H(\tau_{b})-H(\tau_{a})]/2}\xi^{1/2}(\tau_{b})+\frac{1}{2}\int_{\tau_{a}}^{\tau_{b}}e^{[H(\tau)-H(\tau_{a})]/2}g(\tau)d\tau.
  \end{equation}
  Note that for all $\tau_{a}\leq\tau\leq \tau_{c}$, 
  \[
  H(\tau)-H(\tau_{a})\leq c_{0}(\tau-\tau_{a})+C_{a},
  \]
  where $C_{a}$ has the dependence stated in connection with (\ref{eq:LbDbfAunconditionalestimateEi}). Moreover, if the conditions of
  Remark~\ref{remark:basichigherorderenergyestimateimp} are satisfied, $c_{0}$ can be set to zero, at the expense of demanding that the constant
  $C_{a}$, additionally, depend on $d_{q}$ and $d_{\coeff}$. Combining this observation with (\ref{eq:firstenestxihalfest}) yields
  \begin{equation*}
    \begin{split}
      \hE_{k}^{1/2}(\tau_{a};\tau_{c}) \leq & C_{a}e^{c_{0}(\tau_{b}-\tau_{a})/2}\hE_{k}^{1/2}(\tau_{b};\tau_{c})\\
      & +C_{a}\int_{\tau_{a}}^{\tau_{b}}e^{c_{0}(\tau-\tau_{a})/2}\ldr{\tau}^{(k+1)\cweight}\ldr{\tau-\tau_{c}}^{3\iota_{b}/2}\hE_{k-1}^{1/2}(\tau;\tau_{c})d\tau.
    \end{split}
  \end{equation*}
  Combining this estimate with the inductive assumption yields the conclusion that the inductive assumption holds with
  \begin{align*}
    a_{k,m} = & a_{k-1,m}+k+1,\\
    b_{k,m} = & b_{k-1,m}+3\iota_{b}/2,\\
    c_{k,m} = & c_{k-1,m}+1
  \end{align*}
  for all $m\leq k-1$. Moreover, $a_{k,k}=b_{k,k}=c_{k,k}=0$. Combining the above observations yields the conclusions of the lemma, as well as those
  of Remark~\ref{remark:basichigherorderenergyestimateimp}.
\end{proof}

\section{Weighted Sobolev embedding}\label{section:weightedSobembulsupass}

When deriving asymptotics of solutions, the estimate (\ref{eq:basichigherorderenergyestimate}) is a natural starting point. However, we also wish to derive
$C^{k}$-estimates. To this end, we need Sobolev embedding estimates. However, the estimates we need are not completely standard. This is due to the
fact that, in the energies, there is a time and space dependent weight; cf. (\ref{eq:hEktaudefEi}). In fact, we are integrating with respect to the measure 
$\mutgc$ instead of with respect to the measure $\mu_{\bge_{\refer}}$. This necessitates a slight variation of the standard Sobolev estimates. 
To begin with, it is of interest to express $\mutgc$ in terms of $\mu_{\bge_{\refer}}$. Note, to this end, that (\ref{eq:mutgreformulation}) and
(\ref{eq:mutgdef}) yield the conclusion that
\[
\mutgc=\tvarphi_{c}^{-1}\tvarphi\mu_{\bge_{\refer}}.
\]
Note also that Lemma~\ref{lemma:thetavarrhorelqconvtonmotwo} yields an estimate of $|\ln\tvarphi-\ln\tvarphi_{c}|$. 
Combining these observations with Sobolev embedding yields the following conclusion. 
\begin{lemma}
  Let $\kappa_{0}$ be the smallest integer which is strictly larger than $n/2$. Assume that the conditions of Lemma~\ref{lemma:CkestofvarrhoEi} are 
  fulfilled with $l=\kappa_{0}+1$. Assume, moreover, that
  \[
  \|\ln\theta\|_{C^{\bfk_{1}}_{\weight_{0}}(\bM)}+\|q\|_{C^{\kappa_{0}}_{\weight_{0}}(\bM)}\leq C_{\theta,\kappa_{0}}
  \]
  for all $\tau\leq 0$, where $\bfk_{1}=(1,\kappa_{0}+1)$. Then, if $\psi$ is a smooth function on $\bM$ and $w:=\tvarphi_{c}^{-1/2}\tvarphi^{1/2}$, 
  \begin{equation}\label{eq:weightedsobembedding}
    \|\psi\|_{\infty,w}
    \leq C\left(\int_{\bM}\textstyle{\sum}_{m=0}^{\kappa_{0}}\sum_{|\bfI|=m}
    \ldr{\tau}^{2(\kappa_{0}-m)\cweight}\ldr{\tau-\tau_{c}}^{2(\kappa_{0}-m)}|E_{\bfI}\psi|^{2}\mutgc\right)^{1/2}
  \end{equation}
  for all $\tau\leq \tau_{c}$, where $C$ only depends on $c_{\robas}$, $c_{\chi,\kappa_{0}+2}$, $C_{\rorel,\bfk_{1}}$, $C_{\theta,\kappa_{0}}$ and $(\bM,\bge_{\refer})$.
  Here
  \[
  \|\psi\|_{\infty,w}:=\|\psi w\|_{C^{0}(\bM)}.
  \]
\end{lemma}
\begin{remark}\label{remark:bDlnwestimate}
  The arguments presented in the proof also yield the conclusion that if the conditions of Lemma~\ref{lemma:CkestofvarrhoEi} are 
  fulfilled with $l=2$; and
  \[
  \|\ln\theta\|_{C^{\bfm_{1}}_{\weight_{0}}(\bM)}+\|q\|_{C^{1}_{\weight_{0}}(\bM)}\leq C_{\theta,1}
  \]
  for all $\tau\leq 0$, where $\bfm_{1}=(1,2)$, then
  \[
  |\bD\ln w|_{\bge_{\refer}}\leq C_{a}\ldr{\tau}^{\cweight}\ldr{\tau-\tau_{c}}
  \]
  for all $\tau\leq \tau_{c}$, where $C_{a}$ only depends on $c_{\robas}$, $c_{\chi,3}$, $C_{\rorel,\bfm_{1}}$, $C_{\theta,1}$ and $(\bM,\bge_{\refer})$.
\end{remark}
\begin{proof}
  Note, to begin with, that if $\kappa_{0}$ is the smallest integer which is strictly larger than $n/2$, then 
  \begin{equation}\label{eq:psiwSobolevEstimatefs}
    \|\psi w\|_{C^{0}(\bM)}
    \leq C\left(\int_{\bM}\textstyle{\sum}_{|\bfI|\leq \kappa_{0}}|E_{\bfI}(w\psi)|^{2}\mu_{\bge_{\refer}}\right)^{1/2}.
  \end{equation}
  On the other hand, $|E_{\bfI}(\psi w)|$ can be estimated by a linear combination of terms of the form
  \begin{equation}\label{eq:derlnwwestimate}
    |E_{\bfI_{1}}(\ln w)|\cdots |E_{\bfI_{k}}(\ln w)|\cdot |E_{\bfI_{0}}\psi|w,
  \end{equation}
  where $\bfI_{i}\neq 0$, $i=1,\dots,k$, and $|\bfI_{0}|+\dots+|\bfI_{k}|=|\bfI|$. In order to estimate $E_{\bfI}\ln w$, it is convenient to note that
  combining (\ref{eq:hUnlnthetamomqbas}), (\ref{eq:hUvarrhoident}) and (\ref{eq:dtaulnvarphi}) yields
  \begin{equation}\label{eq:dtaulnvarphiexpanded}
    \d_{\tau}\ln\tvarphi=-\tN [q-(n-1)]/n+\tN\hN^{-1}\rodiv_{\bge_{\refer}}\chi+\tN\hN^{-1}\chi\ln\tvarphi.
  \end{equation}
  At this stage, we wish to estimate the expressions that result when applying $E_{\bfI}$ to the right hand side. In order to estimate $E_{\bfI}$
  applied to the first term on the right hand side of (\ref{eq:dtaulnvarphiexpanded}), note that it is sufficient to estimate expressions of the form
  \[
  \tN\cdot E_{\bfI_{1}}\ln\hN\cdots E_{\bfI_{k}}\ln\hN\cdot E_{\bfJ}q
  \]
  where $|\bfI_{1}|+\dots+|\bfI_{k}|+|\bfJ|=|\bfI|$ and $\bfI_{j}\neq 0$. However, due to the assumptions, such expressions can be estimated by
  $C_{a}\ldr{\tau}^{|\bfI|\cweight}$ for all $\tau\leq 0$ and $|\bfI|\leq \kappa_{0}$, where $C_{a}$ only depends on $C_{\rorel,\bfk_{1}}$, $C_{\theta,\kappa_{0}}$
  and $(\bM,\bge_{\refer})$. In order to estimate the second term on the right hand side of (\ref{eq:dtaulnvarphiexpanded}), note that
  $\rodiv_{\bge_{\refer}}\chi=\omega^{i}(\bD_{E_{i}}\chi)$. It is thus sufficient to estimate expressions of the form
  \[
  \tN\hN^{-1}(\bD_{\bfJ}\omega^{i})(\bD_{\bfK}\bD_{E_{i}}\chi),
  \]
  where $|\bfJ|+|\bfK|=|\bfI|$. Due to (\ref{eq:DeltavarrhorelvariationEi}), (\ref{eq:hNtaudotequivEi}), (\ref{eq:chimClClest}) and the assumptions,
  such expressions can be estimated by $C_{b}\ldr{\tau}^{(|\bfI|+1)\cweight}e^{\eSpe\tau}$ for all $\tau\leq 0$ and $|\bfI|\leq \kappa_{0}$, where $C_{b}$ only
  depends on $c_{\robas}$, $c_{\chi,\kappa_{0}+1}$ and $(\bM,\bge_{\refer})$. In order to estimate the last term on the right hand side of
  (\ref{eq:dtaulnvarphiexpanded}), note that 
  \begin{equation}\label{eq:lnchvarphiEbfIpest}
    |E_{\bfI}(\ln\tvarphi)|\leq |E_{\bfI}(\varrho)|+|E_{\bfI}(\ln\theta)|\leq C_{a}\ldr{\varrho}^{|\bfI|\cweight+1}
  \end{equation}
  for all $\tau\leq 0$ and $1\leq |\bfI|\leq\kappa_{0}+1$, where we appealed to Lemma~\ref{lemma:CkestofvarrhoEi} and the assumptions. Here $C_{a}$ only
  depends on $c_{\robas}$, $c_{\chi,\kappa_{0}+2}$, $C_{\rorel,\bfk_{1}}$, $C_{\theta,\kappa_{0}}$ and $(\bM,\bge_{\refer})$. On the other hand, applying
  $E_{\bfI}$ to the last term on the right hand side of (\ref{eq:dtaulnvarphiexpanded}) yields expressions of the form
  \[
  \tN\hN^{-1}(\bD_{\bfJ}\omega^{i})(\bD_{\bfK}\chi)\bD_{\bfL}E_{i}\ln\tvarphi.
  \]
  Due to (\ref{eq:DeltavarrhorelvariationEi}), (\ref{eq:hNtaudotequivEi}), (\ref{eq:chimClClest}), (\ref{eq:lnchvarphiEbfIpest}) and the assumptions,
  such expressions can be estimated by $C_{c}\ldr{\tau}^{(|\bfI|+1)\cweight+1}e^{\eSpe\tau}$ for all $\tau\leq 0$ and $|\bfI|\leq \kappa_{0}$, where $C_{c}$ only
  depends on $c_{\robas}$, $c_{\chi,\kappa_{0}+2}$, $C_{\rorel,\bfk_{1}}$, $C_{\theta,\kappa_{0}}$ and $(\bM,\bge_{\refer})$. Summing up the above
  estimates yields the conclusion that
  \begin{equation}\label{eq:dtauEbfIlntvarphiestimate}
    |\d_{\tau}E_{\bfI}\ln\tvarphi|\leq C_{a}\ldr{\tau}^{|\bfI|\cweight}+C_{b}\ldr{\tau}^{(|\bfI|+1)\cweight+1}e^{\eSpe\tau}
  \end{equation}
  for all $\tau\leq 0$ and all $|\bfI|\leq \kappa_{0}$, where $C_{a}$ only depends on $C_{\rorel,\bfk_{1}}$, $C_{\theta,\kappa_{0}}$ and $(\bM,\bge_{\refer})$;
  and $C_{b}$ only depends on $c_{\robas}$, $c_{\chi,\kappa_{0}+2}$, $C_{\rorel,\bfk_{1}}$, $C_{\theta,\kappa_{0}}$ and $(\bM,\bge_{\refer})$. Integrating this
  estimate from $\tau$ to $\tau_{c}$ yields
  \[
  |E_{\bfI}\ln w|\leq C_{a}\ldr{\tau}^{|\bfI|\cweight}\ldr{\tau-\tau_{c}}+C_{b}\ldr{\tau_{c}}^{(|\bfI|+1)\cweight+1}e^{\eSpe\tau_{c}}
  \leq C_{b}\ldr{\tau}^{|\bfI|\cweight}\ldr{\tau-\tau_{c}}
  \]
  for all $\tau\leq\tau_{c}\leq 0$, where $C_{a}$ and $C_{b}$ have the same dependence as in the case of (\ref{eq:dtauEbfIlntvarphiestimate}).
  Combining this estimate with (\ref{eq:psiwSobolevEstimatefs}) and (\ref{eq:derlnwwestimate}) yields the conclusion of the lemma. 
\end{proof}

\section{Estimates of the weighted $C^{k}$ energy density}\label{section:Ckestimatesweightedenergydensity}

Next, we turn to the problem of estimating $\me_{k}$. 

\begin{lemma}\label{lemma:Ckestofweightedenden}
  Let $\kappa_{0}$ be the smallest integer strictly larger than $n/2$, $0\leq\cweight\in\ro$, $\bfk:=(1,\kappa_{0})$, $\bfk_{1}:=(1,\kappa_{0}+1)$,
  $\weight_{0}:=(0,\cweight)$ and $\weight:=(\cweight,\cweight)$. Assume that the conditions of Lemma~\ref{lemma:taurelvaryingbxEi} as well as the
  $(\cweight,\kappa_{0})$-supremum assumptions are satisfied. Then, if $0\leq k\in\zo$ and $w_{2}:=\tvarphi_{c}^{-1}\tvarphi=w^{2}$, 
  \begin{equation}\label{eq:mekinfwtwoestgeneral}
    \|\me_{k}(\cdot,\tau)\|_{\infty,w_{2}}\leq C_{a}\textstyle{\sum}_{m=0}^{\kappa_{0}}\ldr{\tau}^{2(\kappa_{0}-m)\cweight}
    \ldr{\tau-\tau_{c}}^{2(\kappa_{0}-m)}\hE_{k+m}(\tau;\tau_{c})
  \end{equation}
  for all $\tau\leq\tau_{c}\leq 0$, where $C_{a}$ only depends on $c_{\cweight,\kappa_{0}}$, $k$, $(\bM,\bge_{\refer})$ and a lower bound on $\theta_{0,-}$. 

  Next, let $0\leq k\in\zo$, $l:=k+\kappa_{0}$, and assume, in addition to the above, the $(\cweight,l)$-supremum assumptions to be satisfied;
  (\ref{eq:coefflassumptions}) to hold; and $u$ to be a solution to (\ref{eq:theequation}) with vanishing right hand side. Then, for all
  $\tau\leq \tau_{b}\leq\tau_{c}\leq 0$, 
  \begin{equation}\label{eq:meksupestimate}
    \begin{split}
      & \|\me_{k}(\cdot,\tau)\|_{\infty,w_{2}}\\
      \leq & C_{l}\textstyle{\sum}_{m=0}^{\kappa_{0}}\textstyle{\sum}_{j=0}^{m+k}\ldr{\tau}^{2\ba_{k,m,j}\cweight}\ldr{\tau-\tau_{c}}^{2\bb_{k,m,j}}
      \ldr{\tau-\tau_{b}}^{2\bc_{k,m,j}}e^{c_{0}(\tau_{b}-\tau)}\hE_{j}(\tau_{b};\tau_{c}),
    \end{split}    
  \end{equation}
  where $C_{l}$ only depends on $c_{\cweight,l}$, $c_{\coeff,l}$, $m_{\ros}$, $d_{\a}$ (in case $\iota_{b}\neq 0$), $(\bM,\bge_{\refer})$ and a lower bound
  on $\theta_{0,-}$. Moreover 
  \begin{align*}
    \ba_{k,m,j} = & (k+m+j+3)(m+k-j)/2+\kappa_{0}-m,\\
    \bb_{k,m,j} = & 3(m+k-j)\iota_{b}/2+\kappa_{0}-m,\\
    \bc_{k,m,j} = & k+m-j
  \end{align*}
  for all $0\leq m\leq \kappa_{0}$ and $0\leq j\leq m+k$.
\end{lemma}
\begin{remark}\label{remark:puttingcztozinLinfest}
  If, in addition to the assumptions of the lemma, all the conditions of Corollary~\ref{cor:basicenergyestimate} are satisfied, the estimate
  (\ref{eq:meksupestimate}) can be improved in the sense that the factor $e^{c_{0}(\tau_{b}-\tau)}$ can be removed. On the other hand, the constant $C_{l}$
  appearing in (\ref{eq:meksupestimate}) then also depends on $d_{\coeff}$, $d_{q}$ and $d_{\a}$. Finally, note that, in this setting,
  (\ref{eq:lntvarphimlntvarphicimp}) holds, so that $\tvarphi_{c}^{-1}\tvarphi$ can be bounded from above and below by strictly positive constants. 
\end{remark}
\begin{proof}
  The idea of the proof is to appeal to (\ref{eq:weightedsobembedding}) with $\psi$ replaced by $\hU E_{\bfJ}u$, $e^{-\mu_{A}}X_{A}E_{\bfJ}u$ and 
  $E_{\bfJ}u$. However, this necessitates interchanging the order of $\hU$ and $E_{\bfI}$, as well as the order of $e^{-\mu_{A}}X_{A}$ and $E_{\bfI}$. 
  
  \textbf{Commuting with $\hU$.} Note that 
  \[
  |E_{\bfI}\hU E_{\bfJ}u|\leq |[E_{\bfI},\hU]E_{\bfJ}u|+|\hU E_{\bfI}E_{\bfJ}u|.
  \]
  Combining this inequality with Remark~\ref{remark:ZzeqtoIdEi} yields, assuming $i=|\bfI|$ and $j=|\bfJ|$,
  \begin{equation*}
    \begin{split}
      |E_{\bfI}\hU E_{\bfJ}u| \leq & \sqrt{2}\me_{i+j}^{1/2}+C_{a}\textstyle{\sum}_{m=0}^{i-1}\ldr{\varrho}^{(i-m)\cweight}\me_{m+j}^{1/2}\\
      & +C_{a}\textstyle{\sum}_{m=0}^{i}\ldr{\varrho}^{(i-m+1)\cweight}\ldr{\tau-\tau_{c}}^{3\iota_{b}/2}e^{\e_{\Spe}\varrho}\me_{m+j}^{1/2}
    \end{split}    
  \end{equation*}
  on $M_{-}$, where $C_{a}$ only depends on $c_{\cweight,i}$ and $(\bM,\bge_{\refer})$. In particular, 
  \begin{equation}\label{eq:hUcommsupestfs}
    \begin{split}
      & \int_{\bM_{\tau}}\ldr{\tau}^{2(\kappa_{0}-i)\cweight}\ldr{\tau-\tau_{c}}^{2(\kappa_{0}-i)}|E_{\bfI}\hU E_{\bfJ}u|^{2}\mutgc\\
      \leq & 3\ldr{\tau}^{2(\kappa_{0}-i)\cweight}\ldr{\tau-\tau_{c}}^{2(\kappa_{0}-i)}\hE_{i+j}(\tau;\tau_{c})\\
      & +C_{b}\textstyle{\sum}_{m=0}^{i-1}\ldr{\tau}^{2(\kappa_{0}-m)\cweight}\ldr{\tau-\tau_{c}}^{2(\kappa_{0}-i)}\hE_{m+j}(\tau;\tau_{c})\\
      & +C_{b}\textstyle{\sum}_{m=0}^{i}\ldr{\tau}^{2(\kappa_{0}-m+1)\cweight}\ldr{\tau-\tau_{c}}^{2(\kappa_{0}-i)+3\iota_{b}}e^{2\eSpe\tau}\hE_{m+j}(\tau;\tau_{c})\\
      \leq & C_{b}\textstyle{\sum}_{m=0}^{i}\ldr{\tau}^{2(\kappa_{0}-m)\cweight}\ldr{\tau-\tau_{c}}^{2(\kappa_{0}-i)}\hE_{m+j}(\tau;\tau_{c})
    \end{split}    
  \end{equation}
  for all $\tau\leq\tau_{c}$, where $C_{b}$ only depends on $c_{\cweight,i}$ and $(\bM,\bge_{\refer})$.

  \textbf{Commuting with $e^{-\mu_{A}}X_{A}$.} Next, note that 
  \[
  E_{\bfI}(e^{-\mu_{A}}X_{A}E_{\bfJ}u)=[E_{\bfI},e^{-\mu_{A}}X_{A}]E_{\bfJ}u+e^{-\mu_{A}}X_{A}E_{\bfI}E_{\bfJ}u.
  \]
  Combining this equality with Lemma~\ref{lemma:emmuAXAEbfIcomm} yields, assuming $i=|\bfI|$ and $j=|\bfJ|$,
  \begin{equation*}
    \begin{split}
      \left|E_{\bfI}\left(e^{-\mu_{A}}X_{A}E_{\bfJ}u\right)\right| \leq & \sqrt{2}\me_{i+j}^{1/2}
      +C_{b}\ldr{\varrho}^{i(2\cweight+1)}e^{\e_{\Spe}\varrho}\textstyle{\sum}_{1\leq |\bfK|\leq i}|E_{\bfK}E_{\bfJ}u|
    \end{split}
  \end{equation*}
  on $M_{-}$, where $C_{b}$ only depends on $c_{\cweight,i}$, $(\bM,\bge_{\refer})$ and a lower bound on $\theta_{0,-}$.
  Thus
  \begin{equation*}
    \begin{split}
      & \int_{\bM_{\tau}}\ldr{\tau}^{2(\kappa_{0}-i)\cweight}\ldr{\tau-\tau_{c}}^{2(\kappa_{0}-i)}\left|E_{\bfI}\left(e^{-\mu_{A}}X_{A}E_{\bfJ}u\right)\right|^{2}\mutgc\\
      \leq & C_{b}\ldr{\tau}^{2(\kappa_{0}-i)\cweight}\ldr{\tau-\tau_{c}}^{2(\kappa_{0}-i)}\hE_{i+j}(\tau;\tau_{c}).
    \end{split}    
  \end{equation*}
  Combining this estimate with (\ref{eq:hUcommsupestfs}) and (\ref{eq:weightedsobembedding}) yields (\ref{eq:mekinfwtwoestgeneral}).
  Combining (\ref{eq:mekinfwtwoestgeneral}) with (\ref{eq:basichigherorderenergyestimate}) yields (\ref{eq:meksupestimate}).
\end{proof}

\chapter{Higher order energy estimates, part II}\label{chapter:hoeepII}

In the previous chapter, we derive estimates for $\hE_{k}$, and, via Sobolev embedding, also for $\me_{k}$. The derivation is based on 
$(\cweight,l)$-supremum assumptions. In the present chapter, the idea is to estimate $[E_{\bfI},L]u$ in $L^{2}$ using Moser type estimates
and $(\cweight,l)$-Sobolev assumptions. However, in order for this to be possible, we need to control $u$ and its first
derivatives in $C^{0}$. For that reason, we assume the $(\cweight,\kappa_{1})$-supremum assumptions to be satisfied, where $\kappa_{1}$ is the smallest
integer strictly larger than $n/2+1$. This gives us the desired control of $u$ and its first derivatives. A second problem which arises when appealing
to the Moser estimates is the one of relating expressions of the form
\begin{equation}\label{eq:reorderingspatialderivativesXA}
  \int_{\bM_{\tau}}|E_{\bfI}(e^{-\mu_{A}}X_{A}u)|^{2}\mutgc,\ \ \
  \int_{\bM_{\tau}}|e^{-\mu_{A}}X_{A}E_{\bfI}u|^{2}\mutgc. 
\end{equation}
The reason for this is that the first expression is of a type that naturally results when appealing to the Moser estimates, and the second
expression is of the type that appears in the energies. 

We begin the chapter in Section~\ref{section:reorderingderivatives} by deriving estimates that, e.g., relate the expressions appearing in
(\ref{eq:reorderingspatialderivativesXA}). The proofs are based on Moser estimates obtained in
Section~\ref{section:applicationsofgagliardonirenbergestimates}. Given the results concerning the reordering of derivatives, we then proceed to an estimate
of commutators in Section~\ref{section:commutatorshigherorderenergiesMoserest}. These estimates are based on $(\cweight,1)$-supremum assumptions as well as
$(\cweight,l)$-Sobolev assumptions. However, the right hand sides of the estimates contain supremum norms of up to one derivative of the unknown, and these
expressions will later need to be estimated by appealing to the $(\cweight,\kappa_{1})$-supremum assumptions. When estimating commutators involving the
coefficients of the equations we, needless to say, need to impose analogous assumptions concerning the coefficients.
In some of the commutator estimates, $E_{\bfK}\hU^{2}u$ appears on the right hand side. Estimating this expression requires a separate argument,
which we provide in Section~\ref{section:estimatinghUsquulsobas}. Given the above, we are in a position to estimate the commutator with $L$, and
we do so in Section~\ref{section:commutatorwithLulSobass}. Combining these conclusions with the zeroth order energy estimate and an inductive argument,
higher order energy estimates can now immediately be derived; cf. Section~\ref{section:energyestimatesstepII}. We end the chapter by illustrating
the consequences of the estimates in the case of the Klein-Gordon equation. We also illustrate that it is possible to derive more detailed asymptotic
information in case $q-(n-1)$ converges to zero exponentially; cf. Proposition~\ref{prop:asvelocityKGlikeeq}. 

\section{Reordering derivatives}\label{section:reorderingderivatives}

In the arguments to follow, we appeal to Corollary~\ref{cor:mixedmoserestweight}. When doing so, one of the weights will be
\begin{equation}\label{eq:wdef}
  w : = \tvarphi_{c}^{-1/2}\tvarphi^{1/2},
\end{equation}
\index{$\a$Aa@Notation!Functions!$w$}%
where $\tvarphi$ and $\tvarphi_{c}$ are defined by (\ref{eq:tvarphidefinition}) and (\ref{eq:tvarphicdef}) respectively; from now on $t_{c}$, and
the corresponding $\tau_{c}=\tau(t_{c})$, used to define $\tvarphi_{c}$ will be considered to be fixed. We therefore need to estimate
\index{$\a$Aa@Notation!Functions!$\tga$}%
\begin{equation}\label{eq:chgammadef}
  \tga(t):=1+\sup_{\bx\in\bM}|\bD w(\bx,t)|_{\bge_{\refer}}.
\end{equation}
\begin{lemma}\label{lemma:chgammaestimate}
  Let $0\leq \cweight\in\ro$, $\weight_{0}=(0,\cweight)$ and $\weight=(\cweight,\cweight)$. Assume that the conditions of 
  Lemma~\ref{lemma:taurelvaryingbxEi} as well as the $(\cweight,1)$-supremum assumptions are satisfied. Then there is a constant $C_{\g}$ such that
  \begin{equation}\label{eq:chgammaestimate}
    \tga(t)\leq C_{\g}\ldr{\tau(t)}^{\cweight}\ldr{\tau(t)-\tau_{c}}
  \end{equation}
  for all $t\leq t_{c}$, where $C_{\gamma}$ only depends on $c_{\cweight,1}$ and $(\bM,\bge_{\refer})$.
\end{lemma}
\begin{remark}
  The choice of assumptions is motivated by the assumptions we make in the applications; the conclusion of the lemma holds under weaker
  conditions. 
\end{remark}
\begin{proof}
  The statement follows from Remark~\ref{remark:bDlnwestimate} and the assumptions. 
\end{proof}
Below, we use the following notation for $1\leq p<\infty$ and families $\mt$ of tensor fields on $\bM$, where $w$ is defined by
(\ref{eq:wdef}):
\begin{align}
  \|\mt(\cdot,t)\|_{p,w} := & \left(\int_{\bM}|\mt(\cdot,t)|_{\bge_{\refer}}^{p}w^{p}(\cdot,t)\mu_{\bge_{\refer}}\right)^{1/p},\label{eq:Lpwweightest}\\
  \|\mt(\cdot,t)\|_{\infty,w} := & \sup_{\bx\in\bM}|\mt(\bx,t)|_{\bge_{\refer}}w(\bx,t).\label{eq:Linftyweightest}
\end{align}
Moreover,
\[
\|\bD^{1}_{\bbE}u\|_{\infty,w}:=\textstyle{\sum}_{i=1}^{n}\|E_{i}u\|_{\infty,w}.
\]
In order to relate the expressions appearing in (\ref{eq:reorderingspatialderivativesXA}), note that the following holds. 

\begin{lemma}\label{lemma:EbfIemmuAXAestimate}
  Let $0\leq \cweight\in\ro$, $\weight_{0}=(0,\cweight)$ and $\weight=(\cweight,\cweight)$. Assume that the conditions of 
  Lemma~\ref{lemma:taurelvaryingbxEi} as well as the $(\cweight,1)$-supremum assumptions are satisfied. Then, if $0\leq m\in\zo$ and
  $|\bfI|\leq m$,
  \begin{equation}\label{eq:EbfIemmuAXAestimate}
    \begin{split}
      & \|E_{\bfI}(e^{-\mu_{A}}X_{A}u)\|_{2,w}\\
      \leq & \sqrt{2}\hE_{m}^{1/2}+C_{a}\theta_{0,-}^{-1}\ldr{\tau}^{\a_{m}\cweight+\b_{m}}e^{\eSpe\tau}\|\bD^{1}_{\bbE}u\|_{\infty,w}
           [\|\mK\|_{H^{\bfm}_{\weight_{0}}(\bM)}+\|\mu_{A}\|_{H^{\bfm}_{\weight}(\bM)}]\\
           & +C_{a}\theta_{0,-}^{-1}\ldr{\tau}^{\a_{m}\cweight+\b_{m}}e^{\eSpe\tau}\hE_{m}^{1/2}
    \end{split}
  \end{equation}
  for all $\tau\leq \tau_{c}$ (no summation on $A$), where $\bfm:=(1,m)$, $C_{a}$ only depends on $c_{\cweight,1}$, $m$ and $(\bM,\bge_{\refer})$;
  and $\a_{m}$, $\b_{m}$ only depend on $m$. Moreover, the second and third terms on the right hand side should be omitted when $m=0$.
  
  If, in addition, the $(\cweight,l)$-Sobolev assumptions are satisfied for some $1\leq l\in\zo$ and
  $|\bfI|\leq m\leq l$, then 
  \begin{equation}\label{eq:EbfIemmuAXAestimateaddass}
    \begin{split}
      \|E_{\bfI}(e^{-\mu_{A}}X_{A}u)\|_{2,w}
      \leq & C_{a}\hE_{m}^{1/2}+C_{b}\ldr{\tau}^{\a_{m}\cweight+\b_{m}}e^{\eSpe\tau}\|\bD^{1}_{\bbE}u\|_{\infty,w}           
    \end{split}
  \end{equation}
  for all $\tau\leq \tau_{c}$, where $C_{a}$ only depends on $c_{\cweight,1}$, $m$, $(\bM,\bge_{\refer})$ and a lower bound on $\theta_{0,-}$;
  and $C_{b}$ only depends on $c_{\cweight,1}$, $s_{\cweight,m}$, $(\bM,\bge_{\refer})$ and a lower bound on $\theta_{0,-}$.  
\end{lemma}
\begin{remark}
  In this lemma, and what follows, $\hE_{k}$ means $\hE_{k}(\cdot;\tau_{c})$.
\end{remark}
\begin{remark}
  Due to the proof, $\|[E_{\bfI},e^{-\mu_{A}}X_{A}]u\|_{2,w}$, can be estimated by the sum of the last two terms on the right hand side of
  (\ref{eq:EbfIemmuAXAestimate}). 
\end{remark}
\begin{proof}
  To begin with, 
  \[
  |E_{\bfI}(e^{-\mu_{A}}X_{A}u)|\leq |e^{-\mu_{A}}X_{A}E_{\bfI}u|+|[E_{\bfI},e^{-\mu_{A}}X_{A}]u|
  \]
  (no summation on $A$). On the other hand, the second term on the right hand side can be estimated by appealing to
  Lemma~\ref{lemma:bdbfAGAXGcommformEi}. In fact, 
  \[
  |[E_{\bfI},e^{-\mu_{A}}X_{A}]u|\leq \textstyle{\sum}_{1\leq |\bfJ|\leq |\bfI|}|H_{\bfI,\bfJ}|\cdot |E_{\bfJ}u|,
  \]
  where
  \[
  |H_{\bfI,\bfJ}|\leq C_{a}\textstyle{\sum}_{k_{a}+|\bfK|\leq l_{b}}\mfP_{\mK,k_{a}}|E_{\bfK}(e^{-\mu_{A}})|
  \]
  $l_{b}:=|\bfI|-|\bfJ|+1$ and $C_{a}$ only depends on $\mKsup$, $\e_{\rond}$, $|\bfI|$, $n$ and $(\bM,\bge_{\refer})$. In practice, we thus wish to 
  estimate 
  \[
  e^{-\mu_{A}}|\bD^{m_{1}}\mK|_{\bge_{\refer}}\cdots |\bD^{m_{r}}\mK|_{\bge_{\refer}}|E_{\bfK_{1}}\mu_{A}|\cdots |E_{\bfK_{p}}\mu_{A}||E_{\bfJ}u|
  \]
  in $L^{2}$ (with weight $w$), where $m_{j}\neq 0$, $\bfK_{j}\neq 0$ and $m_{\rotot}:=m_{1}+\dots+m_{r}+|\bfK_{1}|+\dots+|\bfK_{p}|\leq l_{b}$. 

  To this end, we first estimate $e^{-\mu_{A}}$ by appealing to (\ref{eq:muminmainlowerbound}) and (\ref{eq:eSpevarrhoeelowtaurelEi}). If $m_{\rotot}=0$,
  we obtain
  \[
  \int_{\bM}e^{-2\mu_{A}}|E_{\bfJ}u|^{2}\mutgc\leq C_{a}\theta_{0,-}^{-2}\ldr{\tau}^{3\iota_{b}}e^{2\eSpe\tau}\hE_{k_{a}}
  \]
  for $\tau\leq \tau_{c}$ and $|\bfJ|\leq k_{a}$. Here $C_{a}$ only depends on $c_{\robas}$ and $m$. Assume now that $m_{\rotot}>0$. Then $r+p\geq 1$.
  Moreover, we rewrite $E_{\bfK_{i}}\mu_{A}=E_{\bfK_{i,a}}E_{\bfK_{i,b}}\mu_{A}$ and $E_{\bfJ}u=E_{\bfJ_{a}}E_{\bfJ_{b}}u$, where it is understood that $|\bfK_{i,b}|=1$
  and $|\bfJ_{b}|=1$. Again, we estimate $e^{-\mu_{A}}$ by appealing to (\ref{eq:muminmainlowerbound}) and (\ref{eq:eSpevarrhoeelowtaurelEi}) and then
  appeal to Corollary~\ref{cor:mixedmoserestweight}. Note, when doing so, that $q=0$, $s=p+1$, $u_{j}=1$, $g_{j}=1$, $h_{m}=1$, and $v_{m}=1$ for
  $m=1,\dots,p$. Moreover, $h_{s}=1$ and $v_{s}=w$, where $w$ is defined by (\ref{eq:wdef}). In addition, $\mt_{j}=\bD\mK$, $\mU_{m}=E_{\bfK_{m,b}}\mu_{A}$
  for $m=1,\dots,p$, and $\mU_{s}=E_{\bfJ_{b}}u$. Let
  \[
  k_{\rotot}:=m_{\rotot}+|\bfJ|-r-p-1\leq |\bfI|-r-p.
  \]
  Then
  \begin{equation}\label{eq:ganiintermediatecommspatial}
    \begin{split}
      & \left(\int_{\bM}e^{-2\mu_{A}}|\bD^{m_{1}}\mK|_{\bge_{\refer}}^{2}\cdots |\bD^{m_{r}}\mK|_{\bge_{\refer}}^{2}|E_{\bfK_{1}}\mu_{A}|^{2}\cdots
      |E_{\bfK_{p}}\mu_{A}|^{2}|E_{\bfJ}u|^{2}\mutgc\right)^{1/2}\\
      \leq & C_{a}\theta_{0,-}^{-1}e^{\eSpe\tau}\sum_{k\leq k_{\rotot}}\|\bD^{k+1}\mK\|_{2}\|\bD\mK\|_{\infty}^{r-1}
      \prod_{i=1}^{p}\|E_{\bfK_{i,b}}\mu_{A}\|_{\infty}\|E_{\bfJ_{b}}u\|_{\infty,v_{s}}\\
      & +C_{a}\theta_{0,-}^{-1}e^{\eSpe\tau}\sum_{i=1}^{p}\sum_{|\bfK|\leq k_{\rotot}}\|\bD\mK\|_{\infty}^{r}\|E_{\bfK}E_{\bfK_{i,b}}\mu_{A}\|_{2}
      \prod_{j\neq i}\|E_{\bfK_{j,b}}\mu_{A}\|_{\infty}\|E_{\bfJ_{b}}u\|_{\infty,v_{s}}\\
      & +C_{a}\theta_{0,-}^{-1}e^{\eSpe\tau}\sum_{|\bfK|\leq k_{\rotot}}\tga^{k_{\rotot}-|\bfK|}\|\bD\mK\|_{\infty}^{r}\prod_{i=1}^{p}\|E_{\bfK_{i,b}}\mu_{A}\|_{\infty}
      \|E_{\bfK}E_{\bfJ_{b}}u\|_{2,v_{s}}
    \end{split}
  \end{equation}
  for all $\tau\leq \tau_{c}$, where $C_{a}$ only depends on $c_{\robas}$, $k_{\rotot}$ and $(\bM,\bge_{\refer})$. Moreover, $\tga$ is given by 
  (\ref{eq:chgammadef}). Combining (\ref{eq:ganiintermediatecommspatial}) with (\ref{eq:chgammaestimate}), Remark~\ref{remark:CkestofmuAEi} 
  and the assumptions yields
  \begin{equation}\label{eq:ganisecondarycommspatial}
    \begin{split}
      & \left(\int_{\bM}e^{-2\mu_{A}}|\bD^{m_{1}}\mK|_{\bge_{\refer}}^{2}\cdots |\bD^{m_{r}}\mK|_{\bge_{\refer}}^{2}|E_{\bfK_{1}}\mu_{A}|^{2}\cdots
      |E_{\bfK_{p}}\mu_{A}|^{2}|E_{\bfJ}u|^{2}\mutgc\right)^{1/2}\\
      \leq & C_{b}\theta_{0,-}^{-1}\ldr{\tau}^{p\cweight+(k_{\rotot}+r+p)\cweight+p-1}e^{\eSpe\tau}\|\bD^{1}_{\bbE}u\|_{\infty,w}
      [\ldr{\tau}\|\mK\|_{H^{\bka}_{\weight_{0}}(\bM)}+\|\mu_{A}\|_{H^{\bka}_{\weight}(\bM)}]\\
      & +C_{b}\theta_{0,-}^{-1}e^{\eSpe\tau}\sum_{|\bfK|\leq k_{\rotot}}\ldr{\tau}^{p\cweight+(k_{\rotot}+r+p-|\bfK|)\cweight+k_{\rotot}+p+3\iota_{b}/2-|\bfK|}\hE_{|\bfK|+1}^{1/2}
    \end{split}
  \end{equation}
  for all $\tau\leq\tau_{c}$, where $C_{b}$ only depends on $c_{\cweight,1}$, $k_{\rotot}$ and $(\bM,\bge_{\refer})$. Moreover, 
  $\bka=(1,k_{\rotot}+1)$. Thus (\ref{eq:EbfIemmuAXAestimate}) holds. Combining this estimate with (\ref{eq:HlmHlequiv}) and the conclusions of 
  Remark~\ref{remark:muAmhlestimate} and yields (\ref{eq:EbfIemmuAXAestimateaddass}). The lemma follows.
\end{proof}

\subsection{Reordering involving the normal derivative}

Next, we wish to relate expressions of the form $\|E_{\bfI}\hU u\|_{2,w}$ and $\|\hU E_{\bfI}u\|_{2,w}$. The following lemma serves this purpose.

\begin{lemma}\label{lemma:EbfIhUuestbfIgeneralorder}
  Let $0\leq \cweight\in\ro$, $\weight_{0}=(0,\cweight)$ and $\weight=(\cweight,\cweight)$. Assume that the conditions of 
  Lemma~\ref{lemma:taurelvaryingbxEi} as well as the $(\cweight,1)$-supremum assumptions are satisfied. Let $c_{\chi,2}$ be defined 
  as in the statement of Lemma~\ref{lemma:respvarvarrhoEi}. Then, if $|\bfI|=1$, 
  \begin{equation}\label{eq:EbfIhUubfIorderone}
    \|E_{\bfI}\hU u\|_{2,w}\leq C\hE_{1}^{1/2}
  \end{equation}
  for all $\tau\leq\tau_{c}$, where $C$ only depends on $c_{\robas}$, $c_{\chi,2}$ and $(\bM,\bge_{\refer})$. Fix $l$
  as in Definition~\ref{def:sobklassumptions} and assume that the $(\cweight,l)$-Sobolev assumptions are satisfied. Then, if $2\leq m\leq l$ and 
  $|\bfI|=m$, 
  \begin{equation}\label{eq:EbfIhUuestbfIgeneralorder}
    \begin{split}
      \|E_{\bfI}\hU u\|_{2,w}
      \leq & C_{a}\hE_{m}^{1/2}+C_{a}\ldr{\tau}^{\a_{m}\cweight}\ldr{\tau-\tau_{c}}^{\b_{m}}\hE_{m-1}^{1/2}\\
      & +C_{b}\ldr{\tau}^{\a_{m}\cweight}\ldr{\tau-\tau_{c}}^{\b_{m}}[\|\hU u\|_{\infty,w}+e^{\eSpe\tau}\|\bD^{1}_{\bbE}u\|_{\infty,w}]
    \end{split}
  \end{equation}
  for all $\tau\leq\tau_{c}$. Here $\a_{m}$ and $\b_{m}$ are constants depending only on $m$. Moreover, $C_{a}$ only depends on $c_{\cweight,1}$, $m$, 
  and $(\bM,\bge_{\refer})$; and $C_{b}$ only depends on $c_{\cweight,1}$, $s_{\cweight,m}$ and $(\bM,\bge_{\refer})$.
\end{lemma}
\begin{proof}
  Note that 
  \begin{equation}\label{eq:EbfIhUuest}
    |E_{\bfI}(\hU u)|\leq |\hU E_{\bfI}u|+|[E_{\bfI},\hU]u|.
  \end{equation}
  The second term on the right hand side can be estimated by appealing to Lemma~\ref{lemma:bdbfAchthhUcommformEireverse}. This yields
  \begin{equation}\label{eq:EbfIhUuestcommutator}
    |[E_{\bfI},\hU]u|\leq \textstyle{\sum}_{|\bfJ|\leq |\bfI|-1}|\bGe_{\bfI,\bfJ}^{1}|\cdot |E_{\bfJ}\hU u|
    +\textstyle{\sum}_{1\leq |\bfJ|\leq |\bfI|}|\bGe_{\bfI,\bfJ}^{0}|\cdot |E_{\bfJ}u|
  \end{equation}
  where
  \begin{align}
    |\bGe_{\bfI,\bfJ}^{1}| \leq & C_{a}\textstyle{\sum}_{k_{a}\leq l_{a}}\mfP_{N,k_{a}},\label{eq:bGebfIbfJzeroestimate}\\
    |\bGe_{\bfI,\bfJ}^{0}| \leq & C_{a}\textstyle{\sum}_{k_{a}+|\bfK|\leq l_{a}}\textstyle{\sum}_{i,k}\mfP_{N,k_{a}}
    |E_{\bfK}(A_{i}^{k})|\nonumber,
  \end{align}
  $l_{a}:=|\bfI|-|\bfJ|$ and $C_{a}$ only depends on $|\bfI|$, $n$ and $(\bM,\bge_{\refer})$.

  \textbf{Step 1.} Note that if $|\bfI|=1$, then (\ref{eq:EbfIhUuestcommutator}) yields
  \begin{equation}\label{eq:EbfIhUcommutatorstepone}
    |[E_{\bfI},\hU]u|\leq C_{a}|\hU u|+C_{b}\ldr{\tau}^{\cweight}e^{\eSpe\tau}\textstyle{\sum}_{|\bfJ|=1}|E_{\bfJ}u|
  \end{equation}
  for all $\tau\leq 0$, where $C_{a}$ only depends on $C_{\rorel}$, $n$ and $(\bM,\bge_{\refer})$; and $C_{b}$ only depends on $c_{\robas}$,
  $c_{\chi,2}$ and $(\bM,\bge_{\refer})$. In order to obtain this estimate we appealed to 
  Lemma~\ref{lemma:Aiksupest}. Combining (\ref{eq:EbfIhUcommutatorstepone}) with (\ref{eq:EbfIhUuest}) yields
  \[
  |E_{\bfI}(\hU u)|\leq \sqrt{2}\me_{1}^{1/2}+C_{a}\me^{1/2}
  +C_{b}\ldr{\tau}^{\cweight+3\iota_{b}/2}e^{\eSpe\tau}\me_{1}^{1/2}
  \]
  for all $\tau\leq 0$, where $C_{a}$ only depends on $C_{\rorel}$, $n$ and $(\bM,\bge_{\refer})$; and $C_{b}$ only depends on $c_{\robas}$, 
  $c_{\chi,2}$ and $(\bM,\bge_{\refer})$. In particular, 
  \begin{equation}\label{eq:changingorderhUabsbfIeqone}
    \begin{split}
      \|E_{\bfI}\hU u\|_{2,w} \leq & 2\hE_{1}^{1/2}+C_{a}\hE^{1/2}
      +C_{b}\ldr{\tau}^{\cweight+3\iota_{b}/2}e^{\eSpe\tau}\hE_{1}^{1/2}\leq C_{c}\hE_{1}^{1/2}
    \end{split}
  \end{equation}
  for all $\tau\leq\tau_{c}$, where $C_{a}$ only depends on $C_{\rorel}$, $n$ and $(\bM,\bge_{\refer})$; and $C_{b}$ and $C_{c}$ only depend on $c_{\robas}$, 
  $c_{\chi,2}$ and $(\bM,\bge_{\refer})$. Thus (\ref{eq:EbfIhUubfIorderone}) holds. 

  \textbf{Step 2.} Next, we carry out an inductive argument. We begin by estimating the second term on the right hand side of
  (\ref{eq:EbfIhUuestcommutator}) for general $\bfI$. If $|\bfI|=|\bfJ|$ and $|\bfI|\leq m$, then
  \[
  \int_{\bM}|\bGe_{\bfI,\bfJ}^{0}|^{2}|E_{\bfJ}u|^{2}\mutgc\leq C_{a}\ldr{\tau}^{2\cweight}e^{2\eSpe\tau}\int_{\bM}|E_{\bfJ}u|^{2}\mutgc
  \]
  for $\tau\leq\tau_{c}$, where $C_{a}$ only depends on $c_{\robas}$, $m$, $c_{\chi,2}$ and $(\bM,\bge_{\refer})$. In general, let
  $\bfJ_{a}$ and $\bfJ_{b}$ be such that $E_{\bfJ}u=E_{\bfJ_{a}}E_{\bfJ_{b}}u$ and $|\bfJ_{b}|=1$. Then we wish to estimate
  \[
  \left(\int_{\bM}\mfP_{N,k_{a}}^{2}|E_{\bfK}(A_{i}^{k})|^{2}|E_{\bfJ_{a}}E_{\bfJ_{b}}u|^{2}\mutgc\right)^{1/2}.
  \]
  To do so, we proceed as in the proof of Lemma~\ref{lemma:EbfIemmuAXAestimate}. Assuming $2\leq |\bfI|\leq m$, this expression can be estimated by
  \begin{equation*}
    \begin{split}
      &C_{a}\ldr{\tau}^{m\cweight}e^{\eSpe\tau}\|\ln\hN\|_{H^{\bfm}_{\weight_{0}}(\bM)}\|\bD^{1}_{\bbE}u\|_{\infty,w}\\
      &+C_{b}\ldr{\tau}^{m\cweight}\|A_{i}^{k}\|_{\mH^{m-1}_{\bbE,\weight}(\bM)}\|\bD^{1}_{\bbE}u\|_{\infty,w}
      +C_{c}\ldr{\tau}^{m\cweight+m+3\iota_{b}/2-1}e^{\eSpe\tau}\hE_{m}^{1/2}
    \end{split}
  \end{equation*}
  for $\tau\leq\tau_{c}$, where $C_{a}$ and $C_{c}$ only depend on $c_{\cweight,1}$, $m$ and $(\bM,\bge_{\refer})$; and $C_{b}$ only 
  depends on $C_{\rorel}$, $\cweight$, $m$, $n$ and $(\bM,\bge_{\refer})$. Here $\bfm:=(1,m-1)$ and $w:=(\tvarphi/\tvarphi_{c})^{1/2}$. 

  \textbf{Step 3.} Next, consider the first term on the right hand side of (\ref{eq:EbfIhUuestcommutator}) for general $\bfI$. Keeping 
  (\ref{eq:bGebfIbfJzeroestimate}) in mind, there are two cases to consider. If $k_{a}\leq 1$, then 
  \[
  \left(\int_{\bM}\mfP_{N,k_{a}}^{2}|E_{\bfJ}\hU u|^{2}\mutgc\right)^{1/2}\leq C_{a}\left(\int_{\bM}|E_{\bfJ}\hU u|^{2}\mutgc\right)^{1/2}
  \]
  for $\tau\leq\tau_{c}$, where $C_{a}$ only depends on $C_{\rorel}$. In this case, the idea is to estimate the right hand side by appealing to an 
  inductive assumption, since $|\bfJ|\leq |\bfI|-1$. In case $k_{a}\geq 1$, we can proceed as above: if $k\geq 1$, we rewrite factors of the 
  form $|\bD^{k}\ln\hN|_{\bge_{\refer}}$ in $\mfP_{N,k_{a}}$ as $|\bD^{k_{0}+1}\ln\hN|_{\bge_{\refer}}$ and then appeal to 
  Corollary~\ref{cor:mixedmoserestweight}. Assuming $|\bfI|\leq m$, the corresponding expression can be estimated by 
  \begin{equation*}
    \begin{split}
      & C_{a}\ldr{\tau}^{m\cweight}\|\ln\hN\|_{H^{\bfm_{1}}_{\weight_{0}}(\bM)}\|\hU u\|_{\infty,w}\\
      & +C_{a}\textstyle{\sum}_{l=0}^{m-1}\textstyle{\sum}_{|\bfK|=l}\ldr{\tau}^{(m-1-l)\cweight}\ldr{\tau-\tau_{c}}^{m-1-l}\|E_{\bfK}\hU u\|_{2,w}
    \end{split}
  \end{equation*}  
  for all $\tau\leq\tau_{c}$, where $C_{a}$ only depends on $c_{\cweight,1}$, $m$ and $(\bM,\bge_{\refer})$. Moreover, $\bfm_{1}:=(1,m)$. Again, the idea is 
  to estimate the second term by appealing to an inductive assumption. 

  \textbf{Step 4.} Note that (\ref{eq:changingorderhUabsbfIeqone}) holds in case $|\bfI|=1$. Let us therefore assume $2\leq |\bfI|\leq m$. 
  Combining (\ref{eq:EbfIhUuest}) and (\ref{eq:EbfIhUuestcommutator}) with the estimates resulting from steps 2 and 3 then yields
  \begin{equation}\label{eq:inductivestepchordderEbfIhU}
    \begin{split}
       \|E_{\bfI}\hU u\|_{2,w}
      \leq & \|\hU E_{\bfI}u\|_{2,w}+C_{a}\ldr{\tau}^{m\cweight+m+3\iota_{b}/2-1}e^{\eSpe\tau}\hE_{m}^{1/2}\\
      & +C_{a}\ldr{\tau}^{m\cweight}\left[e^{\eSpe\tau}\|\ln\hN\|_{H^{\bfm}_{\weight_{0}}(\bM)}
      +\textstyle{\sum}_{i,k}\|A_{i}^{k}\|_{\mH^{m-1}_{\bbE,\weight}(\bM)}\right]\|\bD^{1}_{\bbE}u\|_{\infty,w}\\
      & +C_{a}\ldr{\tau}^{m\cweight}\|\ln\hN\|_{H^{\bfm_{1}}_{\weight_{0}}(\bM)}\|\hU u\|_{\infty,w}\\
      & +C_{a}\textstyle{\sum}_{l=0}^{m-1}\textstyle{\sum}_{|\bfK|=l}\ldr{\tau}^{(m-1-l)\cweight}\ldr{\tau-\tau_{c}}^{m-1-l}\|E_{\bfK}\hU u\|_{2,w}
    \end{split}
  \end{equation}
  for all $\tau\leq\tau_{c}$, where $C_{a}$ only depends on $c_{\cweight,1}$, $m$ and $(\bM,\bge_{\refer})$. On the other hand, the conditions of 
  Lemma~\ref{lemma:mWAhUASobestimates} are fulfilled, so that (\ref{eq:AikmHlestsobassumpEi}) holds. Combining this observation with the 
  fact that the $(\cweight,l)$-Sobolev assumptions are satisfied and the fact that (\ref{eq:inductivestepchordderEbfIhU}) holds yields the 
  conclusion that if $2\leq |\bfI|\leq m$ and $m\leq l$, 
  \begin{equation*}
    \begin{split}
      \|E_{\bfI}\hU u\|_{2,w}
      \leq & C_{a}\hE_{m}^{1/2}+C_{b}\ldr{\tau}^{m\cweight}e^{\eSpe\tau}\|\bD^{1}_{\bbE}u\|_{\infty,w}
      +C_{b}\ldr{\tau}^{m\cweight}\|\hU u\|_{\infty,w}\\
      & +C_{a}\textstyle{\sum}_{l=0}^{m-1}\textstyle{\sum}_{|\bfK|=l}\ldr{\tau}^{(m-1-l)\cweight}\ldr{\tau-\tau_{c}}^{m-1-l}\|E_{\bfK}\hU u\|_{2,w}
    \end{split}
  \end{equation*}
  for all $\tau\leq\tau_{c}$, where $C_{a}$ only depends on $c_{\cweight,1}$, $m$ and $(\bM,\bge_{\refer})$; and $C_{b}$ only depends on $s_{\cweight,m}$, 
  $c_{\cweight,1}$ and $(\bM,\bge_{\refer})$. Combining this estimate with (\ref{eq:changingorderhUabsbfIeqone}) and an inductive argument yields the 
  conclusion of the lemma. 
\end{proof}

\section{Commutators}\label{section:commutatorshigherorderenergiesMoserest}

Next, we estimate $[L,E_{\bfI}]u$ in $L^{2}$, just as in the previous chapter. However, we here impose conditions on 
weighted $L^{2}$-based norms of the foliation quantities. This necessitates the derivation of different estimates. 

\begin{lemma}\label{lemma:EbfIcommhUsqLtwotwowest}
  Let $0\leq \cweight\in\ro$, $\weight_{0}=(0,\cweight)$ and $\weight=(\cweight,\cweight)$. Assume that the conditions of 
  Lemma~\ref{lemma:taurelvaryingbxEi} as well as the $(\cweight,1)$-supremum assumptions are satisfied. Fix $l$
  as in Definition~\ref{def:sobklassumptions} and assume that the $(\cweight,l)$-Sobolev assumptions are satisfied. Then, if $1\leq m\leq l$,
  $|\bfI|=m$ and $w$ is given by (\ref{eq:wdef}), 
  \begin{equation}\label{eq:EbfIcommhUsqLtwotwowest}
    \begin{split}
      \|[E_{\bfI},\hU^{2}]u\|_{2,w} \leq & C_{a}\ldr{\tau}^{\a_{m}\cweight}\ldr{\tau-\tau_{c}}^{\b_{m}}e^{\eSpe\tau}\hE_{m}^{1/2}
      +C_{a}\ldr{\tau}^{\a_{m}\cweight}\ldr{\tau-\tau_{c}}^{\b_{m}}\hE_{m-1}^{1/2}\\
      & +C_{a}\ldr{\tau}^{(m-1)\cweight}\ldr{\tau-\tau_{c}}^{m-1}\textstyle{\sum}_{|\bfK|\leq |\bfI|-1}\|E_{\bfK}\hU^{2}u\|_{2,w}\\
      & +C_{b}\ldr{\tau}^{\a_{m}\cweight}\ldr{\tau-\tau_{c}}^{\b_{m}}[\|\hU u\|_{\infty,w}+e^{\eSpe\tau}\|\bD^{1}_{\bbE}u\|_{\infty,w}]\\
      & +C_{c}\ldr{\tau}^{m\cweight}\|\hU^{2}u\|_{\infty,w}
    \end{split}
  \end{equation}
  for all $\tau\leq\tau_{c}$. Here $C_{a}$ only depends on $c_{\cweight,1}$, $m$ and $(\bM,\bge_{\refer})$; $C_{b}$ only depends on
  $s_{\cweight,m}$, $c_{\cweight,1}$ and $(\bM,\bge_{\refer})$; and $C_{c}$ only depends on $s_{\cweight,m}$ and $(\bM,\bge_{\refer})$. Moreover, $\a_{m}$
  and $\b_{m}$ only depend on $m$. 
\end{lemma}
\begin{proof}
  In order to estimate $[\hU^{2},E_{\bfI}]u$ in $L^{2}$, we appeal to Lemma~\ref{lemma:bdbfAhUsqcommformEioo}.

  \textbf{The case of two normal derivatives.} To begin with, we estimate the second sum on the right hand side of
  (\ref{eq:bdbfAhUsqcommformEioo}). Due to (\ref{eq:CIJtwoestoo}), it is sufficient to estimate expressions of the form 
  \[
  \left(\int_{\bM}|\bD^{m_{1}+1}\ln\hN|_{\bge_{\refer}}^{2}\cdots |\bD^{m_{k}+1}\ln\hN|_{\bge_{\refer}}^{2}|E_{\bfJ}\hU^{2}u|^{2}\mutgc\right)^{1/2}.
  \]
  Here $m_{1}+\dots+m_{k}+k+|\bfJ|\leq |\bfI|$. Moreover, due to (\ref{eq:bdbfAhUsqcommformEioo}) and (\ref{eq:CIJtwoestoo}), if equality holds, 
  then $k\geq 1$. Combining these observations with an argument similar to the derivation of (\ref{eq:ganiintermediatecommspatial}) yields 
  \begin{equation*}
    \begin{split}
      \|\bC_{\bfI,\bfJ}^{2}E_{\bfJ}\hU^{2}u\|_{2,w} \leq & 
      C_{a}\ldr{\tau}^{m\cweight}\|\ln\hN\|_{H^{\bfm}_{\weight_{0}}(\bM)}\|\hU^{2}u\|_{\infty,w}\\
      & +C_{b}\ldr{\tau}^{(m-1)\cweight}\ldr{\tau-\tau_{c}}^{m-1}\textstyle{\sum}_{|\bfK|\leq |\bfI|-1}\|E_{\bfK}\hU^{2}u\|_{2,w}
    \end{split}
  \end{equation*}
  for $|\bfI|\leq m$ and $|\bfJ|\leq |\bfI|-1$, where $\bfm:=(1,m)$; $C_{a}$ only depends on $C_{\rorel}$, $\cweight$, $m$, $n$ and $(\bM,\bge_{\refer})$;
  and $C_{b}$ only depends on $c_{\cweight,1}$, $m$ and $(\bM,\bge_{\refer})$. In particular, 
  \begin{equation*}
    \begin{split}
      \|\bC_{\bfI,\bfJ}^{2}E_{\bfJ}\hU^{2}u\|_{2,w} \leq & C_{c}\ldr{\tau}^{m\cweight}\|\hU^{2}u\|_{\infty,w}\\
      & +C_{b}\ldr{\tau}^{(m-1)\cweight}\ldr{\tau-\tau_{c}}^{m-1}\textstyle{\sum}_{|\bfK|\leq |\bfI|-1}\|E_{\bfK}\hU^{2}u\|_{2,w}
    \end{split}
  \end{equation*}
  where $C_{b}$ has the same dependence as before and $C_{c}$ only depends on $s_{\cweight,m}$ and $(\bM,\bge_{\refer})$. 

  \textbf{The case of one normal derivative.} Next, we estimate the terms arising from the first sum on the right hand side of
  (\ref{eq:bdbfAhUsqcommformEioo}). In particular, we are interested in the case that $k=1$. Due to (\ref{eq:CIJoneestoo}), there are two
  types of terms that we need to estimate, corresponding to the two sums on the right hand side of (\ref{eq:CIJoneestoo}).

  \textit{Terms of the first type.} In order to estimate a term of the first type, we can proceed as before, and we conclude, assuming 
  $|\bfI|\leq m$, that it can be bounded by 
  \begin{equation*}
    \begin{split}
      & C_{a}\ldr{\tau}^{(m+1)\cweight}\textstyle{\sum}_{i,k}\|A_{i}^{k}\|_{C^{0}_{\weight}(\bM)}\|\hU u\|_{\infty,w}
      \|\ln\hN\|_{H^{\bfm}_{\weight_{0}}(\bM)}\\
      & +C_{a}\ldr{\tau}^{(m+1)\cweight}\|\hU u\|_{\infty,w}\textstyle{\sum}_{i,k}\|A_{i}^{k}\|_{H^{m}_{\weight}(\bM)}\\
      & +C_{b}\ldr{\tau}^{(m+1)\cweight}\ldr{\tau-\tau_{c}}^{m}
      \textstyle{\sum}_{i,k}\|A_{i}^{k}\|_{C^{0}_{\weight}(\bM)}\textstyle{\sum}_{|\bfK|\leq m}\|E_{\bfK}\hU u\|_{2,w}
    \end{split}
  \end{equation*}
  for all $\tau\leq\tau_{c}$. Here $C_{a}$ only depends on $C_{\rorel}$, $\cweight$, $m$, $n$ and $(\bM,\bge_{\refer})$; and $C_{b}$ only depends on 
  $c_{\cweight,1}$, $m$ and $(\bM,\bge_{\refer})$. Combining this estimate with (\ref{eq:AikCzest}) and  Lemma~\ref{lemma:mWAhUASobestimates} and the
  assumptions yields the conclusion that the relevant terms can be estimated by 
  \begin{equation*}
    \begin{split}
      & C_{a}\ldr{\tau}^{(m+1)\cweight}\ldr{\tau-\tau_{c}}^{m}e^{\eSpe\tau}\textstyle{\sum}_{|\bfK|\leq m}\|E_{\bfK}\hU u\|_{2,w}
      +C_{b}\ldr{\tau}^{(m+1)\cweight}e^{\eSpe\tau}\|\hU u\|_{\infty,w}
    \end{split}
  \end{equation*}
  for all $\tau\leq\tau_{c}$. Here $C_{a}$ only depends on $c_{\cweight,1}$, $m$ and $(\bM,\bge_{\refer})$; and $C_{b}$ only depends on $s_{\cweight,m}$ and 
  $(\bM,\bge_{\refer})$. Combining this estimate with Lemma~\ref{lemma:EbfIhUuestbfIgeneralorder} yields the conclusion that
  terms of the first type can be estimated by 
  \begin{equation*}
    \begin{split}
      & \ldr{\tau}^{\a_{m}\cweight}\ldr{\tau-\tau_{c}}^{\b_{m}}e^{\eSpe\tau}[C_{a}\hE_{m}^{1/2}+C_{b}(\|\hU u\|_{\infty,w}+e^{\eSpe\tau}\|\bD^{1}_{\bbE}u\|_{\infty,w})]
    \end{split}
  \end{equation*}
  for all $\tau\leq\tau_{c}$. Here $C_{a}$ only depends on $c_{\cweight,1}$, $m$ and $(\bM,\bge_{\refer})$; and $C_{b}$ only depends on $s_{\cweight,m}$, 
  $c_{\cweight,1}$ and $(\bM,\bge_{\refer})$. Moreover, $\a_{m}$ and $\b_{m}$ only depend on $m$. 

  \textit{Terms of the second type.} In the second type of term appearing in (\ref{eq:CIJoneestoo}), the lower bound in the sum is $1$.
  This means that there must be a factor of the form $|\bD^{m_{1}+1}\ln\hN|_{\bge_{\refer}}$ or a factor of the form
  $|E_{\bfK}\hU(\ln\hN)|$ with $\bfK\neq 0$. In the first case, we rewrite the factor as $|\bD^{m_{1}}(\bD\ln\hN)|_{\bge_{\refer}}$ when appealing
  to Corollary~\ref{cor:mixedmoserestweight}. In the second case, we rewrite the relevant factor as 
  $|E_{\bfK}\hU(\ln\hN)|=|E_{\bfK_{a}}E_{\bfK_{b}}\hU(\ln\hN)|$, where $|\bfK_{b}|=1$. The effect of this reformulation is that the total number of 
  derivatives (denoted $l$ in the statement of Corollary~\ref{cor:mixedmoserestweight}) is bounded from above by $m-1$. Thus a term of the 
  second type can be estimated by, assuming $|\bfI|\leq m$,  
  \begin{equation*}
    \begin{split}
      & C_{a}\ldr{\tau}^{(m+1)\cweight}\|\hU\ln\hN\|_{C^{0}_{\weight}(\bM)}\|\hU u\|_{\infty,w}
      \|\ln\hN\|_{H^{\bfm}_{\weight_{0}}(\bM)}\\
      & +C_{a}\ldr{\tau}^{(m+1)\cweight}\|\hU u\|_{\infty,w}\|\hU\ln\hN\|_{H^{m}_{\weight}(\bM)}\\
      & +C_{b}\ldr{\tau}^{(m+1)\cweight}\ldr{\tau-\tau_{c}}^{m-1}\|\hU\ln\hN\|_{C^{1}_{\weight}(\bM)}\textstyle{\sum}_{|\bfK|\leq m-1}\|E_{\bfK}\hU u\|_{2,w}
    \end{split}
  \end{equation*}
  for all $\tau\leq\tau_{c}$. Here $C_{a}$ only depends on $C_{\rorel}$, $\cweight$, $m$, $n$ and $(\bM,\bge_{\refer})$; and $C_{b}$ only depends on 
  $c_{\cweight,1}$, $m$ and $(\bM,\bge_{\refer})$. Combining this estimate with the assumptions yields the conclusion that a term of the 
  second type can be estimated by, assuming $|\bfI|\leq m$,  
  \begin{equation*}
    \begin{split}
      & C_{a}\ldr{\tau}^{(m+1)\cweight}\ldr{\tau-\tau_{c}}^{m-1}\textstyle{\sum}_{|\bfK|\leq m-1}\|E_{\bfK}\hU u\|_{2,w}+C_{b}\ldr{\tau}^{(m+1)\cweight}\|\hU u\|_{\infty,w}
    \end{split}
  \end{equation*}
  for all $\tau\leq\tau_{c}$. Here $C_{a}$ only depends on $c_{\cweight,1}$, $m$ and $(\bM,\bge_{\refer})$; and $C_{b}$ only depends on $s_{\cweight,m}$ and 
  $(\bM,\bge_{\refer})$. Combining this estimate with Lemma~\ref{lemma:EbfIhUuestbfIgeneralorder} results in terms of the form 
  \begin{equation*}
    \begin{split}
      & C_{a}\ldr{\tau}^{\a_{m}\cweight}\ldr{\tau-\tau_{c}}^{\b_{m}}\hE_{m-1}^{1/2}
      +C_{b}\ldr{\tau}^{\a_{m}\cweight}\ldr{\tau-\tau_{c}}^{\b_{m}}[\|\hU u\|_{\infty,w}+e^{\eSpe\tau}\|\bD^{1}_{\bbE}u\|_{\infty,w}]
    \end{split}
  \end{equation*}
  for all $\tau\leq\tau_{c}$. Here $C_{a}$ only depends on $c_{\cweight,1}$, $m$ and $(\bM,\bge_{\refer})$; and $C_{b}$ only depends on $s_{\cweight,m}$, 
  $c_{\cweight,1}$ and $(\bM,\bge_{\refer})$. Moreover, $\a_{m}$ and $\b_{m}$ only depend on $m$. Summing up yields the conclusion that 
  \begin{equation*}
    \begin{split}
      \|\bC_{\bfI,\bfJ}^{1}E_{\bfJ}\hU u\|_{2,w} \leq & C_{a}\ldr{\tau}^{\a_{m}\cweight}\ldr{\tau-\tau_{c}}^{\b_{m}}e^{\eSpe\tau}\hE_{m}^{1/2}
      +C_{a}\ldr{\tau}^{\a_{m}\cweight}\ldr{\tau-\tau_{c}}^{\b_{m}}\hE_{m-1}^{1/2}\\
      & +C_{b}\ldr{\tau}^{\a_{m}\cweight}\ldr{\tau-\tau_{c}}^{\b_{m}}[\|\hU u\|_{\infty,w}+e^{\eSpe\tau}\|\bD^{1}_{\bbE}u\|_{\infty,w}]
    \end{split}
  \end{equation*}
  for all $\tau\leq\tau_{c}$. Here $C_{a}$ only depends on $c_{\cweight,1}$, $m$ and $(\bM,\bge_{\refer})$; and $C_{b}$ only depends on $s_{\cweight,m}$, 
  $c_{\cweight,1}$ and $(\bM,\bge_{\refer})$. Moreover, $\a_{m}$ and $\b_{m}$ only depend on $m$. 
  
  \textbf{The case of no normal derivatives.} Next, we are interested in the case that $k=0$ in the first sum on the right hand side of
  (\ref{eq:bdbfAhUsqcommformEioo}). We then have to estimate $\bC_{\bfI,\bfJ}^{0}E_{\bfJ}u$ in a weighted $L^{2}$-space. Before doing so, note that 
  $\bC_{\bfI,\bfJ}^{0}$ vanishes if $\bfJ=0$. In the estimates to follow, it is therefore natural to rewrite $E_{\bfJ}u=E_{\bfJ_{a}}E_{\bfJ_{b}}u$, where
  $|\bfJ_{b}|=1$. The corresponding arguments are similar to before, and the result is, assuming $|\bfI|\leq m$, 
  \begin{equation*}
    \begin{split}
      & \|\bC_{\bfI,\bfJ}^{0}E_{\bfJ}u\|_{2,w}\\
      \leq & C_{a}\ldr{\tau}^{(m+1)\cweight}
      \textstyle{\sum}_{i,k}\left[\|\hU(A_{i}^{k})\|_{C^{0}_{\weight_{1}}(\bM)}+\|A_{i}^{k}\|_{C^{0}_{\weight}(\bM)}\|\hU\ln\hN\|_{C^{0}_{\weight}(\bM)}\right.\\
      & \left. +\textstyle{\sum}_{p,q}\|A_{i}^{k}\|_{C^{0}_{\weight}(\bM)}\|A_{p}^{q}\|_{C^{0}_{\weight}(\bM)}\right]
      \|\bD^{1}_{\bbE}u\|_{\infty,w}\|\ln\hN\|_{H^{\bfm_{-}}_{\weight_{0}}(\bM)}\\
      & +C_{a}\ldr{\tau}^{(m+1)\cweight}\|\bD^{1}_{\bbE}u\|_{\infty,w}
      \textstyle{\sum}_{i,k}\left[\|\hU(A_{i}^{k})\|_{H^{m-1}_{\weight_{1}}(\bM)}+\|\hU\ln\hN\|_{C^{0}_{\weight}(\bM)}\|A_{i}^{k}\|_{H^{m-1}_{\weight}(\bM)}\right.\\
      & \left.+\|\hU\ln\hN\|_{H^{m-1}_{\weight}(\bM)}\|A_{i}^{k}\|_{C^{0}_{\weight}(\bM)}
      +\textstyle{\sum}_{p,q}\|A_{i}^{k}\|_{C^{0}_{\weight}(\bM)}\|A_{p}^{q}\|_{H^{m-1}_{\weight}(\bM)}\right]\\
      & +C_{b}\ldr{\tau}^{(m+1)\cweight}\ldr{\tau-\tau_{c}}^{m-1}
      \textstyle{\sum}_{i,k}\left[\|\hU(A_{i}^{k})\|_{C^{0}_{\weight_{1}}(\bM)}+\|A_{i}^{k}\|_{C^{0}_{\weight}(\bM)}\|\hU\ln\hN\|_{C^{0}_{\weight}(\bM)}\right.\\
      & \left. +\textstyle{\sum}_{p,q}\|A_{i}^{k}\|_{C^{0}_{\weight}(\bM)}\|A_{p}^{q}\|_{C^{0}_{\weight}(\bM)}\right]
      \textstyle{\sum}_{|\bfK|\leq m}\|E_{\bfK}u\|_{2,w}
    \end{split}
  \end{equation*}
  for all $\tau\leq\tau_{c}$, where $\bfm_{-}=(1,m-1)$; in case $m=1$, all the terms on the right hand side but the last one should be set to zero. Here 
  $C_{a}$ only depends on $C_{\rorel}$, $\cweight$, $m$, $n$ and $(\bM,\bge_{\refer})$; and $C_{b}$ only depends on
  $c_{\cweight,1}$, $m$ and $(\bM,\bge_{\refer})$. Combining this estimate with Lemmas~\ref{lemma:Aiksupest}, \ref{lemma:hUAiksupest} and
  \ref{lemma:mWAhUASobestimates} as well as the assumptions yields
  \begin{equation*}
    \begin{split}
      \|\bC_{\bfI,\bfJ}^{0}E_{\bfJ}u\|_{2,w} \leq & \ldr{\tau}^{(m+1)\cweight}e^{\eSpe\tau}\left[C_{a}\|\bD^{1}_{\bbE}u\|_{\infty,w}
      +C_{b}\ldr{\tau-\tau_{c}}^{m-1}\textstyle{\sum}_{|\bfK|\leq m}\|E_{\bfK}u\|_{2,w}\right],
    \end{split}
  \end{equation*}  
  where $C_{a}$ only depends on $s_{\cweight,m}$ and $(\bM,\bge_{\refer})$, and $C_{b}$ only depends on $c_{\cweight,1}$, $m$ and $(\bM,\bge_{\refer})$. 
\end{proof}

\subsection{Commutator with $e^{-2\mu_{A}}X_{A}^{2}$}

Next, we wish to estimate the commutator with $e^{-2\mu_{A}}X_{A}^{2}$.

\begin{lemma}
  Let $0\leq \cweight\in\ro$, $\weight_{0}=(0,\cweight)$ and $\weight=(\cweight,\cweight)$. Assume that the conditions of 
  Lemma~\ref{lemma:taurelvaryingbxEi} as well as the $(\cweight,1)$-supremum assumptions are satisfied. Fix $l$
  as in Definition~\ref{def:sobklassumptions} and assume that the $(\cweight,l)$-Sobolev assumptions are satisfied. Then, if $1\leq m\leq l$ and 
  $|\bfI|=m$, 
  \begin{equation}\label{eq:emtmuAXAsqcommtbesttotal}
    \begin{split}
      \|[E_{\bfI},e^{-2\mu_{A}}X_{A}^{2}]u\|_{2,w} \leq & \ldr{\tau}^{\a_{m}\cweight+\b_{m}}e^{\eSpe\tau}\left(C_{a}\hE_{m}^{1/2}
      +C_{b}\textstyle{\sum}_{i}\|e^{-\mu_{A}}X_{A}E_{i}u\|_{\infty,w}\right)\\
      & +C_{b}\ldr{\tau}^{\a_{m}\cweight+\b_{m}}e^{2\eSpe\tau}\|\bD^{1}_{\bbE}u\|_{\infty,w}
    \end{split}
  \end{equation}  
  for all $\tau\leq\tau_{c}$, where $C_{a}$ only depends on $c_{\cweight,1}$, $m$, $(\bM,\bge_{\refer})$ and a lower bound on $\theta_{0,-}$; and $C_{b}$ only
  depends on $c_{\cweight,1}$, $s_{\cweight,m}$, $(\bM,\bge_{\refer})$ and a lower bound on $\theta_{0,-}$. Here $\a_{m}$ and $\b_{m}$ only depend on $m$. 
\end{lemma}
\begin{proof}
  Due to Lemma~\ref{lemma:EbfIemmuAXAsqcommoo}, we wish to estimate the right hand side of (\ref{eq:EbfIemmuAXAsqcommoo}) in $L^{2}$ with respect
  to the measure $\mutgc$. We consider the two terms on the right hand side separately. 

  \textbf{The first term on the right hand side of (\ref{eq:EbfIemmuAXAsqcommoo}).} In case $|\bfJ|=|\bfI|$,
  \[
  |\bD^{A}_{\bfI,\bfJ}|\leq C\ldr{\tau}^{2\cweight+1}
  \]
  for all $\tau\leq 0$, where $C$ only depends on $c_{\cweight,1}$, $|\bfI|$ and $(\bM,\bge_{\refer})$. In order to obtain this estimate, we appealed to
  Remark~\ref{remark:CkestofmuAEi}. Combining this observation with (\ref{eq:EbfIemmuAXAestimateaddass}) yields the conclusion that if $1\leq m\leq l$
  and $|\bfI|=|\bfJ|=m$, 
  \begin{equation}\label{eq:bDAbfIbfJetcttbestpI}
    \begin{split}
      \|\bD^{A}_{\bfI,\bfJ}e^{-\mu_{A}}E_{\bfJ}(e^{-\mu_{A}}X_{A}u)\|_{2,w} \leq & C_{a}\ldr{\tau}^{\a_{m}\cweight+\b_{m}}e^{\eSpe\tau}\hE_{m}^{1/2}\\
      & +C_{b}\ldr{\tau}^{\a_{m}\cweight+\b_{m}}e^{2\eSpe\tau}\|\bD^{1}_{\bbE}u\|_{\infty,w}
    \end{split}
  \end{equation}  
  for all $\tau\leq\tau_{c}$, where $C_{a}$ only depends on $c_{\cweight,1}$, $m$, $(\bM,\bge_{\refer})$ and a lower bound on $\theta_{0,-}$; and $C_{b}$ only
  depends on $c_{\cweight,1}$, $s_{\cweight,m}$, $(\bM,\bge_{\refer})$ and a lower bound on $\theta_{0,-}$. Next, consider the case that
  $1\leq |\bfJ|\leq |\bfI|-1$. Then, in order to estimate the first term on the right hand of (\ref{eq:EbfIemmuAXAsqcommoo}), it is sufficient to
  estimate expressions of the form 
  \begin{equation}\label{eq:bDAbfIbfJetcttbest}
    e^{-\mu_{A}}\textstyle{\prod}_{i=1}^{p}|\bD^{m_{i}+1}\mK|_{\bge_{\refer}}\textstyle{\prod}_{j=1}^{r}|\bD^{k_{j}+1}\mu_{A_{j}}|_{\bge_{\refer}}
    |E_{\bfJ_{a}}E_{\bfJ_{b}}(e^{-\mu_{A}}X_{A}u)|
  \end{equation}
  in $L^{2}$ with weight $w$. Here $|\bfJ_{b}|=1$,
  \[
  l_{\rotot}:=m_{1}+\dots+m_{p}+k_{1}+\dots+k_{r}+|\bfJ_{a}|\leq |\bfI|-p-r
  \]
  and if the far left hand side equals the far right hand side, then $p+r\geq 1$. At this stage, the factor $e^{-\mu_{A}}$ can be estimated by appealing
  to (\ref{eq:muminmainlowerbound}) and the remainder can be estimated by appealing to Corollary~\ref{cor:mixedmoserestweight}. To conclude,
  (\ref{eq:bDAbfIbfJetcttbest}) can, in $L^{2}$ with weight $w$, be estimated by
  \begin{equation}\label{eq:bDAbfIbfJetcttbestgenterm}
    \begin{split}
      & C_{a}\ldr{\tau}^{\a_{m}\cweight+\b_{m}}e^{\eSpe\tau}\|\bD^{1}_{\bbE}(e^{-\mu_{A}}X_{A}u)\|_{\infty,w}\\
      & +C_{b}\ldr{\tau}^{\a_{m}\cweight+\b_{m}}e^{\eSpe\tau}\textstyle{\sum}_{1\leq |\bfK|\leq |\bfI|}\|E_{\bfK}(e^{-\mu_{A}}X_{A}u)\|_{2,w}
    \end{split}
  \end{equation}
  for all $\tau\leq\tau_{c}$ and $|\bfI|\leq m$, where $C_{a}$ only depends on $c_{\cweight,1}$, $s_{\cweight,m}$, $(\bM,\bge_{\refer})$ and a lower bound on
  $\theta_{0,-}$; and $C_{b}$ only depends on $c_{\cweight,1}$, $m$, $(\bM,\bge_{\refer})$ and a lower bound on $\theta_{0,-}$. In order to obtain this
  conclusion, we appealed to Remark~\ref{remark:muAmhlestimate}, Remark~\ref{remark:CkestofmuAEi} and the assumptions. In order to express
  the terms appearing in (\ref{eq:bDAbfIbfJetcttbestgenterm}) in a form more useful for future estimates, note that
  \[
  |E_{i}(e^{-\mu_{A}}X_{A}u)|\leq |e^{-\mu_{A}}X_{A}E_{i}u|+|[E_{i},e^{-\mu_{A}}X_{A}]u|.
  \]
  In order to estimate the second term on the right hand side, we can appeal to Lemma~\ref{lemma:bdbfAGAXGcommformEi}. This yields
  \[
  |[E_{i},e^{-\mu_{A}}X_{A}]u|\leq \textstyle{\sum}_{k}|H_{i,k}||E_{k}u|,
  \]
  where 
  \[
  |H_{i,k}|\leq C_{a}\textstyle{\sum}_{k_{a}+|\bfK|\leq 1}\mfP_{\mK,k_{a}}|E_{\bfK}(e^{-\mu_{A}})|
  \]
  and $C_{a}$ only depends on $\mKsup$, $\e_{\rond}$, $n$ and $(\bM,\bge_{\refer})$. Summing up the above yields the conclusion that
  \[
  \|\bD^{1}_{\bbE}(e^{-\mu_{A}}X_{A}u)\|_{\infty,w}\leq C_{a}\textstyle{\sum}_{i}\|e^{-\mu_{A}}X_{A}E_{i}u\|_{\infty,w}
  +C_{b}\ldr{\tau}^{2\cweight+1}e^{\eSpe\tau}\|\bD^{1}_{\bbE}u\|_{\infty,w},
  \]
  where $C_{a}$ only depends on $n$ and $C_{b}$ only depends on $c_{\cweight,1}$, $(\bM,\bge_{\refer})$ and a lower bound on $\theta_{0,-}$. In order to
  estimate the second term appearing in (\ref{eq:bDAbfIbfJetcttbestgenterm}), it is sufficient to appeal to (\ref{eq:EbfIemmuAXAestimateaddass}).
  Summing up the above yields the conclusion that if $1\leq |\bfJ|\leq |\bfI|-1$, then 
  \begin{equation}\label{eq:bDAbfIbfJetcttbestpII}
    \begin{split}
      \|\bD^{A}_{\bfI,\bfJ}e^{-\mu_{A}}E_{\bfJ}(e^{-\mu_{A}}X_{A}u)\|_{2,w} \leq & C_{a}\ldr{\tau}^{\a_{m}\cweight+\b_{m}}e^{\eSpe\tau}\hE_{m}^{1/2}\\
      & +C_{b}\ldr{\tau}^{\a_{m}\cweight+\b_{m}}e^{\eSpe\tau}\textstyle{\sum}_{i}\|e^{-\mu_{A}}X_{A}E_{i}u\|_{\infty,w}\\
      & +C_{b}\ldr{\tau}^{\a_{m}\cweight+\b_{m}}e^{2\eSpe\tau}\|\bD^{1}_{\bbE}u\|_{\infty,w}
    \end{split}
  \end{equation}  
  for all $\tau\leq\tau_{c}$, where $C_{a}$ only depends on $c_{\cweight,1}$, $m$, $(\bM,\bge_{\refer})$ and a lower bound on $\theta_{0,-}$; and $C_{b}$ only
  depends on $c_{\cweight,1}$, $s_{\cweight,m}$, $(\bM,\bge_{\refer})$ and a lower bound on $\theta_{0,-}$. Noting that (\ref{eq:bDAbfIbfJetcttbestpI})
  holds in case $|\bfJ|=|\bfI|$, it is clear that (\ref{eq:bDAbfIbfJetcttbestpII}) holds if $1\leq |\bfJ|\leq |\bfI|$. 

  \textbf{The second term on the right hand side of (\ref{eq:EbfIemmuAXAsqcommoo}).} In case $|\bfI|=|\bfJ|$,
  \begin{equation}\label{eq:bFbfIbfJAsimpest}
    \|\bF_{\bfI,\bfJ}^{A}e^{-2\mu_{A}}E_{\bfJ}u\|_{2,w}\leq C_{a}\ldr{\tau}^{4\cweight+2+3\iota_{b}/2}e^{2\eSpe\tau}\hE_{m}^{1/2}
  \end{equation}
  for all $\tau\leq\tau_{c}$, where $C_{a}$ only depends on $c_{\cweight,1}$, $m$, $(\bM,\bge_{\refer})$ and a lower bound on $\theta_{0,-}$. In order to
  obtain this conclusion, we appealed to (\ref{eq:muminmainlowerbound}), Remark~\ref{remark:CkestofmuAEi} and the assumptions. Consider
  (\ref{eq:FIJAestoo}). For terms on the right hand side
  of (\ref{eq:FIJAestoo}) such that $m_{1}+m_{2}\leq 2$, we can proceed as above, and the relevant terms can be bounded by the right hand side of 
  (\ref{eq:bFbfIbfJAsimpest}). Let us therefore assume that $m_{1}+m_{2}>2$ in (\ref{eq:FIJAestoo}). We then need to estimate 
  \begin{equation}\label{eq:bFAbfIbfJetcttbest}
    e^{-2\mu_{A}}\textstyle{\prod}_{i=1}^{p}|\bD^{k_{i}+1}\mK|_{\bge_{\refer}}\textstyle{\prod}_{j=1}^{r}|\bD^{q_{j}+1}\mu_{A_{j}}|_{\bge_{\refer}}
    |E_{\bfJ_{a}}E_{\bfJ_{b}}u|
  \end{equation}
  in $L^{2}$ with weight $w$. Here $|\bfJ_{b}|=1$ and 
  \begin{align}
    k_{1}+\dots+k_{p}+q_{1}+\dots+q_{r}+|\bfJ_{a}| \leq & |\bfI|+1-p-r,\label{eq:kisumqisumbfJaest}\\
    q_{i}+|\bfJ_{a}| \leq & |\bfI|-1\label{eq:qisumbfJaest}
  \end{align}
  for $i=1,\dots,r$. 
  This means that if equality holds in the first inequality and if $p+r=1$, then $p=1$. This means that there are three cases 
  to consider. The first possibility is that equality does not hold in (\ref{eq:kisumqisumbfJaest}). Since we, by the above, can assume that 
  $p+r\geq 1$, this means that 
  \begin{equation}\label{eq:ltotdefbF}
    l_{\rotot}:=k_{1}+\dots+k_{p}+q_{1}+\dots+q_{r}+|\bfJ_{a}|\leq |\bfI|-1.
  \end{equation}
  The second possibility is that equality holds in (\ref{eq:kisumqisumbfJaest}), but that $p+r\geq 2$. In that case, (\ref{eq:ltotdefbF}) still 
  holds. The third possibility is that equality holds in (\ref{eq:kisumqisumbfJaest}) and $p+r=1$. Then $p=1$ and $k_{1}\geq 2$, and we need to 
  estimate
  \begin{equation}\label{eq:bFAbfIbfJetcttbestto}
    e^{-2\mu_{A}}|\bD^{k_{1}-1}\bD^{2}\mK|_{\bge_{\refer}}|E_{\bfJ_{a}}E_{\bfJ_{b}}u|
  \end{equation}
  in $L^{2}$ with weight $w$. In this case, we define $l_{\rotot}$ to equal $k_{1}-1+|\bfJ_{a}|\leq |\bfI|-1$. In the first two cases, the factor 
  $e^{-2\mu_{A}}$ can be estimated by appealing to (\ref{eq:muminmainlowerbound}) and the remainder can be estimated by appealing to 
  Corollary~\ref{cor:mixedmoserestweight}. Moreover, the $l$ appearing in the statement of Corollary~\ref{cor:mixedmoserestweight} should be 
  replaced by $l_{\rotot}$ given by (\ref{eq:ltotdefbF}). Assuming $1\leq m\leq l$ and $|\bfI|=m$, the resulting expressions can be estimated by
  \begin{equation*}
    \begin{split}
      C_{a}\ldr{\tau}^{\a_{m}\cweight+\b_{m}}e^{2\eSpe\tau}\|\bD^{1}_{\bbE}u\|_{\infty,w}+C_{b}\ldr{\tau}^{\a_{m}\cweight+\b_{m}}e^{2\eSpe\tau}\hE_{m}^{1/2}
    \end{split}
  \end{equation*}
  for all $\tau\leq \tau_{c}$, where $C_{a}$ only depends on $c_{\cweight,1}$, $s_{\cweight,m}$, $(\bM,\bge_{\refer})$ and a lower bound on $\theta_{0,-}$; and
  $C_{b}$ only depends on $c_{\cweight,1}$, $m$, $(\bM,\bge_{\refer})$ and a lower bound on $\theta_{0,-}$. In the third case, 
  $l_{\rotot}:=k_{1}-1+|\bfJ_{a}|$. Moreover, if $1\leq m\leq l$ and $|\bfI|=m$, then $l_{\rotot}\leq m-1$. Appealing to 
  (\ref{eq:muminmainlowerbound}) and Corollary~\ref{cor:mixedmoserestweight} we conclude that (\ref{eq:bFAbfIbfJetcttbestto}) can be 
  estimated in $L^{2}$ with weight $w$ by
  \[
  C_{a}\ldr{\tau}^{\a_{m}\cweight+\b_{m}}e^{2\eSpe\tau}\|\bD^{1}_{\bbE}u\|_{\infty,w}+C_{b}\ldr{\tau}^{\a_{m}\cweight+\b_{m}}e^{2\eSpe\tau}\hE_{m}^{1/2}
  \]
  for all $\tau\leq\tau_{c}$, where $C_{a}$ only depends on $c_{\cweight,1}$, $s_{\cweight,m}$, $(\bM,\bge_{\refer})$ and a lower bound on $\theta_{0,-}$; and
  $C_{b}$ only depends on $c_{\cweight,1}$, $m$, $(\bM,\bge_{\refer})$ and a lower bound on $\theta_{0,-}$. Summing up the above yields the conclusion
  of the lemma. 
\end{proof}

\subsection{Commutator with $Z^{0}\hU$}

Next, we wish to estimate the commutator with $Z^{0}\hU$. To this end, we appeal to Lemma~\ref{lemma:bdbfAchthhUcommformEireverse}. Note, in 
the application of this lemma, that $Z^{0}$ is given by (\ref{eq:ZzdefEi}), where 
\[
\hmcY^{0}=n^{-1}[q-(n-1)];
\]
cf. (\ref{eq:chthexpressionhUlntheta}) and (\ref{eq:hmcYzdefEi}). In what follows, we, in analogy with (\ref{eq:coefflassumptions}),
impose the condition that (\ref{eq:Sobcoefflassumptions}) holds, where $l$, $\weight_{0}$ and $\weight$ have the properties stated in
Definition~\ref{def:sobklassumptions}.

\begin{lemma}
  Let $0\leq \cweight\in\ro$, $\weight_{0}=(0,\cweight)$ and $\weight=(\cweight,\cweight)$. Assume that the conditions of 
  Lemma~\ref{lemma:taurelvaryingbxEi} as well as the $(\cweight,1)$-supremum assumptions are satisfied. Fix $l$
  as in Definition~\ref{def:sobklassumptions} and assume that the $(\cweight,l)$-Sobolev assumptions are satisfied. Assume, finally, that 
  there are constants $c_{\coeff,1}$ and $s_{\coeff,l}$ such that (\ref{eq:coefflassumptions}) is satisfied with $l$ replaced by $1$ and 
  (\ref{eq:Sobcoefflassumptions}) is satisfied. Then, if $1\leq m\leq l$ and 
  $|\bfI|=m$, 
  \begin{equation}\label{eq:ZzhUcommEbfISobest}
    \begin{split}
      \|[E_{\bfI},Z^{0}\hU]u\|_{2,w} \leq & C_{a}\ldr{\tau}^{\a_{m}\cweight}\ldr{\tau-\tau_{c}}^{\b_{m}}e^{\eSpe\tau}\hE_{m}^{1/2}
      +C_{a}\ldr{\tau}^{\a_{m}\cweight}\ldr{\tau-\tau_{c}}^{\b_{m}}\hE_{m-1}^{1/2}\\
      & +C_{b}\ldr{\tau}^{\a_{m}\cweight}\ldr{\tau-\tau_{c}}^{\b_{m}}\left[\|\hU u\|_{\infty,w}+e^{\eSpe\tau}\textstyle{\sum}_{|\bfI|\leq 1}\|E_{\bfI}u\|_{\infty,w}\right]
    \end{split}
  \end{equation}  
  for all $\tau\leq\tau_{c}$, where $C_{a}$ only depends on $c_{\cweight,1}$, $c_{\coeff,1}$, $m$, $m_{\ros}$ and $(\bM,\bge_{\refer})$; and $C_{b}$ only depends on 
  $s_{\cweight,m}$, $s_{\coeff,m}$, $c_{\cweight,1}$, $c_{\coeff,1}$, $m_{\ros}$ and $(\bM,\bge_{\refer})$. Here $\a_{m}$ and $\b_{m}$ only depend on $m$. 
\end{lemma}
\begin{proof}
  Due to Lemma~\ref{lemma:bdbfAchthhUcommformEireverse}, we need to estimate the terms on the right hand side of (\ref{eq:bDbfAchthhUcommEireverse}),
  applied to $u$, in $L^{2}$ with weight $w$. In order to estimate the first sum on the right hand side, it is sufficient to estimate expressions of 
  the form 
  \[
  \textstyle{\prod}_{i=1}^{p}|\bD^{k_{i}+1}\ln\hN|_{\bge_{\refer}}\|E_{\bfK}Z^{0}\|\cdot |E_{\bfJ}\hU u|,
  \]
  where $l_{\rotot}:=k_{1}+\dots+k_{p}+|\bfK|+|\bfJ|\leq |\bfI|-p$ and $|\bfJ|\leq |\bfI|-1$. If $p\geq 1$, we can appeal to 
  Corollary~\ref{cor:mixedmoserestweight} with $l$ replaced by $l_{\rotot}$. This leads to the conclusion that if $1\leq m\leq l$ and $|\bfI|=m$, then
  the relevant expressions can be estimated by 
  \begin{equation*}
    \begin{split}
      & C_{a}\ldr{\tau}^{m\cweight}\|\hU u\|_{\infty,w}+C_{b}\ldr{\tau}^{(m-1)\cweight}\|\hU u\|_{\infty,w}\\
      & +C_{c}\ldr{\tau}^{(m-1)\cweight}\ldr{\tau-\tau_{c}}^{m-1}\textstyle{\sum}_{|\bfL|\leq m-1}\|E_{\bfL}\hU u\|_{2,w}
    \end{split}
  \end{equation*}  
  for all $\tau\leq\tau_{c}$, where $C_{a}$ only depends on $s_{\cweight,m}$, $c_{\coeff,0}$, $m_{\ros}$ and $(\bM,\bge_{\refer})$; $C_{b}$ only depends on
  $s_{\coeff,m-1}$, $s_{\cweight,m-1}$, $m_{\ros}$ and $(\bM,\bge_{\refer})$; and $C_{c}$ only depends on $c_{\cweight,1}$, $c_{\coeff,0}$, $m$, $m_{\ros}$ and
  $(\bM,\bge_{\refer})$. In case $p=0$ and 
  $|\bfK|+|\bfJ|\leq |\bfI|-1$, we obtain the same estimate. What remains to be considered is the case that $p=0$ and $|\bfK|+|\bfJ|=|\bfI|$. Since
  $|\bfJ|\leq |\bfI|-1$, this means that $|\bfK|\geq 1$. We thus need to estimate
  \[
  \|E_{\bfK_{a}}E_{\bfK_{b}}Z^{0}\|\cdot |E_{\bfJ}\hU u|
  \]
  in $L^{2}$ with weight $w$, where $|\bfK_{b}|=1$. In this case, we let $l_{\rotot}:=|\bfK_{a}|+|\bfJ|\leq |\bfI|-1$. If $1\leq m\leq l$ and $|\bfI|=m$,
  we obtain the following bound by appealing to Corollary~\ref{cor:mixedmoserestweight}:
  \[
  C_{b}\ldr{\tau}^{m\cweight}\|\hU u\|_{\infty,w}+C_{c}\ldr{\tau}^{m\cweight}\ldr{\tau-\tau_{c}}^{m-1}\textstyle{\sum}_{|\bfL|\leq m-1}\|E_{\bfL}\hU u\|_{2,w}
  \]
  where $C_{b}$ only depends on $s_{\coeff,m}$, $s_{\cweight,m}$, $m_{\ros}$ and $(\bM,\bge_{\refer})$; and $C_{c}$ only depends on $c_{\cweight,1}$, $c_{\coeff,1}$,
  $m$, $m_{\ros}$ 
  and $(\bM,\bge_{\refer})$. Combining the above estimates with Lemma~\ref{lemma:EbfIhUuestbfIgeneralorder} yields the conclusion that if $1\leq m\leq l$,
  $|\bfI|=m$ and $|\bfJ|\leq |\bfI|-1$, then
  \begin{equation*}
    \begin{split}
      \|\bGe_{\bfI,\bfJ}^{0}E_{\bfJ}\hU u\|_{2,w} \leq & C_{a}\ldr{\tau}^{\a_{m}\cweight}\ldr{\tau-\tau_{c}}^{\b_{m}}\hE_{m-1}^{1/2}\\
      &+C_{b}\ldr{\tau}^{\a_{m}\cweight}\ldr{\tau-\tau_{c}}^{\b_{m}}[\|\hU u\|_{\infty,w}+e^{\eSpe\tau}\|\bD^{1}_{\bbE}u\|_{\infty,w}]
    \end{split}
  \end{equation*}  
  for all $\tau\leq\tau_{c}$, where $C_{a}$ only depends on $c_{\cweight,1}$, $c_{\coeff,1}$, $m$, $m_{\ros}$ and $(\bM,\bge_{\refer})$; and $C_{b}$ only depends on 
  $s_{\cweight,m}$, $s_{\coeff,m}$, $c_{\cweight,1}$, $c_{\coeff,1}$, $m_{\ros}$ and $(\bM,\bge_{\refer})$. Moreover, $\a_{m}$ and $\b_{m}$ are constants depending only 
  on $m$. 

  Next, we need to estimate the expressions that arise from the second term on the right hand side of (\ref{eq:bDbfAchthhUcommEireverse}). In this
  case, it is possible to directly apply Corollary~\ref{cor:mixedmoserestweight} in order to conclude that 
  \begin{equation*}
    \begin{split}
      \|\bGe_{\bfI,\bfJ}^{1}E_{\bfJ}u\|_{2,w} \leq & C_{a}\ldr{\tau}^{\a_{m}\cweight}\ldr{\tau-\tau_{c}}^{\b_{m}}e^{\eSpe\tau}\hE_{m}^{1/2}
      +C_{b}\ldr{\tau}^{\a_{m}\cweight}\ldr{\tau-\tau_{c}}^{\b_{m}}e^{\eSpe\tau}\|u\|_{\infty,w}
    \end{split}
  \end{equation*}
  for all $\tau\leq\tau_{c}$, where $C_{a}$ only depends on $c_{\cweight,0}$, $c_{\coeff,0}$, $m$, $m_{\ros}$ and $(\bM,\bge_{\refer})$; and $C_{b}$ only depends on 
  $s_{\coeff,m}$, $s_{\cweight,m}$, $c_{\coeff,0}$, $m_{\ros}$ and $(\bM,\bge_{\refer})$. 
\end{proof}

\subsection{Commutator with $Z^{A}X_{A}$}

Next, we wish to estimate the commutator with $Z^{A}X_{A}$. To this end, we appeal to Lemma~\ref{lemma:bdbfAGAXGcommformEi}. Note, in 
the application of this lemma, that $Z^{A}$ is given by (\ref{eq:ZAdefEi}), where $\hmcY^{A}$ is given by (\ref{eq:hmcYAdefEi}). Before 
estimating the commutator, it is convenient to derive Sobolev estimates for $Z^{A}$.

\begin{lemma}
  Let $0\leq \cweight\in\ro$, $\weight_{0}=(0,\cweight)$ and $\weight=(\cweight,\cweight)$. Assume that the conditions of 
  Lemma~\ref{lemma:taurelvaryingbxEi} as well as the $(\cweight,1)$-supremum assumptions are satisfied. Fix $l$
  as in Definition~\ref{def:sobklassumptions} and assume that the $(\cweight,l)$-Sobolev assumptions are satisfied. Then
  \begin{equation}\label{eq:hmcYAHlestimate}
    \|\hmcY^{A}\|_{H^{l}(\bM)}\leq C_{a}\ldr{\tau}^{(l+1)(2\cweight+1)}e^{2\eSpe\tau}
  \end{equation}
  for all $\tau\leq 0$, where $C_{a}$ only depends on $s_{\cweight,l}$, $c_{\cweight,1}$, $(\bM,\bge_{\refer})$ and a lower bound on $\theta_{0,-}$.
  Assume, in addition, that (\ref{eq:coefflassumptions}) holds with $l$ replaced by $0$ and that (\ref{eq:Sobcoefflassumptions}) holds.
  Then 
  \begin{equation}\label{eq:hmcXAijHlestimate}
    \|\hmcX^{A}_{ij}\|_{H^{l}(\bM)}\leq C_{a}\ldr{\tau}^{l\cweight}e^{\eSpe\tau}
  \end{equation}
  for all $\tau\leq 0$, where $C_{a}$ only depends on $s_{\cweight,l}$, $s_{\coeff,l}$, $c_{\coeff,0}$, $(\bM,\bge_{\refer})$ and a lower bound on
  $\theta_{0,-}$. In particular, 
  \begin{equation}\label{eq:ZAHlestimateprel}
    \|Z^{A}\|_{H^{l}(\bM)}\leq C_{a}\ldr{\tau}^{l\cweight}e^{\eSpe\tau}
  \end{equation}
  for all $\tau\leq 0$, where $C_{a}$ only depends on $s_{\cweight,l}$, $s_{\coeff,l}$, $c_{\cweight,1}$, $c_{\coeff,0}$, $m_{\ros}$, $(\bM,\bge_{\refer})$
  and a lower bound on $\theta_{0,-}$.
\end{lemma}
\begin{remark}
  If one, in addition to the assumptions of the lemma, requires the existence of a constant $c_{\coeff,1}$ such that (\ref{eq:coefflassumptions}) holds with
  $l$ replaced by $1$, then
  \[
  \|Z^{A}\|_{C^{1}_{\weight_{0}}(\bM)}\leq C_{a}e^{\eSpe\tau}
  \]
  for all $\tau\leq 0$, where $C_{a}$ only depends on $c_{\cweight,1}$, $c_{\rocoeff,1}$, $m_{\ros}$, $(\bM,\bge_{\refer})$ and a lower bound on $\theta_{0,-}$.
  This follows from Lemma~\ref{lemma:HABestEi}, Remark~\ref{remark:emuAhmcYAest} and (\ref{eq:eSpevarrhoeelowtaurelEi}). 
\end{remark}
\begin{proof}
  We begin by estimating $\hmcY^{A}$. Note, to this end, that (\ref{eq:EbfKhmcYAest}) holds, where we use the notation introduced in 
  Definition~\ref{def:mfPmKmuhN}. To begin with, we wish to estimate the first term on the right hand side of (\ref{eq:EbfKhmcYAest}).
  To this end, it is sufficient to estimate
  \begin{equation}\label{eq:hmcYAfp}
    e^{-2\mu_{A}}\textstyle{\prod}_{i=1}^{p}|\bD^{k_{i}+1}\mK|_{\bge_{\refer}}\textstyle{\prod}_{j=1}^{q}|\bD^{l_{j}+1}\mu_{A_{j}}|_{\bge_{\refer}}
    |\bD^{m_{b}+1}\ln\theta|_{\bge_{\refer}},
  \end{equation}
  where $l_{\rotot}:=k_{1}+\dots+k_{p}+l_{1}+\dots+l_{q}+m_{b}\leq k-p-q$ and $k:=|\bfK|$. In case $p+q\geq 1$, we appeal to (\ref{eq:muminmainlowerbound}),
  (\ref{eq:eSpevarrhoeelowtaurelEi}), Remark~\ref{remark:CkestofmuAEi}, Corollary~\ref{cor:mixedmoserestweight} and the assumptions in order to conclude
  that (\ref{eq:hmcYAfp}) can, in $L^{2}$, be estimated by
  \[
  C_{a}\ldr{\tau}^{k(2\cweight+1)+\cweight}e^{2\eSpe\tau}
  \]
  for all $\tau\leq 0$, where $C_{a}$ only depends on $s_{\cweight,k}$, $c_{\cweight,1}$, $(\bM,\bge_{\refer})$ and a lower bound on $\theta_{0,-}$; here
  $k:=|\bfK|$. In case $p+q=0$, we need only appeal to (\ref{eq:muminmainlowerbound}), (\ref{eq:eSpevarrhoeelowtaurelEi}) and the assumptions in
  order to obtain a better bound. Turning to the second term on the right hand side of (\ref{eq:EbfKhmcYAest}), we need to estimate
  \begin{equation}\label{eq:hmcYAsp}
    e^{-2\mu_{A}}\textstyle{\prod}_{h=1}^{p}|\bD^{k_{h}+1}\mK|_{\bge_{\refer}}\textstyle{\prod}_{i=1}^{q}|\bD^{l_{i}+1}\mu_{A_{i}}|_{\bge_{\refer}}
    \textstyle{\prod}_{j=1}^{r}|\bD^{m_{j}+1}\ln\hN|_{\bge_{\refer}}
  \end{equation}
  where $l_{\rotot}:=k_{1}+\dots+k_{p}+l_{1}+\dots+l_{q}+m_{1}+\dots+m_{r}\leq k+1-p-q-r$. Appealing to (\ref{eq:muminmainlowerbound}),
  (\ref{eq:eSpevarrhoeelowtaurelEi}), Remark~\ref{remark:CkestofmuAEi}, Corollary~\ref{cor:mixedmoserestweight} and the assumptions, we conclude
  that (\ref{eq:hmcYAsp}) can, in $L^{2}$, be estimated by
  \[
  C_{a}\ldr{\tau}^{(k+1)(2\cweight+1)}e^{2\eSpe\tau}
  \]
  for all $\tau\leq 0$, where $C_{a}$ only depends on $s_{\cweight,k}$, $c_{\cweight,1}$, $(\bM,\bge_{\refer})$ and a lower bound on $\theta_{0,-}$. Thus
  (\ref{eq:hmcYAHlestimate}) holds.

  Next, we wish to estimate $E_{\bfI}[\hmcX^{A}_{ij}]$ for $|\bfI|\leq l$. This expression can be written as a linear combination of terms of the form
  (\ref{eq:EbfIhmcXAttbeest}). Combining this observation with (\ref{eq:bDbfAellAetcpteststmtEi}) yields the conclusion that it is sufficient to
  estimate expressions of the form
  \[
  \textstyle{\prod}_{i=1}^{p}|\bD^{k_{i}+1}\mK|_{\bge_{\refer}}|\bD_{\bfJ}\hmcX_{ij}^{\perp}|_{\bge_{\refer}}
  \]
  where $l_{\rotot}:=k_{1}+\dots+k_{p}+|\bfJ|\leq |\bfI|-p$. Appealing to 
  (\ref{eq:chimHlHlesthc}), (\ref{eq:chimClClesthc}), Corollary~\ref{cor:mixedmoserestweight} and the assumptions, we conclude that this
  expression can, in $L^{2}$, be estimated by
  \[
  C_{a}\ldr{\tau}^{l\cweight}e^{\eSpe\tau}
  \]
  for all $\tau\leq 0$, where $C_{a}$ only depends on $s_{\cweight,l}$, $s_{\coeff,l}$, $c_{\coeff,0}$, $(\bM,\bge_{\refer})$ and a lower bound on
  $\theta_{0,-}$. Thus (\ref{eq:hmcXAijHlestimate}) holds, and the lemma follows. 
\end{proof}

\begin{lemma}
  Let $0\leq \cweight\in\ro$, $\weight_{0}=(0,\cweight)$ and $\weight=(\cweight,\cweight)$. Assume that the conditions of 
  Lemma~\ref{lemma:taurelvaryingbxEi} as well as the $(\cweight,1)$-supremum assumptions are satisfied. Fix $l$
  as in Definition~\ref{def:sobklassumptions} and assume that the $(\cweight,l)$-Sobolev assumptions are satisfied. Assume, finally, that
  (\ref{eq:coefflassumptions}) holds with $l$ replaced by $1$ and that (\ref{eq:Sobcoefflassumptions}) holds. Then, if $0\leq |\bfI|\leq l$, 
  \begin{equation}\label{eq:EbfIZAXAcommtwowest}
    \begin{split}
      \|[E_{\bfI},Z^{A}X_{A}]u\|_{2,w} \leq & C_{a}\ldr{\tau}^{\a_{l}\cweight}\ldr{\tau-\tau_{c}}^{\b_{l}}e^{\eSpe\tau}\hE_{l}^{1/2}\\
      & +C_{b}\ldr{\tau}^{\a_{l}\cweight}\ldr{\tau-\tau_{c}}^{\b_{l}}e^{\eSpe\tau}\|\bD^{1}_{\bbE}u\|_{\infty,w}
    \end{split}
  \end{equation}    
  for all $\tau\leq\tau_{c}$, where $C_{a}$ only depends on $c_{\cweight,1}$, $c_{\coeff,1}$, $l$, $m_{\ros}$, $(\bM,\bge_{\refer})$ and a lower bound on
  $\theta_{0,-}$; and $C_{b}$ only depends on $s_{\cweight,l}$, $s_{\coeff,l}$, $c_{\cweight,1}$, $c_{\coeff,1}$, $m_{\ros}$, $(\bM,\bge_{\refer})$ and a lower
  bound on $\theta_{0,-}$. 
\end{lemma}
\begin{proof}
  Due to Lemma~\ref{lemma:bdbfAGAXGcommformEi}, we need to estimate expressions of the form
  \begin{equation}\label{eq:ZAXAprodestSob}
    \textstyle{\prod}_{i=1}^{p}|\bD^{k_{i}+1}\mK|_{\bge_{\refer}}\|E_{\bfK}Z^{A}\|\cdot |E_{\bfJ_{a}}E_{\bfJ_{b}}u|,
  \end{equation}
  where $l_{\rotot}:=k_{1}+\dots+k_{p}+|\bfK|+|\bfJ_{a}|\leq |\bfI|-p$ and $|\bfJ_{b}|=1$. In case $p\geq 1$, we can directly appeal to
  Corollary~\ref{cor:mixedmoserestweight} to conclude that (\ref{eq:ZAXAprodestSob}) can be estimated in $L^{2}$ with weight $w$ by
  \[
  C_{a}\ldr{\tau}^{l\cweight}e^{\eSpe\tau}\|\bD^{1}_{\bbE}u\|_{\infty,w}
  +C_{b}\ldr{\tau}^{l\cweight}\ldr{\tau-\tau_{c}}^{l+3\iota_{b}/2}e^{\eSpe\tau}\hE_{l}^{1/2}
  \]
  for all $\tau\leq\tau_{c}$, where $C_{a}$ only depends on $s_{\cweight,l}$, $s_{\coeff,l}$, $c_{\cweight,1}$, $c_{\coeff,1}$, $m_{\ros}$, $(\bM,\bge_{\refer})$ and
  a lower bound on $\theta_{0,-}$; and $C_{b}$ only depends on $c_{\cweight,1}$, $c_{\coeff,1}$, $l$, $m_{\ros}$, $(\bM,\bge_{\refer})$ and a lower bound on
  $\theta_{0,-}$.

  In case $p=0$ and $|\bfK|+|\bfJ_{a}|\leq |\bfI|-1$, we can proceed as above. However, if $p=0$ and $|\bfK|+|\bfJ_{a}|=|\bfI|$, then, since
  $|\bfJ_{a}|\leq |\bfI|-1$, we have to have $|\bfK|\geq 1$. In that case, we rewrite $E_{\bfK}=E_{\bfK_{a}}E_{\bfK_{b}}$, where $|\bfK_{b}|=1$. Then we
  need to estimate
  \[
  \|E_{\bfK_{a}}E_{\bfK_{b}}Z^{A}\|\cdot |E_{\bfJ_{a}}E_{\bfJ_{b}}u|
  \]
  in $L^{2}$ with weight $w$, where $l_{\rotot}:=|\bfK_{a}|+|\bfJ_{a}|\leq |\bfI|-1$. Appealing to Corollary~\ref{cor:mixedmoserestweight}, we obtain the
  bound
  \[
  C_{a}\ldr{\tau}^{l\cweight}\ldr{\tau-\tau_{c}}^{l+3\iota_{b}/2}e^{\eSpe\tau}\hE_{l}^{1/2}+C_{b}\ldr{\tau}^{l\cweight}e^{\eSpe\tau}\|\bD^{1}_{\bbE}u\|_{\infty,w}
  \]
  for all $\tau\leq\tau_{c}$, where $C_{a}$ only depends on $c_{\cweight,1}$, $c_{\coeff,1}$, $l$, $m_{\ros}$, $(\bM,\bge_{\refer})$ and a lower bound on
  $\theta_{0,-}$; and $C_{b}$ only depends on $s_{\cweight,l}$, $s_{\coeff,l}$, $c_{\cweight,1}$, $c_{\coeff,0}$, $m_{\ros}$, $(\bM,\bge_{\refer})$ and a lower
  bound on $\theta_{0,-}$. 
\end{proof}

\subsection{Commutator with $\hal$}

\begin{lemma}
  Let $0\leq \cweight\in\ro$, $\weight_{0}=(0,\cweight)$ and $\weight=(\cweight,\cweight)$. Assume that the conditions of 
  Lemma~\ref{lemma:taurelvaryingbxEi} as well as the $(\cweight,1)$-supremum assumptions are satisfied. Assume, finally, that
  (\ref{eq:coefflassumptions}) holds with $l$ replaced by $1$ and that (\ref{eq:Sobcoefflassumptions}) holds. Then, if $1\leq |\bfI|\leq l$,
  \begin{equation}\label{eq:EbfIhalcommSobestimate}
    \|[E_{\bfI},\hal]u\|_{2,w}\leq C_{a}\ldr{\tau}^{l\cweight}\|u\|_{\infty,w}+C_{b}\ldr{\tau}^{l\cweight}\ldr{\tau-\tau_{c}}^{l+3\iota_{b}/2}\hE_{l-1}^{1/2}
  \end{equation}
  for all $\tau\leq\tau_{c}$, where $C_{a}$ only depends on $c_{\robas}$, $s_{\coeff,l}$, $l$, $m_{\ros}$ and $(\bM,\bge_{\refer})$; and $C_{b}$ only
  depends on $c_{\cweight,1}$, $c_{\coeff,1}$, $l$, $m_{\ros}$ and $(\bM,\bge_{\refer})$.  
\end{lemma}
\begin{proof}
  Note that $[E_{\bfI},\hal]u$ can be written as a linear combination of terms of the form $E_{\bfJ}\hal\cdot E_{\bfK}u$, where $|\bfJ|+|\bfK|=|\bfI|$
  and $|\bfJ|\geq 1$. Rewrite $E_{\bfJ}=E_{\bfJ_{a}}E_{\bfJ_{b}}$ with $|\bfJ_{b}|=1$, let $1\leq m\leq l$ and assume that $|\bfI|=m$. Then we can appeal
  to Corollary~\ref{cor:mixedmoserestweight} to conclude that $E_{\bfJ_{a}}E_{\bfJ_{b}}\hal\cdot E_{\bfK}u$ can, in $L^{2}$ with weight $w$, be estimated by
  \[
  C_{a}\ldr{\tau}^{m\cweight}\|u\|_{\infty,w}+C_{b}\ldr{\tau}^{m\cweight}\ldr{\tau-\tau_{c}}^{m+3\iota_{b}/2}\hE_{m-1}^{1/2}
  \]
  for all $\tau\leq\tau_{c}$, where $C_{a}$ only depends on $c_{\robas}$, $s_{\coeff,m}$, $m$, $m_{\ros}$ and $(\bM,\bge_{\refer})$; and $C_{b}$ only depends
  on $c_{\cweight,1}$, $c_{\coeff,1}$, $m$, $m_{\ros}$ and $(\bM,\bge_{\refer})$. The lemma follows.
\end{proof}

\section{Estimating $\hU^{2}u$}\label{section:estimatinghUsquulsobas}

At this stage, we need to return to (\ref{eq:EbfIcommhUsqLtwotwowest}). In particular, we need to estimate $\hU^{2}u$, both in weighted Sobolev
spaces and in a weighted $C^{0}$-space. In order to obtain such estimates, we need to assume $u$ to satisfy the equation (\ref{eq:theequation}).
Making this assumption, the desired weighted $C^{0}$-estimate follows from (\ref{eq:hUsquestimate}). In order to obtain the desired weighted
Sobolev estimate, we make the following observation.

\begin{lemma}
  Let $0\leq \cweight\in\ro$, $\weight_{0}=(0,\cweight)$ and $\weight=(\cweight,\cweight)$. Assume that the conditions of 
  Lemma~\ref{lemma:taurelvaryingbxEi} as well as the $(\cweight,1)$-supremum assumptions are satisfied. Fix $l$
  as in Definition~\ref{def:sobklassumptions} and assume that the $(\cweight,l)$-Sobolev assumptions are satisfied. Assume, moreover, that 
  there are constants $c_{\coeff,1}$ and $s_{\coeff,l}$ such that (\ref{eq:coefflassumptions}) is satisfied with $l$ replaced by $1$ and 
  (\ref{eq:Sobcoefflassumptions}) is satisfied. Assume, finally, that (\ref{eq:theeqreformEi}) is satisfied. Then, if $|\bfK|\leq l$,
  \begin{equation*}
    \begin{split}
      \|E_{\bfK}\hU^{2}u\|_{2,w} \leq & C_{a}\ldr{\tau}^{\a_{k}\cweight+\b_{k}}e^{\eSpe\tau}\hE_{k+1}^{1/2}
      +C_{a}\ldr{\tau}^{\a_{k}\cweight}\ldr{\tau-\tau_{c}}^{\b_{k}}\hE_{k}^{1/2}\\
      & +C_{b}\ldr{\tau}^{\a_{k}\cweight+\b_{k}}e^{\eSpe\tau}\|\me_{1}\|_{\infty,w_{2}}^{1/2}\\
      & +C_{b}\ldr{\tau}^{\a_{k}\cweight}\ldr{\tau-\tau_{c}}^{\b_{k}}\|\me_{0}\|_{\infty,w_{2}}^{1/2}+\|E_{\bfK}\hf\|_{2,w}
    \end{split}
  \end{equation*}
  for all $\tau\leq\tau_{c}$, where $k:=|\bfK|$; $C_{a}$ only depends on $c_{\cweight,1}$, $c_{\coeff,1}$, $d_{\a}$ (in case $\iota_{b}\neq 0$), $k$, $m_{\ros}$,
  $(\bM,\bge_{\refer})$ and a lower bound on $\theta_{0,-}$; and $C_{b}$ only depends on $s_{\cweight,k}$, $s_{\coeff,k}$, $c_{\cweight,1}$, $c_{\coeff,1}$, $m_{\ros}$
  $(\bM,\bge_{\refer})$ and a lower bound on $\theta_{0,-}$. Moreover, $w_{2}:=w^{2}$ and $\a_{k}$ and $\b_{k}$ only depend on $k$. 
\end{lemma}
\begin{remark}
  Combining the conclusion of the lemma with (\ref{eq:hUsquestimate}) and Lemma~\ref{lemma:EbfIcommhUsqLtwotwowest} yields the following estimate: if
  $1\leq m\leq l$ and $|\bfI|=m$, 
  \begin{equation}\label{eq:EbfIcommhUsqLtwotwowestfinal}
    \begin{split}
      \|[E_{\bfI},\hU^{2}]u\|_{2,w} \leq & C_{a}\ldr{\tau}^{\a_{m}\cweight+\b_{m}}e^{\eSpe\tau}\hE_{m}^{1/2}
      +C_{a}\ldr{\tau}^{\a_{m}\cweight}\ldr{\tau-\tau_{c}}^{\b_{m}}\hE_{m-1}^{1/2}\\
      & +C_{b}\ldr{\tau}^{\a_{m}\cweight+\b_{m}}e^{\eSpe\tau}\|\me_{1}\|_{\infty,w_{2}}^{1/2}
      +C_{b}\ldr{\tau}^{\a_{m}\cweight}\ldr{\tau-\tau_{c}}^{\b_{m}}\|\me_{0}\|_{\infty,w_{2}}^{1/2}\\
      & +C_{c}\ldr{\tau}^{m\cweight}\|\hf\|_{\infty,w}+C_{d}\ldr{\tau}^{(m-1)\cweight}\ldr{\tau-\tau_{c}}^{m-1}\textstyle{\sum}_{|\bfK|\leq m-1}\|E_{\bfK}\hf\|_{2,w}
    \end{split}
  \end{equation}
  for all $\tau\leq\tau_{c}$. Here $C_{a}$ only depends on $c_{\cweight,1}$, $c_{\coeff,1}$, $d_{\a}$ (in case $\iota_{b}\neq 0$), $m$, $m_{\ros}$,
  $(\bM,\bge_{\refer})$ and a lower bound on $\theta_{0,-}$; $C_{b}$ only depends on $s_{\cweight,m}$, $s_{\coeff,m-1}$, $c_{\cweight,1}$, $c_{\coeff,1}$,
  $m_{\ros}$, $d_{\a}$ (in case $\iota_{b}\neq 0$), $(\bM,\bge_{\refer})$ and a lower bound on $\theta_{0,-}$; $C_{c}$ only depends on $s_{\cweight,m}$ and
  $(\bM,\bge_{\refer})$; and $C_{d}$ only depends on $c_{\cweight,1}$, $m$ and $(\bM,\bge_{\refer})$. Moreover, $\a_{m}$ and $\b_{m}$ only depend on $m$. 
\end{remark}
\begin{proof}
  Due to (\ref{eq:hUsquestimate}), we know that
  \[
  \|\hU^{2}u\|_{2,w}\leq C_{a}e^{\eSpe\tau}\hE_{1}^{1/2}+C_{b}\hE_{0}^{1/2}+\|\hf\|_{2,w}
  \]
  for all $\tau\leq\tau_{c}$, where $C_{a}$ only depends on $c_{\robas}$, $(\bM,\bge_{\refer})$ and a lower bound on $\theta_{0,-}$; and $C_{b}$ only depends on
  $c_{\cweight,0}$, $c_{\rocoeff,0}$, $d_{\a}$ (in case $\iota_{b}\neq 0$), $(\bM,\bge_{\refer})$ and a lower bound on $\theta_{0,-}$. Next, assume that
  $|\bfK|\geq 1$ and note that
  \begin{equation}\label{eq:EbfKhUsquexp}
    E_{\bfK}\hU^{2}u=\textstyle{\sum}_{A}E_{\bfK}(e^{-2\mu_{A}}X_{A}^{2}u)+E_{\bfK}(Z^{0}\hU u)+E_{\bfK}(Z^{A}X_{A}u)+E_{\bfK}(\hal u)-E_{\bfK}\hf.
  \end{equation}
  \textbf{The first term.} In order to estimate the first term, note that
  \begin{equation}\label{eq:EbfKemtwomuAXAsq}
    \begin{split}
      \|E_{\bfK}(e^{-2\mu_{A}}X_{A}^{2}u)\|_{2,w} \leq & \|[E_{\bfK},e^{-2\mu_{A}}X_{A}^{2}]u\|_{2,w}+\|e^{-2\mu_{A}}X_{A}^{2}E_{\bfK}u\|_{2,w}.
    \end{split}
  \end{equation}
  The first term on the right hand side can be estimated by the right hand side of (\ref{eq:emtmuAXAsqcommtbesttotal}). In order to estimate the
  second term on the right hand side of (\ref{eq:EbfKemtwomuAXAsq}), we appeal to (\ref{eq:emtwomuAXAsquest}) with $u$ replaced by $E_{\bfK}u$. This
  yields
  \[
  \|e^{-2\mu_{A}}X_{A}^{2}E_{\bfK}u\|_{2,w}\leq C_{a}e^{\eSpe\tau}\hE_{k+1}^{1/2}
  \]
  for all $\tau\leq\tau_{c}$, where $C_{a}$ only depends on $c_{\robas}$, $(\bM,\bge_{\refer})$ and a lower bound on $\theta_{0,-}$. Summing up,
  \begin{equation}\label{eq:EbfKemtwomuAXAsqfin}
    \begin{split}
      \|E_{\bfK}(e^{-2\mu_{A}}X_{A}^{2}u)\|_{2,w} \leq & \ldr{\tau}^{\a_{k}\cweight+\b_{k}}e^{\eSpe\tau}[C_{a}\hE_{k+1}^{1/2}
      +C_{b}\|\me_{1}\|_{\infty,w_{2}}^{1/2}]
    \end{split}
  \end{equation}
  for all $\tau\leq\tau_{c}$, where $C_{a}$ only depends on $c_{\cweight,1}$, $k$, $(\bM,\bge_{\refer})$ and a lower bound on $\theta_{0,-}$; and $C_{b}$ only
  depends on $c_{\cweight,1}$, $s_{\cweight,k}$, $(\bM,\bge_{\refer})$ and a lower bound on $\theta_{0,-}$.

  \textbf{The second term.} Turning to the second term on the right hand side of (\ref{eq:EbfKhUsquexp}),
  \begin{equation}\label{eq:SecondTermhUsqest}
    \|E_{\bfK}(Z^{0}\hU u)\|_{2,w}\leq \|[E_{\bfK},Z^{0}\hU] u\|_{2,w}+\|Z^{0}\hU E_{\bfK}u\|_{2,w}.
  \end{equation}
  The first term on the right hand side can be estimated by appealing to (\ref{eq:ZzhUcommEbfISobest}). In order to estimate the second term on
  the right hand side, it is sufficient to note that $\|Z^{0}\|$ is bounded by a constant depending only on $c_{\coeff,0}$, $n$, $m_{\ros}$ and $c_{\cweight,0}$.
  Adding up yields 
  \begin{equation}\label{eq:EbfKZzhUuSobest}
    \begin{split}
      \|E_{\bfK}(Z^{0}\hU u)\|_{2,w} \leq & C_{a}\ldr{\tau}^{\a_{k}\cweight}\ldr{\tau-\tau_{c}}^{\b_{k}}\hE_{k}^{1/2}\\
      & +C_{b}\ldr{\tau}^{\a_{k}\cweight}\ldr{\tau-\tau_{c}}^{\b_{k}}\left[e^{\eSpe\tau}\|\me_{1}\|_{\infty,w_{2}}^{1/2}+\|\me_{0}\|_{\infty,w_{2}}^{1/2}\right]
    \end{split}
  \end{equation}  
  for all $\tau\leq\tau_{c}$, where $C_{a}$ only depends on $c_{\cweight,1}$, $c_{\coeff,1}$, $k$, $m_{\ros}$ and $(\bM,\bge_{\refer})$; and $C_{b}$ only depends on 
  $s_{\cweight,k}$, $s_{\coeff,k}$, $c_{\cweight,1}$, $c_{\coeff,1}$, $m_{\ros}$ and $(\bM,\bge_{\refer})$. Here $\a_{k}$ and $\b_{k}$ are constants depending only on
  $k$.

  \textbf{The third term.} Next,
  \[
  \|E_{\bfK}(Z^{A}X_{A}u)\|_{2,w}\leq \|[E_{\bfK},Z^{A}X_{A}] u\|_{2,w}+\|Z^{A}X_{A}E_{\bfK}u\|_{2,w}.
  \]
  In this case, the first term on the right hand side can be estimated by appealing to (\ref{eq:EbfIZAXAcommtwowest}). The second term on the right
  hand side can be estimated by appealing to (\ref{eq:ZACznormestimate}). Summing up yields
  \begin{equation}\label{eq:EbfKZAXAuSobest}
    \begin{split}
      \|E_{\bfK}(Z^{A}X_{A}u)\|_{2,w} \leq & C_{a}\ldr{\tau}^{\a_{k}\cweight}\ldr{\tau-\tau_{c}}^{\b_{k}}e^{\eSpe\tau}\hE_{k+1}^{1/2}\\
      & +C_{b}\ldr{\tau}^{\a_{k}\cweight}\ldr{\tau-\tau_{c}}^{\b_{k}}e^{\eSpe\tau}\|\me_{1}\|_{\infty,w_{2}}^{1/2}
    \end{split}    
  \end{equation}
  for all $\tau\leq\tau_{c}$, where $C_{a}$ only depends on $c_{\cweight,1}$, $c_{\coeff,1}$, $k$, $m_{\ros}$, $(\bM,\bge_{\refer})$ and a lower bound on
  $\theta_{0,-}$; and $C_{b}$ only depends on $s_{\cweight,k}$, $s_{\coeff,k}$, $c_{\cweight,1}$, $c_{\coeff,1}$, $m_{\ros}$, $(\bM,\bge_{\refer})$ and a lower
  bound on $\theta_{0,-}$. 

  \textbf{The fourth term.} Finally,
  \begin{equation}\label{eq:EbfKhaluSobest}
    \begin{split}
      \|E_{\bfK}(\hal u)\|_{2,w} \leq & \|[E_{\bfK},\hal]u\|_{2,w}+\|\hal E_{\bfK}u\|_{2,w}\\
      \leq & C_{a}\ldr{\tau}^{k\cweight}\ldr{\tau-\tau_{c}}^{k+3\iota_{b}/2}\hE_{k}^{1/2}+C_{b}\ldr{\tau}^{k\cweight}\|u\|_{\infty,w}
    \end{split}
  \end{equation}
  for all $\tau\leq\tau_{c}$, where $C_{a}$ only depends on $c_{\cweight,1}$, $c_{\coeff,1}$, $k$, $m_{\ros}$ and $(\bM,\bge_{\refer})$; and $C_{b}$ only depends on
  $c_{\robas}$, $s_{\coeff,k}$, $k$, $m_{\ros}$ and $(\bM,\bge_{\refer})$.

  \textbf{Summing up.} Summing up the above estimates yields the conclusion of the lemma. 
\end{proof}

\section{Commutator with $L$}\label{section:commutatorwithLulSobass}

Summing up the above estimates, we are in a position to bound the commutator of $L$ with $E_{\bfI}$.

\begin{lemma}
  Let $0\leq \cweight\in\ro$, $\weight_{0}=(0,\cweight)$ and $\weight=(\cweight,\cweight)$. Assume that the conditions of 
  Lemma~\ref{lemma:taurelvaryingbxEi} as well as the $(\cweight,1)$-supremum assumptions are satisfied. Fix $l$
  as in Definition~\ref{def:sobklassumptions} and assume that the $(\cweight,l)$-Sobolev assumptions are satisfied. Assume, moreover, that 
  there are constants $c_{\coeff,1}$ and $s_{\coeff,l}$ such that (\ref{eq:coefflassumptions}) is satisfied with $l$ replaced by $1$ and 
  (\ref{eq:Sobcoefflassumptions}) is satisfied. Assume, finally, that (\ref{eq:theeqreformEi}) is satisfied. Then, if $1\leq m\leq l$ and
  $|\bfI|=m$,
  \begin{equation}\label{eq:EbfIcommLPDELtwotwowestfinal}
    \begin{split}
      \|[E_{\bfI},L]u\|_{2,w} \leq & C_{a}\ldr{\tau}^{\a_{m}\cweight+\b_{m}}e^{\eSpe\tau}\hE_{m}^{1/2}
      +C_{a}\ldr{\tau}^{\a_{m}\cweight}\ldr{\tau-\tau_{c}}^{\b_{m}}\hE_{m-1}^{1/2}\\
      & +C_{b}\ldr{\tau}^{\a_{m}\cweight+\b_{m}}e^{\eSpe\tau}\|\me_{1}\|_{\infty,w_{2}}^{1/2}
      +C_{b}\ldr{\tau}^{\a_{m}\cweight}\ldr{\tau-\tau_{c}}^{\b_{m}}\|\me_{0}\|_{\infty,w_{2}}^{1/2}\\
      & +C_{c}\ldr{\tau}^{m\cweight}\|\hf\|_{\infty,w}+C_{d}\ldr{\tau}^{(m-1)\cweight}\ldr{\tau-\tau_{c}}^{m-1}\textstyle{\sum}_{|\bfK|\leq m-1}\|E_{\bfK}\hf\|_{2,w}
    \end{split}
  \end{equation}
  for all $\tau\leq\tau_{c}$. Here $C_{a}$ only depends on $c_{\cweight,1}$, $c_{\coeff,1}$, $d_{\a}$ (in case $\iota_{b}\neq 0$), $m$, $m_{\ros}$,
  $(\bM,\bge_{\refer})$ and a lower bound on $\theta_{0,-}$; $C_{b}$ only depends on $s_{\cweight,m}$, $s_{\coeff,m}$, $c_{\cweight,1}$, $c_{\coeff,1}$, $m_{\ros}$,
  $d_{\a}$ (in case $\iota_{b}\neq 0$), $(\bM,\bge_{\refer})$ and a lower bound on $\theta_{0,-}$; $C_{c}$ only depends on $s_{\cweight,m}$ and
  $(\bM,\bge_{\refer})$; and $C_{d}$ only depends on $c_{\cweight,1}$, $m$ and $(\bM,\bge_{\refer})$. Moreover, $\a_{m}$ and $\b_{m}$ only depend on $m$.     
\end{lemma}
\begin{remark}\label{remark:EbfIcommLPDELtwotwowestfinalinclCkest}
  Assuming, in addition to the conditions of the lemma, that the conditions of Lemma~\ref{lemma:Ckestofweightedenden} are satisfied with $k=1$,
  we conclude that
  \begin{equation}\label{eq:EbfIcommLPDELtwotwowestfinalinclCkest}
    \begin{split}
      \|[E_{\bfI},L]u\|_{2,w} \leq & C_{a}\ldr{\tau}^{\a_{m}\cweight+\b_{m}}e^{\eSpe\tau}\hE_{m}^{1/2}
      +C_{a}\ldr{\tau}^{\a_{m}\cweight}\ldr{\tau-\tau_{c}}^{\b_{m}}\hE_{m-1}^{1/2}\\
      & +C_{b}\ldr{\tau}^{\a_{m,n}\cweight+\b_{m,n}}e^{\eSpe\tau}e^{c_{0}(\tau_{c}-\tau)/2}\hE_{\kappa_{1}}^{1/2}(\tau_{c};\tau_{c})\\
      & +C_{b}\ldr{\tau}^{\a_{m,n}\cweight}\ldr{\tau-\tau_{c}}^{\b_{m,n}}e^{c_{0}(\tau_{c}-\tau)/2}\hE_{\kappa_{0}}^{1/2}(\tau_{c};\tau_{c})      
    \end{split}
  \end{equation}
  for all $\tau\leq\tau_{c}$. Here $C_{a}$ only depends on $c_{\cweight,1}$, $c_{\coeff,1}$, $d_{\a}$ (in case $\iota_{b}\neq 0$), $m$, $m_{\ros}$,
  $(\bM,\bge_{\refer})$ and a lower bound on $\theta_{0,-}$; and $C_{b}$ only depends on $s_{\cweight,m}$, $s_{\coeff,m}$, $c_{\cweight,\kappa_{1}}$,
  $c_{\coeff,\kappa_{1}}$, $m_{\ros}$, $d_{\a}$ (in case $\iota_{b}\neq 0$), $(\bM,\bge_{\refer})$ and a lower bound on $\theta_{0,-}$. Moreover, $c_{0}$ is
  given by (\ref{eq:czCbdef}); $\a_{m}$ and $\b_{m}$ only depend on $m$; and $\a_{m,n}$ and $\b_{m,n}$ only depend on $n$ and $m$. Finally, $\kappa_{0}$
  is the smallest integer strictly larger than $n/2$ and $\kappa_{1}:=\kappa_{0}+1$. 
\end{remark}
\begin{remark}\label{remark:commczzest}
  Assume that the conditions of the lemma and all the conditions of Corollary~\ref{cor:basicenergyestimate} are satisfied. Assume, moreover, that the
  conditions of Lemma~\ref{lemma:Ckestofweightedenden} are satisfied with $k=1$. Then (\ref{eq:EbfIcommLPDELtwotwowestfinalinclCkest}) holds with
  $c_{0}=0$. However, in that case, $C_{b}$ also depends on $d_{q}$, $d_{\coeff}$ and $d_{\a}$. This conclusion is a consequence of the above and
  Remark~\ref{remark:puttingcztozinLinfest}. 
\end{remark}
\begin{proof}
  Combining (\ref{eq:emtmuAXAsqcommtbesttotal}), (\ref{eq:ZzhUcommEbfISobest}), (\ref{eq:EbfIZAXAcommtwowest}), (\ref{eq:EbfIhalcommSobestimate}) and
  (\ref{eq:EbfIcommhUsqLtwotwowestfinal}) yields the estimate stated in the lemma. 
\end{proof}

\section{Energy estimates}\label{section:energyestimatesstepII}

Combining the above conclusions with (\ref{eq:elbasenergyestimateEi}), we can derive energy estimates.

\begin{prop}\label{prop:EnergyEstimateSobolevAssumptions}
  Let $0\leq \cweight\in\ro$, $\weight_{0}=(0,\cweight)$ and $\weight=(\cweight,\cweight)$. Assume that the conditions of
  Lemma~\ref{lemma:taurelvaryingbxEi} are fulfilled and let $\kappa_{1}$ be the smallest integer strictly larger than $n/2+1$. Assume the
  $(\cweight,\kappa_{1})$-supremum
  assumptions to be satisfied; and that there is a constant $c_{\coeff,\kappa_{1}}$ such that (\ref{eq:coefflassumptions}) holds with $l$ replaced by
  $\kappa_{1}$. Fix $l$ as in Definition~\ref{def:sobklassumptions} and assume the $(\cweight,l)$-Sobolev assumptions to be
  satisfied. Assume, moreover, that there is a constant $s_{\coeff,l}$ such that (\ref{eq:Sobcoefflassumptions}) holds. Assume, finally, that
  (\ref{eq:theeqreformEi}) is satisfied with vanishing right hand side. Then
  \begin{equation}\label{eq:TheEnergyEstimate}
    \begin{split}
      \hE_{l}(\tau;\tau_{c}) \leq & C_{a}e^{c_{0}(\tau_{c}-\tau)}\hE_{l}(\tau_{c};\tau_{c})
      +C_{a}\ldr{\tau}^{2\a_{l,n}\cweight}\ldr{\tau-\tau_{c}}^{2\b_{l,n}}e^{c_{0}(\tau_{c}-\tau)}\hE_{l-1}(\tau_{c};\tau_{c})\\
      & +C_{b}\ldr{\tau}^{2\a_{l,n}\cweight}\ldr{\tau-\tau_{c}}^{2\b_{l,n}}e^{c_{0}(\tau_{c}-\tau)}\hE_{\kappa_{1}}(\tau_{c};\tau_{c})
    \end{split}    
  \end{equation}
  for all $\tau\leq\tau_{c}\leq 0$. Here $c_{0}$ is the constant defined by (\ref{eq:czCbdef}); the second and third terms on the right hand side vanish
  in case $l=0$; $\a_{l,n}$ and $\b_{l,n}$ only depend on $n$ and $l$; $C_{a}$ only depends on $c_{\cweight,1}$, $c_{\coeff,1}$, $d_{\a}$
  (in case $\iota_{b}\neq 0$), $l$, $m_{\ros}$, $(\bM,\bge_{\refer})$ and a lower bound on $\theta_{0,-}$; and $C_{b}$ only depends on $s_{\cweight,l}$,
  $s_{\coeff,l}$, $c_{\cweight,\kappa_{1}}$, $c_{\coeff,\kappa_{1}}$, $m_{\ros}$, $d_{\a}$ (in case $\iota_{b}\neq 0$), $(\bM,\bge_{\refer})$ and a lower bound on
  $\theta_{0,-}$.
\end{prop}
\begin{remark}\label{remark:EnergyEstimateSobolevAssumptionsImpCon}
  If, in addition to the assumptions of the lemma, all the conditions of Corollary~\ref{cor:basicenergyestimate} are satisfied, then
  (\ref{eq:TheEnergyEstimate}) can be improved in the sense that $c_{0}$ can be set to zero. On the other hand, the constants $C_{a}$ and $C_{b}$ then,
  additionally, depend on $d_{q}$, $d_{\coeff}$ and $d_{\a}$. The reason for this is the following. First, (\ref{eq:elbasenergyestimateEi}) holds. Second,
  due to Corollary~\ref{cor:basicenergyestimate}, the $\kappa$ appearing in this estimate is integrable. Third, due to Remark~\ref{remark:commczzest},
  (\ref{eq:EbfIcommLPDELtwotwowestfinalinclCkest}) holds with $c_{0}=0$. Combining these observations with an argument similar to the proof below
  yields the desired conclusion.
\end{remark}
\begin{proof}
  Combining Remark~\ref{remark:EbfIcommLPDELtwotwowestfinalinclCkest} with (\ref{eq:elbasenergyestimateEi}) yields
  \begin{equation}\label{eq:elbasenergyestimateEiX}
    \begin{split}
      \hE_{k}(\tau;\tau_{c}) \leq &  \hE_{k}(\tau_{c};\tau_{c})+\int_{\tau}^{\tau_{c}}\kappa(s)\hE_{k}(s;\tau_{c})ds
      +C_{a}\int_{\tau}^{\tau_{c}}\ldr{s}^{\a_{k}\cweight+\b_{k}}e^{\eSpe s}\hE_{k}(s;\tau_{c})ds\\
      & +C_{a}\int_{\tau}^{\tau_{c}}\ldr{s}^{\a_{k}\cweight}\ldr{s-\tau_{c}}^{\b_{k}}\hE_{k-1}^{1/2}(s;\tau_{c})\hE_{k}^{1/2}(s;\tau_{c})ds\\
      & +C_{b}\int_{\tau}^{\tau_{c}}\ldr{s}^{\a_{k,n}\cweight+\b_{k,n}}e^{\eSpe s}e^{c_{0}(\tau_{c}-s)/2}\hE_{\kappa_{1}}^{1/2}(\tau_{c};\tau_{c})\hE_{k}^{1/2}(s;\tau_{c})ds\\
      & +C_{b}\int_{\tau}^{\tau_{c}}\ldr{s}^{\a_{k,n}\cweight}\ldr{s-\tau_{c}}^{\b_{k,n}}e^{c_{0}(\tau_{c}-s)/2}\hE_{\kappa_{0}}^{1/2}(\tau_{c};\tau_{c})
      \hE_{k}^{1/2}(s;\tau_{c})ds   
    \end{split}
  \end{equation}
  for all $\tau\leq\tau_{c}$, where $C_{a}$ only depends on $c_{\cweight,1}$, $c_{\coeff,1}$, $d_{\a}$ (in case $\iota_{b}\neq 0$), $k$, $m_{\ros}$,
  $(\bM,\bge_{\refer})$ and a lower bound on $\theta_{0,-}$; and $C_{b}$ only depends on $s_{\cweight,k}$, $s_{\coeff,k}$, $c_{\cweight,\kappa_{1}}$,
  $c_{\coeff,\kappa_{1}}$, $d_{\a}$ (in case $\iota_{b}\neq 0$), $m_{\ros}$, $(\bM,\bge_{\refer})$ and a lower bound on
  $\theta_{0,-}$. Assume, inductively, that there are constants
  $\g_{m,n}$ and $\de_{m,n}$, depending only on $m$ and $n$, such that
  \begin{equation}\label{eq:hEmLtbasedestimatesindass}
    \begin{split}
      \hE_{m}(\tau;\tau_{c}) \leq & C_{a}e^{c_{0}(\tau_{c}-\tau)}\hE_{m}(\tau_{c};\tau_{c})
      +C_{a}\ldr{\tau}^{2\g_{m,n}\cweight}\ldr{\tau-\tau_{c}}^{2\de_{m,n}}e^{c_{0}(\tau_{c}-\tau)}\hE_{m-1}(\tau_{c};\tau_{c})\\
      & +C_{b}\ldr{\tau}^{2\g_{m,n}\cweight}\ldr{\tau-\tau_{c}}^{2\de_{m,n}}e^{c_{0}(\tau_{c}-\tau)}\hE_{\kappa_{1}}(\tau_{c};\tau_{c})
    \end{split}
  \end{equation}  
  for all $\tau\leq \tau_{c}$. Here $C_{a}$ and $C_{b}$ have the same dependence as in the case of (\ref{eq:elbasenergyestimateEiX}) (with $k$ replaced by
  $m$); and the second and third terms on the right hand side of (\ref{eq:hEmLtbasedestimatesindass}) should be set to zero in case $m=0$. We know this
  assumption to be true for $m=0$; cf. (\ref{eq:hEtauataubest}). Moreover, the relevant constant only depends on $c_{\cweight,0}$,
  $d_{\a}$ (in case $\iota_{b}\neq 0$), $(\bM,\bge_{\refer})$ and a lower bound on $\theta_{0,-}$. Assume that the inductive hypothesis holds for
  $0\leq m\leq k-1$ and that $k\leq l$. In order to demonstrate that it holds for $k$, we proceed as in the proof of Lemma~\ref{lemma:hEkenergyestimate}. To begin with, let $\xi(\tau)$ be
  defined by the right hand side of (\ref{eq:elbasenergyestimateEiX}). Then
  \[
  \xi'\geq -H'\xi-g\xi^{1/2},
  \]
  where 
  \begin{align*}
    H'(\tau) = & \kappa(\tau)+C_{a}\ldr{\tau}^{\a_{k}\cweight+\b_{k}}e^{\eSpe \tau},\\
    g(\tau) = & C_{a}\ldr{\tau}^{\a_{k}\cweight}\ldr{\tau-\tau_{c}}^{\b_{k}}\hE_{k-1}^{1/2}(\tau;\tau_{c})
    +C_{b}\ldr{\tau}^{\a_{k,n}\cweight}\ldr{\tau-\tau_{c}}^{\b_{k,n}}e^{c_{0}(\tau_{c}-\tau)/2}\hE_{\kappa_{1}}^{1/2}(\tau_{c};\tau_{c}).
  \end{align*}
  With this notation, it can be verified that (\ref{eq:firstenestxihalfest}) holds with $\tau_{a}=\tau$ and  $\tau_{b}=\tau_{c}$. Combining this
  estimate with the inductive assumption yields the conclusion that the inductive assumption holds with $k-1$ replaced by $k$. The lemma follows. 
\end{proof}

\section{The Klein-Gordon equation}

In the interest of illustrating the consequences of the above estimates, we apply them in the case of the Klein-Gordon equation. In this case,
we are interested in analysing the asymptotics of solutions to
\[
\Box_{g}u-\mKG^{2}u=0,
\]
where $\mKG$ is a constant. Comparing this equation with (\ref{eq:theequation}), it is clear that $\hal=-\mKG^{2}\theta^{-2}$. On the other hand, due to
(\ref{eq:hUnlnthetamomqbas}) and the fact that $q\geq n\e_{\Spe}$, cf. Remark~\ref{remark:qlwbd}, it is clear that $\theta$ tends to infinity
exponentially; cf. (\ref{eq:lnthetalowbd}). Combining this with, say, $(\cweight,l)$-Sobolev assumptions yields exponential decay of $\hal$ in suitable
weighted Sobolev spaces. In fact, we have the following estimate.

\begin{lemma}\label{lemma:halestimateKG}
  Let $0\leq \cweight\in\ro$, $\weight_{0}=(0,\cweight)$ and $\weight=(\cweight,\cweight)$. Assume that the conditions of 
  Lemma~\ref{lemma:taurelvaryingbxEi} and the estimate (\ref{eq:Kthetaoneestimate}) are satisfied. Then, for $1\leq l\in\zo$ and
  $\bfl=(1,l)$,
  \begin{equation}\label{eq:thetamtwoHbflest}
    \|\theta^{-2}\|_{H^{\bfl}_{\weight_{0}}(\bM)}\leq C_{a}\theta_{0,-}^{-2}e^{2\eSpe\tau}\|\ln\theta\|_{H^{\bfl}_{\weight_{0}}(\bM)}
  \end{equation}
  for all $\tau\leq 0$, where $C_{a}$ only depends on $\bDlnhNsup$, $\cweight$, $c_{\theta,1}$, $l$, $n$ and $(\bM,\bge_{\refer})$. 
\end{lemma}
\begin{remark}\label{remark:KGhalexpdec}
  Note that a $C^{0}$-estimate for $\theta^{-2}$ follows immediately from (\ref{eq:lnthetalowbd}). In particular, if $\hal=-\mKG^{2}\theta^{-2}$,
  then
  \[
  \|\hal\|_{C^{0}(\bM)}\leq C_{a}\theta_{0,-}^{-2}e^{2\eSpe\tau},\ \ \
  \|\hal\|_{H^{\bfl}_{\weight_{0}}(\bM)}\leq C_{b}\theta_{0,-}^{-2}e^{2\eSpe\tau}\|\ln\theta\|_{H^{\bfl}_{\weight_{0}}(\bM)}
  \]
  for all $\tau\leq 0$, where $C_{a}$ only depends on $\mKG$; and $C_{b}$ only depends on $c_{\theta,1}$, $\bDlnhNsup$, $\cweight$, $l$, $n$, $\mKG$ and
  $(\bM,\bge_{\refer})$.
\end{remark}
\begin{remark}
  If we, in addition to the conditions of the lemma, demand that (\ref{eq:qconvergence}) hold, then we obtain a better estimate of $\theta^{-2}$
  by appealing to Lemma~\ref{lemma:thetavarrhorelqconvtonmo}. 
\end{remark}
\begin{proof}
  Due to (\ref{eq:lnthetalowbd})
  \begin{equation}\label{eq:thetageqcn}
    \theta\geq e^{-2}\theta_{0,-}\exp[-(\e_{\Spe}+1/n)\varrho]
  \end{equation}
  for all $t\leq t_{0}$. In particular, appealing to
  (\ref{eq:eSpevarrhoeelowtaurelEi}), it is clear that $|E_{\bfI}\theta^{-2}|$ can be estimated by a linear combination of terms of the form
  \[
  \theta_{0,-}^{-2}e^{2\eSpe\tau}\textstyle{\prod}_{j}|E_{\bfI_{j}}\ln\theta|,
  \]
  where $|\bfI_{1}|+\dots+|\bfI_{k}|=|\bfI|$ and $|\bfI_{j}|\neq 0$. Combining this observation with Lemma~\ref{lemma:basequiv} and
  Corollary~\ref{cor:mixedmoserestweight} yields the conclusion of the lemma.
\end{proof}

In the case of the Klein-Gordon equation, Remark~\ref{remark:KGhalexpdec} makes it clear that $\|\hmcX^{0}\|$, $\|\hmcX^{\perp}\|_{\chg}$ and
$\|\hal\|$ all decay exponentially. For that reason we, from now on, focus on the somewhat more general situation that these expressions
decay to zero exponentially. In other words, we assume that there are constants $d_{\coe}$ and $\e_{\coe}>0$ such that
\begin{align}
  \sup_{\bx\in\bM}[\|\hmcX^{0}(\bx,t)\|+\|\hmcX^{\perp}(\bx,t)\|_{\chg}+\|\hal(\bx,t)\|] \leq & d_{\coe}e^{\e_{\coe}\tau(t)}\label{eq:Czcoeffexpdecassumptions}
\end{align}
for all $t\leq t_{0}$. Considering Lemma~\ref{lemma:basicenergyestimatequant}, it is clear that, under these circumstances, the only term that
contributes to the growth of the zeroth order energy is $q-(n-1)$. However, in what follows, we assume (\ref{eq:qconvergence}) to be satisfied.
Under these circumstances, we might as well use time independent measures in the definitions of the energies; cf. Remark~\ref{remark:tgbgerefequiv}.
For this reason, it is convenient to introduce the notation
\begin{equation}\label{eq:hGekdef}
  \hGe_{k}[u](\tau):=\int_{\bM_{\tau}}\me_{k}[u]\mu_{\bge_{\refer}}.
\end{equation}
\index{$\a$Aa@Notation!Energies!$\hGe_{k}[u]$}%
Note also that, assuming (\ref{eq:Czcoeffexpdecassumptions}) to hold, $\iota_{a}=0$ and $\iota_{b}=1$ in the definition of $\me_{k}$; cf.
(\ref{eq:mektaudefEi}). Under these circumstances, we obtain the following conclusions.

\begin{prop}\label{prop:asvelocityKGlikeeq}
  Let $0\leq \cweight\in\ro$, $\weight_{0}=(0,\cweight)$, $\weight=(\cweight,\cweight)$ and $\kappa_{1}$ be the smallest integer
  strictly larger than $n/2+1$. Assume that the conditions of Lemma~\ref{lemma:taurelvaryingbxEi} are fulfilled. Assume the
  $(\cweight,\kappa_{1})$-supremum assumptions to be satisfied; and that there is a constant $c_{\coeff,\kappa_{1}}$ such that (\ref{eq:coefflassumptions})
  holds with $l$ replaced by $\kappa_{1}$. Fix $l$ as in Definition~\ref{def:sobklassumptions} and assume the
  $(\cweight,l)$-Sobolev assumptions to be satisfied. Assume, moreover, that there is a constant $s_{\coeff,l}$ such that (\ref{eq:Sobcoefflassumptions})
  holds and that (\ref{eq:theeqreformEi}) is satisfied with vanishing right hand side. Assume, finally, that (\ref{eq:qconvergence}), (\ref{eq:haldecay})
  and (\ref{eq:coeffconvergence}) hold and let $\tau_{c}=0$. Then, if $l\geq \kappa_{1}$,
  \begin{align}
    \hGe_{l}(\tau) \leq & C_{a}\ldr{\tau}^{2\a_{l,n}\cweight+2\b_{l,n}}\hGe_{l}(0),\label{eq:TheEnergyEstimateKG}\\
    \|\me_{1}(\cdot,\tau)\|_{C^{0}(\bM)} \leq & C_{b}\ldr{\tau}^{\a_{n}\cweight+\b_{n}}\hGe_{\kappa_{1}}(0)\label{eq:meksupestimateKG}
  \end{align}
  for all $\tau\leq 0$. Here $\a_{l,n}$ and $\b_{l,n}$ only depend on $n$ and $l$; and $C_{a}$ only depends on $s_{\cweight,l}$, $s_{\coeff,l}$,
  $c_{\cweight,\kappa_{1}}$, $c_{\coeff,\kappa_{1}}$, $d_{\a}$, $d_{q}$, $d_{\coeff}$, $m_{\ros}$, $(\bM,\bge_{\refer})$ and a lower bound on $\theta_{0,-}$. Moreover,
  $\a_{n}$ and $\b_{n}$ only depend on $n$; and $C_{b}$ only depends on $c_{\cweight,\kappa_{1}}$, $c_{\coeff,\kappa_{1}}$, $d_{\a}$, $d_{q}$, $d_{\coeff}$,
  $m_{\ros}$, $(\bM,\bge_{\refer})$ and a lower bound on $\theta_{0,-}$.

  Assume, in addition, that (\ref{eq:Czcoeffexpdecassumptions}) holds, and that there are constants $\de_{q}$ and $\e_{q}>0$ such that 
  \begin{equation}\label{eq:qmnmoexpdecest}
    \|q(\cdot,t)-(n-1)\|_{C^{0}(\bM)} \leq \de_{q}e^{\e_{q}\tau(t)}
  \end{equation}
  for all $t\leq t_{0}$. Let $\e_{\roacc}:=\min\{\e_{\coe},\e_{q},\eSpe\}$. Then there is a $v_{\infty}\in C^{0}(\bM)$ such that
  \begin{align}
    \|(\hU u)(\cdot,\tau)-v_{\infty}\|_{C^{0}(\bM)} \leq & C_{\roacc}\ldr{\tau}^{\a_{n}\cweight+\b_{n}}e^{\e_{\roacc}\tau}\hGe_{\kappa_{1}}^{1/2}(0),\label{eq:hUvinflim}\\
    \|v_{\infty}\|_{C^{0}(\bM)}\leq & C_{\roacc}\hGe_{\kappa_{1}}^{1/2}(0)\label{eq:vinfCzbd}
  \end{align}
  for all $\tau\leq 0$, where $C_{\roacc}$ only depends on $c_{\cweight,\kappa_{1}}$, $c_{\coeff,\kappa_{1}}$, $d_{\alpha}$, $d_{q}$, $d_{\coeff}$, $\de_{q}$, $d_{\coe}$,
  $\e_{\coe}$, $\e_{q}$, $m_{\ros}$, $(\bM,\bge_{\refer})$ and a lower bound on $\theta_{0,-}$. Moreover, $\a_{n}$ and $\b_{n}$ only depend on $n$.
\end{prop}
\begin{remark}
  If (\ref{eq:Czcoeffexpdecassumptions}) is fulfilled, it follows that (\ref{eq:haldecay}) and (\ref{eq:coeffconvergence}) hold with $\tau_{c}=0$.
  Moreover, $d_{\a}$ and $d_{\coeff}$ then only depend on $d_{\coe}$ and $\e_{\coe}$.
\end{remark}
\begin{remark}\label{remark:higherorderasvel}
  In the lemma we impose $C^{0}$ assumptions on the coefficients and $q-(n-1)$; cf. (\ref{eq:Czcoeffexpdecassumptions}) and
  (\ref{eq:qmnmoexpdecest}). This leads to the $C^{0}$-estimates expressed in (\ref{eq:hUvinflim}) and (\ref{eq:vinfCzbd}). If one would impose
  stronger assumptions on the coefficients and $q-(n-1)$ ($C^{k}$-estimates for some $k\geq 1$ or Sobolev estimates) as well as, possibly, on the
  remaining components of the geometry, it should be possible to prove analogous estimates where $C^{0}$ is replaced by $C^{k_{1}}$ or $H^{k_{1}}$
  for some suitable $k_{1}\geq 1$. The arguments necessary should be similar to the arguments of the proof below combined with arguments already
  presented in these notes. However, for the sake of brevity, we do not attempt to prove such statements here. 
\end{remark}
\begin{remark}\label{remark:vinftyuinftyexistence}
  If one would have, say, higher order $C^{k}$-estimates analogous to (\ref{eq:hUvinflim}) and (\ref{eq:vinfCzbd})
  (cf. Remark~\ref{remark:higherorderasvel}), the asymptotic information could be improved. In order to justify this statement, assume that there
  is a $v_{\infty}\in C^{1}(\bM)$ such that (\ref{eq:hUvinflim}) and (\ref{eq:vinfCzbd}) hold with $C^{0}$ replaced by $C^{1}$ and $\kappa_{1}$
  replaced by $\kappa_{1}+1$. Given this assumption, let us sketch how to derive more detailed asymptotics. Compute 
  \begin{equation}\label{eq:hUumvinfvarrho}
    \hU(u-v_{\infty}\varrho)=\hU u-v_{\infty}+v_{\infty}[1-\hU(\varrho)]-\hU(v_{\infty})\varrho.
  \end{equation}
  The sum of the first two terms on the right hand side decay exponentially in $C^{0}$ due to (\ref{eq:hUvinflim}). In order to estimate the
  second term on the right hand side, note that (\ref{eq:hUvarrhoident}) yields
  \[
  |v_{\infty}[1-\hU(\varrho)]|=|v_{\infty}\hN^{-1}\rodiv_{\bge_{\refer}}\chi|\leq C_{a}e^{\eSpe\tau}\hGe_{\kappa_{1}}^{1/2}(0)
  \]
  for all $\tau\leq 0$, where we appealed to (\ref{eq:rodivchiestimpr}), (\ref{eq:eSpevarrhoeelowtaurelEi}) and (\ref{eq:vinfCzbd}). Moreover,
  $C_{a}$ has the same dependence as $C_{\roacc}$ in (\ref{eq:vinfCzbd}). Finally, let us estimate the third term on the right hand side of
  (\ref{eq:hUumvinfvarrho}). Note, to this end, that
  \[
  |\hU(v_{\infty})|=\hN^{-1}|\chi(v_{\infty})|\leq C_{b}e^{\eSpe\tau}|\bD v_{\infty}|_{\bge_{\refer}}
  \]
  for all $\tau\leq 0$, where we, in the last step, combined Remark~\ref{remark:chiclvarrhodecay} and (\ref{eq:eSpevarrhoeelowtaurelEi}). Here $C_{b}$
  only depends on $c_{\cweight,0}$ and $(\bM,\bge_{\refer})$. Assuming (\ref{eq:vinfCzbd}) to hold with $C^{0}$ replaced by $C^{1}$ and $\kappa_{1}$ replaced
  by $\kappa_{1}+1$,
  \[
  |\hU(v_{\infty})\varrho|\leq C_{c}\ldr{\tau}e^{\eSpe\tau}\hGe_{\kappa_{1}+1}^{1/2}(0),
  \]
  where $C_{c}$ only depends on $c_{\cweight,\kappa_{1}+1}$, $c_{\coeff,\kappa_{1}+1}$, $d_{\a,1}$, $d_{q,1}$, $d_{\coeff,1}$, $\de_{q,1}$, $d_{\coe,1}$, $m_{\ros}$,
  $(\bM,\bge_{\refer})$ and a lower bound on $\theta_{0,-}$. Here $d_{\a,1}$, $d_{q,1}$, $d_{\coeff,1}$, $\de_{q,1}$ and $d_{\coe,1}$ correspond to assumptions
  on the coefficients and $q$ that have to be imposed in order to obtain the $C^{1}$ version of the estimates (\ref{eq:hUvinflim}) and
  (\ref{eq:vinfCzbd}). Summarising the above estimates yields
  \begin{equation}\label{eq:hUumvinfvarrhoest}
    \|[\hU(u-v_{\infty}\varrho)](\cdot,\tau)\|_{C^{0}(\bM)}\leq C_{\roacc,1}\ldr{\tau}^{\a_{n}\cweight+\b_{n}}e^{\e_{\roacc}\tau}\hGe_{\kappa_{1}+1}^{1/2}(0)
  \end{equation}
  for all $\tau\leq 0$, where $C_{\roacc,1}$ only depends on $c_{\cweight,\kappa_{1}+1}$, $c_{\coeff,\kappa_{1}+1}$, $d_{\a,1}$, $d_{q,1}$, $d_{\coeff,1}$, $\de_{q,1}$,
  $d_{\coe,1}$, $m_{\ros}$, $(\bM,\bge_{\refer})$ and a lower bound on $\theta_{0,-}$; and $\a_{n}$ and $\b_{n}$ only depend on $n$. In analogy with the proof
  of (\ref{eq:hUvinflim}) (see below) this yields the existence of a $u_{\infty}\in C^{0}(\bM)$ such that
  \[
  \|(u-v_{\infty}\varrho-u_{\infty})(\cdot,\tau)\|_{C^{0}(\bM)}\leq K_{\roacc,1}\ldr{\tau}^{\a_{n}\cweight+\b_{n}}e^{\e_{\roacc}\tau}\hGe_{\kappa_{1}+1}^{1/2}(0)
  \]
  for all $\tau\leq 0$, where $K_{\roacc,1}$ only depends on $c_{\cweight,\kappa_{1}+1}$, $c_{\coeff,\kappa_{1}+1}$, $d_{\a,1}$, $d_{q,1}$, $d_{\coeff,1}$, $\de_{q,1}$,
  $d_{\coe,1}$, $\e_{\coe}$, $\e_{q}$, $m_{\ros}$, $(\bM,\bge_{\refer})$ and a lower bound on $\theta_{0,-}$. 
\end{remark}
\begin{proof}
  Note that all the conditions of Corollary~\ref{cor:basicenergyestimate} are satisfied. Due to the assumptions of the proposition, the conditions of
  Proposition~\ref{prop:EnergyEstimateSobolevAssumptions} are also satisfied with $\tau_{c}=0$, so that
  Remark~\ref{remark:EnergyEstimateSobolevAssumptionsImpCon} applies. Since $l\geq \kappa_{1}$, this means that (\ref{eq:TheEnergyEstimateKG}) holds for
  all $\tau\leq 0$, but with $\hGe$ replaced by $\hE(\cdot;0)$, and the same dependence of the constant. Combining this estimate with
  (\ref{eq:lntvarphimlntvarphicimp}), (\ref{eq:mutgreformulation}) and (\ref{eq:mutgdef}) yields
  (\ref{eq:TheEnergyEstimateKG}). Moreover, the assumptions stated in
  Remark~\ref{remark:puttingcztozinLinfest} apply with $k=1$, so that (\ref{eq:meksupestimateKG}) holds.

  If, in addition, (\ref{eq:Czcoeffexpdecassumptions}) and (\ref{eq:qmnmoexpdecest}) hold, then (\ref{eq:hUsquestimateprel}) holds with $f=0$ and an
  $\eta$ (introduced in (\ref{eq:etadef})) satisfying 
  \[
  \|\eta(\cdot,\tau)\|_{C^{0}(\bM)}\leq C_{\eta}\ldr{\tau}^{2(\cweight+1)}e^{\e_{\roacc}\tau}
  \]
  for all $\tau\leq 0$, where $\e_{\roacc}:=\min\{\e_{\coe},\e_{q},\eSpe\}$ and $C_{\eta}$ only depends on $c_{\cweight,0}$, $\de_{q}$,
  $d_{\coe}$, $(\bM,\bge_{\refer})$ and a lower bound on $\theta_{0,-}$. Combining this estimate with (\ref{eq:hUsquestimateprel}),
  (\ref{eq:meksupestimateKG}) and the fact that $f=0$ yields 
  \begin{equation}\label{eq:hUsqufirstczest}
    \|(\hU^{2}u)(\cdot,\tau)\|_{C^{0}(\bM)}\leq C_{\roacc}\ldr{\tau}^{\a_{n}\cweight+\b_{n}}e^{\e_{\roacc}\tau}\hGe_{\kappa_{1}}^{1/2}(0)
  \end{equation}
  for all $\tau\leq 0$, where $C_{\roacc}$ only depends on $c_{\cweight,\kappa_{1}}$, $c_{\coeff,\kappa_{1}}$, $d_{\a}$, $d_{q}$, $d_{\coeff}$, $\de_{q}$, $d_{\coe}$,
  $m_{\ros}$, $(\bM,\bge_{\refer})$ and a lower bound on $\theta_{0,-}$. Moreover, $\a_{n}$ and $\b_{n}$ only depend on $n$. Before proceeding,
  note that
  \begin{equation}\label{eq:hUsqsplitting}
    \hU^{2}u=\hN^{-1}\d_{t}\hU u-\hN^{-1}\chi \hU u.
  \end{equation}
  It is of interest to estimate the the second term in $C^{0}(\bM)$. Due to Remark~\ref{remark:chiclvarrhodecay} and (\ref{eq:eSpevarrhoeelowtaurelEi}),
  \begin{equation}\label{eq:hNinvchihUuestimate}
    |\hN^{-1}\chi \hU u|\leq \hN^{-1}|\chi|_{\bge_{\refer}}|\bD\hU u|_{\bge_{\refer}}\leq C_{a}e^{\eSpe\tau}|\bD\hU u|_{\bge_{\refer}}
  \end{equation}
  for all $\tau\leq 0$, where $C_{a}$ only depends on $c_{\cweight,0}$ and $(\bM,\bge_{\refer})$. On the other hand,
  \[
  |E_{i}\hU u|\leq |\hU E_{i}u|+|[E_{i},\hU]u|\leq C_{a}\me^{1/2}_{1},
  \]
  where we appealed to (\ref{eq:EbfIhUcommutatorstepone}) and $C_{a}$ only depends on $c_{\cweight,0}$ and $(\bM,\bge_{\refer})$.
  Combining this estimate with (\ref{eq:meksupestimateKG}) and (\ref{eq:hNinvchihUuestimate}) yields
  \[
  \|(\hN^{-1}\chi \hU u)(\cdot,\tau)\|_{C^{0}(\bM)}\leq C_{b}\ldr{\tau}^{\a_{n}\cweight+\b_{n}}e^{\eSpe\tau}\hGe_{\kappa_{1}}^{1/2}(0)
  \]
  on $M_{-}$, where $\a_{n}$ and $\b_{n}$ only depend on $n$; and $C_{b}$ only depends on $c_{\cweight,\kappa_{1}}$, $c_{\coeff,\kappa_{1}}$, $d_{\a}$, $d_{q}$,
  $d_{\coeff}$, $m_{\ros}$, $(\bM,\bge_{\refer})$ and a lower bound on $\theta_{0,-}$. Combining this estimate with (\ref{eq:hUsqufirstczest}),
  (\ref{eq:hUsqsplitting}) and (\ref{eq:hNtaudotequivEi}) yields the conclusion that 
  \begin{equation}\label{eq:dtauhUufirstczest}
    \|(\d_{\tau}\hU u)(\cdot,\tau)\|_{C^{0}(\bM)}\leq C_{\roacc}\ldr{\tau}^{\a_{n}\cweight+\b_{n}}e^{\e_{\roacc}\tau}\hGe_{\kappa_{1}}^{1/2}(0)
  \end{equation}
  for all $\tau\leq 0$, where $C_{\roacc}$ only depends on $c_{\cweight,\kappa_{1}}$, $c_{\coeff,\kappa_{1}}$, $d_{\a}$, $d_{q}$, $d_{\coeff}$,
  $\de_{q}$, $d_{\coe}$, $m_{\ros}$, $(\bM,\bge_{\refer})$ and a lower bound on $\theta_{0,-}$. Moreover, $\a_{n}$ and $\b_{n}$ only depend on $n$. Integrating
  (\ref{eq:dtauhUufirstczest}) from $\tau_{a}$ to $\tau_{b}$, where $\tau_{a}\leq \tau_{b}\leq 0$ yields the conclusion that
  \[
  \|(\hU u)(\cdot,\tau_{b})-(\hU u)(\cdot,\tau_{a})\|_{C^{0}(\bM)}\leq K_{\roacc}\ldr{\tau_{b}}^{\a_{n}\cweight+\b_{n}}e^{\e_{\roacc}\tau_{b}}\hGe_{\kappa_{1}}^{1/2}(0),
  \]
  where $K_{\roacc}$ only depends on $c_{\cweight,\kappa_{1}}$, $c_{\coeff,\kappa_{1}}$, $d_{\a}$, $d_{q}$, $d_{\coeff}$, $\de_{q}$, $d_{\coe}$, $\e_{\coe}$, $\e_{q}$,
  $m_{\ros}$, $(\bM,\bge_{\refer})$ and a lower bound on $\theta_{0,-}$. Thus there is a function $v_{\infty}\in C^{0}(\bM)$ such that
  \[
  \|(\hU u)(\cdot,\tau)-v_{\infty}\|_{C^{0}(\bM)}\leq K_{\roacc}\ldr{\tau}^{\a_{n}\cweight+\b_{n}}e^{\e_{\roacc}\tau}\hGe_{\kappa_{1}}^{1/2}(0)
  \]
  for all $\tau\leq 0$. In particular,
  \[
  \|v_{\infty}\|_{C^{0}(\bM)}\leq \|(\hU u)(\cdot,0)\|_{C^{0}(\bM)}+K_{\roacc}\hGe_{\kappa_{1}}^{1/2}(0)\leq C_{a}\hGe_{\kappa_{1}}^{1/2}(0),
  \]
  where $C_{a}$ has the same dependence as $K_{\roacc}$, and we appealed to (\ref{eq:meksupestimateKG}) in the last step. The lemma follows.
\end{proof}

\chapter{Localising the analysis}\label{chapter:localisingtheanalysis}

In the previous two chapters, we derive energy estimates based on various assumptions. Unfortunately, the estimates are quite crude in that
they only yield the conclusion that the energies do not grow faster than exponentially in the direction of the singularity. Moreover, the
information concerning the rate of growth is not very detailed. However, an extremely important feature of the estimates is that the rate of
growth does not depend on the order of the energy. Combining this fact with the silence allows us to derive more detailed asymptotic information
in causally localised regions. The purpose of the present chapter is to take the first step in carrying out such a derivation. 

In what follows, we derive asymptotics in regions that are roughly speaking of the form $J^{+}(\g)$, where $\g$ is an inextendible causal curve
in the spacetime (in the end it turns out to be convenient to consider slightly larger regions, denoted $A^{+}(\g)$ and introduced below). To begin with,
we therefore analyse the causal structure in the direction of the singularity. This is the subject of Section~\ref{section:causalstructure}.
In this section, we also analyse the spatial variation of $\varrho$ in $A^{+}(\g)$ and the behaviour of the weight appearing in the energy
estimates. Beyond analysing the causal structure, the main goal of the present chapter is to derive a model equation for the asymptotic behaviour
in $A^{+}(\g)$; cf. the heuristic discussions in Sections~\ref{section:resultsintrointo} and \ref{section:energyestimatesincausallylocalisedregions}.
We begin this derivation in Section~\ref{section:localisingeqfirstder} by estimating the difference between $\d_{\tau}\psi$ and $\hU\psi$. We also
estimate $\d_{\tau}\hU E_{\bfI}u-\hU^{2}E_{\bfI}u$. However, the main difficulty is to estimate differences such as $\d_{\tau}^{2}\psi-\d_{\tau}\hU\psi$.
This is the purpose of Section~\ref{section:localisingtheequsecondderivatives}. Unfortunately, the required arguments are quite technical. However,
in the end they result in a model equation; cf. Corollary~\ref{cor:rhsreplcfderfullylocal}. 

\section{Causal structure}\label{section:causalstructure}

Let $\g:(s_{-},s_{+})\rightarrow \bM\times I$ be a future pointing and past inextendible causal curve. We begin by providing conditions ensuring that
the spatial component of $\g(s)$ converges to a point in $\bM$ as $s\rightarrow s_{-}$. 

\begin{lemma}\label{lemma:pastinextendiblecausalcurveloc}
  Given that the conditions of Lemma~\ref{lemma:taurelvaryingbxEi} and the basic assumptions, cf. Definition~\ref{def:basicassumptions}, are satisfied,
  let $\tau$ be defined by (\ref{eq:taudefinitionEi}). Let $\g:(s_{-},s_{+})\rightarrow \bM\times I$ be a future pointing and past inextendible causal
  curve. Writing $\g(s)=[\bga(s),\g^{0}(s)]$, where $\bga(s)\in \bM$
  \[
  \frac{d\g^{0}}{ds}>0,\ \ \
  \lim_{s\rightarrow s_{-}+}\g^{0}(s)=t_{-}.
  \]
  Reparametrising $\g$ so that it is a function of $\tau$, there is a constant $C_{a}$ such that
  \begin{equation}\label{eq:dbgadtaubgerefestloc}
    \left|\frac{d\bga}{d\tau}(\tau)\right|_{\bge_{\refer}}\leq C_{a}\theta_{0,-}^{-1}e^{\eSpe\tau}
  \end{equation}
  for $\tau\leq 0$, where $C_{a}$ only depends on $c_{\robas}$ and $(\bM,\bge_{\refer})$. 
\end{lemma}
\begin{remark}\label{remark:bgaconvtobxbgaloc}
  Note that $s\rightarrow s_{-}+$ corresponds to $t\rightarrow t_{-}+$ which corresponds to $\tau\rightarrow \tau_{-}$, where $\tau_{-}\geq -\infty$.
  Combining this observation with the estimate (\ref{eq:dbgadtaubgerefestloc}) and the observation that $(\bM,\bge_{\refer})$ is complete yields the
  conclusion that $\bga(s)$ converges to a point $\bx_{\g}$ as $s\rightarrow s_{-}+$. Moreover,
  \begin{equation}\label{eq:dbgasbxgdistance}
  d(\bga(s),\bx_{\g})\leq C_{a}\theta_{0,-}^{-1}\eSpe^{-1}e^{\eSpe\tau\circ\g^{0}(s)}
  \end{equation}
  for all $s$ such that $\tau\circ\g^{0}(s)\leq 0$. Here $d$ is the topological metric induced on $\bM$ by $\bge_{\refer}$. 
\end{remark}
\begin{proof}
  Represent the tangent vector of $\g$ by
  \begin{equation}\label{eq:dotgorthodecomploc}
    \dot{\g}=v^{0}\hU+v^{A}X_{A},
  \end{equation}
  where $v^{0}>0$, since $\g$ is future pointing. Due to the causality of $\g$,
  \begin{equation}\label{eq:vAvzestloc}
    0\geq \hg(\dot{\g},\dot{\g})=-(v^{0})^{2}+\textstyle{\sum}_{A}e^{2\mu_{A}}(v^{A})^{2}.
  \end{equation}
  Due to (\ref{eq:futurenormalNchiexpr}), (\ref{eq:dotgorthodecomploc}) and the fact that $v^{0}>0$, it is clear that
  \begin{equation}\label{eq:dgzdshNinvvzexprloc}
    \frac{d\g^{0}}{ds}=\hN^{-1}v^{0}>0.
  \end{equation}
  Using (\ref{eq:futurenormalNchiexpr}) and (\ref{eq:dotgorthodecomploc}),
  it can also be deduced that
  \[
  \dot{\bga}=(v^{A}-\hN^{-1}\chi^{A}v^{0})X_{A}.
  \]
  In particular, there is a constant $C$, depending only on $n$, such that
  \begin{equation}\label{eq:dotbgabgereffirstestim}
  |\dot{\bga}|_{\bge_{\refer}}\leq \textstyle{\sum}_{A}(|v^{A}|+\hN^{-1}|\chi^{A}|v^{0})\leq Ce^{-\mu_{\min}}v^{0},
  \end{equation}
  where we appealed to (\ref{eq:dttimelikecondition}) and (\ref{eq:vAvzestloc}). Combining this estimate with (\ref{eq:muminmainlowerbound}) and
  (\ref{eq:eSpevarrhoeelowtaurelEi}) yields
  \begin{equation}\label{eq:dotbgaprelestloc}
    |\dot{\bga}|_{\bge_{\refer}}\leq C\theta_{0,-}^{-1}e^{\eSpe\tau}v^{0},
  \end{equation}
  where $C$ only depends on $c_{\robas}$. On the other hand, due to (\ref{eq:dgzdshNinvvzexprloc}),
  \begin{equation}\label{eq:dbgadtauformula}
  \frac{d\bga}{d\tau}=\left(\frac{d\g^{0}}{ds}\right)^{-1}\left(\frac{d\tau}{dt}\right)^{-1}\frac{d\bga}{ds}
  =\frac{\hN}{v^{0}}\left(\frac{d\tau}{dt}\right)^{-1}\frac{d\bga}{ds}.
  \end{equation}
  Combining this observation with (\ref{eq:hNtaudotequivEi}) and (\ref{eq:dotbgaprelestloc}) yields (\ref{eq:dbgadtaubgerefestloc}). 
\end{proof}

From now on, we are going to fix one curve $\g$ and assume that $\bx_{0}=\bx_{\g}$. In that situation, the estimate (\ref{eq:dbgadtaubgerefestloc})
can be improved slightly.

\begin{cor}\label{cor:pastinextendiblecausalcurveloc}
  Given that the conditions of Lemma~\ref{lemma:taurelvaryingbxEi} and the basic assumptions, cf. Definition~\ref{def:basicassumptions}, are satisfied,
  let $\tau$ be defined by (\ref{eq:taudefinitionEi}) and
  $\g:(s_{-},s_{+})\rightarrow \bM\times I$ be a future pointing and past inextendible causal curve. Let $\bx_{\g}$ be defined as in
  Remark~\ref{remark:bgaconvtobxbgaloc} and assume $\bx_{0}$ to have been chosen so that $\bx_{0}=\bx_{\g}$. Then, reparametrising
  $\g$ so that it is a function of $\tau$, there is a constant $C_{\rocau}$ such that
  \begin{equation}\label{eq:dbgadtaubgerefestimp}
    \left|\frac{d\bga}{d\tau}(\tau)\right|_{\bge_{\refer}}\leq C_{\rocau}\theta_{0,-}^{-1}e^{\e_{\Spe}\tau}
  \end{equation}
  for $\tau\leq 0$, where $C_{\rocau}$ only depends on $c_{\robas}$, $c_{\chi,2}$, $(\bM,\bge_{\refer})$ and a lower bound on $\theta_{0,-}$.
\end{cor}
\begin{remark}\label{remark:bgaconvtobxbgaimp}
  With $d$, $\g$ and $\bx_{\g}$ as in Remark~\ref{remark:bgaconvtobxbgaloc}, the estimate (\ref{eq:dbgadtaubgerefestimp}) yields
  \begin{equation}\label{eq:dbgasbxgdistanceimp}
  d(\bga(s),\bx_{\g})\leq C_{\rocau}\theta_{0,-}^{-1}\e_{\Spe}^{-1}e^{\e_{\Spe}\tau\circ\g^{0}(s)}
  \end{equation}
  for all $s$ such that $\tau\circ\g^{0}(s)\leq 0$. 
\end{remark}
\begin{proof}
  Combining (\ref{eq:muminmainlowerbound}), (\ref{eq:hNtaudotequivEi}), (\ref{eq:dotbgabgereffirstestim}) and (\ref{eq:dbgadtauformula}) yields
  \begin{equation}\label{eq:dbgadtauprelsharpest}
  \left|\frac{d\bga}{d\tau}(\tau)\right|_{\bge_{\refer}}\leq C_{a}\theta_{0,-}^{-1}e^{\e_{\Spe}\varrho\circ\g(\tau)}
  \end{equation}
  for all $\tau\leq 0$, where $C_{a}$ only depends on $c_{\robas}$ and $(\bM,\bge_{\refer})$. On the other hand,
  \begin{equation*}
    \begin{split}
      |\tau-\varrho\circ\g(\tau)| = & |\varrho(\bx_{0},\g^{0}(\tau))-\varrho(\bga(\tau),\g^{0}(\tau))|\\
      \leq & C_{b}\ldr{\tau}d(\bx_{0},\bga(\tau)),
    \end{split}
  \end{equation*}
  where $C_{b}$ only depends on $c_{\robas}$, $c_{\chi,2}$ and $(\bM,\bge_{\refer})$, and we appealed to (\ref{eq:bDvarrhobdEi})
  and (\ref{eq:DeltavarrhorelvariationEi}). Combining this estimate with (\ref{eq:dbgasbxgdistance}) and (\ref{eq:dbgadtauprelsharpest}) yields
  the conclusion of the corollary. 
\end{proof}

Given assumptions and notation as in the statement of Corollary~\ref{cor:pastinextendiblecausalcurveloc} and Remark~\ref{remark:bgaconvtobxbgaimp},
let
\[
K_{A}:=C_{\rocau}\theta_{0,-}^{-1}\e_{\Spe}^{-1}
\]
\index{$\a$Aa@Notation!Constants!$K_{A}$}%
and define, using the notation $M_{-}:=\bM\times I_{-}$, 
\begin{equation}\label{eq:Aplusgammadef}
  A^{+}(\g):=\{(\bx,t)\in M_{-}: d(\bx,\bx_{\g})\leq K_{A}e^{\e_{\Spe}\tau(t)}\}. 
\end{equation}
\index{$\a$Aa@Notation!Sets!$A^{+}(\g)$}%
Then Corollary~\ref{cor:pastinextendiblecausalcurveloc} yields the conclusion that $J^{+}(\g)\cap J^{-}(\bM_{t_{0}})\subseteq A^{+}(\g)$. Moreover, due
to an argument similar to the proof Corollary~\ref{cor:pastinextendiblecausalcurveloc},
\begin{equation}\label{eq:varrhominustauestimate}
|\varrho(\bx,t)-\tau(t)|\leq C_{b}\theta_{0,-}^{-1}\ldr{\tau(t)}e^{\e_{\Spe}\tau(t)}
\end{equation}
for all $(\bx,t)\in A^{+}(\g)$, where $C_{b}$ only depends on $c_{\robas}$, $c_{\chi,2}$, $(\bM,\bge_{\refer})$ and a lower bound on $\theta_{0,-}$.

At this stage, it is also of interest to estimate $w$, defined by (\ref{eq:wdef}), in $A^{+}(\g)$. 
\begin{lemma}\label{lemma:lnwintqminusnminusoneestimate}
  Assume that the conditions of Lemma~\ref{lemma:taurelvaryingbxEi} and the basic assumptions, cf. Definition~\ref{def:basicassumptions}, are fulfilled,
  let $\g$ and $\bx_{\g}$ be as in Remark~\ref{remark:bgaconvtobxbgaloc}, and assume that $\bx_{0}=\bx_{\g}$. Assume, moreover, that there is a constant
  $c_{q}$ such that $|q|\leq c_{q}$ on $M_{-}$ and that (\ref{eq:Kthetaoneestimate}) holds. Then
  \begin{equation}\label{eq:lnwalongAplusgamma}
    \left|(\ln w)(\bx,\tau_{a})-\frac{1}{2n}\int_{\tau_{a}}^{\tau_{c}}[q(\bx_{0},\tau)-(n-1)]d\tau\right|
    \leq C_{a}\ldr{\tau_{c}}^{\bcweight}e^{\e_{\Spe}\tau_{c}}
  \end{equation}
  for all $\tau_{a}\leq\tau_{c}\leq 0$ and $\bx\in\bM$ such that $(\bx,\tau_{a})$ corresponds to an element of $A^{+}(\g)$. Here $C_{a}$ only depends
  on $c_{\robas}$, $c_{\theta,1}$, $c_{q}$, $c_{\chi,2}$, $(\bM,\bge_{\refer})$ and a lower bound on $\theta_{0,-}$. Moreover, $\bcweight:=\max\{\cweight,1\}$.
\end{lemma}
\begin{remark}
  As already pointed out, $q-(n-1)$ converges to zero exponentially in many situations of interest. In that setting, (\ref{eq:lnwalongAplusgamma})
  yields the conclusion that $w$ is essentially constant. However, in oscillatory settings (such as Bianchi VIII and IX), the difference
  $q-(n-1)$ does not converge to zero. On the other hand, it is very small on average. 
\end{remark}
\begin{proof}
  Note that
  \[
  2\ln w(\bx_{0},\tau)=\ln\tvarphi(\bx_{0},\tau)-\ln\tvarphi(\bx_{0},\tau_{c})=\tau-\tau_{c}+\ln\theta(\bx_{0},\tau)-\ln\theta(\bx_{0},\tau_{c}),
  \]
  where we used the fact that $\varrho(\bx_{0},\tau)=\tau$. Next, note that
  \[
  \d_{\tau}\ln\theta=(\d_{t}\tau)^{-1}\hN\hN^{-1}\d_{t}\ln\theta=\tN(\hU+\hN^{-1}\chi)\ln\theta,
  \]
  where $\tN:=\hN/\d_{t}\tau$. On the other hand,
  \[
  |\tN(\bx_{0},\cdot)-1|=\tN(\bx_{0},\cdot)|1-\tN^{-1}(\bx_{0},\cdot)|\leq 3|1-\tN^{-1}(\bx_{0},\cdot)|/2,
  \]
  where we appealed to (\ref{eq:tNestimatepreliminary}). On the other hand, due to (\ref{eq:tNinverseeqonepluserror}),
  \[
  |1-\tN^{-1}(\bx_{0},\cdot)|\leq |[\hN^{-1}\chi(\varrho)](\bx_{0},\cdot)|+|[\hN^{-1}\rodiv_{\bge_{\refer}}\chi](\bx_{0},\cdot)|.
  \]
  However, the first term on the right hand side can be estimated by appealing to (\ref{eq:firststepvarrhospatialvar}) and the
  second term on the right hand side can be estimated by appealing to (\ref{eq:rodivchiestimpr}). To conclude
  \begin{equation}\label{eq:tNinvbxzminusoneestimates}
    |\tN^{-1}(\bx_{0},\cdot)-1|+|\tN(\bx_{0},\cdot)-1|\leq C_{a}\ldr{\tau}e^{\e_{\Spe}\tau}
  \end{equation}
  for all $\tau\leq 0$, where $C_{a}$ only depends on $c_{\robas}$, $c_{\chi,2}$ and $(\bM,\bge_{\refer})$. Next, note that by an argument similar
  to (\ref{eq:firststepvarrhospatialvar}),
  \[
  \hN^{-1}|\chi\ln\theta|\leq n^{1/2}e^{-\mu_{\min}}|\chi|_{\rohy}|\bD\ln\theta|.
  \]
  Evaluating this estimate in $(\bx_{0},\cdot)$ and appealing to (\ref{eq:Kthetaoneestimate}) yields
  \[
    [\hN^{-1}|\chi\ln\theta|](\bx_{0},\tau)\leq C_{b}\ldr{\tau}^{\cweight}e^{\e_{\Spe}\tau}
  \]
  for all $\tau\leq 0$, where $C_{b}$ only depends on $c_{\robas}$, $c_{\theta,1}$, $c_{\chi,2}$ and $(\bM,\bge_{\refer})$. Finally, $|\hU(\ln\theta)|$ is
  bounded by a constant depending only on $c_{q}$ and $n$. Combining the above estimates yields the conclusion that
  \[
  |\d_{\tau}\ln\theta-\hU(\ln\theta)|(\bx_{0},\tau)\leq C_{a}\ldr{\tau}^{\bcweight}e^{\e_{\Spe}\tau}
  \]
  for all $\tau\leq 0$, where $C_{a}$ only depends on $c_{\robas}$, $c_{\theta,1}$, $c_{q}$, $c_{\chi,2}$ and $(\bM,\bge_{\refer})$. Combining this estimate
  with (\ref{eq:hUnlnthetamomqbas}) and the fact that $\tau=\varrho(\bx_{0},\tau)$ yields
  \[
  |(\d_{\tau}\ln\tvarphi)(\bx_{0},\tau)+[q(\bx_{0},\tau)-(n-1)]/n|\leq C_{a}\ldr{\tau}^{\bcweight}e^{\e_{\Spe}\tau}
  \]
  for all $\tau\leq 0$, where $C_{a}$ only depends on $c_{\robas}$, $c_{\theta,1}$, $c_{q}$, $c_{\chi,2}$ and $(\bM,\bge_{\refer})$. In particular,
  \[
  \left|(\ln\tvarphi)(\bx_{0},\tau_{c})-(\ln\tvarphi)(\bx_{0},\tau_{a})+\frac{1}{n}\int_{\tau_{a}}^{\tau_{c}}[q(\bx_{0},\tau)-(n-1)]d\tau\right|
  \leq C_{a}\ldr{\tau_{c}}^{\bcweight}e^{\e_{\Spe}\tau_{c}}
  \]
  for all $\tau_{a}\leq\tau_{c}\leq 0$, where $C_{a}$ only depends on $c_{\robas}$, $c_{\theta,1}$, $c_{q}$, $c_{\chi,2}$ and $(\bM,\bge_{\refer})$. Thus
  \[
  \left|(\ln w)(\bx_{0},\tau_{a})-\frac{1}{2n}\int_{\tau_{a}}^{\tau_{c}}[q(\bx_{0},\tau)-(n-1)]d\tau\right|\leq C_{a}\ldr{\tau_{c}}^{\bcweight}e^{\e_{\Spe}\tau_{c}}
  \]
  for all $\tau_{a}\leq\tau_{c}\leq 0$, where $C_{a}$ only depends on $c_{\robas}$, $c_{\theta,1}$, $c_{q}$, $c_{\chi,2}$ and $(\bM,\bge_{\refer})$.
  Combining this estimate with (\ref{eq:bDlntvarphicestpre}) yields the conclusion of the lemma. 
\end{proof}

\section{Localising the equation, first derivatives}\label{section:localisingeqfirstder}

In what follows, we wish to replace every occurrence of $\hU$ in $L$ with $\d_{\tau}$. In the end, this will allow us to replace the PDE with an 
ODE when analysing the asymptotics. In the present section, we begin by replacing one occurrence of $\hU$. 

\begin{lemma}\label{lemma:dhtaupsirelhUpsiloc}
  Given that the conditions of Lemma~\ref{lemma:taurelvaryingbxEi} and the basic assumptions, cf. Definition~\ref{def:basicassumptions}, are fulfilled, 
  \begin{equation}\label{eq:dhtaupsihUpsiestloc}
    |\d_{\tau}\psi|\leq C_{a}\left(|\hU(\psi)|^{2}+\textstyle{\sum}_{A}e^{-2\mu_{A}}|X_{A}(\psi)|^{2}\right)^{1/2}
  \end{equation}
  on $M_{-}$, where $C_{a}$ only depends on $C_{\rorel}$ and $(\bM,\bge_{\refer})$. Let $\g$ and $\bx_{\g}$ be as in Remark~\ref{remark:bgaconvtobxbgaloc},
  and assume that $\bx_{0}=\bx_{\g}$. Then
  \begin{equation}\label{eq:dhtaupsiminushUpsiestloc}
    \begin{split}
      |\d_{\tau}\psi-\hU\psi| \leq & C_{b}\ldr{\tau}e^{\e_{\Spe}\tau}\left(|\hU(\psi)|^{2}+\textstyle{\sum}_{A}e^{-2\mu_{A}}|X_{A}(\psi)|^{2}\right)^{1/2}\\
      & +\frac{1}{2}\left(\textstyle{\sum}_{A}e^{-2\mu_{A}}|X_{A}(\psi)|^{2}\right)^{1/2}
    \end{split}
  \end{equation}
  on $A^{+}_{c}(\g)$, where $A^{+}_{c}(\g)$ is the subset of $A^{+}(\g)$ corresponding to $\tau\leq\tau_{c}$. Here $C_{b}$ only depends on $c_{\robas}$,
  $c_{\chi,2}$, $(\bM,\bge_{\refer})$ and a lower bound on $\theta_{0,-}$.
\end{lemma}
\begin{remark}
  One particular consequence of (\ref{eq:dhtaupsihUpsiestloc}) is that 
  \begin{equation}\label{eq:dhtaubDbfAuloc}
    |\d_{\tau}E_{\bfI}u|^{2}\leq C_{a} \me_{l}[u]
  \end{equation}
  for all $\tau\leq 0$ and vector field multiindices $\bfI$ satisfying $|\bfI|\leq l$. Here $C_{a}$ only depends on $C_{\rorel}$ and 
  $(\bM,\bge_{\refer})$. One particular consequence of (\ref{eq:dhtaupsiminushUpsiestloc}) is that 
  \begin{equation}\label{eq:dhtaubDbfAuminushUbDbfAuloc}
    |\d_{\tau}E_{\bfI}u-\hU E_{\bfI}u|\leq C_{b}\ldr{\tau}\ldr{\tau-\tau_{c}}^{3\iota_{b}/2}e^{\e_{\Spe}\tau}\me_{l+1}^{1/2}[u]
  \end{equation}
  on $A^{+}_{c}(\g)$, where $C_{b}$ only depends on $c_{\robas}$, $c_{\chi,2}$, $(\bM,\bge_{\refer})$ and a lower bound on $\theta_{0,-}$.
\end{remark}
\begin{proof}
  By assumption, the conditions of Lemma~\ref{lemma:epsilonlowdefEi} are fulfilled, so that
  \[
  |\d_{\tau}\psi|\leq |\d_{t}\tau|^{-1}|\d_{t}\psi|\leq 2K_{\rovar}\hN^{-1}|\d_{t}\psi|\leq 2K_{\rovar}(|\hU\psi|+\hN^{-1}|\chi\psi|),
  \]
  where we appealed to (\ref{eq:hNtaudotequivEi}). Next, note that 
  \begin{equation}\label{eq:hNinvchipsiestloc}
    \begin{split}
      \hN^{-1}|\chi\psi| \leq & \left(\textstyle{\sum}_{A}\hN^{-2}e^{2\mu_{A}}(\chi^{A})^{2}\right)^{1/2}
      \left(\textstyle{\sum}_{A}e^{-2\mu_{A}}[X_{A}(\psi)]^{2}\right)^{1/2}\\
      \leq & N^{-1}|\chi|_{\bge}\left(\textstyle{\sum}_{A}e^{-2\mu_{A}}|X_{A}(\psi)|^{2}\right)^{1/2}
      \leq \frac{1}{2}\left(\textstyle{\sum}_{A}e^{-2\mu_{A}}|X_{A}(\psi)|^{2}\right)^{1/2},
    \end{split}
  \end{equation}
  where we appealed to (\ref{eq:dttimelikecondition}) in the last step (note that (\ref{eq:dttimelikecondition}) is a consequence of
  (\ref{eq:basbaschibd})). Combining the last two estimates yields (\ref{eq:dhtaupsihUpsiestloc}). In order to prove the second estimate, note that
  \begin{equation}\label{eq:dtaupsiminushUpsiidentity}
    \d_{\tau}\psi-\hU\psi=(\d_{t}\tau)^{-1}\d_{t}\psi-\hN^{-1}\d_{t}\psi+\hN^{-1}\chi(\psi).
  \end{equation}
  The last term on the right hand side can be estimated by appealing to (\ref{eq:hNinvchipsiestloc}). It is therefore of interest to consider
  \begin{equation}\label{eq:oneminhNinvdttauid}
    \begin{split}
      1-\hN^{-1}(\bx,t)\d_{t}\tau(t) = & 1-\hN^{-1}(\bx,t)\hN(\bx_{0},t)\\
      & +\hN^{-1}(\bx,t)\hN(\bx_{0},t)[1-\hN^{-1}(\bx_{0},t)\d_{t}\tau(t)].
    \end{split}    
  \end{equation}
  On the other hand,
  \[
  |\ln[\hN^{-1}(\bx,t)\hN(\bx_{0},t)]|\leq C_{\rorel}d(\bx_{0},\bx)\leq C_{\rorel}K_{A}e^{\e_{\Spe}\tau}
  \]
  for all $(\bx,t)\in A^{+}(\g)$. In particular,
  \begin{equation}\label{eq:oneminhNinvdttauidlocpar}
    |1-\hN^{-1}(\bx,t)\hN(\bx_{0},t)|\leq C_{a}e^{\e_{\Spe}\tau}
  \end{equation}
  for all $(\bx,t)\in A^{+}(\g)$, where $C_{a}$  only depends on $c_{\robas}$, $c_{\chi,2}$, $(\bM,\bge_{\refer})$ and a lower bound on
  $\theta_{0,-}$. Next, (\ref{eq:tNinvbxzminusoneestimates}), (\ref{eq:oneminhNinvdttauid}) and (\ref{eq:oneminhNinvdttauidlocpar}) yield
  \begin{equation}\label{eq:hprelestimate}
  |1-\hN^{-1}(\bx,t)\d_{t}\tau(t)|\leq C_{d}\ldr{\tau}e^{\e_{\Spe}\tau}
  \end{equation}
  for all $(\bx,t)\in A^{+}(\g)$, where $C_{d}$  only depends on $c_{\robas}$, $c_{\chi,2}$, $(\bM,\bge_{\refer})$ and a lower bound on
  $\theta_{0,-}$. Combining this estimate with (\ref{eq:dtaupsiminushUpsiidentity}) yields
  \begin{equation*}
    \begin{split}
      |\d_{\tau}\psi-\hU\psi| \leq & |1-\hN^{-1}(\bx,t)\d_{t}\tau(t)||\d_{\tau}\psi|+\hN^{-1}|\chi(\psi)|\\
      \leq & C_{e}\ldr{\tau}e^{\e_{\Spe}\tau}\left(|\hU(\psi)|^{2}+\textstyle{\sum}_{A}e^{-2\mu_{A}}|X_{A}(\psi)|^{2}\right)^{1/2}\\
      & +\frac{1}{2}\left(\textstyle{\sum}_{A}e^{-2\mu_{A}}|X_{A}(\psi)|^{2}\right)^{1/2}
    \end{split}
  \end{equation*}
  for all $(\bx,t)\in A^{+}(\g)$, where we appealed to (\ref{eq:dhtaupsihUpsiestloc}) and (\ref{eq:hNinvchipsiestloc}), and $C_{e}$ only depends
  on $c_{\robas}$, $c_{\chi,2}$, $(\bM,\bge_{\refer})$ and a lower bound on $\theta_{0,-}$. The lemma follows. 
\end{proof}

Next, we wish to replace $\hU^{2}$ with $\d_{\tau}\hU$. 

\begin{lemma}\label{lemma:dhtauhUbDbfAumhUsqetcloc}
  Fix $l$, $\bfl_{1}$, $\cweight$, $\weight_{0}$ and $\weight$ as in Definition~\ref{def:supmfulassumptions}.
  Then, given that the assumptions of Lemma~\ref{lemma:taurelvaryingbxEi} as well as the $(\cweight,l)$-supremum assumptions are satisfied,
  assume (\ref{eq:coefflassumptions}) to hold. Let $L$ be defined by (\ref{eq:LuformulaEi}) and assume $u$ to be a smooth solution to $Lu=0$.
  Let $\g$ and $\bx_{\g}$ be as in Remark~\ref{remark:bgaconvtobxbgaloc}, and assume that $\bx_{0}=\bx_{\g}$. Then, for all $m=|\bfI|\leq l$, 
  \begin{equation}\label{eq:dhtauhUbDbfAumhUsqetcloc}
    |\d_{\tau}\hU E_{\bfI}u-\hU^{2}E_{\bfI}u|\leq C_{a}\ldr{\tau}^{(m+1)\cweight+1}\ldr{\tau-\tau_{c}}^{3\iota_{b}/2}e^{\e_{\Spe}\tau}\me_{m+1}^{1/2}
  \end{equation}
  on $A^{+}_{c}(\g)$, where $C_{a}$ only depends on $c_{\cweight,l}$, $c_{\rocoeff,l}$, $m_{\ros}$, $d_{\a}$ (in case $\iota_{b}\neq 0$), $(\bM,\bge_{\refer})$ and
  a lower bound on $\theta_{0,-}$.
\end{lemma}
\begin{remark}
  An additional consequence of the proof is that for $m=|\bfI|\leq l$ and $\tau\leq \tau_{c}$,  
  \begin{equation}\label{eq:dhtauhUbDbfAuestloc}
    \begin{split}
      |\d_{\tau}\hU E_{\bfI}u|^{2} \leq & C_{a}\ldr{\tau}^{4\cweight+2}\ldr{\tau-\tau_{c}}^{3\iota_{b}}e^{2\e_{\Spe}\tau}\me_{m+1}\\
      & +C_{b}\ldr{\tau}^{2(m+1)\cweight}\ldr{\tau-\tau_{c}}^{3\iota_{b}}\me_{m-1}+C_{c}\me_{m},
    \end{split}    
  \end{equation}
  where the second term on the right hand side can be omitted in case $m=0$. Here $C_{c}$ only depends on $c_{\cweight,0}$, $c_{\rocoeff,0}$,
  $m_{\ros}$, $d_{\a}$ (in case $\iota_{b}\neq 0$) and $(\bM,\bge_{\refer})$; and $C_{a}$ and $C_{b}$ only depend on $c_{\cweight,l}$, $c_{\rocoeff,l}$, $m_{\ros}$,
  $d_{\a}$ (in case $\iota_{b}\neq 0$), $(\bM,\bge_{\refer})$ and a lower bound on $\theta_{0,-}$.
\end{remark}
\begin{proof}
  Since $Lu=0$,
  \begin{equation}\label{eq:LbDbfAuloc}
    L(E_{\bfI}u)=[L,E_{\bfI}]u.
  \end{equation}
  Moreover, since the conditions of Lemma~\ref{lemma:LbDbfAunconditionalestimateEi} are satisfied,
  \begin{equation}\label{eq:LbDbfAunconditionalestimatefzloc}
    |[L,E_{\bfI}]u|^{2}\leq C_{a}\ldr{\varrho}^{4\cweight+2}\ldr{\tau-\tau_{c}}^{3\iota_{b}}e^{2\e_{\Spe}\varrho}\me_{m}
      +C_{b}\ldr{\varrho}^{2(m+1)\cweight}\ldr{\tau-\tau_{c}}^{3\iota_{b}}\me_{m-1}
  \end{equation} 
  for all $\tau\leq \tau_{c}$ and $m=|\bfI|\leq l$, where $C_{a}$ and $C_{b}$ only depend on $c_{\cweight,l}$, $c_{\rocoeff,l}$, $m_{\ros}$,
  $d_{\a}$ (in case $\iota_{b}\neq 0$),
  $(\bM,\bge_{\refer})$ and a lower bound on $\theta_{0,-}$. Note also that if $m=0$, the estimate (\ref{eq:LbDbfAunconditionalestimatefzloc})
  holds with a vanishing right hand side. Next, let us consider the terms appearing in $L(E_{\bfI}u)$. Appealing to
  (\ref{eq:ZACznormestimate}) and (\ref{eq:emtwomuAXAsquest}) with $u$ replaced by $E_{\bfI}u$ yields, with $m=|\bfI|$, 
  \begin{equation}\label{eq:spatialderivativesexpdecayestimateloc}
    |e^{-2\mu_{A}}X_{A}^{2}E_{\bfI}u|^{2}+|Z^{A}X_{A}E_{\bfI}u|^{2}\leq C_{a}\theta_{0,-}^{-2}\ldr{\tau-\tau_{c}}^{3\iota_{b}}e^{2\e_{\Spe}\varrho}\me_{m+1}
  \end{equation}
  for all $\tau\leq\tau_{c}$, where $C_{a}$ only depends on $c_{\cweight,0}$, $c_{\rocoeff,0}$, $m_{\ros}$, $(\bM,\bge_{\refer})$ and a lower bound on $\theta_{0,-}$.
  Combining this estimate with (\ref{eq:LbDbfAuloc}) and (\ref{eq:LbDbfAunconditionalestimatefzloc}) yields, with $m=|\bfI|$, 
  \begin{equation}\label{eq:hUderlhsestloc}
    \begin{split}
      & |-\hU^{2}E_{\bfI}u+Z^{0}\hU E_{\bfI}u+\hal E_{\bfI}u|^{2}\\
      \leq & C_{a}\ldr{\varrho}^{4\cweight+2}\ldr{\tau-\tau_{c}}^{3\iota_{b}}e^{2\e_{\Spe}\varrho}\me_{m+1}
      +C_{b}\ldr{\varrho}^{2(m+1)\cweight}\ldr{\tau-\tau_{c}}^{3\iota_{b}}\me_{m-1}
    \end{split}    
  \end{equation}
  for all $\tau\leq\tau_{c}$, where $Z^{0}$ is introduced in (\ref{eq:ZzdefEi}). Here the second term on the right hand side vanishes if $m=0$.
  Moreover, $C_{a}$ and $C_{b}$ only depend on $c_{\cweight,l}$, $c_{\rocoeff,l}$, $m_{\ros}$, $d_{\a}$ (in case $\iota_{b}\neq 0$), $(\bM,\bge_{\refer})$ and a lower
  bound on $\theta_{0,-}$. Note that one particular consequence of this estimate is that, if $m=|\bfI|$, 
  \begin{equation}\label{eq:hUsqbDbfAusqestloc}
    \begin{split}
      |\hU^{2}E_{\bfI}u|^{2} \leq & C_{a}\ldr{\varrho}^{4\cweight+2}\ldr{\tau-\tau_{c}}^{3\iota_{b}}e^{2\e_{\Spe}\varrho}\me_{m+1}\\
      & +C_{b}\ldr{\varrho}^{2(m+1)\cweight}\ldr{\tau-\tau_{c}}^{3\iota_{b}}\me_{m-1}+C_{c}\me_{m}
    \end{split}    
  \end{equation}
  for all $\tau\leq\tau_{c}$, where we appealed to (\ref{eq:chthexpressionhUlntheta}), (\ref{eq:Czbalancedhat}), (\ref{eq:hmcYzdefEi}), (\ref{eq:ZzdefEi})
  and and the assumptions; note that (\ref{eq:Czbalancedhat}) follows from (\ref{eq:coefflassumptions}) and that $q$ is bounded due to 
  Definition~\ref{def:supmfulassumptions}. Moreover, the second term on the right hand side vanishes if $m=0$; $C_{c}$ only depends on
  $c_{\cweight,0}$, $c_{\rocoeff,0}$, $m_{\ros}$ and $d_{\a}$ (in case $\iota_{b}\neq 0$); and $C_{a}$ and $C_{b}$ only depend on $c_{\cweight,l}$, $c_{\rocoeff,l}$,
  $m_{\ros}$, $d_{\a}$ (in case $\iota_{b}\neq 0$), $(\bM,\bge_{\refer})$ and a lower bound on $\theta_{0,-}$. Moreover, (\ref{eq:dhtaubDbfAuminushUbDbfAuloc})
  yields, with $m=|\bfI|$, 
  \[
  |Z^{0}\hU E_{\bfI}u-Z^{0}\d_{\tau} E_{\bfI}u|\leq C\ldr{\tau}\ldr{\tau-\tau_{c}}^{3\iota_{b}/2}e^{\e_{\Spe}\tau}\me_{m+1}^{1/2}
  \]
  on $A^{+}_{c}(\g)$, where $C$ only depends on $c_{\cweight,0}$, $c_{\rocoeff,0}$, $m_{\ros}$, $(\bM,\bge_{\refer})$ and a lower bound on $\theta_{0,-}$.
  Combining this estimate with (\ref{eq:hUderlhsestloc}) yields the conclusion that, if $m=|\bfI|$, 
  \begin{equation}\label{eq:rhsreplcfderloc}
    \begin{split}
      & |-\hU^{2}E_{\bfI}u+Z^{0}\d_{\tau} E_{\bfI}u+\hal E_{\bfI}u|^{2}\\
      \leq & C_{a}\ldr{\tau}^{4\cweight+2}\ldr{\tau-\tau_{c}}^{3\iota_{b}}e^{2\e_{\Spe}\tau}\me_{m+1}
      +C_{b}\ldr{\tau}^{2(m+1)\cweight}\ldr{\tau-\tau_{c}}^{3\iota_{b}}\me_{m-1}
    \end{split}    
  \end{equation}
  holds on $A^{+}_{c}(\g)$. Again, the second term on the right hand side vanishes if $m=0$, and $C_{a}$ and $C_{b}$ only depend on $c_{\cweight,l}$, 
  $c_{\rocoeff,l}$, $m_{\ros}$, $d_{\a}$ (in case $\iota_{b}\neq 0$), $(\bM,\bge_{\refer})$ and a lower bound on $\theta_{0,-}$. Applying
  (\ref{eq:dhtaupsihUpsiestloc}) with $\psi=\hU E_{\bfI}u$ yields
  \begin{equation}\label{eq:dhtauhUbDbfAuestprelloc}
    |\d_{\tau}\hU E_{\bfI}u|\leq C\left(|\hU^{2}E_{\bfI}u|^{2}+\textstyle{\sum}_{A}e^{-2\mu_{A}}|X_{A}\hU E_{\bfI}u|^{2}\right)^{1/2}
  \end{equation}
  on $M_{-}$, where $C$ only depends on $C_{\rorel}$ and $(\bM,\bge_{\refer})$. In order to estimate the second term inside the paranthesis, note
  that (\ref{eq:hUEicomm}) yields
  \begin{equation*}
    \begin{split}
      E_{i}\hU E_{\bfI}u = & [E_{i},\hU]E_{\bfI}u+\hU E_{i}E_{\bfI}u\\
      = & -A_{i}^{k}E_{k}E_{\bfI}u-E_{i}(\ln\hN)\hU E_{\bfI}u+\hU E_{i}E_{\bfI}u,
    \end{split}
  \end{equation*}
  where $A_{i}^{k}$ and $A_{i}^{0}$ are given by (\ref{eq:Aialphadef}). Due to Lemma~\ref{lemma:Aiksupest} and (\ref{eq:bDlnNbDlnthetabd}), it
  follows that if $m=|\bfI|\leq l$,
  \begin{equation}\label{eq:XAhUbDbfAuestloc}
    |X_{A}\hU E_{\bfI}u|\leq C\me_{m+1}^{1/2}
  \end{equation}
  on $M_{-}$, where $C$ only depends on $c_{\cweight,0}$ and $(\bM,\bge_{\refer})$. In order to estimate the
  first term inside the paranthesis on the right hand side of (\ref{eq:dhtauhUbDbfAuestprelloc}), it is sufficient to appeal to
  (\ref{eq:hUsqbDbfAusqestloc}). Summing up, we conclude that (\ref{eq:dhtauhUbDbfAuestloc}) holds. Appealing to (\ref{eq:dhtaupsiminushUpsiestloc})
  with $\psi=\hU E_{\bfI}u$ yields
  \begin{equation*}
    \begin{split}
      |\d_{\tau}\hU E_{\bfI}u-\hU^{2}E_{\bfI}u| \leq & C\ldr{\tau}e^{\e_{\Spe}\tau}\left(|\hU^{2}E_{\bfI}u|^{2}
      +\textstyle{\sum}_{A}e^{-2\mu_{A}}|X_{A}\hU E_{\bfI}u|^{2}\right)^{1/2}\\
      & +\left(\textstyle{\sum}_{A}e^{-2\mu_{A}}|X_{A}\hU E_{\bfI}u|^{2}\right)^{1/2}
    \end{split}
  \end{equation*}
  on $A^{+}(\g)$, where $C$ only depends on $c_{\robas}$, $c_{\chi,2}$, $(\bM,\bge_{\refer})$ and a lower bound on $\theta_{0,-}$.
  Combining this estimate with (\ref{eq:hUsqbDbfAusqestloc}) and (\ref{eq:XAhUbDbfAuestloc}) yields the conclusion that
  (\ref{eq:dhtauhUbDbfAumhUsqetcloc}) holds.
\end{proof}

\section{Localising the equation, second derivatives}\label{section:localisingtheequsecondderivatives}

Next, we wish to replace $\hU^{2}$ with $\d_{\tau}^{2}$. Note, to this end, that (\ref{eq:dtaupsiminushUpsiidentity}) and (\ref{eq:oneminhNinvdttauid}) 
yield
\[
\d_{\tau}\psi-\hU\psi=h\d_{\tau}\psi+\hN^{-1}\chi(\psi),
\]
where
\begin{equation}\label{eq:hdef}
  \begin{split}
    h(\bx,t) := & 1-\hN^{-1}(\bx,t)\d_{t}\tau(t)\\
    = & 1-\hN^{-1}(\bx,t)\hN(\bx_{0},t)+\hN^{-1}(\bx,t)\hN(\bx_{0},t)[1-\hN^{-1}(\bx_{0},t)\d_{t}\tau(t)].
  \end{split}
\end{equation}
Thus
\[
\d_{\tau}^{2}\psi-\d_{\tau}\hU\psi=\d_{\tau}h\d_{\tau}\psi+h\d_{\tau}^{2}\psi+\d_{\tau}[\hN^{-1}\chi(\psi)].
\]
In particular,
\[
(1-h)(\d_{\tau}^{2}\psi-\d_{\tau}\hU\psi)=h\d_{\tau}\hU\psi+\d_{\tau}h\d_{\tau}\psi+\d_{\tau}[\hN^{-1}\chi(\psi)].
\]
Combining this equality with (\ref{eq:hNtaudotequivEi}) yields
\begin{equation}\label{eq:dtausqdtauhUdiffestfs}
|\d_{\tau}^{2}\psi-\d_{\tau}\hU\psi|\leq 2K_{\rovar}[|h\d_{\tau}\hU\psi|+|\d_{\tau}h\d_{\tau}\psi|+|\d_{\tau}[\hN^{-1}\chi(\psi)]|]. 
\end{equation}
Note that (\ref{eq:hprelestimate}) gives an estimate for $h$. To estimate $\d_{\tau}\hU\psi$ in the context of greatest interest here, it is
sufficient to appeal to (\ref{eq:dhtauhUbDbfAuestloc}). Combining these observations yields an estimate for the first term inside the parenthesis
on the right hand side of (\ref{eq:dtausqdtauhUdiffestfs}). In order to estimate $\d_{\tau}h$, we begin by making the following observation.

\subsection{The spatial variation of the $\tau$-derivative of $\hN$}

In order to estimate $\d_{\tau}h$, it is natural to begin by estimating the $\tau$-derivative of the second term on the far right hand side of 
(\ref{eq:hdef}). 

\begin{lemma}
  Assume that the conditions of Lemma~\ref{lemma:taurelvaryingbxEi} as well as the $(\cweight,1)$-supremum assumptions are satisfied. 
  Let $\g$ and $\bx_{\g}$ be as in Remark~\ref{remark:bgaconvtobxbgaloc}, and assume that $\bx_{0}=\bx_{\g}$. Finally, let
  $\hN_{0}:=\hN(\bx_{0},\cdot)$. Then
  \begin{equation}\label{eq:dtauhNinvhNzeroest}
    |\d_{\tau}(\hN^{-1}\hN_{0})|\leq C\ldr{\tau}^{2\cweight}e^{\e_{\Spe}\tau}
  \end{equation}
  on $A^{+}(\g)$, where $C$ only depends on $c_{\cweight,1}$, $(\bM,\bge_{\refer})$ and a lower bound on $\theta_{0,-}$.
\end{lemma}
\begin{proof}
  Compute
  \begin{equation}\label{eq:dtauhNinvhNzeroid}
    \d_{\tau}(\hN^{-1}\hN_{0})=\hN^{-1}\hN_{0}(\d_{\tau}\ln\hN_{0}-\d_{\tau}\ln\hN).
  \end{equation}
  Next, note that (\ref{eq:dhtaupsihUpsiestloc}) yields 
  \begin{equation}\label{eq:EidtaulnhN}
    |E_{i}\d_{\tau}\ln\hN|=|\d_{\tau}E_{i}\ln\hN|\leq C_{a}\left(|\hU E_{i}\ln\hN|^{2}+\textstyle{\sum}_{A}e^{-2\mu_{A}}|X_{A}E_{i}\ln\hN|^{2}\right)^{1/2}.
  \end{equation}
  In order to estimate the right hand side, note that (\ref{eq:hUEicomm}) and (\ref{eq:Aialphadef}) yield
  \begin{equation*}
    \begin{split}
      |\hU E_{i}\ln\hN| \leq & |[\hU,E_{i}]\ln\hN|+|E_{i}\hU\ln\hN|\\
      \leq & |E_{i}\ln\hN|\cdot |\hU\ln\hN|+\textstyle{\sum}_{k}|A_{i}^{k}||E_{k}\ln\hN|+|E_{i}\hU\ln\hN|\leq C\ldr{\tau}^{2\cweight},
    \end{split}
  \end{equation*}
  where $C$ only depends on $c_{\cweight,1}$ and $(\bM,\bge_{\refer})$. In order to obtain this
  estimate, we appealed to the assumptions and (\ref{eq:AikCzest}). Combining this estimate with (\ref{eq:muminmainlowerbound}), (\ref{eq:EidtaulnhN})
  and the assumptions yields 
  \[
  |E_{i}\d_{\tau}\ln\hN|\leq C\ldr{\tau}^{2\cweight}
  \]
  on $M_{-}$, where $C$ only depends on $c_{\cweight,1}$, $(\bM,\bge_{\refer})$ and a lower bound on $\theta_{0,-}$. Combining this estimate with 
  (\ref{eq:dtauhNinvhNzeroid}) yields (\ref{eq:dtauhNinvhNzeroest}). 
\end{proof}

\subsection{Estimating the contribution from the shift vector field}

Considering (\ref{eq:hdef}), it is clear that what remains to be estimated is the $\tau$-derivative of the right hand side of
\begin{equation}\label{eq:hUdtaudiscrinApl}
  1-\hN^{-1}(\bx_{0},t)\d_{t}\tau(t)=-[\hN^{-1}\chi(\varrho)+\hN^{-1}\rodiv_{\bge_{\refer}}\chi](\bx_{0},t);
\end{equation}
this equality follows from (\ref{eq:tNinverseeqonepluserror}). Returning to (\ref{eq:dtausqdtauhUdiffestfs}), it is clear that we need to estimate
\[
\d_{\tau}[\hN^{-1}\chi(\psi)],\ \ \
\d_{\tau}[\hN^{-1}\chi(\varrho)],\ \ \
\d_{\tau}[\hN^{-1}\rodiv_{\bge_{\refer}}\chi]. 
\]
On the other hand, the last two expressions we only need to estimate along $(\bx_{0},t)$. Next, note that $A_{i}^{k}$ introduced in (\ref{eq:Aialphadef})
satisfies
\[
A^{k}_{i}=-\hN^{-1}\omega^{k}(\ml_{\chi}E_{i})=-\hN^{-1}\omega^{k}(\bD_{\chi}E_{i})+\hN^{-1}\omega^{k}(\bD_{E_{i}}\chi). 
\]
Taking the trace of this equality yields
\begin{equation}\label{eq:hNinvrodivbgereferchiAiiformloc}
\hN^{-1}\rodiv_{\bge_{\refer}}\chi=\textstyle{\sum}_{i}A^{i}_{i}+\hN^{-1}\chi^{j}\omega^{i}(\bD_{E_{j}}E_{i}).
\end{equation}
Due to the above and (\ref{eq:dhtaupsihUpsiestloc}), it is of interest to estimate the result when applying $\hU$ and $X_{A}$ to $A_{i}^{i}$, as well as to
\[
\hN^{-1}\chi\psi,\ \ \
\hN^{-1}\chi(\varrho),\ \ \
\hN^{-1}\chi^{j}\omega^{i}(\bD_{E_{j}}E_{i}).
\]
Moreover, with the exception of $\hN^{-1}\chi\psi$, we only need to estimate these expressions along $(\bx_{0},t)$. 

\begin{lemma}\label{lemma:hUwnfestloc}
  Assume that the conditions of Lemma~\ref{lemma:taurelvaryingbxEi} as well as the $(\cweight,0)$-supremum assumptions are satisfied. 
  Let $\g$ and $\bx_{\g}$ be as in Remark~\ref{remark:bgaconvtobxbgaloc}, and assume that $\bx_{0}=\bx_{\g}$. Then
  \begin{equation}\label{eq:hUhNinvchipsiestloc}
    |\hU[\hN^{-1}\chi(\psi)]|\leq C_{a}\ldr{\tau}^{\cweight}e^{\e_{\Spe}\tau}\textstyle{\sum}_{i}\left(|\hU E_{i}\psi|+|E_{i}\psi|\right)
  \end{equation}
  on $A^{+}(\g)$ for all smooth $\psi$ on $\bM\times I$, where $C_{a}$ only depends on $c_{\cweight,0}$, $(\bM,\bge_{\refer})$ and a lower bound on
  $\theta_{0,-}$. Moreover,
  \begin{equation}\label{eq:hUhNinvchivarrhoestloc}
    |\hU[\hN^{-1}\chi(\varrho)]|\leq C_{a}\ldr{\tau}^{\cweight+1}e^{\e_{\Spe}\tau}
  \end{equation}
  on $A^{+}(\g)$, where $C_{a}$ only depends on $c_{\cweight,0}$, $(\bM,\bge_{\refer})$ and a lower bound on $\theta_{0,-}$. Finally,
  \begin{equation}\label{eq:hUhNinvchijzetajestloc}
    |\hU[\hN^{-1}\chi^{j}\omega^{i}(\bD_{E_{j}}E_{i})]|\leq C_{a}\ldr{\tau}^{\cweight}e^{\e_{\Spe}\tau}
  \end{equation}
  on $A^{+}(\g)$, where $C_{a}$ only depends on $c_{\cweight,0}$, $(\bM,\bge_{\refer})$ and a lower bound on $\theta_{0,-}$.  
\end{lemma}
\begin{proof}
  Note that
  \begin{equation}\label{eq:hUhNinvchifestloc}
    |\hU[\hN^{-1}\chi(\psi)]|\leq |\hU(\ln\hN)|\cdot |\hN^{-1}\chi(\psi)|+|\hN^{-1}\hU\chi(\psi)|.
  \end{equation}
  Before estimating the second term on the right hand side of (\ref{eq:hUhNinvchifestloc}), note that
   \begin{equation}\label{eq:hUchipsiidenloc}
    \hU\chi(\psi)=\hU(\chi^{i})E_{i}(\psi)+\chi^{i}\hU E_{i}(\psi).
  \end{equation}
  On the other hand, (\ref{eq:dotchirelations}) yields
  \[
  \hU(\chi^{i})=\omega^{i}(\dotchi)-\chi^{k}A_{k}^{i}.
  \]
  This means that
  \begin{equation*}
    \begin{split}
      \hN^{-1}|\hU(\chi^{i})| \leq & \hN^{-1}|\dotchi|_{\bge_{\refer}}+\hN^{-1}|\chi|_{\bge_{\refer}}\textstyle{\sum}_{i,k}|A_{k}^{i}|\\
      \leq & C_{a}\ldr{\tau}^{\cweight}e^{\e_{\Spe}\tau}\left(1+e^{\eSpe\tau}\right)
    \end{split}
  \end{equation*}
  in $A^{+}(\g)$, where we appealed to Remark~\ref{remark:chiclvarrhodecay}, (\ref{eq:AikCzest}), 
  (\ref{eq:varrhominustauestimate}) and the assumptions. Moreover, the constant $C_{a}$ only depends on  $c_{\cweight,0}$, $(\bM,\bge_{\refer})$ 
  and a lower bound on $\theta_{0,-}$. The first term
  on the right hand side of (\ref{eq:hUhNinvchifestloc}) and the second term on the right hand side of (\ref{eq:hUchipsiidenloc}) can be estimated
  by similar arguments. Summarising yields (\ref{eq:hUhNinvchipsiestloc}). Next, we wish to apply this estimate with $\psi=\varrho$. Note, to this
  end, that (\ref{eq:bDvarrhobdEi}) and (\ref{eq:DeltavarrhorelvariationEi}) yield
  \begin{equation}\label{eq:Eivarrhobasestiloc}
    |E_{i}\varrho|\leq C_{a}\ldr{\tau}
  \end{equation}
  on $M_{-}$, where $C_{a}$ only depends on $c_{\cweight,0}$ and $(\bM,\bge_{\refer})$. Next, note that
  \[
  |\hU E_{i}(\varrho)|\leq |[\hU,E_{i}](\varrho)|+|E_{i}\hU(\varrho)|
  \leq C_{\rorel}|\hU(\varrho)|+\textstyle{\sum}_{k}|A_{i}^{k}|\cdot |E_{k}(\varrho)|+|E_{i}\hU(\varrho)|,
  \]
  where we appealed to (\ref{eq:hUEicomm}) and (\ref{eq:Aialphadef}). Due to (\ref{eq:hUvarrhoident}), (\ref{eq:rodivchiestimpr}) and
  (\ref{eq:eSpevarrhoeelowtaurelEi}),
  \[
  |\hU(\varrho)|\leq 1+e^{\eSpe\tau}
  \]
  on $M_{-}$. Moreover,
  \[
  |E_{i}\hU(\varrho)|=|E_{i}[\hN^{-1}\rodiv_{\bge_{\refer}}\chi]|\leq C_{\rorel}e^{\eSpe\tau}+C_{b}\ldr{\tau}^{2\cweight}e^{\eSpe\tau},
  \]
  where we appealed to (\ref{eq:Eirodivchiestimate}) and $C_{b}$ only depends on $c_{\cweight,0}$ and $(\bM,\bge_{\refer})$. Combining the
  above observations with (\ref{eq:AikCzest}) yields
  \begin{equation}\label{eq:hUEivarrhoestloc}
  |\hU E_{i}(\varrho)|\leq C_{a}
  \end{equation}
  on $M_{-}$, where $C_{a}$ only depends on $c_{\cweight,0}$ and $(\bM,\bge_{\refer})$. Combining 
  (\ref{eq:hUhNinvchipsiestloc}), (\ref{eq:Eivarrhobasestiloc}) and (\ref{eq:hUEivarrhoestloc}) yields (\ref{eq:hUhNinvchivarrhoestloc}). 
  Finally, the estimate (\ref{eq:hUhNinvchijzetajestloc}) follows by arguments similar to the above. 
\end{proof}

Next, we derive similar estimates for $X_{A}[\hN^{-1}\chi(\psi)]$. 

\begin{lemma}\label{lemma:XAwnfestloc}
  Assume that the conditions of Lemma~\ref{lemma:taurelvaryingbxEi} as well as the $(\cweight,1)$-supremum assumptions are satisfied. Let $\g$ 
  and $\bx_{\g}$ be as in Remark~\ref{remark:bgaconvtobxbgaloc}, and assume that $\bx_{0}=\bx_{\g}$. Then, if $\psi$ is a smooth function on $\bM\times I$,  
  \begin{align}
      |X_{A}[\hN^{-1}\chi(\psi)]| \leq & C_{a}\ldr{\tau}^{\cweight}e^{\e_{\Spe}\tau}\textstyle{\sum}_{i}|E_{i}(\psi)|+
      \frac{1}{2}\left(\textstyle{\sum}_{B,i}e^{-2\mu_{B}}|X_{B}E_{i}(\psi)|^{2}\right)^{1/2},\label{eq:XAhNinvchipsiestimatloc}\\
      |X_{A}[\hN^{-1}\chi^{j}\omega^{i}(\bD_{E_{j}}E_{i})]| \leq & C_{a}\ldr{\tau}^{\cweight}e^{\e_{\Spe}\tau},\label{eq:XAhNinvchijzetajestimatloc}\\
      |X_{A}[\hN^{-1}\chi(\varrho)]| \leq & C_{a}\ldr{\tau}^{2\cweight+1}e^{\e_{\Spe}\tau}\label{eq:XAhNinvchivarrhoestimatloc}
  \end{align}
  on $A^{+}(\g)$, where $C_{a}$ only depends on $c_{\cweight,1}$, $(\bM,\bge_{\refer})$ and a lower bound on $\theta_{0,-}$.
\end{lemma}
\begin{proof}
  To begin with,
  \begin{equation}\label{eq:XAhNinvchifloc}
    |X_{A}[\hN^{-1}\chi(\psi)]|\leq |X_{A}(\ln\hN)|\cdot |\hN^{-1}\chi(\psi)|+|\hN^{-1}X_{A}\chi(\psi)|.
  \end{equation}
  The first term on the right hand side can be estimated by appealing (\ref{eq:muminmainlowerbound}), (\ref{eq:varrhominustauestimate})
  and (\ref{eq:hNinvchipsiestloc}). This yields
  \[
  |X_{A}(\ln\hN)|\cdot |\hN^{-1}\chi(\psi)|\leq Ce^{\e_{\Spe}\tau}\textstyle{\sum}_{i}|E_{i}\psi|
  \]
  on $A^{+}(\g)$, where $C$ only depends on $c_{\cweight,0}$, $(\bM,\bge_{\refer})$ and a lower bound on $\theta_{0,-}$. 
  In order to estimate the second term on the right hand side of (\ref{eq:XAhNinvchifloc}), note that
  \begin{equation*}
    \begin{split}
      |\hN^{-1}X_{A}\chi(\psi)| \leq & \left(\textstyle{\sum}_{i}|\hN^{-1}E_{i}\chi(\psi)|^{2}\right)^{1/2}\\
      \leq & \left(\textstyle{\sum}_{i}|\hN^{-1}(\ml_{E_{i}}\chi)(\psi)|^{2}\right)^{1/2}
      +\left(\textstyle{\sum}_{i}|\hN^{-1}\chi E_{i}(\psi)|^{2}\right)^{1/2}.
    \end{split}
  \end{equation*}
  On the other hand,
  \begin{equation*}
    \begin{split}
      |\hN^{-1}\ml_{E_{i}}\chi|_{\bge_{\refer}} \leq & |\hN^{-1}\bD_{E_{i}}\chi|_{\bge_{\refer}}+|\hN^{-1}\bD_{\chi}E_{i}|_{\bge_{\refer}}
      \leq C_{a}\ldr{\tau}^{\cweight}e^{\e_{\Spe}\tau}
    \end{split}
  \end{equation*}
  on $A^{+}(\g)$, where $C_{a}$ only depends on $c_{\cweight,0}$, $(\bM,\bge_{\refer})$ and a lower bound on $\theta_{0,-}$. To obtain this estimate, 
  we appealed to Remark~\ref{remark:chiclvarrhodecay} and (\ref{eq:varrhominustauestimate}). Next, note that (\ref{eq:hNinvchipsiestloc}) yields
  \begin{equation*}
    \begin{split}
      |\hN^{-1}\chi E_{i}(\psi)| \leq & \frac{1}{2}\left(\textstyle{\sum}_{B}e^{-2\mu_{B}}|X_{B}E_{i}(\psi)|^{2}\right)^{1/2}.
    \end{split}
  \end{equation*}
  To summarise, (\ref{eq:XAhNinvchipsiestimatloc}) holds. The proof of (\ref{eq:XAhNinvchijzetajestimatloc}) is similar but less involved. 

  Next, applying (\ref{eq:XAhNinvchipsiestimatloc}) with $\psi=\varrho$, it is clear that we wish to estimate up to two derivatives of $\varrho$.
  To estimate one derivative of $\varrho$, it is sufficient to appeal to (\ref{eq:Eivarrhobasestiloc}). In order to obtain an estimate of the second
  order derivatives of $\varrho$, we appeal to Lemma~\ref{lemma:CkestofvarrhoEi}. 
\end{proof}

At this stage we return to (\ref{eq:dtausqdtauhUdiffestfs}).

\begin{lemma}\label{lemma:dtausqmdtauhUestimate}
  Assume that the conditions of Lemma~\ref{lemma:taurelvaryingbxEi} as well as the $(\cweight,1)$-supremum assumptions are satisfied. Let $\g$ 
  and $\bx_{\g}$ be as in Remark~\ref{remark:bgaconvtobxbgaloc}, and assume that $\bx_{0}=\bx_{\g}$. Then, if $\psi$ is a smooth function on $\bM\times I$
  and $\bcweight:=\max\{\cweight,1\}$, 
  \begin{equation}\label{eq:dtausqmdtauhUestimate}
    \begin{split}
      & |\d_{\tau}^{2}\psi-\d_{\tau}\hU\psi|\\
      \leq & C_{a}\ldr{\tau}e^{\e_{\Spe}\tau}|\d_{\tau}\hU\psi|
      +C_{a}\ldr{\tau}^{\bcweight+\cweight}e^{\e_{\Spe}\tau}\left(|\hU(\psi)|^{2}+\textstyle{\sum}_{A}e^{-2\mu_{A}}|X_{A}(\psi)|^{2}\right)^{1/2}\\
      & +C_{a}\ldr{\tau}^{\cweight}e^{\e_{\Spe}\tau}\textstyle{\sum}_{i}(|\hU E_{i}\psi|+|E_{i}\psi|)
      +C_{a}e^{\e_{\Spe}\tau}\left(\textstyle{\sum}_{B,i}e^{-2\mu_{B}}|X_{B}E_{i}(\psi)|^{2}\right)^{1/2}
    \end{split}
  \end{equation}
  on $A^{+}(\g)$, where $C_{a}$ only depends on $c_{\cweight,1}$, $(\bM,\bge_{\refer})$ and a lower bound on $\theta_{0,-}$. 
\end{lemma}
\begin{proof}
  Due to (\ref{eq:hprelestimate}), the first term in the parenthesis of the right hand side of (\ref{eq:dtausqdtauhUdiffestfs}) can be estimated by
  \[
  |h\d_{\tau}\hU\psi|\leq C_{a}\ldr{\tau}e^{\e_{\Spe}\tau}|\d_{\tau}\hU\psi|
  \]
  on $A^{+}(\g)$, where $C_{a}$  only depends on $c_{\cweight,0}$, $(\bM,\bge_{\refer})$ and a lower bound on
  $\theta_{0,-}$. Next, let us estimate $\d_{\tau}h$. Consider, to this end, (\ref{eq:hdef}). Combining this equality with (\ref{eq:hprelestimate})
  and (\ref{eq:dtauhNinvhNzeroest}) yields
  \[
  |\d_{\tau}h|\leq C\ldr{\tau}^{2\cweight}e^{\e_{\Spe}\tau}+\hN^{-1}\hN_{0}|\d_{\tau}[1-\hN^{-1}_{0}\d_{t}\tau]|
  \]
  on $A^{+}(\g)$, where $C$ only depends on $c_{\cweight,1}$, $(\bM,\bge_{\refer})$ and a lower bound on $\theta_{0,-}$. In order to estimate the last 
  term on the right hand side, we appeal to (\ref{eq:hUdtaudiscrinApl}). Due to this equality, we need to estimate the $\tau$-derivative of   
  \begin{equation}\label{eq:trickyparttodiff}
    \hN^{-1}\chi(\varrho)+\hN^{-1}\rodiv_{\bge_{\refer}}\chi
    =\hN^{-1}\chi(\varrho)+\textstyle{\sum}_{i}A^{i}_{i}+\hN^{-1}\chi^{j}\omega^{i}(\bD_{E_{j}}E_{i})
  \end{equation}
  at $(\bx_{0},t)$, where we appealed to (\ref{eq:hNinvrodivbgereferchiAiiformloc}) in the last step. In order to estimate the $\tau$-derivative of
  the first and last terms on the right hand side of (\ref{eq:trickyparttodiff}), it is sufficient to appeal to Lemmas~\ref{lemma:dhtaupsirelhUpsiloc},
  \ref{lemma:hUwnfestloc} and \ref{lemma:XAwnfestloc}. This yields
  \begin{align*}
    |\d_{\tau}[\hN^{-1}\chi(\varrho)](\bx_{0},t)| \leq & C_{a}\ldr{\tau(t)}^{\cweight+1}e^{\e_{\Spe}\tau(t)},\\
    |\d_{\tau}[\hN^{-1}\chi^{j}\omega^{i}(\bD_{E_{j}}E_{i})](\bx_{0},t)| \leq & C_{a}\ldr{\tau(t)}^{\cweight}e^{\e_{\Spe}\tau(t)}
  \end{align*}
  for $t\leq t_{0}$, 
  where $C_{a}$ only depends on $c_{\cweight,1}$, $(\bM,\bge_{\refer})$ and a lower bound on $\theta_{0,-}$. Next, in order to estimate the $\tau$-derivative
  of the second term on the right hand side of (\ref{eq:trickyparttodiff}), we appeal to Remark~\ref{remark:mWAhUAClestimates} and
  Lemma~\ref{lemma:dhtaupsirelhUpsiloc}. This yields
  \[
  |(\d_{\tau}A^{i}_{i})(\bx_{0},t)|\leq C_{a}\ldr{\tau(t)}^{2\cweight}e^{\e_{\Spe}\tau(t)}
  \]
  for $t\leq t_{0}$, where $C_{a}$ only depends on $c_{\cweight,1}$, $(\bM,\bge_{\refer})$ and a lower bound on $\theta_{0,-}$. Summing up the above estimates
  leads to the conclusion that if $\bcweight:=\max\{\cweight,1\}$, then 
  \[
  |\d_{\tau}h|\leq C_{a}\ldr{\tau}^{\bcweight+\cweight}e^{\e_{\Spe}\tau}
  \]
  on $A^{+}(\g)$, where $C_{a}$ only depends on $c_{\cweight,1}$, $(\bM,\bge_{\refer})$ and a lower bound on $\theta_{0,-}$. Next,
  Lemmas~\ref{lemma:dhtaupsirelhUpsiloc}, \ref{lemma:hUwnfestloc} and \ref{lemma:XAwnfestloc} yield
  \begin{equation*}
    \begin{split}
      |\d_{\tau}[\hN^{-1}\chi(\psi)]| \leq & C_{a}\ldr{\tau}^{\cweight}e^{\e_{\Spe}\tau}\textstyle{\sum}_{i}(|\hU E_{i}\psi|+|E_{i}\psi|)\\
      & +C_{a}e^{\e_{\Spe}\tau}\left(\textstyle{\sum}_{B,i}e^{-2\mu_{B}}|X_{B}E_{i}(\psi)|^{2}\right)^{1/2}
    \end{split}
  \end{equation*}  
  on $A^{+}(\g)$, where $C_{a}$ only depends on $c_{\cweight,1}$, $(\bM,\bge_{\refer})$ and a lower bound on $\theta_{0,-}$. Combining the above estimates
  with (\ref{eq:dtausqdtauhUdiffestfs}) and Lemma~\ref{lemma:dhtaupsirelhUpsiloc} yields the conclusion of the lemma. 
\end{proof}

At this point, we can combine (\ref{eq:dhtauhUbDbfAumhUsqetcloc}) and (\ref{eq:dtausqmdtauhUestimate}) in order to draw the following conclusion.

\begin{lemma}\label{lemma:dtausqmhUsqEbfIestimate}
  Fix $l\geq 1$, $\bfl_{1}$, $\cweight$, $\weight_{0}$ and $\weight$ as in Definition~\ref{def:supmfulassumptions}.
  Then, given that the assumptions of Lemma~\ref{lemma:taurelvaryingbxEi} as well as the $(\cweight,l)$-supremum assumptions are satisfied,
  assume (\ref{eq:coefflassumptions}) to hold. Let $L$ be defined by (\ref{eq:LuformulaEi}) and assume $u$ to be a smooth solution to $Lu=0$.
  Let $\g$ and $\bx_{\g}$ be as in Remark~\ref{remark:bgaconvtobxbgaloc}, and assume that $\bx_{0}=\bx_{\g}$. Then, for all $m=|\bfI|\leq l$, 
  \begin{equation}\label{eq:dtausqmhUsqEbfIestimate}
    |\d_{\tau}^{2}E_{\bfI}u-\hU^{2}E_{\bfI}u|\leq C_{a}\ldr{\tau}^{(m+2)\cweight+1}\ldr{\tau-\tau_{c}}^{3\iota_{b}/2}e^{\e_{\Spe}\tau}\me_{m+1}^{1/2}
  \end{equation}
  on $A^{+}_{c}(\g)$, where $C_{a}$ only depends on $c_{\cweight,l}$, $c_{\rocoeff,l}$, $m_{\ros}$, $d_{\a}$ (in case $\iota_{b}\neq 0$), $(\bM,\bge_{\refer})$ and
  a lower bound on $\theta_{0,-}$.
\end{lemma}
\begin{remark}
  Combining (\ref{eq:dtausqmhUsqEbfIestimate}) with (\ref{eq:rhsreplcfderloc}) yields
  \begin{equation}\label{eq:rhsreplcfderlocaux}
    \begin{split}
      & |-\d_{\tau}^{2}E_{\bfI}u+Z^{0}\d_{\tau} E_{\bfI}u+\hal E_{\bfI}u|\\
      \leq & C_{a}\ldr{\tau}^{(m+2)\cweight+1}\ldr{\tau-\tau_{c}}^{3\iota_{b}/2}e^{\e_{\Spe}\tau}\me_{m+1}^{1/2}
      +C_{b}\ldr{\tau}^{(m+1)\cweight}\ldr{\tau-\tau_{c}}^{3\iota_{b}/2}\me_{m-1}^{1/2}
    \end{split}    
  \end{equation}
  on $A^{+}_{c}(\g)$. Here, the second term on the right hand side vanishes in case $m=0$. Moreover, $C_{a}$ and $C_{b}$ only depend on $c_{\cweight,l}$,
  $c_{\rocoeff,l}$, $m_{\ros}$, $d_{\a}$ (in case $\iota_{b}\neq 0$), $(\bM,\bge_{\refer})$ and a lower bound on $\theta_{0,-}$.
\end{remark}
\begin{proof}
  Combining (\ref{eq:dhtauhUbDbfAumhUsqetcloc}), (\ref{eq:dhtauhUbDbfAuestloc}) and (\ref{eq:dtausqmdtauhUestimate}) yields the conclusion
  of the lemma. 
\end{proof}

In what follows, we use (\ref{eq:rhsreplcfderlocaux}) to derive estimates. However, it is convenient to simplify the expressions that appear on the
left hand side additionally. Introduce, to this end, 
\begin{equation}\label{eq:Zzerolochalloc}
Z^{0}_{\roloc}(t):=Z^{0}(\bx_{0},t),\ \ \
\hal_{\roloc}(t):=\hal(\bx_{0},t). 
\end{equation}
\index{$\a$Aa@Notation!Functions!$Z^{0}_{\roloc}$}%
\index{$\a$Aa@Notation!Functions!$\hal_{\roloc}$}%
With this notation, the following holds.
\begin{cor}\label{cor:rhsreplcfderfullylocal}
  Fix $l\geq 1$, $\bfl_{1}$, $\cweight$, $\weight_{0}$ and $\weight$ as in Definition~\ref{def:supmfulassumptions}.
  Then, given that the assumptions of Lemma~\ref{lemma:taurelvaryingbxEi} as well as the $(\cweight,l)$-supremum assumptions are satisfied,
  assume (\ref{eq:coefflassumptions}) to hold. Let $L$ be defined by (\ref{eq:LuformulaEi}) and assume $u$ to be a smooth solution to $Lu=0$.
  Let $\g$ and $\bx_{\g}$ be as in Remark~\ref{remark:bgaconvtobxbgaloc}, and assume that $\bx_{0}=\bx_{\g}$. Then, for all $m=|\bfI|\leq l$, 
  \begin{equation}\label{eq:rhsreplcfderfullylocal}
    \begin{split}
      & |-\d_{\tau}^{2}E_{\bfI}u+Z^{0}_{\roloc}\d_{\tau} E_{\bfI}u+\hal_{\roloc} E_{\bfI}u|\\
      \leq & C_{a}\ldr{\tau}^{(m+2)\cweight+1}\ldr{\tau-\tau_{c}}^{3\iota_{b}/2}e^{\e_{\Spe}\tau}\me_{m+1}^{1/2}
      +C_{b}\ldr{\tau}^{(m+1)\cweight}\ldr{\tau-\tau_{c}}^{3\iota_{b}/2}\me_{m-1}^{1/2}
    \end{split}    
  \end{equation}
  holds on $A^{+}_{c}(\g)$. Here, the second term on the right hand side vanishes in case $m=0$. Moreover, $C_{a}$ and $C_{b}$ only depend
  on $c_{\cweight,l}$, $c_{\rocoeff,l}$, $m_{\ros}$, $d_{\a}$ (in case $\iota_{b}\neq 0$), $(\bM,\bge_{\refer})$ and a lower bound on $\theta_{0,-}$.
\end{cor}
\begin{proof}
  Note, first of all, that (\ref{eq:rhsreplcfderlocaux}) holds. Next, note that (\ref{eq:coefflassumptions}) holds with $l=1$. Moreover,
  Definition~\ref{def:supmfulassumptions} yields a bound on the weighted $C^{1}$-norm of $q$. Combining these observations with
  (\ref{eq:chthexpressionhUlntheta}), (\ref{eq:hmcYzdefEi}) and (\ref{eq:ZzdefEi}) yields
  \begin{align}
    \|Z^{0}(\bx,t)-Z^{0}_{\roloc}(t)\| \leq & C_{a}\theta_{0,-}^{-1}\ldr{\tau(t)}^{\cweight}e^{\e_{\Spe}\tau(t)},\label{eq:ZzerominusZetazeroloc}\\
    \|\hal(\bx,t)-\hal_{\roloc}(t)\| \leq & C_{a}\theta_{0,-}^{-1}\ldr{\tau(t)}^{\cweight}e^{\e_{\Spe}\tau(t)}\label{eq:halminushalloc}
  \end{align}
  for all $(\bx,t)\in A^{+}(\g)$, where $C_{a}$ only depends on $c_{\cweight,1}$, $c_{\rocoeff,1}$, $m_{\ros}$, $(\bM,\bge_{\refer})$ and a lower bound on
  $\theta_{0,-}$. Combining these estimates with (\ref{eq:rhsreplcfderlocaux}) and (\ref{eq:dhtaupsihUpsiestloc}) yields the conclusion of the corollary. 
\end{proof}

\chapter{Energy estimates in causally localised regions}\label{chapter:energyestimatescausallyloc}

Due to the estimates of the previous chapter, we have a model equation for the asymptotic behaviour in $A^{+}_{c}(\g)$; cf. (\ref{eq:modelintrointro}).
The model equation is a system of second order ODE's. Since the only assumptions we make concerning the coefficients of this system is that they are
smooth and bounded, we cannot in general derive the asymptotic behaviour of solutions to the model equation. For this reason, we need to make assumptions
concerning the behaviour of solutions to the model equation and then compare these assumptions with the behaviour of solutions to the actual
equation. Since the model equation can be phrased as a first order system of ODE's, and since the behaviour of the corresponding solutions is completely
described by the associated flow, we phrase the assumptions in terms of the flow. We do so at the beginning of Section~\ref{section:loceqasymp};
cf. (\ref{eq:Phinormbasassloc}). Given these assumptions, we derive energy estimates in $A^{+}_{c}(\g)$ in
Theorem~\ref{thm:asgrowthofenergy}. In the end, we prove that the energy, up to polynomial factors, asymptotically behaves as well as we assume the
solutions to the model equation to behave. In order to improve the rate of growth/decay of the energy, we need to sacrifice derivatives. In fact, the loss
of derivatives typically tends to infinity as $\e_{\Spe}$ tends to $0$. In some situations, the functions $Z^{0}_{\roloc}$ and $\hal_{\roloc}$ converge in the
direction of the singularity. In that setting, if the convergence is fast enough, the asymptotic behaviour is characterized by a matrix $A_{0}$. In fact,
we can then prove estimates of the form (\ref{eq:Phinormbasassloc}), where $d_{A}$ and $\varpi_{A}$ can be calculated in terms of $A_{0}$;
$\varpi_{A}$ is the smallest real part of an eigenvalue of $A_{0}$ and $d_{A}+1$ is the largest dimension of a corresponding Jordan block. We justify these
statements in Section~\ref{section:approximationslocalisedenergies}.

\section{Localised equation and improved energy estimates}\label{section:loceqasymp}

Due to Corollary~\ref{cor:rhsreplcfderfullylocal}, we can improve the energy estimates in $A^{+}_{c}(\g)$. Introduce, to this end, the notation
\begin{equation}\label{eq:Psiihtwodef}
  \Psi_{1}:=E_{\bfI}u,\ \ \
  \Psi_{2}:=\d_{\tau} E_{\bfI}u,\ \ \
  h_{2}:=\d_{\tau}^{2}E_{\bfI}u-Z^{0}_{\roloc}\d_{\tau} E_{\bfI}u-\hal_{\roloc}E_{\bfI}u. 
\end{equation}
Then 
\begin{equation}\label{eq:localsystematbfxz}
  \d_{\tau}\Psi=A\Psi+H,
\end{equation}
where 
\begin{equation}\label{eq:PsiAHdef}
  \Psi:=\left(\begin{array}{c} \Psi_{1}\\ \Psi_{2}\end{array}\right),\ \ \
  A:=\left(\begin{array}{cc} 0 & \Id \\ \hal_{\roloc} & Z^{0}_{\roloc}\end{array}\right),\ \ \
  H:=\left(\begin{array}{c} 0\\ h_{2}\end{array}\right). 
\end{equation}
Let $\Phi$ be the flow associated with $A$. In other words, 
\begin{equation}\label{eq:Phidef}
  \d_{\tau}\Phi=A\Phi,\ \ \
  \Phi(\tau;\tau)=\Id. 
\end{equation}
Assume now that there are constants $C_{A}$, $d_{A}$ and $\varpi_{A}$ such that if $s_{1}\leq s_{2}\leq 0$, 
then 
\begin{equation}\label{eq:Phinormbasassloc}
 \|\Phi(s_{1};s_{2})\|\leq C_{A}\ldr{s_{2}-s_{1}}^{d_{A}}e^{\varpi_{A}(s_{1}-s_{2})}. 
\end{equation}
Clearly, $C_{A}$, $d_{A}$ and $\varpi_{A}$ depend on $\bx_{0}$. Fix $\tau_{c}\leq 0$ as before and introduce $\Xi(\tau):=e^{-\varpi_{A}(\tau-\tau_{c})}\Psi(\tau)$,
$\hA:=A-\varpi_{A}\mathrm{Id}$ and $\hH(\tau):=e^{-\varpi_{A}(\tau-\tau_{c})}H(\tau)$. Then 
\[
\d_{\tau}\Xi=\hA\Xi+\hH.
\]
Defining $\hPhi$ as in (\ref{eq:Phidef}) but with $A$ replaced by $\hA$ yields 
\[
\hPhi(\tau;\tau_{a})=e^{-\varpi_{A}(\tau-\tau_{a})}\Phi(\tau;\tau_{a}).
\]
In particular, 
\begin{equation}\label{eq:hPhinormbasassloc}
  \|\hPhi(s_{1};s_{2})\|\leq C_{A}\ldr{s_{2}-s_{1}}^{d_{A}}
\end{equation}
for all $s_{1}\leq s_{2}\leq 0$. On the other hand,
\[
\Xi(\bx,\tau)=\hPhi(\tau;\tau_{a})\Xi(\bx,\tau_{a})+\int_{\tau_{a}}^{\tau}\hPhi(\tau;s)\hH(\bx,s)ds. 
\]
In particular, 
\begin{equation}\label{eq:Psibasestloc}
|\Xi(\bx,\tau)|\leq \|\hPhi(\tau;\tau_{a})\|\cdot |\Xi(\bx,\tau_{a})|+\left|\int_{\tau_{a}}^{\tau}\|\hPhi(\tau;s)\|\cdot |\hH(\bx,s)|ds\right|; 
\end{equation}
note that we are mainly interested in the case that $\tau$ is smaller than $\tau_{a}$. 

Next, we improve the energy estimate in $A^{+}_{c}(\g)$. Recall, to this end, the notation introduced in (\ref{eq:mektaudefEi}) and
(\ref{eq:hGekdef}).

\begin{thm}\label{thm:asgrowthofenergy}
  Let $0\leq \cweight\in\ro$, $\weight_{0}=(0,\cweight)$ and $\weight=(\cweight,\cweight)$. Assume that the conditions of
  Lemma~\ref{lemma:taurelvaryingbxEi} are fulfilled. Let $\kappa_{0}$ be the smallest integer which is strictly larger than $n/2$;
  $\kappa_{1}=\kappa_{0}+1$; $\kappa_{1}\leq k\in\zo$; and $l=k+\kappa_{0}$. Assume the
  $(\cweight,k)$-supremum and the $(\cweight,l)$-Sobolev assumptions to be satisfied; and that there are constants $c_{\coeff,k}$ and
  $s_{\coeff,l}$ such that (\ref{eq:Sobcoefflassumptions}) holds and such that (\ref{eq:coefflassumptions}) holds with $l$ replaced by $k$.
  Let $\g$ and $\bx_{\g}$ be as in Remark~\ref{remark:bgaconvtobxbgaloc}, and assume that $\bx_{0}=\bx_{\g}$. Assume, finally, that
  (\ref{eq:theeqreformEi}) is satisfied with vanishing right hand side; and that if $A$ is defined by (\ref{eq:PsiAHdef}) and $\Phi$ is defined
  by (\ref{eq:Phidef}), then there are constants $C_{A}$, $d_{A}$ and $\varpi_{A}$ such that (\ref{eq:Phinormbasassloc}) holds.
  Let $c_{0}$ be defined by
  (\ref{eq:czCbdef}) and $\tc_{0}$ be defined by
  \begin{equation}\label{eq:tczdef}
    \tc_{0}:=c_{0}+1-1/n-\e_{\Spe}.
  \end{equation}
  Let $m_{0}$ be the smallest integer greater than or equal to 
  \[
  \max\left\{1,\frac{2\varpi_{A}+\tc_{0}}{2\e_{\Spe}}+\frac{1}{2}\right\}.
  \]
  Assuming $k\geq m_{0}$, the estimate 
  \begin{equation}\label{eq:melindassfslocfinalstmt}
    \begin{split}
      \me_{m}^{1/2} \leq & C_{m,a}\ldr{\tau-\tau_{c}}^{\kappa_{m,a}}\ldr{\tau}^{\lambda_{m,a}}e^{\varpi_{A}(\tau-\tau_{c})}\hGe_{m+\kappa_{0}}^{1/2}(\tau_{c})\\
      & +C_{m,b}\ldr{\tau-\tau_{c}}^{\kappa_{m,b}}\ldr{\tau}^{\lambda_{m,b}}e^{\varpi_{A}(\tau-\tau_{c})}\ldr{\tau_{c}}^{\zeta_{m}}
      \textstyle{\sum}_{j=1}^{m_{0}}e^{j\e_{\Spe}\tau_{c}}\hGe_{m+j+\kappa_{0}}^{1/2}(\tau_{c})
    \end{split}    
  \end{equation}
  holds on $A^{+}_{c}(\g)$ for $0\leq m\leq k-m_{0}$, where $C_{m,a}$ and $C_{m,b}$ only depend on $s_{\cweight,l}$, $s_{\coeff,l}$, $c_{\cweight,k}$, $c_{\coeff,k}$,
  $m_{\ros}$, $d_{\a}$ (in case $\iota_{b}\neq 0$), $C_{A}$, $d_{A}$, $(\bM,\bge_{\refer})$ and a lower bound on $\theta_{0,-}$; $\kappa_{m,a}$ and $\kappa_{m,b}$
  only depend on $d_{A}$, $n$, $m$ and $k$; $\lambda_{m,a}$, $\lambda_{m,b}$ and $\zeta_{m}$ only depend on $\cweight$, $n$, $m$ and $k$; and $\hGe_{l}$ is
  introduced in (\ref{eq:hGekdef}). Moreover, $\kappa_{0,a}=\kappa_{0,b}=d_{A}$ and $\lambda_{0,a}=\lambda_{0,b}=0$. 
\end{thm}
\begin{remark}
  One particular consequence of the statement is that the growth of $|u_{\tau}|^{2}+|u|^{2}$ is exactly the one you would expect by replacing the equation
  with the system of ODE's given by (\ref{eq:modelintrointro}). 
\end{remark}
\begin{proof}
  Note, to begin with, that the conditions of Proposition~\ref{prop:EnergyEstimateSobolevAssumptions} are fulfilled. Thus (\ref{eq:TheEnergyEstimate})
  holds. Combining this estimate with (\ref{eq:mekinfwtwoestgeneral}) and the fact that $l=k+\kappa_{0}$ yields
  \begin{equation}\label{eq:mekinfwtwoestasymp}
    \begin{split}
      \|\me_{j}(\cdot,\tau)\|_{\infty,w_{2}} \leq & C_{a}e^{c_{0}(\tau_{c}-\tau)}\hE_{j+\kappa_{0}}(\tau_{c};\tau_{c})\\
      & +C_{b}\ldr{\tau}^{2\a_{j,n}\cweight}\ldr{\tau-\tau_{c}}^{2\b_{j,n}}e^{c_{0}(\tau_{c}-\tau)}\hE_{k_{j}}(\tau_{c};\tau_{c})
    \end{split}    
  \end{equation}
  for all $\tau\leq \tau_{c}$ and all $j\leq k$. Here $k_{j}:=\max\{\kappa_{1},j+\kappa_{0}-1\}$; $c_{0}$ is the constant defined by (\ref{eq:czCbdef});
  $\a_{j,n}$ and $\b_{j,n}$ only depend on $n$ and $j$; $C_{a}$ only depends on $c_{\cweight,\kappa_{0}}$, $c_{\coeff,1}$, $m_{\ros}$, $l$, $d_{\a}$ (in case
  $\iota_{b}\neq 0$), $(\bM,\bge_{\refer})$ and a lower bound on $\theta_{0,-}$; and $C_{b}$ only depends on $s_{\cweight,l}$, $s_{\coeff,l}$,
  $c_{\cweight,\kappa_{1}}$, $c_{\coeff,\kappa_{1}}$, $m_{\ros}$, $d_{\a}$ (in case $\iota_{b}\neq 0$), $(\bM,\bge_{\refer})$ and a lower bound on $\theta_{0,-}$.
  Combining (\ref{eq:mekinfwtwoestasymp}) with (\ref{eq:lnwalongAplusgamma}) and the fact that $q\geq n\e_{\Spe}$ (cf. Remark~\ref{remark:qlwbd}) yields
  \begin{equation}\label{eq:mekhGelzest}
    \begin{split}
      \me_{j} \leq & C_{a}e^{\tc_{0}(\tau_{c}-\tau)}\hGe_{j+\kappa_{0}}(\tau_{c})
      +C_{b}\ldr{\tau}^{2\a_{j,n}\cweight}\ldr{\tau-\tau_{c}}^{2\b_{j,n}}e^{\tc_{0}(\tau_{c}-\tau)}\hGe_{k_{j}}(\tau_{c})
    \end{split}    
  \end{equation}
  on $A^{+}_{c}(\g)$, where the constants have the same dependence as the constants with the same names appearing in (\ref{eq:mekinfwtwoestasymp});
  $\tc_{0}$ is defined by (\ref{eq:tczdef}); and the notation $\hGe_{l}$ is introduced in (\ref{eq:hGekdef}). Here $A^{+}_{c}(\g)$ denotes the subset
  of $A^{+}(\g)$ corresponding to $t\leq t_{c}$. Let 
  \[
  \mG_{j}:=\frac{1}{2}\textstyle{\sum}_{|\bfI|=j}\left[|\d_{\tau}E_{\bfI}u|^{2}+|E_{\bfI}u|^{2}\right].
  \]
  Due to (\ref{eq:dhtaupsihUpsiestloc}),
  \begin{equation}\label{eq:mGjmEjdominance}
    \mG_{j}\leq C\ldr{\tau-\tau_{c}}^{3\iota_{b}}\me_{j}
  \end{equation}
  on $M_{-}$, where $C$ only depends on $C_{\rorel}$ and $(\bM,\bge_{\refer})$. In what follows, it is also of interest to keep in mind that
  \begin{equation}\label{eq:mGkhGelSobembedding}
    \|\mG_{j}(\cdot,\tau)\|_{\infty}\leq C_{a}\ldr{\tau-\tau_{c}}^{3\iota_{b}}\hGe_{j+\kappa_{0}}(\tau)
  \end{equation}
  for all $\tau\leq\tau_{c}$, where $C_{a}$ only depends on $C_{\rorel}$, $j$ and $(\bM,\bge_{\refer})$, and we appealed to
  (\ref{eq:dhtaupsihUpsiestloc}). 

  Due to (\ref{eq:mekhGelzest}),
  \begin{equation}\label{eq:meketazest}
    \begin{split}
      \me_{m} \leq & C_{a}e^{2\nu_{0}(\tau-\tau_{c})}\hGe_{m+\kappa_{0}}(\tau_{c})
      +C_{b}\ldr{\tau}^{2d_{m}}\ldr{\tau-\tau_{c}}^{2c_{m}}e^{2\nu_{0}(\tau-\tau_{c})}\hGe_{k_{m}}(\tau_{c})
    \end{split}    
  \end{equation}
  on $A^{+}_{c}(\g)$ for all $m\leq k$. Here
  \begin{equation}\label{eq:dketazdef}
    d_{m}:=\a_{m,n}\cweight,\ \ \  c_{m}:=\b_{m,n},\ \ \ \nu_{0}:=-\tc_{0}/2,
  \end{equation}
  where $\tc_{0}$ is defined by (\ref{eq:tczdef}). Moreover, the remaining constants have the same dependence as in the case of
  (\ref{eq:mekinfwtwoestasymp}). For technical reasons, it will be convenient to deteriorate the estimate (\ref{eq:meketazest}) slightly.
  Let $p_{0}$ be the largest integer $\leq 0$ such that
  \begin{equation}\label{eq:pzdef}
    p_{0}\e_{\Spe}-\e_{\Spe}/2\leq\nu_{0}-\varpi_{A}
  \end{equation}
  (note that $m_{0}$ introduced in the statement of the theorem is related to $p_{0}$ via $m_{0}=-p_{0}+1$) and define $\varkappa_{j}$ by
  \begin{equation}\label{eq:varkappajdef}
    \varkappa_{0}:=\varpi_{A}+p_{0}\e_{\Spe}-\e_{\Spe}/2,\ \ \
    \varkappa_{j}:=\varkappa_{0}+j\e_{\Spe}.
  \end{equation}
  Then $\varkappa_{0}\leq \nu_{0}$, so that (\ref{eq:meketazest}) holds with $\nu_{0}$ replaced by $\varkappa_{0}$. Moreover, for all $j$,
  \begin{equation}\label{eq:varkappajminusvarpiAlowerbound}
    |\varkappa_{j}-\varpi_{A}|\geq\e_{\Spe}/2.
  \end{equation}
  Let us now assume, inductively, that there are functions $f_{m,j}$ and $g_{m,j}$ that
  are polynomials in $\ldr{\tau}$ and $\ldr{\tau-\tau_{c}}$ with non-negative coefficients such that 
  \begin{equation}\label{eq:melindassloc}
    \me_{m}^{1/2}\leq f_{m,j}e^{\varpi_{A}(\tau-\tau_{c})}+g_{m,j}e^{\varkappa_{j}(\tau-\tau_{c})}    
  \end{equation}
  on $A^{+}_{c}(\g)$ for $m\leq k-j$. Here the properties of the functions $f_{m,j}$ and $g_{m,j}$ remain to be determined. Due to
  (\ref{eq:meketazest}), we know this estimate to hold for $j=0$ with $f_{m,0}=0$ and
  \begin{equation}\label{eq:gmzerodefinition}
    g_{m,0}(\tau)=c_{m,0}\ldr{\tau}^{d_{m}}\ldr{\tau-\tau_{c}}^{c_{m}}\hGe_{p_{m}}^{1/2}(\tau_{c}),
  \end{equation}
  where $p_{m}:=\max\{\kappa_{1},m+\kappa_{0}\}$ and $c_{m,0}$ only depends on $s_{\cweight,l}$, $s_{\coeff,l}$, $c_{\cweight,\kappa_{1}}$, $c_{\coeff,\kappa_{1}}$,
  $m_{\ros}$, $d_{\a}$ (in case $\iota_{b}\neq 0$), $(\bM,\bge_{\refer})$ and a lower bound on $\theta_{0,-}$. The idea of the proof is to  improve
  (\ref{eq:melindassloc}) inductively. The improvement consists in the increase of $\varkappa_{j}$. However, there is additional structure in the
  estimate which will become apparent in what follows.

  \textbf{General observations.} In the argument below, we appeal to the following two observations without further comment. First, if
  $0\leq a,b\in\ro$, $0<\b\in\ro$ and $\tau\leq \tau_{c}\leq 0$, then $\ldr{\tau}\leq 2^{1/2}\ldr{\tau-\tau_{c}}\ldr{\tau_{c}}$ and 
  \begin{equation}\label{eq:ldrtauldrtaumtaucexpest}
    \ldr{\tau}^{a}\ldr{\tau-\tau_{c}}^{b}e^{\b\tau}
    \leq 2^{a/2}\ldr{\tau_{c}}^{a}e^{\b\tau_{c}}\ldr{\tau-\tau_{c}}^{a+b}e^{\b(\tau-\tau_{c})}\leq C\ldr{\tau_{c}}^{a}e^{\b\tau_{c}},
  \end{equation}
  where $C$ only depends on $a$, $b$ and $\b$. In particular, if $f$ is a polynomial in $\ldr{\tau}$ and $\ldr{\tau-\tau_{c}}$ with non-negative
  coefficients, then
  \[
  f(\tau)e^{\b\tau}\leq Cf(\tau_{c})e^{\b\tau_{c}}
  \]
  for $\tau\leq\tau_{c}\leq 0$, where $C$ only depends on $\b$ and the powers of $\ldr{\tau}$ and $\ldr{\tau-\tau_{c}}$ appearing in $f$.
  Second, similar arguments imply that if $0<\b\in\ro$ and $f$ is as above, then
  \[
  \int_{\tau}^{\tau_{c}}f(s)e^{\b s}ds\leq Cf(\tau_{c})e^{\b\tau_{c}}
  \]
  for $\tau\leq\tau_{c}\leq 0$, where $C$ only depends on $\b$ and the powers of $\ldr{\tau}$ and $\ldr{\tau-\tau_{c}}$ appearing in $f$. 

  \textbf{Prototype improvement.} In order to improve (\ref{eq:melindassloc}), note that for $m\leq k$, the estimate
  (\ref{eq:rhsreplcfderfullylocal}) yields
  \begin{equation}\label{eq:Hestfirststep}
    |H|\leq \pi_{m}(\tau)[e^{\e_{\Spe}\tau}\me_{m+1}^{1/2}(\tau)+\upsilon_{m}\me_{m-1}^{1/2}(\tau)]
  \end{equation}
  on $A^{+}_{c}(\g)$, where $|\bfI|=m$ in the definition of $H$. Here $\upsilon_{m}=0$ if $m=0$ and $\upsilon_{m}=1$ if $m\geq 1$. Moreover, 
  \[
  \pi_{m}(\tau):=C_{a}\ldr{\tau}^{(m+2)\cweight+1}\ldr{\tau-\tau_{c}}^{3\iota_{b}/2},
  \]
  where $C_{a}\geq 1$ only depends on $c_{\cweight,k}$, $c_{\rocoeff,k}$, $m_{\ros}$, $d_{\a}$ (in case $\iota_{b}\neq 0$), $(\bM,\bge_{\refer})$ and a lower bound on
  $\theta_{0,-}$. 

  Assume that (\ref{eq:melindassloc}) holds for some $j\geq 0$ and all $m\leq k-j$, and that either $m=0$ or that an improved estimate holds for $m-1$;
  i.e., that (\ref{eq:melindassloc}) holds with $m$ replaced by $m-1$ and $j$ replaced by $j+1$. Combining these assumptions with (\ref{eq:Hestfirststep})
  yields
  \begin{equation*}
    \begin{split}
      |\hH| \leq &  \pi_{m}(\tau)[e^{\e_{\Spe}\tau}f_{m+1,j}(\tau)+\upsilon_{m}f_{m-1,j+1}(\tau)]\\
      & +\pi_{m}(\tau)[g_{m+1,j}(\tau)e^{\e_{\Spe}\tau_{c}}+\upsilon_{m}g_{m-1,j+1}(\tau)]e^{(\varkappa_{j+1}-\varpi_{A})(\tau-\tau_{c})}
    \end{split}
  \end{equation*}
  on $A^{+}_{c}(\g)$. Combining this estimate with (\ref{eq:hPhinormbasassloc}) and (\ref{eq:Psibasestloc}) yields
  \begin{equation}\label{eq:mGmonehalfest}
    \begin{split}
      \mG_{m}^{1/2} \leq & F_{m,j+1}(\tau)e^{\varpi_{A}(\tau-\tau_{c})}+G_{m,j+1}(\tau)e^{\varkappa_{j+1}(\tau-\tau_{c})}
    \end{split}
  \end{equation}
  on $A^{+}_{c}(\g)$, where $F_{m,j+1}$ and $G_{m,j+1}$ are defined as follows (recall that (\ref{eq:varkappajminusvarpiAlowerbound}) holds). 

  \textit{Case 1.} If $\varkappa_{j+1}-\varpi_{A}\leq -\e_{\Spe}/2$, then
  \begin{align}
      F_{m,j+1}(\tau) := & C_{a}\ldr{\tau-\tau_{c}}^{d_{A}}\hGe^{1/2}_{m+\kappa_{0}}(\tau_{c})
      +C_{b}\ldr{\tau-\tau_{c}}^{d_{A}}\pi_{m}(\tau_{c})f_{m+1,j}(\tau_{c})e^{\e_{\Spe}\tau_{c}}\label{eq:Fjpomcaseone}\\
      & +\upsilon_{m}C_{c}\ldr{\tau-\tau_{c}}^{d_{A}+1}\pi_{m}(\tau)f_{m-1,j+1}(\tau),\nonumber\\
      G_{m,j+1}(\tau) := & C_{d}\ldr{\tau-\tau_{c}}^{d_{A}}\pi_{m}(\tau)[g_{m+1,j}(\tau)e^{\e_{\Spe}\tau_{c}}+\upsilon_{m}g_{m-1,j+1}(\tau)],\label{eq:Gjpomcaseone}
  \end{align}  
  where $C_{a}$ only depends on $C_{A}$, $\bDlnhNsup$, $m$ and $(\bM,\bge_{\refer})$; $C_{b}$ only depends on $C_{A}$, $\cweight$, $m$, $n$, $\e_{\Spe}$
  and the powers of $\ldr{\tau}$ and $\ldr{\tau-\tau_{c}}$ appearing in $f_{m+1,j}$; $C_{c}$ only depends on $C_{A}$, $m$ and $n$; and
  $C_{d}$ only depends on $C_{A}$, $\e_{\Spe}$, $m$ and $n$. Moreover, we appealed to (\ref{eq:mGkhGelSobembedding}). 
  
  \textit{Case 2.} If $\varkappa_{j+1}-\varpi_{A}\geq \e_{\Spe}/2$, then $G_{m,j+1}=0$ and 
  \begin{align}
      F_{m,j+1}(\tau) := & C_{a}\ldr{\tau-\tau_{c}}^{d_{A}}\hGe^{1/2}_{m+\kappa_{0}}(\tau_{c})
      +C_{b}\ldr{\tau-\tau_{c}}^{d_{A}}\pi_{m}(\tau_{c})f_{m+1,j}(\tau_{c})e^{\e_{\Spe}\tau_{c}}\label{eq:Fjpomcasetwo}\\
      & +\upsilon_{m}C_{c}\ldr{\tau-\tau_{c}}^{d_{A}+1}\pi_{m}(\tau)f_{m-1,j+1}(\tau)\nonumber\\
      & +C_{d}\ldr{\tau-\tau_{c}}^{d_{A}}\pi_{m}(\tau_{c})[g_{m+1,j}(\tau_{c})e^{\e_{\Spe}\tau_{c}}+\upsilon_{m}g_{m-1,j+1}(\tau_{c})]\nonumber
  \end{align}  
  where $C_{a}$ only depends on $C_{A}$, $\bDlnhNsup$, $m$ and $(\bM,\bge_{\refer})$; $C_{b}$ only depends on $C_{A}$, $\cweight$, $m$, $n$, $\e_{\Spe}$
  and the powers of $\ldr{\tau}$ and $\ldr{\tau-\tau_{c}}$ appearing in $f_{m+1,j}$; $C_{c}$ only depends on $C_{A}$, $m$ and $n$; and
  $C_{d}$ only depends on $C_{A}$, $\cweight$, $\e_{\Spe}$, $m$, $n$, the powers of $\ldr{\tau}$ and $\ldr{\tau-\tau_{c}}$ appearing in $g_{m+1,j}$
  and (if $m\geq 1$) the powers of $\ldr{\tau}$ and $\ldr{\tau-\tau_{c}}$ appearing in $g_{m-1,j+1}$. 

  In order to take the step from the estimate (\ref{eq:mGmonehalfest}) to an improvement of (\ref{eq:melindassloc}), note that if $|\bfI|=m$,
  (\ref{eq:dhtaubDbfAuminushUbDbfAuloc}) yields
  \begin{equation}\label{eq:hUumGzmeoneestimate}
    |\hU E_{\bfI}u|\leq |\d_{\tau}E_{\bfI}u|+|\d_{\tau}E_{\bfI}u-\hU E_{\bfI}u|\leq |\d_{\tau}E_{\bfI}u|
    +C_{a}\ldr{\tau}\ldr{\tau-\tau_{c}}^{3\iota_{b}/2}e^{\e_{\Spe}\tau}\me_{m+1}^{1/2}
  \end{equation}
  on $A^{+}_{c}(\g)$, where $C_{a}$ only depends on $c_{\cweight,0}$, $(\bM,\bge_{\refer})$ and a lower bound on $\theta_{0,-}$. Moreover, 
  \begin{equation}\label{eq:emmuAXaumeoneest}
    e^{-\mu_{A}}|X_{A}E_{\bfI}u|\leq C_{a}e^{\e_{\Spe}\tau}\ldr{\tau-\tau_{c}}^{3\iota_{b}/2}\me_{m+1}^{1/2}
  \end{equation}
  on $A^{+}_{c}(\g)$, where we appealed to (\ref{eq:muminmainlowerbound}) and (\ref{eq:varrhominustauestimate}), and $C_{a}$ only depends
  on $c_{\cweight,0}$, $(\bM,\bge_{\refer})$ and a lower bound on $\theta_{0,-}$. Thus
  \begin{equation*}
    \begin{split}
      \me_{m}^{1/2} \leq & \upsilon_{m}\me_{m-1}^{1/2}+2^{-1/2}\textstyle{\sum}_{|\bfI|=m}[|\hU E_{\bfI}u|+\sum_{A}e^{-\mu_{A}}|X_{A}E_{\bfI}u|\\
        & +\iota_{a}|E_{\bfI}u|+\iota_{b}\ldr{\tau-\tau_{c}}^{-3/2}|E_{\bfI}u|]\\
      \leq & \upsilon_{m}\me_{m-1}^{1/2}+K_{a}\mG_{m}^{1/2}+K_{b}\ldr{\tau}\ldr{\tau-\tau_{c}}^{3\iota_{b}/2}e^{\e_{\Spe}\tau}\me_{m+1}^{1/2}
    \end{split}
  \end{equation*}
  on $A^{+}_{c}(\g)$, where $K_{a}$ only depends on $m$ and $n$; and $K_{b}$ only depends on $c_{\cweight,0}$, $m$, $(\bM,\bge_{\refer})$ and a lower bound
  on $\theta_{0,-}$. Combining the above estimates yields
  \[
  \me_{m}^{1/2}\leq \fb_{m,j+1}e^{\varpi_{A}(\tau-\tau_{c})}+\bge_{m,j+1}e^{\varkappa_{j+1}(\tau-\tau_{c})},
  \]
  where
  \begin{align}
    \fb_{m,j+1}(\tau) = & \upsilon_{m}f_{m-1,j+1}(\tau)+K_{a}F_{m,j+1}(\tau)+L_{b}\ldr{\tau_{c}}f_{m+1,j}(\tau_{c})e^{\e_{\Spe}\tau_{c}},\label{eq:fmjponedef}\\
    \bge_{m,j+1}(\tau) = & \upsilon_{m}g_{m-1,j+1}(\tau)+K_{a}G_{m,j+1}(\tau)+K_{b}\ldr{\tau}\ldr{\tau-\tau_{c}}^{3\iota_{b}/2}g_{m+1,j}(\tau)e^{\e_{\Spe}\tau_{c}},
    \label{eq:gmjponedef}
  \end{align}
  where $K_{a}$ only depends on $m$ and $n$; and $K_{b}$ only depends on $c_{\cweight,0}$, $(\bM,\bge_{\refer})$ and a lower bound on $\theta_{0,-}$;
  and $L_{b}$ only depends on $c_{\cweight,0}$, $m$, $(\bM,\bge_{\refer})$, a lower bound on $\theta_{0,-}$ and the powers of $\ldr{\tau}$ and
  $\ldr{\tau-\tau_{c}}$ appearing in $f_{m+1,j}$. Thus (\ref{eq:melindassloc}) holds with $j$ replaced by $j+1$, where $f_{m,j+1}$ and $g_{m,j+1}$
  are determined as follows:

  \textit{Case 1.} If $\varkappa_{j+1}-\varpi_{A}\leq -\e_{\Spe}/2$,
  \begin{align}
    f_{m,j+1}(\tau) := & \wp_{0,0}(\tau)\hGe_{m+\kappa_{0}}^{1/2}(\tau_{c})+\wp_{m,+}(\tau)f_{m+1,j}(\tau_{c})e^{\e_{\Spe}\tau_{c}}
    +\upsilon_{m}\wp_{m,-}(\tau)f_{m-1,j+1}(\tau),\label{eq:fmjponedefinition}\\
    g_{m,j+1}(\tau) := & \wp_{m}(\tau)(\upsilon_{m}g_{m-1,j+1}(\tau)+g_{m+1,j}(\tau)e^{\e_{\Spe}\tau_{c}}),\label{eq:gmjponedefinition}
  \end{align}
  where $\wp_{0,0}(\tau):=C_{a}\ldr{\tau-\tau_{c}}^{d_{A}}$,
  \begin{align*}
    \wp_{m,+}(\tau) := & C_{b}\ldr{\tau-\tau_{c}}^{d_{A}}\ldr{\tau_{c}}^{(m+2)\cweight+1},\ \ \
    \wp_{m,-}(\tau) := C_{c}\ldr{\tau-\tau_{c}}^{d_{A}+1+3\iota_{b}/2}\ldr{\tau}^{(m+2)\cweight+1}\\
    \wp_{m}(\tau) := & C_{d}\ldr{\tau}^{(m+2)\cweight+1}\ldr{\tau-\tau_{c}}^{d_{A}+3\iota_{b}/2}
  \end{align*}
  and $C_{a}$ only depends on $C_{A}$, $\bDlnhNsup$, $m$ and $(\bM,\bge_{\refer})$; $C_{b}$ only depends on $c_{\cweight,k}$, $c_{\coeff,k}$, $C_{A}$, $d_{A}$,
  $m_{\ros}$, $d_{\a}$ (in case $\iota_{b}\neq 0$), $(\bM,\bge_{\refer})$, a lower bound on $\theta_{0,-}$ and the powers of $\ldr{\tau}$ and
  $\ldr{\tau-\tau_{c}}$ appearing in $f_{m+1,j}$; $C_{c}$ only depends on $c_{\cweight,k}$, $c_{\coeff,k}$, $C_{A}$, $m_{\ros}$, $d_{\a}$ (in case
  $\iota_{b}\neq 0$), $(\bM,\bge_{\refer})$ and a lower bound on $\theta_{0,-}$; and $C_{d}$ only depends on $C_{A}$, $c_{\cweight,k}$, $c_{\rocoeff,k}$, $m_{\ros}$,
  $d_{\a}$ (in case $\iota_{b}\neq 0$), $(\bM,\bge_{\refer})$ and a lower bound on $\theta_{0,-}$. To be able to use (\ref{eq:fmjponedefinition})
  to obtain an estimate of $f_{m,j}$, we first need to determine the dependence of the powers of $\ldr{\tau}$ and $\ldr{\tau-\tau_{c}}$ appearing
  in $f_{m+1,j}$. However, due to (\ref{eq:fmjponedefinition}) and the fact that $f_{m,0}=0$, it can inductively be verified that
  these powers only depend on $d_{A}$, $m$, $\cweight$, $j$ and $n$. Since $m$ and $j$ are bounded by $k$, the dependence of the powers can thus be
  ignored.

  \textit{Case 2.} If $\varkappa_{j+1}-\varpi_{A}=\e_{\Spe}/2$, then $g_{m,j+1}=0$ and 
  \begin{equation}\label{eq:fmjponedefinitionfinal}
    \begin{split}
      f_{m,j+1}(\tau) := & \wp_{0,0}(\tau)\hGe_{m+\kappa_{0}}^{1/2}(\tau_{c})+\wp_{m,+}(\tau)f_{m+1,j}(\tau_{c})e^{\e_{\Spe}\tau_{c}}
      +\upsilon_{m}\wp_{m,-}(\tau)f_{m-1,j+1}(\tau)\\
      & +\wp_{m,\fin}(\tau)g_{m+1,j}(\tau_{c})e^{\e_{\Spe}\tau_{c}},
    \end{split}
  \end{equation}
  where $\wp_{0,0}$, $\wp_{m,+}$ and $\wp_{m,-}$ are defined in case 1. Moreover,
  \[
  \wp_{m,\fin}(\tau):=C_{e}\ldr{\tau-\tau_{c}}^{d_{A}}\ldr{\tau_{c}}^{(m+2)\cweight+1},
  \]
  where $C_{e}$ only depends on $c_{\cweight,k}$, $c_{\rocoeff,k}$, $C_{A}$, $d_{A}$, $m_{\ros}$, $d_{\a}$ (in case $\iota_{b}\neq 0$), $(\bM,\bge_{\refer})$, a
  lower bound on $\theta_{0,-}$ and the powers of $\ldr{\tau}$ and $\ldr{\tau-\tau_{c}}$ appearing in $g_{m+1,j}$. To be able to use
  (\ref{eq:fmjponedefinitionfinal}) to obtain an estimate of $f_{m,j+1}$, we first need to determine the dependence of the powers of
  $\ldr{\tau}$ and $\ldr{\tau-\tau_{c}}$ appearing in $g_{m+1,j}$. However, due to (\ref{eq:gmjponedefinition}) and the fact that
  (\ref{eq:gmzerodefinition}) holds, it can inductively be verified that these powers only depend on $d_{A}$, $m$, $\cweight$, $j$ and $n$. Since
  $m$ and $j$ are bounded by $k$, the dependence of the powers can thus be ignored.

  \textbf{Conclusions.} Our starting point is the estimate (\ref{eq:melindassloc}). We know this estimate to hold on $A^{+}_{c}(\g)$ with $j=0$, where
  $\varkappa_{0}$ is given by (\ref{eq:varkappajdef}). Moreover, we know that if it holds for some $j$ and
  $\varkappa_{j}-\varpi_{A}\leq -\e_{\Spe}/2$, we can improve it. The improvement consists in a replacement of $\varkappa_{j}$ by
  $\varkappa_{j+1}$. By induction, we obtain (\ref{eq:melindassloc}) on $A^{+}_{c}(\g)$ for all $m,j\geq 0$ such that $m+j\leq k$, as long as
  $\varkappa_{j}-\varpi_{A}\leq \e_{\Spe}/2$. Assuming $k$ to be large enough (corresponding to $k\geq m_{0}$ in the statement of the theorem),
  $\varkappa_{j}-\varpi_{A}$ will, for $j=m_{0}$, equal $\e_{\Spe}/2$. At this stage, the improvements terminate, since the second term on the
  right hand side of (\ref{eq:melindassloc}) then vanishes. This leads to the desired conclusion, modulo the detailed structure of the polynomials
  involved in the estimates. 

  \textbf{The structure of the polynomials, step 1.} Assuming $\varkappa_{j+1}-\varpi_{A}\leq -\e_{\Spe}/2$, (\ref{eq:gmjponedefinition}) yields
  \[
  g_{m,j+1}=\textstyle{\sum}_{r=0}^{m}\left(\prod_{p=r}^{m}\wp_{p}\right)g_{r+1,j}e^{\e_{\Spe}\tau_{c}},
  \]
  assuming $m+j+1\leq k$. Combining this observation with (\ref{eq:gmzerodefinition}) yields
  \begin{equation}\label{eq:gmjestimatecaseone}
    g_{m,j}(\tau)\leq \mQ_{m,j}(\tau)e^{j\e_{\Spe}\tau_{c}}\hGe_{m+j+\kappa_{0}}^{1/2}(\tau_{c})
  \end{equation}
  for $j\geq 1$ and $m+j\leq k$. Here
  \[
  \mQ_{m,j}(\tau):=K_{m,j}\ldr{\tau-\tau_{c}}^{r_{m,j}}\ldr{\tau}^{s_{m,j}},
  \]
  where $K_{m,j}$ only depends on $s_{\cweight,l}$, $s_{\coeff,l}$, $c_{\cweight,k}$, $c_{\coeff,k}$, $C_{A}$, $m_{\ros}$, $d_{\a}$ (in case $\iota_{b}\neq 0$),
  $(\bM,\bge_{\refer})$ and a lower bound on $\theta_{0,-}$; $r_{m,j}$ only depends on $m$, $n$, $j$ and $d_{A}$; and $s_{m,j}$ only depends on $m$, $n$,
  $j$ and $\cweight$. 

  \textbf{The structure of the polynomials, step 2.} Next, we estimate $f_{m,j+1}$ under the assumption that $\varkappa_{j+1}-\varpi_{A}\leq -\e_{\Spe}/2$.
  To this end, we appeal to (\ref{eq:fmjponedefinition}). Since $f_{m,0}=0$, (\ref{eq:fmjponedefinition}) yields
  \[
  f_{m,1}(\tau)\leq \wp_{0,0}\hGe_{m+\kappa_{0}}^{1/2}(\tau_{c})+\upsilon_{m}\mR_{m,1}(\tau)\hGe_{m+\kappa_{0}-1}^{1/2}(\tau_{c})
  \]
  for $m\leq k-1$, where
  \[
  \mR_{m,1}(\tau):=L_{m,1}\ldr{\tau}^{p_{m,1}}\ldr{\tau-\tau_{c}}^{q_{m,1}}.
  \]
  Here $L_{m,1}$ only depends on $c_{\cweight,k}$, $c_{\coeff,k}$, $C_{A}$, $m_{\ros}$, $d_{\a}$ (in case $\iota_{b}\neq 0$), $(\bM,\bge_{\refer})$ and a lower
  bound on $\theta_{0,-}$; $p_{m,1}$ only depends on $m$ and $\cweight$; and $q_{m,1}$ only depends on $m$ and $d_{A}$.

  In general, for $j\geq 1$, an inductive argument yields the conclusion that
  \begin{equation}\label{eq:fmjgeneralestimate}
    \begin{split}
      f_{m,j}(\tau) \leq & \wp_{0,0}\hGe_{m+\kappa_{0}}^{1/2}(\tau_{c})+\upsilon_{m}\mR_{m,j}(\tau)\hGe_{m+\kappa_{0}-1}^{1/2}(\tau_{c})\\
      & +\upsilon_{j-1}\ldr{\tau-\tau_{c}}^{d_{A}}\mS_{m,j}(\tau_{c})\textstyle{\sum}_{l=1}^{j}e^{l\e_{\Spe}\tau_{c}}\hGe_{m+l+\kappa_{0}}^{1/2}(\tau_{c})\\
      & +\upsilon_{m}\upsilon_{j-1}\mR_{m,j}(\tau)\mS_{m,j}(\tau_{c})\textstyle{\sum}_{l=0}^{j-1}e^{(l+1)\e_{\Spe}\tau_{c}}\hGe_{m+l+\kappa_{0}}^{1/2}(\tau_{c})
    \end{split}
  \end{equation}  
  for $m+j\leq k$ (in fact, a better estimate holds, but the corresponding improvement does not result in an improvement of the final result), where 
  \begin{equation}\label{eq:mQmjmRmjdef}
    \mR_{m,j}(\tau) := L_{m,j}\ldr{\tau}^{p_{m,j}}\ldr{\tau-\tau_{c}}^{q_{m,j}},\ \ \
    \mS_{m,j}(\tau) := M_{m,j}\ldr{\tau}^{r_{m,j}}.
  \end{equation}  
  Here $L_{m,j}$ and $M_{m,j}$ only depend on $c_{\cweight,k}$, $c_{\coeff,k}$, $C_{A}$, $d_{A}$, $m_{\ros}$, $d_{\a}$ (in case $\iota_{b}\neq 0$),
  $(\bM,\bge_{\refer})$ and a lower bound on $\theta_{0,-}$; $p_{m,j}$ and $r_{m,j}$ only depend on $m$, $j$ and $\cweight$; and $q_{m,j}$ only depends
  on $m$, $j$ and $d_{A}$. More precisely, if the estimate (\ref{eq:fmjgeneralestimate}) holds for $m+j\leq k$, where $j\geq 1$, then it is
  preserved by the formula (\ref{eq:fmjponedefinition}) (for $m+j+1\leq k$). This observation is of importance in the next step, since the first
  three terms on the right hand side of (\ref{eq:fmjponedefinitionfinal}) coincide with the right hand side of (\ref{eq:fmjponedefinition}).

  \textit{Case 2.} Say now that $\varkappa_{j+1}-\varpi_{A}=\e_{\Spe}/2$; this happens for $j=-p_{0}$, where $p_{0}$ is introduced in connection with
  (\ref{eq:pzdef}). Since $g_{m,j+1}=0$, we only need to estimate $f_{m,j+1}$. To this end we appeal to (\ref{eq:fmjponedefinitionfinal}). However,
  there are two cases to consider. Either $p_{0}=0$ or $p_{0}\leq -1$. In case $p_{0}=0$, (\ref{eq:fmjponedefinitionfinal}) can be used to deduce that
  (\ref{eq:fmjgeneralestimate}) and (\ref{eq:mQmjmRmjdef}) still hold with $j=1$, but with the following modifications: First, $\upsilon_{j-1}$ should be
  removed from the right hand side of (\ref{eq:fmjgeneralestimate}). Second, $L_{m,1}$ and $M_{m,1}$ are, additionally, allowed to depend on $s_{\cweight,l}$
  and $s_{\coeff,l}$; and $p_{m,1}$ and $r_{m,1}$ are allowed to depend on $n$.
  Assume now that $p_{0}\leq -1$. Then we know that (\ref{eq:fmjgeneralestimate}) and (\ref{eq:mQmjmRmjdef}) hold and that $j\geq 1$. In the case
  of $m=0$, (\ref{eq:fmjponedefinitionfinal}), (\ref{eq:gmjestimatecaseone}) and (\ref{eq:fmjgeneralestimate}) yield the conclusion that
  (\ref{eq:fmjgeneralestimate}) holds with $m=0$ and $j$ replaced with $j+1$. However, $M_{0,j+1}$ is, additionally, allowed to depend on $s_{\cweight,l}$
  and $s_{\coeff,l}$; and $r_{0,j+1}$ is additionally allowed to depend on $n$. Next, by an inductive argument, it can be demonstrated that if $m\geq 1$ and
  $m+j+1\leq k$, then (\ref{eq:fmjgeneralestimate}) holds with $j$ replaced by $j+1$. The proof of this is largely the same as the proof of
  (\ref{eq:fmjgeneralestimate}). The only difference is the contribution (in the inductive argument) of the last term on the right hand side of
  (\ref{eq:fmjponedefinitionfinal}). However, appealing to (\ref{eq:gmjestimatecaseone}), it can be estimated that
  \begin{equation*}
    \begin{split}
      \wp_{m,\fin}(\tau)g_{m+1,j}(\tau_{c})e^{\e_{\Spe}\tau_{c}}
      \leq & C_{e}\ldr{\tau-\tau_{c}}^{d_{A}}\ldr{\tau_{c}}^{\g_{m,j+1}}e^{(j+1)\e_{\Spe}\tau_{c}}\hGe^{1/2}_{m+j+1+\kappa_{0}}(\tau_{c}),
    \end{split}
  \end{equation*}
  where $C_{e}$ only depends on $s_{\cweight,l}$, $s_{\coeff,l}$, $c_{\cweight,k}$, $c_{\coeff,k}$, $C_{A}$, $d_{A}$, $m_{\ros}$, $d_{\a}$ (in case $\iota_{b}\neq 0$),
  $(\bM,\bge_{\refer})$ and a lower bound on $\theta_{0,-}$; and $\g_{m,j}$ only depends on $m$, $n$, $j$ and $\cweight$. This expression is of a form
  compatible with (\ref{eq:fmjgeneralestimate}). However, the dependence of the constants has to be modified; $L_{m,j}$ and $M_{m,j}$ are, additionally,
  allowed to depend on $s_{\cweight,l}$ and $s_{\coeff,l}$; and $p_{m,j}$ and $r_{m,j}$ are allowed to depend on $n$. This finishes the inductive argument
  and the desired conclusion follows.
\end{proof}

\section{Approximations}\label{section:approximationslocalisedenergies}

Sometimes, the behaviour of $A$, introduced in (\ref{eq:PsiAHdef}), simplifies asymptotically. In particular, $A$ could converge to a constant
matrix. In that setting, it is of interest to make the following observation.

\begin{lemma}\label{lemma:PhiestfromPhizest}
  Let $A_{i}\in C^{0}[I,\Mn{k}{\ro}$, $i=0,1$, where $I$ is an open interval containing $(-\infty,0]$. Let $A=A_{0}+A_{1}$ and $\Phi$ be defined as
  in (\ref{eq:Phidef}). Let $\Phi_{0}$ be defined as in (\ref{eq:Phidef}), where $A$ is replaced by $A_{0}$. Assume that there are constants $d_{A}$,
  $C_{0}$ and $\varpi_{A}$ such that if $s_{1}\leq s_{2}\leq 0$, then
  \begin{equation}\label{eq:Phizeronormbasassloc}
    \|\Phi_{0}(s_{1};s_{2})\|\leq C_{0}\ldr{s_{2}-s_{1}}^{d_{A}}e^{\varpi_{A}(s_{1}-s_{2})}.
  \end{equation}
  Let $\xi(s):=\ldr{s}^{d_{A}}\|A_{1}(s)\|$ and assume $\|\xi\|_{1}:=\|\xi\|_{L^{1}(-\infty,0]}<\infty$. Then
  \begin{equation}\label{eq:Phinormbasasslocder}
    \|\Phi(s_{1};s_{2})\|\leq C_{B}\ldr{s_{2}-s_{1}}^{d_{A}}e^{\varpi_{A}(s_{1}-s_{2})},
  \end{equation}
  where $C_{B}$ only depends on $C_{0}$ and $\|\xi\|_{1}$. 
\end{lemma}
\begin{proof}
  Introducing $\hA_{0}:=A_{0}-\varpi_{A}\mathrm{Id}$, the associated flow $\hPhi_{0}$ satisfies an estimate analogous to (\ref{eq:Phizeronormbasassloc}),
  with $\varpi_{A}$ set to zero; cf. the argument leading to (\ref{eq:hPhinormbasassloc}). Let $\hA:=A-\varpi_{A}\mathrm{Id}$, and consider a solution
  to $\dot{x}=\hA x$. Then
  \[
  x(\tau)=\hPhi_{0}(\tau;\tau_{0})x(\tau_{0})+\int_{\tau_{0}}^{\tau}\hPhi_{0}(\tau;s)A_{1}(s)x(s)ds,
  \]
  so that, for all $\tau\leq \tau_{0}\leq 0$,
  \[
  |x(\tau)|\leq C_{0}\ldr{\tau-\tau_{0}}^{d_{A}}|x(\tau_{0})|+C_{0}\int_{\tau}^{\tau_{0}}\ldr{\tau-s}^{d_{A}}\|A_{1}(s)\|\cdot |x(s)|ds.
  \]
  Introducing $\zeta(\tau):=\ldr{\tau-\tau_{0}}^{-d_{A}}|x(\tau)|$, it follows that
  \[
  \zeta(\tau)\leq C_{0}\zeta(\tau_{0})+C_{0}\int_{\tau}^{\tau_{0}}\ldr{s-\tau_{0}}^{d_{A}}\|A_{1}(s)\| \zeta(s)ds.
  \]
  A Gr\"{o}nwall's lemma argument yields the conclusion that
  \[
  \zeta(\tau)\leq C_{B}\zeta(\tau_{0}),\ \ \
  |x(\tau)|\leq C_{B}\ldr{\tau-\tau_{0}}^{d_{A}}|x(\tau_{0})|,
  \]
  where $C_{B}$ only depends on $C_{0}$ and $\|\xi\|_{1}$. Thus, for $s_{1}\leq s_{2}\leq 0$,
  \[
  \|\hPhi(s_{1};s_{2})\|\leq C_{B}\ldr{s_{2}-s_{1}}^{d_{A}},\ \ \
  \|\Phi(s_{1};s_{2})\|\leq C_{B}\ldr{s_{2}-s_{1}}^{d_{A}}e^{\varpi_{A}(s_{1}-s_{2})},
  \]
  where $\hPhi$ is the flow associated with $\hA$. 
\end{proof}

One particular case of interest is when $A$ converges to a constant matrix. Before stating the relevant result, it is convenient to introduce
the following notation; cf. \cite[Definition~4.3, p.~47]{finallinsys}.

\begin{definition}\label{def:SpRspdef}
  Given $A\in\Mn{k}{\co}$, let $\Spe A$ denote the set of eigenvalues of $A$. Moreover, let
  \[
  \varpi_{\max}(A):=\sup\{\mathrm{Re}\lambda\ |\ \lambda\in \Spe A\},\ \ \
  \varpi_{\min}(A):=\inf\{\mathrm{Re}\lambda\ |\ \lambda\in \Spe A\}.
  \]
  In addition, if $\varpi\in \{\mathrm{Re}\lambda\ |\ \lambda\in \Spe A\}$, then $d_{\max}(A,\varpi)$ is defined to be the largest dimension of a Jordan
  block corresponding to an eigenvalue of $A$ with real part $\varpi$. 
\end{definition}

\begin{cor}\label{cor:Asymptoticallyconstant}
  Let $A\in C^{0}[I,\Mn{k}{\ro}]$, where $I$ is an open interval containing $(-\infty,0]$. Assume that there is an $A_{0}\in \Mn{k}{\ro}$
  such that $A(s)\rightarrow A_{0}$ as $s\rightarrow -\infty$. Let $\varpi_{A}=\varpi_{\min}(A_{0})$ and $d_{A}:=d_{\max}(A_{0},\varpi_{A})-1$.
  Let $\xi(s):=\ldr{s}^{d_{A}}\|A(s)-A_{0}\|$. If $\|\xi\|_{1}:=\|\xi\|_{L^{1}(-\infty,0]}<\infty$,
  \[
  \|\Phi(s_{1};s_{2})\|\leq C_{A}\ldr{s_{2}-s_{1}}^{d_{A}}e^{\varpi_{A}(s_{1}-s_{2})}
  \]
  for all $s_{1}\leq s_{2}\leq 0$, where $C_{A}$ only depends on $A_{0}$ and $\|\xi\|_{1}$. 
\end{cor}
\begin{proof}
  The statement is an immediate consequence of Lemma~\ref{lemma:PhiestfromPhizest} and the fact that
  \[
  \|e^{A_{0}(s_{1}-s_{2})}\|\leq C_{0}\ldr{s_{1}-s_{2}}^{d_{A}}e^{\varpi_{A}(s_{1}-s_{2})}
  \]
  for all $s_{1}\leq s_{2}\leq 0$, where $d_{A}$ and $\varpi_{A}$ are defined as in the statement of the corollary and
  $C_{0}$ only depends on $A_{0}$.
\end{proof}

\chapter{Deriving asymptotics}\label{chapter:derivingasymptotics}

In order to derive detailed asymptotics, we need to make stronger assumptions than the ones made in the previous chapter. In the present chapter, we
therefore assume $Z^{0}_{\roloc}$ and $\hal_{\roloc}$ to converge exponentially. In that setting, we can replace the model equation with a constant
coefficient equation. For solutions to the latter equation, we of course know what the asymptotics are. However, even though we can hope to extract the
leading order behaviour from the constant coefficient equation, at a lower level, the error terms might begin to dominate. At the beginning of
Section~\ref{section:detailedasymptoticsderas}, we therefore introduce terminology that makes it possible to quantify the level to which solutions to the
constant coefficient equation describe the asymptotics of solutions to the actual equation. Moreover, we state and prove a general result concerning the
asymptotics of solutions to equations of the form $\xi_{\tau}=B\xi+H$, where $B$ is a matrix and $H$ is a vector valued function satisfying appropriate
asymptotic estimates. Given this result, we are then in a position to derive the leading order asymptotics of $u$ and $\hU u$ in $A^{+}(\g)$, where $u$ is a
solution to the actual equation; cf. Theorem~\ref{thm:leadingorderasymptoticszero} below. Before proceeding to the asymptotics of the higher order
derivatives, we need to derive a model equation for them. This is the subject of the beginning of
Section~\ref{section:asymptoticshigherorderderivatives}. The cause of the difficulties is that the commutator of $\hU$ and $E_{i}$ cannot be ignored. On
the other hand, there is a hierarchy in the sense that one can derive the asymptotics up to a certain order, and then the correction terms (relative
to the constant coefficient model equation for the zeroth order spatial derivatives) appearing in the equation for the order above can be calculated in
terms of the coefficients, the geometry and the lower order asymptotics. Note, in particular, that in order to derive the leading order asymptotics for
the higher order derivatives, we only need to assume that $Z^{0}$ and $\hal$ converge along the causal curve $\g$. We do not need to assume that the
spatial derivatives of these coefficients converge along the causal curve. Given the model equation for the higher order spatial derivatives, we
derive the asymptotics using an inductive argument on the order of the spatial derivatives; cf. Theorem~\ref{thm:leadingorderasymptoticszeroho} below.

\section{Detailed asymptotics}\label{section:detailedasymptoticsderas}

In the situation considered in Corollary~\ref{cor:Asymptoticallyconstant}, more detailed asymptotics can be derived in case $A$ converges to $A_{0}$
exponentially. In order to state the relevant result, we first need to introduce additional terminology; cf. \cite[Definition~4.7, p.~48]{finallinsys}.

\begin{definition}\label{def:defofgeneigensp}
  Let $1\leq k\in\zo$, $B\in\Mn{k}{\co}$ and $P_{B}(X)$ be the characteristic polynomial of $B$. Then
  \[
  P_{B}(X)=\prod_{\lambda\in\Spe B}(X-\lambda)^{k_{\lambda}},
  \]
  where $1\leq k_{\lambda}\in\zo$. Moreover, given $\lambda\in \Spe B$, the \textit{generalised eigenspace of $B$ corresponding
  to $\lambda$}, denoted $E_{\lambda}$, is defined by
  \begin{equation}\label{eq:Elambdadef}
    E_{\lambda}:=\ker (B-\lambda\Id_{k})^{k_{\lambda}},
  \end{equation}
  where $\Id_{k}$ denotes the $k\times k$-dimensional identity matrix. If $J\subseteq\ro$ is an interval, then the
  $J$-\textit{generalised eigenspace of} $B$, denoted $E_{B,J}$, is the subspace of $\cn{k}$ defined to be the direct sum of the generalised eigenspaces
  of $B$ corresponding to eigenvalues with real parts belonging to $J$ (in case there are no eigenvalues with real part belonging to $J$, then $E_{B,J}$
  is defined to be $\{0\}$). Finally, given $0<\b\in\ro$, the \textit{first generalised eigenspace in the $\b$, $B$-decomposition of $\cn{k}$}, denoted
  $E_{B,\b}$, is defined to be $E_{B,J_{\b}}$, where $J_{\b}:=(\varpi-\b,\varpi]$ and $\varpi:=\varpi_{\max}(B)$; cf. Definition~\ref{def:SpRspdef}.
\end{definition}
\begin{remark}\label{remark:EBJreal}
  In case $B\in\Mn{k}{\ro}$, the vector spaces $E_{B,J}$ have bases consisting of vectors in $\rn{k}$. The reason for this is that if $\lambda$ is an
  eigenvalue of $B$ with $\mathrm{Re}\lambda\in J$, then $\lambda^{*}$ (the complex conjugate of $\lambda$) is an eigenvalue of $B$  with
  $\mathrm{Re}\lambda^{*}\in J$. Moreover, if $v\in E_{\lambda}$, then $v^{*}\in E_{\lambda^{*}}$. Combining bases of $E_{\lambda}$ and $E_{\lambda^{*}}$, we
  can thus construct a basis of the direct sum of these two vector spaces which consists of vectors in $\rn{k}$. 
\end{remark}

Before turning to the particular equations of interest here, it is convenient to make a technical observation concerning systems of ODE's.

\begin{lemma}\label{lemma:asymptoticsmodelODE}
  Let $B\in\Mn{k}{\ro}$ and $H\in C^{\infty}(I,\rn{k})$, where $I$ is an open interval containing $(-\infty,0]$. Let $\xi\in C^{\infty}(I,\rn{k})$
  be a solution to
  \begin{equation}\label{eq:modelODEas}
    \xi_{\tau}=B\xi+H.
  \end{equation}
  Let $\varpi_{B}:=\varpi_{\min}(B)$, $\b>0$ and assume that there are constants $C_{H}>0$ and $\eta_{H}\geq 0$ such that
  \[
  |H(\tau)|\leq C_{H}\ldr{\tau-\tau_{c}}^{\eta_{H}}e^{(\varpi_{B}+\b)(\tau-\tau_{c})}
  \]
  for all $\tau\leq\tau_{c}$ and some $\tau_{c}\leq 0$. Let $J_{a}:=[\varpi_{B},\varpi_{B}+\b)$, $J_{b}:=[\varpi_{B}+\b,\infty)$, $E_{a}:=E_{B,J_{a}}$
  and $E_{b}:=E_{B,J_{b}}$; cf. Definition~\ref{def:defofgeneigensp}. Then there is a unique division of $\xi$ as $\xi=\xi_{a}+\xi_{b}$, where
  $\xi_{a}\in C^{\infty}(I,E_{a})$ and $\xi_{b}\in C^{\infty}(I,E_{b})$. Moreover, there is a unique $\xi_{\infty,a}\in E_{a}$ such that
  \begin{equation}\label{eq:ximinuseBtauxiinftya}
    \begin{split}
      |\xi(\tau)-e^{B(\tau-\tau_{c})}\xi_{\infty,a}| \leq & C_{B}\ldr{\tau-\tau_{c}}^{\eta_{B}}e^{(\varpi_{B}+\b)(\tau-\tau_{c})}|\xi_{b}(\tau_{c})|\\
      & +KC_{H}\ldr{\tau-\tau_{c}}^{\eta_{H}+\eta_{B}}e^{(\varpi_{B}+\b)(\tau-\tau_{c})}
    \end{split}
  \end{equation}  
  for all $\tau\leq\tau_{c}$, where $K$ only depends on $B$, $\eta_{H}$ and $\b$; and $C_{B}$ and $\eta_{B}$ only depend on $B$. In addition,
  $\xi_{\infty,a}\in\rn{k}$ and there is a $\xi_{\infty}\in\rn{k}$, given by $\xi_{\infty}=\xi_{\infty,a}+\xi_{b}(\tau_{c})$, such that 
  \begin{equation}\label{eq:ximinuseBtauxiinfty}
    |\xi(\tau)-e^{B(\tau-\tau_{c})}\xi_{\infty}|\leq KC_{H}\ldr{\tau-\tau_{c}}^{\eta_{H}+\eta_{B}}e^{(\varpi_{B}+\b)(\tau-\tau_{c})}
  \end{equation}
  for all $\tau\leq\tau_{c}$, where $K$ and $\eta_{B}$ have the same dependence as in (\ref{eq:ximinuseBtauxiinftya}). Finally,
  \begin{equation}\label{eq:xiinftyaxiinftyestimate}
    |\xi_{\infty,a}|\leq |\xi_{a}(\tau_{c})|+KC_{H},\ \ \
    |\xi_{\infty}|\leq C_{B}|\xi(\tau_{c})|+KC_{H},
  \end{equation}
  where $K$ and $C_{B}$ have the same dependence as in (\ref{eq:ximinuseBtauxiinftya}). 
\end{lemma}
\begin{remark}
  Due to Remark~\ref{remark:EBJreal}, $\xi_{a}$ and $\xi_{b}$ are $\rn{k}$-valued.
\end{remark}
\begin{proof}
  Note that $\cn{k}$ is the direct sum of the generalised eigenspaces of $B$. Given a vector $v\in\cn{k}$, there are thus uniquely determined
  $v_{\lambda}\in E_{\lambda}$, $\lambda\in\Spe(B)$, such that
  \begin{equation}\label{eq:vgeneigenvaluedecompositionODE}
    v=\textstyle{\sum}_{\lambda\in\Spe(B)}v_{\lambda};
  \end{equation}
  here $E_{\lambda}$ is defined by (\ref{eq:Elambdadef}). In particular, we can write $H$
  as a sum of functions $H_{\lambda}$, $\lambda\in\Spe(B)$, where $H_{\lambda}$ is a smooth function which takes its values in $E_{\lambda}$. Since
  $B$ maps $E_{\lambda}$ into $E_{\lambda}$, the equation (\ref{eq:modelODEas}) can be decomposed into
  \[
  \d_{\tau}\xi_{\lambda}=B\xi_{\lambda}+H_{\lambda},
  \]
  where the definition of $\xi_{\lambda}$ is analogous to the definition of $H_{\lambda}$. In particular,
  \[
  \d_{\tau}(e^{-B(\tau-\tau_{c})}\xi_{\lambda})=e^{-B(\tau-\tau_{c})}H_{\lambda}.
  \]
  Let $\tau_{a}\leq\tau_{b}\leq\tau_{c}$ and integrate this equality from $\tau_{a}$ to $\tau_{b}$. This yields
  \begin{equation}\label{eq:integratedequationlambdaODE}
    e^{-B(\tau_{b}-\tau_{c})}\xi_{\lambda}(\tau_{b})-e^{-B(\tau_{a}-\tau_{c})}\xi_{\lambda}(\tau_{a})
    =\int_{\tau_{a}}^{\tau_{b}}e^{-B(\tau-\tau_{c})}H_{\lambda}(\tau)d\tau.
  \end{equation}
  However, the right hand side can be estimated by 
  \begin{equation*}
    \begin{split}
      \left|\int_{\tau_{a}}^{\tau_{b}}e^{-B(\tau-\tau_{c})}H_{\lambda}(\tau)d\tau\right| \leq &
      \int_{\tau_{a}}^{\tau_{b}}C_{\lambda}\ldr{\tau-\tau_{c}}^{k_{\lambda}-1}e^{-\mathrm{Re}{\lambda}(\tau-\tau_{c})}|H_{\lambda}(\tau)|d\tau\\
      \leq & K_{B}C_{H}
      \int_{\tau_{a}}^{\tau_{b}}\ldr{\tau-\tau_{c}}^{\eta_{H}+k_{\lambda}-1}e^{(\varpi_{B}+\beta-\mathrm{Re}\lambda)(\tau-\tau_{c})}d\tau,
    \end{split}
  \end{equation*}
  where $K_{B}$ only depends on $B$ and $k_{\lambda}$ is the algebraic multiplicity of $\lambda$. Let $S_{a}$ be the set of $\lambda\in\Spe(B)$ such that
  $\roRe(\lambda)\in J_{a}$, and let $S_{b}$ be the set of $\lambda\in\Spe(B)$ be such that $\roRe(\lambda)\in J_{b}$. Then $\xi_{a}$ and $\xi_{b}$,
  defined in the statement of the theorem, can be written
  \[
  \xi_{a}=\textstyle{\sum}_{\lambda\in S_{a}}\xi_{\lambda},\ \ \
  \xi_{b}=\textstyle{\sum}_{\lambda\in S_{b}}\xi_{\lambda}.
  \]
  Using the fact that $\varpi_{B}+\beta-\mathrm{Re}\lambda\geq \b_{\rem}>0$ for all $\lambda\in S_{a}$, we conclude that
  \begin{equation}\label{eq:xiaCauchyconvergencelambda}
    \begin{split}
      & \left|e^{-B(\tau_{b}-\tau_{c})}\xi_{\lambda}(\tau_{b})-e^{-B(\tau_{a}-\tau_{c})}\xi_{\lambda}(\tau_{a})\right|\\
      \leq & KC_{H}\ldr{\tau_{b}-\tau_{c}}^{\eta_{H}+k_{\lambda}-1}e^{(\varpi_{B}+\beta-\mathrm{Re}\lambda)(\tau_{b}-\tau_{c})}
    \end{split}    
  \end{equation}
  for all $\tau_{a}\leq\tau_{b}\leq\tau_{c}$ and $\lambda\in S_{a}$, where $K$ only depends on $B$, $\eta_{H}$ and $\b$. Thus, for $\lambda\in S_{a}$,
  the limit
  \begin{equation}\label{eq:xilambdainftydef}
    \xi_{\lambda,\infty}:=\lim_{\tau\rightarrow-\infty}e^{-B(\tau-\tau_{c})}\xi_{\lambda}(\tau)
  \end{equation}
  exists. Moreover, letting $\tau_{a}$ tend to $-\infty$ and choosing $\tau_{b}=\tau$ in (\ref{eq:xiaCauchyconvergencelambda}) yields the conclusion
  that 
  \begin{equation}\label{eq:xiaCauchyconvergencelambdalimit}
    \begin{split}
      \left|e^{-B(\tau-\tau_{c})}\xi_{\lambda}(\tau)-\xi_{\lambda,\infty}\right|
      \leq & KC_{H}\ldr{\tau-\tau_{c}}^{\eta_{H}+k_{\lambda}-1}e^{(\varpi_{B}+\beta-\mathrm{Re}\lambda)(\tau-\tau_{c})}
    \end{split}    
  \end{equation}
  for all $\tau\leq \tau_{c}$ and $\lambda\in S_{a}$, where $K$ has the same dependence as in the case of (\ref{eq:xiaCauchyconvergencelambda}). Thus
  \begin{equation*}
    \begin{split}
      & \left|\xi_{\lambda}(\tau)-e^{B(\tau-\tau_{c})}\xi_{\lambda,\infty}\right|\\
      \leq & C_{\lambda}\ldr{\tau-\tau_{c}}^{k_{\lambda}-1}e^{\mathrm{Re}\lambda (\tau-\tau_{c})}
      KC_{H}\ldr{\tau-\tau_{c}}^{\eta_{H}+k_{\lambda}-1}e^{(\varpi_{B}+\beta-\mathrm{Re}\lambda)(\tau-\tau_{c})}
    \end{split}    
  \end{equation*}
  for all $\tau\leq \tau_{c}$ and $\lambda\in S_{a}$. Summing up over all $\lambda\in S_{a}$ yields
  \begin{equation*}
    \begin{split}
      \left|\xi_{a}(\tau)-e^{B(\tau-\tau_{c})}\xi_{a,\infty}\right|
      \leq & KC_{H}\ldr{\tau-\tau_{c}}^{\eta_{H}+\eta_{B}}e^{(\varpi_{B}+\beta)(\tau-\tau_{c})}
    \end{split}    
  \end{equation*}
  for $\tau\leq\tau_{c}$, where $\xi_{a,\infty}:=\textstyle{\sum}_{\lambda\in S_{a}}\xi_{\lambda,\infty}$, $\eta_{B}$ only depends on $B$ and $K$ has the same
  dependence as in the case of (\ref{eq:xiaCauchyconvergencelambda}). Letting $\tau=\tau_{c}$ in this estimate yields
  \begin{equation}\label{eq:xiainftyestimate}
    \left|\xi_{a,\infty}\right|\leq |\xi_{a}(\tau_{c})|+KC_{H}.
  \end{equation}
  Thus the first estimate in (\ref{eq:xiinftyaxiinftyestimate}) holds. Next, letting $\tau_{b}=\tau_{c}$ and $\tau_{a}=\tau$ in
  (\ref{eq:integratedequationlambdaODE}) yields
  \[
  \xi_{\lambda}(\tau)=e^{B(\tau-\tau_{c})}\xi_{\lambda}(\tau_{c})-\int_{\tau}^{\tau_{c}}e^{B(\tau-s)}H_{\lambda}(s)ds.
  \]
  In particular,
  \[
  |\xi_{\lambda}(\tau)|\leq C_{\lambda}\ldr{\tau-\tau_{c}}^{k_{\lambda}-1}e^{\mathrm{Re}\lambda (\tau-\tau_{c})}|\xi_{\lambda}(\tau_{c})|
  +\int_{\tau}^{\tau_{c}}C_{\lambda}\ldr{\tau-s}^{k_{\lambda}-1}e^{\mathrm{Re}\lambda (\tau-s)}|H_{\lambda}(s)|ds.
  \]
  Due to the assumptions and the definition of $S_{b}$, it follows that
  \[
  |\xi_{b}(\tau)|\leq K_{B}\ldr{\tau-\tau_{c}}^{\eta_{B}}e^{(\varpi_{B}+\b)(\tau-\tau_{c})}|\xi_{b}(\tau_{c})|
  +K_{B}C_{H}\ldr{\tau-\tau_{c}}^{\eta_{H}+\eta_{B}}e^{(\varpi_{B}+\b)(\tau-\tau_{c})}
  \]
  for all $\tau\leq \tau_{c}$, where $\xi_{b}:=\textstyle{\sum}_{\lambda\in S_{b}}\xi_{\lambda}$ and $K_{B}$ and $\eta_{B}$ only depend on $B$. This estimate
  can be refined to 
  \[
  |\xi_{b}(\tau)-e^{B(\tau-\tau_{c})}\xi_{b}(\tau_{c})|\leq
  K_{B}C_{H}\ldr{\tau-\tau_{c}}^{\eta_{H}+\eta_{B}}e^{(\varpi_{B}+\b)(\tau-\tau_{c})}
  \]
  for all $\tau\leq \tau_{c}$. Combining the above estimates yields the conclusions that (\ref{eq:ximinuseBtauxiinftya}) and
  (\ref{eq:ximinuseBtauxiinfty}) hold, where $\xi_{\infty}:=\xi_{a,\infty}+\xi_{b}(\tau_{c})$. Since $\xi_{a,\infty}$ satisfies the estimate
  (\ref{eq:xiainftyestimate}) we also conclude that the second estimate in (\ref{eq:xiinftyaxiinftyestimate}) holds. What remains to be demonstrated
  is that $\xi_{\infty,a}$ is unique. Let us, to this end, assume that there are $\xi_{i}\in E_{a}$, $i=1,2$, such that (\ref{eq:ximinuseBtauxiinftya}) holds
  with $\xi_{\infty,a}$ replaced by $\xi_{i}$, $i=1,2$. This means that there are constants $C$ and $\eta$ such that
  \[
  |e^{B(\tau-\tau_{c})}(\xi_{1}-\xi_{2})|\leq C\ldr{\tau-\tau_{c}}^{\eta}e^{(\varpi_{B}+\b)(\tau-\tau_{c})}
  \]
  for all $\tau\leq\tau_{c}$. If $\xi_{1}\neq\xi_{2}$, then the left hand side becomes larger than the right hand side as $\tau\rightarrow-\infty$ due
  to the fact that $\xi_{1}-\xi_{2}\in E_{a}$. The lemma follows. 
\end{proof}

\begin{thm}\label{thm:leadingorderasymptoticszero}
  Let $0\leq \cweight\in\ro$, $\weight_{0}=(0,\cweight)$ and $\weight=(\cweight,\cweight)$. Assume that the conditions of
  Lemma~\ref{lemma:taurelvaryingbxEi} are fulfilled. Let $\kappa_{0}$ be the smallest integer which is strictly larger than $n/2$;
  $\kappa_{1}=\kappa_{0}+1$; $\kappa_{1}\leq k\in\zo$; and $l=k+\kappa_{0}$. Assume the
  $(\cweight,k)$-supremum and the $(\cweight,l)$-Sobolev assumptions to be satisfied; and that there are constants $c_{\coeff,k}$ and
  $s_{\coeff,l}$ such that (\ref{eq:Sobcoefflassumptions}) holds and such that (\ref{eq:coefflassumptions}) holds with $l$ replaced by $k$.
  Assume, moreover, that (\ref{eq:theeqreformEi}) is satisfied with vanishing right hand side and that $\varrho(\bx_{0},t)\rightarrow -\infty$
  as $t$ tends to the left endpoint of $I_{-}$; cf. (\ref{eq:Iminusdef}). Let $\g$ and $\bx_{\g}$ be as in
  Remark~\ref{remark:bgaconvtobxbgaloc}, and assume that $\bx_{0}=\bx_{\g}$. Assume, finally, that there are
  $Z^{0}_{\infty},\hal_{\infty}\in\Mn{m_{\ros}}{\ro}$ and constants $\e_{A}>0$, $c_{\rem}\geq 0$ such that
  \begin{equation}\label{eq:Zzerohalexpconvergence}
    [\|Z^{0}_{\roloc}(\tau)-Z^{0}_{\infty}\|^{2}+\|\hal_{\roloc}(\tau)-\hal_{\infty}\|^{2}]^{1/2}\leq c_{\rem}e^{\e_{A}\tau}
  \end{equation}
  for all $\tau\leq 0$, where $Z^{0}_{\roloc}$ and $\hal_{\roloc}$ are introduced in (\ref{eq:Zzerolochalloc}). Let 
  \begin{equation}\label{eq:AzeroAremdef}
    A_{0}:=\left(\begin{array}{cc} 0 & \Id \\ \hal_{\infty} & Z^{0}_{\infty}\end{array}\right),\ \ \
    A_{\rem}:=A-A_{0},
  \end{equation}
  where $A$ is defined in (\ref{eq:PsiAHdef}). Let, moreover, $\varpi_{A}:=\varpi_{\min}(A_{0})$ and $d_{A}:=d_{\max}(A_{0},\varpi_{A})-1$. Then
  (\ref{eq:Phinormbasassloc}) is satisfied for all $s_{1}\leq s_{2}\leq 0$, where $\Phi$ is defined by (\ref{eq:Phidef}) and $C_{A}$
  only depends on $A_{0}$, $c_{\rem}$ and $\e_{A}$. Let $m_{0}$ be defined as in the statement of Theorem~\ref{thm:asgrowthofenergy}
  and assume $k>m_{0}$. Let, moreover, $\b:=\min\{\e_{A},\e_{\Spe}\}$, $J_{a}:=[\varpi_{A},\varpi_{A}+\b)$, $E_{a}:=E_{A_{0},J_{a}}$ and 
  \begin{equation}\label{eq:Vdef}
    V:=\left(\begin{array}{c} u \\ \hU u\end{array}\right).
  \end{equation}
  Then, given $\tau_{c}\leq 0$, there is a unique $V_{\infty,a}\in E_{a}$ with $V_{\infty,a}\in\rn{2m_{s}}$ such that
  \begin{equation}\label{eq:VAplusestimate}
    \left|V-e^{A_{0}(\tau-\tau_{c})}V_{\infty,a}\right|
    \leq  C_{a}\ldr{\tau_{c}}^{\eta_{b}}\hGe_{l}^{1/2}(\tau_{c})\ldr{\tau-\tau_{c}}^{\eta_{a}}e^{(\varpi_{A}+\b)(\tau-\tau_{c})}
  \end{equation}
  on $A^{+}_{c}(\g)$, where $C_{a}$ only depends on $s_{\cweight,l}$, $s_{\coeff,l}$, $c_{\cweight,k}$, $c_{\coeff,k}$, $d_{\a}$ (in case $\iota_{b}\neq 0$), $A_{0}$,
  $c_{\rem}$, $\e_{A}$, $(\bM,\bge_{\refer})$ and a lower bound on $\theta_{0,-}$; and $\eta_{a}$, $\eta_{b}$ only depend on $\cweight$, $A_{0}$, $n$ and $k$.
  Moreover,
  \begin{equation}\label{eq:Vinfestimatefinalot}
    |V_{\infty,a}| \leq C_{a}\ldr{\tau_{c}}^{\eta_{b}}\hGe_{l}^{1/2}(\tau_{c}),
  \end{equation}
  where $C_{a}$ and $\eta_{b}$ have the same dependence as in the case of (\ref{eq:VAplusestimate}). 
\end{thm}
\begin{remark}
  Due to the proof, the function $V$ appearing in (\ref{eq:VAplusestimate}) can be replaced by $\Psi$ introduced in (\ref{eq:PsiAHdef}), where $\Psi_{i}$,
  $i=1,2$, is defined by (\ref{eq:Psiihtwodef}) and we here assume $\bfI=0$. 
\end{remark}
\begin{remark}\label{remark:Vinftyimprovement}
  The estimate (\ref{eq:VAplusestimate}) can be improved in that there is a $V_{\infty}\in \rn{2m_{s}}$ such that
  \begin{equation}\label{eq:VAplusestimateimproved}
    \left|V-e^{A_{0}(\tau-\tau_{c})}V_{\infty}\right|
    \leq  C_{a}\ldr{\tau_{c}}^{\eta_{b}}e^{\b\tau_{c}}\hGe_{l}^{1/2}(\tau_{c})\ldr{\tau-\tau_{c}}^{\eta_{a}}e^{(\varpi_{A}+\b)(\tau-\tau_{c})}
  \end{equation}
  on $A^{+}_{c}(\g)$, where $C_{a}$, $\eta_{a}$ and $\eta_{b}$ have the same dependence as in the case of (\ref{eq:VAplusestimate}). However, the
  corresponding $V_{\infty}$ is not unique. Nevertheless, $V_{\infty}$ can be chosen so that it satisfies (\ref{eq:Vinfestimatefinalot}) with $V_{\infty,a}$
  replaced by $V_{\infty}$. 
\end{remark}
\begin{proof}
  The first statement of the theorem, i.e., that (\ref{eq:Phinormbasassloc}) is satisfied for all $s_{1}\leq s_{2}\leq 0$, where $\Phi$ is defined by
  (\ref{eq:Phidef}), is an immediate consequence of Corollary~\ref{cor:Asymptoticallyconstant}. Letting $m_{0}$ be defined as in the statement of
  Theorem~\ref{thm:asgrowthofenergy} and assuming $k>m_{0}$, the assumptions of Theorem~\ref{thm:asgrowthofenergy} are fulfilled. In particular, the
  estimate (\ref{eq:melindassfslocfinalstmt}) yields the conclusion that
  \begin{equation}\label{eq:melindassfslocfinalstmttauceqzero}
    \begin{split}
      \me_{m}^{1/2} \leq & C_{a}\ldr{\tau_{c}}^{\eta_{b}}\ldr{\tau-\tau_{c}}^{\eta_{a}}e^{\varpi_{A}(\tau-\tau_{c})}\hGe_{l}^{1/2}(\tau_{c})
    \end{split}    
  \end{equation}
  holds on $A^{+}_{c}(\g)$ for $0\leq m\leq k-m_{0}$. Here $C_{a}$ only depends on $s_{\cweight,l}$, $s_{\coeff,l}$, $c_{\cweight,k}$, $c_{\coeff,k}$, $d_{\a}$ (in case
  $\iota_{b}\neq 0$), $A_{0}$, $c_{\rem}$, $\e_{A}$, $(\bM,\bge_{\refer})$ and a lower bound on $\theta_{0,-}$; and $\eta_{a}$ and $\eta_{b}$ only depend on
  $\cweight$, $A_{0}$, $n$, $m$ and $k$. Next, note that (\ref{eq:localsystematbfxz}) holds. In this equation, we are only interested in estimating
  $\Psi$ for $\bx=\bx_{0}$ and $|\bfI|=0$. For that reason, we here assume $\bx=\bx_{0}$ in (\ref{eq:localsystematbfxz}) and abuse notation in that we, most
  of the time, omit the argument $\bx_{0}$ in what follows. By assumption, $A=A_{0}+A_{\rem}$, where
  $\|A_{\rem}(\tau)\|\leq c_{\rem}e^{\e_{A}\tau_{c}}e^{\e_{A}(\tau-\tau_{c})}$ for all $\tau\leq \tau_{c}$. Here $c_{\rem}$ and $\e_{A}$ are the
  constants appearing in the statement of the theorem. In order to estimate $H$, we appeal to (\ref{eq:rhsreplcfderfullylocal}) with $m=0$ and to
  (\ref{eq:melindassfslocfinalstmttauceqzero}) with $m=1$. This yields
  \begin{equation}\label{eq:mHbxzuptauest}
    |H(\tau)|\leq C_{a}\ldr{\tau_{c}}^{\eta_{b}}e^{\e_{\Spe}\tau_{c}}\ldr{\tau-\tau_{c}}^{\eta_{a}}e^{(\varpi_{A}+\e_{\Spe})(\tau-\tau_{c})}\hGe_{l}^{1/2}(\tau_{c})
  \end{equation}
  for all $\tau\leq \tau_{c}$, where $C_{a}$, $\eta_{a}$ and $\eta_{b}$ have the same dependence as in (\ref{eq:melindassfslocfinalstmttauceqzero}).
  Next, due to (\ref{eq:dhtaupsihUpsiestloc}), (\ref{eq:melindassfslocfinalstmttauceqzero}) and the definition of the energy,
  \begin{equation}\label{eq:Psiroughestimate}
    |\Psi|\leq C_{a}\ldr{\tau_{c}}^{\eta_{b}}\ldr{\tau-\tau_{c}}^{\eta_{a}}e^{\varpi_{A}(\tau-\tau_{c})}\hGe_{l}^{1/2}(\tau_{c})
  \end{equation}
  for all $\tau\leq \tau_{c}$, where $C_{a}$, $\eta_{a}$ and $\eta_{b}$ have the same dependence as in (\ref{eq:melindassfslocfinalstmttauceqzero}).
  Combining this estimate with (\ref{eq:localsystematbfxz}), (\ref{eq:mHbxzuptauest}) and the above estimates for $A_{\rem}$ yields
  the conclusion that
  \begin{equation}\label{eq:PsiequmHver}
    \d_{\tau}\Psi=A_{0}\Psi+\mH,
  \end{equation}
  where
  \begin{equation}\label{eq:mHestimatebelowtauc}
    |\mH(\tau)|\leq C_{a}\ldr{\tau_{c}}^{\eta_{b}}e^{\b\tau_{c}}\ldr{\tau-\tau_{c}}^{\eta_{a}}e^{(\varpi_{A}+\b)(\tau-\tau_{c})}\hGe_{l}^{1/2}(\tau_{c})
  \end{equation}
  for all $\tau\leq \tau_{c}$, where $\b:=\min\{\e_{A},\e_{\Spe}\}$ and $C_{a}$, $\eta_{a}$ and $\eta_{b}$ have the same dependence as in
  (\ref{eq:melindassfslocfinalstmttauceqzero}).

  At this stage we can appeal to Lemma~\ref{lemma:asymptoticsmodelODE}. In fact, the conditions of this lemma are fulfilled with $\xi=\Psi$; $B=A_{0}$;
  $H=\mH$; $k=2m_{\ros}$; $\varpi_{B}=\varpi_{A}$; $\b$ defined as in the statement of the theorem; $\eta_{H}=\eta_{a}$; and
  \begin{equation}\label{eq:CHdef}
    C_{H}=C_{a}\ldr{\tau_{c}}^{\eta_{b}}e^{\b\tau_{c}}\hGe_{l}^{1/2}(\tau_{c}).
  \end{equation}
  Defining $E_{a}$ and $E_{b}$ as in the statement of Lemma~\ref{lemma:asymptoticsmodelODE}, there is then a unique $\Psi_{\infty,a}\in E_{a}$
  such that
  \begin{equation*}
    \begin{split}
      |\Psi-e^{A_{0}(\tau-\tau_{c})}\Psi_{\infty,a}| \leq & C_{B}\ldr{\tau-\tau_{c}}^{\eta_{B}}e^{(\varpi_{A}+\b)(\tau-\tau_{c})}|\Psi_{b}(\tau_{c})|\\
      & +KC_{H}\ldr{\tau-\tau_{c}}^{\eta_{H}+\eta_{B}}e^{(\varpi_{A}+\b)(\tau-\tau_{c})}
    \end{split}
  \end{equation*}  
  for all $\tau\leq\tau_{c}$, where $K$ only depends on $A_{0}$, $\eta_{a}$ and $\b$; and $C_{B}$ and $\eta_{B}$ only depend on $A_{0}$. Combining this
  estimate with (\ref{eq:Psiroughestimate}) and (\ref{eq:CHdef}) yields
  \[
  |\Psi-e^{A_{0}(\tau-\tau_{c})}\Psi_{\infty,a}|\leq C_{a}\ldr{\tau_{c}}^{\eta_{b}}\ldr{\tau-\tau_{c}}^{\eta_{a}}e^{(\varpi_{A}+\b)(\tau-\tau_{c})}\hGe_{l}^{1/2}(\tau_{c})
  \]
  for all $\tau\leq \tau_{c}$, where $\b:=\min\{\e_{A},\e_{\Spe}\}$ and $C_{a}$, $\eta_{a}$ and $\eta_{b}$ have the same dependence as in
  (\ref{eq:melindassfslocfinalstmttauceqzero}). Combining Lemma~\ref{lemma:asymptoticsmodelODE} with similar arguments yields the conclusion that
  $\Psi_{\infty}\in \rn{2m_{\ros}}$ such that 
  \begin{equation}\label{eq:PsiminuseAzPsiinfest}
    |\Psi-e^{A_{0}(\tau-\tau_{c})}\Psi_{\infty}|\leq C_{a}\ldr{\tau_{c}}^{\eta_{b}}e^{\b\tau_{c}}
    \ldr{\tau-\tau_{c}}^{\eta_{a}}e^{(\varpi_{A}+\b)(\tau-\tau_{c})}\hGe_{l}^{1/2}(\tau_{c})
  \end{equation}
  for all $\tau\leq \tau_{c}$, where $\b:=\min\{\e_{A},\e_{\Spe}\}$ and $C_{a}$, $\eta_{a}$ and $\eta_{b}$ have the same dependence as in
  (\ref{eq:melindassfslocfinalstmttauceqzero}). Note also that if $\Psi_{b}(\tau_{c})=0$, then $\Psi_{\infty}$ appearing on the left hand side of
  (\ref{eq:PsiminuseAzPsiinfest}) can be replaced by $\Psi_{\infty,a}$. Finally, combining Lemma~\ref{lemma:asymptoticsmodelODE} with similar arguments yields
  \[
  |\Psi_{\infty,a}|+|\Psi_{\infty}|\leq C_{a}\ldr{\tau_{c}}^{\eta_{b}}\hGe_{l}^{1/2}(\tau_{c}),
  \]
  where $C_{a}$ and $\eta_{b}$ have the same dependence as in (\ref{eq:melindassfslocfinalstmttauceqzero}).
 
  \textbf{Estimating the spatial variation.} At this stage, we wish to replace $\Psi$ with $V$; cf. (\ref{eq:Vdef}). We therefore need to estimate
  $(\d_{\tau}u)(\bx,\tau)-(\d_{\tau}u)(\bx_{0},\tau)$ for $\bx$ such that $d(\bx,\bx_{0})\leq K_{A}e^{\e_{\Spe}\tau}$; cf. the definition (\ref{eq:Aplusgammadef})
  of $A^{+}(\g)$. However, (\ref{eq:dhtaupsihUpsiestloc}) yields the conclusion that
  \[
  |E_{i}\d_{\tau}u|\leq C_{b}\me_{1}^{1/2}\leq C_{a}\ldr{\tau_{c}}^{\eta_{b}}\hGe_{l}^{1/2}(\tau_{c})\ldr{\tau-\tau_{c}}^{\eta_{a}}e^{\varpi_{A}(\tau-\tau_{c})}
  \]
  on $A^{+}_{c}(\g)$, where we appealed to (\ref{eq:melindassfslocfinalstmttauceqzero}) and $C_{a}$, $\eta_{a}$ and $\eta_{b}$ have the same dependence as
  in the case of (\ref{eq:melindassfslocfinalstmttauceqzero}). Combining the above observations,
  \[
  |(\d_{\tau}u)(\bx,\tau)-(\d_{\tau}u)(\bx_{0},\tau)|\leq
  C_{a}\ldr{\tau_{c}}^{\eta_{b}}e^{\e_{\Spe}\tau_{c}}\hGe_{l}^{1/2}(\tau_{c})\ldr{\tau-\tau_{c}}^{\eta_{a}}e^{(\varpi_{A}+\e_{\Spe})(\tau-\tau_{c})}
  \]
  for all $(\bx,\tau)\in A^{+}_{c}(\g)$. The argument concerning the spatial variation of $u$ in $A^{+}_{c}(\g)$ is similar but simpler. In particular, we
  can replace $\Psi(\bx_{0},\tau)$ with $\Psi(\bx,\tau)$ for $(\bx,\tau)\in A^{+}_{c}(\g)$. Next, we wish to replace $\d_{\tau}u$ with $\hU u$. However,
  that this is allowed is an immediate consequence of (\ref{eq:dhtaubDbfAuminushUbDbfAuloc}) and (\ref{eq:melindassfslocfinalstmttauceqzero}). Finally,
  the uniqueness of $V_{\infty,a}$ follows by the same argument as at the end of the proof of Lemma~\ref{lemma:asymptoticsmodelODE}. The theorem follows. 
\end{proof}

\section{Asymptotics of higher order derivatives}\label{section:asymptoticshigherorderderivatives}

\textbf{Preliminary equation.} Assume $u$ to be a solution to (\ref{eq:theeqreformEi}) with a vanishing right hand side; i.e.,
\begin{equation}\label{eq:maindominant}
  -\hU^{2}u+Z^{0}\hU u+\hal u=\mfS u,
\end{equation}
where
\begin{equation}\label{eq:mfSudef}
  \mfS u:=-\textstyle{\sum}_{A}e^{-2\mu_{A}}X_{A}^{2}u-Z^{A}X_{A}u. 
\end{equation}
Setting $\mfS u$ to zero yields a model equation. In some sense, this model equation corresponds to ``dropping the spatial derivatives'' in the original
equation, an idea that goes back to BKL, and which has been refined in the works of many authors; cf., e.g., \cite{aeta,dhn,huar,isamo,chisamo,els} and
references cited therein. A related notion is that of asymptotically velocity term dominated (AVTD) solutions. Due to
Theorem~\ref{thm:leadingorderasymptoticszero},
we know the leading order behaviour of $u$ and $\hU u$ in $A^{+}(\g)$. Combining this knowledge with (\ref{eq:maindominant}) yields the leading order
behaviour of $\hU^{2}u$ in $A^{+}(\g)$. However, it is also of interest to determine the asymptotics of $\hU^{m}E_{\bfI}u$ in $A^{+}(\g)$ for $m=0,1,2$.
Let us begin by giving a heuristic description of how this is to be achieved. First, we commute (\ref{eq:maindominant}) with $E_{\bfI}$. When doing so,
we ignore all resulting terms that contain a factor of the form $E_{\bfK}(A^{i}_{j})$ or $E_{\bfK}[\hU(A^{i}_{j})]$. Note that this corresponds to dropping
the second term on the right hand side of (\ref{eq:hUEicomm}). This results in an equation of the form
\[
-\hU^{2}E_{\bfI}u+Z^{0}\hU E_{\bfI}u+\hal E_{\bfI}u=L_{\pre,\bfI}u+\dots,
\]
where the dots signify the terms that we have ignored. In what follows, we assume $Z^{0}$ and $\hal$ to converge exponentially in the sense that 
(\ref{eq:Zzerohalexpconvergence}) holds. Moreover, as before, we can, effectively, replace $\hU$ with $\d_{\tau}$. This yields the equation
\[
-\d_{\tau}^{2}E_{\bfI}u+Z^{0}_{\infty}\d_{\tau} E_{\bfI}u+\hal_{\infty}E_{\bfI}u=L_{\pre,\bfI}u+\dots.
\]
Again, the dots signify the terms that we have ignored. Moreover, $L_{\pre,\bfI}u$ can be written in the form
\begin{equation}\label{eq:LbfIcompdefinitionpre}
  L_{\pre,\bfI}u=\textstyle{\sum}_{|\bfJ|<|\bfI|}\textstyle{\sum}_{m=0}^{2}L_{\pre,\bfI,\bfJ}^{m}\hU^{m}E_{\bfJ}u;
\end{equation}
cf. the proof of Theorem~\ref{thm:leadingorderasymptoticszeroho} below, in particular (\ref{eq:LmbfIbfJitoPGandb}), for a more detailed explanation
of how to compute $L_{\pre,\bfI}$. When it comes to deriving asymptotics, there is no problem in using $L_{\pre,\bfI}$ as the basis for our arguments.
However, when specifying asymptotics, we have to take into account that the different $E_{\bfI}u$ are not independent. In fact, $E_{\bfI}u$ can be
expressed in terms of $E_{\omega}u$ for $\rn{n}$-multiindices $\omega$ satisfying $|\omega|\leq |\bfI|$; if $\omega$ is an $\rn{n}$-multiindex, we here
use the notation
\[
E_{\omega}u:=E_{1}^{\omega_{1}}\cdots E_{n}^{\omega_{n}}u.
\]
\textbf{Removing redundancies.} In what follows, it is convenient to define, for every vector field multiindex $\bfI$, an associated $\rn{n}$-multiindex.
\begin{definition}\label{def:upomegadef}
  Given a vector field multiindex $\bfI=(I_{1},\dots,I_{p})$, let $\upomega(\bfI)\in\nn{n}$ be the vector whose components, written
  $\upomega_{i}(\bfI)$, $i=1,\dots,n$, are given as follows: $\upomega_{i}(\bfI)$ equals the number of times $I_{q}=i$, $q=1,\dots,p$. 
\end{definition}
Given a vector field multiindex $\bfI$, let $\omega:=\upomega(\bfI)$. Then
\begin{equation}\label{eq:EbFIminusEomegapsi}
  E_{\bfI}\psi-E_{\omega}\psi=\textstyle{\sum}_{|\xi|<|\bfI|}\mfC_{\bfI,\xi}E_{\xi}\psi,
\end{equation}
where $\mfC_{\bfI,\xi}$ are functions depending only on $\bfI$, $\xi$ and the frame $\{E_{i}\}$; and $\xi$ are $\rn{n}$-multiindices. It is straightforward
to prove this for $|\bfI|\leq 2$. In order to prove the statement in general, let $2\leq m\in\zo$, and assume that it holds for $|\bfI|\leq m$. Let
$\bfI=(I_{1},\dots,I_{p})$ with $p=m+1$. Note that if $\bfJ$ is obtained from $\bfI$ by permuting two adjacent indices, then
\[
E_{\bfI}\psi-E_{\bfJ}\psi=\textstyle{\sum}_{|\bfK|<|\bfI|}\mfD_{\bfI,\bfJ,\bfK}E_{\bfK}\psi
\]
for some functions $\mfD_{\bfI,\bfJ,\bfK}$ depending only on $\bfI$, $\bfJ$, $\bfK$ and the frame $\{E_{i}\}$. However, due to the inductive assumption,
$E_{\bfK}\psi$ can, up to
functions depending only on $\bfK$, $\xi$ and the frame $\{E_{i}\}$, be written as a sum of terms of the form $E_{\xi}\psi$ for $\rn{n}$-multiindices
$\xi$ satisfying $|\xi|\leq |\bfK|$. To conclude, permuting two adjacent indices in $\bfI$ is harmless due to the inductive assumption. On the other hand, a
finite number of such permutations takes us from $\bfI$ to $\upomega(\bfI)$. To conclude, (\ref{eq:EbFIminusEomegapsi}) holds.

Consider (\ref{eq:LbfIcompdefinitionpre}). Due to (\ref{eq:EbFIminusEomegapsi}), $E_{\bfJ}u$ can be rewritten in terms of $E_{\xi}u$, $|\xi|\leq |\bfI|$,
with coefficients depending only $\bfI$, $\xi$ and the frame $\{E_{i}\}$. Moreover, if a $\hU$ hits one of these coefficients, the resulting term is an
error term. In the end, we thus conclude that 
\[
-\d_{\tau}^{2}E_{\bfI}u+Z^{0}_{\infty}\d_{\tau} E_{\bfI}u+\hal_{\infty}E_{\bfI}u=L_{\bfI}u+\dots,
\]
where 
\begin{equation}\label{eq:LbfIcompdefinition}
  L_{\bfI}u=\textstyle{\sum}_{|\omega|<|\bfI|}\textstyle{\sum}_{m=0}^{2}L_{\bfI,\omega}^{m}\hU^{m}E_{\omega}u
\end{equation}
and $\omega$ are $\rn{n}$-multiindices; cf. (\ref{eq:LbfIdef}) and (\ref{eq:LbfIcoeffdef}) for a more detailed explanation of how to compute
$L_{\bfI}u$ and its coefficients. 

\textbf{Inductive argument.} When deriving the asymptotics of the higher order derivatives, it is important to note that the sum in
(\ref{eq:LbfIcompdefinition}) ranges over $|\omega|<|\bfI|$. Due to this fact, it is possible to proceed inductively. To begin with, appealing
to Theorem~\ref{thm:leadingorderasymptoticszero}, we control the leading order behaviour of $\hU u$ and $u$. Combining this knowledge with the equation
yields the behaviour of $\hU^{2}u$. It is therefore meaningful to assume, inductively, that for some $1\leq j\in\zo$, there are functions $U_{\bfJ,m}$
for $|\bfJ|<j$ and $m=0,1,2$, depending only on $\tau$, such that the difference between $\hU^{m}E_{\bfJ}u$ and $U_{\bfJ,m}$ is small. Localising,
additionally, the coefficients of $L_{\bfI}$, it is natural to introduce
\begin{equation}\label{eq:sfLbfIdef}
  \sfL_{\bfI}(\tau):=\textstyle{\sum}_{|\omega|<|\bfI|}\textstyle{\sum}_{m=0}^{2}L_{\bfI,\omega}^{m}(\bx_{0},\tau)U_{\omega,m}(\tau).
\end{equation}
As a part of the inductive argument, it can be demonstrated that this expression captures the leading order behaviour of $L_{\bfI}u$. In the end, the
equation can be written
\begin{equation}\label{eq:somesteptowardsODE}
  -\d_{\tau}^{2}E_{\bfI}u+Z^{0}_{\infty}\d_{\tau} E_{\bfI}u+\hal_{\infty}E_{\bfI}u=\sfL_{\bfI}+\dots.
\end{equation}
To conclude, the model equation is the following ODE: 
\[
-\d_{\tau}^{2}U_{\bfI}+Z^{0}_{\infty}\d_{\tau} U_{\bfI}+\hal_{\infty}U_{\bfI}=\sfL_{\bfI}.
\]
The solutions to this equation can be written
\[
\left(\begin{array}{c} U_{\bfI}(\tau) \\ (\d_{\tau}U_{\bfI})(\tau) \end{array}\right)
=e^{A_{0}(\tau-\tau_{c})}X_{\bfI}+\int_{\tau}^{\tau_{c}}e^{A_{0}(\tau-s)}\left(\begin{array}{c} 0 \\ \sfL_{\bfI}(s) \end{array}\right)ds,
\]
where $X_{\bfI}\in\rn{2m_{\ros}}$. For this reason, the goal in the present section is to prove, inductively, that, for a suitable choice of
$X_{\bfI}$, the difference 
\[
\left(\begin{array}{c} E_{\bfI}u \\ \hU E_{\bfI}u \end{array}\right)
-e^{A_{0}(\tau-\tau_{c})}X_{\bfI}-\int_{\tau}^{\tau_{c}}e^{A_{0}(\tau-s)}\left(\begin{array}{c} 0 \\ \sfL_{\bfI}(s) \end{array}\right)ds
\]
is small in $A^{+}_{c}(\g)$. In the process of deriving the corresponding estimates, we also obtain estimates with $\hU E_{\bfI}u$ replaced by
$\d_{\tau}E_{\bfI}u$. Once such estimates have been derived, we can immediately read off $U_{\bfI,m}$ for $m=0,1$. Combining this knowledge with
(\ref{eq:dtausqmhUsqEbfIestimate}) and (\ref{eq:somesteptowardsODE}) yields $U_{\bfI,2}$. This reproduces the inductive assumption and completes the
argument.

\begin{thm}\label{thm:leadingorderasymptoticszeroho}
  Let $0\leq \cweight\in\ro$, $\weight_{0}=(0,\cweight)$ and $\weight=(\cweight,\cweight)$. Assume that the conditions of
  Lemma~\ref{lemma:taurelvaryingbxEi} are fulfilled. Let $\kappa_{0}$ be the smallest integer which is strictly larger than $n/2$;
  $\kappa_{1}=\kappa_{0}+1$; $\kappa_{1}\leq k\in\zo$; and $l=k+\kappa_{0}$. Assume the
  $(\cweight,k)$-supremum and the $(\cweight,l)$-Sobolev assumptions to be satisfied; and that there are constants $c_{\coeff,k}$ and
  $s_{\coeff,l}$ such that (\ref{eq:Sobcoefflassumptions}) holds and such that (\ref{eq:coefflassumptions}) holds with $l$ replaced by $k$.
  Assume, moreover, that (\ref{eq:theeqreformEi}) is satisfied with vanishing right hand side and that $\varrho(\bx_{0},t)\rightarrow -\infty$
  as $t$ tends to the left endpoint of $I_{-}$; cf. (\ref{eq:Iminusdef}). Let $\g$ and $\bx_{\g}$ be as in
  Remark~\ref{remark:bgaconvtobxbgaloc}, and assume that $\bx_{0}=\bx_{\g}$. Assume, finally, that there are
  $Z^{0}_{\infty},\hal_{\infty}\in\Mn{m_{\ros}}{\ro}$ and constants $\e_{A}>0$, $c_{\rem}\geq 0$ such that (\ref{eq:Zzerohalexpconvergence}) holds
  for all $\tau\leq 0$. Let $A_{0}$ be defined by (\ref{eq:AzeroAremdef}) and $A$ be defined by (\ref{eq:PsiAHdef}). Let, moreover,
  $\varpi_{A}:=\varpi_{\min}(A_{0})$ and $d_{A}:=d_{\max}(A_{0},\varpi_{A})-1$. Then
  (\ref{eq:Phinormbasassloc}) is satisfied for all $s_{1}\leq s_{2}\leq 0$, where $\Phi$ is defined by (\ref{eq:Phidef}) and $C_{A}$
  only depends on $A_{0}$, $c_{\rem}$ and $\e_{A}$. Let $m_{0}$ be defined as in the statement of Theorem~\ref{thm:asgrowthofenergy}
  and assume $k>m_{0}+1$. Let, moreover, $\b:=\min\{\e_{A},\e_{\Spe}\}$, $J_{a}:=[\varpi_{A},\varpi_{A}+\b)$, $E_{a}:=E_{A_{0},J_{a}}$, $V$ be defined by
  (\ref{eq:Vdef}) and 
  \[
  V_{\bfI}:=\left(\begin{array}{c} E_{\bfI}u \\ \hU E_{\bfI}u\end{array}\right).
  \]
  Fix $\tau_{c}\leq 0$, let $V_{\infty,a}$ be defined as in the statement of Theorem~\ref{thm:leadingorderasymptoticszero} and define
  $U_{0,m}\in C^{\infty}(\ro,\rn{m_{\ros}})$, $m=0,1,2$, by
  \begin{equation}\label{eq:Uzeromdef}
    \left(\begin{array}{c} U_{0,0}(\tau) \\ U_{0,1}(\tau)\end{array}\right):=e^{A_{0}(\tau-\tau_{c})}V_{\infty,a},\ \ \
    U_{0,2}(\tau):=Z^{0}_{\infty}U_{0,1}(\tau)+\hal_{\infty}U_{0,0}(\tau).
  \end{equation}
  Let $1\leq j\leq k-m_{0}-1$ and assume that $U_{\bfJ,m}$ has been defined for $|\bfJ|<j$ and $m=0,1,2$ (for $\bfJ=0$, these functions are defined by
  (\ref{eq:Uzeromdef}) and for $|\bfJ|>0$, they are defined inductively by (\ref{eq:UbfIzeroandonedef}) and (\ref{eq:UbfItwodef}) below). Let $\bfI$
  be such that $|\bfI|=j$ and define $\sfL_{\bfI}$ by (\ref{eq:sfLbfIdef}). Then there is a unique $V_{\bfI,\infty,a}\in E_{a}$ with
  $V_{\bfI,\infty,a}\in\rn{2m_{s}}$ such that
  \begin{equation}\label{eq:VAplusestimateho}
    \begin{split}
      & \left|V_{\bfI}-e^{A_{0}(\tau-\tau_{c})}V_{\bfI,\infty,a}-\int_{\tau}^{\tau_{c}}e^{A_{0}(\tau-s)}\left(\begin{array}{c} 0 \\ \sfL_{\bfI}(s)\end{array}\right)ds\right|\\
      \leq & C_{a}\ldr{\tau_{c}}^{\eta_{b}}\hGe_{l}^{1/2}(\tau_{c})\ldr{\tau-\tau_{c}}^{\eta_{a}}e^{(\varpi_{A}+\b)(\tau-\tau_{c})}
    \end{split}    
  \end{equation}
  on $A^{+}_{c}(\g)$, where $C_{a}$ only depends on $s_{\cweight,l}$, $s_{\coeff,l}$, $c_{\cweight,k}$, $c_{\coeff,k}$, $d_{\a}$ (in case $\iota_{b}\neq 0$), $A_{0}$,
  $c_{\rem}$, $\e_{A}$, $(\bM,\bge_{\refer})$ and a lower bound on $\theta_{0,-}$; and $\eta_{a}$ and $\eta_{b}$ only depend on $\cweight$, $A_{0}$, $n$, 
  and $k$. Moreover,
  \begin{equation}\label{eq:Vinfestimatefinalotho}
    |V_{\bfI,\infty,a}| \leq C_{a}\ldr{\tau_{c}}^{\eta_{b}}\hGe_{l}^{1/2}(\tau_{c}),
  \end{equation}
  where $C_{a}$ and $\eta_{b}$ have the same dependence as in the case of (\ref{eq:VAplusestimateho}). Given $V_{\bfI,\infty,a}$ as above,
  define $U_{\bfI,m}$, $m=0,1,2$, by
  \begin{align}
    \left(\begin{array}{c} U_{\bfI,0}(\tau) \\ U_{\bfI,1}(\tau)\end{array}\right)
    := & e^{A_{0}(\tau-\tau_{c})}V_{\bfI,\infty,a}+\int_{\tau}^{\tau_{c}}e^{A_{0}(\tau-s)}
    \left(\begin{array}{c} 0 \\ \sfL_{\bfI}(s)\end{array}\right)ds,\label{eq:UbfIzeroandonedef}\\
    U_{\bfI,2}(\tau) := & Z^{0}_{\infty}U_{\bfI,1}(\tau)+\hal_{\infty}U_{\bfI,0}(\tau)-\sfL_{\bfI}(\tau).\label{eq:UbfItwodef}
  \end{align}
  Proceeding inductively as above yields $U_{\bfI,m}$ and $V_{\bfI,\infty,a}$ for $|\bfI|\leq k-m_{0}-1$ and $m=0,1,2$ such that (\ref{eq:VAplusestimateho})
  holds. 
\end{thm}
\begin{remark}\label{remark:higherorderestimatesimprovedestimates}
  It is possible to improve the estimates. First, define $V_{\infty}$ as in Remark~\ref{remark:Vinftyimprovement}. This yields
  (\ref{eq:VAplusestimateimproved}). Defining $U_{0,m}$, $m=0,1,2$, by (\ref{eq:Uzeromdef}) with $V_{\infty,a}$ replaced by $V_{\infty}$, we can proceed
  inductively as in the statement of the theorem. In particular, a $V_{\bfI,\infty}\in\rn{2m_{s}}$ can be constructed such that (\ref{eq:VAplusestimateho})
  is improved to
  \begin{equation}\label{eq:VAplusestimatehoimproved}
    \begin{split}
      & \left|V_{\bfI}-e^{A_{0}(\tau-\tau_{c})}V_{\bfI,\infty}-\int_{\tau}^{\tau_{c}}e^{A_{0}(\tau-s)}\left(\begin{array}{c} 0 \\ \sfL_{\bfI}(s)\end{array}\right)ds\right|\\
      \leq & C_{a}\ldr{\tau_{c}}^{\eta_{b}}e^{\b\tau_{c}}\hGe_{l}^{1/2}(\tau_{c})\ldr{\tau-\tau_{c}}^{\eta_{a}}e^{(\varpi_{A}+\b)(\tau-\tau_{c})}
    \end{split}    
  \end{equation}
  on $A^{+}_{c}(\g)$, where $C_{a}$, $\eta_{a}$ and $\eta_{b}$ have the same dependence as in (\ref{eq:VAplusestimateho}). Defining $U_{\bfI,m}$ as in
  (\ref{eq:UbfIzeroandonedef}) and (\ref{eq:UbfItwodef}) with $V_{\bfI,\infty,a}$ replaced by $V_{\bfI,\infty}$, and modifying $\sfL_{\bfI}$ accordingly, it
  can be demonstrated that (\ref{eq:VAplusestimatehoimproved}) holds for $|\bfI|\leq k-m_{0}-1$. Note that the advantage here is that by taking
  $\tau_{c}$ close enough to $-\infty$, the factor $C_{a}\ldr{\tau_{c}}^{\eta_{b}}e^{\b\tau_{c}}$ can be chosen to be as small as we wish. The disadvantage of
  the estimate is that $V_{\bfI,\infty}$ is not unique. However, $V_{\bfI,\infty}$ satisfies (\ref{eq:Vinfestimatefinalotho}) with $V_{\bfI,\infty,a}$ replaced
  by $V_{\bfI,\infty}$. 
\end{remark}
\begin{proof}
  The conditions of Theorem~\ref{thm:leadingorderasymptoticszero} are satisfied, and this theorem and Remark~\ref{remark:Vinftyimprovement} immediately
  yield the existence of $V_{\infty,a}$ and $V_{\infty}$ and imply that (\ref{eq:Phinormbasassloc}) holds. 

  \textbf{Preliminary equation.}
  The goal of the proof is to determine the asymptotics of $\hU^{m}E_{\bfI}u$ in $A^{+}_{c}(\g)$ for $m=0,1,2$. As described prior to the statement of the
  theorem, we need, to this end, to commute (\ref{eq:maindominant}) with $E_{\bfI}$ and to keep the leading order terms. Due to the proof of
  Lemma~\ref{lemma:bdbfAhUsqcommformEi},
  \begin{equation}\label{eq:bdbfAhUsqcommformEiPrelEqn}
    [\hU^{2},E_{\bfI}]\psi = \textstyle{\sum}_{|\bfJ|<|\bfI|}\textstyle{\sum}_{m=1}^{2}P_{\bfI,\bfJ}^{m}\hU^{m}E_{\bfJ}\psi+\mfR_{\bfI}^{2}\psi,
  \end{equation}
  where $\mfR^{2}_{\bfI}u$ collects all the terms that contain a factor of the form $E_{\bfK}(A_{i}^{j})$. To be more precise, $P_{\bfI,\bfJ}^{2}$ is a linear
  combination of terms of the form (\ref{eq:CbfIbfJtwoterms}) (with $k$ replaced by $r$), where $|\bfI_{1}|+\dots+|\bfI_{r}|=|\bfI|-|\bfJ|$, $r\geq 1$
  and $\bfI_{j}\neq 0$; and $P_{\bfI,\bfJ}^{1}$ is a linear combination of terms of the form (\ref{eq:hConeterms}) (with $k$ replaced by $r$), where
  $|\bfI_{1}|+\dots+|\bfI_{r}|+|\bfK|=|\bfI|-|\bfJ|$, $\bfI_{j}\neq 0$. Moreover,
  \[
  \mfR_{\bfI}^{2}\psi=\textstyle{\sum}_{|\bfJ|\leq |\bfI|}\textstyle{\sum}_{m=0}^{1}\mfR_{\bfI,\bfJ}^{m}\hU^{m}E_{\bfJ}\psi.
  \]
  Here $\mfR_{\bfI,\bfJ}^{1}$ is a linear combination of terms of the form (\ref{eq:mfConeterms}) (with $k$ replaced by $r$), where
  $|\bfI_{1}|+\dots+|\bfI_{r}|+|\bfK|=|\bfI|-|\bfJ|$, $\bfI_{j}\neq 0$; and $\mfR_{\bfI,\bfJ}^{0}$ is a linear combination of terms of the form
  (\ref{eq:CzerotermsA})--(\ref{eq:CzerotermsC}) (with $k$ replaced by $r$), where $|\bfI_{1}|+\dots+|\bfI_{r}|+|\bfK|=|\bfI|-|\bfJ|$ in
  (\ref{eq:CzerotermsA}); $|\bfI_{1}|+\dots+|\bfI_{r}|+|\bfJ_{1}|+|\bfJ_{2}|=|\bfI|-|\bfJ|$ in (\ref{eq:CzerotermsB}) and (\ref{eq:CzerotermsC});
  $\bfI_{j}\neq 0$; and $r+|\bfJ_{2}|\geq 1$ in (\ref{eq:CzerotermsC}).

  Next, due to Lemma~\ref{lemma:bdbfAchthhUcommformEi}, and with the notation $\mfG_{\bfI,\bfJ}^{0}=G_{\bfI,\bfJ}^{0}$,
  \[
  [E_{\bfI},Z^{0}\hU]=\textstyle{\sum}_{|\bfJ|<|\bfI|}G_{\bfI,\bfJ}^{1}\hU E_{\bfJ}
  +\textstyle{\sum}_{1\leq |\bfJ|\leq |\bfI|}\mfG_{\bfI,\bfJ}^{0}E_{\bfJ}.
  \]
  Here $G_{\bfI,\bfJ}^{1}$ is a linear combination of terms of the form (\ref{eq:commZzeroEbfIfirsttype}) (with $k$ replaced by $r$), where $\bfI_{j}\neq 0$
  and $|\bfI_{1}|+\dots+|\bfI_{r}|+|\bfK|=|\bfI|-|\bfJ|$; and $\mfG_{\bfI,\bfJ}^{0}$ is a linear combination of terms of the form
  (\ref{eq:commZzeroEbfIsecondtype}) (with $k$ replaced by $r$), where $\bfI_{j}\neq 0$ and $|\bfI_{1}|+\dots+|\bfI_{r}|+|\bfJ_{1}|+|\bfJ_{2}|=|\bfI|-|\bfJ|$.
  Finally,
  \[
  [E_{\bfI},\hal]=\textstyle{\sum}_{|\bfJ|<|\bfI|}b_{\bfI,\bfJ}E_{\bfJ},
  \]
  where $b_{\bfI,\bfJ}$ is a linear combination of terms of the form $E_{\bfK}\hal$, where $|\bfK|=|\bfI|-|\bfJ|$. 

  Combining the above observations yields the conclusion that $E_{\bfI}u$ satisfies the equation
  \begin{equation}\label{eq:EbfImaindominantpre}
    -\hU^{2}E_{\bfI}u+Z^{0}\hU E_{\bfI}u+\hal E_{\bfI}u=L_{\pre,\bfI}u+\mfR_{\pre,\bfI}u.
  \end{equation}
  Here
  \begin{align}
    L_{\pre,\bfI}u = & \textstyle{\sum}_{|\bfJ|<|\bfI|}\textstyle{\sum}_{m=1}^{2}P_{\bfI,\bfJ}^{m}\hU^{m}E_{\bfJ}u
    -\textstyle{\sum}_{|\bfJ|<|\bfI|}G_{\bfI,\bfJ}^{1}\hU E_{\bfJ}u-\textstyle{\sum}_{|\bfJ|<|\bfI|}b_{\bfI,\bfJ}E_{\bfJ}u,\label{eq:LbfIdef}\\
    \mfR_{\pre,\bfI}u = & \textstyle{\sum}_{|\bfJ|\leq |\bfI|}\textstyle{\sum}_{m=0}^{1}\mfR_{\bfI,\bfJ}^{m}\hU^{m}E_{\bfJ}u
    -\textstyle{\sum}_{1\leq |\bfJ|\leq |\bfI|}\mfG_{\bfI,\bfJ}^{0}E_{\bfJ}u+E_{\bfI}\mfS u.\label{eq:RbfIdef}
  \end{align}
  Comparing (\ref{eq:LbfIcompdefinitionpre}) with (\ref{eq:LbfIdef}) yields
  \begin{equation}\label{eq:LmbfIbfJitoPGandb}
    L^{2}_{\pre,\bfI,\bfJ}=P^{2}_{\bfI,\bfJ},\ \ \
    L^{1}_{\pre,\bfI,\bfJ}=P^{1}_{\bfI,\bfJ}-G^{1}_{\bfI,\bfJ},\ \ \
    L^{0}_{\pre,\bfI,\bfJ}=-b_{\bfI,\bfJ}.
  \end{equation}

  \textbf{Removing redundancies.} Recalling (\ref{eq:EbFIminusEomegapsi}),
  \[
  L^{m}_{\pre,\bfI,\bfJ}\hU^{m}E_{\bfJ}u=L^{m}_{\pre,\bfI,\bfJ}\textstyle{\sum}_{|\xi|\leq |\bfJ|}\hU^{m}(\mfC_{\bfJ,\xi}E_{\xi}u),
  \]
  where we define $\mfC_{\bfJ,\upomega(\bfJ)}=1$; $\mfC_{\bfJ,\xi}=0$ if $|\xi|=|\bfJ|$ and $\xi\neq\upomega(\bfJ)$; and
  $\mfC_{\bfJ,\xi}=0$ if $|\xi|>|\bfJ|$. Thus
  \begin{equation}\label{eq:EbfImaindominant}
    -\hU^{2}E_{\bfI}u+Z^{0}\hU E_{\bfI}u+\hal E_{\bfI}u=L_{\bfI}u+\mfR_{\bfI}u,
  \end{equation}
  where
  \begin{align}
    L_{\bfI}u := & \textstyle{\sum}_{|\xi|<|\bfI|}\textstyle{\sum}_{m=0}^{2}L_{\bfI,\xi}^{m}\hU^{m}E_{\xi}u,\label{eq:LbfIdefaux}\\
    L^{m}_{\bfI,\xi} := & \textstyle{\sum}_{|\bfJ|<|\bfI|}L^{m}_{\pre,\bfI,\bfJ}\mfC_{\bfJ,\xi}.\label{eq:LbfIcoeffdef}
  \end{align}
  Moreover,
  \begin{align*}
    \mfR_{\bfI}u := & \mfR_{\pre,\bfI}u+\textstyle{\sum}_{|\bfJ|<|\bfI|}\textstyle{\sum}_{|\xi|<|\bfJ|}\mfR_{\rocor,\bfI,\bfJ,\xi}u,\\
    \mfR_{\rocor,\bfI,\bfJ,\xi}u := & 2L^{2}_{\pre,\bfI,\bfJ}\hU(\mfC_{\bfJ,\xi})\hU E_{\xi}u
    +[L^{2}_{\pre,\bfI,\bfJ}\hU^{2}(\mfC_{\bfJ,\xi})+L^{1}_{\pre,\bfI,\bfJ}\hU(\mfC_{\bfJ,\xi})]E_{\xi}u.
  \end{align*}

  \textbf{Inductive argument.} Combining (\ref{eq:EbfImaindominant}) with an inductive argument, it is possible to derive the leading order asymptotics
  of $\hU^{m}E_{\bfI}u$ in $A^{+}_{c}(\g)$ for $m=0,1,2$. The rough structure of the argument is the following. To begin with, due to
  Theorem~\ref{thm:leadingorderasymptoticszero} and Remark~\ref{remark:Vinftyimprovement},
  we know the leading order asymptotics of $u$ and $\hU u$ in $A^{+}_{c}(\g)$. Combining this information with (\ref{eq:maindominant}) yields the leading
  order asymptotics of $\hU^{2}u$. Let $\bfI$ be such that $|\bfI|\neq 0$ and assume that we know the leading order asymptotics of $\hU^{m}E_{\bfJ}u$ in
  $A^{+}_{c}(\g)$ for $m=0,1,2$ and $|\bfJ|<|\bfI|$. Inserting this information into (\ref{eq:EbfImaindominant}) and proceeding, roughly speaking, as in the
  proof of Theorem~\ref{thm:leadingorderasymptoticszero} yields the leading order asymptotics of $\hU^{m}E_{\bfI}u$ in $A^{+}_{c}(\g)$ for $m=0,1,2$.

  \textbf{Deriving the ODE.} In order to derive an ODE for $E_{\bfI}u$, let us begin by appealing to Lemma~\ref{lemma:dtausqmhUsqEbfIestimate} and
  (\ref{eq:melindassfslocfinalstmttauceqzero}). This yields 
  \begin{equation}\label{eq:dtausqEbfIuminushUsqEbfIu}
    |\d_{\tau}^{2}E_{\bfI}u-\hU^{2}E_{\bfI}u|\leq C_{a}\ldr{\tau-\tau_{c}}^{\eta_{a}}e^{(\varpi_{A}+\e_{\Spe})(\tau-\tau_{c})}
    \ldr{\tau_{c}}^{\eta_{b}}e^{\e_{\Spe}\tau_{c}}\hGe_{l}^{1/2}(\tau_{c})
  \end{equation}
  on $A^{+}_{c}(\g)$ for $0\leq |\bfI|\leq k-m_{0}-1$. Here $C_{a}$ only depends on $s_{\cweight,l}$, $s_{\coeff,l}$, $c_{\cweight,k}$, $c_{\coeff,k}$, $d_{\a}$ (in case
  $\iota_{b}\neq 0$), $A_{0}$, $c_{\rem}$, $\e_{A}$, $(\bM,\bge_{\refer})$ and a lower bound on $\theta_{0,-}$; and $\eta_{a}$ and $\eta_{b}$ only depend on
  $\cweight$, $A_{0}$, $n$, $m$ and $k$. Next, combining (\ref{eq:dhtaubDbfAuminushUbDbfAuloc}), (\ref{eq:ZzerominusZetazeroloc}),
  (\ref{eq:Zzerohalexpconvergence}) and (\ref{eq:melindassfslocfinalstmttauceqzero}) yields
  \[
  |Z^{0}\hU E_{\bfI}u-Z^{0}_{\infty}\d_{\tau}E_{\bfI}u|\leq C_{a}\ldr{\tau-\tau_{c}}^{\eta_{a}}e^{(\varpi_{A}+\b)(\tau-\tau_{c})}
  \ldr{\tau_{c}}^{\eta_{b}}e^{\b\tau_{c}}\hGe_{l}^{1/2}(\tau_{c})
  \]
  on $A^{+}_{c}(\g)$ for $0\leq |\bfI|\leq k-m_{0}-1$. Here $C_{a}$, $\eta_{a}$ and $\eta_{b}$ have the same dependence as in the case of
  (\ref{eq:dtausqEbfIuminushUsqEbfIu}). Combining the above estimates with (\ref{eq:halminushalloc}), (\ref{eq:Zzerohalexpconvergence}) and
   (\ref{eq:melindassfslocfinalstmttauceqzero}) yields 
  \begin{equation}\label{eq:ascoeffhUtaureplestimatesi}
    \begin{split}
      & |-\d_{\tau}^{2}E_{\bfI}u+Z^{0}_{\infty}\d_{\tau}E_{\bfI}u+\hal_{\infty}E_{\bfI}u+\hU^{2}E_{\bfI}u-Z^{0}\hU E_{\bfI}u-\hal E_{\bfI}u|\\
      \leq & C_{a}\ldr{\tau-\tau_{c}}^{\eta_{a}}e^{(\varpi_{A}+\b)(\tau-\tau_{c})}\ldr{\tau_{c}}^{\eta_{b}}e^{\b\tau_{c}}\hGe_{l}^{1/2}(\tau_{c})
    \end{split}
  \end{equation}
  on $A^{+}_{c}(\g)$ for $0\leq |\bfI|\leq k-m_{0}-1$. Here $C_{a}$, $\eta_{a}$ and $\eta_{b}$ have the same dependence as in the case of
  (\ref{eq:dtausqEbfIuminushUsqEbfIu}).

  Next, we need to estimate $E_{\bfI}\mfS u$; cf. (\ref{eq:maindominant}) and (\ref{eq:mfSudef}). Due to (\ref{eq:bDbfAemtmuAXAsqcommEi}),
  (\ref{eq:ZAXAcommestEi}), (\ref{eq:spatialderivativesexpdecayestimateloc}) and (\ref{eq:melindassfslocfinalstmttauceqzero})
  \[
  |E_{\bfI}\mfS u|\leq C_{a}\ldr{\tau-\tau_{c}}^{\eta_{a}}e^{(\varpi_{A}+\e_{\Spe})(\tau-\tau_{c})}
    \ldr{\tau_{c}}^{\eta_{b}}e^{\e_{\Spe}\tau_{c}}\hGe_{l}^{1/2}(\tau_{c})
  \]
  on $A^{+}_{c}(\g)$ for $0\leq |\bfI|\leq k-m_{0}-1$. Here $C_{a}$, $\eta_{a}$ and $\eta_{b}$ have the same dependence as in the case of
  (\ref{eq:dtausqEbfIuminushUsqEbfIu}). 

  In order to estimate the first term on the right hand side of (\ref{eq:RbfIdef}), it is sufficient to
  estimate the contribution from the first term on the right hand side of (\ref{eq:CIJoneest}) as well as the right hand side of
  (\ref{eq:CIJzeroest}). This is done in Lemma~\ref{lemma:CkbfBbfAClaestEi}, and the contributions correspond to the first term on the right hand
  side of (\ref{eq:ConebfIbfJaE}) and the right hand side of (\ref{eq:CzerobfIbfJaE}). This yields
  \[
  \left|\textstyle{\sum}_{|\bfJ|\leq |\bfI|}\textstyle{\sum}_{m=0}^{1}\mfR_{\bfI,\bfJ}^{m}\hU^{m}E_{\bfJ}u\right|
  \leq C_{a}\ldr{\tau-\tau_{c}}^{\eta_{a}}e^{(\varpi_{A}+\e_{\Spe})(\tau-\tau_{c})}
    \ldr{\tau_{c}}^{\eta_{b}}e^{\e_{\Spe}\tau_{c}}\hGe_{l}^{1/2}(\tau_{c})
  \]
  on $A^{+}_{c}(\g)$ for $0\leq |\bfI|\leq k-m_{0}$. Here $C_{a}$, $\eta_{a}$ and $\eta_{b}$ have the same dependence as in the case of
  (\ref{eq:dtausqEbfIuminushUsqEbfIu}). In order to estimate the second term on the right hand side of (\ref{eq:RbfIdef}), it is sufficient to
  appeal to (\ref{eq:wGonebfIbfJest}). This yields
  \[
  \left|\textstyle{\sum}_{1\leq |\bfJ|\leq |\bfI|}\mfG_{\bfI,\bfJ}^{0}E_{\bfJ}u\right|
  \leq C_{a}\ldr{\tau-\tau_{c}}^{\eta_{a}}e^{(\varpi_{A}+\e_{\Spe})(\tau-\tau_{c})}
    \ldr{\tau_{c}}^{\eta_{b}}e^{\e_{\Spe}\tau_{c}}\hGe_{l}^{1/2}(\tau_{c})
  \]
  on $A^{+}_{c}(\g)$ for $0\leq |\bfI|\leq k-m_{0}$. Here $C_{a}$, $\eta_{a}$ and $\eta_{b}$ have the same dependence as in the case of
  (\ref{eq:dtausqEbfIuminushUsqEbfIu}). Combining the above estimates yields an estimate for $\mfR_{\pre,\bfI}u$.

  Next, we wish to estimate $\mfR_{\rocor,\bfI,\bfJ,\xi}u$. Before doing so, note that
  \[
  |\hU(\mfC_{\bfI,\xi})|=\hN^{-1}|\chi(\mfC_{\bfI,\xi})|\leq C_{a}e^{\e_{\Spe}\tau}
  \]
  in $A^{+}(\g)$, where we appealed to (\ref{eq:muminmainlowerbound}), (\ref{eq:varrhominustauestimate}) and (\ref{eq:hNinvchipsiestloc}); and
  $C_{a}$ only depends on $|\bfI|$, $c_{\robas}$, $c_{\chi,2}$, $(\bM,\bge_{\refer})$ and a lower bound on $\theta_{0,-}$. Next, note that
  \[
  \hU^{2}(\mfC_{\bfI,\xi})=\hU(\ln\hN)\hN^{-1}\chi(\mfC_{\bfI,\xi})-\hN^{-1}(\ml_{\hU}\chi)(\mfC_{\bfI,\xi})+\hN^{-1}\chi[\hN^{-1}\chi(\mfC_{\bfI,\xi})].
  \]
  Appealing to (\ref{eq:Aialphadef}), (\ref{eq:dotchirelations}), (\ref{eq:muminmainlowerbound}), (\ref{eq:varrhominustauestimate}),
  Remark~\ref{remark:chiclvarrhodecay} and the assumptions, it can thus be demonstrated that
  \[
  |\hU^{2}(\mfC_{\bfI,\xi})|\leq C_{a}\ldr{\tau}^{\cweight}e^{\e_{\Spe}\tau}
  \]
  in $A^{+}(\g)$, where $C_{a}$ only depends on $|\bfI|$, $c_{\cweight,0}$, $(\bM,\bge_{\refer})$ and a lower bound on $\theta_{0,-}$. Combining these
  estimates with the above estimates for $\mfR_{\pre,\bfI}u$; the definition of $\mfR_{\rocor,\bfI,\bfJ,\xi}$; and the assumptions yields the conclusion
  that
  \begin{equation}\label{eq:mfRbfIuestimate}
    |\mfR_{\bfI}u|\leq C_{a}\ldr{\tau-\tau_{c}}^{\eta_{a}}e^{(\varpi_{A}+\e_{\Spe})(\tau-\tau_{c})}
    \ldr{\tau_{c}}^{\eta_{b}}e^{\e_{\Spe}\tau_{c}}\hGe_{l}^{1/2}(\tau_{c})
  \end{equation}
  on $A^{+}_{c}(\g)$ for $0\leq |\bfI|\leq k-m_{0}-1$. Here $C_{a}$, $\eta_{a}$ and $\eta_{b}$ have the same dependence as in the case of
  (\ref{eq:dtausqEbfIuminushUsqEbfIu}).

  \textbf{Inductive assumptions.} 
  Next, we wish to simplify $L_{\bfI}u$ by imposing two inductive assumptions, one corresponding to the statement of the theorem and one corresponding
  to the statement of Remark~\ref{remark:higherorderestimatesimprovedestimates}. Fix $1\leq j\leq k-m_{0}-1$. The inductive assumption is 
  that there are functions $U_{\bfJ,m}$ for $|\bfJ|<j$ and $m=0,1,2$, depending only on $\tau$, such that one of the following estimates hold:
  \begin{align}
    |\hU^{m}E_{\bfJ}u-U_{\bfJ,m}(\tau)| \leq & C_{a}\ldr{\tau_{c}}^{\eta_{b}}\ldr{\tau-\tau_{c}}^{\eta_{a}}e^{(\varpi_{A}+\b)(\tau-\tau_{c})}
    \hGe_{l}^{1/2}(\tau_{c}),\label{eq:hUmEbfJuminusUbfJm}\\
    |\hU^{m}E_{\bfJ}u-U_{\bfJ,m}(\tau)| \leq & C_{a}\ldr{\tau_{c}}^{\eta_{b}}e^{\b\tau_{c}}\ldr{\tau-\tau_{c}}^{\eta_{a}}e^{(\varpi_{A}+\b)(\tau-\tau_{c})}
    \hGe_{l}^{1/2}(\tau_{c}),\label{eq:hUmEbfJuminusUbfJmRemark}
  \end{align}
  on $A^{+}_{c}(\g)$ for $m=0,1,2$ and $0\leq |\bfJ|<j$. Here $C_{a}$, $\eta_{a}$ and $\eta_{b}$ have the same dependence as in the case of
  (\ref{eq:dtausqEbfIuminushUsqEbfIu}). Moreover, the first assumption corresponds to the statement of the theorem and the second corresponds to
  the statement of Remark~\ref{remark:higherorderestimatesimprovedestimates}. We also assume, inductively, that
  \begin{equation}\label{eq:UbfJmestimate}
    |U_{\bfJ,m}(\tau)|\leq C_{a}\ldr{\tau_{c}}^{\eta_{b}}\ldr{\tau-\tau_{c}}^{\eta_{a}}e^{\varpi_{A}(\tau-\tau_{c})}\hGe_{l}^{1/2}(\tau_{c})
  \end{equation}
  for $\tau\leq \tau_{c}$, $m=0,1,2$ and $0\leq |\bfJ|\leq j$. Here $C_{a}$, $\eta_{a}$ and $\eta_{b}$ have the same dependence as in the case of
  (\ref{eq:dtausqEbfIuminushUsqEbfIu}). Note that by combining (\ref{eq:UbfJmestimate}) with either (\ref{eq:hUmEbfJuminusUbfJm}) or
  (\ref{eq:hUmEbfJuminusUbfJmRemark}) yields (\ref{eq:UbfJmestimate}) with $U_{\bfJ,m}$ replaced by $\hU^{m}E_{\bfJ}u$. To begin with, it is of interest to
  verify that the inductive assumption is satisfied for $j=1$. Note to this end, that by defining $U_{0,m}$, $m=0,1,2$, as in the statement of the theorem,
  (\ref{eq:hUmEbfJuminusUbfJm}), (\ref{eq:hUmEbfJuminusUbfJmRemark}) and (\ref{eq:UbfJmestimate}) are satisfied for $\bfJ=0$ and $m=0,1$. This is
  an immediate consequence of Theorem~\ref{thm:leadingorderasymptoticszero} and Remark~\ref{remark:Vinftyimprovement}. That (\ref{eq:UbfJmestimate}) holds
  for $\bfJ=0$ and $m=2$ follows from the definition of $U_{0,2}$, cf. (\ref{eq:Uzeromdef}), and the fact that (\ref{eq:UbfJmestimate}) holds for
  $\bfJ=0$ and $m=0,1$. Finally, in order to verify that (\ref{eq:hUmEbfJuminusUbfJm}) and (\ref{eq:hUmEbfJuminusUbfJmRemark}) hold for $\bfJ=0$ and $m=2$,
  it is sufficient to appeal to the fact that they hold for $\bfJ=0$ and $m=0,1$; the equation (\ref{eq:maindominant}); and arguments similar to the above. 

  \textbf{Inductive step.} In order to take the inductive step, let $L_{\bfI}u=\sfL_{\bfI}+\mfL_{\bfI}$, where
  \[
  \sfL_{\bfI}(\tau):=\textstyle{\sum}_{|\xi|<|\bfI|}\textstyle{\sum}_{m=0}^{2}L_{\bfI,\xi}^{m}(\bx_{0},\tau)U_{\xi,m}(\tau),\ \ \
  \mfL_{\bfI}:=L_{\bfI}u-\sfL_{\bfI}
  \]
  and $L_{\bfI,\xi}^{m}$ is given by (\ref{eq:LmbfIbfJitoPGandb}) and (\ref{eq:LbfIcoeffdef}). In other words, we have localised the coefficients of
  $L_{\bfI}u$ as in (\ref{eq:Zzerolochalloc}). Note that we can equally well localise the coefficients along the causal curve $\g$. Combining
  (\ref{eq:hUmEbfJuminusUbfJm}), (\ref{eq:UbfJmestimate}) and the assumptions yields
  \begin{equation}\label{eq:mfLbfIestim}
    |\mfL_{\bfI}| \leq  C_{a}\ldr{\tau_{c}}^{\eta_{b}}\ldr{\tau-\tau_{c}}^{\eta_{a}}e^{(\varpi_{A}+\b)(\tau-\tau_{c})}\hGe_{l}^{1/2}(\tau_{c})
  \end{equation}
  on $A^{+}_{c}(\g)$ for $0\leq |\bfI|\leq k-m_{0}-1$. Combining (\ref{eq:hUmEbfJuminusUbfJmRemark}), (\ref{eq:UbfJmestimate}) and the assumptions yields
  \begin{equation}\label{eq:mfLbfIestimRemark}
    |\mfL_{\bfI}| \leq  C_{a}\ldr{\tau_{c}}^{\eta_{b}}e^{\b\tau_{c}}\ldr{\tau-\tau_{c}}^{\eta_{a}}e^{(\varpi_{A}+\b)(\tau-\tau_{c})}\hGe_{l}^{1/2}(\tau_{c})
  \end{equation}
  on $A^{+}_{c}(\g)$ for $0\leq |\bfI|\leq k-m_{0}-1$. In both of these estimates, $C_{a}$, $\eta_{a}$ and $\eta_{b}$ have the same dependence as in the case of
  (\ref{eq:dtausqEbfIuminushUsqEbfIu}). Combining (\ref{eq:EbfImaindominant}), (\ref{eq:ascoeffhUtaureplestimatesi}) and (\ref{eq:mfRbfIuestimate})
  with (\ref{eq:mfLbfIestim}) or (\ref{eq:mfLbfIestimRemark}) yields the conclusion that
  \begin{equation}\label{eq:modelODEbfI}
    -\d_{\tau}^{2}E_{\bfI}u+Z^{0}_{\infty}\d_{\tau}E_{\bfI}u+\hal_{\infty}E_{\bfI}u=\sfL_{\bfI}+\sfR_{\bfI}.
  \end{equation}
  Here
  \begin{equation}\label{eq:sfRbfIestimate}
    |\sfR_{\bfI}|\leq C_{a}\ldr{\tau_{c}}^{\eta_{b}}\ldr{\tau-\tau_{c}}^{\eta_{a}}e^{(\varpi_{A}+\b)(\tau-\tau_{c})}\hGe_{l}^{1/2}(\tau_{c})
  \end{equation}
  on $A^{+}_{c}(\g)$ for $0\leq |\bfI|\leq k-m_{0}-1$, assuming (\ref{eq:mfLbfIestim}) is the relevant estimate. Moreover,
  \begin{equation}\label{eq:sfRbfIestimateRemark}
    |\sfR_{\bfI}|\leq C_{a}\ldr{\tau_{c}}^{\eta_{b}}e^{\b\tau_{c}}\ldr{\tau-\tau_{c}}^{\eta_{a}}e^{(\varpi_{A}+\b)(\tau-\tau_{c})}\hGe_{l}^{1/2}(\tau_{c})
  \end{equation}
  on $A^{+}_{c}(\g)$ for $0\leq |\bfI|\leq k-m_{0}-1$, assuming (\ref{eq:mfLbfIestimRemark}) is the relevant estimate. In the case of both estimates, $C_{a}$,
  $\eta_{a}$ and $\eta_{b}$ have the same dependence as in the case of (\ref{eq:dtausqEbfIuminushUsqEbfIu}). At this stage, we can evaluate the equation
  (\ref{eq:modelODEbfI}) at $(\bx_{0},\tau)$ in order to obtain an ODE for $(E_{\bfI}u)(\bx_{0},\tau)$. The resulting equation can be written 
  \begin{equation}\label{eq:PsiequationbfIcase}
    \d_{\tau}\Psi=A_{0} \Psi-H_{1}-H_{2},
  \end{equation}
  where $A_{0}$ is given by (\ref{eq:AzeroAremdef}) and
  \[
  \Psi(\tau):=\left(\begin{array}{c} (E_{\bfI}u)(\bx_{0},\tau) \\ (\d_{\tau}E_{\bfI}u)(\bx_{0},\tau)\end{array}\right),\ \
  H_{1}(\tau):=\left(\begin{array}{c} 0 \\ \sfL_{\bfI}(\tau) \end{array}\right),\ \
  H_{2}(\tau):=\left(\begin{array}{c} 0 \\ \sfR_{\bfI}(\bx_{0},\tau) \end{array}\right).
  \]
  \textbf{Analysing the asymptotics.} Introducing 
  \begin{equation}\label{eq:tPsidef}
    \tPsi(\tau):=\Psi(\tau)-\int_{\tau}^{\tau_{c}}e^{A_{0}(\tau-s)}H_{1}(s)ds,
  \end{equation}
  the equation (\ref{eq:PsiequationbfIcase}) yields the conclusion that $\d_{\tau}\tPsi=A_{0} \tPsi-H_{2}$. Due to the definition of $H_{2}$, it is clear
  that $|H_{2}|$ can be estimated by the right hand side of either (\ref{eq:sfRbfIestimate}) or
  (\ref{eq:sfRbfIestimateRemark}), depending on the assumptions. At this stage, we can appeal to Lemma~\ref{lemma:asymptoticsmodelODE} with
  $B=A_{0}$; $k=2m_{\ros}$; $H=-H_{2}$; $\xi=\tPsi$; $\varpi_{B}=\varpi_{A}$; $\eta_{H}=\eta_{a}$; and $C_{H}$ given by one of
  \[
  C_{a}\ldr{\tau_{c}}^{\eta_{b}}\hGe_{l}^{1/2}(\tau_{c}),\ \ \
  C_{a}\ldr{\tau_{c}}^{\eta_{b}}e^{\b\tau_{c}}\hGe_{l}^{1/2}(\tau_{c}).
  \]
  Here $C_{H}$ is given by the first expression in case (\ref{eq:sfRbfIestimate}) is satisfied and by the second in case (\ref{eq:sfRbfIestimateRemark})
  is satisfied. In particular, there are thus $\Psi_{\bfI,\infty,a}\in E_{a}$ and $\Psi_{\bfI,\infty}\in\rn{2m_{\ros}}$ such that
  \begin{align}
    \left|\tPsi(\tau)-e^{A_{0}(\tau-\tau_{c})}\Psi_{\bfI,\infty,a}\right|
    \leq & C_{a}\ldr{\tau_{c}}^{\eta_{b}}\hGe_{l}^{1/2}(\tau_{c})\ldr{\tau-\tau_{c}}^{\eta_{a}}e^{(\varpi_{A}+\b)(\tau-\tau_{c})},\nonumber\\
    \left|\tPsi(\tau)-e^{A_{0}(\tau-\tau_{c})}\Psi_{\bfI,\infty}\right|
    \leq & C_{a}\ldr{\tau_{c}}^{\eta_{b}}e^{\b\tau_{c}}\hGe_{l}^{1/2}(\tau_{c})\ldr{\tau-\tau_{c}}^{\eta_{a}}e^{(\varpi_{A}+\b)(\tau-\tau_{c})},\label{eq:tPsiminuseAPsibfIinf}
  \end{align}
  where $\Psi_{\bfI,\infty}=\Psi_{\bfI,\infty,a}+\Psi_{b}(\tau_{c})$ and the latter estimate holds only in case (\ref{eq:sfRbfIestimateRemark}) is
  satisfied. Moreover, $C_{a}$, $\eta_{a}$ and $\eta_{b}$ have the same dependence as in the case of (\ref{eq:dtausqEbfIuminushUsqEbfIu}). In order to obtain
  these conclusions, we appealed to Lemma~\ref{lemma:asymptoticsmodelODE} and the fact that an estimate of the form (\ref{eq:Psiroughestimate}) holds in
  the present setting. We also obtain the conclusion that
  \[
  |\Psi_{\bfI,\infty,a}|+|\Psi_{\bfI,\infty}|\leq C_{a}\ldr{\tau_{c}}^{\eta_{b}}\hGe_{l}^{1/2}(\tau_{c}),
  \]
  where $C_{a}$ and $\eta_{b}$ have the same dependence as in the case of (\ref{eq:dtausqEbfIuminushUsqEbfIu}). Combining these estimates with observations
  concerning the spatial variation of the solution in $A^{+}_{c}(\g)$ (as in the end of the proof of Theorem~\ref{thm:leadingorderasymptoticszero}) yields the
  conclusion that
  \begin{equation*}
    \begin{split}
      & \left|\left(\begin{array}{c} E_{\bfI}u \\ \hU E_{\bfI}u\end{array}\right)-e^{A_{0}(\tau-\tau_{c})}\Psi_{\bfI,\infty,a}
        -\int_{\tau}^{\tau_{c}}e^{A_{0}(\tau-s)}\left(\begin{array}{c} 0 \\ \sfL_{\bfI}(s) \end{array}\right)ds\right|\\
        \leq & C_{a}\ldr{\tau_{c}}^{\eta_{b}}\hGe_{l}^{1/2}(\tau_{c})\ldr{\tau-\tau_{c}}^{\eta_{a}}e^{(\varpi_{A}+\b)(\tau-\tau_{c})}
    \end{split}
  \end{equation*}  
  on $A^{+}_{c}(\g)$ for all $0\leq |\bfI|\leq k-m_{0}-1$, where $C_{a}$, $\eta_{a}$ and $\eta_{b}$ have the same dependence as in the case of
  (\ref{eq:dtausqEbfIuminushUsqEbfIu}). Similarly, in case (\ref{eq:sfRbfIestimateRemark}) holds, 
  \begin{equation*}
    \begin{split}
      & \left|\left(\begin{array}{c} E_{\bfI}u \\ \hU E_{\bfI}u\end{array}\right)-e^{A_{0}(\tau-\tau_{c})}\Psi_{\bfI,\infty}
        -\int_{\tau}^{\tau_{c}}e^{A_{0}(\tau-s)}\left(\begin{array}{c} 0 \\ \sfL_{\bfI}(s) \end{array}\right)ds\right|\\
        \leq & C_{a}\ldr{\tau_{c}}^{\eta_{b}}e^{\b\tau_{c}}\hGe_{l}^{1/2}(\tau_{c})\ldr{\tau-\tau_{c}}^{\eta_{a}}e^{(\varpi_{A}+\b)(\tau-\tau_{c})}
    \end{split}
  \end{equation*}  
  on $A^{+}_{c}(\g)$ for all $0\leq |\bfI|\leq k-m_{0}-1$, where $C_{a}$, $\eta_{a}$ and $\eta_{b}$ have the same dependence as in the case of
  (\ref{eq:dtausqEbfIuminushUsqEbfIu}).

  Define $V_{\bfI,\infty,a}:=\Psi_{\bfI,\infty,a}$; define $V_{\bfI,\infty}:=\Psi_{\bfI,\infty}$ in case (\ref{eq:sfRbfIestimateRemark}) holds; and define
  $U_{\bfI,m}$, $m=0,1,2$, as in the statement of the theorem (or as in Remark~\ref{remark:higherorderestimatesimprovedestimates}). Due to the inductive
  assumption and the definitions, it can be verified that (\ref{eq:hUmEbfJuminusUbfJm}) (or (\ref{eq:hUmEbfJuminusUbfJmRemark})) and
  (\ref{eq:UbfJmestimate}) hold with $\bfJ$ replaced by $\bfI$ and $m=0,1$. Combining this information with the inductive assumption and the definitions,
  it also follows that (\ref{eq:UbfJmestimate}) holds with $\bfJ$ replaced by $\bfI$ and $m=2$. Finally, in order to prove that (\ref{eq:hUmEbfJuminusUbfJm})
  (or (\ref{eq:hUmEbfJuminusUbfJmRemark})) holds with $\bfJ$ replaced by $\bfI$ and $m=2$, it is sufficient to appeal to (\ref{eq:modelODEbfI}); the conclusions
  we have already derived for $\d_{\tau}^{m}E_{\bfI}u$, $m=0,1$; and (\ref{eq:dtausqEbfIuminushUsqEbfIu}). In order prove the uniqueness of
  $V_{\bfI,\infty,a}$, it is sufficient to proceed inductively and to appeal to arguments similar to the ones presented at the end of
  Lemma~\ref{lemma:asymptoticsmodelODE}.
\end{proof}

\chapter{Specifying the asymptotics}\label{chapter:specifyingtheasymptotics}

The final goal of these notes is to prove that we can specify the leading order asymptotics, given exponential convergence of $Z^{0}$ and $\hal$ along a
causal curve. This is the purpose of the present chapter. The idea of the proof is to define a set of initial data which has the same dimension as the
set of asymptotic data one wishes to specify. The evolution associated with the equation then defines a linear map from this set of initial data to the
set of asymptotic data. Given good enough estimates, one can then prove that this linear map between vector spaces of the same dimension is injective.
However, this also means that it is surjective and demonstrates that we can specify the leading order asymptotics. 

\section{Specifying the asymptotics}\label{section:specifyingtheasymptotics}

Our next goal is to prove that we can specify the leading order asymptotics of $E_{\omega}u$ and $\hU E_{\omega}u$ for $\rn{n}$-multiindices $\omega$
satisfying $|\omega|\leq k-m_{0}-1$. 

\begin{thm}\label{thm:specifyingtheasymptoticdata}
  Assume that the conditions of Theorem~\ref{thm:leadingorderasymptoticszeroho} are satisfied. Then, using the notation of
  Theorem~\ref{thm:leadingorderasymptoticszeroho}, the following holds. Fix vectors $v_{\omega}\in E_{a}$ for $\rn{n}$-multiindices $\omega$ satisfying
  $|\omega|\leq k-m_{0}-1$. Then, given $\tau_{c}$ close enough to $-\infty$, there is a solution to (\ref{eq:theeqreformEi}) with vanishing right
  hand side such that if $V_{\bfI,\infty,a}$ are the vectors uniquely determined by the solution as in the statement of
  Theorem~\ref{thm:leadingorderasymptoticszeroho}, then $V_{\bfI_{\omega},\infty,a}=v_{\omega}$, where $\bfI_{\omega}=(I_{1},\dots,I_{p})$ is the vector field
  multiindex such that $I_{j}\leq I_{j+1}$ for $j=1,\dots,p-1$ and such that $\upomega(\bfI_{\omega})=\omega$. 
\end{thm}
\begin{remark}
  Here $\upomega$ is given by Definition~\ref{def:upomegadef}.
\end{remark}
\begin{remark}
  The bound $\tau_{c}$ has to satisfy in order for the conclusions to hold is of the form $\tau_{c}\leq T_{c}$, where $T_{c}$ only depends on $s_{\cweight,l}$,
  $s_{\coeff,l}$, $c_{\cweight,k}$, $c_{\coeff,k}$, $d_{\a}$ (in case $\iota_{b}\neq 0$), $A_{0}$, $c_{\rem}$, $\e_{A}$, $(\bM,\bge_{\refer})$, a lower bound on
  $\theta_{0,-}$, a choice of local coordinates on $\bM$ around $\bx_{0}$ and a choice of a cut-off function near $\bx_{0}$.
\end{remark}
\begin{remark}
  The solutions constructed in the theorem are such that
  \begin{equation}\label{eq:VAplusestimatehoimprovedconstr}
    \begin{split}
      & \sum_{|\bfI|\leq k-m_{0}-1}\left|V_{\bfI}-e^{A_{0}(\tau-\tau_{c})}V_{\bfI,\infty,a}-\int_{\tau}^{\tau_{c}}
      e^{A_{0}(\tau-s)}\left(\begin{array}{c} 0 \\ \sfL_{\bfI}(s)\end{array}\right)ds\right|\\
      \leq & C_{a}\ldr{\tau_{c}}^{\eta_{b}}e^{\b\tau_{c}}\ldr{\tau-\tau_{c}}^{\eta_{a}}e^{(\varpi_{A}+\b)(\tau-\tau_{c})}
      \textstyle{\sum}_{|\omega|\leq k-m_{0}-1}|v_{\omega}|
    \end{split}    
  \end{equation}
  on $A^{+}_{c}(\g)$, where $C_{a}$ only depends on $s_{\cweight,l}$, $s_{\coeff,l}$, $c_{\cweight,k}$, $c_{\coeff,k}$, $d_{\a}$ (in case $\iota_{b}\neq 0$), $A_{0}$,
  $c_{\rem}$, $\e_{A}$, $(\bM,\bge_{\refer})$, a lower bound on $\theta_{0,-}$, a choice of local coordinates on $\bM$ around $\bx_{0}$ and a choice of a cut-off
  function near $\bx_{0}$; and $\eta_{a}$ and $\eta_{b}$ only depend on $\cweight$, $A_{0}$, $n$, and $k$. Note, in particular, that by choosing $\tau_{c}$
  close enough to
  $-\infty$, the factor $C_{a}\ldr{\tau_{c}}^{\eta_{b}}e^{\b\tau_{c}}$ appearing on the right hand side of (\ref{eq:VAplusestimatehoimprovedconstr}) can be chosen
  to be as small as we wish. 
\end{remark}
\begin{proof}
  Most of the arguments necessary to prove that we can specify the asymptotics are already present in the proof of
  Theorem~\ref{thm:leadingorderasymptoticszeroho}. In particular, Theorem~\ref{thm:leadingorderasymptoticszeroho} yields a linear map from initial data at
  $\tau_{c}$ to the asymptotic data. Restricting this map to a suitable finite dimensional subspace, it is, in the end, possible to demonstrate that the map
  is bijective, which gives the desired conclusion. The main difference in comparison with earlier results is that it is here of crucial importance to 
  fix a $\tau_{c}$ close to $-\infty$. The reason we need to choose $\tau_{c}$ close to $-\infty$ is that the constants appearing in the estimates are of the
  form
  \begin{equation}\label{eq:modelconstantspecasymptotics}
    C_{a}\ldr{\tau_{c}}^{\eta_{b}}e^{\b\tau_{c}}\hGe_{l}^{1/2}(\tau_{c}).
  \end{equation}
  The point here is that the initial data we specify at $\tau_{c}$ are such that $\hGe_{l}^{1/2}(\tau_{c})\leq C_{b}|v|$, where $v$ corresponds to the size
  of the initial data (where we have restricted the initial data to a finite dimensional subspace, and $v$ corresponds to an element in this subspace).
  In particular, $\hGe_{l}(\tau_{c})$ can be bounded by a constant independent of the choice of $\tau_{c}$. Thus, given $\e>0$, letting $\tau_{c}$ be close
  enough to $-\infty$, the constant (\ref{eq:modelconstantspecasymptotics}) can be assumed to be bounded by $\e |v|$. It is this kind of estimate which
  will allow us to prove bijectivity of the linear map mentioned above. 
  
  \textit{Choosing a finite dimensional subspace of initial data.} From the above, it is clear that we need to specify a suitable finite dimensional
  subspace of initial data. Let, to this end, $(\msU,\bsfx)$ be local coordinates on $\bM$ such that $\bx_{0}\in\msU$, $\bsfx(\bx_{0})=0$ and such that
  \[
  \d_{\bsfx^{i}}|_{\bx_{0}}=E_{i}|_{\bx_{0}}.
  \]
  Let $\phi$ be a smooth function on $\bM$ such that $\phi(\bx)=1$ for $\bx$ in a neighbourhood of $\bx_{0}$ and such that $\phi$ has support contained in
  $\msU$. Let $\omega$ be an $\rn{n}$-multiindex, $v\in\rn{2m_{\ros}}$ and define
  \[
  \phi_{\omega,v}(\bx)=(\omega_{1}!\cdots\omega_{n}!)^{-1}\phi(\bx)\bsfx^{\omega}(\bx)v.
  \]
  Here 
  \[
  \bsfx^{\omega}(\bx):=\textstyle{\prod}_{m=1}^{n}[\bsfx^{m}(\bx)]^{\omega_{m}}.
  \]
  Then $(E_{\bfI}\phi_{\omega,v})(\bx_{0})=v$ if $\omega=\upomega(\bfI)$ and $(E_{\bfI}\phi_{\omega,v})(\bx_{0})=0$ if $|\upomega(\bfI)|\leq|\omega|$  and
  $\upomega(\bfI)\neq\omega$ (note that for an $\rn{n}$-multiindex $\omega$, $|\omega|$ denotes the sum of the components of $\omega$). Let $\msX_{j}$ be the
  subspace of $C^{\infty}(\bM,\rn{2m_{\ros}})$ spanned by $\phi_{\omega,v}$ for $|\omega|=j$ and $v\in \rn{2m_{\ros}}$; and let $\msX_{j,a}$ be the subspace
  of $C^{\infty}(\bM,\rn{2m_{\ros}})$ spanned by $\phi_{\omega,v}$ for $|\omega|=j$ and $v\in E_{a}$. Note that $E_{a}$ and $\msX_{0,a}$ are
  isomorphic. The isomorphism is given by the map $\msT_{0}:E_{a}\rightarrow \msX_{0,a}$ defined by $\msT_{0}(v)=\phi_{0,v}$.

  \textit{Definition of the linear map.} Define a map $\msL_{c,0}:\msX_{0,a}\rightarrow E_{a}$ as follows. Given $\psi\in \msX_{0,a}$, let
  \begin{equation}\label{eq:uutauindataattauceqpsi}
    \left(\begin{array}{c} u(\cdot,\tau_{c}) \\ u_{\tau}(\cdot,\tau_{c})\end{array}\right)=\psi.
  \end{equation}
  Solving the equation with this initial data yields $\msL_{c,0}\psi:=V_{\infty,a}$. Since the equation is linear and homogeneous, and since $V_{\infty,a}$
  is uniquely determined by the solution, the map $\msL_{c,0}$ is linear. In what follows, we wish to prove that
  $\msL_{c,0}\circ \msT_{0}:E_{a}\rightarrow E_{a}$ is an isomorphism. However, due to (\ref{eq:PsiminuseAzPsiinfest}), the remarks made immediately below
  this estimate and the fact that $\Psi_{b}(\tau_{c})=0$ in our setting, the following estimate holds:
  \[
  |\Psi-e^{A_{0}(\tau-\tau_{c})}V_{\infty,a}|\leq C_{a}\ldr{\tau_{c}}^{\eta_{b}}e^{\b\tau_{c}}\ldr{\tau-\tau_{c}}^{\eta_{a}}e^{(\varpi_{A}+\b)(\tau-\tau_{c})}\hGe_{l}^{1/2}(\tau_{c});
  \]
  note that $\Psi_{\infty,a}=V_{\infty,a}$. Putting $\tau=\tau_{c}$ in this estimate yields
  \begin{equation}\label{eq:mcVataucminusmcVainfty}
  |\Psi(\tau_{c})-V_{\infty,a}|\leq C_{a}\ldr{\tau_{c}}^{\eta_{b}}e^{\b\tau_{c}}\hGe_{l}^{1/2}(\tau_{c}).
  \end{equation}
  Since $\Psi(\tau_{c})=\Psi_{a}(\tau_{c})$, we can of course replace $\Psi(\tau_{c})$ with $\Psi_{a}(\tau_{c})$ on the left hand side. If we can prove that
  $\msL_{c,0}\circ\msT_{0}$ is injective for a suitable choice of $\tau_{c}$, then it follows that $\msL_{c,0}$ is surjective.

  \textit{Proving injectivity.} In order to prove injectivity, let us begin by estimating $\hGe_{l}(\tau_{c})$. Assuming $\omega$ to be an
  $\rn{n}$-multiindex with $|\omega|\leq k-m_{0}-1$ and $v\in E_{a}$, let $\psi=\phi_{\omega,v}$. Specifying the initial data at $\tau_{c}$ by
  (\ref{eq:uutauindataattauceqpsi}), we wish to prove that
  \begin{equation}\label{eq:hGelutaucestimatevomega}
    \hGe_{l}^{1/2}[u](\tau_{c})\leq C_{a}|v|.
  \end{equation}
  Note, to this end, that if $|\bfK|\leq l+1$, then
  \begin{equation}\label{eq:dtauebfKuEbfKuestimate}
    |(\d_{\tau}E_{\bfK}u)(\cdot,\tau_{c})|+|(E_{\bfK}u)(\cdot,\tau_{c})|\leq C_{a}|v|,
  \end{equation}
  where $C_{a}$ only depends on $l$, $(\bM,\bge_{\refer})$, $\phi$ and the local coordinates. Consider (\ref{eq:dhtaupsiminushUpsiestloc}) with
  $\tau=\tau_{c}$. Assume $\tau_{c}$ to be sufficiently close to $-\infty$ that $C_{b}\ldr{\tau_{c}}e^{\e_{\Spe}\tau_{c}}\leq 1/2$, where $C_{b}$ is the constant
  appearing in (\ref{eq:dhtaupsiminushUpsiestloc}). Then, for a smooth function $\varphi$, 
  \[
  |\hU(\varphi)|\leq |\d_{\tau}\varphi|+|\hU(\varphi)-\d_{\tau}\varphi|\leq |\d_{\tau}\varphi|+\frac{1}{2}|\hU(\varphi)|
  +\left(\textstyle{\sum}_{A}e^{-2\mu_{A}}|X_{A}\varphi|^{2}\right)^{1/2}
  \]
  on $A^{+}_{c}(\g)$. In particular,
  \[
  |\hU(\varphi)|\leq 2|\d_{\tau}\varphi|+2\left(\textstyle{\sum}_{A}e^{-2\mu_{A}}|X_{A}\varphi|^{2}\right)^{1/2}
  \]
  on $A^{+}_{c}(\g)$. 
  Combining this inequality with $\varphi$ replaced by $E_{\bfJ}u$ with (\ref{eq:muminmainlowerbound}) and (\ref{eq:dtauebfKuEbfKuestimate}) yields
  the conclusion that (\ref{eq:hGelutaucestimatevomega}) holds, where $C_{a}$ only depends on $l$, $c_{\robas}$, $(\bM,\bge_{\refer})$, a lower bound
  on $\theta_{0,-}$, $\phi$ and the local coordinates. Combining (\ref{eq:hGelutaucestimatevomega}) and (\ref{eq:mcVataucminusmcVainfty}) with
  $\psi=\phi_{0,v}$ yields
  \[
  \left|v-V_{\infty,a}\right|\leq C_{a}\ldr{\tau_{c}}^{\eta_{b}}e^{\b\tau_{c}}\left|v\right|.
  \]
  Assuming $\tau_{c}$ to be such that the factor in front of the absolute value on the right hand side is bounded from above by $1/2$, it follows that
  \[
  \left|v\right|\leq 2\left|V_{\infty,a}\right|=2|\msL_{c,0}\circ\msT_{0}(v)|.
  \]
  This demonstrates injectivity of $\msL_{c,0}\circ\msT_{0}$, and thereby the surjectivity of $\msL_{c,0}$.  

  \textbf{Estimating the quality of the approximation.} Assume the initial data at $\tau_{c}$ to be given by (\ref{eq:uutauindataattauceqpsi}), where
  $\psi$ belongs to a direct sum of $\msX_{j,a}$'s. Then $E_{\bfI}\psi$ takes all its values in $E_{a}$. As a consequence, $V_{\infty,a}=V_{\infty}$ and
  $V_{\bfI,\infty,a}=V_{\bfI,\infty}$. This is due to the fact that, with these initial data, the $\Psi$'s appearing in the proofs of
  Theorems~\ref{thm:leadingorderasymptoticszero}
  and \ref{thm:leadingorderasymptoticszeroho} are such that $\Psi_{b}(\tau_{c})=0$, and the fact that the construction of $V_{\infty,a}$, $V_{\infty}$,
  $V_{\bfI,\infty,a}$ and $V_{\bfI,\infty}$ is based on an application of Lemma~\ref{lemma:asymptoticsmodelODE}; note that the relation between $\xi_{\infty}$
  and $\xi_{\infty,a}$ in Lemma~\ref{lemma:asymptoticsmodelODE} is given by $\xi_{\infty}=\xi_{\infty,a}+\xi_{b}(\tau_{c})$. Due to
  Remarks~\ref{remark:Vinftyimprovement} and \ref{remark:higherorderestimatesimprovedestimates}, the estimates (\ref{eq:VAplusestimate}) and
  (\ref{eq:VAplusestimateho}) can then be improved, in that an extra factor $e^{\b\tau_{c}}$ can be inserted on the right hand side in each of these
  estimates. In fact, due to the proofs, (\ref{eq:PsiminuseAzPsiinfest}) holds with $\Psi_{\infty}$ replaced by $V_{\infty,a}$, and
  (\ref{eq:tPsiminuseAPsibfIinf}) holds with $\Psi_{\bfI,\infty}$ replaced by $V_{\bfI,\infty,a}$.
  Inductively, it can also be demonstrated that $U_{\bfI,m}$, $m=0,1,2$, depends linearly on the initial data. The inductive step consists
  of the observation that if $U_{\bfJ,m}$, $m=0,1,2$, depends linearly on the initial data for $|\bfJ|<k$, then $\sfL_{\bfI}$ depends linearly on the initial
  data for $|\bfI|=k$, so that $\tPsi$ introduced in (\ref{eq:tPsidef}) depends linearly on the initial data. Since $V_{\bfI,\infty,a}$ is defined linearly
  in terms of $\tPsi$, it follows that $V_{\bfI,\infty,a}$ depends linearly on the initial data. Inserting this information into the definition of
  $U_{\bfI,m}$ yields the conclusion that $U_{\bfI,m}$, $m=0,1,2$, depends linearly on the initial data.

  \textbf{Specifying the asymptotic data.} Evaluating (\ref{eq:PsiminuseAzPsiinfest}) and (\ref{eq:tPsiminuseAPsibfIinf}) at $\tau_{c}$ and keeping the
  above observations in mind yields
  \begin{equation}\label{eq:psibxzeroVinftyadiffest}
    |\psi(\bx_{0})-V_{\infty,a}|+|(E_{\bfI}\psi)(\bx_{0})-V_{\bfI,\infty,a}|\leq C_{a}\ldr{\tau_{c}}^{\eta_{b}}e^{\b\tau_{c}}\hGe_{l}^{1/2}(\tau_{c})
  \end{equation}
  for all $|\bfI|\leq k-m_{0}-1$. 

  \textit{Choosing a finite dimensional subspace of initial data.} At this stage, note that there is a linear map from initial data at $\tau_{c}$ to
  $V_{\infty,a}$ and $V_{\bfI,a,\infty}$. In order to prove that we can specify the asymptotic data, we need, as in the case of $\omega=0$, to
  choose a convenient finite dimensional subspace of initial data. Let $W_{j}=E_{a}^{q_{j}}$, where $q_{j}$ denotes the number of distinct $\rn{n}$-multiindices
  $\omega$ with $|\omega|\leq j$; and let $Y_{j}$ be the direct sum of $\msX_{q,a}$ for $q\leq j$ (where $\msX_{q,a}$ is defined as above). Then we can define
  $\msL_{c,j}:Y_{j}\rightarrow W_{j}$ as follows. Given $\psi\in Y_{j}$, let $u$ be the solution to the equation corresponding to initial data given
  by (\ref{eq:uutauindataattauceqpsi}). Then the zeroth component of $\msL_{c,j}(\psi)$ is given by $V_{\infty,a}$, and if $|\omega|\leq j$, the component of
  $\msL_{c,j}(\psi)$ corresponding to $\omega$ is given by $V_{\omega,\infty,a}$ (strictly speaking by $V_{\bfI_{\omega},\infty,a}$). Due to the above arguments, it
  is clear that these components depend linearly on $\psi$. Let $\msT_{j}:W_{j}\rightarrow Y_{j}$ be defined by the condition that it takes
  $v_{\omega}\in E_{a}$, $|\omega|\leq j$, to
  \[
  \textstyle{\sum}_{|\omega|\leq j}\phi_{\omega,v_{\omega}}.
  \]
  To prove that $\msL_{c,j}$ is surjective, it is sufficient to prove that $\msL_{c,j}\circ\msT_{j}$ is an isomorphism.

  \textit{Proving surjectivity, basic estimates.} Given $w\in W_{j}$, corresponding to $v_{\omega}\in E_{a}$, $|\omega|\leq j$, let $u$ be the solution to
  the equation corresponding to initial data given by (\ref{eq:uutauindataattauceqpsi}), where $\psi=\msT_{j}(w)$. To begin with, it is of interest to
  verify that, for $\tau_{c}$ close enough to $-\infty$,
  \begin{equation}\label{eq:hGelutaucestimatevomegaaux}
    \hGe_{l}^{1/2}[u](\tau_{c})\leq C_{a}\textstyle{\sum}_{|\omega|\leq j}|v_{\omega}|.
  \end{equation}
  However, this estimate follows from the fact that (\ref{eq:hGelutaucestimatevomega}) holds in case the initial data
  $\psi$ in (\ref{eq:uutauindataattauceqpsi}) are given by $\phi_{\omega,v}$. Note also that $C_{a}$ only depends on $l$, $c_{\robas}$, $(\bM,\bge_{\refer})$,
  a lower bound on $\theta_{0,-}$, $\phi$ and the choice of local coordinates. Combining (\ref{eq:hGelutaucestimatevomegaaux}) with
  (\ref{eq:psibxzeroVinftyadiffest}) yields the conclusion that
  \begin{equation}\label{eq:theestimateleadingtoinjectivity}
    \textstyle{\sum}_{|\omega|\leq j}|(E_{\omega}\psi)(\bx_{0})-V_{\omega,\infty,a}|
    \leq C_{a}\ldr{\tau_{c}}^{\eta_{b}}e^{\b\tau_{c}}\textstyle{\sum}_{|\omega|\leq j}|v_{\omega}|.
  \end{equation}

  \textit{Proving surjectivity.} As mentioned above, it is sufficient to prove that $\msL_{c,j}\circ\msT_{j}$ is an isomorphism. Thus, since
  $\msL_{c,j}\circ\msT_{j}$ is a linear map from $W_{j}$ (a finite dimensional vector space) to itself, it is sufficient to prove that this map is injective.
  Assume, to this end, that $w\in W_{j}$ is such that $\msL_{c,j}\circ\msT_{j}(w)=0$. Combining this assumption with
  (\ref{eq:theestimateleadingtoinjectivity}) yields
  \begin{equation}\label{eq:mcVomegaaestinjectarg}
    \textstyle{\sum}_{|\omega|\leq j}|(E_{\omega}\psi)(\bx_{0})|\leq C_{a}\ldr{\tau_{c}}^{\eta_{b}}e^{\b\tau_{c}}\textstyle{\sum}_{|\omega|\leq j}|v_{\omega}|.
  \end{equation}
  Note that there is a bijection taking $w\in W_{j}$ to $(E_{\omega}\psi)(\bx_{0})$ for $|\omega|\leq j$. Moreover, $v_{0}=\psi(\bx_{0})$; and if
  $1\leq |\omega|\leq j$,
  then
  \[
  v_{\omega}=(E_{\omega}\psi)(\bx_{0})-\textstyle{\sum}_{|\xi|<|\omega|}q_{\omega,\xi}v_{\xi},
  \]
  where $q_{\omega,\xi}$ can be calculated in terms of functions that are independent of $\tau_{c}$ (so that, in particular, $q_{\omega,\xi}$ is independent
  of $\tau_{c}$). By an inductive argument, it follows that there are constants $r_{\omega,\xi}$ (depending only on $\phi$ and the choice of coordinates
  $\bsfx$) such that 
  \[
  v_{\omega}=(E_{\omega}\psi)(\bx_{0})-\textstyle{\sum}_{|\xi|<|\omega|}r_{\omega,\xi}(E_{\xi}\psi)(\bx_{0}).
  \]
  Inserting this information into (\ref{eq:mcVomegaaestinjectarg}) yields the conclusion that
  \[
  \textstyle{\sum}_{|\omega|\leq j}|(E_{\omega}\psi)(\bx_{0})|\leq C_{a}\ldr{\tau_{c}}^{\eta_{b}}e^{\b\tau_{c}}\textstyle{\sum}_{|\omega|\leq j}
  |(E_{\omega}\psi)(\bx_{0})|.
  \]
  Letting $\tau_{c}$ be close enough to $-\infty$, so that $C_{a}\ldr{\tau_{c}}^{\eta_{b}}e^{\b\tau_{c}}\leq 1/2$, it follows that $(E_{\omega}\psi)(\bx_{0})=0$
  for all $|\omega|\leq j$. This implies that $v_{\omega}=0$ for all $\omega$ with $|\omega|\leq j$. Thus $w=0$, and the map is injective.
\end{proof}

\part{Appendices}

\appendix

\chapter{Terminology and justification of assumptions}

The purpose of the present chapter is to introduce some of the terminology we use in these notes. We also provide a more detailed motivation for some of the
assumptions stated in the introduction. We begin, in Section~\ref{section:globalframefinitecover}, by proving that if $\mK$ has distinct eigenvalues but
does not have a global
frame, then it is sufficient to take a finite covering space of $\bM$ in order for the expansion normalised Weingarten map on the resulting spacetime
to have a global frame. In Section~\ref{section:timederivativeofmK}, we then define $\hml_{U}\mK$. To end the chapter, we describe how the conditions on
the relative spatial variation of $\theta$ in some situations essentially follow from the assumption that the blow up of the mean curvature is synchronized
and assumptions on the deceleration parameter and the lapse function. This is the subject of Section~\ref{section:syncblowupmc}.

\section{Existence of a global frame}\label{section:globalframefinitecover}

As pointed out in Remark~\ref{remark:existenceglobalframe}, the non-degeneracy of $\mK$ is not sufficient to guarantee the existence of a global
frame. However, the existence of a frame can be ensured by taking a finite cover of $\bM$, as we now demonstrate. The proof consists of a simple
application of basic ideas from algebraic topology. However, since the subject of these notes is the asymptotics of solutions to partial differential
equations, we write out the details here. 

\begin{lemma}
  Let $(M,g)$ be a time oriented Lorentz manifold. Assume it to have an expanding partial pointed foliation and $\mK$ to be non-degenerate on $I$.
  Assuming $\bM$ to be connected, there is a connected finite covering space $\tM$ of $\bM$ with covering map $\pi_{a}:\tM\rightarrow \bM$. Letting
  $\pi_{b}:\tM\times I\rightarrow\bM\times I$ be defined by $\pi_{b}(\tx,t)=[\pi_{a}(\tx),t]$, then $\pi_{b}$ is also a covering map. Letting
  $\tg=\pi_{b}^{*}g$, $\pi_{b}$ is a local isometry. Moreover, the expansion normalised Weingarten map associated with $\tg$ and the foliation
  $\tM\times I$ has a global frame. 
\end{lemma}
\begin{remark}
  The notion of a global frame is introduced in Definition~\ref{def:XAellA}; on $\tM$ we take it to be understood that the reference metric
  is $\pi_{a}^{*}\bge_{\refer}$. 
\end{remark}
\begin{proof}
  Let $\ell_{1}<\cdots<\ell_{n}$ denote the distinct eigenvalues of $\mK$. Let $t\in I$, $\bx\in \bM$, $p=(\bx,t)$ and
  $A\in\{1,\dots,n\}$. Then there are two tangent vectors to $\bM$ at $\bx$, say $\xi^{\pm}_{A,p}$ such that $\xi^{\pm}_{A,p}$ is an eigenvector of
  $\mK|_{p}$ corresponding to $\ell_{A}(p)$ with norm one relative to $\bge_{\refer}$. Let
  \[
  N:=\{(\xi^{i_{1}}_{1,p},\dots,\xi^{i_{n}}_{n,p})\times \{t\}\ :\ t\in I,\ \bx\in\bM,\ p=(\bx,t),\ i_{j}\in \{+,-\},\ j=1,\dots,n\}
  \]
  and define $\pi:N\rightarrow\bM\times I$ by $\pi(\xi^{i_{1}}_{1,p},\dots,\xi^{i_{n}}_{n,p},t)=p$. To begin with, we prove that $N$ has the
  structure of a smooth manifold and that $\pi$ is a covering map. 

  Let $q\in N$ with $(\bx,t)=\pi(q)$. Then there is an open neighbourhood $U_{q}$ of $\bx\in\bM$ and an interval $I_{q}\subset I$, open relative to
  $I$ and containing $t$,
  such that on $U_{q}\times I_{q}$, there is a unique collection $\{X_{A}\}$, $A=1,\dots,n$, of smooth vector fields tangent to the leaves of the foliation
  which, for $A\in\{1,\dots,n\}$, is such that $\mK X_{A}=\ell_{A}X_{A}$ (no summation); $|X_{A}|_{\bge_{\refer}}=1$; and $X_{A}|_{(\bx,t)}=\xi^{i_{A}}_{A,p}$.
  We can think of $U_{q}$ as being the domain of some coordinates $\psi_{q}:U_{q}\rightarrow \rn{n}$ on $\bM$, and, when convenient, we can
  assume $U_{q}$ and $I_{q}$ to be members of a countable basis of $\bM$ and $I$ respectively. Define
  \[
  V_{q}:=\{[X_{1}(\by,s),\dots,X_{n}(\by,s),s]:\by\in U_{q},\ s\in I_{q}\}
  \]
  and $\Psi_{q}:V_{q}\rightarrow\rn{n+1}$ (or $\mathbb{H}^{n+1}$) by $\Psi[X_{1}(\by,s),\dots,X_{n}(\by,s),s]=[\psi_{q}(\by),s]$. Note that $\Psi_{q}$ is
  one-to-one. In fact, all the conditions of \cite[Proposition~42, p.~23]{oneill} are satisfied (note also that this proposition can be generalised
  to the case of manifolds with boundary). Thus, due to \cite[Proposition~42, p.~23]{oneill}, demanding that $\Psi_{q}$ be
  homeomorphisms endows $N$ with a unique Hausdorff topology. Moreover, there is a
  complete smooth atlas on $N$ such that each of the $(\Psi_{q},V_{q})$ are coordinate neighbourhoods. Finally, the manifold $N$ is second countable.
  Next, note that $\pi$ is a covering map; cf., e.g., \cite[Definition~7, p.~443]{oneill}. 

  Next, let $\tM:=\pi^{-1}(\bM\times \{t_{0}\})$ and let $\pi_{a}:=p_{1}\circ\pi|_{\tM}$, where $p_{1}:\bM\times I\rightarrow\bM$ is defined by
  $p_{1}(\bx,t)=\bx$. Then $\pi_{a}:\tM\rightarrow\bM$ is a smooth covering map. Define $\xi:\tM\times I\rightarrow \bM\times I$ by
  $\xi(\tx,t)=[\pi_{a}(\tx),t]$. Note that $\xi$ is homotopy equivalent to $\xi_{0}$ defined by $\xi_{0}(\tx,s)=\pi(\tx)$. In particular,
  \[
  \xi_{*}=\xi_{0*}:\pi_{1}(\tM\times I)\rightarrow\pi_{1}(\bM\times I).
  \]
  On the other hand, $\xi_{0}$ factors through $N$ by $\xi_{0}(\tx,s)=\pi\circ \psi_{1}(\tx,s)$, where $\psi_{1}(\tx,s)=\tx$. This means that
  \[
  \xi_{*}[\pi_{1}(\tM\times I)]=\pi_{*}\circ \psi_{1*}[\pi_{1}(\tM\times I)]\subseteq\pi_{*}(N).
  \]
  In particular, there is a unique lift of $\xi$ to a map $\Xi:\tM\times I\rightarrow N$ such that $\xi=\pi\circ\Xi$ and such that the restriction of
  $\Xi$ to $\tM\times \{t_{0}\}$ is given by $\Xi(\tx,t_{0})=\iota (\tx)$, where $\iota:\tM\rightarrow N$ is the inclusion.

  In order to define a map from $N$ to $\tM\times I$, let $q=(\xi^{i_{1}}_{1,p},\dots,\xi^{i_{n}}_{n,p})\times \{t\}\in N$, where $p=(\bx,t)$ and
  $\bx\in\bM$. Let $\g(s)=[\bx,(1-s)t+st_{0}]$. Then $\pi(q)=\g(0)$. This means that $\g$ has a unique lift $\tga:[0,1]\rightarrow N$ such that
  $\tga(0)=q$ and $\pi\circ\tga=\g$. Define $\rho:N\rightarrow\tM\times I$ by $\rho(q)=[\tga(1),t]$. Compute $\xi\circ\rho(q)=\pi(q)$. This means
  that $\xi\circ\rho$ has a unique lift to a map from $N$ to $N$ such that it is the identity on $\tM$. Note that $\mathrm{Id}:N\rightarrow N$ is
  one such lift. On the other hand, $\Xi\circ\rho$ is a lift of $\xi\circ\rho$ to a map from $N$ to $N$. Next, let $q\in\tM$. Then
  $\Xi\circ\rho(q)=\Xi(q,t_{0})=q$. Thus $\mathrm{Id}:N\rightarrow N$ and $\Xi\circ\rho:N\rightarrow N$ have to coincide due to the uniquness of the
  lifts. In particular, $\Xi$ is surjective and $\rho$ is injective. 

  Next, note that $\rho$ is surjective. In order to prove this statement, let $(\tx,t)\in \tM\times I$. Then the curve $\g(s)=[\bx,(1-s)t_{0}+st]$,
  where $\pi(\tx)=(\bx,t_{0})$, has a unique lift $\tga:[0,1]\rightarrow N$ such that $\tga(0)=\tx$. From the definition of $\rho$, it is clear that
  $\rho[\tga(1)]=(\tx,t)$. In other words, $\rho$ is surjective. Since $\rho\circ\Xi\circ\rho=\rho$, we conclude that $\rho\circ\Xi=\mathrm{Id}$.
  In particular, there is a bijection from $N$ to $\tM\times I$.

  Next, fix $(\tx,t)\in\tM\times I$ and let $q:=\Xi(\tx,t)$. Then there is a neighbourhood $U$ of $(\tx,t)$ such that $\xi|_{U}$ is a diffeomorphism
  onto its image. Moreover, there is an open neighbourhood $V$ of $q$ such that $\pi|_{V}$ is a diffeomorphism onto its image. Let $W=U\cap \Xi^{-1}(V)$.
  Then $\pi\circ\Xi=\xi$, and restricting this equality to $W$, $\pi$ and $\xi$ are local diffeomorphisms. This means that $\Xi$ is a local
  diffeomorphism. To conclude, $\Xi$ is a global bijection which is also a local diffeomorphism. Thus $\Xi$ and $\rho$ are diffeomorphisms.

  To conclude, we can think of $N$ as having the form $\tM\times I$. Moreover, since it is sufficient to consider a connected component of $\tM$,
  we can assume $\tM$ to be connected. Since $\tM\times I$ is a covering space, we can of course pull back $g$ to a Lorentz metric on $\tM\times I$.
  Since the projection to $\bM\times I$ is a local isometry, all the geometric quantities on $\tM\times I$ are locally the same as the corresponding
  geometric quantities on $\bM\times I$. We can also pull back the coefficients of a system of wave equations on $\bM\times I$.

  Finally, we wish to verify that the expansion normalised Weingarten map has a global frame on $N\cong \tM\times I$. Note, to this end, that if $q\in N$, then
  $q=(\xi^{i_{1}}_{1,p},\dots,\xi^{i_{n}}_{n,p})\times \{t\}$. However, $\xi^{i_{1}}_{1,p},\dots,\xi^{i_{n}}_{n,p}$ is here a basis of eigenvectors of $\mK$
  at $p$. Since $\pi$ is a local diffeomorphism, this basis corresponds to a unique basis of the expansion normalised Weingarten map at $q$. 
\end{proof}

\section{Defining the expansion normalised normal derivative of $\mK$}\label{section:timederivativeofmK}

Next, we define the notion of a normal derivative of the expansion normalised Weingarten map. We do so in several steps. 

\begin{definition}\label{def:tildebardefgeometric}
Let $(M,g)$ be a time oriented Lorentz manifold. Assume that it has a partial pointed foliation. If $\psi$ is a family of functions on 
$\bM$ (for $t\in I$), then $\psi$ can be thought of as a function on $\bM\times I$, say $\tilde{\psi}$. Inversely, if $\psi$ is a function 
on $\bM\times I$, then it can be interpreted as a family of functions on $\bM$ (for $t\in I$). This family is denoted by $\bar{\psi}$. 
If $X$ is a family of vector fields on $\bM$ (for $t\in I$), then $X$ can be thought of as a vector field on $\bM\times I$, say $\tX$, 
defined by
\[
\tX(\psi):=\widetilde{X(\bar{\psi})}
\]
for every $\psi\in C^{\infty}(\bM\times I)$. Next, if $\eta$ is a family of one-form fields on $\bM$ (for $t\in I$), then $\eta$ can be extended to a 
one-form field, say $\teta$, on $\bM\times I$ by demanding that $\teta(U)=0$ and 
\[
\teta(\tX)=\widetilde{\eta(X)}
\]
for every family $X$ of vector fields on $\bM$ (for $t\in I$).  Moreover, if $\eta$ is a one form field on $\bM\times I$, then there is 
an associated family of one-form fields on $\bM$. This family is denoted by $\overline{\eta}$ and is defined by 
\[
\overline{\eta}(X)=\overline{\eta(\tX)}
\]
for every family $X$ of vector fields on $\bM$ (for $t\in I$). Finally, if $X$ is a vector field on $\bM\times I$, then there is an associated 
family of vector fields on $\bM$, denoted $\bX$, defined by the condition that 
\[
X-\widetilde{\bX}\perp\bM_{t}
\]
for all $t\in I$; i.e., $X-\widetilde{\bX}$ is parallel to $U$. 
\end{definition}
\begin{remark}
In what follows, it is necessary to be precise concerning the different notions of regularity. Here we focus on the smooth setting. Let
$\psi$ be a family of functions on $\bM$ (for $t\in I$). Then $\psi$ is a map from $\bM\times I\rightarrow \ro$. Moreover, $\psi$ is said to be 
smooth if this map is smooth; i.e., if $\tilde{\psi}$ is smooth. Next, let $X$ be a family of vector fields on $\bM$ (for all $t\in I$). Then $X$ 
is said to be smooth if, for every smooth family $\psi$ of functions on $\bM$ (for $t\in I$), the expression $X(\psi)$ is a smooth family of 
functions on $\bM$ (for $t\in I$). Finally, let $\eta$ be a family of one-form fields on $\bM$ (for $t\in I$). Then $\eta$ is said to be smooth 
if $\eta(X)$ is a smooth family of functions on $\bM$ (for $t\in I$) for every smooth family $X$ of vector fields on $\bM$ (for all $t\in I$).
\end{remark}

Given the notation introduced in Definition~\ref{def:tildebardefgeometric}, we are in a position to introduce the Lie derivative of
a family $\mt$ of $(1,1)$-tensor fields on $\bM$ (for $t\in I$) with respect to the future pointing unit normal $U$. 

\begin{definition}
Let $\mt$ be a family of $(1,1)$-tensor fields on $\bM$ (for $t\in I$). Then $\ml_{U}\mt$ is defined by
\begin{equation}\label{eq:mlUchKdefgeometric}
(\ml_{U}\mt)(\eta,X):=\overline{U[\widetilde{\eta(\mt X)}]}-\mt(\overline{\ml_{U}\teta},X)-\mt(\eta,\overline{\ml_{U}\tX}),
\end{equation}
\index{$\a$Aa@Notation!Operators!$\ml_{U}$}%
where $\eta$ is a family of one-form fields on $\bM$ (for $t\in I$) and $X$ is a family of vector fields on $\bM$ (for $t\in I$).
\end{definition}
In order to 
justify that the definition (\ref{eq:mlUchKdefgeometric}) is meaningful, we need to prove that $\ml_{U}\mt$ is a family of $(1,1)$-tensor fields on 
$\bM$ (for $t\in I$). In other words, we need to verify that $\ml_{U}\mt$ is linear over families of functions on $\bM$ (for $t\in I$) in both 
$\eta$ and $X$. We leave the verification of this statement to the reader.

Introducing $\{\omega^{i}\}$ and $\{E_{i}\}$ as in Remark~\ref{remark:framenondegenerate}, it is of interest to calculate the constituents of
(\ref{eq:mlUchKdefgeometric}) for $\eta=\omega^{i}$ and $X=E_{j}$. To begin with,
\begin{equation}\label{eq:mlUEkform}
\overline{\ml_{U}\widetilde{E}_{k}}=\overline{[U,\widetilde{E}_{k}]}=-\frac{1}{N}\ml_{\chi}E_{k},
\end{equation}
since the components of $\widetilde{E}_{k}$ with respect to a fixed coordinate system on $\bM$ are independent of $t$. Next, 
\[
\overline{\ml_{U}\widetilde{\omega}^{i}}(E_{k})=\overline{\ml_{U}\widetilde{\omega}^{i}(\widetilde{E}_{k})}
=\overline{\ml_{U}[\widetilde{\omega}^{i}(\widetilde{E}_{k})]-\widetilde{\omega}^{i}(\ml_{U}\widetilde{E}_{k})}=\frac{1}{N}\omega^{i}(\ml_{\chi}E_{k}).
\]
Thus
\[
\overline{\ml_{U}\widetilde{\omega}^{i}}=\frac{1}{N}\omega^{i}(\ml_{\chi}E_{k})\omega^{k}=-\frac{1}{N}\ml_{\chi}\omega^{i}. 
\]
Introducing the notation
\[
(\ml_{U}\mt)^{i}_{\phantom{i}j}:=(\ml_{U}\mt)(\omega^{i},E_{j}),\ \ \
\mt^{i}_{\phantom{i}j}:=\mt(\omega^{i},E_{j})
\]
and omitting the overlines and the twiddles, the definition (\ref{eq:mlUchKdefgeometric}) implies that 
\begin{equation}\label{eq:mlUmKincoordinates}
\begin{split}
(\ml_{U}\mt)^{i}_{\phantom{i}j} = & U(\mt^{i}_{\phantom{i}j})+\frac{1}{N}\mt(\ml_{\chi}\omega^{i},E_{j})+\frac{1}{N}\mt(\omega^{i},\ml_{\chi}E_{j})\\
 = & \frac{1}{N}\d_{t}(\mt^{i}_{\phantom{i}j})-\frac{1}{N}(\ml_{\chi}\mt)^{i}_{\phantom{i}j},
\end{split}
\end{equation}
where
\[
(\ml_{\chi}\mt)^{i}_{\phantom{i}j}:=(\ml_{\chi}\mt)(\omega^{i},E_{j})
\]
In other words, 
\[
\ml_{U}\mt=N^{-1}[\d_{t}(\mt^{i}_{\phantom{i}j})-(\ml_{\chi}\mt)^{i}_{\phantom{i}j}]E_{i}\otimes\omega^{j}. 
\]
In practice, we are mainly interested in $\hml_{U}\mt$, defined by 
\begin{equation}\label{eq:hmlUmtinfixedspatialcoord}
  \hml_{U}\mt:=\theta^{-1}\ml_{U}\mt=\hN^{-1}[\d_{t}(\mt^{i}_{\phantom{i}j})-(\ml_{\chi}\mt)^{i}_{\phantom{i}j}] E_{i}\otimes\omega^{j}, 
\end{equation}
\index{$\a$Aa@Notation!Operators!$\hml_{U}$}%
where $\hN$ is introduced in Definition~\ref{def:lapseandshift}. In what follows, it is convenient to note if $\mS$ and $\mt$ are two 
families of $(1,1)$-tensor fields on $\bM$ (for $t\in I$) and $\psi\in C^{\infty}(\bM\times I)$, then 
\begin{equation}\label{eq:hmlULeibnizrule}
\hml_{U}(\mS\mt)=\hml_{U}(\mS)\mt+\mS\hml_{U}(\mt),\ \ \
\hml_{U}(\psi\mt)=\hU(\psi)\mt+\psi\hml_{U}(\mt).
\end{equation}

\section{Synchronised blow up of the mean curvature}\label{section:syncblowupmc}

In these notes, we are interested in foliations such that there is a $t_{-}$ with the property that, for all $\bx\in\bM$,
$\theta(\bx,t)\rightarrow \infty$ as $t\rightarrow t_{-}+$. 
In other words, the blow up occurs at the same ``time'' for all spatial points; below we speak of a synchronised blow up. Foliations with this property 
are quite special, as the observations below illustrate. Even though we are interested in more general situations, we here restrict our attention to 
situations in which $\ln N$ is bounded and $\chi=0$. 

\begin{lemma}\label{lemma:blowuprategradientoftheta}
Let $(M,g)$ be a time oriented Lorentz manifold. Assume it to have an expanding partial pointed foliation and $\chK$ to have a silent upper bound on $I$; 
cf. Definition~\ref{def:silenceandnondegeneracy}. Assume, finally, that $\chi=0$ and that there are constants $C_{N}$ and $C_{q}$ such that 
$|\ln N|\leq C_{N}$ and $|q|\leq C_{q}$ on $M_{-}$. Then $t_{-}>-\infty$ and either $\theta(\cdot,t)$ converges uniformly as $t\rightarrow t_{-}$, or there 
is an $\bx\in\bM$ such that 
\begin{equation}\label{eq:pointwisedivergtheta}
\lim_{t\rightarrow t_{-}}\theta(\bx,t)=\infty.
\end{equation}
Moreover, there is a constant $C_{0}\geq 1$, depending only on $C_{N}$, $C_{q}$ and $n$, such that 
\begin{equation}\label{eq:thetaupper}
\theta(\bx,t)\leq C_{0}|t-t_{-}|^{-1}
\end{equation}
for all $\bx\in\bM$ and all $t\in (t_{-},t_{0}]$. This $C_{0}$ is also such that 
\begin{equation}\label{eq:thetalower}
\theta(\bx,t)\geq C_{0}^{-1}|t-t_{-}|^{-1}.
\end{equation}
for all $\bx\in\bM$ such that (\ref{eq:pointwisedivergtheta}) holds and all $t\in (t_{-},t_{0}]$.

If there are $\bx_{a},\bx_{b}\in\bM$ such that $\theta(\bx_{a},t)\rightarrow\infty$ and $\theta(\bx_{b},t)\nrightarrow\infty$ as $t\rightarrow t_{-}$, then,
for each $1\leq m\in\zo$, there is a sequence $(\bx_{k},t_{k})\in \bM\times I$ and a constant $c_{m}>0$ such that $t_{k}\rightarrow t_{-}$ and such that 
\begin{equation}\label{eq:gradthetalowerbd}
|(\theta^{-m-1}\grad\theta)(\bx_{k},t_{k})|_{\bge_{\refer}}\geq c_{m},
\end{equation}
where $\grad$ denotes the gradient of $\theta$ (considered as a function on $\bM$) with respect to the metric $\bge_{\refer}$. 
\end{lemma}
\begin{remark}
If the best estimate we are allowed to assume is that the left hand side of (\ref{eq:gradthetalowerbd}) is bounded, then it is quite hard to derive 
any conclusions concerning the asymptotics. However, below we demonstrate that if we combine the assumption of synchronised blow up with assumptions 
concerning $N$ and $q$, then we can deduce much better bounds on the spatial variation of $\ln\theta$. 
\end{remark}
\begin{proof}
Due to (\ref{eq:hUnlnthetamomqbas}), Remark~\ref{remark:qlwbd}, the definition of $\hU$ and the fact that $\chi=0$, 
\begin{equation}\label{eq:dtthetainv}
\d_{t}\theta^{-1}=-\theta^{-1}\d_{t}\ln\theta=-n^{-1}N\hU(n\ln\theta)=n^{-1}N(1+q)\geq n^{-1}(1+n\e_{\Spe})e^{-C_{N}}.
\end{equation}
This means that $\theta^{-1}(\bx,\cdot)$ reaches zero in finite time, starting at $t_{0}$, unless $t$ reaches $t_{-}$ first. Say now that 
$\theta^{-1}(\bx,\cdot)\rightarrow 0$ as $t\rightarrow t_{1}+$, where $t_{-}\leq t_{1}<t_{0}$. Then $t_{1}$ must equal $t_{-}$ (since $\theta(\bx,t_{1})$ 
would otherwise be bounded). Thus $t_{1}=t_{-}$ and $t_{-}>-\infty$. Next, note that 
\begin{equation}\label{eq:thetainvgeneralform}
\theta^{-1}(\bx,t_{0})-\theta^{-1}(\bx,t_{-})=\int_{t_{-}}^{t_{0}}n^{-1}[N(1+q)](\bx,s)ds,
\end{equation}
where the second term on the left hand side should be interpreted as the limit of $\theta^{-1}(\bx,t)$ as $t\rightarrow t_{-}$; since $\theta^{-1}$ is bounded from
below by $0$ and monotonically decreasing to the past, this limit exists. The  first term on the left hand side defines a continuous function of $\bx$. The same 
is true of the right hand side; this follows from the fact that $t_{-}>-\infty$ and the fact that $N$ and $q$ are bounded. Thus $\theta^{-1}(\cdot,t_{-})$ is a 
continuous function and it is the uniform limit of continuous functions. If it is strictly positive, it is clear that $\theta(\cdot,t_{-})$ is a well defined 
continuous function which is the uniform limit of $\theta(\cdot,t)$. In case $\theta^{-1}(\bx,t_{-})=0$ for some $\bx\in\bM$, we also have 
\begin{equation}\label{eq:thetainvblowupform}
\theta^{-1}(\bx,t)=\int_{t_{-}}^{t}n^{-1}[N(1+q)](\bx,s)ds.
\end{equation}
In this case, there is a constant $C_{0}\geq 1$, depending only on $C_{N}$, $C_{q}$ and $n$, such that 
\begin{equation}\label{eq:thetablowuprate}
C_{0}^{-1}|t-t_{-}|\leq \frac{1}{\theta(\bx,t)}\leq C_{0}|t-t_{-}|.
\end{equation}
Note that $C_{0}$ is the same for all $\bx$ such that $\theta(\bx,t)\rightarrow\infty$ as $t\rightarrow t_{-}$. Note, moreover, that the lower bound holds for 
all $\bx$. This yields (\ref{eq:thetaupper}) and (\ref{eq:thetalower}). 

Given that there are $\bx_{a}$ and $\bx_{b}$ as in the statement of the lemma, let $\g:[0,1]\rightarrow\bM$ be a length minimising geodesic with respect to 
$\bge_{\refer}$ connecting $\bx_{a}$ and $\bx_{b}$. Then 
\[
|\theta^{-m}(\bx_{b},t)-\theta^{-m}(\bx_{a},t)|=\left|\int_{0}^{1}[\bd\theta^{-m}(\cdot,t)](\dot{\g}(s))ds\right|
\leq d_{\refer}(\bx_{b},\bx_{a})\sup_{s\in [0,1]}|\bd\theta^{-m}[\g(s),t]|_{\bge_{\refer}},
\]
where $\bd$ is the standard operator on differential forms on $\bM$ and $d_{\refer}$ is the topological metric on $\bM$ induced by $\bge_{\refer}$.
Combining the above observations, it is possible to construct a sequence $(\bx_{k},t_{k})$
with the properties stated in the lemma. In particular, such that (\ref{eq:gradthetalowerbd}) holds. 
\end{proof}

Considering (\ref{eq:thetainvgeneralform}), it is clear that if, given $\bx$, $\theta^{-1}(\bx,t_{-})=0$, then the value of the right hand side is determined 
by $\theta(\bx,t_{0})$. This is clearly a very special situation. Moreover, if $\theta^{-1}(\bx,t_{-})=0$ for all $\bx\in\bM$, then (\ref{eq:thetainvblowupform})
holds for all $\bx\in\bM$. In general, this formula cannot be expected to yield any bounds on the gradient of $\theta$. However, we are not interested in
situations with uncontrolled gradients of $N$ and $q$. In analogy with the weighted norms we impose on $\mK$, we here restrict our attention to the case that 
analogous norms of $\ln N$ and $\ln(1+q)$ are bounded; recall that we are here interested in situations where $q\geq 0$, $N>0$ and $N^{-1}$ is bounded.
In order to be able to draw conclusions from these assumptions, we need to relate $\varrho$ to $t-t_{-}$. 

\begin{lemma}\label{lemma:varrholnthetalntmtmequiv}
Let $(M,g)$ be a time oriented Lorentz manifold. Assume it to have an expanding partial pointed foliation and $\chK$ to have a silent upper bound on $I$; 
cf. Definition~\ref{def:silenceandnondegeneracy}. Assume, moreover, that $\chi=0$ and that there is a constant $C_{q}$ such that $|q|\leq C_{q}$ on $M_{-}$. 
Then there is a constant $c_{a}\geq 1$ such that 
\begin{equation}\label{eq:varrholnthetarelation}
c_{a}^{-1}\leq \frac{\ldr{\varrho}}{\ldr{\ln\theta}}\leq c_{a}
\end{equation}
for all $t\leq t_{0}$. Moreover, $c_{a}$ only depends on $C_{q}$, $\theta_{0,\pm}$ and $n$, where $\theta_{0,-}$ and $\theta_{0,+}$ are defined in
(\ref{eq:thetazdef}). If, in addition, there is a constant $C_{N}$ such that $|\ln N|\leq C_{N}$; and (\ref{eq:pointwisedivergtheta}) holds for all
$\bx\in\bM$, then there is a constant $c_{b}\geq 1$ such that 
\begin{equation}\label{eq:varrholntmintminrelation}
c_{b}^{-1}\leq \frac{\ldr{\varrho}}{\ldr{\ln|t-t_{-}|}}\leq c_{b}
\end{equation}
for all $t\leq t_{0}$. Finally, $c_{b}$ only depends on $C_{q}$, $C_{N}$, $\theta_{0,\pm}$ and $n$.
\end{lemma}
\begin{proof}
Note that (\ref{eq:hUnlnthetamomqbas}) and (\ref{eq:hUvarrhoident}) (in the case that $\chi=0$) imply that
\[
\hU(\varrho+n\ln\theta)=-q\leq 0;
\]
recall Remark~\ref{remark:qlwbd}. This means, in particular, that 
\[
\varrho+n\ln\theta\geq n\ln\theta_{0,-}
\]
for all $t\leq t_{0}$, where $\theta_{0,-}$ is defined by (\ref{eq:thetazdef}); recall that $\varrho(\bx,t_{0})=0$ by definition. Given that there is a $C_{q}$ with 
the properties stated in Lemma~\ref{lemma:blowuprategradientoftheta}, 
\[
\hU[(C_{q}+1)\varrho+n\ln\theta]=C_{q}-q\geq 0. 
\]
Thus
\[
(C_{q}+1)\varrho+n\ln\theta\leq n\ln\theta_{0,+}
\]
for all $t\leq t_{0}$. To summarise, there is a constant $c_{a}\geq 1$ such that (\ref{eq:varrholnthetarelation}) holds. Moreover, $c_{a}$ has the
stated dependence. 

Assuming, in addition, that there is a constant $C_{N}$ such that $|\ln N|\leq C_{N}$ and that (\ref{eq:pointwisedivergtheta})
holds for all $\bx\in\bM$, it follows from (\ref{eq:thetaupper}) and (\ref{eq:thetalower}) that $\ldr{\ln\theta}$ is equivalent to $\ldr{\ln|t-t_{-}|}$. 
This yields a $c_{b}\geq 1$ such that (\ref{eq:varrholntmintminrelation}) holds. Finally, $c_{b}$ has the stated dependence. 
\end{proof}

\begin{prop}
Let $(M,g)$ be a time oriented Lorentz manifold. Assume it to have an expanding partial pointed foliation and $\chK$ to have a silent upper bound on $I$; 
cf. Definition~\ref{def:silenceandnondegeneracy}. Assume, moreover, that $\chi=0$; that there are constants $C_{N}$ and $C_{q}$ such that 
$|\ln N|\leq C_{N}$ and $|q|\leq C_{q}$ on $M_{-}$; and that (\ref{eq:pointwisedivergtheta}) holds for all $\bx\in\bM$. Let $0\leq \cweight\in\ro$ and assume that
there is a $1\leq k\in\zo$ and constants $C_{N,k}$ and $C_{q,k}$ such that 
\begin{equation}\label{eq:lnNlnoneplqweightCkest}
\textstyle{\sum}_{j=1}^{k}\ldr{\varrho}^{-j\cweight}|\bD^{j}\ln N|_{\bge_{\refer}}\leq C_{N,k},\ \ \
\textstyle{\sum}_{j=1}^{k}\ldr{\varrho}^{-j\cweight}|\bD^{j}\ln (1+q)|_{\bge_{\refer}}\leq C_{q,k}
\end{equation}
on $M_{-}$. Then there is a constant $C_{\theta,k}$ such that 
\begin{equation}\label{eq:bDjlnthetaCzbd}
\textstyle{\sum}_{j=1}^{k}\ldr{\varrho}^{-j\cweight}|\bD^{j}\ln \theta|_{\bge_{\refer}}\leq C_{\theta,k}
\end{equation}
on $M_{-}$, where $C_{\theta,k}$ only depends on $n$, $C_{N}$, $C_{q}$, $C_{N,k}$, $C_{q,k}$, $\cweight$, $\theta_{0,\pm}$ and $(\bM,\bge_{\refer})$. 
\end{prop}
\begin{remark}
The estimates (\ref{eq:bDjlnthetaCzbd}) should be contrasted with (\ref{eq:gradthetalowerbd}). Whereas a bound on the left hand side of 
(\ref{eq:gradthetalowerbd}) is not very useful in the arguments, the bound (\ref{eq:bDjlnthetaCzbd}) is sufficient to yield several interesting 
conclusions. 
\end{remark}
\begin{proof}
Let $\{E_{i}\}$ be a frame of the form described in Remark~\ref{remark:framenondegenerate}. Since (\ref{eq:lnNlnoneplqweightCkest}) holds, and since all the 
assumptions stated in Lemma~\ref{lemma:varrholnthetalntmtmequiv} are satisfied, we can appeal to (\ref{eq:thetainvblowupform}) in order to conclude that 
\[
-\theta^{-2}E_{i}\theta=\int_{t_{-}}^{t}n^{-1}[E_{i}\ln N+E_{i}\ln (1+q)]N(1+q)ds.
\]
Thus
\[
|\theta^{-2}E_{i}\theta|\leq C\int_{t_{-}}^{t}\ldr{\ln|s-t_{-}|}^{\cweight}ds\leq C\ldr{\ln|t-t_{-}|}^{\cweight}|t-t_{-}|,
\]
where $C$ only depends on $n$, $C_{N}$, $C_{q}$, $C_{N,1}$, $C_{q,1}$, $\cweight$ and $\theta_{0,\pm}$. Combining this estimate with (\ref{eq:thetaupper}) and 
(\ref{eq:varrholntmintminrelation}) yields the conclusion that (\ref{eq:bDjlnthetaCzbd}) holds for $k=1$, where $C$ only depends on $n$, $C_{N}$, $C_{q}$, 
$C_{N,1}$, $C_{q,1}$, $\cweight$ and $\theta_{0,\pm}$. 

Assume now, inductively, that (\ref{eq:bDjlnthetaCzbd}) holds with $k$ replaced by an $m$ satisfying $1\leq m\leq k-1$. Let
$E_{\bfI}:=E_{i_{1}}\cdots E_{i_{m+1}}$, where $\bfI=(i_{1},\dots,i_{m+1})$. Then applying $E_{\bfI}$ to (\ref{eq:thetainvblowupform}) yields an equality where
the left hand side is a linear combination of terms of the form 
\[
\theta^{-1}E_{\bfI_{1}}\ln\theta\cdots E_{\bfI_{p}}\ln\theta,
\]
where $|\bfI_{1}|+\dots+|\bfI_{p}|=|\bfI|$ and $|\bfI_{j}|\geq 1$.
If $p\geq 2$, this term is bounded after multiplying with $\theta\ldr{\varrho}^{-|\bfI|\cweight}$; this is a consequence of the inductive assumption combined
with Lemma~\ref{lemma:bDbfAbDkequiv}. Note, however, that the resulting constant then depends on $(\bM,\bge_{\refer})$. The only term that is not controlled 
by the inductive assumption is $-\theta^{-1}E_{\bfI}\ln\theta$. The right hand side that 
results when applying $E_{\bfI}$ to (\ref{eq:thetainvblowupform})  is a linear combination of terms of the form 
\[
\int_{t_{-}}^{t}n^{-1}E_{\bfI_{1}}[\ln N+\ln (1+q)]\cdots E_{\bfI_{p}}[\ln N+\ln (1+q)]N(1+q)ds.
\]
However, multiplying this expression with $\theta\ldr{\varrho}^{-|\bfI|\cweight}$ yields a bounded expression due to the assumptions combined with 
Lemma~\ref{lemma:bDbfAbDkequiv}. Combining these observations yields the conclusion that 
\[
\ldr{\varrho}^{-|\bfI|\cweight}|\bD^{|\bfI|}\ln\theta|\leq C,
\]
where we appealed to the inductive assumption combined with Lemma~\ref{lemma:bDbfAbDkequiv}. Combining this estimate with the inductive assumption
proves that the inductive assumption holds with $m$ replaced by $m+1$. The statement of the lemma follows. 
\end{proof}

Consider an expanding partial pointed foliation. Since the interval of the foliation does not necessarily reach the points at which $\theta$ blows up,
it is 
not natural to assume synchronised blow up. However, due to the above examples, it is natural to assume bounds of the form (\ref{eq:bDjlnthetaCzbd}).
For that reason, we typically assume such bounds, or analogous $H^{l}$-bounds. Since we also assume $\bD\ln N$ to be bounded in suitable weighted $C^{l}$
and $H^{l}$-spaces, it is clear that $\bD\ln\hN$ is also bounded in suitable weighted $C^{l}$ and $H^{l}$-spaces. 

\chapter{Gagliardo-Nirenberg estimates}\label{chapter:gagnir}

The purpose of the present chapter is to generalise the Gagliardo-Nirenberg estimates. In particular, we replace ordinary derivatives with vector fields
(which are allowed to be time dependent and the collection of which need not necessarily be a frame); include a space and time dependent weight; carry out
the analysis on closed manifolds; and derive the estimates for general families of tensor fields. This also leads to a generalisation of Moser estimates.
The resulting conclusions play a central role in the derivation of energy estimates. 

\section{Setup and notation}\label{section:setupnotation}

To begin with, let $(\Sigma,h)$ be a closed
$n$-dimensional Riemannian manifold and $\mathscr{I}$ be an interval. We denote the Levi-Civita connection associated with $h$ by $D$. Let $w$ be 
a smooth, strictly positive function on $\Sigma\times \mathscr{I}$. We refer to $w$ as \textit{the weight}. Finally, let $\{W_{1},\dots,W_{P}\}$ 
be a family of smooth time dependent vector fields on $\Sigma$, where $1\leq P\in\zo$. In other words, the $W_{i}$ are smooth vector fields on 
$\Sigma\times \mathscr{I}$ which are tangent to the leaves $\Sigma_{t}:=\Sigma\times \{t\}$, and we think of them as being a family of vector fields
on $\Sigma$. Note that we do not assume $P$ to equal the dimension of $\Sigma$. In particular, we do not assume $\{W_{i}\}$ to constitute a frame. 
In analogy with Definition~\ref{def:multiindexnotation}, we introduce the following notation. 

\begin{definition}\label{def:multiindexnotationgani}
A $W$-\textit{vector field multiindex} is a vector, say $\bfI=(I_{1},\dots,I_{l})$,
where $I_{j}\in \{1,\dots,P\}$. The number $l$ is said to be the \textit{order} of the vector field multiindex, and it is denoted by 
$|\bfI|$. The vector field multiindex corresponding to the empty set is denoted by $\bfz$. Moreover, $|\bfz|=0$. Given a vector field multiindex
$\bfI$, 
\begin{align*}
\bfW_{\bfI} := & (W_{I_{1}},\dots,W_{I_{l}}),\ \ \
D_{\bfI}:=D_{W_{I_{1}}}\cdots D_{W_{I_{l}}}.
\end{align*}
with the special convention that $D_{\bfz}$ is the identity operator, and $\bfW_{\bfz}$ is the empty argument.
\end{definition}
If $\mt$ is a family of smooth tensor fields on $\Sigma$ for $t\in \mathscr{I}$, let 
\[
\|\mt(\cdot,t)\|_{p,w}:=\left(\int_{\Sigma} |\mt(\cdot,t)|_{h}^{p}w^{p}(\cdot,t)\mu_{h}\right)^{1/p},\ \ \
\|\mt(\cdot,t)\|_{\infty,w}:=\|\mt(\cdot,t)w(\cdot,t)\|_{C^{0}(\Sigma)}
\]
for $1\leq p<\infty$. If $\mt$ is a tensor field on $\Sigma$ such that $\|\mt\|_{p,w}<\infty$, then we write $\mt\in L^{p}_{w}(\Sigma)$.
We also use the notation 
\begin{align}
\|D^{l}_{\bbW}\mt(\cdot,t)\|_{p,w} := & \left(\int_{\Sigma} \left(\textstyle{\sum}_{|\bfI|=l}
|(D_{\bfI}\mt)(\cdot,t)|_{h}^{2}\right)^{p/2} w^{p}(\cdot,t)\mu_{h}\right)^{1/p},\label{eq:DlbbWLpwdef}\\
\|D^{l}_{\bbW}\mt(\cdot,t)\|_{\infty,w} := & \sup_{\bx\in\Sigma}\textstyle{\sum}_{|\bfI|=l}|(D_{\bfI}\mt)(\bx,t)|_{h}w(\bx,t). \label{eq:DlbbWLinftywdef}
\end{align}
Let $S,T$ be tensor fields which are covariant of order $l$ and contravariant of order $k$. Then 
\[
\ldr{S,T}_{h}:=h^{i_{1}j_{1}}\cdots h^{i_{l}j_{l}}h_{m_{1}n_{1}}\cdots h_{m_{k}n_{k}}
S^{m_{1}\cdots m_{k}}_{i_{1}\cdots i_{l}}T^{n_{1}\cdots n_{k}}_{j_{1}\cdots j_{l}}.
\]
With this notation $D_{W_{i}}\ldr{S,T}_{h}=\ldr{D_{W_{i}}S,T}_{h}+\ldr{S,D_{W_{i}}T}_{h}$. 

\section{The basic estimate}

The following lemma is the heart of the proof of the Gagliardo-Nirenberg estimates.

\begin{lemma}
Given the assumptions and notation introduced in Section~\ref{section:setupnotation}, let $1\leq i\leq P$ and $\kappa,r\in\mathbb{R}$ 
be such that $1\leq r\leq \kappa$. Then, if $\mt$ is a family of smooth tensor fields on $\Sigma$ for $t\in \mathscr{I}$, 
\begin{equation}\label{eq:stepo}
\|(D_{W_{i}}\mt)(\cdot,t)\|_{2\kappa/r,w}^{2}\leq (2\kappa/r)
\|\mt(\cdot,t)\|_{2\kappa/(r-1),w}
\textstyle{\sum}_{l=1}^{2}\md_{i}^{2-l}(t)\|(D_{W_{i}}^{l}\mt)(\cdot,t)\|_{2\kappa/(r+1),w}
\end{equation}
for all $t\in \mathscr{I}$, where $\md_{i}(t)$ is defined by 
\begin{equation}\label{eq:mdAdef}
\md_{i}(t):=\sup_{\bx\in\Sigma}\left(|(\rodiv_{h}W_{i})(\bx,t)|+|[W_{i}(\ln w)](\bx,t)|\right).
\end{equation}
\end{lemma}
\begin{remark}
The expression $2\kappa/(r-1)$ should be interpreted as $\infty$ when $r=1$. Moreover, $\md_{i}^{0}(t)$ should always be interpreted as equalling $1$ (even
when $\md_{i}(t)=0$).
\end{remark}
\begin{remark}
The assumption that $\Sigma$ be compact is not necessary. In fact, if $(\Sigma,h)$ is a Riemannian manifold without boundary, then the estimate holds,
assuming $\mt$ has compact support. Of course, the estimate is only of interest if $\md_{i}$ introduced in (\ref{eq:mdAdef}) is finite. One particular 
case of interest is of course when $(M,h)$ is $\rn{n}$ with the standard Euclidean metric; $\mt$ is a smooth function with compact support; $w=1$; 
and $\{W_{i}\}$ is the standard frame $\{\d_{i}\}$. In that case, $\md_{i}=0$ and the conclusion reduces to the first step in the standard derivation 
of the Gagliardo-Nirenberg estimates. 
\end{remark}
\begin{proof}
Let $2\leq q\in\mathbb{R}$ and consider $\phi_{i}$, defined by
\[
\phi_{i}(\cdot,t)=w^{q}(\cdot,t)\langle \mt(\cdot,t),D_{W_{i}}\mt(\cdot,t)\rangle_{h}
\langle D_{W_{i}}\mt(\cdot,t),D_{W_{i}}\mt(\cdot,t)\rangle_{h}^{\frac{q-2}{2}}.
\]
Here the last factor should be interpreted as $1$ if $q=2$. If $q=2$, it is clear that $\phi_{i}$ is smooth. 
Let us consider the case that $q>2$. If $\xi$ is such that $(D_{W_{i}}\mt)(\xi,t)\neq 0$, then $\phi_{i}$ is smooth
in a neighbourhood of $(\xi,t)$. Consider a $(\xi,t)$ such that $(D_{W_{i}}\mt)(\xi,t)=0$. Let 
$\psi_{i}(\cdot,t)=\langle D_{W_{i}}\mt(\cdot,t),D_{W_{i}}\mt(\cdot,t)\rangle_{h}$. Then 
$\psi_{i}$ is smooth and has a zero of order $2$ in $(\xi,t)$. Thus $[\psi_{i}(\cdot,t)]^{(q-1)/2}$ has a zero of order 
$q-1>1$ in $\xi$, so that 
\[
|\phi_{i}(\cdot,t)|\leq w^{q}(\cdot,t)|\mt(\cdot,t)|_{h}[\psi_{i}(\cdot,t)]^{1/2}
[\psi_{i}(\cdot,t)]^{\frac{q-2}{2}}=w^{q}(\cdot,t)|\mt(\cdot,t)|_{h}[\psi_{i}(\cdot,t)]^{\frac{q-1}{2}}
\]
has a zero of order $q-1>1$ in $\xi$. To conclude, $\phi_{i}(\cdot,t)$ is differentiable at $\xi$ and the derivative is zero. 
If $(D_{W_{i}}\mt)(\cdot,t)\neq 0$, we can differentiate $\phi_{i}$ with respect to any vector field $X$ on $\Sigma$ in order to obtain
\begin{equation}\label{eq:gnaux}
\begin{split}
(D_{X}\phi_{i})(\cdot,t) = & qX[\ln w(\cdot,t)]\phi_{i}(\cdot,t)\\
 & +w^{q}(\cdot,t)\langle (D_{X}\mt)(\cdot,t),(D_{W_{i}}\mt)(\cdot,t)\rangle_{h}
[\psi_{i}(\cdot,t)]^{\frac{q-2}{2}}\\
& +w^{q}(\cdot,t)\langle \mt(\cdot,t),(D_{X}D_{W_{i}}\mt)(\cdot,t)\rangle_{h} 
[\psi_{i}(\cdot,t)]^{\frac{q-2}{2}}\\
& +(q-2)w^{q}(\cdot,t)\langle \mt(\cdot,t),(D_{W_{i}}\mt)(\cdot,t)\rangle_{h} \\
& \cdot
\langle (D_{W_{i}}\mt)(\cdot,t),(D_{X}D_{W_{i}}\mt)(\cdot,t)\rangle_{h} 
[\psi_{i}(\cdot,t)]^{\frac{q-4}{2}}.
\end{split}
\end{equation}
Note that if $q>2$, $(D_{W_{i}}\mt)(\bx_{l},t)\neq 0$ and $\bx_{l}\rightarrow \xi$ with $(D_{W_{i}}\mt)(\xi,t)=0$, then $(D_{X}\phi_{i})(\bx_{l},t)\rightarrow 0$.
In other words, $\phi_{i}$ is continuously differentiable with respect to the spatial variables. Next, note that if 
\[
\omega_{i}:=\phi_{i}(\cdot,t)\mu_{h},
\]
then Cartan's magic formula (i.e., $\ml_{X}=d\iota_{X}+\iota_{X}d$) yields 
\[
d[\iota_{W_{i}}\omega_{i}]=\ml_{W_{i}}\omega_{i}; 
\]
note that $\omega_{i}$ is an $n$-form on an $n$-manifold. Integrating this equality over $\Sigma$ yields
\begin{equation}\label{eq:relevantinteq}
0=\int_{\Sigma}\ml_{W_{i}}\omega_{i}=\int_{\Sigma}(D_{W_{i}}\phi_{i})(\cdot,t)\mu_{h}+\int_{\Sigma}\phi_{i}(\cdot,t)\ml_{W_{i}}\mu_{h}.
\end{equation}
Since $\ml_{W_{i}}\mu_{h}=(\rodiv_{h}W_{i})\mu_{h}$, this equality implies that 
\begin{equation}\label{eq:relevantinteqvtwo}
\int_{\Sigma}(D_{W_{i}}\phi_{i})(\cdot,t)\mu_{h}=-\int_{\Sigma}\phi_{i}(\cdot,t)(\rodiv_{h}W_{i})(\cdot,t)\mu_{h}.
\end{equation}
Combining this equality with (\ref{eq:gnaux}) (with $X$ replaced by $W_{i}$) yields
\begin{equation}\label{eq:gnfsat}
\begin{split}
 & \int_{\Sigma} |(D_{W_{i}}\mt)(\cdot,t)|_{h}^{q} w^{q}(\cdot,t)\mu_{h}\\
 \leq & q\md_{i}(t)
\int_{\Sigma} |\mt(\cdot,t)|_{h}|(D_{W_{i}}\mt)(\cdot,t)|_{h}|(D_{W_{i}}\mt)(\cdot,t)|_{h}^{q-2} w^{q}(\cdot,t)\mu_{h}\\
 &  +(q-1)\int_{\Sigma} |\mt(\cdot,t)|_{h}|(D_{W_{i}}^{2}\mt)(\cdot,t)|_{h}|(D_{W_{i}}\mt)(\cdot,t)|_{h}^{q-2} w^{q}(\cdot,t)\mu_{h},
\end{split}
\end{equation}
where $\md_{i}$ is defined by (\ref{eq:mdAdef}). For $q=2$, we obtain the same result if we interpret $|(D_{W_{i}}\mt)(\cdot,t)|_{h}^{q-2}$ 
as $1$. On the other hand
\begin{align*}
\int_{\Sigma} |\mt(\cdot,t)|_{h}|(D_{W_{i}}^{2}\mt)(\cdot,t)|_{h} w^{2}(\cdot,t)\mu_{h} \leq & \|\mt(\cdot,t)\|_{2\kappa/(r-1),w}
\|(D_{W_{i}}^{2}\mt)(\cdot,t)\|_{2\kappa/(r+1),w},\\
\int_{\Sigma} |\mt(\cdot,t)|_{h}|(D_{W_{i}}\mt)(\cdot,t)|_{h} w^{2}(\cdot,t)\mu_{h} \leq & \|\mt(\cdot,t)\|_{2\kappa/(r-1),w}
\|(D_{W_{i}}\mt)(\cdot,t)\|_{2\kappa/(r+1),w},
\end{align*}
assuming $\kappa=r\geq 1$, where we appealed to H\"{o}lder's inequality. In particular, 
\[
\|(D_{W_{i}}\mt)(\cdot,t)\|_{2,w}^{2}\leq \|\mt(\cdot,t)\|_{2\kappa/(r-1),w}
\textstyle{\sum}_{l=1}^{2}[2\md_{i}(t)]^{2-l}\|(D_{W_{i}}^{l}\mt)(\cdot,t)\|_{2\kappa/(r+1),w}
\]
for all $t\in \mathscr{I}$. Thus (\ref{eq:stepo}) holds when $\kappa=r\geq 1$. In case $1\leq r<\kappa$, let 
\[
q=\frac{2\kappa}{r},\ \ \
q_{1}=\frac{2\kappa}{r-1},\ \ \
q_{2}=\frac{2\kappa}{r+1},\ \ \
q_{3}=\frac{q}{q-2},
\]
Then $1/q_{1}+1/q_{2}+1/q_{3}=1$, so that we can apply H\"{o}lder's
inequality to (\ref{eq:gnfsat}) in order to obtain
\begin{equation*}
\begin{split}
 & \|(D_{W_{i}}\mt)(\cdot,t)\|_{q,w}^{q}\\
 \leq & 
q\|\mt(\cdot,t)\|_{2\kappa/(r-1),w}\textstyle{\sum}_{l=1}^{2}\md_{i}^{2-l}(t)\|(D_{W_{i}}^{l}\mt)(\cdot,t)\|_{2\kappa/(r+1),w}
\|(D_{W_{i}}\mt)(\cdot,t)\|_{q,w}^{q-2}
\end{split}
\end{equation*}
for all $t\in \mathscr{I}$. The lemma follows. 
\end{proof}

\section{Iterating the basic estimate}

The second step consists in combining the basic estimate with an inductive argument in order to obtain a more general interpolation estimate. 

\begin{lemma}\label{lemma:ganisest}
Given the assumptions and notation introduced in Section~\ref{section:setupnotation}, let $1\leq j,l,i\in\mathbb{Z}$ and $\kappa,r\in\mathbb{R}$ 
be such that $j\leq r\leq \kappa+1-i$ and $l\geq j$. Then there is a constant $C$ such that if $\mt$ is a family of smooth tensor fields on 
$\Sigma$ for $t\in \mathscr{I}$, 
\begin{equation}\label{eq:stept}
\begin{split}
 & \textstyle{\sum}_{m=0}^{j}\md^{j-m}(t)\|(D^{l-j+m}_{\bbW}\mt)(\cdot,t)\|_{2\kappa/r,w}\\
 \leq & C\left[\|(D^{l-j}_{\bbW}\mt)(\cdot,t)\|_{2\kappa/(r-j),w}
+\textstyle{\sum}_{m=0}^{i+j}\md^{i+j-m}(t)\|(D^{l-j+m}_{\bbW}\mt)(\cdot,t)\|_{2\kappa/(r+i),w}\right],
\end{split}
\end{equation}
where 
\begin{equation}\label{eq:mddef}
\md(t):=\textstyle{\max}_{i\in \{1,\dots,P\}}\md_{i}(t). 
\end{equation}
Moreover, the constant $C$ only depends on $P$ and an upper bound on $\kappa$ and $l+i$. 
\end{lemma}
\begin{remark}
The expression $2\kappa/(r-j)$ should be interpreted as $\infty$ when $r=j$. 
\end{remark}
\begin{proof} 
Define $\md(t)$ by (\ref{eq:mddef}). Then, due to (\ref{eq:stepo}),
\begin{equation}\label{eq:intermediateinterpolationaux}
\begin{split}
 & \|(D^{l}_{\bbW}\mt)(\cdot,t)\|_{2\kappa/r,w}^{2}\\
 \leq & C\|(D^{l-1}_{\bbW}\mt)(\cdot,t)\|_{2\kappa/(r-1),w}
\textstyle{\sum}_{m=0}^{1}\md^{1-m}(t)\|(D^{l+m}_{\bbW}\mt)(\cdot,t)\|_{2\kappa/(r+1),w},
\end{split}
\end{equation}
assuming $l\geq 1$ and $1\leq r\leq \kappa$. Note that the constant only depends on upper bounds on $\kappa$, $P$, $l$. From now on, and for the sake of 
brevity, we omit the arguments $(\cdot,t)$ and $(t)$. Then (\ref{eq:intermediateinterpolationaux}) reads
\begin{equation}\label{eq:intermediateinterpolation}
\begin{split}
\|D^{l}_{\bbW}\mt\|_{2\kappa/r,w}^{2} \leq C\|D^{l-1}_{\bbW}\mt\|_{2\kappa/(r-1),w}
\textstyle{\sum}_{m=0}^{1}\md^{1-m}\|D^{l+m}_{\bbW}\mt\|_{2\kappa/(r+1),w}.
\end{split}
\end{equation} 
Due to (\ref{eq:intermediateinterpolation}), the following estimate holds for all $\e>0$:
\begin{equation}\label{eq:originalestimate}
\begin{split}
\|D^{l}_{\bbW}\mt\|_{2\kappa/r,w} \leq 
C\left[\e\|D^{l-1}_{\bbW}\mt\|_{2\kappa/(r-1),w}+\e^{-1}\textstyle{\sum}_{m=0}^{1}\md^{1-m}\|D^{l+m}_{\bbW}\mt\|_{2\kappa/(r+1),w}\right].
\end{split}
\end{equation}
Before proceeding, note that if $f\in L^{2\kappa/(r-j)}_{w}(\Sigma)$, $1\leq i,j\in\zo$, $j\leq r\in\ro$, $r\leq\kappa\in\ro$ and $\e>0$, then 
\begin{equation}\label{eq:ftwokappabyrHolder}
\begin{split}
\|f\|_{2\kappa/r,w} \leq & \|f\|_{2\kappa/(r-j),w}^{i/(i+j)}\|f\|_{2\kappa/(r+i),w}^{j/(i+j)}\\
 \leq & \e\frac{i}{i+j}\|f\|_{2\kappa/(r-j),w}+\e^{-i/j}\frac{j}{i+j}\|f\|_{2\kappa/(r+i),w};
\end{split}
\end{equation}
this follows from H\"{o}lder's and Young's inequalities. In particular, 
\[
\md\|D_{\bbW}^{l-1}\mt\|_{2\kappa/r,w}\leq \frac{1}{2}\e \|D_{\bbW}^{l-1}\mt\|_{2\kappa/(r-1),w}
+\frac{1}{2}\e^{-1}\md^{2} \|D_{\bbW}^{l-1}\mt\|_{2\kappa/(r+1),w}.
\]
Combining this estimate with (\ref{eq:originalestimate}) yields
\begin{equation}\label{eq:stepzerosumest}
\begin{split}
 & \textstyle{\sum}_{m=0}^{1}\md^{1-m}\|D^{l-1+m}_{\bbW}\mt\|_{2\kappa/r,w}\\
 \leq & C\left[\e\|D^{l-1}_{\bbW}\mt\|_{2\kappa/(r-1),w}
+\e^{-1}\textstyle{\sum}_{m=0}^{2}\md^{2-m}\|D^{l-1+m}_{\bbW}\mt\|_{2\kappa/(r+1),w}\right].
\end{split}
\end{equation}
Assume, inductively, that 
\begin{equation}\label{eq:staux}
\begin{split}
 & \textstyle{\sum}_{m=0}^{j}\md^{j-m}\|D^{l-j+m}_{\bbW}\mt\|_{2\kappa/r,w}\\
 \leq & C\left[\e\|D^{l-j}_{\bbW}\mt\|_{2\kappa/(r-j),w}
+C(\e)\textstyle{\sum}_{m=0}^{i+j}\md^{i+j-m}\|D^{l-j+m}_{\bbW}\mt\|_{2\kappa/(r+i),w}\right]
\end{split}
\end{equation}
for arbitrary $r,\kappa,j,l,i$ satisfying the conditions of the lemma, as well as the condition that $j,i\leq \iota$. Due to 
(\ref{eq:stepzerosumest}), we know the inductive assumption 
to hold for $\iota=1$. Given that it holds for some $1\leq \iota\in\zo$, let us prove it for $\iota+1$. First we prove that we can increase $j$ to 
$j+1$. Assume the conditions of the lemma to be satisfied with $j$ replaced by $j+1$ and that $1\leq i,j\leq\iota$. By the inductive 
hypothesis, applied to $r'=r-j$, $\kappa'=\kappa$, $l'=l-j$, $i'=j$ and $j'=1$, 
\begin{equation}\label{eq:estofmiddleterm}
\begin{split}
 & \textstyle{\sum}_{m=0}^{1}\md^{1-m}\|D^{l-j-1+m}_{\bbW}\mt\|_{2\kappa/(r-j),w}\\
 \leq & C\left[\e_{1}\|D^{l-j-1}_{\bbW}\mt\|_{2\kappa/(r-j-1),w}
+C(\e_{1})\textstyle{\sum}_{m=0}^{j+1}\md^{j+1-m}\|D^{l-j-1+m}_{\bbW}\mt\|_{2\kappa/r,w}\right].
\end{split}
\end{equation}
Note also that (\ref{eq:ftwokappabyrHolder}) yields 
\begin{equation*}
\begin{split}
\md^{j+1}\|D_{\bbW}^{l-j-1}\mt\|_{2\kappa/r,w}
 \leq & \frac{i}{i+j+1}\e \|D_{\bbW}^{l-j-1}\mt\|_{2\kappa/(r-j-1),w} \\
 & +\frac{j+1}{i+j+1}\e^{-i/(j+1)}\md^{i+j+1} \|D_{\bbW}^{l-j-1}\mt\|_{2\kappa/(r+i),w}. 
\end{split}
\end{equation*}
Combining this estimate with (\ref{eq:staux}) yields 
\begin{equation}\label{eq:stauxos}
\begin{split}
 & \textstyle{\sum}_{m=0}^{j+1}\md^{j+1-m}\|D^{l-j-1+m}_{\bbW}\mt\|_{2\kappa/r,w}\\
 \leq & C\left[\e \|D_{\bbW}^{l-j-1}\mt\|_{2\kappa/(r-j-1),w}+\e\|D^{l-j}_{\bbW}\mt\|_{2\kappa/(r-j),w}\right.\\
 & \left.+C(\e)\textstyle{\sum}_{m=0}^{i+j+1}\md^{i+j+1-m}\|D^{l-j-1+m}_{\bbW}\mt\|_{2\kappa/(r+i),w}\right].
\end{split}
\end{equation}
In order to estimate the second term in the parenthesis on the right hand side, we appeal to (\ref{eq:estofmiddleterm}). 
This yields (assuming $\e_{1}\leq 1$), 
\begin{equation*}
\begin{split}
 & \textstyle{\sum}_{m=0}^{j+1}\md^{j+1-m}\|D^{l-j-1+m}_{\bbW}\mt\|_{2\kappa/r,w}\\
 \leq & C\left[\e \|D_{\bbW}^{l-j-1}\mt\|_{2\kappa/(r-j-1),w}
+\e C(\e_{1})\textstyle{\sum}_{m=0}^{j+1}\md^{j+1-m}\|D^{l-j-1+m}_{\bbW}\mt\|_{2\kappa/r,w}\right.\\
 & \left.+C(\e)\textstyle{\sum}_{m=0}^{i+j+1}\md^{i+j+1-m}\|D^{l-j-1+m}_{\bbW}\mt\|_{2\kappa/(r+i),w}\right].
\end{split}
\end{equation*}
Fixing $\e_{1}$ and then assuming $\e$ to be small enough yields the conclusion that $C\e C(\e_{1})\leq 1/2$. Then the second term in 
the parenthesis of the right hand side can be moved to the left hand side. Thus (\ref{eq:staux}) holds for all $r,\kappa,j,l,i$ satisfying 
the conditions of the lemma and $i\leq\iota$, $j\leq \iota+1$. 

Next, assume that the conditions of the lemma are satisfied with $i$ replaced by $i+1$. Assume, moreover, that $1\leq i\leq\iota$ and $j\leq\iota+1$. 
Due to (\ref{eq:staux}) with $r'=r+i$, $\kappa'=\kappa$, $j'=i$, $l'=l+i$ and $i'=1$
\begin{equation}\label{eq:stauxsecstep}
\begin{split}
 & \textstyle{\sum}_{m=0}^{i}\md^{i-m}\|D^{l+m}_{\bbW}\mt\|_{2\kappa/(r+i),w}\\
 \leq & C\left[\e\|D^{l}_{\bbW}\mt\|_{2\kappa/r,w}
+C(\e)\textstyle{\sum}_{m=0}^{1+i}\md^{1+i-m}\|D^{l+m}_{\bbW}\mt\|_{2\kappa/(r+i+1),w}\right].
\end{split}
\end{equation}
On the other hand, 
\begin{equation}\label{eq:stauxaux}
\begin{split}
 & \textstyle{\sum}_{m=0}^{j}\md^{j-m}\|D^{l-j+m}_{\bbW}\mt\|_{2\kappa/r,w}\\
 \leq & C\left[\e\|D^{l-j}_{\bbW}\mt\|_{2\kappa/(r-j),w}
+C(\e)\textstyle{\sum}_{m=0}^{i+j}\md^{i+j-m}\|D^{l-j+m}_{\bbW}\mt\|_{2\kappa/(r+i),w}\right]
\end{split}
\end{equation}
Note that 
\begin{equation}\label{eq:splitsumgn}
\begin{split}
 & \textstyle{\sum}_{m=0}^{i+j}\md^{i+j-m}\|D^{l-j+m}_{\bbW}\mt\|_{2\kappa/(r+i),w}\\ 
= & \textstyle{\sum}_{m=0}^{j-1}\md^{i+j-m}\|D^{l-j+m}_{\bbW}\mt\|_{2\kappa/(r+i),w}+\textstyle{\sum}_{m=0}^{i}\md^{i-m}\|D^{l+m}_{\bbW}\mt\|_{2\kappa/(r+i),w}.
\end{split}
\end{equation}
The second term on the right hand side can be estimated by (\ref{eq:stauxsecstep}). In order to estimate the first term on the right 
hand side, we can use H\"{o}lder's and Young's inequalities. In fact, note that (\ref{eq:ftwokappabyrHolder}) implies that 
\begin{equation*}
\|f\|_{2\kappa/(r+i),w}\leq \|f\|_{2\kappa/r,w}^{1/(i+1)}\|f\|_{2\kappa/(r+i+1),w}^{i/(i+1)}.
\end{equation*}
Note also that 
\[
i+j-m=(j-m)\frac{1}{i+1}+(i+j+1-m)\frac{i}{i+1}. 
\]
Thus, given $\de,\e>0$,
\begin{equation*}
\begin{split}
\de^{i+j-m}\|f\|_{2\kappa/(r+i),w} \leq & (\e\de^{j-m}\|f\|_{2\kappa/r,w})^{1/(i+1)}(\e^{-1/i}\de^{i+j+1-m}\|f\|_{2\kappa/(r+i+1),w})^{i/(i+1)}\\
 \leq & \frac{1}{i+1}\e\de^{j-m}\|f\|_{2\kappa/r,w}+\frac{i}{i+1}\e^{-1/i}\de^{i+j+1-m}\|f\|_{2\kappa/(r+i+1),w}.
\end{split}
\end{equation*}
In particular, if $\e_{1}>0$, 
\begin{equation*}
\begin{split}
 & \textstyle{\sum}_{m=0}^{j-1}\md^{i+j-m}\|D^{l-j+m}_{\bbW}\mt\|_{2\kappa/(r+i),w}\\
 \leq & \frac{1}{i+1}\e_{1}\textstyle{\sum}_{m=0}^{j-1}\md^{j-m}\|D^{l-j+m}_{\bbW}\mt\|_{2\kappa/r,w}\\
 & +\frac{i}{i+1}\e_{1}^{-1/i}\textstyle{\sum}_{m=0}^{j-1}\md^{i+j+1-m}\|D^{l-j+m}_{\bbW}\mt\|_{2\kappa/(r+i+1),w}.
\end{split}
\end{equation*}
Combining this estimate with (\ref{eq:stauxsecstep}) (with $\e=\e_{1}$) and (\ref{eq:splitsumgn}) yields 
\begin{equation*}
\begin{split}
 & \textstyle{\sum}_{m=0}^{i+j}\md^{i+j-m}\|D^{l-j+m}_{\bbW}\mt\|_{2\kappa/(r+i),w}\\
 \leq & C\e_{1}\textstyle{\sum}_{m=0}^{j}\md^{j-m}\|D^{l-j+m}_{\bbW}\mt\|_{2\kappa/r,w}\\
 & +C(\e_{1})\textstyle{\sum}_{m=0}^{i+j+1}\md^{i+j+1-m}\|D^{l-j+m}_{\bbW}\mt\|_{2\kappa/(r+i+1),w}.
\end{split}
\end{equation*}
Combining this estimate with (\ref{eq:stauxaux}) yields
\begin{equation}\label{eq:stauxauxaux}
\begin{split}
 & \textstyle{\sum}_{m=0}^{j}\md^{j-m}\|D^{l-j+m}_{\bbW}\mt\|_{2\kappa/r,w}\\
 \leq & C\left[\e\|D^{l-j}_{\bbW}\mt\|_{2\kappa/(r-j),w}
+C(\e)C\e_{1}\textstyle{\sum}_{m=0}^{j}\md^{j-m}\|D^{l-j+m}_{\bbW}\mt\|_{2\kappa/r,w}\right.\\
 & \left.+C(\e)C(\e_{1})\textstyle{\sum}_{m=0}^{i+j+1}\md^{i+j+1-m}\|D^{l-j+m}_{\bbW}\mt\|_{2\kappa/(r+i+1),w}\right].
\end{split}
\end{equation}
First fixing $\e>0$ and then choosing $\e_{1}$ small enough (depending on $\e$), it can be ensured that the middle term in the paranthesis on 
the right hand side can be moved over to the left hand side. This leads to the desired estimate: 
\begin{equation}\label{eq:stauxjplusone}
\begin{split}
 & \textstyle{\sum}_{m=0}^{j}\md^{j-m}\|D^{l-j+m}_{\bbW}\mt\|_{2\kappa/r,w}\\
 \leq & C\left[\e\|D^{l-j}_{\bbW}\mt\|_{2\kappa/(r-j),w}
+C(\e)\textstyle{\sum}_{m=0}^{i+j+1}\md^{i+j+1-m}\|D^{l-j+m}_{\bbW}\mt\|_{2\kappa/(r+i+1),w}\right].
\end{split}
\end{equation}
Thus the induction hypothesis holds with $\iota$ replaced by $\iota+1$. 
\end{proof}

\section{Gagliardo-Nirenberg estimates}

By a simple rescaling, Lemma~\ref{lemma:ganisest} has the following consequence. 
\begin{cor}
Given the assumptions and notation introduced in Section~\ref{section:setupnotation}, let $1\leq j,l,i\in\mathbb{Z}$ and $\kappa,r\in\mathbb{R}$ 
be such that $j\leq r\leq \kappa+1-i$ and $l\geq j$. Then there is a constant $C$ such that if $\mt$ is a family of smooth tensor fields on 
$\Sigma$ for $t\in \mathscr{I}$,
\begin{equation}\label{eq:stepttransaauxprod}
\begin{split}
 & \textstyle{\sum}_{m=0}^{j}\md^{j-m}(t)\|(D^{l-j+m}_{\bbW}\mt)(\cdot,t)\|_{2\kappa/r,w}\\
 \leq & C\|(D^{l-j}_{\bbW}\mt)(\cdot,t)\|_{2\kappa/(r-j),w}^{i/(i+j)}
\left(\textstyle{\sum}_{m=0}^{i+j}\md^{i+j-m}(t)\|(D^{l-j+m}_{\bbW}\mt)(\cdot,t)\|_{2\kappa/(r+i),w}\right)^{j/(i+j)}. 
\end{split}
\end{equation}
Moreover, the constant $C$ only depends on $P$ and an upper bound on $\kappa$ and $l+i$. 
\end{cor}
\begin{proof}
Let $0<s\in\ro$. We begin by analysing how the estimate (\ref{eq:stept}) rescales when we rescale the underlying metric $h$ to $h_{s}:=s^{2}h$ 
and the vector fields $W_{I}$ to $W_{I,s}:=s^{-1}W_{I}$. Note, to begin with, that $\|D^{l}_{\bbW}\mt(\cdot,t)\|_{p}$ transforms to 
$s^{-l}s^{m-k}s^{n/p}\|D^{l}_{\bbW}\mt(\cdot,t)\|_{p}$, assuming $\mt$ to be covariant of order $k$ and contravariant of order $m$. Moreover, 
$\md(t)$ transforms to $s^{-1}\md(t)$. Summing up, (\ref{eq:stept}) transforms to 
\begin{equation}\label{eq:stepttransaaux}
\begin{split}
 & \textstyle{\sum}_{m=0}^{j}\md^{j-m}(t)\|(D^{l-j+m}_{\bbW}\mt)(\cdot,t)\|_{2\kappa/r,w}\\
 \leq & C\left[s^{a}\|(D^{l-j}_{\bbW}\mt)(\cdot,t)\|_{2\kappa/(r-j),w}
+s^{b}\textstyle{\sum}_{m=0}^{i+j}\md^{i+j-m}(t)\|(D^{l-j+m}_{\bbW}\mt)(\cdot,t)\|_{2\kappa/(r+i),w}\right]
\end{split}
\end{equation}
(after division by a suitable power of $s$), where
\[
a:=-\frac{nj}{2\kappa}+j=j\left(1-\frac{n}{2\kappa}\right),\ \ \
b:=\frac{ni}{2\kappa}-i=-i\left(1-\frac{n}{2\kappa}\right). 
\]
Note that, if $n\neq 2\kappa$, one of $a$ and $b$ is strictly positive and one is strictly negative. Schematically, the estimate (\ref{eq:stepttransaaux})
can be written 
\[
S\leq C(s^{a}Q+s^{b}R).
\]
Assume that $n\neq 2\kappa$. If one of $Q$ and $R$ vanishes, we can let $s$ tend to $0+$ or $\infty$ in order to deduce that $S$ vanishes. If 
both are non-zero, we can choose $s=(R/Q)^{1/(a-b)}$. Then 
\[
S\leq 2CR^{a/(a-b)}Q^{b/(b-a)}.
\]
In our case, 
\[
\frac{a}{a-b}=\frac{j}{i+j},\ \ \ \frac{b}{b-a}=\frac{i}{i+j}. 
\]
In particular, (\ref{eq:stepttransaaux}) implies that (\ref{eq:stepttransaauxprod}) holds if $n\neq 2\kappa$. In order to prove the lemma in case 
$n=2\kappa$, let $\e>0$, $\kappa_{\e}=\kappa+\e$ and $r_{\e}=r+\e$. Then (\ref{eq:stepttransaauxprod}) holds with $\kappa$ and $r$ replaced by $\kappa_{\e}$
and $r_{\e}$ respectively. The final idea is to take the limit $\e\rightarrow 0+$. In order for this to be allowed, we need to verify that 
$\|\mt(\cdot,t)\|_{p}\rightarrow \|\mt(\cdot,t)\|_{q}$ as $p\rightarrow q$ (even in the case that $q=\infty$). Moreover, we need to verify that the 
constant remains bounded in the limit. However, this can be achieved by an argument similar to the proof of \cite[Corollary~6.1]{minbok}. The lemma follows. 
\end{proof}

Consider (\ref{eq:stepttransaauxprod}). The case that $r=j=l$ and $r+i=\kappa$ is of particular interest. Then,
for $l\geq 1$ and $\kappa\geq l$, 
\begin{equation}\label{eq:stepttransaauxprodgana}
\begin{split}
 & \textstyle{\sum}_{m=0}^{l}\md^{l-m}(t)\|(D^{m}_{\bbW}\mt)(\cdot,t)\|_{2\kappa/l,w} \\
 \leq & C\|\mt(\cdot,t)\|_{\infty,w}^{1-l/\kappa}
\left(\textstyle{\sum}_{m=0}^{\kappa}\md^{\kappa-m}(t)\|(D^{m}_{\bbW}\mt)(\cdot,t)\|_{2,w}\right)^{l/\kappa}. 
\end{split}
\end{equation}

\section{Applications of the Gagliardo-Nirenberg estimates}\label{section:applicationsofgagliardonirenbergestimates}

Next, we derive consequences of the Gagliardo-Nirenberg estimates. One immediate consequence is the following. 
\begin{cor}\label{cor:moserest}
Given the assumptions and notation introduced in Section~\ref{section:setupnotation}, assume that $w=1$. Let, moreover, 
$0\leq l_{i}\in\mathbb{Z}$ and $l=l_{1}+\dots+l_{j}$. Then there is a constant $C$ such that if $\mt_{1},\dots,\mt_{j}$ are 
families of smooth tensor fields on $\Sigma$ for $t\in \mathscr{I}$, then 
\begin{equation}\label{eq:mosergagliardonirenberg}
\begin{split}
\left\| |(D^{l_{1}}_{\bbW}\mt_{1})(\cdot,t)|_{h}\cdots |(D^{l_{j}}_{\bbW}\mt_{j})(\cdot,t)|_{h}\right\|_{2}\leq 
C\textstyle{\sum}_{i}\|\mt_{i}(\cdot,t)\|_{\mH^{l}_{\bbW}}\textstyle{\prod}_{m\neq i}\|\mt_{m}(\cdot,t)\|_{\infty},
\end{split}
\end{equation}
where $|(D^{l_{i}}_{\bbW}\mt_{i})(\cdot,t)|_{h}:=\left(\textstyle{\sum}_{|\bfI|=l_{i}}|(D_{\bfI}\mt_{i})(\cdot,t)|_{h}^{2}\right)^{1/2}$ and 
\begin{equation}\label{eq:mHlbbWdef}
\|\mt(\cdot,t)\|_{\mH^{l}_{\bbW}}:=\left(\textstyle{\sum}_{k\leq l}\|D^{k}_{\bbW}\mt(\cdot,t)\|_{2}^{2}\right)^{1/2}.
\end{equation}
Moreover, the constant $C$ only depends on the supremum of $\md(t)$, $P$ and an upper bound on $l$. 
\end{cor}
\begin{proof}
Note that if at most one $l_{i}$ is non-zero, the estimate holds trivially. Moreover, the factors corresponding to $l_{i}$'s that are zero can 
be estimated in $L^{\infty}$ and extracted outside the $L^{2}$-norm. In other words, we can assume all the $l_{i}$ to be non-zero. Let 
$l:=l_{1}+\dots+l_{j}$ and $p_{i}:=l/l_{i}$. Then H\"{o}lder's inequality yields 
\[
\left\| |(D^{l_{1}}_{\bbW}\mt_{1})(\cdot,t)|_{h}\cdots |(D^{l_{j}}_{\bbW}\mt_{j})(\cdot,t)|_{h}\right\|_{2}\leq 
\textstyle{\prod}_{i=1}^{l}\|(D^{l_{i}}_{\bbW}\mt_{i})(\cdot,t)\|_{2l/l_{i}}.
\]
On the other hand, (\ref{eq:stepttransaauxprodgana}) implies that 
\[
\textstyle{\prod}_{i=1}^{l}\|(D^{l_{i}}_{\bbW}\mt_{i})(\cdot,t)\|_{2l/l_{i}}\leq
C\textstyle{\prod}_{i=1}^{l}\|\mt_{i}(\cdot,t)\|_{\infty}^{1-l_{i}/l}\|\mt_{i}(\cdot,t)\|_{\mH^{l}_{\bbW}}^{l_{i}/l},
\]
where the constant depends on the supremum of $\md(t)$. Since $1-l_{i}/l=\sum_{m\neq i}l_{m}/l$, the product on the right hand side can be divided
into $l$ factors of the form 
\[
\left(\|\mt_{i}(\cdot,t)\|_{\mH^{l}_{\bbW}}\textstyle{\prod}_{m\neq i}\|\mt_{m}(\cdot,t)\|_{\infty}\right)^{l_{i}/l}.
\]
Combining this estimate with Young's inequality yields the conclusion of the corollary. 
\end{proof}

In these notes, there are two natural frames: $\{X_{A}\}$ and $\{E_{i}\}$. In case we use the frame $\{X_{A}\}$ and $h=\bge_{\refer}$, we use
the notation $\bD_{\bbA}$ instead of $D_{\bbW}$. In case we use the frame $\{E_{i}\}$ and $h=\bge_{\refer}$, we use the notation $\bD_{\bbE}$ instead of $D_{\bbW}$. 

\begin{cor}\label{cor:mixedmoserestweight}
  Assume $(M,g)$ to be a time oriented Lorentz manifold. Assume that it has an expanding partial pointed foliation. Assume, moreover, $\mK$ to be
  non-degenerate on $I$, to have a global frame and to be $C^{0}$-uniformly bounded on $I_{-}$; i.e., (\ref{eq:mKsupbasest}) to hold. Let
  $0\leq q,r,s\in\zo$. For $1\leq i\leq q$, $1\leq j\leq r$ and $1\leq m\leq s$, let: $w_{i},u_{j},v_{m}$
  be smooth strictly positive functions on $\bM\times I$; $f_{i},g_{j},h_{m}$ be strictly positive functions on $I$; $l_{i}$, $k_{j}$ and $p_{m}$ be
  non-negative integers; and $\ms_{i}$, $\mt_{j}$ and $\mU_{m}$ be families of smooth tensor fields on $\bM$ for $t\in I$. Let $l$ be the sum of the
  $l_{i}$, the $k_{j}$ and the $p_{m}$. Then, assuming $g_{j}\leq 1$ and $h_{m}\leq 1$, 
  \begin{equation}\label{eq:mixedmosergagliardonirenbergweight}
    \begin{split}
      & \left\|\textstyle{\prod}_{i=1}^{q}w_{i}f_{i}^{l_{i}}|\bD^{l_{i}}_{\bbA}\ms_{i}|_{\bge_{\refer}}
      \prod_{j=1}^{r}u_{j}g_{j}^{k_{j}}|\bD^{k_{j}}\mt_{j}|_{\bge_{\refer}}
      \textstyle{\prod}_{m=1}^{s}v_{m}h_{m}^{p_{m}}|\bD^{p_{m}}_{\bbE}\mU_{m}|_{\bge_{\refer}}\right\|_{2}\\
      \leq & C_{a}\textstyle{\sum}_{i}\textstyle{\sum}_{k\leq l}\a_{i}^{l-k}\|w_{i}f_{i}^{k}\bD^{k}_{\bbA}\ms_{i}\|_{2}
      \textstyle{\prod}_{o\neq i}\|\ms_{o}\|_{\infty,w_{o}}\textstyle{\prod}_{j}\|\mt_{j}\|_{\infty,u_{j}}
      \textstyle{\prod}_{m}\|\mU_{m}\|_{\infty,v_{m}}\\
      & +C_{b}\textstyle{\sum}_{j}\textstyle{\sum}_{k\leq l}\b_{j}^{l-k}\|u_{j}g_{j}^{k}\bD^{k}\mt_{j}\|_{2}
      \textstyle{\prod}_{o\neq j}\|\mt_{o}\|_{\infty,u_{o}}\textstyle{\prod}_{i}\|\ms_{i}\|_{\infty,w_{i}}
      \textstyle{\prod}_{m}\|\mU_{m}\|_{\infty,v_{m}}\\
      & +C_{b}\textstyle{\sum}_{m}\textstyle{\sum}_{k\leq l}\g_{m}^{l-k}\|v_{m}h_{m}^{k}\bD^{k}_{\bbE}\mU_{m}\|_{2}
      \textstyle{\prod}_{o\neq m}\|\mU_{o}\|_{\infty,v_{o}}\textstyle{\prod}_{i}\|\ms_{i}\|_{\infty,w_{i}}
      \textstyle{\prod}_{j}\|\mt_{j}\|_{\infty,u_{j}}
    \end{split}
  \end{equation}
  on $I_{-}$, where the constant $C_{a}$ only depends on $\mKsup$, $\e_{\rond}$, $l$ and $n$; $C_{b}$ only depends on $l$, $n$ and $(\bM,\bge_{\refer})$;
  and 
  \begin{align*}
    \a_{i}(t) := & 1+f_{i}(t)\sup_{\bx\in\bM}[|(\bD\mK)(\bx,t)|_{\bge_{\refer}}+|(\bD\ln w_{i})(\bx,t)|_{\bge_{\refer}}],\\
    \b_{j}(t) := & 1+g_{j}(t)\sup_{\bx\in\bM}|(\bD\ln u_{j})(\bx,t)|_{\bge_{\refer}},\\
    \g_{m}(t) := & 1+h_{m}(t)\sup_{\bx\in\bM}|(\bD\ln v_{m})(\bx,t)|_{\bge_{\refer}}.
  \end{align*}
\end{cor}
\begin{remark}
  If $q=0$, there are no $\ms_{i}$-factors on the left hand side of (\ref{eq:mixedmosergagliardonirenbergweight}); the first term on the right hand
  side is absent; and the products of $\ms_{i}$-factors in the second and third terms on the right hand side can be put equal to $1$. Similar
  statements hold in case $r$ or $s$ equal zero. 
\end{remark}
\begin{remark}
  Due to the arguments presented in the proof, it follows that $\bD^{k}\mt_{j}$ on the right hand side can be replaced by $\bD^{k}_{\bbE}\mt_{j}$.
  Similarly, $\bD^{k}_{\bbE}\mU_{m}$ on the right hand side can be replaced by $\bD^{k}\mU_{m}$.
\end{remark}
\begin{proof}
  Consider $|\bD^{k_{j}}\mt_{j}|_{\bge_{\refer}}$ on the left hand side of (\ref{eq:mixedmosergagliardonirenbergweight}). Due to
  Lemma~\ref{lemma:bDbfAbDkequiv}, this expression can be replaced by a linear combination of expressions
  of the form $|\bD^{k}_{\bbE}\mt_{j}|_{\bge_{\refer}}$, where $k\leq k_{j}$. Since $g_{j}\leq 1$ and since a reduction in $k_{j}$ leads to a reduction in
  $l$, it is thus sufficient to prove the lemma with $|\bD^{k_{j}}\mt_{j}|_{\bge_{\refer}}$ replaced by $|\bD^{k}_{\bbE}\mt_{j}|_{\bge_{\refer}}$. Moreover, we
  can assume $k=k_{j}$ in the latter expression. However, the resulting constants depend on $(\bM,\bge_{\refer})$. 
  
  Note that if at most one of $l_{i}$, $k_{j}$ and $p_{m}$ is non-zero, the estimate holds trivially. Moreover, the factors corresponding to the $l_{i}$'s,
  the $k_{j}$'s and the $p_{m}$'s that are zero can be estimated in $L^{\infty}$ and extracted outside the $L^{2}$-norm. In other words, we can assume all
  the $l_{i}$'s, the $k_{j}$'s and the $p_{m}$'s to be non-zero. Let $l$ be defined as in the statement of the corollary, $q_{i}=l/l_{i}$,
  $r_{j}=l/k_{j}$ and $s_{m}=l/p_{m}$. Then H\"{o}lder's inequality yields the conclusion that the left hand side of
  (\ref{eq:mixedmosergagliardonirenbergweight}) is bounded by
  \begin{equation}\label{eq:totalproducttobeestimated}
  \textstyle{\prod}_{i=1}^{q}\|w_{i}f_{i}^{l_{i}}\bD^{l_{i}}_{\bbA}\ms_{i}\|_{2q_{i}}
  \prod_{j=1}^{r}\|u_{j}g_{j}^{k_{j}}\bD^{k_{j}}_{\bbE}\mt_{j}\|_{2r_{j}}
  \textstyle{\prod}_{m=1}^{s}\|v_{m}h_{m}^{p_{m}}\bD^{p_{m}}_{\bbE}\mU_{m}\|_{2s_{m}}.
  \end{equation}
  At this stage we wish to apply (\ref{eq:stepttransaauxprodgana}) to the three products on the right hand side. In order to apply it to one of
  the factors in first product, note that the assumptions introduced at the beginning of the present chapter are fulfilled with $\Sigma=\bM$;
  $h=\bge_{\refer}$; $w=w_{i}$; $\mathscr{I}=I$; $D=\bD$; $P=n$; and with the $W_{A}$ equal to $f_{i}X_{A}$. Applying
  (\ref{eq:stepttransaauxprodgana}) then yields   
  \begin{equation}\label{eq:bDbbAmsiestimate}
  \|w_{i}f_{i}^{l_{i}}\bD^{l_{i}}_{\bbA}\ms_{i}\|_{2q_{i}}\leq
  C\|\ms_{i}\|_{\infty,w_{i}}^{1-1/q_{i}}\left(\textstyle{\sum}_{k\leq l}\md^{l-k}\|w_{i}f_{i}^{k}\bD^{k}_{\bbA}\ms_{i}\|_{2}\right)^{1/q_{i}},
  \end{equation}
  where the constant only depends on $n$ and $l$. In this particular setting, $\md(t)$ is the supremum (over $\bx\in \bM$ and
  $A\in \{1,\dots,n\}$) of
  \[
  f_{i}|\rodiv_{\bge_{\refer}}X_{A}|+f_{i}|X_{A}\ln w_{i}|\leq Cf_{i}|\bD\mK|_{\bge_{\refer}}+f_{i}|\bD\ln w_{i}|_{\bge_{\refer}},
  \]
  where $C$ only depends on $\mKsup$, $\e_{\rond}$ and $n$, and we used the fact that
  \[
  |\rodiv_{\bge_{\refer}}X_{A}|=|Y^{B}(\bD_{X_{B}}X_{A})|\leq C|\bD\mK|_{\bge_{\refer}};
  \]
  cf. Lemma~\ref{lemma:frameinvest} and (\ref{eq:covderofframe}). Defining $\a_{i}$ as in the statement of the lemma, 
  the estimate (\ref{eq:bDbbAmsiestimate}) implies
  \[
  \|w_{i}f_{i}^{l_{i}}\bD^{l_{i}}_{\bbA}\ms_{i}\|_{2q_{i}}\leq
  C\|\ms_{i}\|_{\infty,w_{i}}^{1-1/q_{i}}\left(\textstyle{\sum}_{k\leq l}\a_{i}^{l-k}\|w_{i}f_{i}^{k}\bD^{k}_{\bbA}\ms_{i}\|_{2}\right)^{1/q_{i}},
  \]
  where $C$ only depends on $\mKsup$, $\e_{\rond}$, $l$ and $n$.

  Next, we need to estimate the second product on the right hand side of (\ref{eq:totalproducttobeestimated}). Note, to this end, that
  (\ref{eq:stepttransaauxprodgana}) applies with $\Sigma=\bM$; $h=\bge_{\refer}$; $w=u_{j}$; $\mathscr{I}=I$; $D=\bD$; $P=n$; and with the
  $W_{p}$ equal to the $g_{j}E_{p}$. An argument similar to the above then yields the estimate
  \[
  \|u_{j}g_{j}^{k_{j}}\bD^{k_{j}}_{\bbE}\mt_{j}\|_{2r_{j}}\leq
  C\|\mt_{j}\|_{\infty,u_{j}}^{1-1/r_{j}}\left(\textstyle{\sum}_{k\leq l}\b_{j}^{l-k}\|u_{j}g_{j}^{k}\bD^{k}_{\bbE}\mt_{j}\|_{2}\right)^{1/r_{j}},
  \]
  where $C$ only depends on $l$, $n$ and $(\bM,\bge_{\refer})$. Moreover, $\b_{j}$ is defined as in the statement of the lemma. 
  The estimate for the factors in the third product on the right hand side of (\ref{eq:totalproducttobeestimated}) is the same.
  At this stage, we can group the factors in analogy with the end of the proof of Corollary~\ref{cor:moserest} and apply Young's
  inequality. This yields (\ref{eq:mixedmosergagliardonirenbergweight}) with $\bD^{k}\mt_{j}$ on the right hand side replaced by
  $\bD^{k}_{\bbE}\mt_{j}$. However, appealing to Lemma~\ref{lemma:bDbfAbDkequiv} again, as well
  as the fact that $g_{j}\leq 1$, we can replace $\bD^{k}_{\bbE}\mt_{j}$ with $\bD^{k}\mt_{j}$. The corollary follows. 
\end{proof}

\chapter{Examples}\label{chapter:examples}

The purpose of the present chapter is to compare the assumptions made in these notes with the conditions satisfied by a few families of
solutions for which the asymptotics are known. We begin, in Section~\ref{section:BianchiApp}, by discussing the Bianchi spacetimes. In
Section~\ref{ssection:examples}, we describe results in the absence of symmetry, but where the authors specify data on the singularity.
This is followed by a discussion of results on stable big bang formation; cf. Section~\ref{section:stablebigbangformApp}. Finally, in
Section~\ref{section:t3Gowdy}, we discuss the asymptotics of vacuum $\tn{3}$-Gowdy solutions. 

\section{Bianchi spacetimes}\label{section:BianchiApp}

Let us begin by considering Bianchi spacetimes, where we use the terminology introduced in \cite[Definition~1, p.~600]{KGCos}:

\begin{definition}[\cite{KGCos}, Definition~1, p.~600]\label{def:Bianchispacetime}
  A \textit{Bianchi spacetime} is a Lorentz manifold $(M,g)$, where $M=G\times I$; $I=(t_{-},t_{+})$ is an open interval; $G$ is a
  connected $3$-dimensional Lie group; and $g$ is of the form
  \begin{equation}\label{eq:Bianchimetricdef}
    g=-dt\otimes dt+a_{ij}(t)\xi^{i}\otimes \xi^{j},
  \end{equation}
  where $\{\xi^{i}\}$ is the dual basis of a basis $\{e_{i}\}$ of the Lie algebra $\mfg$ and $a_{ij}\in C^{\infty}(I,\ro)$ are such that
  $a_{ij}(t)$ are the components of a positive definite matrix $a(t)$ for every $t\in I$. 
\end{definition}

In order to be specific, let us here restrict our attention to orthogonal perfect fluids with a linear equation of state. This means that the
stress energy tensor takes the form (\ref{eq:setperfectfluid}) where $U$ is orthogonal to the hypersurfaces of spatial homogeneity. In the
case of metrics of the form (\ref{eq:Bianchimetricdef}), this means that $U=\d_{t}$. The linear equation of state reads $p=(\g-1)\rho$, where
$\g$ is a constant. If $G$ is unimodular/non-unimodular (cf., e.g., \cite[Definition~4, p.~604]{KGCos}), then $(M,g)$ given in
Definition~\ref{def:Bianchispacetime} is said to be of Bianchi class A/Bianchi class B; cf. \cite[Definition~5, p.~604]{KGCos}. 
The basic results we appeal to in the present section are \cite{BianchiIXattr} (for Bianchi class A orthogonal perfect fluid solutions with
$2/3<\g\leq 2$) and \cite{RadermacherNonStiff} and \cite{RadermacherStiff} (for non-exceptional Bianchi class B orthogonal perfect fluid solutions).
In the case of Bianchi class B, some of the results hold for $0\leq\g\leq 2$ and some hold for $0\leq \g<2/3$.

\textbf{Bianchi spacetimes, basic properties.} Excluding Minkowski space and quotients thereof, Bianchi orthogonal perfect fluid solutions
have crushing singularities such that $\varrho\rightarrow-\infty$, cf. \cite[Subsection~3.1, pp.~607--608]{KGCos} and
\cite[Subsection~3.2, pp.~608--609]{KGCos}. Here we assume $2/3<\g\leq 2$ in the case of Bianchi class A, with the exception of Bianchi type IX (in
which case we assume $1\leq\gamma\leq 2$). In the case of Bianchi class B, we restrict ourselves to the non-exceptional case and assume that
$0\leq\g\leq 2$. 

Next, note that $N=1$ and $\chi=0$ in the case of Bianchi spacetimes. Moreover, $\theta$ is independent of the spatial variable. The only conditions
appearing in Chapter~\ref{chapter:assumptions} that need to be verified are thus the ones concerning the boundedness of $q$ and the ones concerning
$\mK$ and its normal derivative. Concerning $q$, note that in the Bianchi class A setting, $q$ is given by
\[
q=\frac{1}{2}(3\g-2)\Omega+2(\Sigma_{+}^{2}+\Sigma_{-}^{2});
\]
cf. the formula at the bottom of \cite[p.~414]{BianchiIXattr}. For all the Bianchi class A types except IX, this expression fulfills a universal bound. This
follows from \cite[(11), p.~415]{BianchiIXattr} and the fact the the expression involving the $N_{i}$ in \cite[(11), p.~415]{BianchiIXattr} is non-negative
for all the Bianchi types except IX. Due to the results of \cite{BianchiIXattr} concerning Bianchi type IX solutions, it also follows that $q$ is bounded in
the direction of the singularity in that case. In the case of non-exceptional Bianchi class B with $\g\in [0,2]$, $q$ takes its values in
$[-1,2]$; cf. \cite[(16), p.~708]{RadermacherNonStiff}. To conclude, the relevant conditions to examine are those concerning $\mK$. 

Next, recall the matrix $\Sigma_{ij}$ introduced in \cite[(10), (11), p.~603]{KGCos} (note that the components are calculated with respect to a
fixed frame $\{e_{i}\}$). Raising one index by means of the metric yields $\Sigma^{i}_{\phantom{i}j}$. These are the components of the trace free part
of the expansion normalised Weingarten map. In other words,
\[
\mK^{i}_{\phantom{i}j}=\Sigma^{i}_{\phantom{i}j}+\frac{1}{3}\de^{i}_{j},\ \ \
\chK^{i}_{\phantom{i}j}=\Sigma^{i}_{\phantom{i}j}-\frac{1}{3}q\de^{i}_{j}.
\]

\textbf{Bianchi class A solutions.} An extremely important observation concerning Bianchi class A orthogonal perfect fluid solutions is that we can choose
a fixed (time-independent) basis of $\mfg$ such that $\mK$ is diagonal (for this reason, the arguments of these notes should go through in this setting
without requiring non-degeneracy; the purpose of demanding non-degeneracy is to obtain a frame diagonalising $\mK$). Moreover, the diagonal components
of $\mK$ (which are also the eigenvalues of
$\mK$) can be computed in terms of $\Sigma_{\pm}$ appearing in the Wainwright-Hsu equations \cite[(9)-(11), pp.~414-415]{BianchiIXattr}. This means,
in particular, that the frame $\{X_{A}\}$ introduced in Definition~\ref{def:XAellA} is fixed (time-independent). Thus we can choose the frame $\{E_{i}\}$ to
coincide with $\{X_{A}\}$. Moreover,
\[
\mK=\mK^{i}_{\phantom{i}j}E_{i}\otimes \omega^{j},\ \ \
\hml_{U}\mK=\frac{1}{3}(\d_{\tau}\mK^{i}_{\phantom{i}j})E_{i}\otimes \omega^{j},
\]
where we appealed to (\ref{eq:mlUmKincoordinates}) and \cite[(137), p.~487]{BianchiIXattr}. Here $\mK^{i}_{\phantom{i}j}$ and $\d_{\tau}\mK^{i}_{\phantom{i}j}$
are bounded in the direction of the singularity for all Bianchi class A orthogonal perfect fluids with $2/3<\g\leq 2$. This means that $\mK$ and
$\hml_{U}\mK$ satisfies all the weighted Sobolev and $C^{k}$-bounds appearing in Definitions~\ref{def:sobklassumptions} and \ref{def:supmfulassumptions}.
In addition, since $(\hml_{U}\mK)(Y^{A},X_{B})=0$ for $A\neq B$, it is clear that $\hml_{U}\mK$ satisfies an off-diagonal exponential bound. 

Turning to silence and non-degeneracy, note that in the case of Bianchi type VIII and IX non-stiff fluids, generic solutions are expected to
be oscillatory. In the case of Bianchi type IX, this is demonstrated in \cite{BianchiIXattr}. In the case of vacuum Bianchi type VIII solutions, it
is demonstrated in \cite{cbu}. Due to the oscillations, the eigenvalues of $\mK$ switch places, and this means that, while the eigenvalues may be
distinct for long periods of time, there is generically a sequence of times, tending to $-\infty$, such that two eigenvalues coincide for each element
of the sequence. In other words, Bianchi type VIII and IX solutions, while non-degenerate for long periods of time, are generically not non-degenerate
on a time interval stretching to $-\infty$. Turning to silence, the $\a$-limit sets of generic Bianchi type VIII and IX solutions are expected to
include all the Taub points. This means that $\chK$ cannot have a silent upper bound on an interval stretching to $-\infty$. On the other hand,
$\chK$ can be expected to have a silent upper bound on large intervals. To conclude, in the oscillatory setting, the conditions of non-degeneracy
and silence can only be expected to hold on large intervals, but not on intervals stretching to $-\infty$. 

Consider generic Bianchi type I, II, VI${}_{0}$ and VII${}_{0}$ orthogonal perfect fluid solutions with $2/3<\g<2$. Then $\mK$ and $\chK$ converge
and $\chK$ is asymptotically negative definite. This follows from \cite[Subsection~15.2]{KGCos} and \cite[Subsection~17.1]{KGCos}. In the
case of Bianchi type VI${}_{0}$, we also need to appeal to \cite[Theorem~1.6, p.~3076]{hog}. The eigenvalues of $\mK$ can be expected to generically
be distinct. However, there is, to the best of our knowledge, no formal proof of this statement. Note also that $q$ converges exponentially to $2$
in the generic setting. Finally, $\d_{\tau}\mK^{i}_{\phantom{i}j}$ converges to zero exponentially in this setting, so that $\hml_{U}\mK$ converges to
zero exponentially with respect to every weighted $C^{k}$ and Sobolev norm. 

Finally, consider the stiff fluid setting. Due to \cite[Subsection~15.1]{KGCos} and \cite[Subsection~17.1]{KGCos}, $\mK$ and $\chK$ converge
and $\chK$ is asymptotically negative definite. Moreover, $q-2$ and $\hml_{U}\mK$ converge to zero exponentially with
respect to every weighted $C^{k}$ and Sobolev norm. 

\textbf{Bianchi class B solutions.} In the case of non-exceptional Bianchi class B solutions, there are results in
\cite{RadermacherNonStiff,RadermacherStiff}.
However, the analysis is in that case carried out with respect to an orthonormal frame which is not necessarily an eigenframe for $\mK$. Moreover,
one of the elements of the orthonormal frame is a time dependent multiple of a fixed element of $\mfg$. However, the remaining two elements of the
orthonormal frame are typically not. This complicates the analysis of the asymptotic behaviour of $\mK$. In fact, the analysis of
\cite{RadermacherNonStiff,RadermacherStiff} does not give the asymptotics of $\{X_{A}\}$. This makes it more difficult to prove that $\mK$ is bounded etc.
We expect it to be possible to prove the relevant bounds. However, the corresponding analysis can be expected to be more lengthy than would be appropriate
for an appendix to these notes. We therefore do not carry it out here. The issue of silence is discussed in \cite[Subsections~15.1, 15.2 and 17.1]{KGCos}. 
Finally, we expect the solutions to generically be non-degenerate asymptotically. 

\section{Specifying data on the singularity}\label{ssection:examples}

Turning to the spatially inhomogeneous setting, we first consider solutions obtained by specifying data on the singularity. Most of the results
in the literature concern classes of solutions with a $2$-dimensional isometry group; cf., e.g., \cite{kar,iak,ren,sta,ABIF,aeta}. However, there are
results in the absence of symmetries; cf., e.g.,
\cite{aarendall,daetal,fal}. The results of \cite{aarendall,daetal} are obtained under circumstances that can be expected to be ``generic'';
one is allowed to specify the ``correct'' number of free functions on the singularity. On the other hand, these results are obtained in
the real analytic setting, which is not so natural in the context of general relativity. The results of \cite{fal} are not expected to correspond to
a generic setting, since the asymptotic states in this result are known to be unstable. In fact, in order to obtain solutions, the authors, roughly speaking,
have to eliminate degrees of freedom on the singularity. In the present section, we focus on the results of \cite{aarendall,fal}. However, in \cite{daetal},
results similar to those of \cite{aarendall} are obtained in the case of higher dimensions and different matter models. The interested reader is therefore
encouraged to carry out arguments similar to the ones below in the situations considered in \cite{daetal}. We begin by discussing the quiescent cosmological
singularities considered by Andersson and Rendall in \cite{aarendall}.

\textbf{Stiff fluids and scalar fields in $3+1$-dimensions.} Consider the spacetimes constructed in \cite{aarendall}. The asymptotics of solutions are
described in the statements of \cite[Theorems~1 and 2, pp.~484--485]{aarendall}. Note that Andersson and Rendall use a Gaussian time coordinate in
\cite{aarendall} (in particular, the lapse function equals one and the shift vector field equals zero) and $t=0$ corresponds to the singularity. Note
also that our sign convention concerning the second fundamental form is the opposite to the one of Andersson and Rendall. From
\cite[Theorems~1 and 2, pp.~484--485]{aarendall} it follows that there are constants $\zeta,C>0$ such that 
\[
|t\theta-1|\leq Ct^{\zeta}.
\]
In particular, it is clear that the singularity is a crushing singularity. For a Gaussian time coordinate, (\ref{eq:hUvarrhoident}) yields
\[
\d_{t}\varrho=\theta=\frac{1}{t}+O(t^{-1+\zeta}).
\]
Integrating this equality yields the conclusion that $\varrho=\ln t+\varrho_{0}+O(t^{\zeta})$. Here $\varrho_{0}$ is a function of the spatial variables
only. In particular $\varrho\rightarrow-\infty$ in the direction of the singularity. According to \cite[Theorems~1 and 2, pp.~484--485]{aarendall},
$\mK^{i}_{\phantom{i}j}$ converges exponentially to the components of a positive definite matrix. Since the trace of this matrix is $1$, it is also clear that
all the eigenvalues converge to values that are strictly between $0$ and $1$. In \cite{aarendall} it is also clearly possible to specify data on the
singularity in such a way that the eigenvalues of $\mK$ are asymptotically distinct. 

In the setting of \cite{aarendall}, (\ref{eq:chKmKthetarelation}) reads 
\begin{equation}\label{eq:chKijexpre}
\chK=\mK+\theta^{-1}(\d_{t}\ln\theta)\mathrm{Id}.
\end{equation}
In order to estimate $\d_{t}\theta$, note that \cite[(3b), p.~481]{aarendall} implies that 
\begin{equation}\label{eq:thetainvdttheta}
\theta^{-2}\d_{t}\theta+1=-\theta^{-2}R-4\pi\theta^{-2}\tr S+12\pi\theta^{-2}\rho,
\end{equation}
where $R$ is the scalar curvature of the spatial metric. Moreover, in the case of a scalar field, $S$ is given by \cite[(5c), p.~481]{aarendall}
and $\rho$ is given by \cite[(5a), p.~481]{aarendall}. In the case of a stiff fluid, $S$ is given by \cite[(8c), p.~482]{aarendall}
and $\rho$ is given by \cite[(8a), p.~482]{aarendall}. Due to \cite[Lemma~6, p.~504]{aarendall}, it follows that $\theta^{-2}R$ converges to 
zero exponentially in $\tau$-time, where $\tau:=\ln t$. In the case of a scalar field, it can be calculated that 
\[
\tr S-3\rho=-2g^{ab}e_{a}(\phi)e_{b}(\phi).
\]
Combining this observation with the argument presented on \cite[p.~505]{aarendall} implies that $\theta^{-2}(\tr S-3\rho)$ converges to zero exponentially. 
In the case of the stiff fluid, 
\[
\tr S-3\rho=-4\mu|u|^{2}. 
\]
Combining this observation with the statements on \cite[p.~505]{aarendall}, it follows that $\theta^{-2}(\tr S-3\rho)$ converges to zero exponentially;
note that the quantity $M_{ab}$ is introduced in \cite[(47), p.~493]{aarendall}. Summing up the above conclusions, it is clear that (\ref{eq:thetainvdttheta})
implies that $\theta^{-2}\d_{t}\theta$ converges to $-1$ exponentially. Combining this observation with (\ref{eq:chKijexpre}) and the fact that the eigenvalues
of $\mK$ belong to $(0,1)$ yields the conclusion that $\chK$ converges to a negative definite matrix. Note also  that the deceleration parameter $q$
converges to $2$ exponentially. 

By arguments similar to the above, it can also be argued that $\hml_{U}\mK$ converges to zero exponentially. We leave the details to the reader. 

The above estimates are only in $C^{0}$, but in the present paper we make assumptions in weighted $C^{k}$- and $H^{k}$-spaces. The question is then if
one can draw conclusions concerning higher order derivatives from \cite[Theorems~1 and 2, pp.~484--485]{aarendall}. The results of \cite{aarendall} build
on \cite{kar}. Consider,
for this reason, \cite[Theorem~3, p.~1350]{kar}. The proof of existence and uniqueness of solutions is based on a fixed point argument. In particular,
the authors prove that a certain map is a contraction; cf. \cite[pp.~1350--1354]{kar}, in particular \cite[Step~3, p.~1353]{kar}. The norm with respect
to which the map is a contraction is $|\cdot|_{a}$ introduced at the bottom of \cite[p.~1351]{kar}. Considering this norm, it is clear that the
estimates that are obtained as a result of the argument are such that they extend a small distance into the complex plane. Combining this observation
with Cauchy's theorem in each spatial variable separately, it is clear that similar estimates hold for any number of spatial derivatives. For this reason,
it should be possible to obtain conclusions for any number of spatial derivatives. Here, we do not attempt to convert this information into the type of
estimates of interest in these notes. However, it is reasonable to expect the estimates derived previously to not only hold in $C^{0}$ but with respect to
any $C^{k}$-norm. 

\textbf{Asymptotically Kasner solutions.} In \cite{fal}, the authors specify data on the singularity for Einstein's vacuum equations. In particular, they
prescribe Kasner-like asymptotics. In \cite[Theorem~1.7]{fal}, they provide asymptotic conditions on the solutions that guarantees uniqueness. In
particular, \cite[(1.10)]{fal} states that
\begin{equation}\label{eq:asymptoticKasner}
  \textstyle{\sum}_{r=0}^{1}\sum_{|\a|\leq 2-r}t^{r}|\d_{t}^{r}\d^{\a}(\bk^{i}_{\phantom{i}j}-t^{-1}\kappa^{i}_{\phantom{i}j})|\leq Ct^{-1+\varepsilon}
\end{equation}
for some constants $C>0$ and $\varepsilon>0$. Here $\kappa$ is a prescribed matrix valued function depending only on the spatial variables (since our
conventions are opposite to those of \cite{fal}, the $\kappa$ appearing here is obtained by multiplying the object with the same name in
\cite{fal} with $-1$). In particular,
$\mathrm{tr}\kappa=1$ here. Due to (\ref{eq:asymptoticKasner}), the estimate $|t\theta-1|\leq Ct^{\varepsilon}$ holds. Thus we have a crushing singularity
and since the time coordinate is Gaussian, we again conclude that $\varrho=\ln t+\varrho_{0}+O(t^{\varepsilon})$. Combining these observations with
(\ref{eq:asymptoticKasner}) yields the conclusion that $\mK^{i}_{\phantom{i}j}$ converges exponentially to $\kappa^{i}_{\phantom{i}j}$. By assumption, the
diagonal components of $\kappa$ are distinct and $\kappa$ is a triangular matrix; cf. \cite[Theorem~1.1]{fal}. In particular, $\mK$ asymptotically has
distinct eigenvalues. Since the time coordinate is Gaussian,
\[
(\hml_{U}\mK)^{i}_{\phantom{i}j}=\theta^{-1}\d_{t}(\bk^{i}_{\phantom{i}j}/\theta)=\theta^{-2}\d_{t}\bk^{i}_{\phantom{i}j}-\theta^{-3}\theta_{t}\bk^{i}_{\phantom{i}j}.
\]
By arguments similar to the above, it follows that this expression converges to zero exponentially with respect to $\varrho$. It can also be demonstrated
that $\theta^{-2}\theta_{t}$ converges exponentially to $-1$, so that $q$ converges exponentially to $2$. Combining this observation with
(\ref{eq:chKijexpre}) and the fact that the eigenvalues of $\mK$ are asymptotically distinct and satisfy the Kasner relations (cf.
\cite[(1), Theorem~1.1, p.~2]{fal}), we conclude that $\chK$ asymptotically has a silent upper bound. Note also that (\ref{eq:asymptoticKasner}) yields
the conclusion that $\theta^{-1}|\d^{\a}\theta|\leq Ct^{\varepsilon}$ for $1\leq |\a|\leq 2$. In particular, the relative spatial variation of $\theta$ converges
to zero asymptotically. Finally, since the time coordinate is Gaussian, $N=1$ and $\chi=0$. 

\section{Stable big bang formation}\label{section:stablebigbangformApp}

As pointed out in Subsection~\ref{ssection:quiescent}, the results contained in \cite{rasql,rasq,rsh,specks3,FRS} yield stable big bang formation in
the case
of stiff fluids, in the case of scalar fields, and in the case of higher dimensions. Here we focus on the results of \cite{rasq}. The main conclusions
concerning the asymptotics are summarised in \cite[Section~1.4, p.~4303--4306]{rasq}. In the present notes, we have the opposite conventions (relative to
\cite{rasq}) concerning the second fundamental form. In what follows, we therefore reinterpret the results of \cite{rasq} accordingly without further
comment. To begin with, \cite[(1.10b), p.~4304]{rasq} yields the conclusion that $\varrho\rightarrow-\infty$ in the direction of the big bang. Moreover,
\cite[(1.10d), p.~4304]{rasq} yields the conclusion that $\theta\rightarrow\infty$ and that $\mK$ converges. Note, finally, that $\chi=0$ and that $N$
converges to $1$ exponentially; cf. \cite[(1.10a), p.~4304]{rasq}. These observations are consistent with the assumptions made in these notes, but they
are clearly not sufficient to verify that the assumptions are satisfied. We encourage the interested reader to refine the results of
\cite{rasql,rasq,rsh,specks3,FRS} in order
to verify that the assumptions made here (except, possibly, for the non-degeneracy) are satisfied. However, we do not attempt to carry out such an analysis
here. 

\section{$\tn{3}$-Gowdy spacetimes}\label{section:t3Gowdy}

Concerning Gowdy symmetric spacetimes, there are several results describing the asymptotics in the direction of the singularity. In the polarised Gowdy
setting, an analysis of the asymptotics is contained in \cite{chisamo}. There are also results in which the authors specify data on the singularity; cf.,
e.g., \cite{kar,ren,sta}. However, the basis for the discussion in the present section is the analysis concerning generic $\tn{3}$-Gowdy vacuum spacetimes
contained in \cite{asvelGowdy,SCCGowdy}. Here we use the areal time foliation. The metric then takes the form
\begin{equation}\label{eq:Gowdymetricarealapp}
  g=t^{-1/2}e^{\lambda/2}(-dt^{2}+d\vartheta^{2})+te^{P}(dx+Qdy)^{2}+te^{-P}dy^{2}
\end{equation}
on $\tn{3}\times (0,\infty)$ (note, however, that the quantity $\lambda$ introduced here has opposite sign relative to the conventions of
\cite{asvelGowdy,SCCGowdy}). Here the functions $P$, $Q$ and $\lambda$ only depend on $t$ and $\vartheta$, so that the metric is invariant under the
action of $\tn{2}$ corresponding to translations in $x$ and $y$. Note that the area of the orbit of $\tn{2}$ is proportional to $t$. This is the
reason we speak of the \textit{areal} time coordinate and foliation. Here we are interested in the asymptotics as $t\rightarrow 0+$. However, in many
contexts, it is convenient to change time coordinate to $\tau=-\ln t$. With respect to this time coordinate, the singularity corresponds to
$\tau\rightarrow\infty$. When we speak of a $\tn{3}$-Gowdy spacetime in what follows, we assume that the metric takes the form
(\ref{eq:Gowdymetricarealapp}) and speak of $t$, $\vartheta$, $x$, $y$, $\tau$, $P$, $Q$ and $\lambda$ without further comment. 

We begin by calculating $\mK$ for the areal foliation of $\tn{3}$-Gowdy vacuum spacetime. 

\subsection{Components of the expansion normalised Weingarten map}

In order to carry out calculations, we appeal to \cite[Appendix~A]{AAR}. In this appendix, the curvatures and connection coefficients of $\tn{2}$-symmetric
spacetimes are calculated. In order to specialise to $\tn{3}$-Gowdy spacetimes, it is sufficient to put $G=H=0$ and $\a=1$ in
\cite[(1.1), p.~1568]{AAR}. In what follows, we use the frame $\{e_{\b}\}$ introduced in \cite[(1.7), p.~1571]{AAR} with $G=H=0$ and $\a=1$ (in all
the references to the formulae in \cite{AAR} that follow, we take this substitution for granted). We also use the dual frame $\{\xi^{\b}\}$ introduced 
on \cite[p.~1634]{AAR}.

We define $\mK$ as at the beginning of these notes. Moreover, we use the notation
\[
\mK^{\vartheta}_{\phantom{\vartheta}\vartheta}=d\vartheta(\mK\d_{\vartheta}),\ \ \
\mK^{\vartheta}_{\phantom{\vartheta}x}=d\vartheta(\mK\d_{x}),\ \ \
\mK^{\vartheta}_{\phantom{\vartheta}y}=d\vartheta(\mK\d_{y}),\ \ \
\mK^{x}_{\phantom{x}\vartheta}=dx(\mK\d_{\vartheta}),\ \ \
\]
etc.
\begin{lemma}\label{lemma:mKijGowdycalc}
  Consider a $\tn{3}$-Gowdy vacuum spacetime. Then the non-zero components of $\mK$ with respect to the frame $\{\d_{\vartheta},\d_{x},\d_{y}\}$ (with dual
  frame $\{d\vartheta,dx,dy\}$) are given by
  \begin{eqnarray*}
    &\mK^{\vartheta}_{\phantom{\vartheta}\vartheta}  = \rho_{0}^{-1}(t\lambda_{t}-1),\ \ \ & \mK^{x}_{\phantom{x}x}=2\rho_{0}^{-1}(1+tP_{t})-2\rho_{0}^{-1}te^{2P}QQ_{t},\\
    &\mK^{x}_{\phantom{x}y}  = 4\rho_{0}^{-1}tP_{t}Q+2\rho_{0}^{-1}(1-e^{2P}Q^{2})tQ_{t},\ \ \ & \mK^{y}_{\phantom{y}x}=2\rho_{0}^{-1}te^{2P}Q_{t},\\
    &\mK^{y}_{\phantom{y}y}  = 2\rho_{0}^{-1}(1-tP_{t})+2\rho_{0}^{-1}te^{2P}QQ_{t},&
  \end{eqnarray*}
  where $\rho_{0}$ is defined by
  \begin{equation}\label{eq:rhozdef}
    \rho_{0}:=t\lambda_{t}+3.
  \end{equation}
  Moreover,
  \begin{equation}\label{eq:thetaGowdy}
    \theta=\textstyle{\frac{1}{4}}t^{-3/4}e^{-\lambda/4}\rho_{0}.
  \end{equation}
\end{lemma}
\begin{remark}
  Due to (\ref{eq:tlambdatidentity}) below, it follows that $t\lambda_{t}$ is non-negative. This means that $\lambda_{\tau}$ is negative and
  that $\rho_{0}\geq 3$. Combining these observations with (\ref{eq:thetaGowdy}) yields the conclusion that $\theta$ tends to infinity
  uniformly and exponentially (in $\tau$) in the direction of the singularity. 
\end{remark}
\begin{remark}\label{remark:eigenvaluesintthreegowdy}
  Let $\bmK$ denote the $2\times 2$-matrix with components $\mK^{x}_{\phantom{x}x}$, $\mK^{x}_{\phantom{x}y}$, $\mK^{y}_{\phantom{y}x}$ and
  $\mK^{y}_{\phantom{y}y}$. Then
  \[
  \mathrm{tr}\bmK=4\rho_{0}^{-1},\ \ \
  \det\bmK=4\rho_{0}^{-2}(1-P_{\tau}^{2}-e^{2P}Q_{\tau}^{2}).
  \]
  Using this information we can calculate the eigenvalues of $\mK$. They are given by
  \begin{equation}\label{eq:lambdait3gowdydef}
  \ell_{1}:=\rho_{0}^{-1}(t\lambda_{t}-1),\ \ \
  \ell_{2}:=2\rho_{0}^{-1}(1-\kappa^{1/2}),\ \ \
  \ell_{3}:=2\rho_{0}^{-1}(1+\kappa^{1/2}),
  \end{equation}
  where
  \begin{equation}\label{eq:kappadefgowdy}
    \kappa=P_{\tau}^{2}+e^{2P}Q_{\tau}^{2}.
  \end{equation}
  Finally, note that combining (\ref{eq:rhozdef}), (\ref{eq:kappadefgowdy}) and (\ref{eq:tlambdatidentity}) below with the formulae for the eigenvalues
  yields the conclusion that the eigenvalues are globally uniformly bounded. 
\end{remark}
\begin{proof}
  Note that
  \[
  \bk_{ij}=\bk(e_{i},e_{j})=\ldr{\nabla_{e_{i}}e_{0},e_{j}}=\G_{i0}^{j},
  \]
  where we use the notation for connection coefficients introduced in \cite[Section~A.2]{AAR}. Due to the calculations carried out on
  \cite[p.~1636]{AAR}, it follows that
  \begin{equation}\label{eq:bkiibkoneAbktwothree}
    \bk_{ii}=-\g^{i}_{0i},\  \ \
    \bk_{1A}=-\frac{1}{2}\g^{A}_{01},\ \ \
    \bk_{23}=-\frac{1}{2}\g^{2}_{03},
  \end{equation}
  where there is no summation in the first equality and $A\in \{2,3\}$ in the second equality. Moreover, the $\g^{\b}_{\de\g}$ are the structure
  constants associated with the frame $\{e_{\b}\}$; cf. \cite[Section~A.1, p.~1634--1635]{AAR}. Combining this observation with the calculations
  carried out in \cite[Section~A.1]{AAR} yields the conclusion that
  \begin{align*}
    \bk_{11} = & \textstyle{\frac{1}{4}}t^{1/4}e^{-\lambda/4}(\lambda_{t}-t^{-1}),\ \ \
    \bk_{22} = \textstyle{\frac{1}{2}}t^{1/4}e^{-\lambda/4}(t^{-1}+P_{t}),\\
    \bk_{33} = & \textstyle{\frac{1}{2}}t^{1/4}e^{-\lambda/4}(t^{-1}-P_{t}),\\
  \end{align*}
  so that, in particular, the mean curvature is given by (\ref{eq:thetaGowdy}). Here, due to \cite[(2.4), p.~1587]{AAR}; the fact that $K=J=0$ (this
  follows from the fact that we are considering Gowdy spacetimes); and the fact that $P_{1}=\Lambda=0$ (this is a consequence of the fact that we are
  considering solutions to Einstein's vacuum equations),
  \begin{equation}\label{eq:tlambdatidentity}
    t\lambda_{t}=t^{2}[P_{t}^{2}+P_{\vartheta}^{2}+e^{2P}(Q_{t}^{2}+Q_{\vartheta}^{2})].
  \end{equation}
  Next, combining (\ref{eq:bkiibkoneAbktwothree}) with \cite[(A.3), p.~1634]{AAR}, \cite[(A.4), p.~1634]{AAR} and the fact that $J=K=0$ yields
  $\bk_{12}=\bk_{13}=0$. Finally, due to (\ref{eq:bkiibkoneAbktwothree}) and \cite[Section~A.1]{AAR},
  \[
  \bk_{23}=\textstyle{\frac{1}{2}}t^{1/4}e^{-\lambda/4}e^{P}Q_{t}.
  \]
  Using the notation (\ref{eq:rhozdef}), we conclude from the above that the non-zero components of $\theta^{-1}\bk$ are
  \begin{align*}
    \theta^{-1}\bk_{11} = & \rho_{0}^{-1}(t\lambda_{t}-1),\ \ \
    \theta^{-1}\bk_{22} = 2\rho_{0}^{-1}(1+tP_{t}),\\
    \theta^{-1}\bk_{33} = & 2\rho_{0}^{-1}(1-tP_{t}),\ \ \
    \theta^{-1}\bk_{23} = 2\rho_{0}^{-1}te^{P}Q_{t}.
  \end{align*}
  Introducing $\mK$ as before, note that
  \[
  \xi^{i}(\mK e_{j})=\ldr{\mK e_{j},e_{i}}=\theta^{-1}\bk_{ij}.
  \]
  Given the above terminology and calculations, it can be demonstrated that the conclusions of the lemma hold. 
\end{proof}

\subsection{The asymptotic limits of the eigenvalues of $\mK$ and $\chK$}\label{subsection:aslimiteigenvalues}

Next, it is of interest to calculate the asymptotic limits of the eigenvalues of $\mK$. Let us, to this end, first note that, given a $\tn{3}$-Gowdy
symmetric solution to Einstein's vacuum equations, and given a $\vartheta_{0}\in\so$, there is a non-negative number $v_{\infty}(\vartheta_{0})$ such that
\[
\lim_{\tau\rightarrow\infty}\kappa(\vartheta_{0},\tau)=v_{\infty}^{2}(\vartheta_{0}),
\]
where $\kappa$ is defined by (\ref{eq:kappadefgowdy}). This statement is an immediate consequence of \cite[Corollary~6.9, p.~1009]{asvelGowdy}. We refer to the
function $v_{\infty}:\so\rightarrow [0,\infty)$ as the \textit{asymptotic velocity}. Next, let
$\md_{\vartheta_{0},\tau}:=[\vartheta_{0}-e^{-\tau},\vartheta_{0}+e^{-\tau}]$. Then \cite[Proposition~1.3, p.~983]{asvelGowdy} yields the conclusion that
\begin{equation}\label{eq:limkappawp}
  \lim_{\tau\rightarrow\infty}\|\kappa(\cdot,\tau)-v_{\infty}^{2}(\vartheta_{0})\|_{C^{0}(\md_{\vartheta_{0},\tau})}=0,\ \ \
  \lim_{\tau\rightarrow\infty}\|\wp(\cdot,\tau)\|_{C^{0}(\md_{\vartheta_{0},\tau})}=0,
\end{equation}
where
\[
\wp:=e^{-2\tau}(P_{\vartheta}^{2}+e^{2P}Q_{\vartheta}^{2}).
\]
Combining this notation with (\ref{eq:rhozdef}), (\ref{eq:tlambdatidentity}) and (\ref{eq:kappadefgowdy}), it follows that $\rho_{0}=3+\kappa+\wp$ and
that $t\lambda_{t}=\kappa+\wp$. Combining these equalities with (\ref{eq:limkappawp}) yields
\begin{equation}\label{eq:rhoztlambdatlim}
  \lim_{\tau\rightarrow\infty}\|\rho_{0}(\cdot,\tau)-v_{\infty}^{2}(\vartheta_{0})-3\|_{C^{0}(\md_{\vartheta_{0},\tau})}=0,\ \ \
  \lim_{\tau\rightarrow\infty}\|(t\lambda_{t})(\cdot,\tau)-v_{\infty}^{2}(\vartheta_{0})\|_{C^{0}(\md_{\vartheta_{0},\tau})}=0.
\end{equation}
The limits of the eigenvalues $\ell_{i}$ introduced in (\ref{eq:lambdait3gowdydef}) are thus given by
\begin{align}
  \lim_{\tau\rightarrow\infty}\left\|\ell_{1}(\cdot,\tau)
  -\frac{v_{\infty}^{2}(\vartheta_{0})-1}{v_{\infty}^{2}(\vartheta_{0})+3}\right\|_{C^{0}(\md_{\vartheta_{0},\tau})} = & 0,\label{eq:ellonelim}\\
  \lim_{\tau\rightarrow\infty}\left\|\ell_{2}(\cdot,\tau)
  -2\frac{1-v_{\infty}(\vartheta_{0})}{v_{\infty}^{2}(\vartheta_{0})+3}\right\|_{C^{0}(\md_{\vartheta_{0},\tau})} = & 0,\label{eq:elltwolim}\\
  \lim_{\tau\rightarrow\infty}\left\|\ell_{3}(\cdot,\tau)
  -2\frac{1+v_{\infty}(\vartheta_{0})}{v_{\infty}^{2}(\vartheta_{0})+3}\right\|_{C^{0}(\md_{\vartheta_{0},\tau})} = & 0.\label{eq:ellthreelim}
\end{align}
Denoting the limits by $\ell_{i,\infty}(\vartheta_{0})$, it can be verified that
\begin{equation}\label{eq:elliinfKasnerrelations}
  \textstyle{\sum}\ell_{i,\infty}(\vartheta_{0})=1,\ \ \
  \textstyle{\sum}\ell_{i,\infty}^{2}(\vartheta_{0})=1.
\end{equation}
In other words, the limits of the eigenvalues satisfy both of the Kasner relations. Next, note that if $\g$ is a past inextendible causal curve,
then the $\vartheta$ coordinate of $\g$ converges in the direction of the singularity. Call the limit $\vartheta_{0}$. Then, if the $\tau$-component
of $\g(s)$ is denoted $\tau(s)$ and the $\vartheta$-component of $\g(s)$ is denoted $\vartheta(s)$, then $\vartheta(s)\in\md_{\vartheta_{0},\tau(s)}$; this is
an immediate consequence of the causal structure. Thus $\ell_{i}$ converges uniformly to $\ell_{i,\infty}$ in $J^{+}(\g)$. In particular, $\ell_{i}$ converges
to $\ell_{i,\infty}$ along $\g$.

\textbf{Stable regime.} Considering (\ref{eq:ellonelim})--(\ref{eq:ellthreelim}), it is clear that there is a conceptual difference between
the case $v_{\infty}(\vartheta_{0})<1$ and the case $v_{\infty}(\vartheta_{0})>1$. The reason is that if $v_{\infty}(\vartheta_{0})<1$, then
$\ell_{1,\infty}(\vartheta_{0})<0<\ell_{2,\infty}(\vartheta_{0})<\ell_{3,\infty}(\vartheta_{0})$, and if $v_{\infty}(\vartheta_{0})>1$, then
$\ell_{2,\infty}(\vartheta_{0})<0$ and $\ell_{1,\infty}(\vartheta_{0})$ and $\ell_{3,\infty}(\vartheta_{0})$ are strictly positive. Moreover, the eigenvector
fields corresponding to $\ell_{2}$ and $\ell_{3}$ commute. To summarise, if $v_{\infty}(\vartheta_{0})<1$, then
there is asymptotically only one negative eigenvalue of $\mK$, and the eigenvector fields corresponding to the remaining eigenvalues commute. This is a
special situation which is due to the assumption of $\tn{3}$-Gowdy symmetry. As will become clear in the accompanying article on geometry,
cf. \cite{RinGeometry}, the corresponding structure is related to the existence of a stable and convergent regime in the case of $\tn{3}$-Gowdy symmetry
for Einstein's vacuum equations; cf. Subsection~\ref{ssection:convergentsettingGowdy} below. 

\textbf{The eigenvalues of $\chK$.} Next, we wish to calculate the eigenvalues of $\chK$. To this end, we first need to calculate the deceleration
parameter, given by $q=-1-\hU(3\ln\theta)$; cf. (\ref{eq:hUnlnthetamomqbas}).

\begin{lemma}\label{lemma:quniformbdGowdy}
  Consider a $\tn{3}$-Gowdy symmetric vacuum spacetime and let $q$ denote the associated deceleration parameter. Then $q$ is uniformly bounded in the
  direction of the singularity. Moreover, if $\vartheta_{0}\in\so$, 
  \begin{equation}\label{eq:qlocallimgowdy}
    \lim_{\tau\rightarrow\infty}\|q(\cdot,\tau)-2\|_{C^{0}(\md_{\vartheta_{0},\tau})}=0.
  \end{equation}
\end{lemma}
\begin{remark}
  One particular consequence of (\ref{eq:qlocallimgowdy}) is that if $\g$ is a past inextendible causal curve, then $q$ converges to $2$ uniformly in
  $J^{+}(\g)$.
\end{remark}
\begin{proof}
  Recalling that $\theta$ is given by (\ref{eq:thetaGowdy}),
  \begin{equation}\label{eq:decelerationparamGowdy}
    q=-1-12\rho_{0}^{-1}t\d_{t}[\ln(t^{-3/4}e^{-\lambda/4}\rho_{0})]=2-12\rho_{0}^{-1}t\d_{t}\ln\rho_{0}.
  \end{equation}
  In order to calculate $t\d_{t}\rho_{0}=t\d_{t}(t\lambda_{t})$, note that \cite[(2.6), p.~1587]{AAR} yields
  \[
  t\d_{t}(t\lambda_{t}-3)=t^{2}\lambda_{\vartheta\vartheta}-t^{2}(P_{t}^{2}+e^{2P}Q_{t}^{2}-P_{\vartheta}^{2}-e^{2P}Q_{\vartheta}^{2})+t\lambda_{t}.
  \]
  Recalling (\ref{eq:tlambdatidentity}) and that, due to \cite[(2.7), p.~1587]{AAR},
  \begin{equation}\label{eq:lambdatheta}
    \lambda_{\vartheta}=2t(P_{t}P_{\vartheta}+e^{2P}Q_{t}Q_{\vartheta}),
  \end{equation}
  we conclude that
  \begin{equation}\label{eq:tdttlambdatGowdy}
    \begin{split}
      t\d_{t}(t\lambda_{t}) = & -2e^{-2\tau}(P_{\tau\vartheta}P_{\vartheta}+P_{\tau}P_{\vartheta\vartheta}+\d_{\vartheta}(e^{2P}Q_{\tau})Q_{\vartheta}
      +e^{2P}Q_{\tau}Q_{\vartheta\vartheta})\\
      & +2e^{-2\tau}(P_{\vartheta}^{2}+e^{2P}Q_{\vartheta}^{2}).
    \end{split}
  \end{equation}
  In order to analyse the boundedness of this expression, note, first of all, that $\kappa$ and $\wp$ are uniformly bounded in the direction of
  the singularity. This is an immediate consequence of, e.g., \cite[Lemma~5.1, p.~1000]{asvelGowdy}. The same lemma also yields the conclusion that
  there is a constant $C<\infty$ such that
  \[
  e^{-\tau}|P_{\tau\vartheta}|+e^{-2\tau}|P_{\vartheta\vartheta}|+e^{P-\tau}|Q_{\tau\vartheta}|+e^{P-2\tau}|Q_{\vartheta\vartheta}|\leq C
  \]
  for all $\tau\geq 0$. Thus $t\d_{t}(t\lambda_{t})$ is uniformly bounded in the direction of the singularity. Combining this observation with
  (\ref{eq:decelerationparamGowdy}) yields the conclusion that $q$ is uniformly bounded in the direction of the singularity. 

  Next, let us consider the behaviour of $q$ along causal curves. Note, to this end, that the second equality in (\ref{eq:limkappawp}) combined with
  \cite[Lemma~5.1, p.~1000]{asvelGowdy} yields
  \begin{align*}
    \lim_{\tau\rightarrow-\infty}[\|e^{-\tau}P_{\tau\vartheta}(\cdot,\tau)\|_{C^{0}(\md_{\vartheta_{0},\tau})}
      +\|e^{-2\tau}P_{\vartheta\vartheta}(\cdot,\tau)\|_{C^{0}(\md_{\vartheta_{0},\tau})}] = & 0,\\
    \lim_{\tau\rightarrow-\infty}[\|(e^{P-\tau}Q_{\tau\vartheta})(\cdot,\tau)\|_{C^{0}(\md_{\vartheta_{0},\tau})}
      +\|(e^{P-2\tau}Q_{\vartheta\vartheta})(\cdot,\tau)\|_{C^{0}(\md_{\vartheta_{0},\tau})}] = & 0.
  \end{align*}
  Summing up the above yields the conclusion that (\ref{eq:qlocallimgowdy}) holds. 
\end{proof}

Next, we consider the eigenvalues of $\chK$. Due to (\ref{eq:chKmKthetarelation}), they are given by $\lambda_{i}=\ell_{i}-(1+q)/3$. Due to
Remark~\ref{remark:eigenvaluesintthreegowdy} and the uniform bound on $q$, it is clear that the $\lambda_{i}$ are uniformly bounded in the
direction of the singularity. Combining (\ref{eq:ellonelim})--(\ref{eq:ellthreelim}) with (\ref{eq:qlocallimgowdy}) and the relation between
$\ell_{i}$ and $\lambda_{i}$ yields the conclusion that
\begin{align}
  \lim_{\tau\rightarrow\infty}\left\|\lambda_{1}(\cdot,\tau)
  +\frac{4}{v_{\infty}^{2}(\vartheta_{0})+3}\right\|_{C^{0}(\md_{\vartheta_{0},\tau})} = & 0,\label{eq:lambdaonelim}\\
  \lim_{\tau\rightarrow\infty}\left\|\lambda_{2}(\cdot,\tau)
  +\frac{[v_{\infty}(\vartheta_{0})+1]^{2}}{v_{\infty}^{2}(\vartheta_{0})+3}\right\|_{C^{0}(\md_{\vartheta_{0},\tau})} = & 0,\label{eq:lambdatwolim}\\
  \lim_{\tau\rightarrow\infty}\left\|\lambda_{3}(\cdot,\tau)
  +\frac{[v_{\infty}(\vartheta_{0})-1]^{2}}{v_{\infty}^{2}(\vartheta_{0})+3}\right\|_{C^{0}(\md_{\vartheta_{0},\tau})} = & 0.\label{eq:lambdathreelim}
\end{align}
In particular, it is clear that if $v_{\infty}(\vartheta_{0})\neq 1$, then $\chK$ is asymptotically negative definite. On the other hand, if
$v_{\infty}(\vartheta_{0})=1$, then the singularity could correspond to a Cauchy horizon. In fact, the flat Kasner solutions can be interpreted
as a $\tn{3}$-Gowdy solution with $Q=0$, $P=\tau$ and $\lambda=-\tau$. In this case $v_{\infty}(\vartheta)=1$ for all $\vartheta\in\so$.  

\subsection{Normal derivatives}

Introducing the notation $z^{1}=\vartheta$, $z^{2}=x$ and $z^{3}=y$, let
\[
\mK^{i}_{\phantom{i}j}=dz^{i}(\mK\d_{z^{j}}).
\]
Then (\ref{eq:hmlUmtinfixedspatialcoord}) yields the conclusion that
\[
(\hml_{U}\mK)^{i}_{\phantom{i}j}=\hU(\mK^{i}_{\phantom{i}j}).
\]
Combining this observation with Lemma~\ref{lemma:mKijGowdycalc} and the fact that
\begin{equation}\label{eq:NhNhU}
  N = t^{-1/4}e^{\lambda/4},\ \ \ \hN = \textstyle{\frac{1}{4}}t^{-1}\rho_{0},\ \ \
  \hU = 4\rho_{0}^{-1}t\d_{t},
\end{equation}
the components of $\hml_{U}\mK$ can be calculated. However, the detailed formulae are not of interest, since we only wish to estimate the asymptotic
behaviour. For future reference, it is also of interest to note that
\begin{equation}\label{eq:hUlnhNformuGowdy}
  \hU(\ln\hN)=4\rho_{0}^{-1}[t\d_{t}\ln\rho_{0}-1].
\end{equation}

\begin{lemma}
  Consider a $\tn{3}$-Gowdy symmetric vacuum spacetime. Then $\hU(\ln\hN)$ is uniformly bounded. Moreover, if $\vartheta_{0}\in\so$, 
  \begin{equation}\label{eq:hUlnhNlocallimgowdy}
    \lim_{\tau\rightarrow\infty}\left\|[\hU(\ln\hN)](\cdot,\tau)+\frac{4}{v_{\infty}^{2}(\vartheta_{0})+3}\right\|_{C^{0}(\md_{\vartheta_{0},\tau})}=0.
  \end{equation}
\end{lemma}
\begin{proof}
  The uniform boundedness of $\hU(\ln\hN)$ follows from (\ref{eq:hUlnhNformuGowdy}) the proof of Lemma~\ref{lemma:quniformbdGowdy}. The equality
  (\ref{eq:hUlnhNlocallimgowdy}) is an immediate consequence of (\ref{eq:rhoztlambdatlim}) and the proof of Lemma~\ref{lemma:quniformbdGowdy}.
\end{proof}

\subsection{The logarithmic volume density}

Due to (\ref{eq:Gowdymetricarealapp}), it can be calculated that
\[
\mu_{\bge}=t^{3/4}e^{\lambda/4}d\vartheta \wedge dx\wedge dy.
\]
Up to a function $\varrho_{0}$, depending only on $\vartheta$, it is thus clear that
\begin{equation}\label{eq:varrhoGowdy}
  \varrho=\lambda/4+3\ln t/4+\varrho_{0}.
\end{equation}
Note also that this means that $t\d_{t}\varrho=\rho_{0}/4$. In particular, $\d_{\tau}\varrho\leq -3/4$, so that $\varrho$ converges uniformly and
linearly (in $\tau$) to $-\infty$. 

\subsection{The low velocity regime}\label{ssection:convergentsettingGowdy}

Next, we want to compare the assumptions of these notes with the asymptotics of generic $\tn{3}$-Gowdy vacuum spacetimes in the direction of the
singularity. Due to \cite[Proposition~3, p.~1190]{SCCGowdy} and \cite[Theorem~2, p.~1190]{SCCGowdy}, for a generic solution, we have $0<v_{\infty}<1$ and
$\lim_{\tau\rightarrow\infty}P_{\tau}(\cdot,\tau)=v_{\infty}$ for all but a finite number of elements of $\sn{1}$. In the present subsection, we therefore focus on
the case that $0<v_{\infty}(\vartheta_{0})<1$ and $\lim_{\tau\rightarrow\infty}P_{\tau}(\vartheta_{0},\tau)=v_{\infty}(\vartheta_{0})$ for some $\vartheta_{0}\in\so$.
Due to \cite[Proposition~2, pp.~1186--1187]{SCCGowdy}, there is then an open interval $I$ containing $\vartheta_{0}$. Moreover, there are smooth functions
$v_{a}$, $\phi$, $r$ and $Q_{\infty}$ on $I$, where $\varepsilon<v_{a}<1-\varepsilon$ (for a constant $\varepsilon>0$), and a constant $\eta>0$ such that the
following estimates hold
\begin{align}
  \|P_{\tau}(\cdot,\tau)-v_{a}\|_{C^{k}(I)}+\|P(\cdot,\tau)-p(\cdot,\tau)\|_{C^{k}(I)} \leq & C_{k}e^{-\eta\tau},\label{eq:Pexp}\\
  \|e^{2p(\cdot,\tau)}Q_{\tau}(\cdot,\tau)-r\|_{C^{k}(I)}+\left\|e^{2p(\cdot,\tau)}[Q(\cdot,\tau)-Q_{\infty}]+r/(2v_{a})\right\|_{C^{k}(I)}
  \leq & C_{k}e^{-\eta\tau},\label{eq:Qexp}
\end{align}
for all $k\in\mathbb{N}$ and $\tau\geq 0$, where $p(\vartheta,\tau):=v_{a}(\vartheta)\tau+\phi(\vartheta)$. Note also that (\ref{eq:tlambdatidentity})
yields the conclusion that
\begin{equation}\label{eq:tlambdattauversion}
  -\lambda_{\tau}=t\lambda_{t}=P_{\tau}^{2}+e^{-2\tau}P_{\vartheta}^{2}+e^{2P}(Q_{\tau}^{2}+e^{-2\tau}Q_{\vartheta}^{2}).
\end{equation}
In particular,
\[
\|t\lambda_{t}(\cdot,t)-v_{a}^{2}\|_{C^{k}(I)}+\|\rho_{0}(\cdot,t)-3-v_{a}^{2}\|_{C^{k}(I)}\leq C_{k}e^{-\eta\tau}
\]
for all $\tau\geq 0$. Integrating this estimate yields a smooth function $\lambda_{\infty}$ on $I$ such that 
\[
\|\lambda(\cdot,\tau)+v_{a}^{2}\tau-\lambda_{\infty}\|_{C^{k}(I)}\leq C_{k}e^{-\eta\tau}
\]
for all $\tau\geq 0$. Combining this estimate with (\ref{eq:varrhoGowdy}) yields the conclusion that there is a smooth function
$\varrho_{\infty}$ on $I$ such that 
\begin{equation}\label{eq:varrhoasapp}
  \left\|\varrho+(v_{a}^{2}+3)\tau/4-\varrho_{\infty}\right\|_{C^{k}(I)}\leq C_{k}e^{-\eta\tau}
\end{equation}
for all $\tau\geq 0$. Combining (\ref{eq:thetaGowdy}) with the above asymptotics, it can also be verified that there is a smooth positive function
$\theta_{\infty}$ on $I$ such that 
\begin{equation}\label{eq:lnthetaasapp}
  \left\|\ln\theta-(v_{a}^{2}+3)\tau/4-\ln\theta_{\infty}\right\|_{C^{k}(I)}\leq C_{k}e^{-\eta\tau}
\end{equation}
for all $\tau\geq 0$. Note also that (\ref{eq:varrhoasapp}) and (\ref{eq:lnthetaasapp}) yield the conclusion that the spatial derivatives of $\ln\theta$
do not grow faster than linearly in $\varrho$. 

\textbf{Convergence of the expansion normalised Weingarten map.} Combining the formulae of Lemma~\ref{lemma:mKijGowdycalc} with the asymptotics given by
(\ref{eq:Pexp}) and (\ref{eq:Qexp}) yields
\begin{align*}
  \left\|\mK^{\vartheta}_{\phantom{\vartheta}\vartheta}(\cdot,t)-\frac{v_{a}^{2}-1}{v_{a}^{2}+3}\right\|_{C^{k}(I)} \leq & C_{k}e^{-\eta\tau},\\
  \left\|\mK^{x}_{\phantom{x}x}(\cdot,t)-\frac{2}{v_{a}^{2}+3}(1-v_{a}+Q_{\infty}r)\right\|_{C^{k}(I)} \leq & C_{k}e^{-\eta\tau},\\
  \left\|\mK^{x}_{\phantom{x}y}(\cdot,t)-\frac{2}{v_{a}^{2}+3}Q_{\infty}(Q_{\infty}r-2v_{a})\right\|_{C^{k}(I)} \leq & C_{k}e^{-\eta\tau},\\
  \left\|\mK^{y}_{\phantom{y}x}(\cdot,t)+\frac{2}{v_{a}^{2}+3}r\right\|_{C^{k}(I)} \leq & C_{k}e^{-\eta\tau},\\  
  \left\|\mK^{y}_{\phantom{y}y}(\cdot,t)-\frac{2}{v_{a}^{2}+3}(1+v_{a}-Q_{\infty}r)\right\|_{C^{k}(I)} \leq & C_{k}e^{-\eta\tau}
\end{align*}
for all $\tau\geq 0$. In particular, $\mK$ converges exponentially to a smooth tensorfield. Since $v_{a}=v_{\infty}$ on $I$, the eigenvalues
converge to the expressions appearing in (\ref{eq:ellonelim})--(\ref{eq:ellthreelim}) with $v_{\infty}(\vartheta_{0})$ replaced by $v_{a}$.
However, the convergence is now exponential in any $C^{k}$-norm on $I$. Since $0<v_{a}<1$, it is clear that the last two
asymptotic eigenvalues
are distinct and strictly positive. Since the first asymptotic eigenvalue is negative, we conclude that the asymptotic eigenvalues are distinct. 

\textbf{Decay of the normal derivative of the expansion normalised Weingarten map.} In order to estimate $\hml_{U}\mK$, it is sufficent to estimate
$\hU$ applied to the components of $\mK$ recorded in Lemma~\ref{lemma:mKijGowdycalc}. Since $\hU$ is given by (\ref{eq:NhNhU}) and since $\rho_{0}$
converges exponentially in any $C^{k}$-norm to a strictly positive function, it is sufficient to apply $t\d_{t}$ to the components of $\mK$. Let us
begin by considering $t\d_{t}$ applied to $t\lambda_{t}$ (and, thereby, to $\rho_{0}$). Combining (\ref{eq:tdttlambdatGowdy}) with (\ref{eq:Pexp}) and
(\ref{eq:Qexp}) and using the fact that $\varepsilon<v_{a}<1-\varepsilon$ yields
\[
\|\hU(\rho_{0})\|_{C^{k}(I)}+\|\hU(t\lambda_{t})\|_{C^{k}(I)}\leq C_{k}e^{-\eta\tau}
\]
for all $\tau\geq 0$. Combining this observation with (\ref{eq:decelerationparamGowdy}) yields the conclusion that
\[
\|q(\cdot,\tau)-2\|_{C^{k}(I)}\leq C_{k}e^{-\eta\tau}
\]
for all $\tau\geq 0$. On the other hand, due to (\ref{eq:chKmKthetarelation}), we know that $\chK=\mK-(1+q)\Id/3$. Since both terms on the right hand
side converge exponentially, the same is true of $\chK$. Moreover, the asymptotic eigenvalues of $\chK$ are
\[
-\frac{4}{v_{a}^{2}+3},\ \ \
-\frac{(v_{a}+1)^{2}}{v_{a}^{2}+3},\ \ \
-\frac{(v_{a}-1)^{2}}{v_{a}^{2}+3}.
\]
In particular, the asymptotic eigenvalues are all strictly negative, so that $\chK$ asymptotically has a silent upper bound. 

Next, note that \cite[(2.5) and (2.12), p.~1587]{AAR} yield
\begin{align}
  t\d_{t}(tP_{t}) = & t^{2}P_{\vartheta\vartheta}+t^{2}e^{2P}(Q_{t}^{2}-Q_{\vartheta}^{2}),\label{eq:tdttdtPeq}\\
  t\d_{t}(te^{2P}Q_{t}) = & t^{2}\d_{\vartheta}(e^{2P}Q_{\vartheta}).\label{eq:tdttetwoPdtQeq}
\end{align}
Combining these observations with the fact that $\varepsilon<v_{a}<1-\varepsilon$ yields the conclusion that
\[
\|\hU(tP_{t})\|_{C^{k}(I)}+\|\hU(te^{2P}Q_{t})\|_{C^{k}(I)}\leq C_{k}e^{-\eta\tau}
\]
for all $\tau\geq 0$. Due to (\ref{eq:tdttetwoPdtQeq}) and the asymptotics, it can also be deduced that 
\[
\|\hU(tQ_{t})\|_{C^{k}(I)}+\|tQ_{t}\|_{C^{k}(I)}\leq C_{k}e^{-\eta\tau}
\]
for all $\tau\geq 0$. Due to the above estimates and the formulae for the components of $\mK$ recorded in Lemma~\ref{lemma:mKijGowdycalc}, it
can be demonstrated that
\[
\|\hml_{U}\mK\|_{C^{k}(I)}\leq C_{k}e^{-\eta\tau}
\]
for all $\tau\geq 0$. For most of the components of $\mK$, this is an immediate consequence of the above estimates. However, let us consider
$\mK^{x}_{\phantom{x}y}$ in greater detail. When $\hU$ hits
$\rho_{0}^{-1}$, the result is an exponentially decaying term; when it hits $tP_{t}$, the result is an exponentially decaying term; and when it
hits the $Q$ appearing in the first term on the right hand side of the formula for $\mK^{x}_{\phantom{x}y}$, the result is the same. What remains is
to estimate 
\[
\hU[(1-e^{2P}Q^{2})tQ_{t}]=\hU(tQ_{t})-\hU(Q^{2})e^{2P}tQ_{t}-Q^{2}\hU(e^{2P}tQ_{t}).
\]
Due to the above estimates, the right hand side consists of exponentially decaying terms.

\textbf{The lapse function.} Due to (\ref{eq:NhNhU}) and (\ref{eq:hUlnhNformuGowdy}), it is clear that $\d_{\vartheta}\ln\hN$ converges exponentially
to a limit in any $C^{k}$-norm and that $\hU(\ln\hN)$ converges exponentially to a limit in any $C^{k}$-norm. 

\textbf{The mean curvature and deceleration parameter.} Due to (\ref{eq:lnthetaasapp}),
\[
\|\d_{\vartheta}\ln\theta\|_{C^{k}(I)}\leq C_{k}\ldr{\tau}
\]
for all $\tau\geq 0$. Combining this estimate with (\ref{eq:varrhoasapp}) yields
\[
\|\ldr{\varrho}^{-1}\d_{\vartheta}^{k+1}\ln\theta\|_{C^{0}(I)}\leq C_{k}
\]
for all $\tau\geq 0$, so that $\d_{\vartheta}\ln\theta$ satisfies the desired bounds. 

\textbf{Summarising.} Due to the above observations and the fact that the shift vector field vanishes, it can be verified that the geometric assumptions
we make in these notes are satisfied in the low velocity regime of $\tn{3}$-Gowdy vacuum spacetimes. 

\subsection{Inversions and false spikes}\label{subsection:falsespikes}

Due to \cite[Proposition~3, p.~1190]{SCCGowdy} and \cite[Theorem~2, p.~1190]{SCCGowdy}, there is, for a generic solution, a finite number of points
(possibly zero) such that $0<v_{\infty}<1$
and $\lim_{\tau\rightarrow\infty}P_{\tau}(\cdot,\tau)=-v_{\infty}$. The goal of the present subsection is to analyse the asymptotic behaviour of the foliation in
a neighbourhood of such a point, say $\vartheta_{0}$. Due to \cite[Proposition~1, pp.~1186]{SCCGowdy}, we know that $(Q_{1},P_{1}):=\mathrm{Inv}(Q,P)$ then
has the property that $P_{1\tau}(\vartheta_{0},\tau)\rightarrow v_{\infty}(\vartheta_{0})$. Moreover, 
$Q_{1}(\vartheta_{0},\tau)$ converges to $0$. Here the inversion of $(Q_{0},P_{0})$, written $\mathrm{Inv}(Q_{0},P_{0})$, is defined to equal $(Q_{1},P_{1})$,
where
\begin{equation}\label{eq:InversionDefinition}
  e^{-P_{1}}=\frac{e^{-P_{0}}}{Q_{0}^{2}+e^{-2P_{0}}},\ \ \
  Q_{1}=\frac{Q_{0}}{Q_{0}^{2}+e^{-2P_{0}}}.
\end{equation}
Note that $\mathrm{Inv}$ is an isometry of the upper half plane, when it is represented by $(\rn{2},g_{R})$, where $g_{R}:=dP^{2}+e^{2P}dQ^{2}$.
Moreover, the equations for $P$ and $Q$ are of wave map type with hyperbolic space as a target, so that isometries of hyperbolic space (such as
inversions) take solutions to solutions; this issue is discussed, e.g., in \cite[p.~2962]{raw}. If $(Q_{0},P_{0})$ is a solution to the $\tn{3}$-Gowdy
symmetric vacuum equations and $(Q_{1},P_{1})=\mathrm{Inv}(Q_{0},P_{0})$, the fact that $\mathrm{Inv}$ is an isometry of hyperbolic space thus implies,
e.g., that $(Q_{1},P_{1})$ is a solution to the equations and that 
\[
P_{1\tau}^{2}+e^{2P_{1}}Q_{1\tau}^{2}=P_{0\tau}^{2}+e^{2P_{0}}Q_{0\tau}^{2},\ \ \
P_{1\vartheta}^{2}+e^{2P_{1}}Q_{1\vartheta}^{2}=P_{0\vartheta}^{2}+e^{2P_{0}}Q_{0\vartheta}^{2}.
\]
In particular, $\kappa$, $\wp$, $\lambda$, $\rho_{0}$, $\theta$, $\varrho$, $\mK^{\vartheta}_{\phantom{\vartheta}\vartheta}$, $\ell_{i}$, $N$, $\hN$, $\hU$ etc.
introduced above are the same for the two solutions $(Q_{0},P_{0})$ and $(Q_{1},P_{1})$. However, it is less clear what happens for the remaining components
of $\mK$ appearing in the statement of Lemma~\ref{lemma:mKijGowdycalc}. In order to analyse the asymptotics of the remaining components, note that
\begin{align}
  e^{P_{1}}Q_{1\tau} = & e^{P_{0}}Q_{0\tau}+\frac{2e^{P_{0}}Q_{0}}{e^{2P_{0}}Q_{0}^{2}+1}(-Q_{0}e^{2P_{0}}Q_{0\tau}+P_{0\tau}),\label{eq:ePoneQonetaufalse}\\
  P_{1\tau} = & -P_{0\tau}+2\frac{Q_{0}e^{2P_{0}}Q_{0\tau}+e^{2P_{0}}Q_{0}^{2}P_{0\tau}}{Q_{0}^{2}e^{2P_{0}}+1}.\label{eq:Ponetaufalse}
\end{align}
Using (\ref{eq:InversionDefinition}), (\ref{eq:ePoneQonetaufalse}) and (\ref{eq:Ponetaufalse}), it can then be computed that
\begin{align}
  -2e^{2P_{1}}Q_{1\tau} = & -4Q_{0}P_{0\tau}-2(1-e^{2P_{0}}Q_{0}^{2})Q_{0\tau},\label{eq:etwoPoneQonetau}\\
  -2P_{1\tau}+2e^{2P_{1}}Q_{1}Q_{1\tau} = & 2P_{0\tau}-2e^{2P_{0}}Q_{0}Q_{0\tau}.\label{eq:mKxymKyxone}
\end{align}
Since $\mathrm{Inv}$ is its own inverse, we can interchange the subscripts $0$ and $1$ in (\ref{eq:etwoPoneQonetau}). This yields
\begin{equation}\label{eq:mKxymKyxtwo}
  -4Q_{1}P_{1\tau}-2(1-e^{2P_{1}}Q_{1}^{2})Q_{1\tau}=-2e^{2P_{0}}Q_{0\tau}.
\end{equation}
Combining (\ref{eq:etwoPoneQonetau}), (\ref{eq:mKxymKyxone}) and (\ref{eq:mKxymKyxtwo}) with the fact that $\rho_{0}$ is the same for the two solutions,
it is clear that the only effect the inversion has on the components of $\mK$ is to interchange $\mK^{x}_{\phantom{x}y}$ with $\mK^{y}_{\phantom{y}x}$ and
$\mK^{x}_{\phantom{x}x}$ with $\mK^{y}_{\phantom{y}y}$. In particular, if $(Q_{0},P_{0})$ is a solution such that $0<v_{\infty}<1$ and
$\lim_{\tau\rightarrow\infty}P_{\tau}(\cdot,\tau)=-v_{\infty}$, and if $(Q_{1},P_{1}):=\mathrm{Inv}(Q,P)$, then it is sufficient to analyse the asymptotics of
$(Q_{1},P_{1})$ in a neighbourhood of $\vartheta_{0}$. However, then $P_{1\tau}(\vartheta_{0},\tau)\rightarrow v_{\infty}(\vartheta_{0})$ and 
$0<v_{\infty}<1$. In other words, we are back in the situation considered in the previous subsection, and the desired conclusions follow.

\subsection{Non-degenerate true spikes}\label{ssection:nondegeneratetruespikes}

Generic $\tn{3}$-Gowdy symmetric vacuum spacetimes have a finite number of so-called non-degenerate true spikes and a finite number of so-called
non-degenerate false spikes; cf. \cite[Definition~4, pp.~1189--1190]{SCCGowdy}, \cite[Proposition~3, p.~1190]{SCCGowdy} and
\cite[Theorem~2, p.~1190]{SCCGowdy}. Beyond the corresponding
finite number of points, the asymptotic behaviour is of the type described in (\ref{eq:Pexp}) and (\ref{eq:Qexp}). For a justification of this statement
and a clarification of the terminology, we refer the reader to \cite[Subsection~1.4, pp.~1188-1191]{SCCGowdy}. It is possible that one could therefore
prove that, in a generic $\tn{3}$-Gowdy symmetric vacuum spacetime, generic causal geodesics going into the singularity avoid the spikes. Considering
systems of wave equations on a generic $\tn{3}$-Gowdy symmetric vacuum spacetime, combining the analysis of
Subsection~\ref{ssection:convergentsettingGowdy} with the results of these notes, it would then be possible to analyse the asymptotics of
solutions restricted to $J^{+}(\g)$ for a generic past inextendible causal geodesic $\g$. Taking this perspective, the issue of the spikes could be
avoided altogether. However, it is of interest to consider the behaviour of solutions in $J^{+}(\g)$ for causal curves whose spatial component converges
to the tip of a spike. In the previous subsection, we provide an analysis in a neighbourhood of a false spike. In the present subsection, we therefore
focus on non-degenerate true spikes.

The natural starting point for discussing spikes is the article \cite{raw}. In what follows, we briefly describe the ideas of
\cite[Section~3, pp.~2963--2967]{raw}. In order to construct a solution with a non-degenerate true spike, we first start with a solution, given by $P_{0}$
and $Q_{0}$, and then perform an inversion; cf. the previous subsection. We then obtain a solution $(Q_{1},P_{1})$, given by (\ref{eq:InversionDefinition}).
Next, we apply the Gowdy to Ernst transformation, obtaining a new solution $P$, $Q$ defined by 
\begin{equation}\label{eq:PtwoQtwodef}
  P=-P_{1}+\tau,\ \ \
  Q_{\tau}=-e^{2(P_{1}-\tau)}Q_{1\vartheta},\ \ \
  Q_{\vartheta}=-e^{2P_{1}}Q_{1\tau};
\end{equation}
cf. \cite[(7), p. 2963]{raw}. 
In order to obtain a non-degenerate true spike, we have to assume the original solution (given by $P_{0}$ and $Q_{0}$) to have expansions such as
(\ref{eq:Pexp}) and (\ref{eq:Qexp}) of a special form. In particular, we assume that $Q_{\infty}(\vartheta_{0})=0$, and $Q_{\infty}'(\vartheta_{0})\neq 0$,
so that $Q_{\infty}$
is non-zero in a punctured neighbourhood of $\vartheta_{0}$. We are mainly interested in analysing the behaviour of solutions in $J^{+}(\g)$, where $\g$
is a past inextendible causal curve whose $\vartheta$-component converges to $\vartheta_{0}$. This means that it is sufficient to analyse the behaviour in 
\[
\msA^{+}(\g):=\{(\vartheta,\tau):|\vartheta-\vartheta_{0}|\leq e^{-\tau}\}.
\]
It is of interest to derive expansions for $e^{P_{0}}Q_{0}$ in this set. Due to (\ref{eq:Qexp}),
\[
Q_{0}=Q_{\infty}-\frac{r}{2v_{a}}e^{-2p}+e^{-2p}f,\ \ \
e^{P_{0}}Q_{0}=e^{P_{0}}Q_{\infty}-\frac{r}{2v_{a}}e^{P_{0}-2p}+e^{P_{0}-2p}f,
\]
where the $C^{k}$ norm of $f$ is $O(e^{-\eta\tau})$ for every $k\in\nn{}$. However, in $\msA^{+}(\g)$,
\[
e^{P_{0}}Q_{\infty}=e^{P_{0}}Q_{\infty}'(\vartheta_{0})(\vartheta-\vartheta_{0})+O(e^{P_{0}-2\tau})=O(e^{P_{0}-\tau})=O(e^{-[1-v_{a}(\vartheta_{0})]\tau}).
\]
In particular,
\[
e^{P_{0}}Q_{0}=O(e^{-[1-v_{a}(\vartheta_{0})]\tau})+O(e^{-v_{a}(\vartheta_{0})\tau})
\]
in $\msA^{+}(\g)$. Next, note that (\ref{eq:ePoneQonetaufalse}) and an analogous formula for the $\vartheta$-derivative hold. This means that
\[
e^{P_{1}}Q_{1\tau}=O(e^{-[1-v_{a}(\vartheta_{0})]\tau})+O(e^{-v_{a}(\vartheta_{0})\tau}),\ \ \
e^{P_{1}-\tau}Q_{1\vartheta}=O(e^{-[1-v_{a}(\vartheta_{0})]\tau})
\]
in $\msA^{+}(\g)$. In fact, the latter equality can be improved to
\begin{equation*}
  \begin{split}
    e^{P_{1}-\tau}Q_{1\vartheta} = & e^{-[1-v_{a}(\vartheta_{0})]\tau}[e^{\phi(\vartheta_{0})}Q_{\infty}'(\vartheta_{0})+O(e^{-\eta\tau})]
  \end{split}
\end{equation*}
in $\msA^{+}(\g)$. Next, note that $P_{1}=-P_{0}+\ln(1+Q_{0}^{2}e^{2P_{0}})$. Moreover, (\ref{eq:Ponetaufalse}) and an analogous formula for the
$\vartheta$-derivative hold. In particular,
\begin{align*}
  P_{1} = & -v_{a}(\vartheta_{0})\tau-\phi(\vartheta_{0})+O(e^{-\eta\tau}),\ \ \ P_{1\tau}+v_{a}(\vartheta_{0})=O(e^{-\eta\tau}),\\
  e^{-\tau}P_{1\vartheta} = & O(\ldr{\tau}e^{-\tau})+O(e^{-2[1-v_{a}(\vartheta_{0})]\tau})
\end{align*}
in $\msA^{+}(\g)$. Combining the above observations with (\ref{eq:PtwoQtwodef}) yields the conclusion that
\[
Q_{\tau}=O(e^{-2\tau}),\ \ \
Q_{\vartheta}=O(e^{-2v_{a}(\vartheta_{0})\tau})+O(e^{-\tau})
\]
in $\msA^{+}(\g)$. In fact, the first equality can be refined to
\[
e^{2P}Q_{\tau}=-e^{2v_{a}(\vartheta_{0})\tau}e^{2\phi(\vartheta_{0})}Q_{\infty}'(\vartheta_{0})[1+O(e^{-\eta\tau})]
\]
in $\msA^{+}(\g)$. Moreover,
\[
e^{P}Q_{\tau}=O(e^{-[1-v_{a}(\vartheta_{0})]\tau}),\ \ \
e^{P-\tau}Q_{\vartheta}=O(e^{-v_{a}(\vartheta_{0})\tau})+O(e^{-[1-v_{a}(\vartheta_{0})]\tau})
\]
in $\msA^{+}(\g)$. On the basis of the above estimates, we also conclude that
\[
t\lambda_{t}=[1+v_{a}(\vartheta_{0})]^{2}+O(e^{-\eta\tau})
\]
in $\msA^{+}(\g)$. If we let $q_{\infty}:=\lim_{\tau\rightarrow\infty}Q(\vartheta_{0},\tau)$, we conclude that
\[
Q-q_{\infty}=O(e^{-[1+2v_{a}(\vartheta_{0})]\tau})+O(e^{-2\tau}),\ \ \
e^{P}(Q-q_{\infty})=O(e^{-v_{a}(\vartheta_{0})\tau})+O(e^{-[1-v_{a}(\vartheta_{0})]\tau}),\ \ \
\]
in $\msA^{+}(\g)$.

In order to obtain a clear picture of the asymptotics, it is convenient to introduce new coordinates
\[
s:=t,\ \ \
\xi:=\vartheta,\ \ \
z:=x+q_{\infty}y,\ \ \
w:=y.
\]
If $\mK$ is the expansion normalised Weingarten map associated with the solution $(P,Q)$, it can then be computed that the non-zero components of
$\mK$ are given by 
\begin{align*}
  \mK^{\xi}_{\phantom{\xi}\xi} = & \rho_{0}^{-1}(t\lambda_{t}-1),\\
  \mK^{z}_{\phantom{z}z} = & 2\rho_{0}^{-1}(1-P_{\tau})+2\rho_{0}^{-1}e^{2P}(Q-q_{\infty})Q_{\tau},\\
  \mK^{z}_{\phantom{z}w} = & -4\rho_{0}^{-1}P_{\tau}(Q-q_{\infty})-2\rho_{0}^{-1}[1-e^{2P}(Q-q_{\infty})^{2}]Q_{\tau},\\
  \mK^{w}_{\phantom{w}z} = & -2\rho_{0}^{-1}e^{2P}Q_{\tau},\\
  \mK^{w}_{\phantom{w}w} = & 2\rho_{0}^{-1}(1+P_{\tau})-2\rho_{0}^{-1}e^{2P}Q_{\tau}(Q-q_{\infty}).  
\end{align*}
Combining these calculations with the above estimates yields
\begin{align*}
  \mK^{\xi}_{\phantom{\xi}\xi} = & \frac{v_{\infty}^{2}(\vartheta_{0})-1}{v_{\infty}^{2}(\vartheta_{0})+3}+O(e^{-\eta\tau}),\\
  \mK^{z}_{\phantom{z}z} = & -\frac{2v_{a}(\vartheta_{0})}{v_{\infty}^{2}(\vartheta_{0})+3}+O(e^{-\eta\tau}),\\
  \mK^{z}_{\phantom{z}w} = & O(e^{-[1+2v_{a}(\vartheta)]\tau})+O(e^{-2\tau}),\\
  \mK^{w}_{\phantom{w}z} = & \frac{2e^{2\phi(\vartheta_{0})}Q_{\infty}'(\vartheta_{0})}{v_{\infty}^{2}(\vartheta_{0})+3}e^{2v_{a}(\vartheta_{0})\tau}[1+O(e^{-\eta\tau})],\\
  \mK^{w}_{\phantom{w}w} = & \frac{2+2v_{\infty}(\vartheta_{0})}{v_{\infty}^{2}(\vartheta_{0})+3}+O(e^{-\eta\tau})
\end{align*}
in $\msA^{+}(\g)$, where $v_{\infty}(\vartheta_{0})=v_{a}(\vartheta_{0})+1$. Note that even though $\mK^{w}_{\phantom{w}z}$ tends to infinity in the direction of
the singularity, the product $\mK^{w}_{\phantom{w}z}\mK^{z}_{\phantom{z}w}$ converges to zero exponentially. Thus the eigenvalues, say $\ell_{i}$, $i=1,2,3$,
converge exponentially to
\[
\frac{v_{\infty}^{2}(\vartheta_{0})-1}{v_{\infty}^{2}(\vartheta_{0})+3},\ \ \
-\frac{2v_{a}(\vartheta_{0})}{v_{\infty}^{2}(\vartheta_{0})+3},\ \ \
\frac{2+2v_{\infty}(\vartheta_{0})}{v_{\infty}^{2}(\vartheta_{0})+3}
\]
in $\msA^{+}(\g)$. Denote the eigenvectors corresponding to $\ell_{A}$ by $X_{A}$. Then $X_{1}$ is proportional to $\d_{\xi}$ and 
\[
X_{A}=X_{A}^{z}\d_{z}+X_{A}^{w}\d_{w}
\]
for $A=2,3$. Normalising the eigenvectors by the requirement that $X_{A}^{w}=1$, it can then be verified that
\[
X_{2}^{z}=O(e^{-2v_{a}(\vartheta_{0})\tau}),\ \ \
X_{3}^{z}=O(e^{-[1+2v_{a}(\vartheta)]\tau})+O(e^{-2\tau})
\]
in $\msA^{+}(\g)$. In the limit, the eigenspaces corresponding to $\ell_{2}$ and $\ell_{3}$ thus coincide. 

\printindex

\end{document}